\documentclass[fleqn,usenatbib]{mnras}
\usepackage[T1]{fontenc}
\usepackage{ae,aecompl}
\usepackage{graphicx}	
\usepackage{amsmath}	
\usepackage{tablefootnote}
\usepackage{amssymb}	

\usepackage{newtxtext,newtxmath}

\title[UWL pulsar census]{A polarization census of bright pulsars using the Ultra-Wideband Receiver on the Parkes radio telescope }

\author[C. Sobey et al.]{
C. Sobey$^{1}$\thanks{E-mail: charlotte.sobey@csiro.au (CS)},
S. Johnston$^{2}$, 
S. Dai$^{2}$, 
M. Kerr$^{3}$,  
R. N. Manchester$^{2}$,
L. S. Oswald$^{4}$,\newauthor
A. Parthasarathy$^{5}$, 
R. M. Shannon$^{6,7}$,  
P. Weltevrede$^{8}$
\\
$^{1}$CSIRO Astronomy and Space Science, PO Box 1130, Bentley, WA 6102, Australia\\
$^{2}$Australia Telescope National Facility, CSIRO Astronomy and Space Science, PO Box 76, Epping, NSW 1710, Australia\\
$^{3}$Space Science Division, Naval Research Laboratory, Washington, DC 20375, USA\\
$^{4}$Department of Astrophysics, University of Oxford, Denys Wilkinson Building, Keble Road, Oxford OX1 3RH, UK\\
$^{5}$Max-Planck-Institut f{\"u}r Radioastronomie, Auf dem H{\"u}gel 69, D-53121 Bonn, Germany\\
$^{6}$Centre for Astrophysics and Supercomputing, Swinburne University of Technology, PO Box 218, Hawthorn, VIC 3122, Australia\\
$^{7}$OzGrav: Australian Research Council Centre of Excellence for Gravitational Wave Discovery, Swinburne University of Technology,\\Mail number H74, PO Box 218, Melbourne, VIC 3122, Australia\\
$^{8}$Jodrell Bank Centre for Astrophysics, The University of Manchester, Alan Turing Building, Manchester M13 9PL, UK
}

\date{Accepted 2021 March 19. Received 2021 February 26; in original form 2020 December 23}
 
\pubyear{2021}

\usepackage{etoolbox}

\begin{document}
\label{firstpage}
\pagerange{\pageref{firstpage}--\pageref{lastpage}}
\maketitle

\begin{abstract}
We present high signal-to-noise, full polarization pulse profiles for 40 bright, `slowly'-rotating (non-recycled) pulsars using the new Ultra-Wideband Low-frequency (UWL; 704--4032\,MHz) receiver on the Parkes radio telescope.
We obtain updated and accurate interstellar medium parameters towards these pulsars (dispersion measures and Faraday rotation measures), and reveal Faraday dispersion towards PSR J1721--3532 caused by interstellar scattering.
We find general trends in the pulse profiles including decreasing fractional linear polarization and increasing degree of circular polarization with increasing frequency, consistent with previous studies, while also revealing new features and frequency evolution.
This demonstrates results that can be obtained using UWL monitoring observations of slow pulsars, which are valuable for improving our understanding of pulsar emission and the intervening interstellar medium.
The calibrated data products are publicly available.
\end{abstract}


\section{Introduction}\label{sec:int}

Observing pulsars (rapidly-rotating, highly-magnetised neutron stars) provides insights into a range of fundamental physics: 
from the behaviour of magnetospheres with incredible field strengths \citep[e.g.][]{Kaspi+2017,Dai+2018}, and the elusive mechanism responsible for their broad spectrum of electromagnetic emission \citep[e.g.][]{Melrose+2021};
to the 3-D structure of the magneto-ionic interstellar medium \citep[ISM; e.g.][]{VanEck+2011,sbg+19,Abbate+2020}.

Monitoring `young' pulsars promises to improve our understanding of neutron stars and pulsars, and the intervening ISM.
Applying the method of pulsar timing to these monitoring observations facilitates precise astrometry \citep[e.g.][]{pjs+20}, measurements of the spectrum of timing noise \citep[e.g.][]{Parthasarathy+2019} and glitches \citep[e.g.][]{Lower+2020}. Timing ephemerides also enable searches for phase-resolved $\gamma$-ray light curves using the \emph{Fermi} Gamma-ray Space Telescope \citep[e.g.][]{fermi,rwjk17,sbc+19} and other high-energy observatories.
We can also study variations in pulse profiles over time, due to intrinsic changes in the radio beam \citep[e.g.][]{Brook+2016}, inhomogeneities in the ISM \citep[e.g.][]{Michilli+2018,Kumamoto+2021}, or precession \citep[e.g.,][]{khjs16,Desvignes+2019}.

In addition, broad-band polarization observations provide essential information about the pulsar radio emission mechanism, beam geometry, and the Galactic magneto-ionic ISM. 
Pulsars are among the most highly polarized radio sources known \citep[e.g.][]{Lyne+1968,Gould+1998}, and the polarization varies with observing frequency \citep[e.g.][]{Manchester+1973,Johnston+2008,Dai+2015}, providing insight into the magnetospheric emission and propagation mechanisms.
In addition, the linear polarization position angles (P.A.s) across pulse phase can constrain the beam size and inclination angles, with respect to the pulsar's rotation axis and our line of sight (LoS). For example, the rotating vector model (RVM) predicts a smooth `S'-shape, due to the projected vectors of the magnetic field lines as they sweep across our LoS \citep[e.g.][]{rc69,lm88,Johnston+2007}.
Many pulsars show more complex P.A. curves, particularly discontinuities with rapid jumps of $\approx$90$^{\circ}$, which suggests the presence of two orthogonal polarization modes \citep[OPM; e.g.][]{Manchester+1975,Backer+1976}.
Furthermore, circular polarization across the pulse is observed to either remain in the same hand or change sense \citep[e.g.][]{Radhakrishnan+1990}, which may be intrinsic to the emission mechanism or due to propagation effects \citep[e.g.][]{Han+1998,Kennett+1998}.
Additional diagnostics of magnetospheric effects have also been investigated, including variations in Faraday rotation measure (RM) and circular polarization across the pulse \citep[e.g.][]{Ramachandran+2004,Karastergiou2009,Noutsos+2009,ijw19}.
Although pulsars were discovered over 50 years ago \citep{Hewish+1968}, it is clear that current understanding and models of the emission mechanism are far from replicating this wide range of observed behaviour, as well as additional emission phenomena such as nulling and mode-changing. 

The impulsive signals from pulsars undergo several effects as they propagate through the ISM, including dispersion and Faraday rotation. Measurements of the observable quantities, dispersion measure (DM) and RM, can be used to estimate the average magnetic field strength and net direction parallel to the LoS, weighted by the thermal electron density, 
\begin{equation} 
\langle B_{\parallel}\rangle=
\frac{\int^{{d}}_{0} n_{\mathrm e} B_{\parallel}\mathrm{d}l}{\int^{{d}}_{0} n_{\mathrm e} \mathrm{d}l}=
1.232~\upmu\mathrm{G} \left( \frac{\mathrm {RM}}{\mathrm {rad~m^{\mathrm -2}}} \right)\left( \frac{\mathrm {DM}}{\mathrm{pc~cm^{\mathrm {-3}}}} \right)^{\mathrm {-1}},
  \label{eq:B}
\end{equation}
where ${n_{\mathrm e}}$ is the electron density, $d$ is the distance to the pulsar, and ${\rm{d}}l$ is the differential distance element. By definition, positive (negative) RMs indicate that the net direction of $\langle B_{\parallel}\rangle$ is towards (away from) the observer. Equation \ref{eq:B} assumes that the electron density and magnetic field are uncorrelated \citep[e.g.][]{Beck+2003}, which is likely a good estimate on Galactic, kpc scales \citep{Seta+2021}.
The large population of pulsars samples the ISM along various lines-of-sight and over a range of distances \citep[e.g.][]{ymw+2017}, and can be used to reconstruct the 3-D structure of the large-scale Galactic magnetic field, which shows reversals in the magnetic field direction between some neighbouring spiral arms \citep[e.g.][]{njkk08,VanEck+2011,Han+2006}.
In addition, DM and RM variations over long timescales (expected as the pulsars traverse the ISM) as well as scattering, diffractive and refractive scintillation, and depolarization can trace the small-scale ISM inhomogeneities dominated by turbulence \citep[e.g.][]{pkj+13,Geyer+2017,Desvignes+2018,kcw+18,Xue+2019,Kumamoto+2021,Johnston+2021}. 
Although the Galactic magnetic field was first measured over 70 years ago \citep[][]{Hall+1949,Hiltner+1949}, our understanding of its structure, generation and amplification remain limited \citep[e.g.][]{Haverkorn+2019}. 
 
There is clearly much to discover about the pulsar emission mechanism and the Galactic magnetic field, and promising inroads are being made by using the suite of Square Kilometre Array (SKA) pathfinders and precursors with cutting-edge sensitivity, bandwidth, and computing resources.
In particular, the Ultra-Wideband Low-frequency (UWL) receiver on the Parkes radio telescope (a technology pathfinder to the SKA) provides unprecedented contiguous broadband information (frequency coverage: 704--4032\,MHz; bandwidth: 3328\,MHz; fractional bandwidth: 1.4; \citealt{Hobbs+2020}). For comparison, the previous receivers frequently used for pulsar observations included the 20cm Multibeam receiver \citep[centre frequency: 1369\,MHz; bandwidth: $\sim$300\,MHz;][]{Staveley-Smith+1996}, the `H-OH' receiver (centre frequency: 1433\,MHz; bandwidth: $\sim$500\,MHz), and 10/50-cm receiver \citep[centre frequencies: 732/3100\,MHz; bandwidths: $\sim$70/1024\,MHz;][]{Granet+2005}, see also, e.g. \citet[][]{Manchester+2013}.
This valuable broadband polarization information is increasingly acknowledged to be important for further understanding the radio emission and the intervening ISM \citep[e.g.][]{Karastergiou+2015,Anderson+2016,Heald+2020}. 

A large observing program using the Parkes radio telescope has been monitoring over 250 young, non-recycled, `slow' pulsars with an approximately monthly cadence for over a decade \citep[project code P574;][]{Weltevrede+2010}.
In this work, we present results from an ultra-broadband polarization study focused on a subset of 40 relatively bright pulsars over the first 14 months of observations using the UWL receiver.
We add the data from all available observing epochs together in order to stabilise and increase the signal-to-noise (S/N) of the pulse profiles through increasing the number of pulses collected and smoothing over the effects of diffractive and refractive scintillation. 
A companion paper \citep{Johnston+2021} presents results focusing on time-variability for the complete set of 276 pulsars observed over 24 months. 

In Section \ref{sec:obs} we describe the selection of the pulsars studied here, the observations using the UWL receiver on the Parkes radio telescope, the data reduction, and data analysis. 
For details on how to obtain the raw and final data products, please see `Data Availability'.
In Section \ref{sec:res} we present the results, namely the DM and RM measurements towards all 40 pulsars, and their ultra-wideband, average polarization pulse profiles.
In Section \ref{sec:dis} we discuss our results, focusing on the ISM in Section \ref{subsec:ISM}, particularly the LoS towards PSR J1721--3532 in Section \ref{subsub:1721}, and explore the general and pulsar-specific polarized emission behaviour in Section \ref{subsec:emn}.
In Section \ref{sec:con} we present our conclusions. 
We provide plots of the Faraday spectrum and the broad-band polarization pulse profile for each pulsar in the supporting information available online.

\section{Observations and data reduction}\label{sec:obs}

\subsection{Source selection}

As part of the large program to study `slow' pulsars for a variety of science goals, we are using the Parkes radio telescope to observe 276 of these objects with an approximately monthly cadence. The full list of these pulsars is provided in a companion paper \citep{Johnston+2021}.
This program commenced in 2007, observing $\sim$180 pulsars with high spin-down luminosities ($\dot{E}>10^{34}$\,erg\,s$^{-1}$) from \cite{sgc+08}. From 2014, additional pulsars with high flux densities and lower spin-down luminosities were included in the observations, and pulsars with lower flux densities were removed \citep[][]{jk18}. 
 
In this work, we conduct a census of 40 of the brightest pulsars from this large sample. 
These were selected to provide high signal-to-noise pulse profiles as an initial illustration of the performance of the UWL observations, data reduction, and science results. 
The pulsars in this census are distributed across the sky (00$^{\rm{h}}<$RA$<$21$^{\rm{h}}$), and are listed in Table \ref{tab:obs}. 
They have pulse periods in the range $0.09<P<1.96$\,s (PSRs J0835--4510 and J2048--1616, respectively); spin-down luminosities between $1.2\times10^{31}<\dot{E}<6.9\times10^{36}$\,erg\,s$^{\rm{-1}}$ (PSRs J0536--7543 and J0835--4510, respectively); characteristic magnetic field strengths in the range $7\times10^{10}<B_{\rm{s}}<2\times10^{13}$\,G (PSRs J1807--0847 and J1740-3015, respectively); and characteristic ages between $1\times10^{4}<\tau_{\rm{c}}<9\times10^{7}$\,yr (PSRs J0835--4510 and J1807--0847, respectively).  

\subsection{Observations} 

The observations presented in this paper were carried out using the UWL receiver on the Parkes radio telescope, which has a system equivalent flux density ranging from 33 to 72\,Jy depending on frequency \citep{Hobbs+2020}.
Each pulsar was observed for $\sim$180\,s at up to 15 epochs (approximately monthly) between 2018 November 18 and 2020 January 3.

The UWL radio frequency band (704--4032\,MHz) is divided into 3328$\times$1\,MHz frequency channels. For each pulsar observation, the data from each channel are coherently dedispersed and folded at the topocentric spin period to form a pulse profile with 1024 phase bins using the pulsar ephemerides from the ATNF Pulsar Catalogue\footnote{\href{http://www.atnf.csiro.au/people/pulsar/psrcat/}{http://www.atnf.csiro.au/people/pulsar/psrcat/}} \citep{Manchester+2005}.  The `Medusa' Graphics Processing Unit cluster processes the four polarization-product data in this way and writes sub-integrations of duration 30\,s modulo the pulse period to disk in {\sc PSRFITS}\footnote{\href{http://www.atnf.csiro.au/research/pulsar/psrfits/}{http://www.atnf.csiro.au/research/pulsar/psrfits/}} fold-mode format \citep{Hotan+2004}. The resulting data rate is $\approx$27\,MB per subintegration \citep{Hobbs+2020}.

Every $\sim$60\,min during the observing sessions, we obtain an 80-s observation of a pulsed, square wave, noise-diode calibration signal at a position offset by $<$1 degree from a pulsar, allowing polarization calibration of the data. 
Separate observations of the radio galaxy Hydra~A (PKS 0915--11) were used for primary flux density calibration; a long-track observation of PSR J0437--4715 was used as an input to the Polarization Calibration Modeling \textit{pcm} \textsc{psrchive}\footnote{\href{http://psrchive.sourceforge.net}{http://psrchive.sourceforge.net}} routine for polarization calibration \citep[e.g.][]{vanStraten+2004,Manchester+2013,Hobbs+2020,Kerr+2020}.

\subsection{Data reduction}\label{subsec:red}

We inspected the calibration signal observations and flagged (zero-weighted) the frequency channels at sub-band boundaries (to remove aliasing effects) and channels affected by radio frequency interference (RFI), before averaging in time.   

We flux and polarization calibrated each pulsar observation using the \textit{pac} routine in {\sc psrchive} \citep{hwm04}. Flux calibration is applied from a solution obtained via observations of Hydra~A. We used the nearest calibration signal observation in time to correct the gains and phases. Instrumental leakage terms were corrected using the solution obtained from the observation of PSR J0437--4715.
This provides the `absolute' polarization position angle of the radiation above the feedhorn at the centre frequency. 
RFI-affected frequency channels in the pulsar observations were then identified and flagged using a median filter applied to the time-averaged data. RFI-affected sub-integrations were also examined and flagged. 

We averaged the pulsar observations in time, before adding data from all available observing epochs together.
For each pulsar, we obtained times of arrival (ToAs) for each epoch's time- and frequency-averaged pulse profile using the {\sc Tempo2} software \citep{hem06} and a pulse profile template centred at 20-cm from previous observations using the Multibeam receiver \citep{jk18}. 
Timing residuals were then obtained using {\sc Tempo2} and the timing ephemeris. 
We rotated each epoch's pulse profile by the corresponding timing residual via the {\sc psrchive} routine \textit{pam} and summed the data using \textit{psradd}.

Summing observations over multiple epochs does not significantly affect the polarization pulse profiles or the ISM measurements \citep[e.g.][]{Ng+2020}. 
The effect of small changes in the DM and RM is negligible at high frequencies as most pulsars do not show secular variations at a level above 10$^{-3}$\,pc\,cm$^{-3}$\,yr$^{-1}$ \citep[e.g.][]{pkj+13,Johnston+2021}, which is a tiny fraction of a phase bin.
The largest change in the RMs for our observations is expected from time- and LoS-dependent ionospheric Faraday rotation. Modelling the range in ionospheric Faraday rotation seen in this work as a Burn slab \citep{Burn1966} with Faraday dispersion 2.28\,rad\,m$^{\rm{-2}}$, see Section \ref{subsec:ana}, we expect negligible depolarization.

We conducted some final interactive RFI flagging on the final pulse profiles averaged over all epochs.
The final data have between 23.98--25.0 per cent of the frequency channels flagged. The median number of frequency channels excised is 24.16 per cent (804/3328).

\subsection{Data analysis}\label{subsec:ana}

\subsubsection{Dispersion measures}

For each pulsar's final time-averaged pulse profile, we measured the DMs and dedispersed the data using this refined measurement. Two methods were explored for obtaining the DM measurements: 
1) the \textit{pdmp} routine in {\sc psrchive} was applied to the full frequency resolution data, which determines the DM that maximizes the total intensity pulse profile signal-to-noise (S/N) ratio. 
2) {\sc Tempo2} and the 20-cm pulse profile templates were used to measure the dispersion delay between the ToAs obtained from the data averaged in frequency to 13 frequency sub-bands.
For both measurement methods used, the DM was the only free parameter. 
The DM measurements using the {\sc Tempo2} method are reported in Table \ref{tab:obs}. We discuss the DM measurements further in Section \ref{subsec:ISM}.

\subsubsection{Faraday rotation measures}

To measure the RMs, we used the technique of RM synthesis \citep{Burn1966,Brentjens+2005}, and the associated deconvolution procedure RM CLEAN \citep{Heald+2009_RMCLEAN}, using a publicly available python package\footnote{\href{https://github.com/gheald/RMtoolkit}{https://github.com/gheald/RMtoolkit}} \citep[e.g.][]{Michilli+2018}.
See, e.g. \citet{Schnitzeler+2015} and \citet{Porayko+2019}, for alternative implementations of RM synthesis.

\begin{figure*}
	\includegraphics[width=0.68\columnwidth]{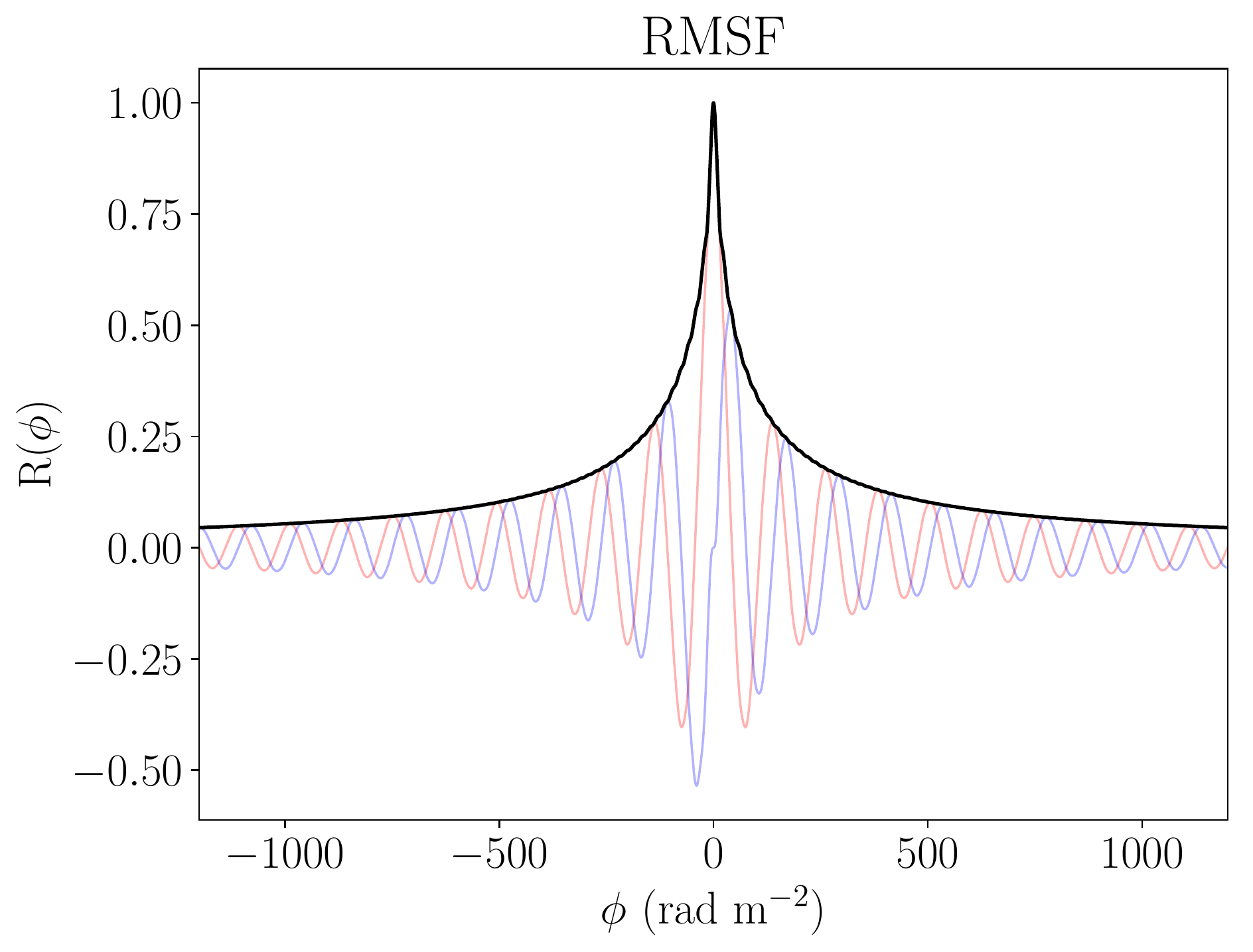}
	\includegraphics[width=0.67\columnwidth]{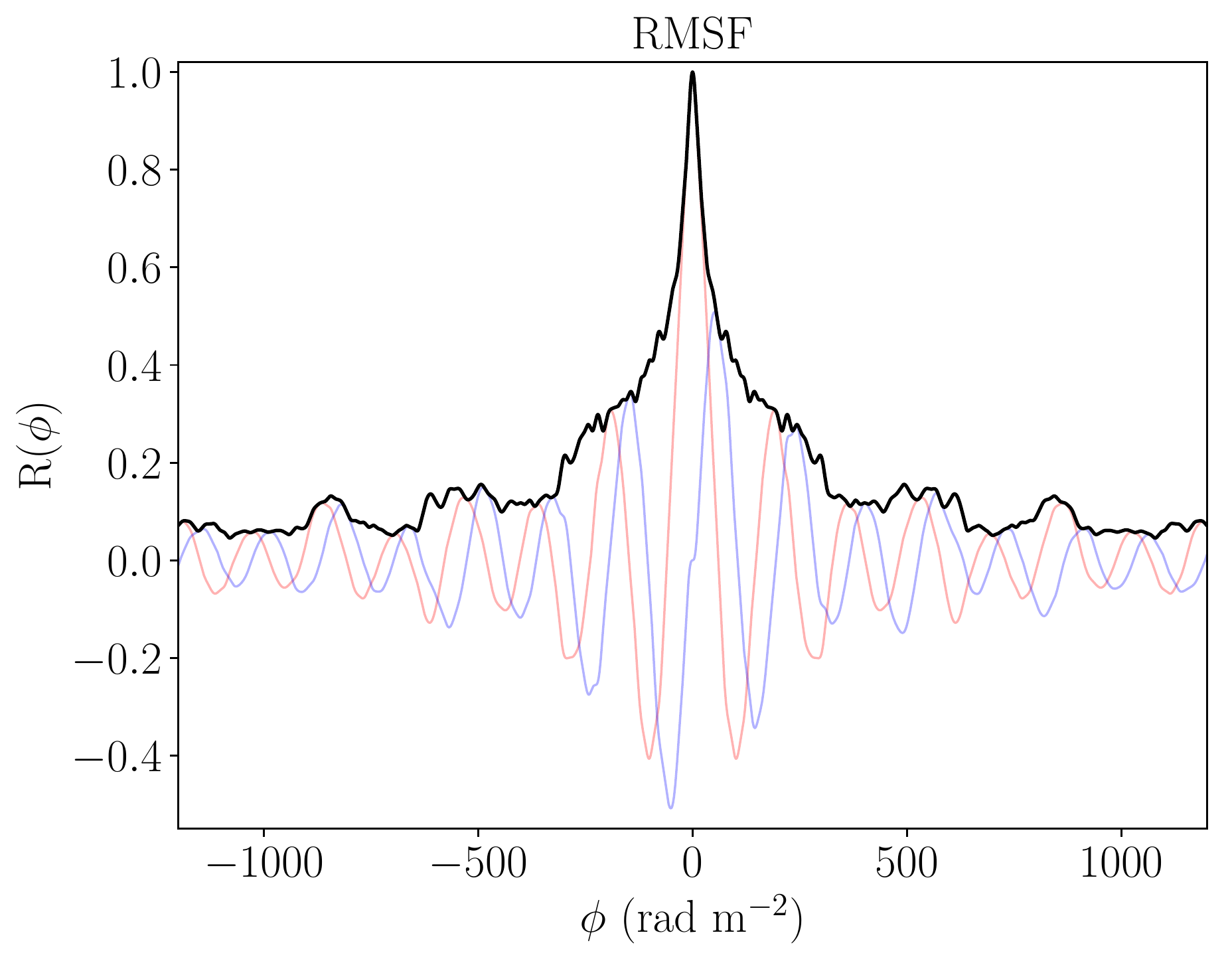}
	\includegraphics[width=0.727\columnwidth]{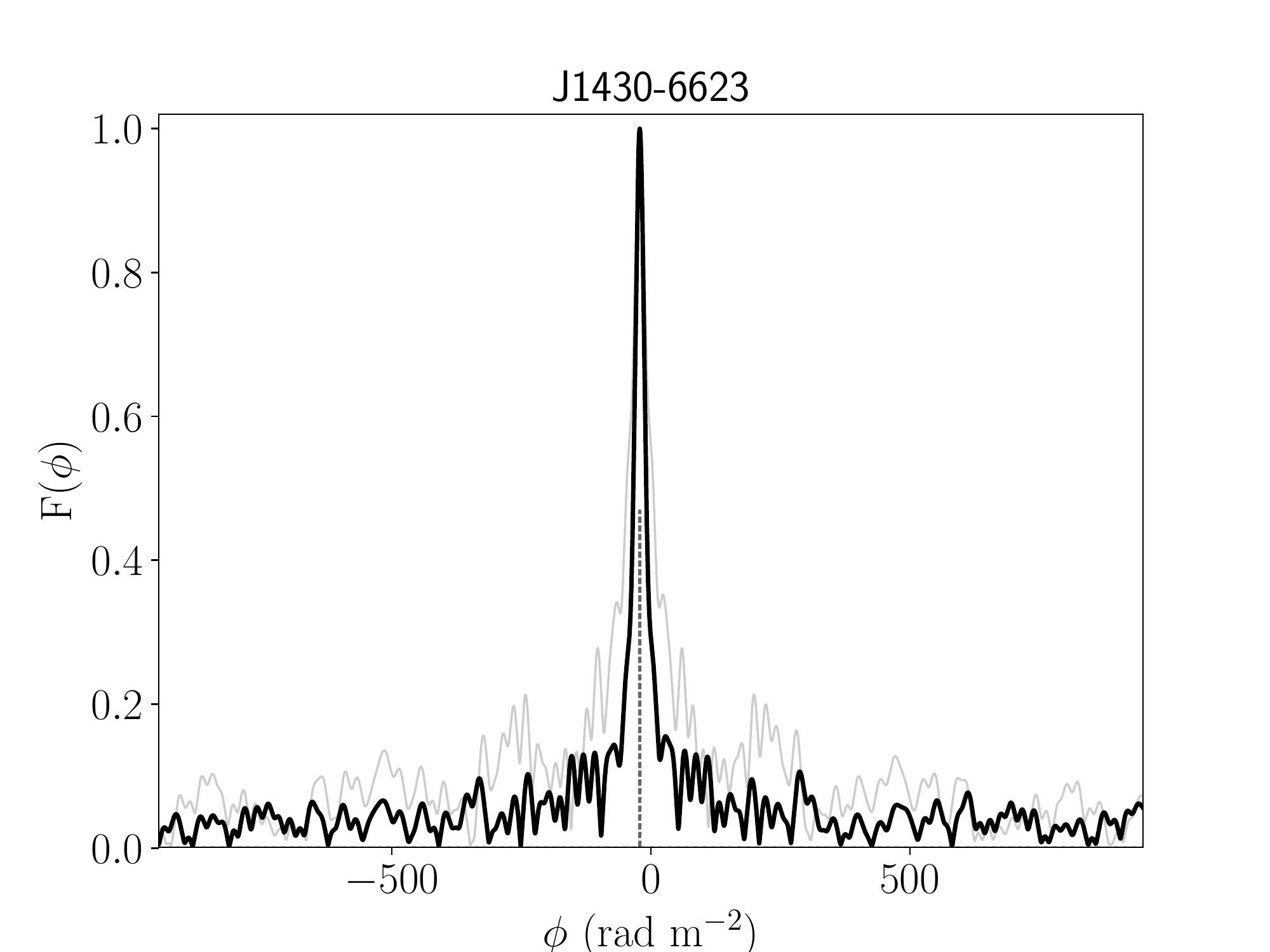}
    \caption{Left and centre: Normalised Faraday rotation measure spread functions (RMSFs) for --1200$<\phi<$1200\,rad\,m$^{\rm{-2}}$. The black lines show the absolute value of the RMSF, and red and blue lines show the real and imaginary values, respectively. The reference wavelength squared is the mean value for the observed wavelengths ($\lambda^{\rm 2}_{\rm 0}$=0.0317\,m$^{\rm{2}}$) and we use uniform weighting across the bandwidth. Left: Theoretical, noiseless RMSF for UWL data at a centre frequency of 2368\,MHz with 3328\,MHz bandwidth and 3328$\times$1\,MHz channels. We measure the $\delta\phi$ of the absolute value of the RMSF to be 94\,rad\,m$^{\rm{-2}}$. For comparison, the previous Multibeam data provide $\delta\phi\approx$208\,rad\,m$^{\rm{-2}}$. Centre: RMSF computed using the frequency information obtained from an UWL observation of PSR J1430--6623 with a reduced number of channels (2524$\times$1\,MHz channels, or 75.8 per cent of the original bandwidth) mostly due to RFI excision. We measure $\delta\phi$=113\,rad\,m$^{\rm{-2}}$. Right: Faraday spectrum computed using PSR J1430--6623 data, shown for --950$<\phi<$950\,rad\,m$^{\rm{-2}}$. The grey and black lines show the absolute values of the original (with FWHM$\approx$52\,rad\,m$^{\rm{-2}}$) and the RM CLEANed spectra (with FWHM$\approx$25\,rad\,m$^{\rm{-2}}$), respectively. The grey dashed line shows the RM CLEAN component at -21.5\,rad\,m$^{\rm{-2}}$. The remaining Faraday spectra for the pulsars in this work are provided in the supporting information. }
    \label{fig:RMSF}
\end{figure*}

We summarise the parameters relevant for RM synthesis in Table \ref{tab:RMsyn}, and present RM synthesis spread functions (RMSFs; equivalent to point spread functions for optical telescopes) in Figure \ref{fig:RMSF}. 
The native resolution in Faraday space (full width at half maximum of the RMSF, $\delta\phi$) reflects the precision of the measurements \citep[][]{Brentjens+2005,Schnitzeler+2009}.
We note that the $\delta\phi$ measurements reported in the Figure \ref{fig:RMSF} caption were produced using uniform weighting for the 1\,MHz channels across the UWL bandwidth, which effectively increases the weighting of the higher frequencies, and is larger than the value in Table \ref{tab:RMsyn}. Since the pulsar spectral indices are $\approx-$1 \citep[e.g.][]{Jankowski+2018}, this down-weights the higher frequencies, by approximately the same amount, and serves to deliver the expected FWHM, see Figure \ref{fig:RMSF}, right.

The RM measurements were obtained using the final time-averaged, DM-corrected pulse profiles and following the method used in \citet{sbg+19}, briefly described here.
We used the \textit{rmfit} routine in \textsc{psrchive} to extract the Stokes $I,Q,U,V$ parameters from the \textsc{PSRFITS} files for between 4 and 58 on-pulse phase bins around the peak in linear polarization ($L=\sqrt{Q^{2}+U^{2}}$; often equivalent to the peak in Stokes $I$; corresponding to an individual polarized pulse profile component up to its FWHM, where possible).
RM synthesis and RM CLEAN were applied to the linear polarizations, with the output Faraday spectra (also referred to as the Faraday dispersion functions; FDFs) computed using 0.1\,rad\,m$^{\rm{-2}}$ steps over the Faraday depth range --1200$\leq\phi\leq$1200\,rad\,m$^{\rm{-2}}$. 
This range is approximately twice the maximum |RM| in the ATNF Pulsar Catalogue for this census (delivering the peak, equivalent to the RM, and the noise in the Faraday spectrum), and is also much lower than the maximum observable Faraday depth, see Table \ref{tab:RMsyn}.
We show an example Faraday spectrum obtained for PSR J1430--6623 in Figure \ref{fig:RMSF}, right.
The Faraday spectra obtained for all pulsars in the census are provided in the supporting information.
The RM measurements were obtained by determining the peak of the Faraday spectrum.
The formal uncertainties were calculated following \citet{Brentjens+2005}: $\delta\phi\div (2\times$S/N), where S/N is the signal to noise in the Faraday spectrum.
The RM measurements and uncertainties for each pulsar are reported in Table \ref{tab:obs}.

\begin{table*}
	\centering
	\caption{Summary of Parkes UWL data used in this work and corresponding theoretical RM-synthesis parameters. See Figure \ref{fig:RMSF} for additional details.}
	\label{tab:RMsyn}
	\begin{tabular}{llr} 
		\hline
		\hline
		Parameter & Symbol & Data \\
		\hline
		Centre frequency & $\nu$ &  2368\,MHz  \\
		Bandwidth & $\Delta\nu$ & 3328\,MHz \\
		Frequency channel width & $\delta\nu$ & 1\,MHz   \\
		Centre wavelength squared & $\lambda^{\rm 2}$  & 0.016\,m$^{\rm 2}$ \\
		Total bandwidth in wavelength squared ($\lambda_{\rm{max}}^{\rm{2}}-\lambda_{\rm{min}}^{\rm{2}}$) & $\Delta(\lambda^{\rm 2})$ & 0.176 (0.181--0.005)\,m$^{\rm 2}$   \\
		 Resolution in Faraday space (FWHM of the RMSF) & $\delta\phi$  & 22\,rad\,m$^{\rm{-2}}$  \\
		 Largest scale in Faraday space to which one is sensitive* & max-scale/$\Delta\phi$ & $\approx$568\,rad\,m$^{\rm{-2}}$  \\
		 Maximum observable Faraday depth using the entire bandwidth & $|\phi_{\rm{max}}|$ & 3368\,rad\,m$^{\rm{-2}}$   \\
		\hline
	\end{tabular}
\end{table*}

\subsubsection{Ionospheric Faraday rotation subtraction}

The ionospheric RM, RM$_{\rm{ion}}$, was computed for all epochs towards the pulsars observed using the publicly available \textsc{ionFR}\footnote{\href{https://github.com/csobey/ionFR}{https://github.com/csobey/ionFR}} package \citep{Sotomayor+2013}. We used publicly available data as inputs: the International GNSS Service global ionospheric total electron content maps\footnote{\href{https://cddis.nasa.gov/archive/gps/products/ionex}{https://cddis.nasa.gov/archive/gps/products/ionex}} \citep[e.g.][]{Hernandez+2009,NOLL2010}; and the International Geomagnetic Reference Field\footnote{\href{https://www.ngdc.noaa.gov/IAGA/vmod/igrf.html}{https://www.ngdc.noaa.gov/IAGA/vmod/igrf.html}} \citep[IGRF-13;][]{Thebault+2015}. 
The RM$_{\rm{ion}}$ towards all pulsars on all epochs ranges between --2.25$\pm$0.19\,rad\,m$^{\rm{-2}}$ and --0.20$\pm$0.04\,rad\,m$^{\rm{-2}}$ with mean --1.05\,rad\,m$^{\rm{-2}}$ (negative in the Southern Hemisphere due to the direction of the Geomagnetic field).
We calculated the mean of the RM$_{\rm{ion}}$ values for each pulsar, and subtracted this from the RM measurements determined from the pulsar data, RM$_{\rm{obs}}$, to determine the RM due to the ISM alone, RM$_{\rm{ISM}}$:
\begin{equation} 
\mathrm{RM_{\rm{ISM}}=RM_{\rm{obs}}-RM_{\rm{ion}}}.
 \label{eq:RM}
\end{equation}
The uncertainty on the mean RM$_{\rm{ion}}$ towards each pulsar was calculated as the square root of the sum of the uncertainties for all RM$_{\rm{ion}}$ epochs squared divided by the number of epochs, all of which rounded up or down to 0.1\,rad\,m$^{\rm{-2}}$. 
See, e.g. \citet{Porayko+2019}, for further discussion of the accuracy of ionospheric RM corrections.
The RM$_{\rm{ion}}$ and RM$_{\rm{ISM}}$ values and associated uncertainties are also reported in Table \ref{tab:obs}.
In addition, we used both the DM and RM$_{\rm{ISM}}$ measurements to estimate $\langle B_{\parallel}\rangle$, following Equation \ref{eq:B}. 
We use standard uncertainty propagation assuming Gaussian distributions, which is a reasonable assumption for the RM measurements obtained from the high S/N data presented in this work \citep[e.g.][]{George+2012,Schnitzeler+2017}.

\subsubsection{Galactic Faraday rotation estimation}

The measurements we obtain towards the pulsars in this work probe a variety of lines-of-sight within the Galaxy over distance ranges of $\sim$0.1--7\,kpc \citep[e.g.][]{ymw+2017}.
For comparison, we calculated the RM expected due to the entire LoS through the Galaxy towards each pulsar using all-sky Faraday Sky maps and associated uncertainties, reconstructed using RM measurements of polarized extragalactic sources, from \citet{Oppermann+2015}\footnote{\href{https://wwwmpa.mpa-garching.mpg.de/ift/faraday/2014/index.html}{HEALPix map available in hdf5 or fits formats}} and \citet{Hutschenreuter+2020}\footnote{\href{https://wwwmpa.mpa-garching.mpg.de/~ensslin/research/data/faraday_revisited.html}{HEALPix map available in hdf5 format}}. 
This comparison is illustrated in Figure \ref{fig:RMcomp}.

\subsubsection{Average wide-band polarization profiles}

We obtained the polarization pulse profiles for each pulsar using the final time-averaged data, which were corrected for dispersion and Faraday rotation using the DM and RM$_{\rm{obs}}$ measurements reported in Table \ref{tab:obs}. 
We obtained average polarization pulse profiles for the entire 3328-MHz bandwidth and for 8$\times$416\,MHz sub-bands.
We used the \textit{pdv} routine in {\sc psrchive} to output the Stokes $I,Q,U,V$ parameters, de-biased linear polarization $L$ \citep[see][]{Wardle+1974}, polarization position angle (P.A.) and uncertainty for each pulse profile phase bin.
Figure \ref{fig:profs} shows example visualizations of these data for PSRs J0738--4042 and J1740--3015. We use the IEEE convention for the P.A.s and circular polarization \citep{vanStraten+2010}, i.e., P.A.s increase anti-clockwise on the sky, and left- (right-)handed circular polarization is shown as positive (negative) values.
The comparison of the 8 sub-bands to the full band shows how the polarization pulse profiles ($I,L,V$), P.A., fractional linear polarization ($L/I$), and fractional circular polarization ($V/I$) change across the wide range in frequency. 
Similar plots for all 40 pulsars explored in this work are provided in the supporting information.
For each pulsar, the centre frequencies for the 8 sub-bands may differ because of the different channels excised due to RFI.

We also used these polarization pulse profiles to compile the fractional linear and circular polarization behaviour with frequency, Figure \ref{fig:fracLV}. 
For each pulsar, the linear, circular, and absolute circular polarization fractions were obtained for integrated on-pulse bin ranges corresponding to the narrowest window with significant ($>$3$\upsigma$) polarization -- usually the highest frequency sub-band -- for all eight sub-bands.  
Identical on-pulse phase bin ranges were used for all sub-bands to provide a fair comparison of corresponding pulse profile component features across the frequency range.
Uncertainties were estimated based on the root mean square (RMS) of the off-pulse baseline and are less than 3 and 4 per cent for linear and circular polarizations, respectively. 

\section{Results}\label{sec:res}

We obtained high S/N polarization pulse profiles for each pulsar across an ultra-wide, contiguous range in observing frequency simultaneously, which has hitherto not been possible. This provides high-quality data, enabling us to measure accurate ISM parameters and explore each pulsar's emission characteristics.

Table \ref{tab:obs} provides a summary of the DM and RM measurements, and $\langle B_{\parallel}\rangle$ estimates towards each pulsar. 

\begingroup
\setlength{\tabcolsep}{5pt}
\begin{table*}
\centering
\caption{Summary of pulsars and measurements. Columns 1-15 show: 1) pulsar B1950 name; 2) pulsar J2000 name; 3) pulse period to 3 decimal places; 4) number of observing epochs (N$_{\rm{epoch}}$) added to achieve the final average pulse profile (see Section \ref{subsec:red}); 5) total integration time after adding all N$_{\rm{epoch}}$ observations; 6) DM measurement and 7) associated uncertainty; 8) RM measurement and 9) associated uncertainty (see Section \ref{subsec:ana}); 10) mean ionospheric RM and 11) associated uncertainty; 12) RM corrected for the ionosphere and 13) propagated uncertainty; 14) average magnetic field strength and net direction parallel to the LoS, weighted by electron density, and 15) propagated uncertainty. *Note: For PSR J0034--0721, we use the original ATNF Pulsar Catalogue DM=10.922$\pm$0.006\,pc\,cm$^{\rm{-3}}$ \citep{srb+2015} for further analysis.}
\label{tab:obs}
\begin{tabular}{l l l r r r r r r r r r r r r} 
\hline
\hline
Name & Jname & P0 & N$_{\rm{epoch}}$ & T$_{\rm{int}}$ & DM & $\pm$ & RM$_{\rm{obs}}$ & $\pm$ & RM$_{\rm{ion}}$ & $\pm$ & RM$_{\rm{ISM}}$ & $\pm$ & $\langle B_{\parallel}\rangle$ & $\pm$ \\
PSR & PSR J & s &  & min & pc\,cm$^{\rm{-3}}$ & & rad\,m$^{\rm{-2}}$ &  & rad\,m$^{\rm{-2}}$  &  & rad\,m$^{\rm{-2}}$  &  & $\upmu$G & \\
\hline
B0031--07 & J0034--0721 & 0.943 & 10 & 32 & 13.3* & 0.3 & 8.2 & 0.8 & --0.9 & 0.1 & 9.0 & 0.8 & 1.021 & 0.098 \\
B0538--75 & J0536--7543 & 1.246 & 12 & 40 & 18.45 & 0.05 & 26.6 & 0.8 & --1.3 & 0.1 & 28.0 & 0.8 & 1.869 & 0.054 \\
B0540+23 & J0543+2329 & 0.246 & 13 & 40 & 77.58 & 0.01 & 3.0 & 0.5 & --0.3 & 0.1 & 3.3 & 0.5 & 0.053 & 0.008 \\
B0628--28 & J0630--2834 & 1.244 & 15 & 50 & 34.84 & 0.06 & 45.5 & 0.8 & --1.1 & 0.1 & 46.6 & 0.8 & 1.647 & 0.028 \\
B0736--40 & J0738--4042 & 0.375 & 15 & 208 & 160.94 & 0.04 & 12.8 & 0.8 & --1.2 & 0.1 & 14.0 & 0.8 & 0.107 & 0.006 \\
B0740--28 & J0742--2822 & 0.167 & 13 & 49 & 73.754 & 0.004 & 150.2 & 0.6 & --1.0 & 0.1 & 151.2 & 0.6 & 2.526 & 0.010 \\
B0833--45 & J0835--4510 & 0.089 & 14 & 44 & 67.771 & 0.009 & 44.1 & 0.7 & --1.2 & 0.1 & 45.3 & 0.7 & 0.823 & 0.012 \\
B0835--41 & J0837--4135 & 0.752 & 13 & 44 & 147.20 & 0.01 & 143.3 & 0.7 & --1.2 & 0.1 & 144.5 & 0.7 & 1.210 & 0.006 \\
B0905--51 & J0907--5157 & 0.254 & 14 & 46 & 103.668 & 0.004 & --24.6 & 1.0 & --1.2 & 0.1 & --23.4 & 1.0 & --0.278 & 0.011 \\
B0940--55 & J0942--5552 & 0.664 & 13 & 43 & 180.16 & 0.02 & --62.9 & 0.6 & --1.1 & 0.1 & --61.8 & 0.6 & --0.422 & 0.004 \\
B1046--58 & J1048--5832 & 0.124 & 14 & 45 & 128.86 & 0.03 & -150.5 & 1.1 & --1.1 & 0.1 & --149.4 & 1.1 & --1.430 & 0.011 \\
B1054--62 & J1056--6258 & 0.422 & 13 & 43 & 320.62 & 0.03 & 7.3 & 0.8 & --1.1 & 0.1 & 8.4 & 0.8 & 0.032 & 0.003 \\
B1133--55 & J1136--5525 & 0.365 & 14 & 46 & 85.111 & 0.009 & 30.8 & 0.7 & --1.1 & 0.1 & 31.8 & 0.7 & 0.461 & 0.011 \\
B1221--63 & J1224--6407 & 0.216 & 14 & 46 & 97.63 & 0.01 & --6.8 & 0.6 & --1.2 & 0.1 & --5.6 & 0.6 & --0.071 & 0.008 \\
B1240--64 & J1243--6423 & 0.388 & 14 & 44 & 297.09 & 0.02 & 160.0 & 0.8 & --1.2 & 0.1 & 161.2 & 0.8 & 0.668 & 0.003 \\
B1323--58 & J1326--5859 & 0.478 & 15 & 49 & 287.17 & 0.02 & --581.6 & 0.8 & --1.1 & 0.1 & --580.4 & 0.8 & --2.490 & 0.004 \\
B1322--66 & J1326--6700 & 0.543 & 14 & 46 & 208.97 & 0.07 & --53.7 & 0.7 & --1.2 & 0.1 & --52.5 & 0.7 & --0.310 & 0.004 \\
B1323--62 & J1327--6222 & 0.530 & 14 & 46 & 318.48 & 0.01 & --323.8 & 0.9 & --1.1 & 0.1 & --322.6 & 0.9 & --1.248 & 0.004 \\
B1353--62 & J1357--62 & 0.456 & 15 & 50 & 416.7 & 0.2 & --590.2 & 0.8 & --1.2 & 0.1 & --589.0 & 0.9 & --1.741 & 0.003 \\
B1356--60 & J1359--6038 & 0.128 & 14 & 46 & 293.78 & 0.02 & 37.2 & 0.6 & --1.1 & 0.1 & 38.3 & 0.6 & 0.161 & 0.003 \\
B1426--66 & J1430--6623 & 0.785 & 12 & 39 & 65.102 & 0.004 & --21.5 & 0.5 & --1.1 & 0.1 & --20.3 & 0.5 & --0.385 & 0.009 \\
B1449--64 & J1453--6413 & 0.179 & 13 & 43 & 71.35 & 0.06 & --22.9 & 0.9 & --1.2 & 0.1 & --21.7 & 0.9 & --0.375 & 0.015 \\
B1451--68 & J1456--6843 & 0.263 & 14 & 46 & 8.613 & 0.004 & --1.4 & 0.6 & --1.3 & 0.1 & --0.1 & 0.6 & --0.021 & 0.091 \\
B1556--44 & J1559--4438 & 0.257 & 13 & 42 & 55.94 & 0.05 & --3.1 & 0.8 & --1.1 & 0.1 & --1.9 & 0.8 & --0.043 & 0.018 \\
B1557--50 & J1600--5044 & 0.193 & 13 & 43 & 262.83 & 0.01 & 137.1 & 0.8 & --1.2 & 0.1 & 138.2 & 0.8 & 0.648 & 0.004 \\
B1601--52 & J1605--5257 & 0.658 & 13 & 42 & 34.90 & 0.06 & 3.7 & 0.7 & --1.1 & 0.1 & 4.9 & 0.7 & 0.171 & 0.025 \\
B1641--45 & J1644--4559 & 0.455 & 14 & 45 & 478.66 & 0.04 & --620.9 & 0.9 & --1.1 & 0.1 & --619.8 & 0.9 & --1.595 & 0.002 \\
B1706--16 & J1709--1640 & 0.653 & 13 & 43 & 24.885 & 0.003 & 0.7 & 0.9 & --0.8 & 0.1 & 1.5 & 0.9 & 0.074 & 0.045 \\
B1706--44 & J1709--4429 & 0.102 & 15 & 49 & 75.593 & 0.003 & --0.6 & 0.6 & --1.1 & 0.1 & 0.5 & 0.6 & 0.008 & 0.009 \\
B1718--35 & J1721--3532 & 0.280 & 11 & 36 & 496.82 & 0.04 & 165.9 & 1.9 & --1.1 & 0.1 & 167.0 & 1.9 & 0.414 & 0.005 \\
B1727--47 & J1731--4744 & 0.830 & 13 & 42 & 122.786 & 0.007 & --444.0 & 0.5 & --1.0 & 0.1 & --443.0 & 0.5 & --4.445 & 0.005 \\
B1737--30 & J1740--3015 & 0.607 & 13 & 42 & 151.80 & 0.04 & --157.0 & 0.7 & --1.0 & 0.1 & --155.9 & 0.7 & --1.265 & 0.006 \\
B1742--30 & J1745--3040 & 0.367 & 13 & 42 & 88.20 & 0.05 & 96.4 & 0.5 & --1.1 & 0.1 & 97.4 & 0.5 & 1.360 & 0.007 \\
B1749--28 & J1752--2806 & 0.563 & 12 & 39 & 50.35 & 0.01 & 95.0 & 1.0 & --1.1 & 0.1 & 96.1 & 1.0 & 2.351 & 0.025 \\
B1804--08 & J1807--0847 & 0.164 & 14 & 46 & 112.364 & 0.004 & 165.2 & 0.8 & --0.8 & 0.1 & 166.0 & 0.8 & 1.820 & 0.008 \\
B1822--09 & J1825--0935 & 0.769 & 12 & 40 & 19.50 & 0.01 & 66.3 & 0.6 & --0.7 & 0.1 & 67.0 & 0.6 & 4.233 & 0.037 \\
B1826--17 & J1829--1751 & 0.307 & 15 & 50 & 216.829 & 0.003 & 304.3 & 1.0 & --0.9 & 0.1 & 305.2 & 1.0 & 1.734 & 0.006 \\
J1852--0635 & J1852--0635 & 0.524 & 13 & 44 & 173.9 & 0.1 & 414.0 & 0.4 & --0.7 & 0.1 & 414.8 & 0.4 & 2.939 & 0.004 \\
B1857--26 & J1900--2600 & 0.612 & 10 & 32 & 38.25 & 0.03 & --9.1 & 0.8 & --1.0 & 0.1 & --8.1 & 0.8 & --0.260 & 0.025 \\
B2045--16 & J2048--1616 & 1.962 & 10 & 32 & 11.41 & 0.02 & --10.2 & 0.8 & --0.9 & 0.1 & --9.3 & 0.8 & --1.002 & 0.092 \\
\hline
\hline
\end{tabular}
\end{table*}
\endgroup

The DMs we measured for these pulsars range from 8.613$\pm$0.004\,pc\,cm$^{\rm{-3}}$ (towards PSR J1456--6843) to 496.82$\pm$0.04\,pc\,cm$^{\rm{-3}}$ (towards PSR J1721--3532). 
The mean of the uncertainties on the DM measurements is 0.04\,pc\,cm$^{\rm{-3}}$. 
For comparison, the mean uncertainty on the DMs from the ATNF Pulsar Catalogue is 0.3\,pc\,cm$^{\rm{-3}}$. 
This factor of 7.5 improvement is largely due to the wider bandwidth UWL observations that we employ.
Furthermore, the median difference between the DMs measured in this work and those from the ATNF Pulsar Catalogue is 1.8$\upsigma$, showing reasonable agreement. 

The absolute RM values we measured for the pulsars range from 0.6$\pm$0.6\,rad\,m$^{\rm{-2}}$ (towards PSR J1709--4429) to 620.9$\pm$0.9\,rad\,m$^{\rm{-2}}$ (towards PSR J1644--4559).
We find that applying the RM synthesis and RM CLEAN methods to the wide-band data provide reliable RM measurements. This is evident from the Faraday spectra in Figure \ref{fig:RMSF} and the supporting information, and the good agreement with the measurements published in the ATNF Pulsar Catalogue ($<1.4\upsigma$, on average).
The mean of the uncertainties we obtain for RM$_{\rm{ISM}}$ is 0.8\,rad\,m$^{\rm{-2}}$.
For comparison, the mean uncertainty on the RMs from the ATNF Pulsar Catalogue is 2\,rad\,m$^{\rm{-2}}$, the majority of which were obtained using older receivers on the Parkes radio telescope and similar observation lengths. 
This reduction in the uncertainties by a factor of $\approx$2.5 is also likely due to the wider bandwidth UWL observations.

The systematic correction for the ionospheric RM is necessary and adds a minor contribution to the RM measurement uncertainty.
The mean ionospheric RM contributes an average of --1\,rad\,m$^{\rm{-2}}$ to the observed RMs, with 0.1\,rad\,m$^{\rm{-2}}$ uncertainty, Table \ref{tab:obs}. 
This level of uncertainty is expected using the method outlined in Section \ref{subsec:ana} \citep[e.g.][]{Sotomayor+2013,Porayko+2019,sbg+19}.
For example, the RM$_{\rm{ISM}}$ for PSR J1709--4429 is in better agreement (0.3$\upsigma$) with the RM from the ATNF Pulsar Catalogue (0.70$\pm$0.07\,rad\,m$^{\rm{-2}}$; measured using 1369\,MHz observations, 256\,MHz bandwidth, and ionospheric Faraday rotation correction; \citealt{Johnston+2005}) compared to the RM$_{\rm{obs}}$ (1.9$\upsigma$).

The largest (absolute) average magnetic field parallel to the LoS we calculate is $\langle B_{\parallel}\rangle$=--4.445$\pm$0.005\,$\upmu$G towards PSR J1731--4744, at an estimated distance of 0.7\,kpc \citep{ymw+2017}.
The smallest absolute average magnetic field parallel to the LoS we calculate is 0.008$\pm$0.009\,$\upmu$G for PSR J1709--4429, at an estimated 2.6\,kpc distance \citep{ymw+2017}.
The mean uncertainty on $\langle B_{\parallel}\rangle$, propagated using the DM and RM uncertainties, is 0.018\,$\upmu$G. The mean and median fractional uncertainties are 19 and 1.5 per cent, respectively, dominated by the uncertainties on the RM measurements.
The mean and median fractional uncertainties on RM$_{\rm{ISM}}$ are 19 per cent (dominated by the small |RM| values) and 1.5 per cent, respectively.
The mean and median fractional uncertainties on the DMs are 0.1 and 0.01 per cent, respectively.

\begin{figure}
	\includegraphics[width=\columnwidth]{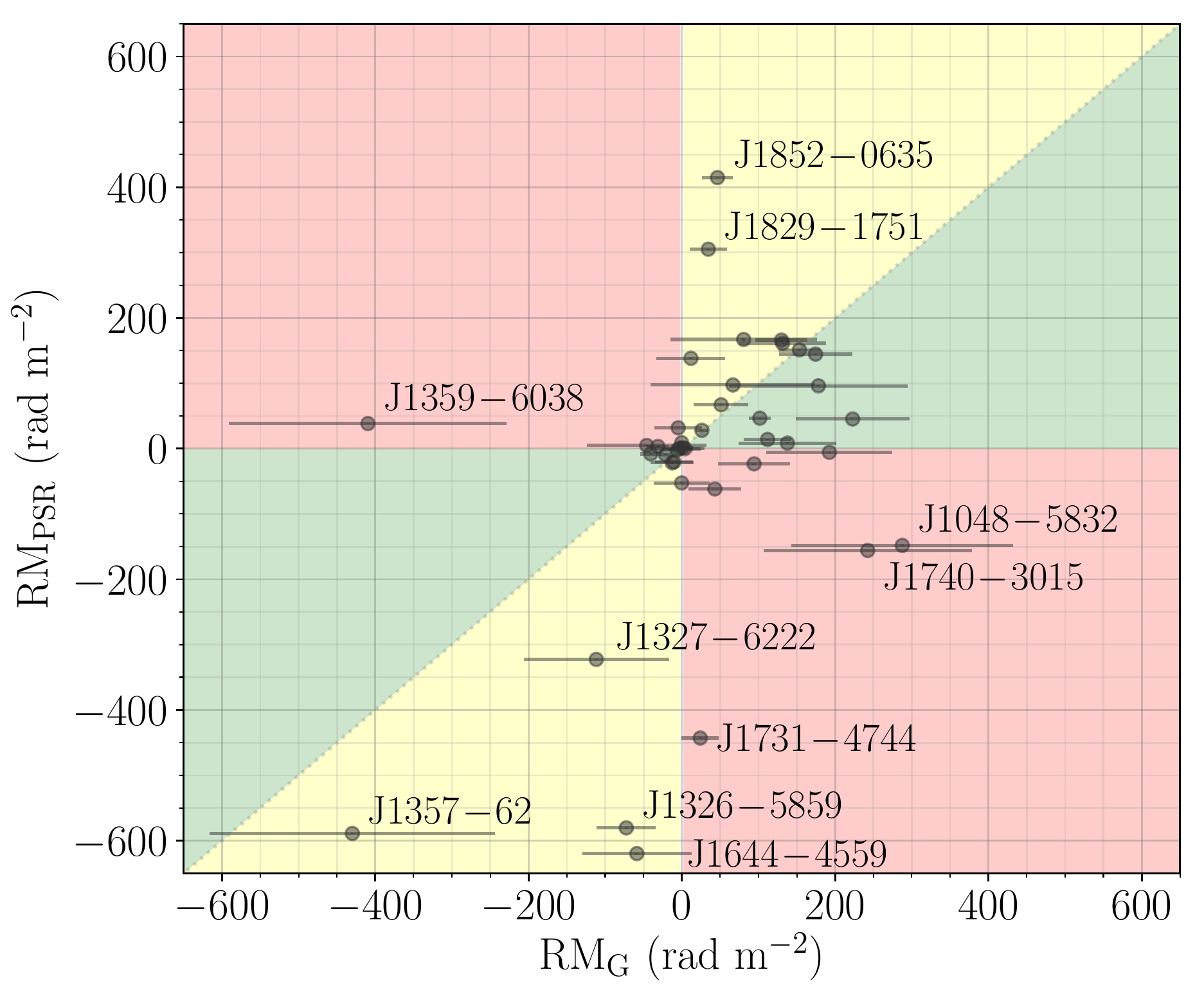}
    \caption{Comparison of the RMs measured in this work towards the pulsars (RM$_{\rm{ISM}}$; here referred to as RM$_{\rm{PSR}}$ for clarity) and the RMs obtained for the entire LoS through the Galaxy (RM$_{\rm{G}}$) obtained from the Faraday sky reconstruction in \citet{Hutschenreuter+2020}, shown by the grey points with error bars representing the uncertainties. The uncertainties on RM$_{\rm{PSR}}$ are not visible at this scale. The shaded regions indicate: green) |RM$_{\rm{PSR}}$| is smaller than |RM$_{\rm{G}}$| and both have the same sign; yellow) |RM$_{\rm{PSR}}$| is larger than |RM$_{\rm{G}}$| and both have the same sign; red) RM$_{\rm{PSR}}$ and RM$_{\rm{G}}$ have opposite signs. We label ten individual pulsars in the yellow and red shaded regions, with large deviations from the dashed line RM$_{\rm{PSR}}$=RM$_{\rm{G}}$. }
    \label{fig:RMcomp}
\end{figure}

We obtained the RMs due to the entire LoS through the Galaxy towards the pulsars in the census using the Galactic Faraday sky reconstructions. A plot comparing the RMs towards the pulsars, probing various distances within the Galaxy, and the RMs expected due to the entire Galactic lines-of-sight from \citet{Hutschenreuter+2020} is shown in Figure \ref{fig:RMcomp}. We further discuss this in Section \ref{subsec:ISM}.

\begin{figure*}
	\includegraphics[width=\columnwidth,trim={-0.5cm 0 0 0},clip]{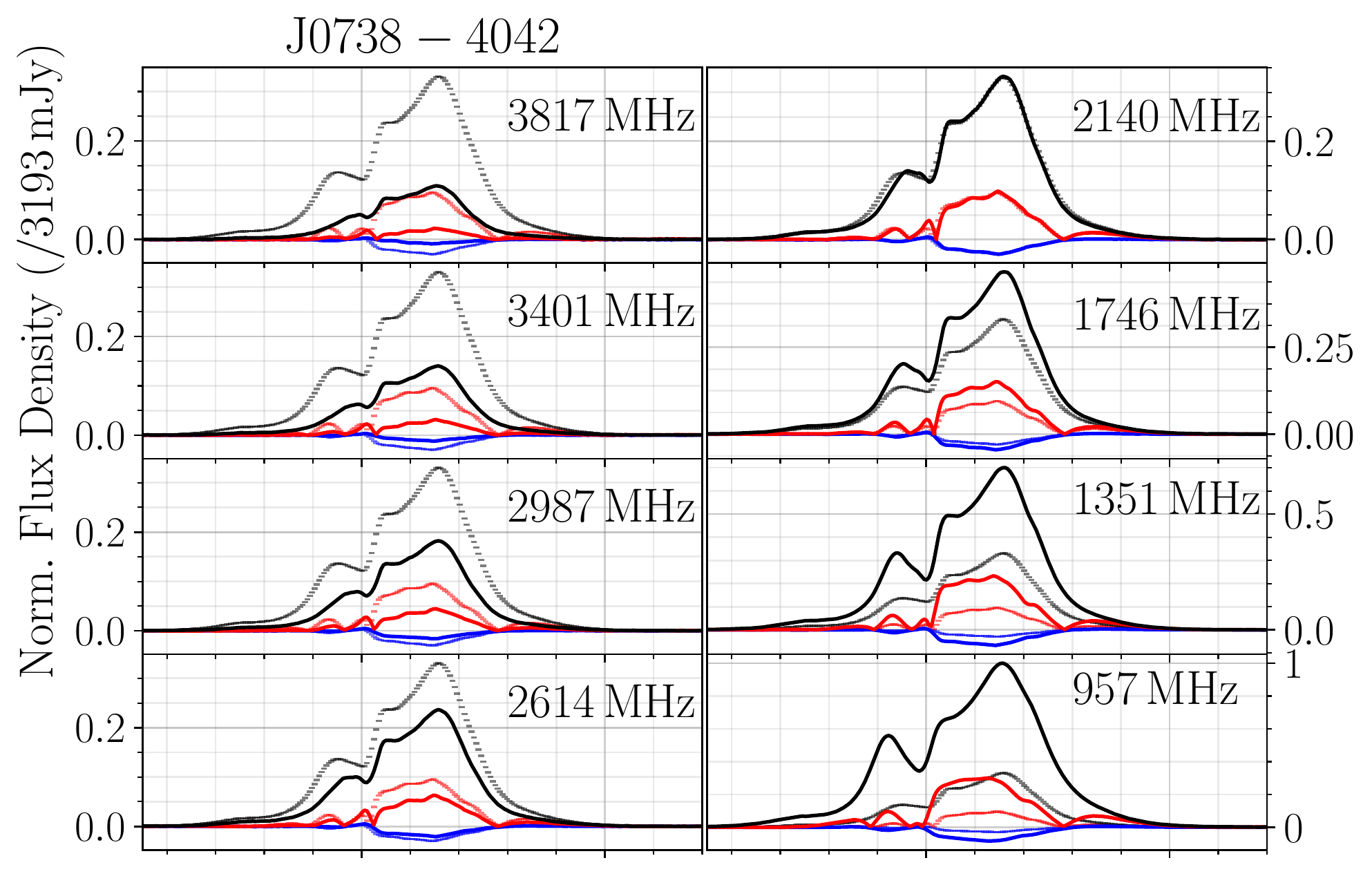}
	\includegraphics[width=\columnwidth,trim={-0.55cm 0 0 0},clip]{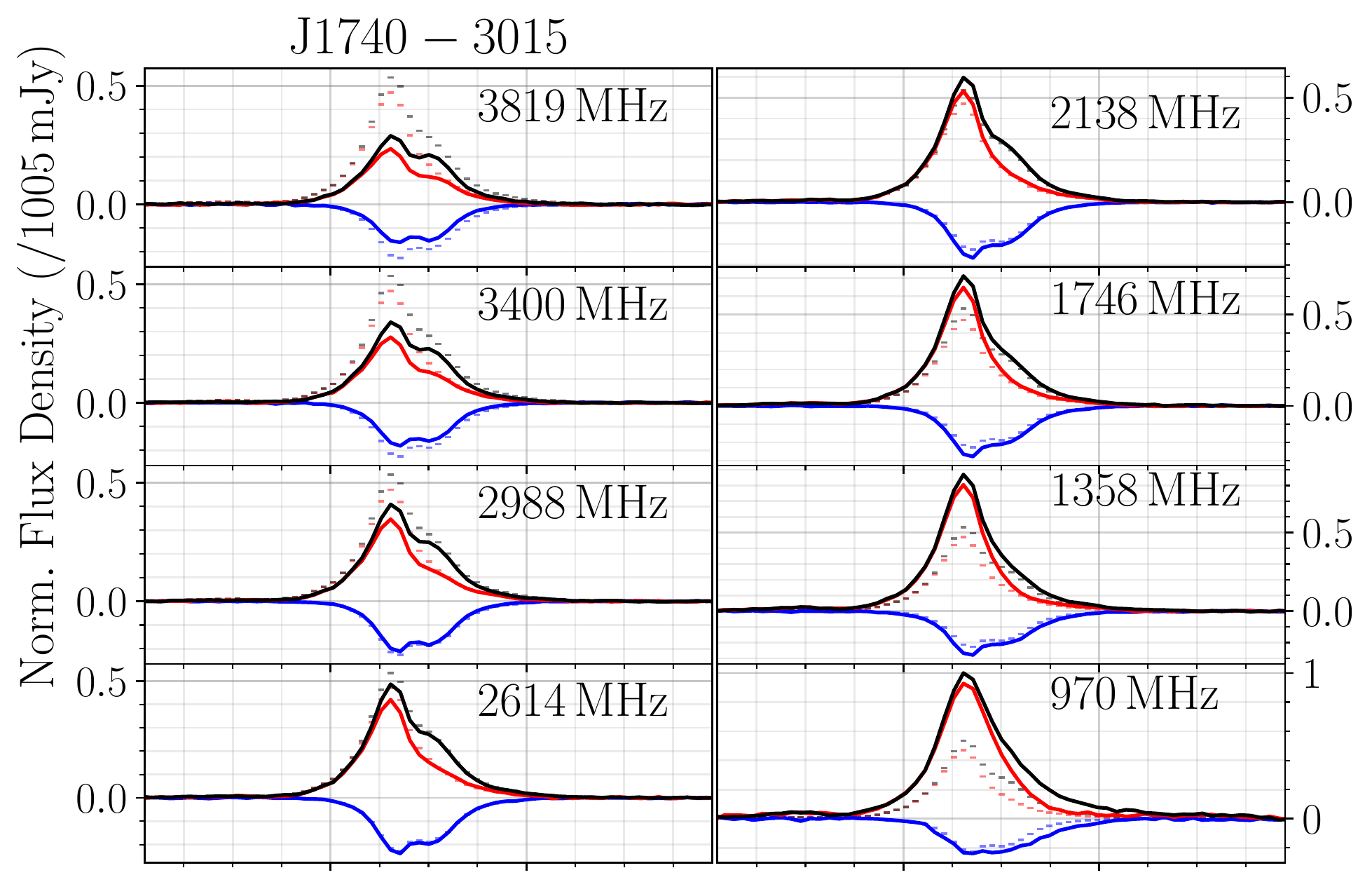}

	\includegraphics[width=\columnwidth,trim={-0.4cm 0 -1.18cm 0},clip]{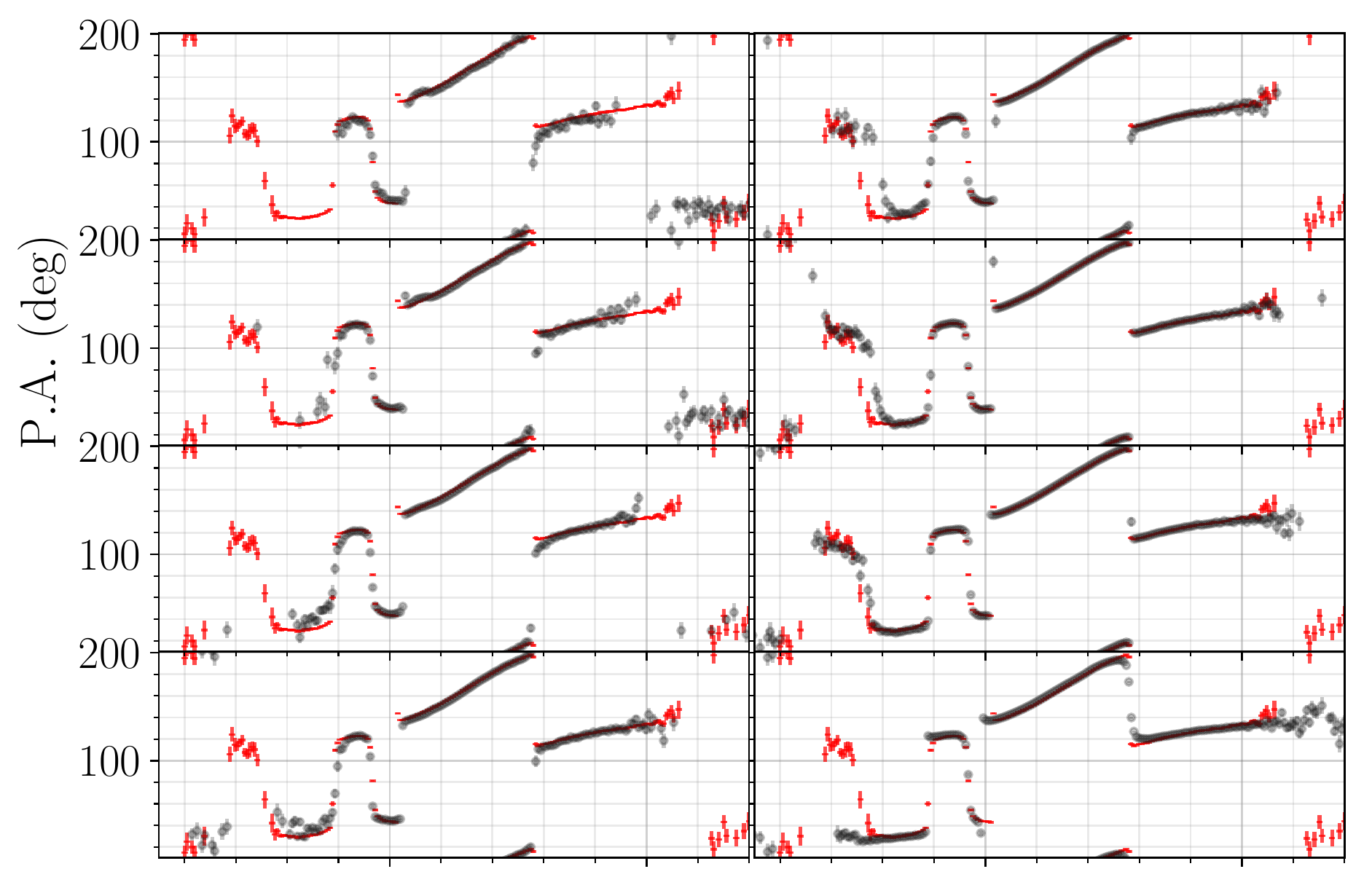}
	\includegraphics[width=\columnwidth,trim={-0.18cm 0 -.85cm 0},clip]{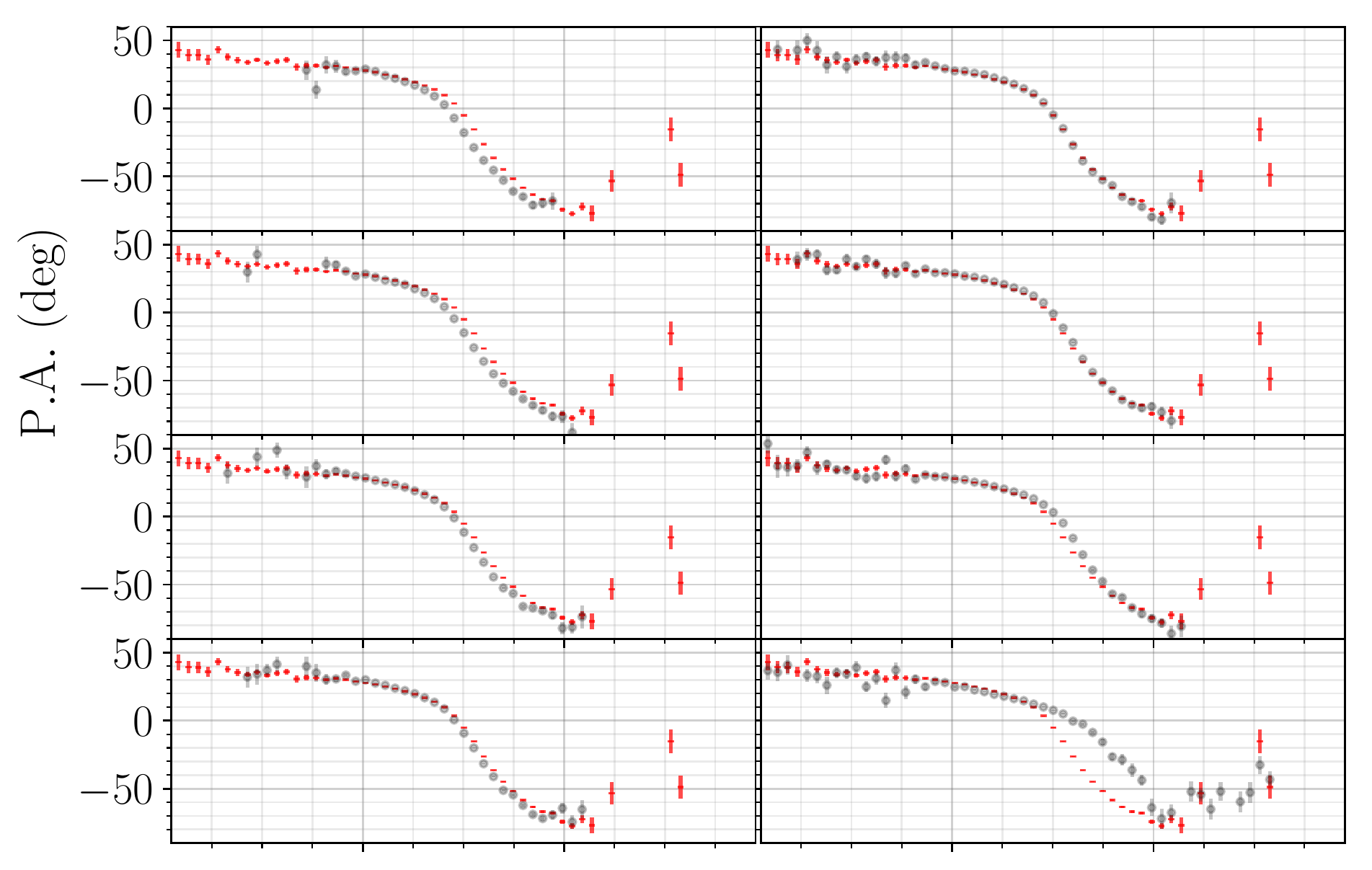}

	\includegraphics[width=\columnwidth,trim={-0.5cm 0 -1.18cm 0},clip]{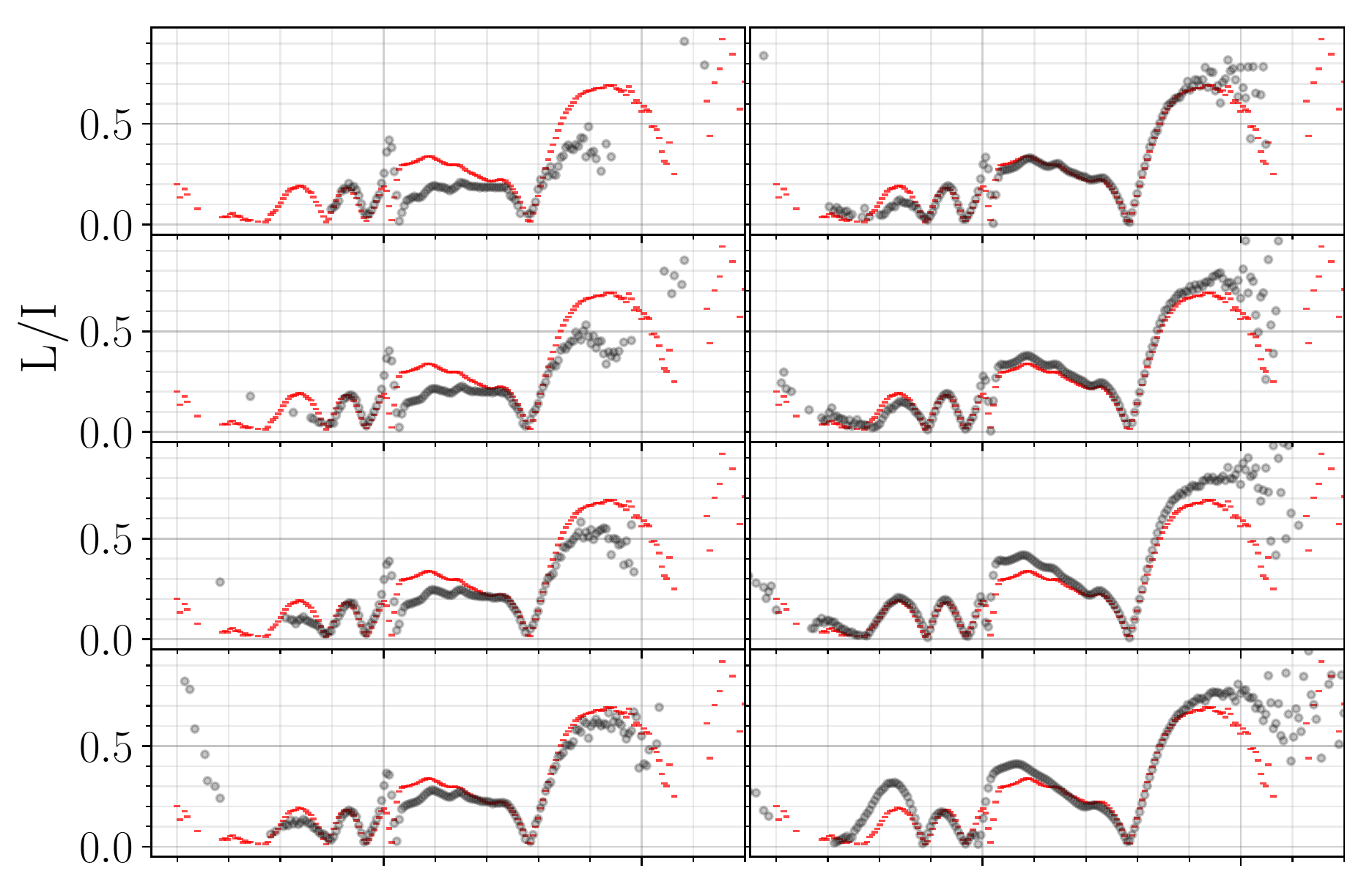}
	\includegraphics[width=\columnwidth,trim={-0.45cm 0 -.8cm 0},clip]{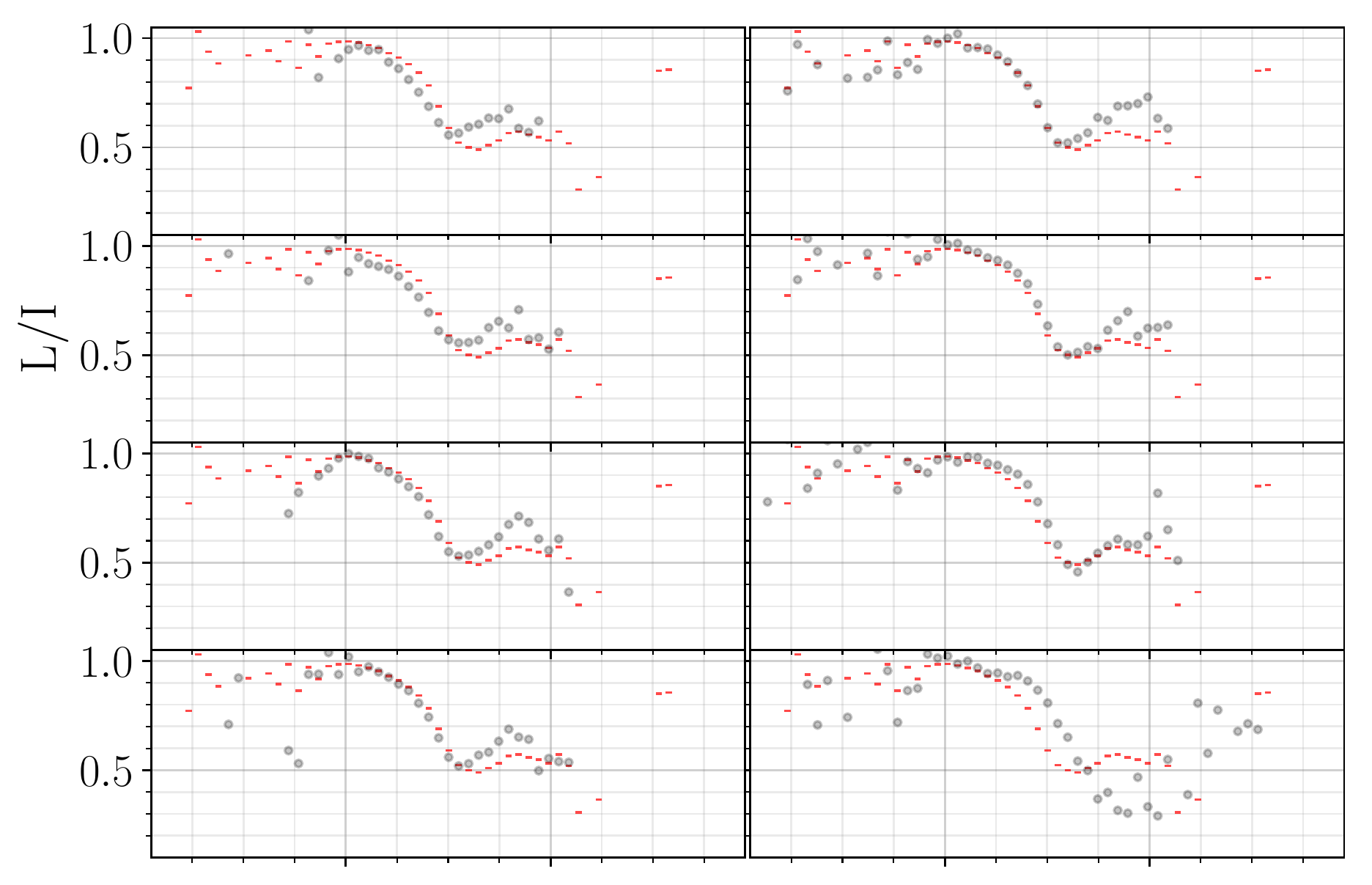}

	\includegraphics[width=\columnwidth,trim={0 0 -1.18cm 0},clip]{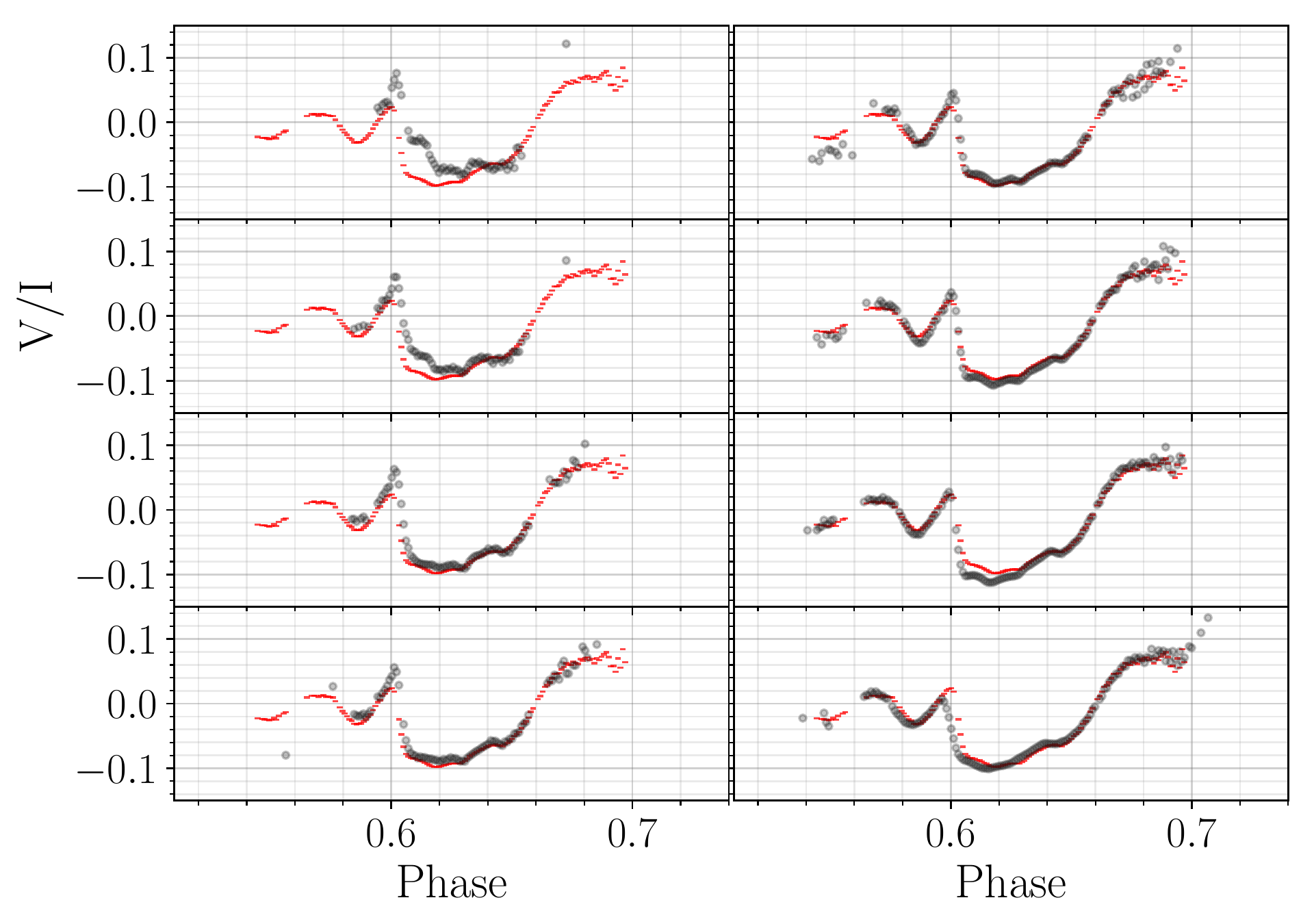}
	\includegraphics[width=\columnwidth,trim={0 0 -.85cm 0},clip]{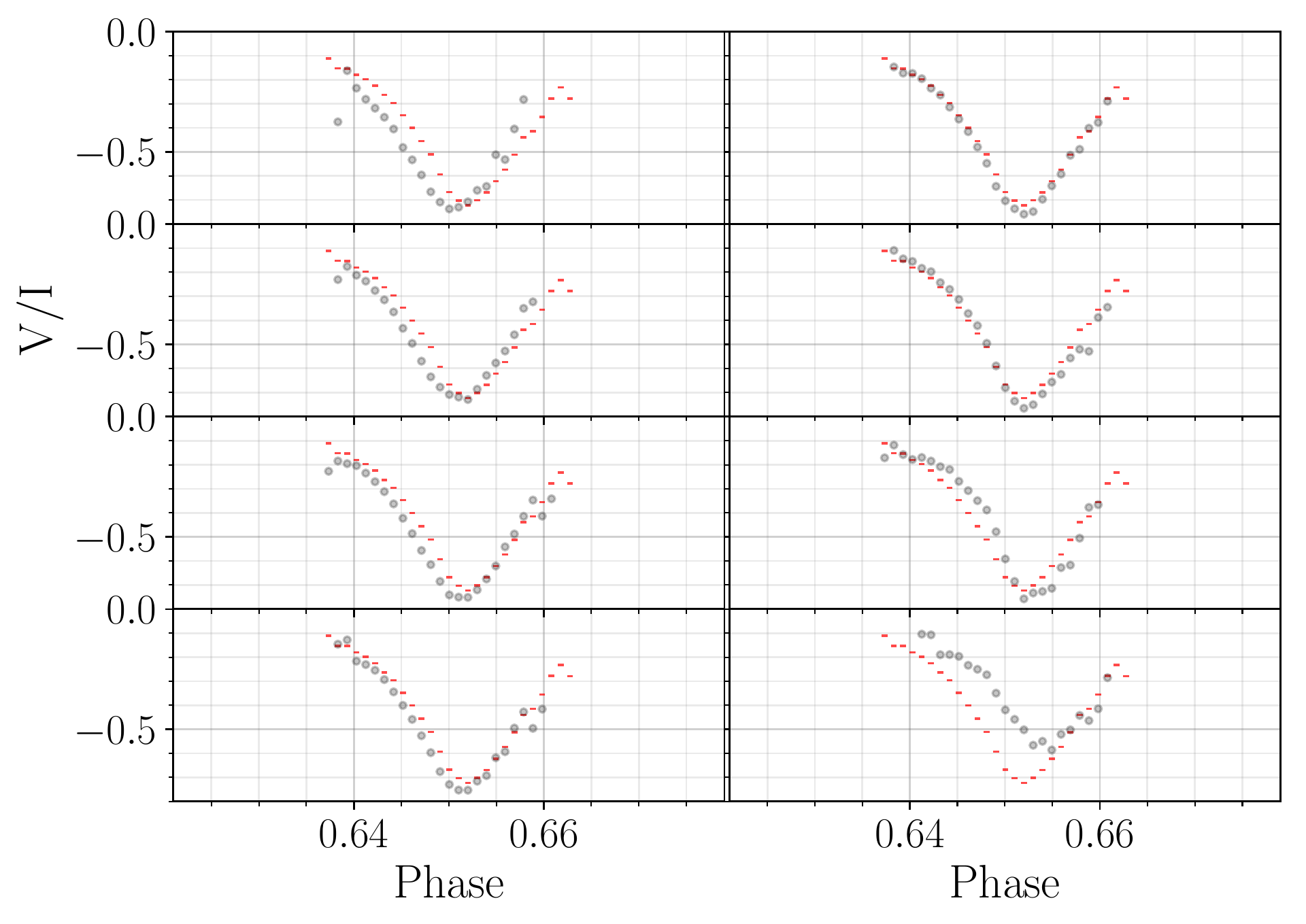}

    \caption{Examples, for PSRs J0738--4042 and J1740--3015, of plots produced for all pulsars, available in the supporting information. Panels (from upper to lower) show: Upper) polarization profiles, zoomed in to emphasize the on-pulse phase bins, normalised to the maximum peak flux density (usually that of the lowest frequency sub-band) shown in the y-axis label, for 8 frequency bands with centre frequencies labelled, plus the frequency-averaged profile across the entire bandwidth, shown for total intensity (Stokes $I$, black lines and grey dashes, respectively), linear polarization (red lines and faint dashes, respectively), and circular polarization (blue lines and faint dashes, respectively). Second from top) polarization position angles (P.A.s) and uncertainties for 8 frequency bands (grey circles) and for the entire bandwidth (red points) for phase bins with $>3\sigma$ linear polarization. Third from top) Fractional linear polarization for 8 frequency bands (grey circles)  and for the entire bandwidth (red dashes) for phase bins with $>3\sigma$ linear polarization. Lower) Fractional circular polarization for 8 labelled frequency bands (grey circles) and for the entire bandwidth (red dashes) for phase bins with $>3\sigma$ circular polarization. }
    \label{fig:profs}
\end{figure*}

We constructed the time-averaged, DM- and RM-corrected, polarization profiles for each of the pulsars, both averaged in frequency across the entire bandwidth, and for 8 sub-bands (each 416\,MHz in bandwidth). 
The example polarization profiles for two pulsars, PSRs J0738--4042 and J1740--3015, are shown in Figure \ref{fig:profs}. 
Similar plots for all 40 pulsars studied in this work are available in the supporting information.
These are further discussed for each pulsar in Section \ref{subsec:emn}.

\begin{figure}
	\includegraphics[width=\columnwidth,trim={-0.45cm 0 0 0},clip]{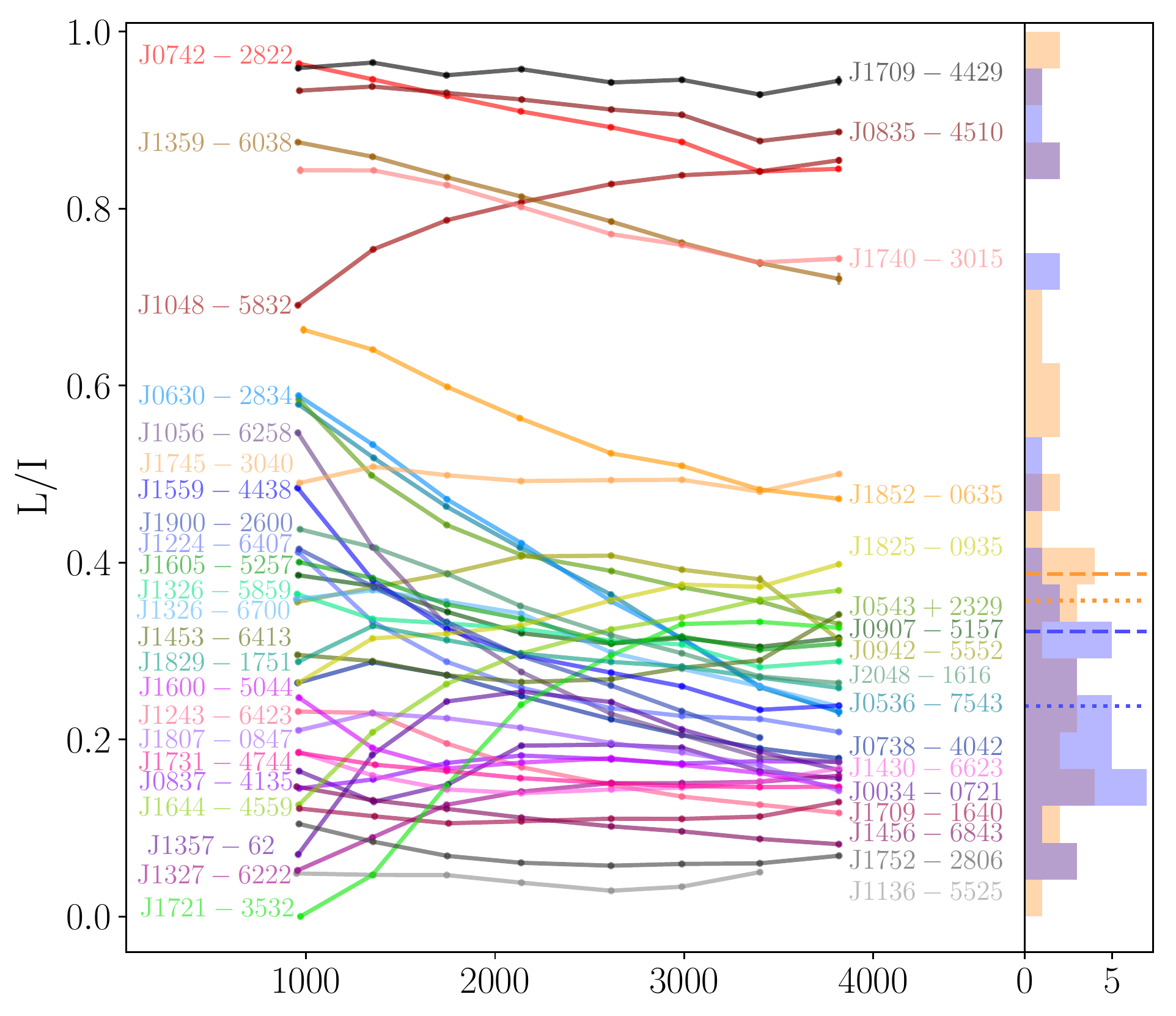}

	\includegraphics[width=\columnwidth]{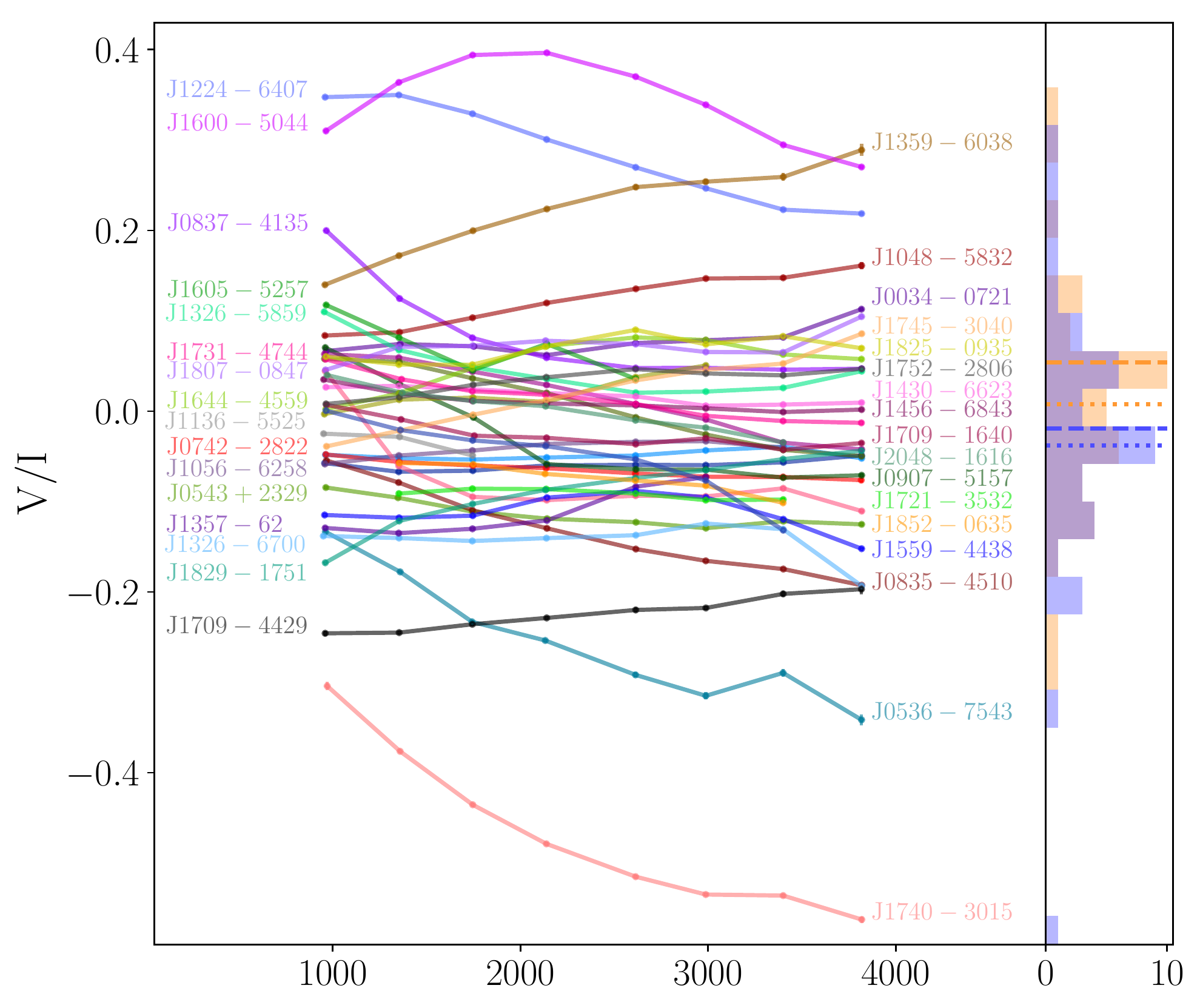}

	\includegraphics[width=\columnwidth,trim={-0.45cm 0 0 0},clip]{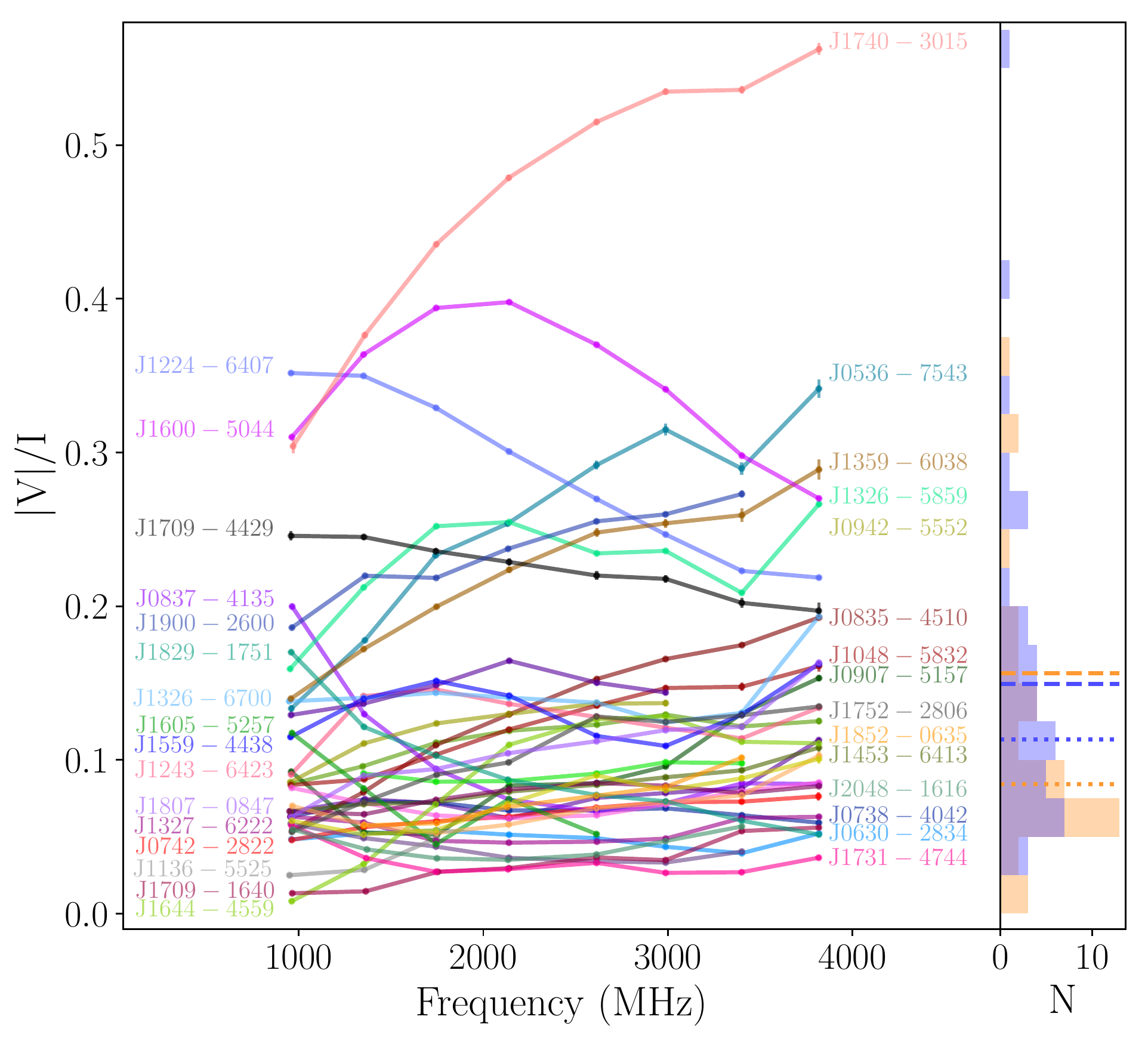}

    \caption{Summary of on-pulse integrated fractional polarization versus frequency for all pulsars, compiled using the average pulse profiles. Upper--lower panels show: linear; circular; and absolute circular fractional polarizations. Left panels: All (most) pulsars are labelled in the upper (middle, lower) plot; each pulsar is shown using the same colour in all panels. Right panels: Histograms, mean (dashed line) and median (dotted line) summarising the data for the lowest (orange) and highest (blue) frequencies.}
    \label{fig:fracLV}
\end{figure}

Figure \ref{fig:fracLV} summarises the variation in the on-pulse integrated linear, circular, and absolute circular polarizations across the 8 sub-bands for each pulsar. This highlights the general trends seen in this pulsar census and is further discussed in Section \ref{subsec:emn}.

\section{Discussion}\label{sec:dis}

\subsection{ISM parameters}\label{subsec:ISM}

We report the DM measurements using the {\sc Tempo2} method here, Table \ref{tab:obs}, because this produced the smallest difference between the measurements and the ATNF Pulsar Catalogue values compared to using the more simple \textit{pdmp} method (0.09\,pc\,cm$^{\rm{-3}}$, vs 0.15\,pc\,cm$^{\rm{-3}}$, on average, respectively), and also smaller uncertainties (0.04\,pc\,cm$^{\rm{-3}}$, vs 0.36\,pc\,cm$^{\rm{-3}}$, on average, respectively). This is likely because {\sc Tempo2} is the preferred method used for measuring the DMs in the ATNF Pulsar Catalogue \citep[e.g.][]{pkj+13,Hobbs+2004,Hobbs+2004b}.
However, consistency in absolute DM measurements obtained using different observing set-ups (e.g. centre frequencies and bandwidths) can be difficult to achieve due to, e.g. pulse profile evolution with frequency -- both intrinsic to the pulsar emission and caused by ISM propagation effects such as interstellar scattering. For example, when using {\sc Tempo2}, the pulse profile templates and, if included, their frequency evolution, often differ between studies, and these are not often made publicly available.

The majority of the DM measurements in the ATNF Pulsar Catalogue were obtained using $\sim$1\,GHz observations and more limited bandwidths in comparison to the ultra-broadband frequency coverage the Parkes UWL receiver now provides \citep[e.g. ][]{McCulloch+1973,Newton+1981,DAlessandro+1993,Hobbs+2004,Hobbs+2004b,pkj+13}. 
There are a handful of pulsars with DMs in the ATNF Pulsar Catalogue from low frequency observations and, therefore, amongst the smallest uncertainties \citep[at 50\,MHz for PSRs J0034--0721, J0543+2329, J1709--1640, J1825--0935; and 150\,MHz for J0543+2329;][respectively]{srb+2015,Bilous+2016}. Most of these pulsars have $>7\upsigma$ DM differences, possibly highlighting systematic (e.g. pulse profile evolution) or physical (e.g. DM variation or frequency-dependent DM) differences between measurements using different observing frequencies and bandwidths \citep[e.g.][]{Cordes+2016,Donner+2019,Lam+2020,Donner+2020}. 
However, for PSR J0034--0721, we use the ATNF Pulsar Catalogue DM for further analysis. 
PSR J0034--0721's pulse profile shows notable evolution with frequency, including some interstellar scattering evident towards the lower frequencies (see the supporting information and, e.g. \citealt{srb+2015}).
The DM we measured for this pulsar (13.3$\pm$0.3\,pc\,cm$^{\rm{-3}}$) is 7$\upsigma$ different from the DM in the ATNF Pulsar Catalogue (10.922$\pm$0.006\,pc\,cm$^{\rm{-3}}$; measured at centre frequency 50\,MHz; \citealt{srb+2015}). 
We produced Faraday spectra for PSR J0034--0721 using both of these DMs, and found that the ATNF Pulsar Catalogue DM provided a `better-behaved' Faraday spectrum (higher S/N and more symmetrical), introducing the possibility of an additional diagnostic check. 
This was not the case for any of the other pulsars in this census.
Alternative and novel methods for aligning the pulse profiles across the frequency range and measuring the resulting DM have been proposed \citep[e.g.][]{pen19}, including using the complementary RM information \citep[e.g.][]{Oswald+2020}, or simultaneously fitting the covariant DM and scattering measures \citep[e.g.][]{Hassall+2012,Geyer+2017}.

One pulsar with a DM measurement in this work, PSR J1056--6258, is also explored in detail in \citet{Oswald+2020}. Our measurement ranges in agreement with the DMs summarised in that work, from 0.2$\upsigma$ \citep[closest to the ATNF Pulsar Catalogue value estimated in][]{Karastergiou+2006}, to 10$\sigma$ with the value reported in \citet{ijw19}. Although the latter value does not have the largest discrepancy between DM values, it does have the smallest uncertainty, indicating that the systematic uncertainties in measuring DMs are likely larger and ideally should be taken into consideration when comparing absolute DM measurements. 
Further investigation of measuring absolute DM values is beyond the scope of this work, and we do not further compare our DM measurements with those published in the ATNF Pulsar Catalogue. 


As was the case for the DMs, there are a handful of pulsars with RMs measured using lower frequency observations in the catalogue and, therefore, amongst the smallest uncertainties. This includes measurements from 150\,MHz for PSR J0034-0721 \citep{Noutsos+2015} and PSR J0543+2329 \citep{sbg+19}; there are also recent measurements at 200\,MHz for PSRs J0630--2834, J0742--2822, J0835--4510, J0837--4135, J0907--5157, J1359--6038, J1453--6413, J1752--2806, J2048--1616 \citep{Riseley+2020}.
These measurements from the literature show excellent agreement with our RMs and are within $1\upsigma$, except for PSRs J0835--4510 and J0837--4135, which are discussed below.
In comparison to the ultra-broadband frequency coverage the Parkes UWL receiver now provides, the majority of the RM measurements in the ATNF Pulsar Catalogue were obtained using high frequency ($\sim$1\,GHz) data and more limited bandwidths \citep[e.g.][]{hl87,Costa+1991,tml93,Qiao+1995,Han+1999,Johnston+2005,Johnston+2007,Han+2006,njkk08,Force+2015,hmvd18}. This is reflected in the reduction in uncertainties by a factor of $>$2, on average, for the measurements reported in this work.
PSRs J1709--4429 and J1456--6843 with small absolute RM values, see Table \ref{tab:obs}, resulting in larger than 100 per cent fractional uncertainties would likely benefit from lower-frequency observations to provide more precise RM measurements \citep[e.g.][]{sbg+19,Xue+2019}.

For some pulsars, the method and on-pulse profile bins used to calculate the RM can affect the absolute RM measured. For example, \citet{Noutsos+2009} and \citet{ijw19} find significant RM variations with pulse phase for PSR J1243--6423, with a range of $\approx$20-60\,rad\,m$^{\rm{-2}}$, see also \citet{Johnston+2021}. Our measurement reported in Table \ref{tab:obs} agrees with the RM reported for the peak linear polarization flux density in \citet{ijw19} and within 3$\upsigma$ of the value published in the ATNF Pulsar Catalogue \citep{tml93}. 
Similarly to DM measurements, systematic uncertainties should be taken into account when comparing RM measurements \citep[e.g.][]{Schnitzeler+2017,Porayko+2019,Johnston+2021}. 
Measuring both DM and RM in a consistent way \citep[e.g.][]{Oswald+2020} may provide the most reliable method for using both measurements towards estimating $\langle B_{\parallel}\rangle$.
Furthermore, to investigate DM and RM variations with time, a consistent measurement method applied to consistent frequency information should ideally be used for all epochs of the monitoring data \citep[e.g.][]{Donner+2020,Johnston+2021}.

There are only two pulsars, PSRs J0835--4510 and J1731--4744, that have significant ($>$13$\upsigma$) differences between the RMs measured in this work and previous measurements in the ATNF Pulsar Catalogue. PSR J0835--4510, the Vela pulsar, is located towards a complex LoS and is known to show DM and RM variations over time \citep[e.g.][Xue et al, in prep.; Sobey et al., in prep.]{Hamilton+1985,pkj+13,Lenc+2017}. PSR J1731--4744 is also located towards a complex LoS and is further discussed below.  

The RMs measured towards the pulsars included in this work are compared to the RMs expected from the entire LoS through the Galaxy, RM$_{\rm{G}}$, see Figure \ref{fig:RMcomp}.
We obtain RM$_{\rm{G}}$ values using the Galactic Faraday sky reconstructions from both \citet{Hutschenreuter+2020}, RM$_{\rm{G}}^{\rm{H20}}$, and \citet{Oppermann+2015}, RM$_{\rm{G}}^{\rm{O15}}$.
The median difference between the pulsar RMs and RM$_{\rm{G}}$ values is 1.7 and 1.6$\sigma$, respectively. 
The signs of the pulsar RMs and RM$_{\rm{G}}$ also agree in 85 and 83 per cent of the LoS, respectively.
We use Figure \ref{fig:RMcomp} primarily as a method to identify noteworthy lines of sight associated with physical structures in the ISM. 
We discuss the pulsars whose RM sign is opposite to that for the entire LoS and identified by name in Figure \ref{fig:RMcomp}, in order of Galactic longitude below.

\textbf{PSR J1048--5832}; $l,b$=287$^{\circ}$.4,0$^{\circ}$.6: 
There is a large negative RM towards this pulsar (--149$\pm$1\,rad\,m$^{\rm{-2}}$), which is the opposite sign to the large positive RM for the entire Galactic LoS, RM$_{\rm{G}}^{\rm{H20}}=$338$\pm$34\,rad\,m$^{\rm{-2}}$ and RM$_{\rm{G}}^{\rm{O15}}=$288$\pm$145\,rad\,m$^{\rm{-2}}$ (14$\upsigma$ and 3$\upsigma$ difference, respectively). The DM towards this pulsar also significantly changes with time (--0.054$\pm$0.004\,pc\,cm$^{\rm{-3}}$\,yr$^{-1}$; \citealt{Johnston+2021}). 
We estimate the average, electron density weighted Galactic magnetic field strength towards the pulsar to be -1.43$\pm$0.01\,$\upmu$G, Table \ref{tab:obs}. In contrast, we estimate that for the entire LoS from the Galaxy (with DM=640\,pc\,cm$^{\rm{-3}}$, using NE2001; \citealt{Cordes+2002}) is $\sim$0.6$\upmu$G. 
PSR J1048--5832 lies in the Galactic plane at a distance of $\sim$1.8--2.9\,kpc \citep{ymw+2017}, possibly near the outer edge of the Carina-Sagittarius spiral arm and the Carina Nebula (NGC 3372; one of the largest, diffuse nebulae and star-forming regions in the sky at $\approx$2.3\,kpc distance). This area shows up as a large positive region in the Faraday sky maps. From the DM changes with time, it would seem that the pulsar is probing an inhomogeneous ISM (although there are no proper motion measurements), possibly associated with the material from the nebula, although the RM sign suggests it may lie in front of the nebula. Therefore, further investigation may provide a limit on the distance to PSR J1048--5832 and the nature of the (local) magnetic field in the foreground region. 

\textbf{PSR J1359--6038}; $l,b=$311$^{\circ}$.2,1$^{\circ}$.1: 
This pulsar has a small positive RM compared to the large negative RM expected due to the entire Galactic LoS (4$\upsigma$ and 2$\upsigma$ difference for RM$_{\rm{G}}^{\rm{H20}}=-$430$\pm$115\,rad\,m$^{\rm{-2}}$ and RM$_{\rm{G}}^{\rm{O15}}=-$410$\pm$181\,rad\,m$^{\rm{-2}}$, respectively).
We estimate $\langle B_{\parallel}\rangle$=0.161$\pm$0.003\,$\upmu$G, Table \ref{tab:obs}. 
In contrast, we estimate that for the entire LoS from the Galaxy (with DM$\approx$842\,pc\,cm$^{\rm{-3}}$, using NE2001; \citealt{Cordes+2002}) is $\sim-$0.6$\upmu$G. 
This pulsar is also located near the Galactic plane at a distance of $\sim$5--5.5\,kpc \citep{ymw+2017}, possibly in or near the Scutum-Centaurus spiral arm, which is known to show a large-scale Galactic magnetic field reversal \citep[e.g.][]{VanEck+2011}.
Our results fit with this general picture of positive magnetic field direction in the spiral arm, but overall negative magnetic field direction from the integrated Galactic LoS, although the average field strength towards the pulsar is low, possibly placing it on the near side of the spiral arm and favouring the lower distance estimate. 

\textbf{PSR J1731--4744}; $l,b=$342$^{\circ}$.6,--7$^{\circ}$.7:
There is a large negative RM towards this pulsar (--443.0$\pm$0.5\,rad\,m$^{\rm{-2}}$), compared to that expected due to the entire Galactic LoS,  
RM$_{\rm{G}}^{\rm{H20}}=$24$\pm$24\,rad\,m$^{\rm{-2}}$ and
RM$_{\rm{G}}^{\rm{O15}}=-$6$\pm$81\,rad\,m$^{\rm{-2}}$ (19 and 5$\upsigma$ difference, respectively).
There is also a 13-$\upsigma$ difference (13.9$\pm$0.7\,rad\,m$^{\rm{-2}}$) between the RM measurement in this work compared to that in ATNF Pulsar Catalogue (--429.1$\pm$0.5\,pc\,cm$^{\rm{-3}}$; \citealt{tml93}), although no significant change was detected over the course of just two years in \citet{Johnston+2021}.
The LoS towards PSR J1731--4744 also has the largest $\langle B_{\parallel}\rangle$=--4.445$\pm$0.005\,$\upmu$G for the pulsars in this census.
In contrast, we estimate that for the entire LoS from the Galaxy (with DM=323\,pc\,cm$^{\rm{-3}}$, using NE2001; \citealt{Cordes+2002}) is $\sim$0.03$\upmu$G. 
The pulsar is located some distance from the Galactic plane with distance estimates in the range 0.7--5.5\,kpc \citep{ymw+2017} and has a proper motion of 151$\pm$19\,mas yr$^{-1}$ \citep[][]{Shternin+2019}.
Its sky position is coincident with an inhomogeneous filamentary structure in H$\upalpha$ emission \citep[e.g.][]{Finkbeiner+2003}, associated with a limb of the RCW 114 nebula.
\citet[][]{Shternin+2019} demonstrate that PSR J1731--4744 may have been created in the supernova that we now see as the evolved SNR G343.0--6.0\footnote{\href{http://www.mrao.cam.ac.uk/surveys/snrs/snrs.info.html}{http://www.mrao.cam.ac.uk/surveys/snrs/snrs.info.html}} \citep[e.g.][]{Green+1984}, and argue that the distance of the pulsar is likely consistent with that of the SNR, $\sim$0.4--1.1\,kpc. 
This structure may explain the discrepancy between the RM towards the pulsar and the Galactic LoS. 
In light of this, careful monitoring of the DM and RM towards this pulsar could probe the small-scale fluctuations in the magnetic field of the supernova remnant shell.  

\textbf{PSR J1740--3015}; $l,b=$358$^{\circ}$.3,0$^{\circ}$.2:
This pulsar also has a relatively large negative RM (--155.9$\pm$0.7\,rad\,m$^{\rm{-2}}$), with an opposite sign, although smaller in absolute value, than that due to the entire Galactic LoS, RM$_{\rm{G}}^{\rm{H20}}=$242$\pm$136\,rad\,m$^{\rm{-2}}$ and 
RM$_{\rm{G}}^{\rm{O15}}=$524$\pm$13\,rad\,m$^{\rm{-2}}$
(3 and 51$\upsigma$ difference, respectively).
There is also a 9-$\upsigma$ difference (12$\pm$1\,rad\,m$^{\rm{-2}}$) between the RM measurement in this work compared to that in ATNF Pulsar Catalogue (--168$\pm$0.7\,pc\,cm$^{\rm{-3}}$; \citealt{njkk08}), although no significant change was detected over the course of just two years in \citet{Johnston+2021}.
The LoS towards PSR J1740--3015 has an estimated
$\langle B_{\parallel}\rangle$=--1.265$\pm$0.006\,$\upmu$G.
In contrast, we estimate that for the entire LoS from the Galaxy (with DM=1553\,pc\,cm$^{\rm{-3}}$, using NE2001; \citealt{Cordes+2002}) $\langle B_{\parallel}\rangle\sim$0.3$\upmu$G.
The pulsar is located in the Galactic plane, in the direction of the Galactic Centre, with distance estimates in the range 0.4--3\,kpc \citep{ymw+2017}. 
Although uncertain, the negative RM may place it in front of the Scutum-Centaurus spiral arm, which is known to show a large-scale Galactic magnetic field reversal and positive magnetic field direction \citep[e.g.][]{VanEck+2011}, in line with $<$3\,kpc distance.
However, local, small-scale structures may also have an influence on the observed ISM parameters.

These pulsars towards these complex lines-of-sight, in particular, demonstrate the importance and complementarity of RM and DM measurements towards pulsars along with measurements towards many extragalactic sources to gain information about many directions, distances, and scales within our Galaxy \citep[as previously demonstrated in, e.g.][]{Han+2006,njkk08,VanEck+2011}.  These measurements can be used in combination with additional observational tracers towards reconstructing the structure of our Galaxy in 3-D, particularly its magnetic field, which plays a role in many astrophysical process over many scales \citep[e.g.][]{Haverkorn+2008,Ordog+2017,Haverkorn+2019}.

\subsubsection{Scattering and depolarization: PSR J1721--3532}\label{subsub:1721} 

%
Seven pulsars show visibly significant interstellar scattering effects, evidenced by the long exponential scattering tails in their pulse profiles at the lowest frequencies observed: PSRs J1326--5859, J1327--6222, J1357--62, J1359--6038, J1600--5044, J1644--4559; J1721--3532 is the most extreme example (see the supporting information). 
Their P.A. curves and fractional polarizations flatten towards the trailing edges of the profiles, likely due to the effect of the interstellar scattering \citep[e.g.][]{Komesaroff+1972,Li+2003,Karastergiou2009}.
PSRs J1327--6222, J1357--62, and J1644--4559 also show generally decreasing fractional linear polarization with decreasing frequency in Figure \ref{fig:fracLV}, also evident in their pulse profiles, where the fractional linear polarization drops to much smaller values (close to $\approx$0) in the scattering tail. 

Only PSR J1721--3532 shows complete depolarization across the entire on-pulse window at 971\,MHz, and shows the greatest decrease in fractional linear polarization with decreasing frequency in Figure \ref{fig:fracLV}. 
Of the pulsars in this census, PSR J1721--3532 has the largest DM, Table \ref{tab:obs}, and scattering time, 113.4$\pm$0.7\,ms at 1\,GHz (Oswald et al., submitted). 
Its position is coincident with an H\,\textsc{ii} region \citep[HRDS G351.691+00.669;][]{Anderson+2011} that is visible from radio \citep[21\,cm; e.g.][]{Tian+2007} to infra-red wavelengths \citep[8\,$\upmu$m;][]{Benjamin+2003}, which appears to be part of a high column density feature \citep{Russeil+2016}. This H\,\textsc{ii} region likely lies in front of the pulsar, since H\,\textsc{i} and OH absorption has been detected against PSR J1721--3532 \citep[][respectively]{Weisberg+1995,Minter+2008}. 
Intervening H\,\textsc{ii} regions have been seen to enhance the DM and RM towards pulsars \citep[e.g.][]{Mitra+2003}. 
This is expected, as the stellar winds from O and B stars expand and sweep up ISM material, increasing the local electron density and affecting the magnetic field through advection and compression  \citep[e.g.][]{Weaver+1977,Costa+2016}.
PSR J1721--3532 has a modest RM for its large DM, Table \ref{tab:obs}. This may be because the integrated net direction of the Galactic magnetic field lies closer to the plane of the sky than parallel to the LoS, or because there is at least one reversal in the net direction of the Galactic magnetic field along the LoS \citep[e.g.][]{VanEck+2011}. 

\begin{figure}
	\includegraphics[width=\columnwidth]{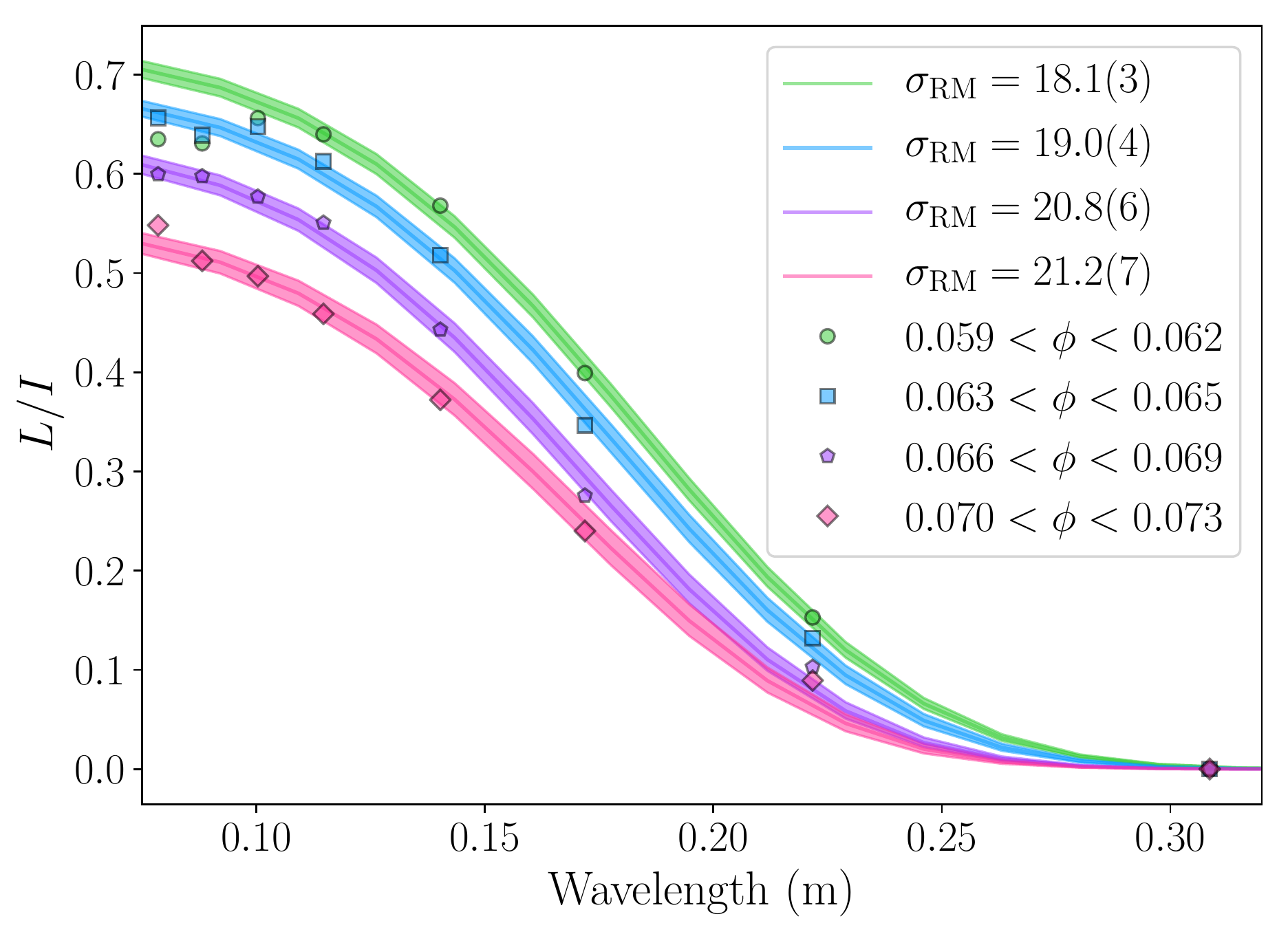}

    \caption{Fractional linear polarization measured for PSR J1721--3532 as a function of wavelength (points), and the best-fitting lines and uncertainties (shaded) using Equation \ref{eq:depol}. The legend shows the ranges in pulse phase, $\phi$, used to measure the fractional polarization and the $\upsigma_{\rm{RM}}$ in \,rad\,m$^{\rm{-2}}$ obtained from the fit and associated formal uncertainties (in brackets, corresponding to the least significant digit). The two shortest wavelength data points for 0.059$<\phi<$0.062 were not included in the fit, where $L/I$ is less than the maximum value reached (likely due to pulse profile evolution), because Equation \ref{eq:depol} assumes increasing $L/I$ values towards shorter wavelengths. }
    \label{fig:depol}
\end{figure}

This H\,\textsc{ii} region is a likely candidate for the location of the dominant interstellar scattering screen, with anisotropies in the electron density and magnetic field causing the depolarization of PSR J1721--3532's emission towards lower frequencies \citep[e.g.][]{Xue+2019}.
Furthermore, when measuring the RM towards PSR J1721--3532, incorporating larger ranges of on-pulse phse bins seems to produce larger RM measurements. For example, 16 bins produces RM=162.1$\pm$2.0\,rad\,m$^{\rm{-2}}$, while 49 bins produces RM=167.3$\pm$1.7\,rad\,m$^{\rm{-2}}$ (differing by 1.4$\upsigma$). 
The RM value shown in Table \ref{tab:obs}, 165.9$\pm$1.9\,rad\,m$^{\rm{-2}}$, uses 30 bins ($w_{50}$ across the peak in linear polarization. \citet{ijw19}, however, do not detect significant variations in RM with pulse phase across the profile, although somewhat of a trend is plausible. 
Moreover, RM CLEANing the Faraday spectrum down to a threshold of 1$\times$RMS produces a significant Faraday dispersion of 17\,rad\,m$^{\rm{-2}}$ (using 30 on-pulse bins), while not affecting the RM measurement. 
Although most pulsar data are expected to be `Faraday thin', pulsars with scattering tails show evidence of Faraday dispersion \citep[e.g.][]{sbg+19,Xue+2019}.
To further investigate this, we fit a depolarization model -- due to an external Faraday screen with a random magnetic field component with Faraday dispersion $\upsigma_{\rm{RM}}$ \citep[][and references therein]{Sokoloff+1998} -- to the mean fractional linear polarization measured over the 8 sub-bands (at wavelengths $\lambda$) across four ranges in on-pulse phase bins:
\begin{equation} 
L/I = (L/I)_{\rm{i}}\mathrm{exp[-2\upsigma_{RM}\lambda^{4}]},
  \label{eq:depol}
\end{equation}
where $(L/I)_{\rm{i}}$ is the intrinsic fractional linear polarization \citep[e.g.][]{OSullivan+2012,Xue+2019}. Figure \ref{fig:depol} shows the data points and best-fitting lines with uncertainties using Equation \ref{eq:depol}. 
We find that the Faraday dispersions obtained from the fitting (18--21\,rad\,m$^{\rm{-2}}$) are similar to that found using RM CLEAN. Also, the Faraday dispersions seem to increase towards later pulse phases, suggesting that increasingly larger areas of the anisotropic, magneto-ionic scattering screen are being probed towards later pulse phases in the scattering tail. This suggests that there are relatively large variations in the magnetic field and electron density within the scattering screen, likely located in the H\,\textsc{ii} region.

\subsection{Pulsar emission}\label{subsec:emn}

Figure \ref{fig:fracLV}, upper panel, summarises the on-pulse fractional linear polarization ($L/I$) behaviour, over the 8 frequency sub-bands, for all of the pulsars in this work. The range in fractional linear polarization for these pulsars' emission is $\approx$4--95 per cent (for PSRs J1136--5525 and J1709--4429, respectively). There are a handful of pulsars with consistently large fractional linear polarization across the bandwidth ($L/I>$84 per cent, although somewhat decreasing with frequency): PSRs J1709--4429, J0835--4510, and J0742--2822.  The majority of the pulsars are between $\sim$10--40 per cent linearly polarized and show decreasing linear polarization fraction with increasing frequency of $\approx$--7 per cent between $\approx$960--3820\,MHz, on average. This trend has been seen previously, and may be due to the higher-frequency emission propagating through a larger distance of the pulsar magnetosphere, according to radius-to-frequency mapping \citep[e.g.][and references therein]{Johnston+2005}.

A handful of pulsars show notably different behaviours to the general trend. 
PSRs J1327--6222, J1357--62, J1644--4559, and particularly J1721--3532 are likely affected by interstellar scattering, discussed in Section \ref{subsub:1721}. 
PSR J1048--5832 also shows unusual behaviour -- increasingly rapid depolarization with decreasing frequencies. This is likely intrinsic to the pulsar's emission, and is further discussed below.

Figure \ref{fig:fracLV}, middle and lower panels, show the on-pulse fractional circular, and absolute circular polarization behaviour with frequency. The colours used for each pulsar are identical in all plots. The fractional circular polarization for these pulsars' emission ranges from $\approx50$ per cent right-handed to $\approx40$ per cent left-handed (for PSRs J1740--3015 and J1600--5044, respectively). On average, the pulsars have a circular polarization degree of $\approx$16\,per cent. 
They also show a general trend of increasing absolute circular polarization with increasing frequency by $\approx$4 per cent between $\approx$960--3820\,MHz, on average. A linear increase in circular polarization with increasing frequency can be expected from a relativistic effect called aberrated backward circular polarization \citep[ABCP;][]{Melrose+2004}, or from `conversion' from linear to circular polarization \citep[][]{vonHoensbroech+1999}, but has only been reported for some pulsars observations \citep[e.g.][]{Han+1998,Johnston+2006}.  

There are a handful of notable pulsars in the circular polarization plots in Figure \ref{fig:fracLV}.
In particular, PSR J1740--3015 shows the largest absolute circular polarization that increases dramatically with frequency. PSRs J1224--6407, J1709--4429, J0837--4135, J1829--1751 (in order of absolute circular polarization at the lowest frequency band) show the greatest declines in absolute circular polarization (between 5--14 per cent) with increasing frequency. This generally appears to be due to more complex pulse profile evolution than average. PSR J1600--5044 also shows interesting non-linear increase and decline in circular polarization across this frequency range, further discussed below. 

We further discuss the pulse profiles for all of the pulsars in the census, shown in the supporting information, individually below. In general, pulse profiles are known to be unique, like a fingerprint, which holds true for the pulsars we study here. Many of the pulse profiles also show magnetospheric and ISM effects (e.g. interstellar scattering, discussed in Section \ref{subsub:1721}).
We found that almost one quarter (9) show significant profile evolution, where the dominant pulse profile component (with the highest peak flux density) at the highest frequency is not equivalent to that at the lowest frequency observed (PSRs J1224--6407, J1326--5859, J1326--6700, J1327--6222, J1357--62, J1559--4438, J1600--5044, 1807--0847, J1829--1751).
In particular, three of these pulsars show somewhat similar behaviour with frequency, where two outer pulse profile components are dominant at higher frequencies (with the peak in flux density occurring at the trailing component) and the inner component is dominant at the lower frequencies (PSRs J1224--6407, J1357--62, J1829--1751).  
We also found 4 pulsars with relatively little pulse profile evolution across the frequency range (excluding the effect of spectral index; PSRs J1605--5257, J1709--4429, J1852--0635, J2048--1616).
Six of the pulsars show a classical `S'-shape P.A. curve, as predicted by the RVM model, PSRs J0536--7543, J0835--4510, J1048--5832, J1600--5044, J1740--3015, J2048--1616. 
Almost half of the pulsars (19) show least one OPM jump, these are described for each pulsar below.
Four pulsars show complex P.A. curves, part of which comprises a swing whose sense changes direction within the observed frequency range: PSRs J1224--6407, J1453--6413, J1456--6843, J1900--2600.
Almost one quarter (9) pulsars show greater fractional circular polarization than linear at some pulse phases (PSRs J0837--4135, J1224--6407, J1326--5859, J1456--6843, J1600--5044, J1740--3015, J1752--2906, J1900--2600, J2048--1616). 



\textbf{PSR B0031--07 (J0034--0721):}
This pulsar is known to display a range of emission phenomena, including nulling, mode-changing, and bright pulses \citep[e.g.][]{Huguenin+1970,Weltevrede+2007,Karuppusamy+2011,McSweeney+2017,ijw19}.
Therefore, our time-averaged pulse profile likely shows the aggregate of all of its emission modes, dominated by the prevailing mode `B' defined in previous work. 
The UWL data shows that PSR J0034--0721 has some notable pulse profile evolution, particularly in linear polarization. At lower frequencies, there are two obvious components in linear polarization, with the latter showing the larger peak flux density, and two orthogonal, $\approx$90$^{\circ}$ jumps in the P.A.s between these components. Towards the higher frequencies, the earlier linear polarization component dominates and the orthogonal P.A. jumps are no longer visible. This is consistent with previous studies \citep[e.g.][]{Johnston+2008,Noutsos+2015}. This pulsar's emission reaches over 25 per cent linear polarization fraction and 10 per cent. It is also consistent with the general trends of decreasing linear polarization and increasing circular polarization with increasing frequency.
Although some RM variations with pulse phase have been seen \citep[][]{ijw19}, interstellar scattering effects may be the primary cause. 

\textbf{PSR B0538--75 (J0536--7543):}
This pulsar is approximately 50 per cent linearly polarized across much of the pulse profile at the lower frequencies and reaches a circular polarization fraction of 40 per cent. It shows a well-defined and classic `S'-shape swing in P.A. across most of the observed frequency range. Previous studies at lower frequencies suggest that the somewhat unusual long leading edge may be the combination of a number of pulse profile components \citep[e.g.][]{Manchester+1998}, although we do not see a swing in the handedness of circular polarization, only right-handed circular polarization across the profile. This pulsar's emission is also consistent with the general trends of decreasing linear polarization and increasing absolute circular polarization with increasing frequency. Although the profile is largely consistent with \citet{ijw19}, we do not see an OPM jump at the trailing edge. 

\textbf{PSR B0540+23 (J0543+2329):}
Showing a single dominant pulse profile component in total intensity, this pulsar is approximately 50 per cent linearly polarized across much of the pulse profile and up to approximately 25 per cent right-handed circularly polarized.
It also follows the general trends of decreasing linear polarization and increasing absolute circular polarization with increasing frequency.
The pulse profile obtained is consistent with previous works at lower and higher frequencies \citep[e.g.][]{Weisberg+1999,Weisberg+2004}, and suggests there is little pulse profile evolution even outside the frequencies used in this work. However, interstellar scattering is evident towards lower frequencies and the pulsar is associated with the supernova remnant IC 443 \citep[e.g.][]{Geyer+2017}. 

\textbf{PSR B0628--28 (J0630--2834):}
Showing a single peak in the total intensity pulse profile, made up of several blended components, this pulsar reaches over 60 per cent linear polarization at the lower frequencies, but this decreases more rapidly than average with increasing frequency, by over 35 per cent. It is also shows modest right-hand circular polarization of approximately 5 per cent. 
There is a smooth and almost linear P.A. curve, with a hint of separate linear polarization components in the leading component at the higher frequencies, consistent with previous studies \citep[e.g.][]{Johnston+2005,Johnston+2006,Johnston+2008}.

\textbf{PSR B0736--40 (J0738--4042):}
The pulse profile for PSR J0738--4042 is complex, with several linear polarization components, up to five $\approx$90$^{\circ}$ OPM jumps, and up to five changes in handedness in the circular polarization, depending on the observing frequency. The fractional linear polarization follows the general trend of decreasing with increasing frequency, although there is little change in the $\approx$6 per cent absolute polarization fraction. This is consistent with previous work \citep[e.g.][]{Johnston+2006,Johnston+2007}, which also shows dramatic interstellar scattering at lower frequencies \citep[e.g.][]{Johnston+2008}.
Systematic changes in the leading components of the pulse profile over hundreds of days have been observed, likely due to competing orthogonal polarization modes \citep[e.g.][]{Karastergiou+2011,Brook+2016}; and RM variations across the pulse profile, with a magnetospheric origin, have been reported \citep{ijw19}.

\textbf{PSR B0740--28 (J0742--2822):}
PSR J0742-2822 is highly linearly polarized, reaching 100 per cent across much of the pulse profile at all frequencies observed. The trailing linear polarization component is approximately 50 per cent polarized at the lowest frequencies observed, decreasing to $\approx$0 at the highest frequencies. The circular polarization fraction also reaches $\approx$10 per cent (right-handed) towards the profile's trailing edge and also increases with frequency. 
This is consistent with previous findings \citep[e.g.][]{Johnston+2005}. 
The pulsar also shows two distinct emission modes, which are correlated with glitch events \citep{Keith+2013}.
PSR J0742--2822 is also notably affected by interstellar scintillation and scattering, caused by the intervening Gum Nebula ISM structure \citep[e.g.][]{Johnston+1998,Johnston+2008,Geyer+2017,Xue+2019}. 
 

\textbf{PSR B0833--45 (J0835--4510):}
Associated with the Vela supernova remnant and commonly known as the Vela pulsar, PSR J0835--4510 is a bright, energetic, well-studied young pulsar known to emit from radio to $\gamma$-rays and exhibit glitches \citep[e.g.][]{Palfreyman+2018}. Its emission is highly polarized: reaching approximately 100 per cent linear polarization fraction across most of its pulse profile and the observed frequency range; and over 20 per cent (right-handed) fractional circular polarization fraction across much of the frequency range, also increasing with increasing frequency.  The linear (circular) polarizations likely decrease (increase) towards even higher frequencies \citep[17 and 24\,GHz][]{Keith+2011}.
The P.A. swing is also a well-defined and classic `S'-shape, although with some possible deviations at the pulse profile edges at some frequencies.
Consistent with previous studies, the pulse profile shows two dominant components, with the leading component increasing in relative flux density and dominating towards the lower frequencies, while the trailing component increases in relative flux density and dominates the profile towards higher frequencies \citep[][]{Johnston+2005,Johnston+2008}.

\textbf{PSR B0835--41 (J0837--4135):}
The relatively steep spectral index \citep[--1.8$\pm$0.2 above 740$\pm$20\,MHz][]{Jankowski+2018} is evident in the UWL data. The central main pulse component dominates the flux density at the lowest frequencies, while the pre- and post-cursor peaks increase in intensity towards higher frequencies, consistent with previous studies \citep[e.g.][]{Johnston+2006,Johnston+2007}.
Despite its relatively simple total intensity profile, PSR J0837--4135 shows more complex linear and circular polarization. 
What seems to be an $\approx$90$^{\circ}$ OPM jump in the leading section of the pulse profile at the lower frequencies, decreases in value and turns into a steep $\approx$50$^{\circ}$-`S'-shape towards the higher frequencies. The linear polarization fraction increases in the pre- and postcursors, showing little evolution with frequency. 
The circular polarization reaches a fraction of over 25 per cent (left-handed), with a swing to $\approx$10 per cent right-handed, and back to left-handed in the postcursor.

\textbf{PSR B0905--51 (J0907--5157):}
This pulsar has a steadily increasing leading component in the main pulse. There is also a relatively low flux-density precursor, which is separate from the main pulse and unpolarized at the higher frequencies and joins with the main pulse and is over 10 per cent polarized at the lower frequencies. The polarization of the precursor component at the lower frequencies does not seem to have been detected previously \citep[e.g.][]{ijw19}.
The profile reaches over 50 per cent fractional linear polarization and generally decreases with increasing frequency. 
The P.A. swing becomes shorter but steeper with increasing frequency. 
We measured RMs separately for the leading and trailing main pulse components because \citet{ijw19} reported changes in RM with pulse phase, although \citet{Noutsos+2009} did not. 
The RM for the leading component (--24.56$\pm$0.95\,rad\,m$^{\rm{-2}}$) is reported in Table \ref{tab:obs} because this corresponds to the peak in fractional linear polarization. 
The RM for the trailing component was measured to be --26.0$\pm$0.9\,rad\,m$^{\rm{-2}}$.
These RMs are in good agreement within uncertainties (0.8$\upsigma$) and de-Faraday rotating the linear polarization using the former RM measurement results in the highest average polarization fraction.
The fractional circular polarization reaches over 20 per cent (left-handed) and switches to right-handed in the trailing component, and the absolute fraction increases with increasing frequency.

\textbf{PSR B0940--55 (J0942--5552):}
This pulsar, known to null \citep[e.g.][]{Biggs+1992}, shows three pulse profile components. All of these are linearly polarized, reaching a fraction of over 75 per cent, which decreases towards later pulse phases. The P.A.s show an $\approx$90$^{\circ}$ OPM jump after the leading component, and also shows an inflection of $\approx$30$^{\circ}$ prior to the trailing component. Only the middle component is right-hand circularly polarized, reaching a fraction of over 10 per cent. 
Significant pulse profile evolution is evident, with the middle component dominating the flux density at the lowest frequencies, consistent with previous studies \citep[e.g.][]{Karastergiou+2006}, and the outer components becoming equivalent in flux density at the highest frequencies. The relatively steep spectral index \citep[--2.3$\pm$0.1 above 1100$\pm$200\,MHz;][]{Jankowski+2018} is also evident, with one of the greatest differences in peak flux density between the highest frequency sub-band and the average profile, for the pulsars in this census.
Despite this, the polarization characteristics show relatively little evolution over the observed frequency range.

\textbf{PSR B1046--58 (J1048--5832):}
The UWL data shows that this pulsar's profile has a trailing, approximately unpolarized component that becomes increasingly prominent at lower frequencies, compared to at least three earlier pulse profile components that are otherwise close to 100 per cent linearly polarized, consistent with previous studies \citep[e.g.][]{Johnston+2005,Johnston+2006}. The P.A. curve shows a classical `S'-shape swing \citep{rc69}. The profile also shows left-handed circular polarization, the fraction of which increases with frequency and reaches up to $\approx$20 per cent.   
This pulsar is the third most energetic in this census ($\dot{E}$=2$\times$10$^{36}$\,erg\,s$^{\rm{-1}}$; \citealt{Manchester+2005}).
Periodic, short-timescale mode-changing in this pulsar's emission has been reported by \citet{Yan+2020}. \citet{Johnston+2021} also note that although a sufficient number of pulses are collected per observing epoch \citep[almost 1500; e.g.][]{Liu+2012}, the pulse profile is not stable when comparing epochs. 

\textbf{PSR B1054--62 (J1056--6258):}
This pulsar has a slowly decreasing trailing section of the pulse profile and appears to consist of at least three pulse profile components that are blended together. This is particularly evident in the variations in fractional linear polarization across the profile, reaching over 50 per cent at the lower frequencies, but decreasing with increasing frequency. At the three highest frequencies, the P.A. curve shows evidence of an $\approx$90$^{\circ}$ OPM jump due to a leading orthogonal polarization mode, consistent with previous studies \citep[e.g.][]{Karastergiou+2006}. The right-hand circular polarization fraction increases with increasing pulse phases and towards lower frequencies, reaching almost 10 per cent. Some pulse profile evolution with frequency is evident, with the leading peak component becoming relatively more dominant towards lower frequencies. As with other pulsars with notable frequency evolution, this may have implications for measuring the (absolute) DM, as discussed in the previous Section \ref{subsec:ISM}, particularly for this pulsar \citep[see also][]{Oswald+2020}.

\textbf{PSR B1133--55 (J1136--5525):}
This pulsar also appears to have a pulse profile consisting of at least four components blended together, with the bridge emission to the trailing component becoming more prominent with decreasing frequencies. There is little significant linear polarization at the highest frequencies, while at the lower frequencies the profile shows larger fractions towards the edges -- up to $\approx$50 per cent. 
There is also evidence for three $\approx$90$^{\circ}$ OPM jumps at 1356\,MHz, with the P.A.s transitioning between two seemingly parallel, approximately linearly increasing tracks. 
This increases the number previously seen, but otherwise the profile shows comparable characteristics to previous studies \citep[e.g.][]{ijw19}.
The profile is also not significantly circularly polarized, except at the lowest two frequencies, where the right-handed circular polarization fraction reaches $\approx$5 per cent towards the leading edge. 

\textbf{PSR B1221--63 (J1224--6407):}
This pulsar has a pulse profile consisting of at least three overlapping components, with notable evolution across the frequency range observed. The two outer total intensity components are prominent at the highest frequencies, with the trailing component most dominant. The inner, middle component increasingly dominates the flux density towards the lower frequencies while the trailing component fades relatively quickly. 
There is a small amount of right-handed circular polarization before swinging to $\approx$50 per cent left-handed fraction in the middle component at the lowest frequencies (greater than in linear polarization).
The P.A. curve becomes more complex towards higher frequencies, with small swings at the lower frequencies turning into $\approx$30$^{\circ}$ and $\approx$80$^{\circ}$ jumps at the highest frequencies.  The leading component shows the highest peak flux in linear polarization across the majority of the higher frequencies, but is narrower than the trailing components. We computed the RM for the leading linear polarization component, --3.4$\pm$1.2\,rad\,m$^{\rm{-2}}$. More bins can be summed for the trailing linear polarization component, and produces a higher S/N in the Faraday spectrum with RM=---6.8$\pm$0.6\,rad\,m$^{\rm{-2}}$, reported in Table \ref{tab:obs}. This is 1.8-$\upsigma$ different compared to the leading component, and is also discrepant with the psrcat value (3$\upsigma$), which is in better agreement with smaller absolute RM from the leading component (within 0.1-$\upsigma$).  PSR J1224--6407 was not included in the study of RM variations with pulse phase in \citet{ijw19}, and may be interesting for further follow-up to investigate the magnetospheric radio emission beam, using both the UWL data and possibly at lower-frequency ($<$300\,MHz) observations using the Murchison Widefield Array \citep[e.g.][]{Tingay+2013,Tremblay+2015,Xue+2019}. 

\textbf{PSR B1240--64 (J1243--6423):}
This pulsar's pulse profile consists of at least two total intensity components and at least three polarized components blended together in the main pulse, with little evolution with frequency. There is also a precursor component that does show greater evolution with frequency, becoming increasingly significant towards the higher frequencies, also seen in previous studies \citep[e.g.][]{Karastergiou+2006,Johnston+2006}. The P.A. shows an approximately `S'-shaped swing across the main pulse, with some deviations, particularly at the higher frequencies. 
There is an $\approx$90$^{\circ}$ OPM jump between the precursor (with approximately flat P.A.s) and the main pulse. The precursor shows the highest fractional linear polarization, reaching $\approx$90 per cent. The integrated main pulse reaches over 40 per cent fractional linear polarization at the leading edge, and decreases with increasing frequency. The circular polarization swings, with deviations, from over 25 per cent right-handed (negative) circular polarization fraction to over 20 per cent left-handed (positive), with the absolute fractional circular polarization increasing with increasing frequencies, particularly at the leading edge. The precursor is not significantly circularly polarized.
In addition, PSR J1243--6423 shows an unusually broad pulse energy distribution \citep{Burke-Spolaor+2012}, $\approx$2 per cent nulling fraction \citep[e.g.][]{Biggs+1992,Wang+2007}, and significant RM variations with pulse phase \citep[][]{Noutsos+2009,ijw19}. 

\textbf{PSR B1323--58 (J1326--5859):} 
This pulsar has a total intensity pulse profile that consists of a pre-cursor component, at least two blended components in the main pulse, and a postcursor component. The bridge emission between the precursor and the main pulse becomes more significant towards lower frequencies. 
The postcursor emission becomes washed out by the exponential interstellar scattering tail visible in the lowest two frequencies (954 and 1352\,MHz). 
The scattering timescale at 1\,GHz was estimated to be 9.5\,ms \citep{Lewandowski+2015}. 
The precursor becomes the most prominent component at the highest frequencies, which continues above 6\,GHz \citep[][]{Johnston+2006}. 
The P.A. curve is somewhat complex, particularly at the higher frequencies. The P.A.s are flat across the precursor (similar to PSR J1243--6423), after which there is a $\approx$90$^{\circ}$ OPM jump. The P.A. curve across the main pulse is steep and shows an approximate `S'-shape with a small inflection between the linearly polarized components. 
Between the main pulse and postcursor, there is another $\approx$90$^{\circ}$ OPM jump, where the P.A.s one again become approximately flat with pulse phase. At the lowest frequencies, the main pulse P.A. curve becomes flatter, likely due to the effect of interstellar scattering, as discussed in Section \ref{subsub:1721}.
Several linearly polarized components are evident in the fractional linear polarization plots. 
We measured RMs using both the precursor and main pulse data, and these agree within 1$\upsigma$ uncertainties. The narrow precursor component, using 15 summed on-pulse bins, provides an RM=--52.5$\pm$0.8\,rad\,m$^{\rm{-2}}$. The peak linear polarization in the main pulse, using 12 on-pulse bins, provides RM=--53.7$\pm$0.7\,rad\,m$^{\rm{-2}}$, reported in Table \ref{tab:obs}.
The main pulse circular polarization shows a classical swing in handedness across pulse phase, expected of emission from a `core' component of the radio beam \citep[e.g.][]{Radhakrishnan+1990}, reaching $\approx$30-50 per cent right-handed fractions at early phases and then $\approx$20-40 per cent left-handed fractions at later phases.

\textbf{PSR B1322--66 (J1326--6700):}
The total intensity pulse profile evolves considerably with frequency -- at the highest four frequencies it is double peaked, and the leading component has the highest flux density, while towards lower frequencies the profile widens, a third intervening component emerges, and the trailing component grows dominant. The linear polarization shows two major components, demonstrated in the P.A. curve with an $\approx$90$^{\circ}$ OPM jump, corresponding to the local $L/I$ minimum. The fractional linear polarization reaches over 75 per cent (at the lowest frequencies) in the leading component and $\approx$50 per cent in the latter component. The fractional circular polarization is over 15 per cent (right-handed) and the peak tends to earlier pulse phases for lower frequencies and later phases for higher frequencies, compared to the average pulse profile. 
This pulsar is known to show magnetospheric phenomena including nulling \citep[e.g.][]{Wang+2007} and mode-changing. The pulse profile presented in the additional information resembles the `normal' emission mode seen in \citet{Wen+2020}.

\textbf{PSR B1323--62 (J1327--6222):} 
This pulsar's pulse profile is double-peaked with at least four blended components at the highest frequencies. The leading component has the highest flux density (and continues to dominate above 8\,GHz; \citealt{Johnston+2006}), while at the lowest frequencies, the trailing component dominates the total intensity and all components merge into an almost single-peaked profile with an exponential scattering tail. \citet[][]{Lewandowski+2015} estimated a scattering time of 2.4\,ms at 1\,GHz.
There is a complex variation in P.A.s with pulse phase, with two OPM jumps: between the first two linear polarization components the jump increases from $\approx$40$^{\circ}$--90$^{\circ}$; between the final two polarization components the jump decreases from $\approx$90$^{\circ}$--60$^{\circ}$, with increasing frequency (excluding the lowest frequency that is most affected by scattering).
As previously discussed in Section \ref{subsub:1721}, the fractional linear polarization also shows the effect of interstellar scattering. The leading feature is over 50 per cent linearly polarized while the trailing two features, $\approx$25 per cent polarized at the higher frequencies, are `washed out' and show little ($<\approx$5 per cent) linear polarization. 
Consequently, the fractional linear polarization in the leading component follows the general trend of decreasing with increasing frequency, while this is the opposite for the trailing components. 
The modest circular polarization, swinging between $\approx$10 per cent right-handed and left-handed fractional polarization, also seems to show this scattering effect, with the lowest frequency remaining constant at $\approx$5 per cent over much of the trailing pulse phases. 
This pulsar was detected as right-hand circularly polarized at 200\,MHz using the MWA \citep[][]{Lenc+2018}, suggesting that interstellar scattering may affect the emission more dramatically at lower frequencies and `level-off' the circular polarization at even earlier pulse phases so it remains right-hand circularly polarized across the entirety of the pulse. 

\textbf{PSR B1353--62 (J1357--62):} 
This pulsar shows significant pulse profile evolution, orthogonal polarization modes, and an interstellar scattering exponential tail. The total intensity pulse profile shows three significant components, with the outer two higher in flux density (and the trailing component dominating) at the higher frequencies, and the middle component becoming the highest in flux density towards the lowest two frequencies (somewhat similar to PSR J1224--6407). There are also four significant linear polarization components, demonstrated by the three $\approx$90$^{\circ}$ OPM jumps between them, corresponding to the local $L/I$ minima, and each reaching a peak in fractional linear polarization over approximately 25 per cent. The circular polarization also swings between $\approx$25 per cent left-handed and right-handed circular polarization. \citet{ijw19} report two of the orthogonal polarization mode jumps and detect changes in RM with pulse phase, although cannot rule out that this is due to interstellar scattering.

\textbf{PSR B1356--60 (J1359--6038):} 
This pulsar has a relatively simple pulse profile, consisting of one dominant component that reaches up to almost 100 per cent fractional linear polarization and almost 50 per cent left-handed circular polarization. 
The profile remains significantly linearly and circularly polarized (although less so) at higher frequencies \citep[$>$8\,GHz][]{Johnston+2006}. 
The P.A. curve is concave -- relatively flat at the leading edge and relatively steep at the trailing edge.
The profile is evidently affected by interstellar scattering and shows an exponential tail at lower frequencies (680\,MHz; Simon Johnston, unpublished), while \citet{Lewandowski+2015} estimate the scattering time to be 1\,ms at 1\,GHz. 
The effects of interstellar scattering can be seen at the lowest frequency in the UWL data -- the P.A. curve, fractional linear and fractional circular polarizations become flatter towards later pulse phases, as discussed in Section \ref{subsub:1721}.
\citet{Noutsos+2009} and \citet{ijw19} find RM variations with pulse phase towards the trailing edge of the profile, possibly due to magnetospheric effects, although scattering may also play a role.
\citet{Brook+2016} also find variations in the pulse profile over time with a short-timescale change in the profile shape that approximately coincides with a drop in the spin-down rate.

\textbf{PSR B1426--66 (J1430--6623):}
This pulsar's total intensity pulse profile has a slowly rising leading edge that increases in flux density relative to the main pulse component and evolves into a clearly separate component towards lower frequencies, and is not visible at $>$8\,GHz \citep{Johnston+2006}.
The fractional linear polarization reaches up to $\approx$50 per cent at the minimum between leading and main total intensity components. The P.A. curve is somewhat complex, with a large range and an $\approx$90$^{\circ}$ OPM jump between the two trailing components. 
The circular polarization also shows a swing from right-handed to left-handed, coincident with the main pulse, reaching a fraction of $\approx$20 per cent towards the trailing edge.

\textbf{PSR B1449--64 (J1453--6413):}
The pulse profile is a combination of at least two total intensity pulse components blended together and shows significant evolution with frequency in linear polarization. There are three distinct linear polarization components associated with the leading, peak, and trailing sections, evident in the P.A. curve and fractional linear polarization. These components are up to $\approx$100, 40 (at the lowest frequencies), and 40 per cent linearly polarized, respectively. While the leading component remains highly polarized at all frequencies, the middle, and trailing components depolarize with increasing frequency, although the middle component more rapidly than the latter. At the five lowest frequencies, there is an $\approx$90$^{\circ}$ OPM jump, which occurs later in pulse phase with increasing frequency, corresponding to the evolution in linear polarization where the most dominant component shifts from that associated with the peak to that associated with the leading component.  This appears to change sense and decrease to $\approx$40$^{\circ}$ at the higher frequencies. There is a second $\approx$90$^{\circ}$ OPM jump between the peak and trailing edge that is more consistent in position and sense, but still decreases with higher frequencies to $\approx$40$^{\circ}$. 
The circular polarization swings from right-handed to left-handed and back across the profile, reaching a maximum of over 10 per cent in both directions.
This pulsar is known to display nulling \citep[e.g.][]{Burke-Spolaor+2012}. RM variations with pulse phase were also reported and may, in part, be due to magnetospheric effects \citep[][]{Noutsos+2009,ijw19}.

\textbf{PSR B1451--68 (J1456--6843):} 
This pulsar has an approximately triangular total intensity pulse profile at most frequencies, clearly a blend of at least 4 components, also shown by the P.A.s and fractional linear (reaching over 30 per cent) and circular polarizations. The P.A. curve is relatively complex, evolving with frequency, and with an $\approx$90$^{\circ}$ OPM jump in the leading edge at the lowest four frequencies. The $\approx$70$^{\circ}$ `S'-shape swing at the leading edge after the OPM jump reverses direction between 2615\,MHz and 2988\,MHz. 
This may reflect the two polarization modes of similar strength, which produce two clear orthogonal polarization angle tracks when observed in single pulses \citep[][]{Dyks+2020}.
The circular polarization fraction swings between $\approx$15 per cent right-handed and $\approx$15 per cent left-handed. 
The absolute circular polarization at the outer edges of the profile increases with increasing frequency (following the general trend), while the central left-handed peak's absolute circular polarization increases towards lower frequencies.
This pulsar is reported to null \citep[][]{Biggs+1992} and show atypically broad pulse energy distributions \citep[][]{Burke-Spolaor+2012}. RM variations across pulse phase of $\approx$20\,rad\,m$^{\rm{-2}}$ were also detected, likely due to magnetospheric effects \citep[][]{ijw19}.

\textbf{PSR B1556--44 (J1559--4438):}
The total intensity pulse profile consists of a precursor and a main pulse comprised of at least four blended components. Significant profile evolution is evident: the main pulse dominates the flux density at the lowest frequencies, while towards high frequencies the ratio between the precursor and main pulse peak flux density approaches 1. The fractional linear polarization reaches almost 100 per cent at the profile edges. The fractional circular polarization also noticeably evolves with frequency; the maximum in right-hand circularly polarization, up to $\approx$20 per cent, shifts to later pulse phases with increasing frequency. The P.A.s show a steep $\approx$70 per cent swing between leading linear polarization components (where the fractional linear polarization reaches its first minimum), and an $\approx$90$^{\circ}$ OPM jump prior to the trailing edge (where the fractional linear polarization reaches its final minimum, and shifting towards earlier pulse phases with increasing frequency). Also, a small downward inflection at the latter part of the central P.A. slope grows in amplitude and moves towards later pulse phases towards the higher frequencies, somewhat merging with the OPM jump.
Although significant changes in RM with pulse phase were detected in \citet{ijw19}, these are likely caused by interstellar scattering \citep[e.g.][]{Johnston+2008}, which may also act to smooth out features in the pulse profile towards lower frequencies, as seen in the UWL data.

\textbf{PSR B1557--50 (J1600--5044):} 
The total intensity pulse profile shows some evolution, with a double peak visible at and above 1745\,MHz \citep[which continues $>$8\,GHz;][]{Johnston+2006}, which merges into a single peak with an exponentially decreasing tail at the lowest frequency, due to interstellar scattering.
The scattering time was estimated to be 5.5\,ms at 1\,GHz \citep[][]{Lewandowski+2015}.
The linear polarization (reaching over $\approx$25 per cent fraction) also evolves with frequency, with the peak drifting to earlier pulse phases with decreasing frequency. This behaviour is also seen in circular polarization, which is greater than $L$ at most pulse phases and at all frequencies, also seen previously \citep[e.g.][]{jk18}. PSR J1600--5044 shows unusual non-linear behaviour in circular polarization across the observing band, Figure \ref{fig:fracLV}. $V/I$ peaks at $\approx$50 per cent the two highest frequencies, increases to $\approx$60 per cent in the intermediate frequencies, and decreases again and levels-off at $\approx$30 per cent at the lowest frequency.
This appears to be caused by the pulse profile evolution towards the higher frequencies, and interstellar scattering effects towards the lower frequencies.
This pulsar also shows a classical `S'-shape P.A. swing at most frequencies, although declining in amplitude and flattening towards the lowest frequencies, also likely due to the effect of the scattering.
Flux density variations are also observed and are likely to be largely due to refractive scintillation \citep[e.g.][]{Brook+2016}, with a timescale longer than 769$\pm$77\,days \citep{Kumamoto+2021}.

\textbf{PSR B1601--52 (J1605--5257):}
Although the pulse profile profile is somewhat complex, including at least four blended total intensity components, it shows little evolution with frequency.
The P.A.s show three $\approx$80$^{\circ}$--100$^{\circ}$ OPM jumps at the lowest three frequencies, while the leading polarization component is not detected towards the higher frequencies, which show the latter two OPM jumps, consistent in angle and pulse phase with the lower frequencies.  
We measured RMs for the three linear polarization components with the highest flux densities, and all of the measurements agreed within the uncertainties. This is consistent with previous studies that find no significant RM variations across the pulse profile \citep{ijw19}. The RM reported in Table \ref{tab:obs} is the value obtained using 35 on-pulse bins across the peak in linear polarization, which reaches a fraction of over 60 per cent. 
The circular polarization appears to show the greatest change with frequency, which declines relatively rapidly with frequency, peaking at $\approx$20 per cent at 964\,MHz and $\approx$10 per cent at 1745\,MHz.

\textbf{PSR B1641--45 (J1644--4559):}
This pulsar shows a relatively simple total intensity pulse profile, with a single-peak plus a weak precursor, which becomes more prominent towards higher frequencies, a trend that continues beyond 8\,GHz \citep[e.g.][]{Johnston+2006,Keith+2011}. An exponentially decreasing scattering tail is seen at the lowest frequency. The scattering time was estimated to be 11\,ms at 1\,GHz \citep[e.g.][]{Lewandowski+2015}.
The precursor is highly polarized (reaching $\approx$90 per cent), and there is an $\approx$90$^{\circ}$ OPM jump between this and the main pulse just before $\approx$0.3 pulse phase. 
The generally positive slope of the P.A.s across the main pulse are interrupted by two positive inflections coincident with the leading linear polarization component's peak, and in the tail of the trailing component. 
The edges of the main pulse show the highest polarization fractions of over 60 per cent.
The circular polarization is relatively complex, with three changes in handedness, starting right-handed and ending left-handed, reaching over 20 per cent in both hands at the highest four frequencies.
The amplitudes of the fractional linear and circular polarizations, inflections in the P.A. curve, and the OPM jump decrease towards the lowest frequency, likely due to the effect of interstellar scattering, as discussed in Section \ref{subsub:1721}. 
RM variations with pulse phase have been detected, but may also be influenced by scattering \citep[][]{Noutsos+2009,ijw19}. 

\textbf{PSR B1706--16 (J1709--1640):}
This pulsar's pulse profile is also relatively simple and shows modest frequency evolution. There is a single peak in total intensity, likely at least two blended components, with one dominant stationary component and one component that moves from leading pule phases at lower frequencies \citep[consistent with $\leq$322\,MHz][]{Johnston+2008} to trailing pulse phases at higher frequencies.
This is the only pulsar that shows an obvious deviation from the total flux density decreasing with increasing frequency (i.e. above the average profile total intensity at 966 and 1364, and 2614 and 2988\,MHz). Since this pulsar has a relatively low DM (24.885$\pm$0.003\,rad\,m$^{\rm{-2}}$, Table \ref{sec:obs}), this is likely due to diffractive scintillation \citep[e.g.][]{Kumamoto+2021}. 
The linear polarization reaches a fraction of $\approx$15 per cent just before the peak in total intensity and increasing even further towards the trailing edge. 
In circular polarization there is a swing from left- to right-handedness, and back, reaching a fraction of $\approx$10 per cent in the trailing edge. 
There is a clear $\approx$90$^{\circ}$ OPM jump corresponding to the trailing linearly polarized component at $\approx$0.965 pulse phase. There is also a smaller $\approx$40$^{\circ}$ jump prior to this at $\approx$0.96 pulse phase visible from $\approx$2139--3400\,MHz, corresponding to the minimum in fractional linear polarization, although this appears to be smoothed at the lowest two frequencies, where this $L/I$ minimum is less extreme. 
This pulsar exhibits nulling on short and long timescales \citep[e.g.][]{Burke-Spolaor+2012,Naidu+2018,Wang+2020}.
PSR J1709-1640's emission was visible in all observing epochs used in this paper, except one. 
On 2019 Nov 3, emission was not detected and the pulsar was likely nulling for at least the duration of the 3.4-min observation. The epoch where the pulsar was nulling was therefore not included in the time-added average pulse profiles produced and analysed in this work. 

\textbf{PSR B1706--44 (J1709--4429):}
The pulse profile is relatively simple and shows little evolution with frequency, with single-peaked total intensity, linear and circular polarization components.  
Of the pulsars studied in this work, PSR J1709--4429 is the most highly linearly polarized across the entire frequency range observed ($L/I>$95 per cent, Figure \ref{fig:fracLV}). The leading edge shows the highest fractional polarization, and remaining so at even higher frequencies \citep[$>8$\,GHz;][]{Johnston+2006}. The pulsar is also highly circularly polarized ($\approx$22 per cent), which peaks towards the trailing edge of the profile. The P.A. curve smoothly increases across the profile with a small curvature. 
For pulsars in this set, PSR J1709--4429 has the second highest spin-down luminosity, $\dot{E}=3.4\times 10^{36}$\,erg\,s$^{-1}$, and the second youngest characteristic age $\tau=1.75\times10^{4}$\,years, after the Vela pulsar, and is also known to glitch \citep[e.g.][]{Yu+2013,Lower+2018}.

\textbf{PSR B1718--35 (J1721--3532):} 
This pulsar's pulse profile is dramatically affected by interstellar scattering, evidenced by the significant exponential scattering tail and depolarization at the lowest three frequencies, discussed in Section \ref{subsub:1721}.
At higher frequencies, the profile consists of at least two blended components.  The leading component shows the highest linear and circular polarization ($\geq$60 per cent and $\geq$10 per cent fractions, respectively, at the highest four frequencies). The P.A. curve rapidly descends over 100$^{\circ}$ with a slight curve, decreasing in amplitude and becoming smoother at 1352-2137\,MHz, and is not significant at the lowest frequency due to the depolarization. 
 
\textbf{PSR B1727--47 (J1731--4744):}
This somewhat complex pulse profile consists of at least 4 blended components in total intensity, showing little pulse profile evolution with frequency, except for the outer two components becoming increasingly prominent towards lower frequencies. The linear polarization fraction peaks in the leading component, has an average fraction of $\approx$15 per cent across the profile and two minima. The circular polarization tends towards more left-handed (reaching over $\approx7$ per cent) at lower frequencies and more right-handed (reaching over $\approx2$ per cent) at higher frequencies. The P.A. curve is flat at the edges of the profile, with a negative slope corresponding to the centre of the profile between the two prominent outer components, and a notable deflection (more prominent at the higher frequencies) approximately corresponding to the first $L/I$ minimum. RM variations across the profile have been detected, but the origin is uncertain \citep[][]{ijw19}. 
This pulsar is also known to glitch \citep[e.g.][and references therein]{Yu+2013}, and has the second highest characteristic magnetic field strength of the pulsars in this census (B$_{\rm{s}}$=1.2$\times$10$^{13}$\,G).

\textbf{PSR B1737--30 (J1740--3015):}
The pulse profile consists of a main pulse with 2 prominent blended components, the leading component of which is the most highly linearly polarized (up to $\approx$100 per cent), and the trailing component of which decreases more rapidly in flux density towards lower frequencies. At even higher frequencies, the trailing component dominates over the leading component and becomes more highly polarized \citep[17\,GHz;][]{Keith+2011}.
There is also a weak precursor component that becomes more prominent towards the lower frequencies, and is also highly linearly polarized ($\approx$90 per cent). 
The P.A. curve shows a somewhat classical `S'-shape, except at 970\,MHz, the cause of which (profile evolution or interstellar scattering) requires further investigation. 
Of the pulsars in this work, PSR J1740--3015 has the highest average circular polarization degree (reaching over 70 per cent right-handed, exceeding the linear polarization fraction at some pulse phases) and also the largest increase with frequency, over 25 per cent. The peak in right-handed fractional circular polarization occurs towards the trailing component, but arrives slightly earlier (later) at higher (lower) frequencies compared to the average.
PSR J1740--3015 is among the most frequently glitching pulsars known \citep[e.g.][and references therein]{Yu+2013}, and has the highest characteristic magnetic field strength of the pulsars in this census.

\textbf{PSR B1742--30 (J1745--3040):}
This pulsar has a relatively complex pulse profile: three prominent blended main pulse components, the latter of which becomes more prominent towards lower frequencies; a precursor component, also with two prominent blended components; and a weak trailing component. The leading edge of the precursor has the highest fractional linear polarization, reaching over 75 per cent; the main and weak trailing components are also over 55 and 25 per cent linearly polarized, respectively. 
The circular polarization changes handedness twice (from right to left, and back) with the peak fraction ($\approx$10 per cent) corresponding to the peak fractional linear polarization in the leading component in the main pulse. 
The P.A. curve is approximately flat at the edges and shows two $\approx$90$^{\circ}$ OPM jumps: between the precursor and main pulse; and between the main pulse and trailing weak components. Across the main pulse, the P.A. is somewhat flat with an upward deflection, approximately corresponding again to the leading component in the main pulse.
RM variations were detected for this pulsar, which may have a magnetospheric origin \citep[][]{Noutsos+2009,ijw19}.
PSR J1745--3040 is also known to null \citep[e.g.][]{Biggs+1992,Burke-Spolaor+2012} and may also show time variability \citep{Johnston+2021}.

\textbf{PSR B1749--28 (J1752--2806):}
The pulse profile is relatively simple with a single main peak consisting of two prominent blended components (the latter of which is most easily seen at the lowest three frequencies) and a weak postcursor that increases in relative flux density towards higher frequencies, which continues above 8\,GHz \citep[][]{Johnston+2006}. 
The P.A. curve is flat at the profile edges, with a steep negative slope and (negative) inflections in the centre, somewhat reminiscent of PSR J1731--4744. \citet{ijw19} report an OPM jump in the central region close to the minimum in circular polarization. This is most consistent with our profile at 1748\,MHz after pulse phase 0.495 (see the supporting information). However, this is not consistently seen in the P.A.s across the other frequencies, which show a steepening gradient towards lower frequencies. 
The fractional polarization is relatively modest, increasing towards the edges of the profile, but with pulse-average fractions of less than 10 per cent linear (generally increasing towards lower frequencies) and less than 14 per cent absolute circular (generally decreasing towards lower frequencies). The circular polarization swings from left to right-handed and back at the lowest four frequencies; and right- to left-handed at the highest four frequencies, since the profile narrows towards higher frequencies. RM variations with pulse phase have been detected, but are likely due to interstellar scattering \citep[][]{ijw19}. \citet{Lewandowski+2015} report a scattering time of 0.003\,ms at 1\,GHz.

\textbf{PSR B1804--08 (J1807--0847):}
The total intensity pulse profile is relatively complex, with four prominent blended components, which significantly evolve with frequency -- the peak flux density occurs at the leading component at the highest frequencies, but at the inner component at the three lowest frequencies, a trend which continues to lower frequencies \citep[$<$350\,MHz][]{Johnston+2008}. The fractional linear polarization increases towards the profile edges, preceded by two minima; the central profile is $\leq$20 per cent linearly polarized and this fraction decreases with increasing frequency. 
The P.A. curve shows a generally shallow negative slope with two $\approx$80$^{\circ}$ OPM jumps corresponding to the two fractional linear polarization minima, and some additional evolution with frequency of the inflections at $\approx$0.155 pulse phase. 
The circular polarization swings from $\approx$18 per cent left-handed to $\approx$25 per cent right-handed in the trailing component.
Possible variations in RM with pulse phase have been detected and may be due to magnetospheric effects \citep[][]{ijw19}. 

\textbf{PSR B1822--09 (J1825--0935):}
This pulsar is the only object in this census to show interpulse emission, plots of which are also provided in the supporting information. 
PSR J1825--0935 is known to show several magnetospheric phenomena including emission mode-changing \citep[e.g.][]{Gil+1994}, which is correlated with changes in the spin-down \citep[][]{Lyne+2010}; and glitches \citep[e.g.][]{Espinoza+2011}.
Therefore, the pulse profile presented here is likely an average of these different emission modes, dominated by the `strong' state \citep[e.g.][]{Yan+2019}. 
The total intensity pulse profile shows two prominent blended components in the main pulse, a leading precursor component, and an interpulse. The main pulse's trailing component and the interpulse increase in relative flux density (and the precursor decreases in relative flux density) towards lower frequencies. Although this trend seems to continue towards lower frequencies ($<350$\,MHz) for the interpulse and precursor, this does not appear to be the case for the trailing component \citep[][]{Johnston+2008}. 
The precursor is highly, $\approx$100 per cent linearly polarized; shows somewhat of an `S'-shape P.A. curve; and is left-hand circularly polarized up to $\approx$40 per cent.
The main pulse is less linearly polarized ($\approx$5-20 per cent) with two clear minima at most frequencies, except at 3399 and 3818\,MHz. These seem to correspond to two inflections (negative and positive, respectively) in the otherwise slowly increasing P.A. curve. There is a small swing from right- to left-handed circular polarization at all but the two lowest frequencies, where it seems to remain left-handed across the profile.
The interpulse shows $\approx$25 per cent linear polsarisation at the three lowest frequencies; an approximately flat P.A. curve; and no significant circular polarization.
RMs were measured for each profile component, although we report the value from the main pulse in Table \ref{tab:obs}, as this had the highest S/N. The RMs from the main pulse and precursor agree within uncertainties. For the interpulse data, with lower in S/N, we obtain RM=70.5$\pm$0.9\,rad\,m$^{\rm{-2}}$, which is 2.8$\upsigma$ different to the RM measured using the main pulse data (66.3$\pm$0.6\,rad\,m$^{\rm{-2}}$). 

\textbf{PSR B1826--17 (J1829--1751):}
The total intensity pulse profile shows three prominent, blended components, and significant profile evolution with frequency -- the outer components are dominant at the higher frequencies, with the latter the most dominant, while the inner component shows the peak in flux density at the lowest three frequencies.
The outer edges of the profile show the highest fractional linear polarization, up to $\approx$75 per cent, with two $L/I$ minima corresponding to two $\approx$90$^{\circ}$ OPM jumps in the P.A.s. 
The P.A. curve across the centre of the profile is approximately flat, but with an inflection (mostly positive between 960-2134\,MHz, negative at higher frequencies) corresponding to the local minimum in linear polarization, and maximum absolute circular polarization fraction ($\approx$25 per cent right-handed).

\textbf{PSR J1852--0635 (J1852--0635):}
This pulsar's pulse profile shows three prominent blended components in total intensity, which becomes wider, and the bridging emission between components decreases, towards lower frequencies. The relatively flat spectral index is evident; this is a known gigahertz-peaked spectrum (GPS) source \citep[e.g.][]{Kijak+2007,Jankowski+2018}. 
The P.A. curve shows two $\approx$90$^{\circ}$ OPM jumps at the $L/I$ minima between the three linear polarization components. Otherwise the P.A.s are subtly curved, and the central component has a steeper gradient compared to those leading and trailing.  
The largest linear polarization fraction occurs in the trailing component, up to $\approx$90 per cent.
Only the central component shows significant circular polarization -- up to $\approx$10 per cent right-handed -- although the absolute fraction decreases towards lower frequencies and is not detected at 987\,MHz.
Each linear polarization component was used to measure the RM, and all agree within the uncertainties. The value obtained using the leading precursor component is reported in Table \ref{tab:obs} because this provided the smallest uncertainty. No RM variations with pulse phase were reported in \citet{ijw19}, although the linear polarization fraction seems to be much larger in this work. 

\textbf{PSR B1857--26 (J1900--2600):}
The total intensity pulse profile appears to be slowly rising in flux density, a combination of at least four blended components. These components are more easily identifiable towards lower frequencies, with the central one dominating the flux density \citep[$<$350\,MHz][]{Johnston+2008}. There are three prominent linear polarization components: the central component shows the highest fraction (over 40 per cent) over much of the frequency range, although the leading component becomes more highly polarized with decreasing frequency, reaching $\approx$50 per cent at 964\,MHz, no significant linear polarization was detected at the highest frequency. 
Across the two leading linearly polarized components, before the deepest $L/I$ minimum, the P.A. curve resembles an `S'-shape. The P.A.s corresponding to the trailing component evolve noticeably with frequency: between 2613-3400\,MHz, the P.A. curve continues, increasing approximately linearly; between 1360-2135\,MHz, there is a steep increasing `S'-shape (also discussed in \citealt{ijw19} as being inconsistent with an OPM jump) and flattening towards the profile edge; while at 964\,MHz, this steep `S'-shape seems to reverse sense, preceded by a further flattening of the P.A.s towards the trailing edge.
The transition in P.A.s either side of the fractional polarization minumum around pulse phase 0.552 occurs increasingly earlier towards lower frequencies, and could be a reason for the large RM variation with pulse phase seen in \citet{ijw19}, although these variations were thought to be caused by interstellar scattering. 
There is a dramatic swing in circular polarization from left- to right-handed, reaching peaks of over 30 per cent in both cases.
This pulsar is also known to show emission mode changing \citep[][]{Burke-Spolaor+2012}. 

\textbf{PSR B2045--16 (J2048--1616):}
The total intensity profile shows three prominent blended components, with the trailing component most dominant. Apart from a somewhat steep spectral index of $-2.6\pm0.1$ \citep{Jankowski+2018}, there is little evolution with frequency over 969--3817\,MHz, although the central component grows to dominate the flux density towards lower frequencies \citep[$<$350\,MHz;][]{Johnston+2008}.
There are two prominent fractional linear polarization features that reach $\approx$70 per cent, associated with the trailing edge of the leading component, and leading edge of the trailing component; the minimum between them is associated with the peak total intensity of the inner component.
The P.A. curve appears to show a classical `S'-shape swing across the profile. 
The circular polarization is reasonably complex: left-handed and reaching $\approx$12 per cent, then swinging rapidly to $\approx$6 per cent right-handed across the minimum in linear polarization, before approaching left-handed values again corresponding to the trailing peak in linear polarization (most dramatic at lower frequencies). 
This pulsar is also known to null with fraction $\approx$10-20 per cent \citep[e.g.][and references therein]{Wang+2020}. Some RM variation with pulse phase is detected, but is likely caused by interstellar scattering \citep[][]{Noutsos+2009,ijw19}. 

\section{Conclusions}\label{sec:con}

We presented flux- and polarization-calibrated, high signal-to-noise, full polarization pulse profiles for a census of 40 bright, `slow' (non-recycled) pulsars using the UWL receiver on the Parkes radio telescope, which provides an unprecedented bandwidth between 704--4032\,MHz.

We updated the DM and RM measurements towards the pulsars, and obtained median uncertainties of 0.016\,pc\,cm$^{\rm{-3}}$ and 0.77\,rad\,m$^{\rm{-2}}$, respectively: an average reduction the uncertainties compared to previous measurements in the ATNF Pulsar Catalogue by factors of 7.5 and 2.5, respectively.
This improvement is useful, for example, towards detecting secular changes in these quantities through monitoring/timing observations over several years, which provides information about small-scale ISM characteristics \citep[e.g.][]{pkj+13,Donner+2020,Johnston+2021}. 
However, systematic uncertainties should be considered when comparing measurements determined using diverse datasets and methods.
We discussed four specific lines of sight where the RMs towards the pulsars have the opposite sign to the RMs expected from the entire Galactic LoS, and discuss the possible influences of intervening ISM structures.
Of the pulsars in this census, we found that the pulse profile of PSR J1721--3532 is most affected by interstellar scattering and becomes completely depolarized at 971\,MHz. We surmise that an intervening H\,\textsc{ii} region is a large contributing factor to the high DM, interstellar scattering, and depolarization, and we measured a Faraday dispersion of $\sim$20\,rad\,m$^{\rm{-2}}$. 

We found general trends in the pulsar emission from the wide-band polarization pulse profiles, consistent with previous studies, where the fractional linear polarization decreases by $\sim$7 per cent, and the degree of circular polarization increases by $\sim$4 per cent from low to high frequencies across the bandwidth.
We also discussed each pulsar's polarization pulse profile, and found them consistent with previous studies, while uncovering some new features and frequency evolution.

This work presents an initial demonstration of the data quality and results that can be obtained using UWL observations of slow pulsars.
The suite of SKA pathfinders and precursors (including the Parkes telescope) are currently obtaining exquisite, complementary pulsar data, particularly in the Southern hemisphere \citep[e.g.][]{mtime,tpa20}.
These will enable us to continue discovering valuable information about our Galaxy's ISM, and the elusive and enigmatic pulsar radio emission mechanism. 

\section*{Acknowledgements}


The Parkes radio telescope, `Murriyang', is part of the Australia Telescope National Facility which is funded by the Australian Government for operation as a National Facility managed by CSIRO. 
We acknowledge the Wiradjuri people as the traditional owners of the Observatory site.
CSIRO acknowledges the Traditional Owners of the land, sea and waters, of the area that we live and work on across Australia. We acknowledge their continuing connection to their culture and we pay our respects to their Elders past and present.
We thank Lawrence Toomey for assistance with the UWL data and pulsar software, and Phil Edwards and the referee for detailed comments that improved the manuscript.
LSO acknowledges funding from the Science and Technology Facilities Council (STFC) Grant Code ST/R505006/1.
RMS acknowledges Australian Research Council future fellowship FT190100155.  
Work at NRL is supported by NASA.
Pulsar research at Jodrell Bank Centre for Astrophysics and Jodrell Bank Observatory is supported by a consolidated grant from the UK Science and Technology Facilities Council (STFC).
This work has made use of the \textsc{Python} plotting library \textsc{Matplotlib} \citep{Hunter:2007};
NASA's Astrophysics Data System Bibliographic Services; 
``Aladin sky atlas'' developed at CDS, Strasbourg Observatory, France \citep[][]{Bonnarel+2000,Boch+2014}.

\section*{Data Availability}

The `raw' (original, uncalibrated) data taken for the P574 project are or will become available from the CSIRO Data Access Portal\footnote{\href{https://data.csiro.au}{https://data.csiro.au}} after the 18-month proprietary period. 

The final data products used in this work (i.e., flux- and polarization-calibrated, full-polarization, time-averaged, DM- and RM-corrected, 1\,MHz frequency resolution, UWL data in \textsc{PSRFITS} format) are available as a collection through the CSIRO Data Access Portal\footnote{\href{https://doi.org/10.25919/gptm-d012}{https://doi.org/10.25919/gptm-d012}} \citep{Sobey+2021_DAP}.
This collection also provides the pulse profile templates (also in \textsc{PSRFITS} format), used as an input to \textsc{Tempo2} to obtain DM measurements, created using 20-cm Multibeam data presented in \citet{jk18}.   

The final polarization pulse profiles, available as PDF plots in the supporting information, will also be made publicly available to the community through the European Pulsar Network (EPN) database\footnote{\href{http://www.epta.eu.org/epndb}{http://www.epta.eu.org/epndb}}.


\bibliographystyle{mnras}
\bibliography{bibliography} 

\newpage

\section*{Supporting Information}

\begin{figure*}
	\includegraphics[width=0.693\columnwidth]{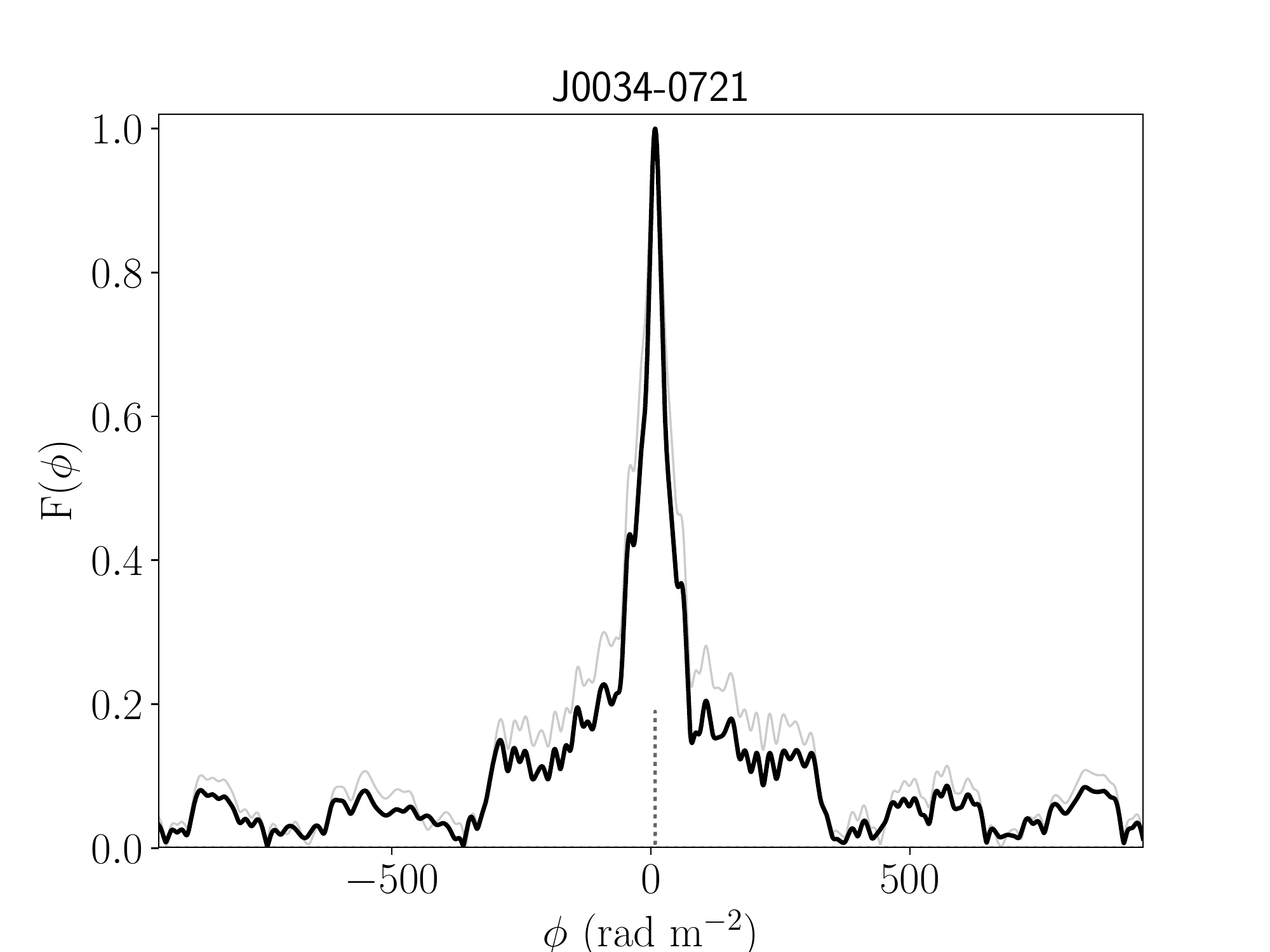}
	\includegraphics[width=0.693\columnwidth]{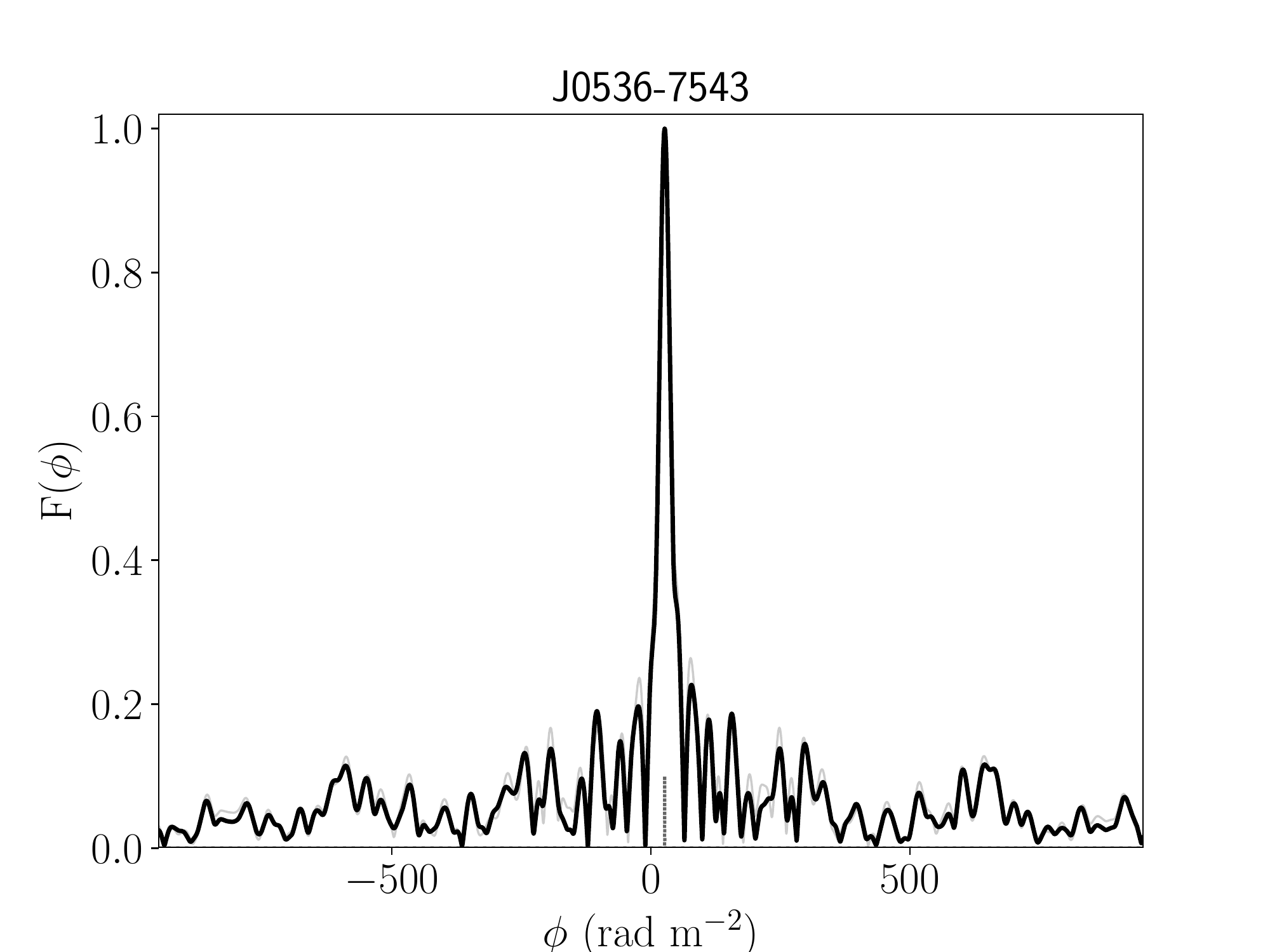}
	\includegraphics[width=0.693\columnwidth]{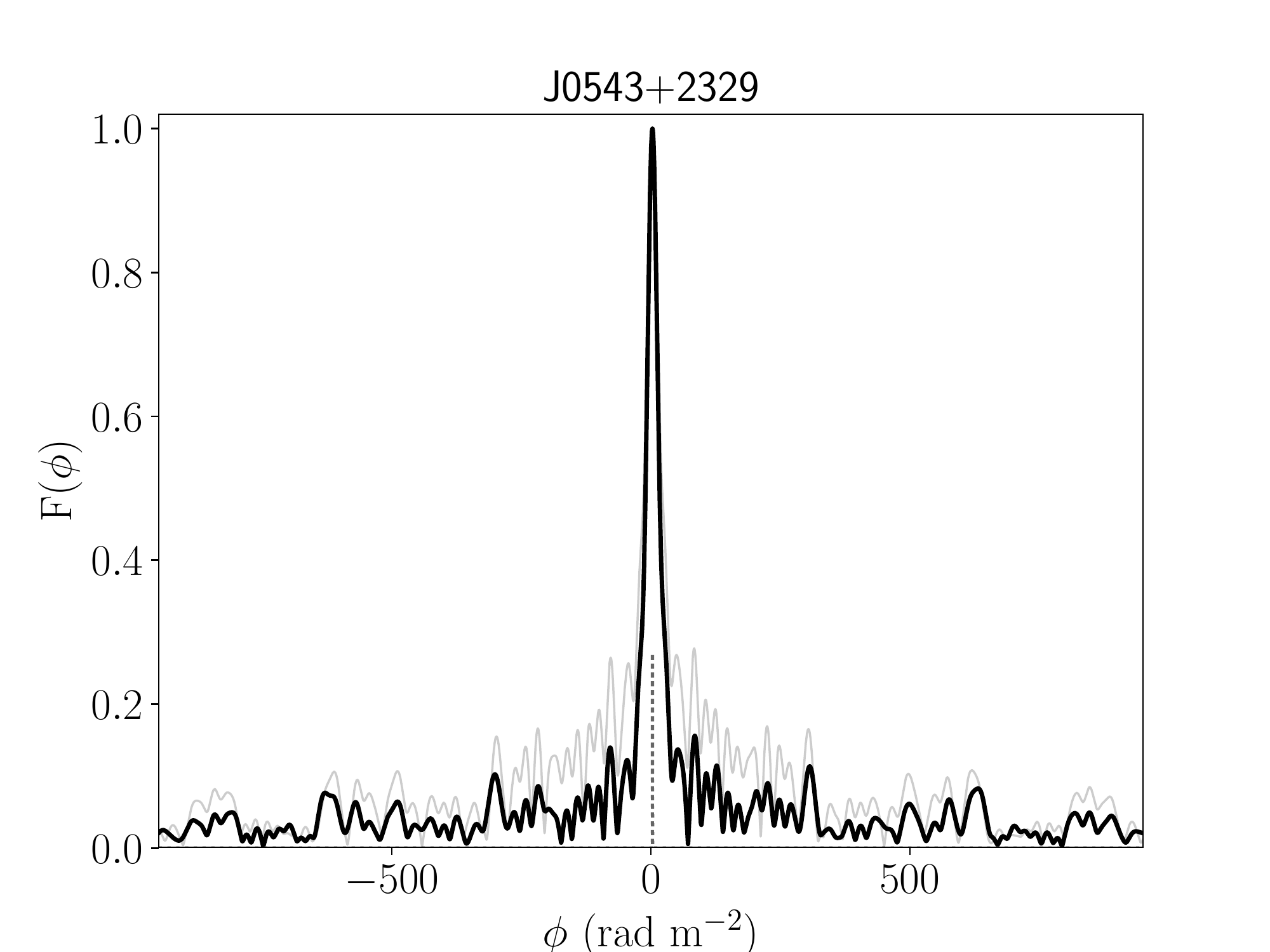}
	
\includegraphics[width=0.693\columnwidth]{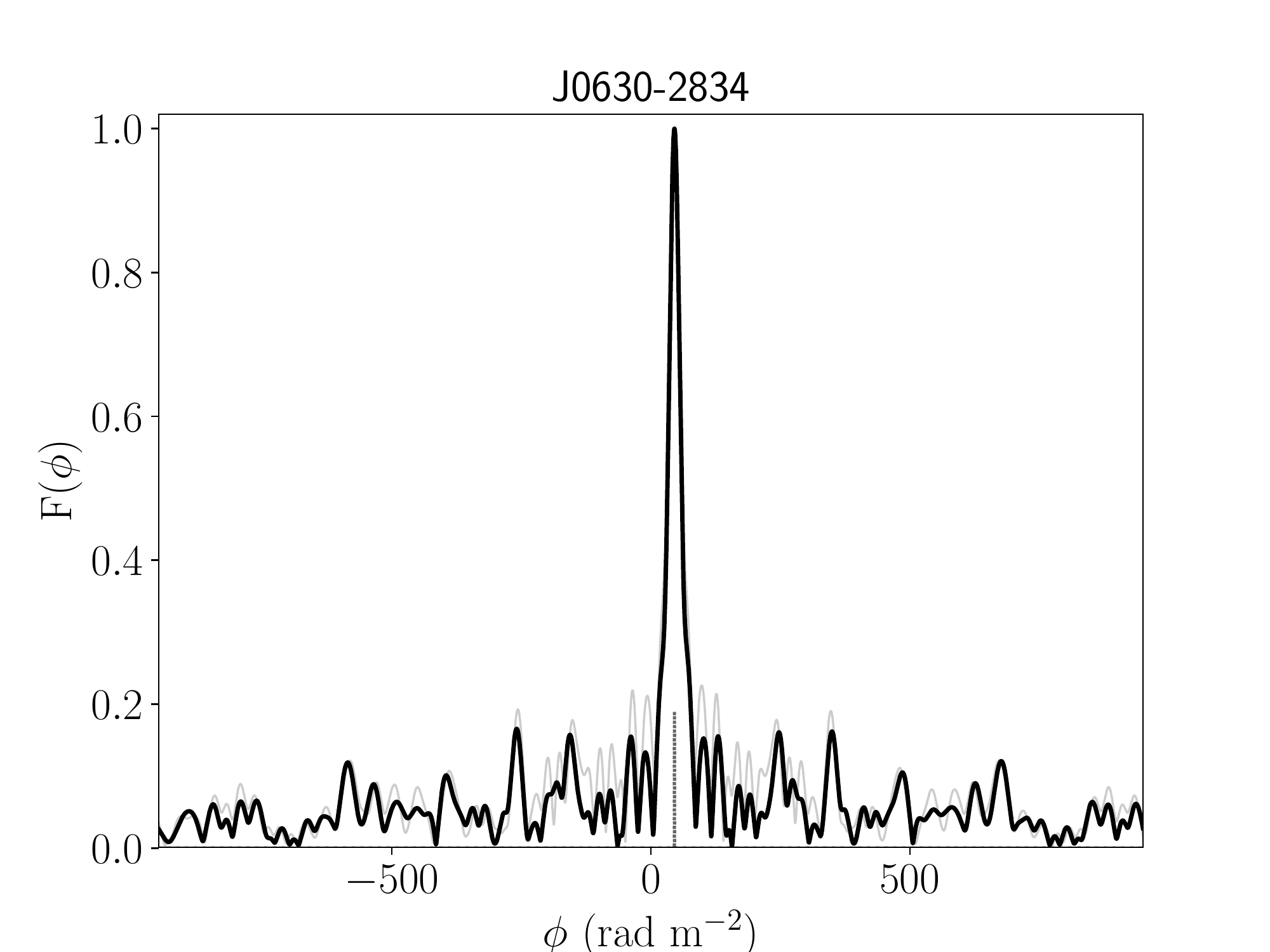}
\includegraphics[width=0.693\columnwidth]{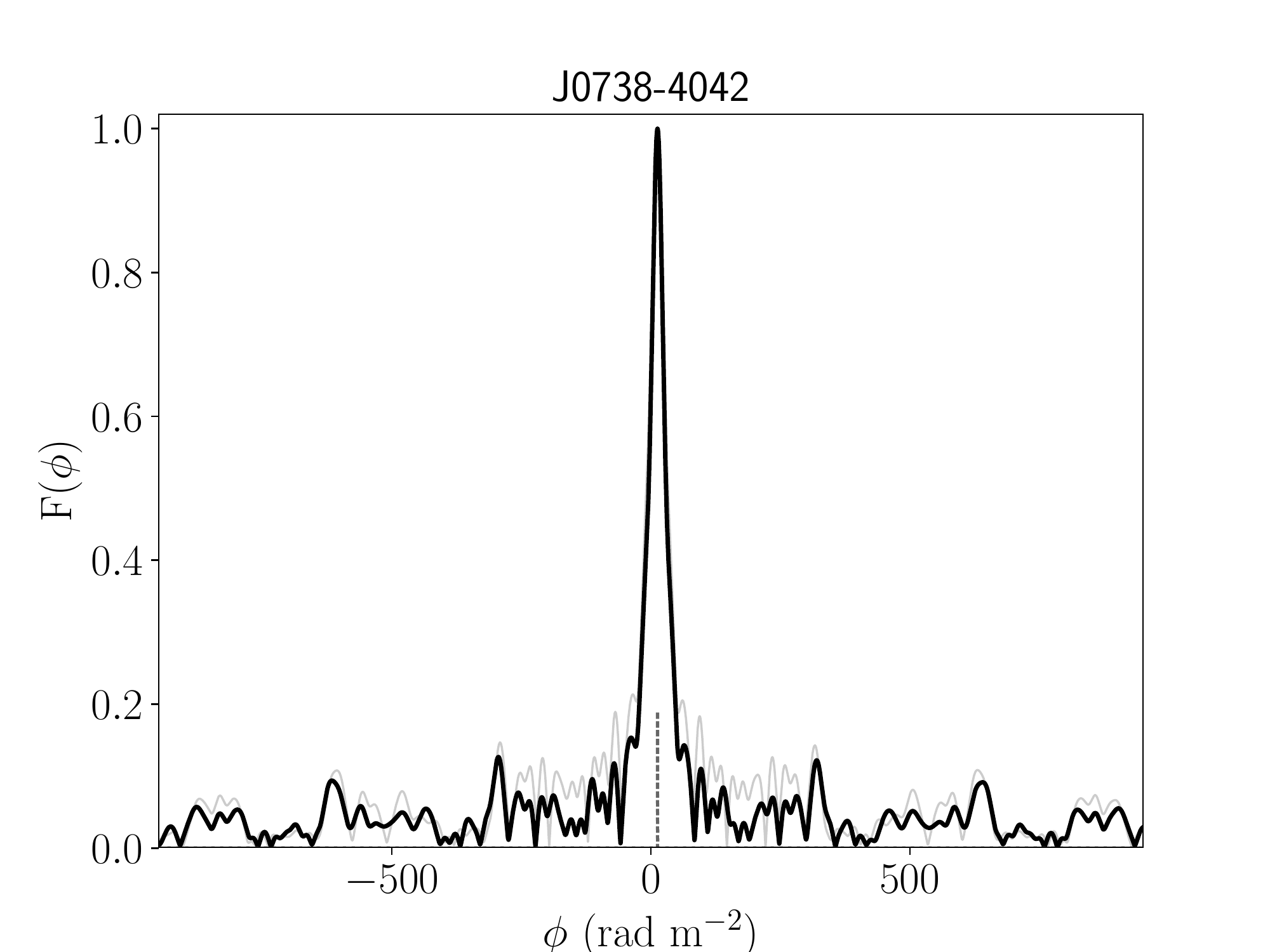}
\includegraphics[width=0.693\columnwidth]{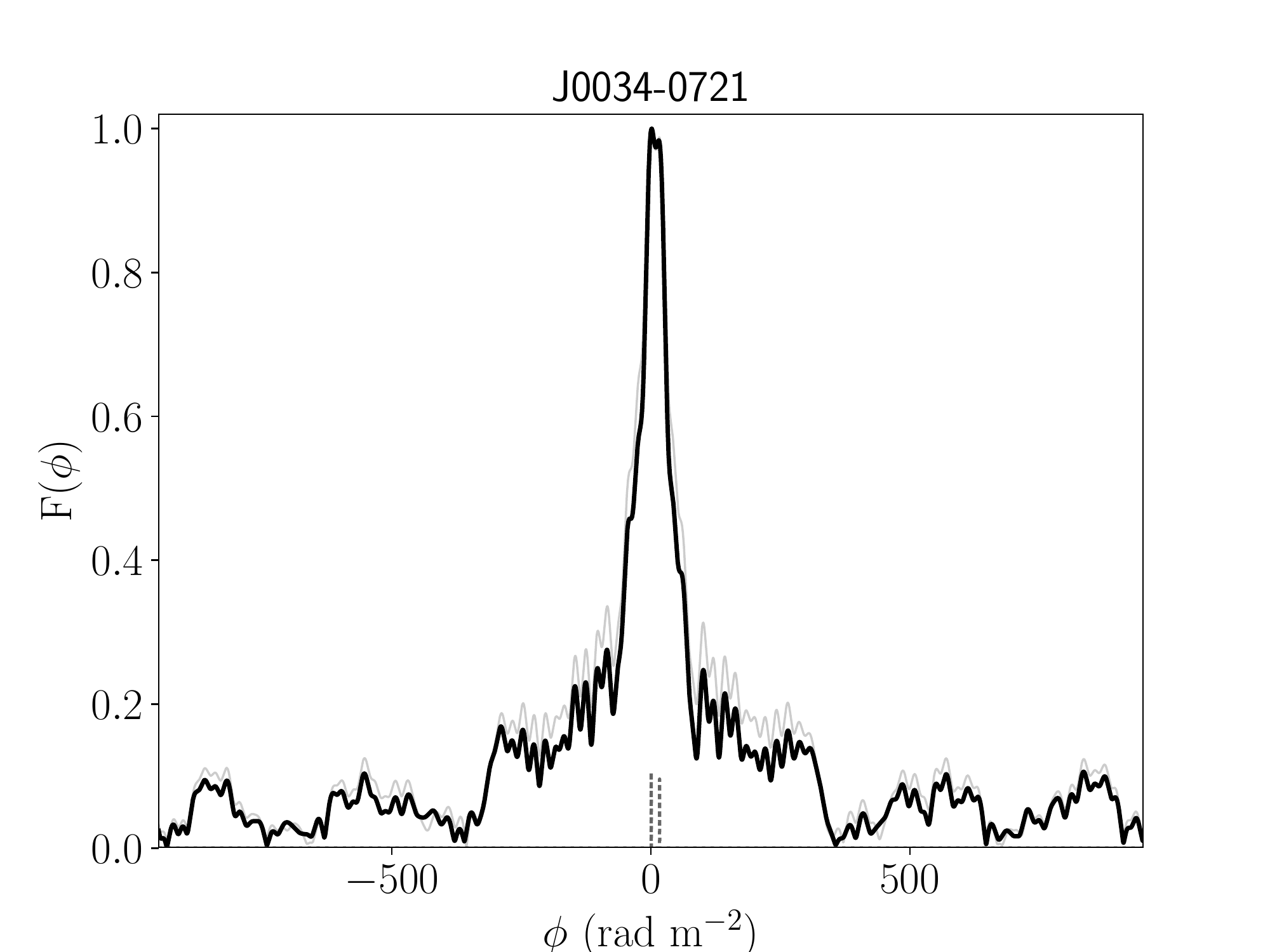}

\includegraphics[width=0.693\columnwidth]{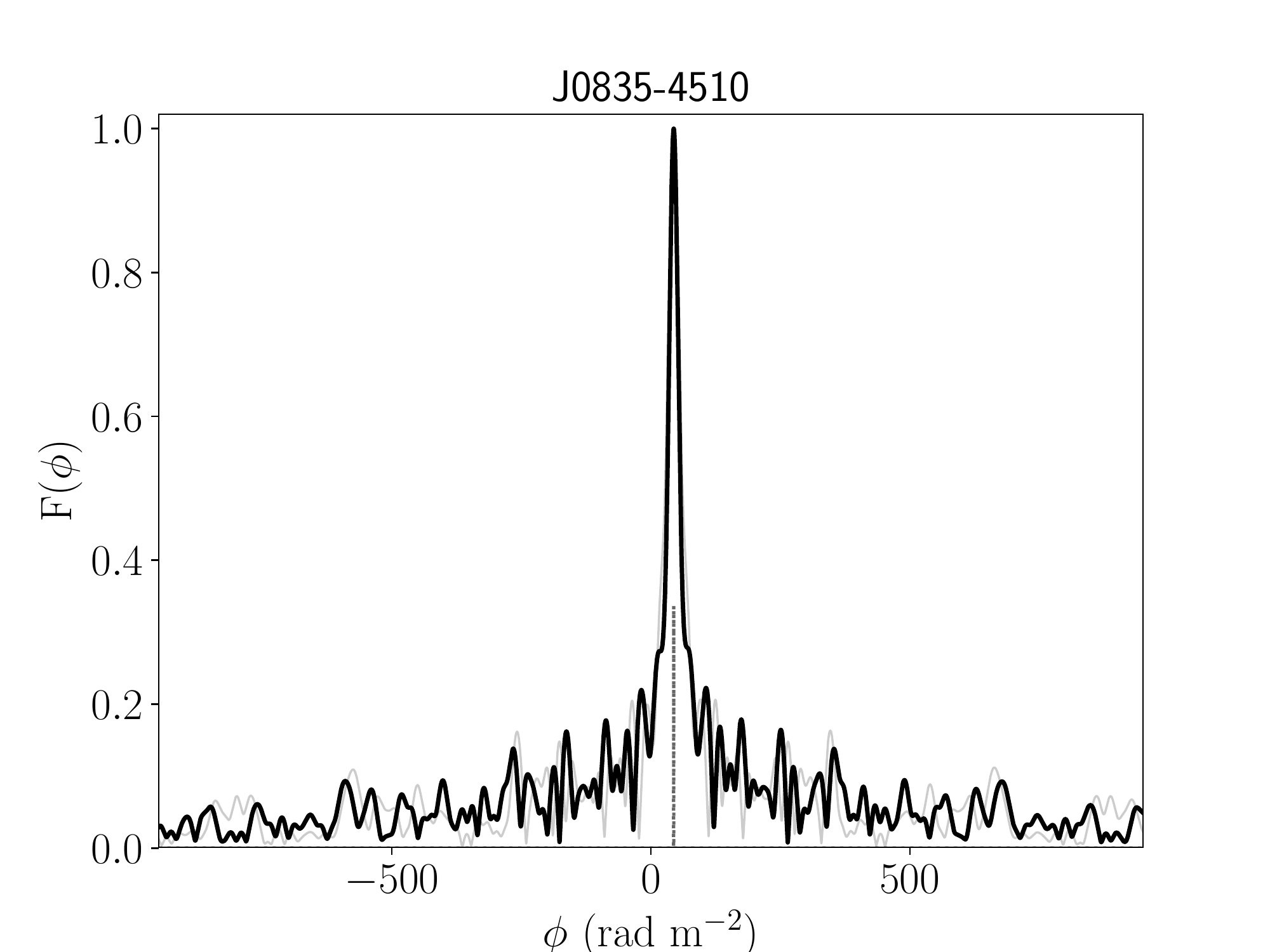}
\includegraphics[width=0.693\columnwidth]{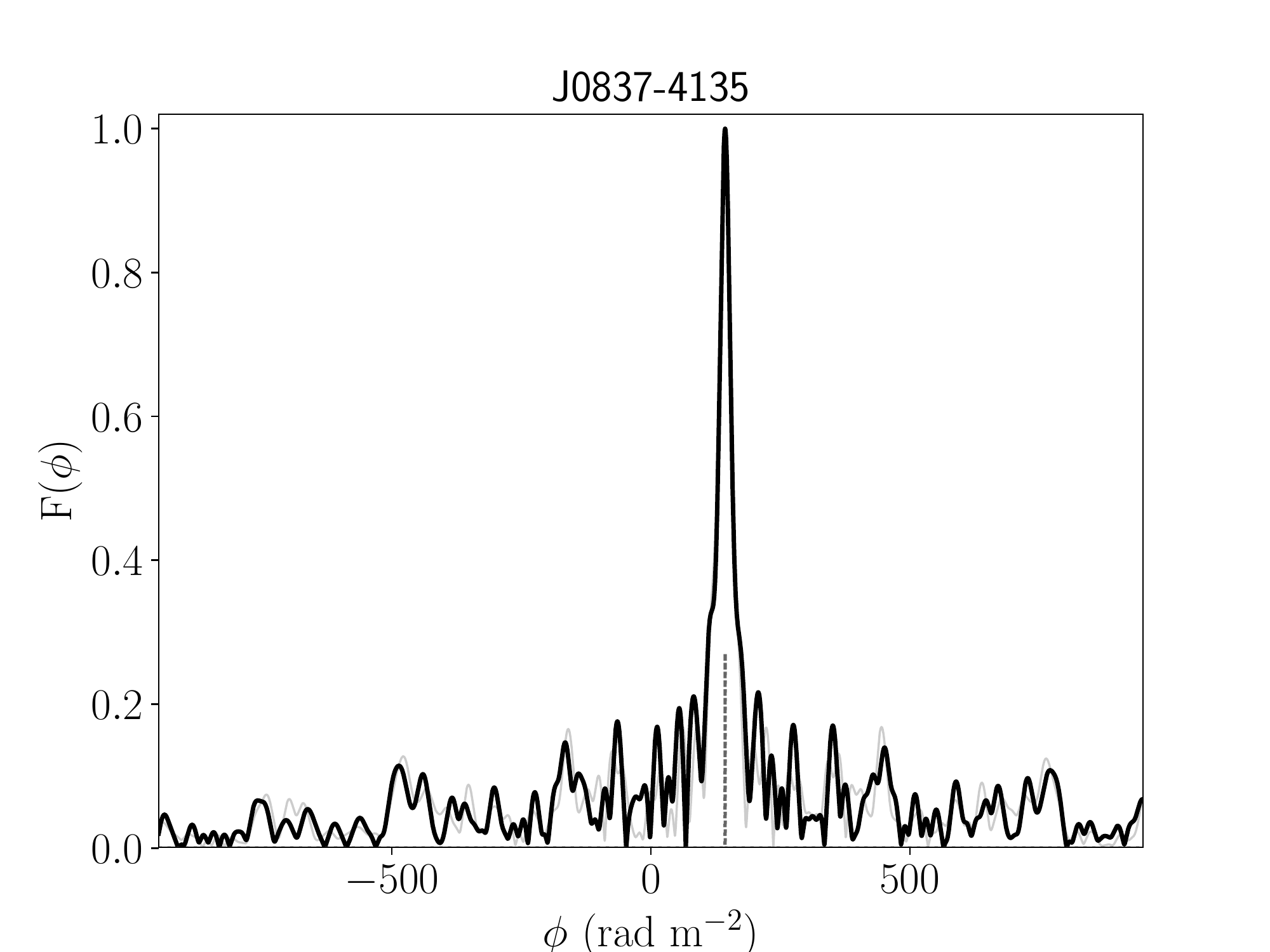}
\includegraphics[width=0.693\columnwidth]{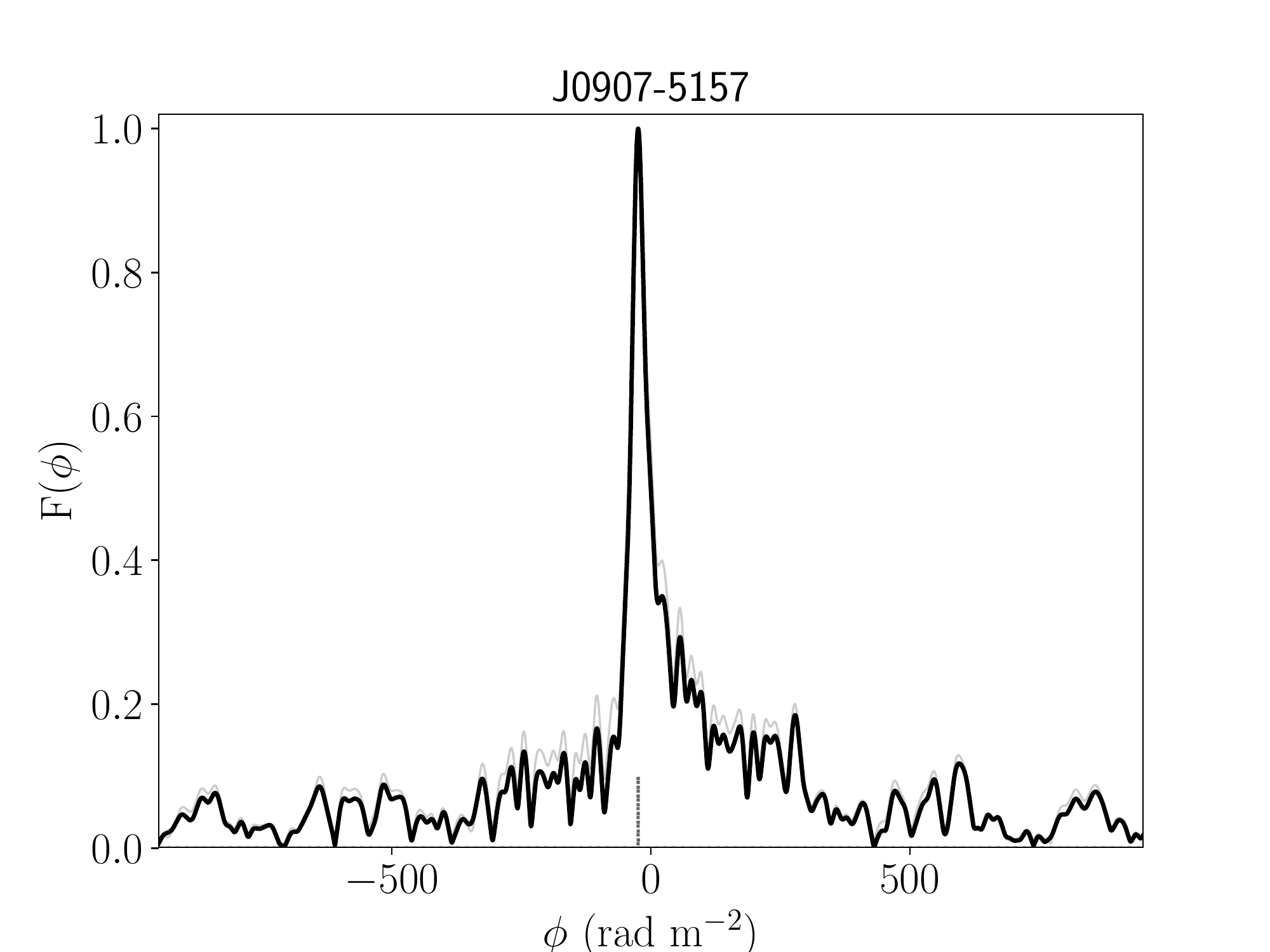}

\includegraphics[width=0.693\columnwidth]{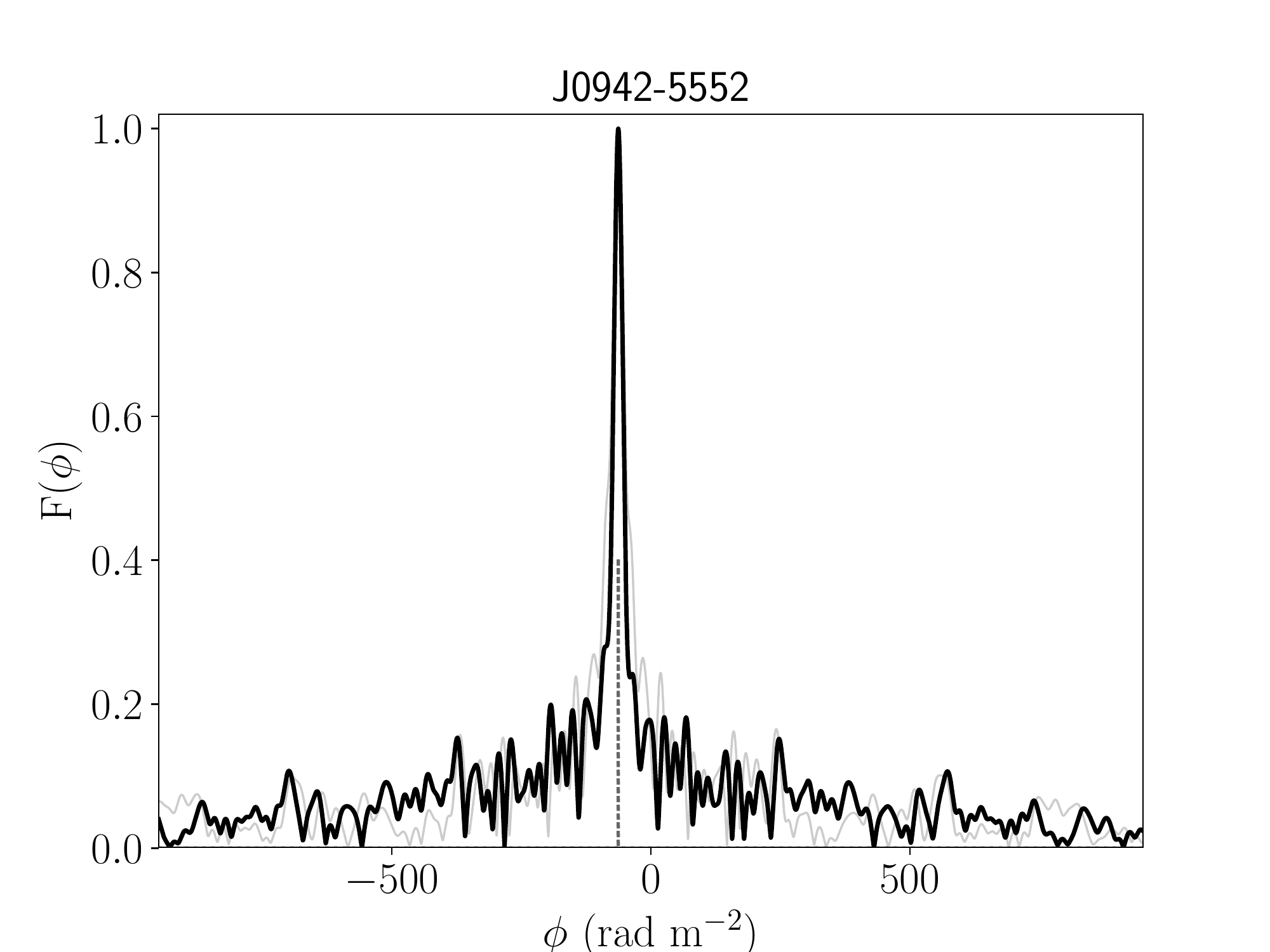}
\includegraphics[width=0.693\columnwidth]{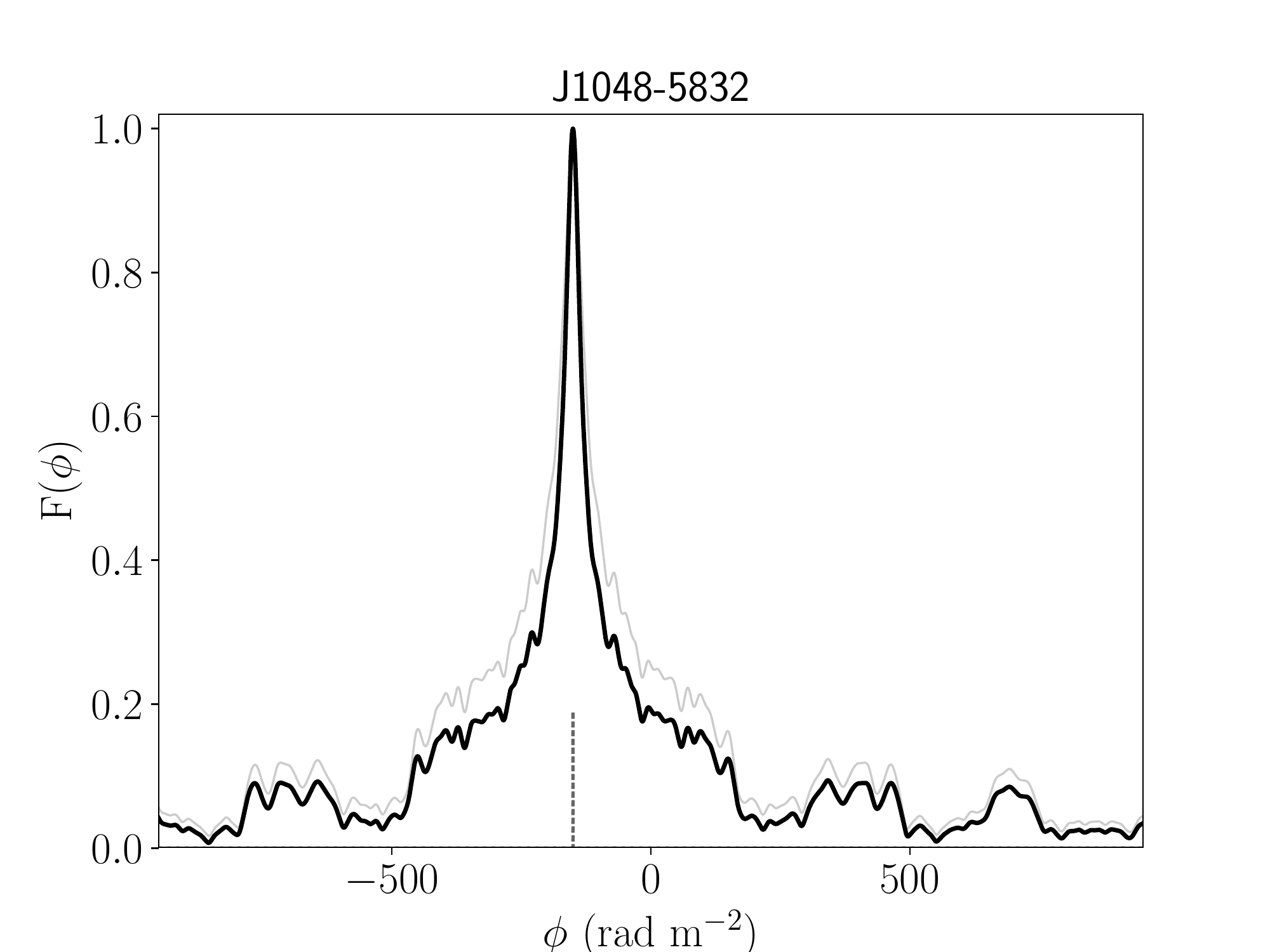}
\includegraphics[width=0.693\columnwidth]{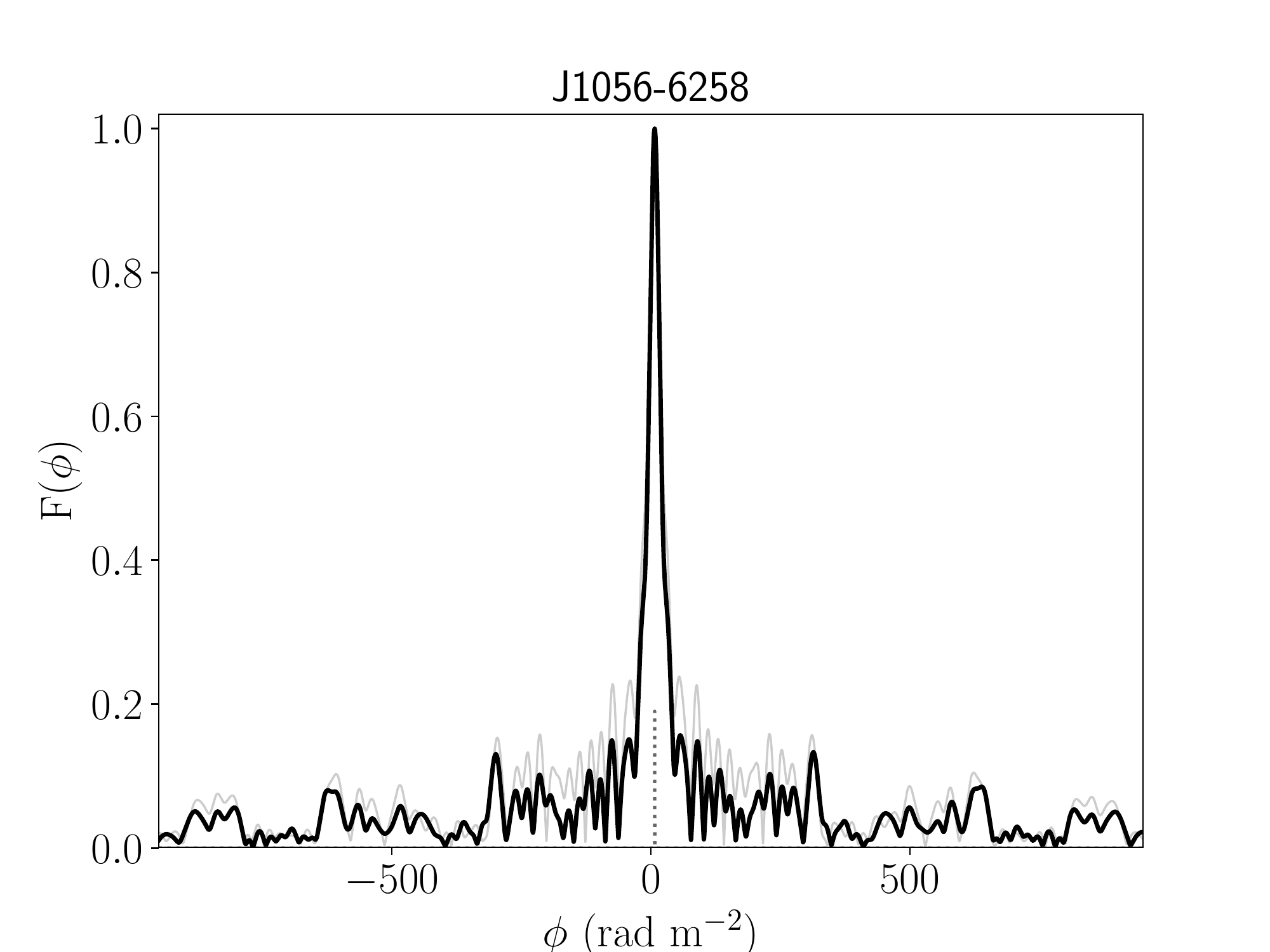}

\includegraphics[width=0.693\columnwidth]{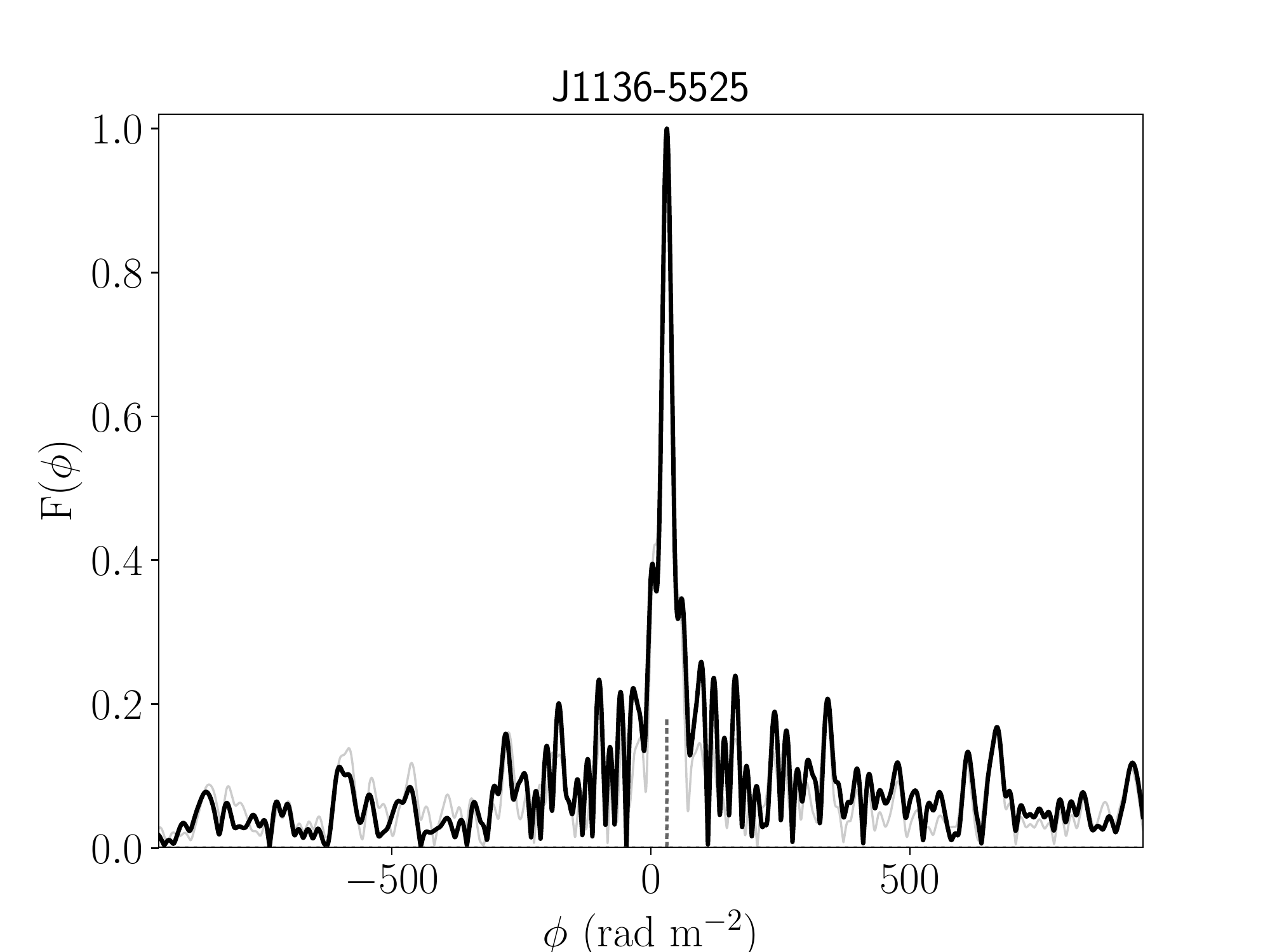}
\includegraphics[width=0.693\columnwidth]{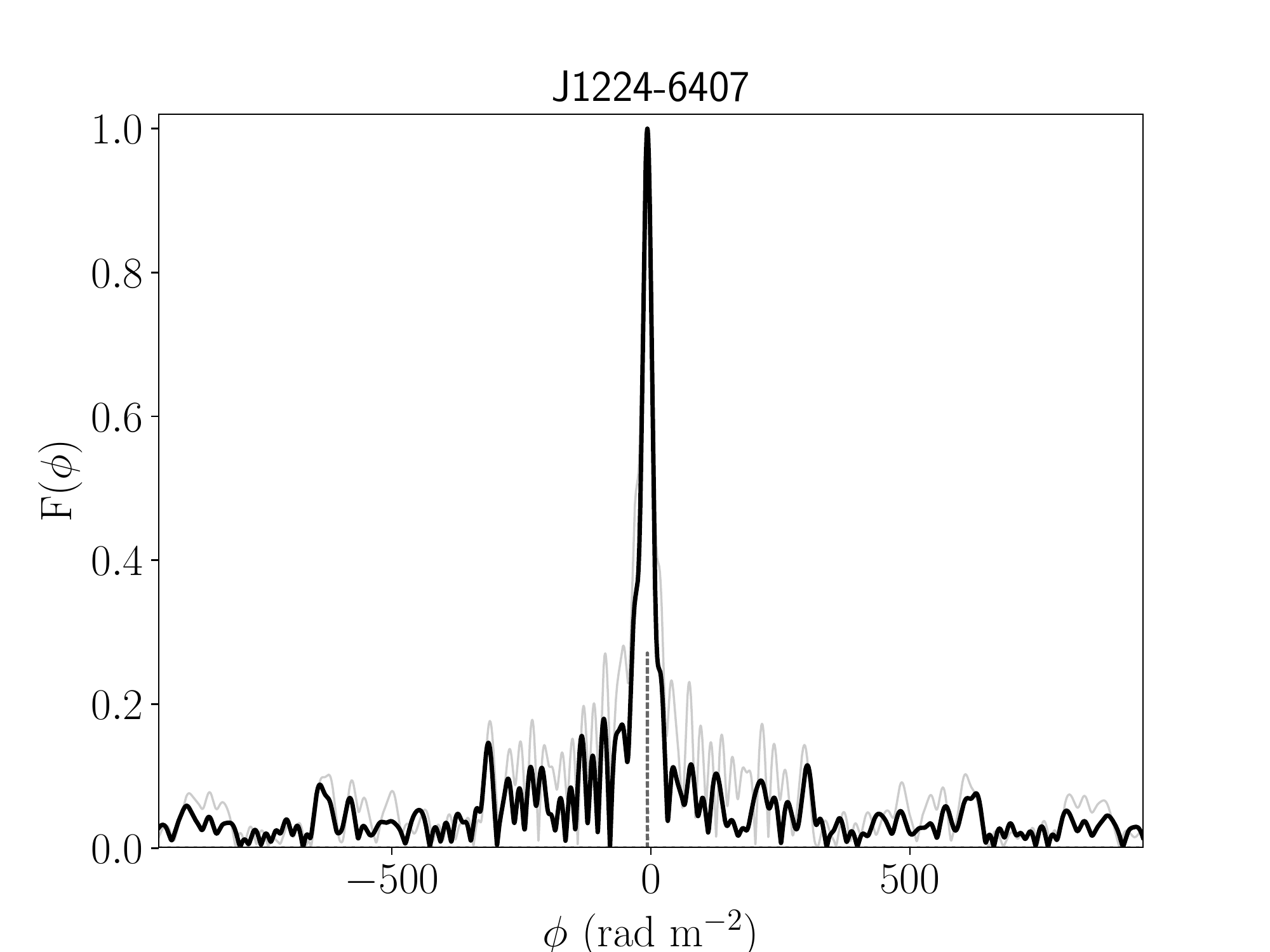}
\includegraphics[width=0.693\columnwidth]{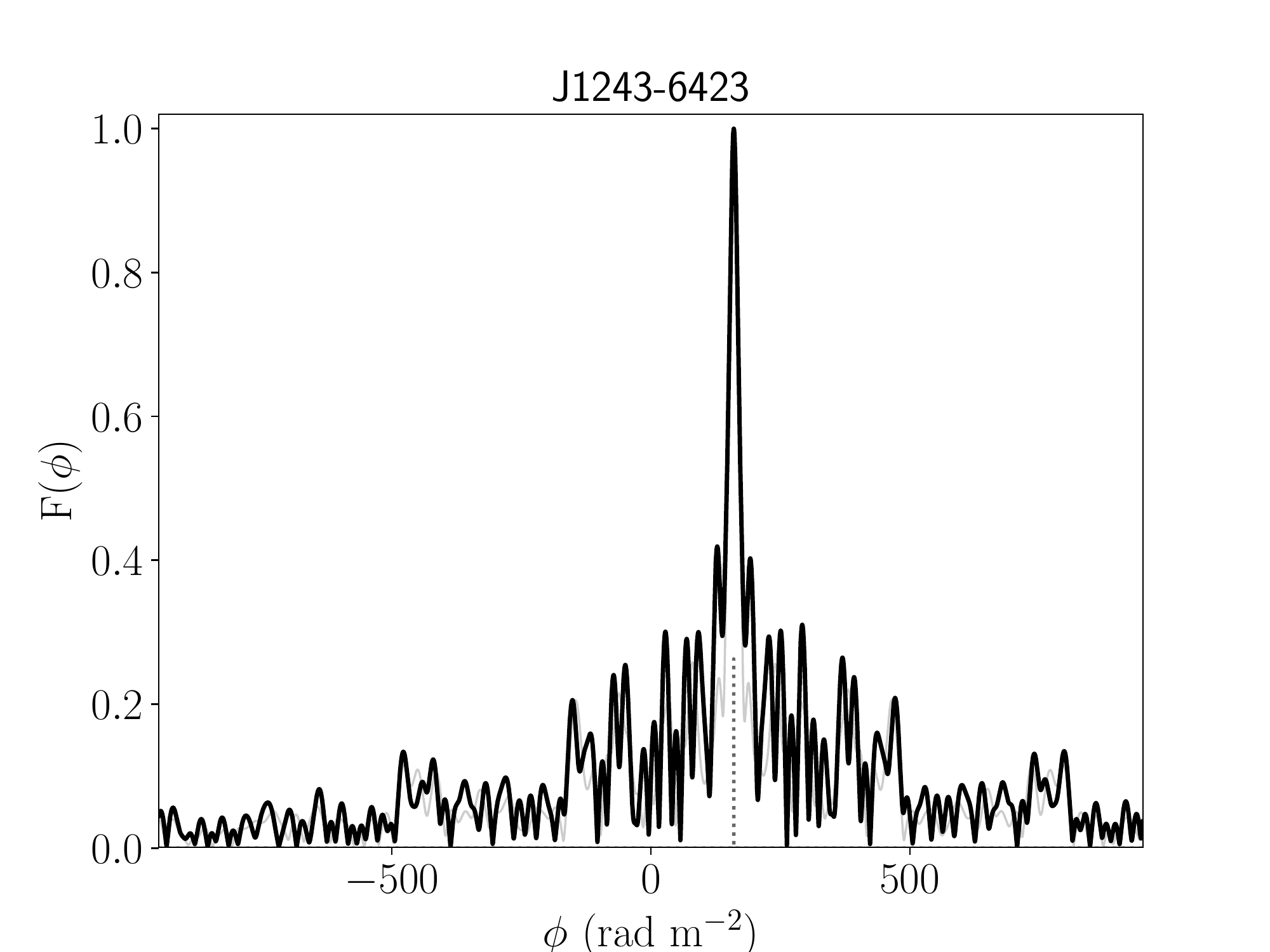}

    \caption{Faraday spectra (FDFs) obtained for each pulsar using RM synthesis. Grey lines show the orignial spectrum, black lines show the RM CLEANed spectrum; grey dashed lines show the position of the CLEAN components at the location corresponding to the measured RM. The spectra are normalised and the Faraday depth range shown is $-950<\phi<950$\,rad\,m$^{\rm{-2}}$.}
    \label{fig:FDFs1}
\end{figure*}

Figure \ref{fig:FDFs1} shows the Faraday spectrum obtained for each pulsar.
An example is shown in the main text in Figure \ref{fig:RMSF}.

\begin{figure*} 
	\includegraphics[width=\columnwidth]{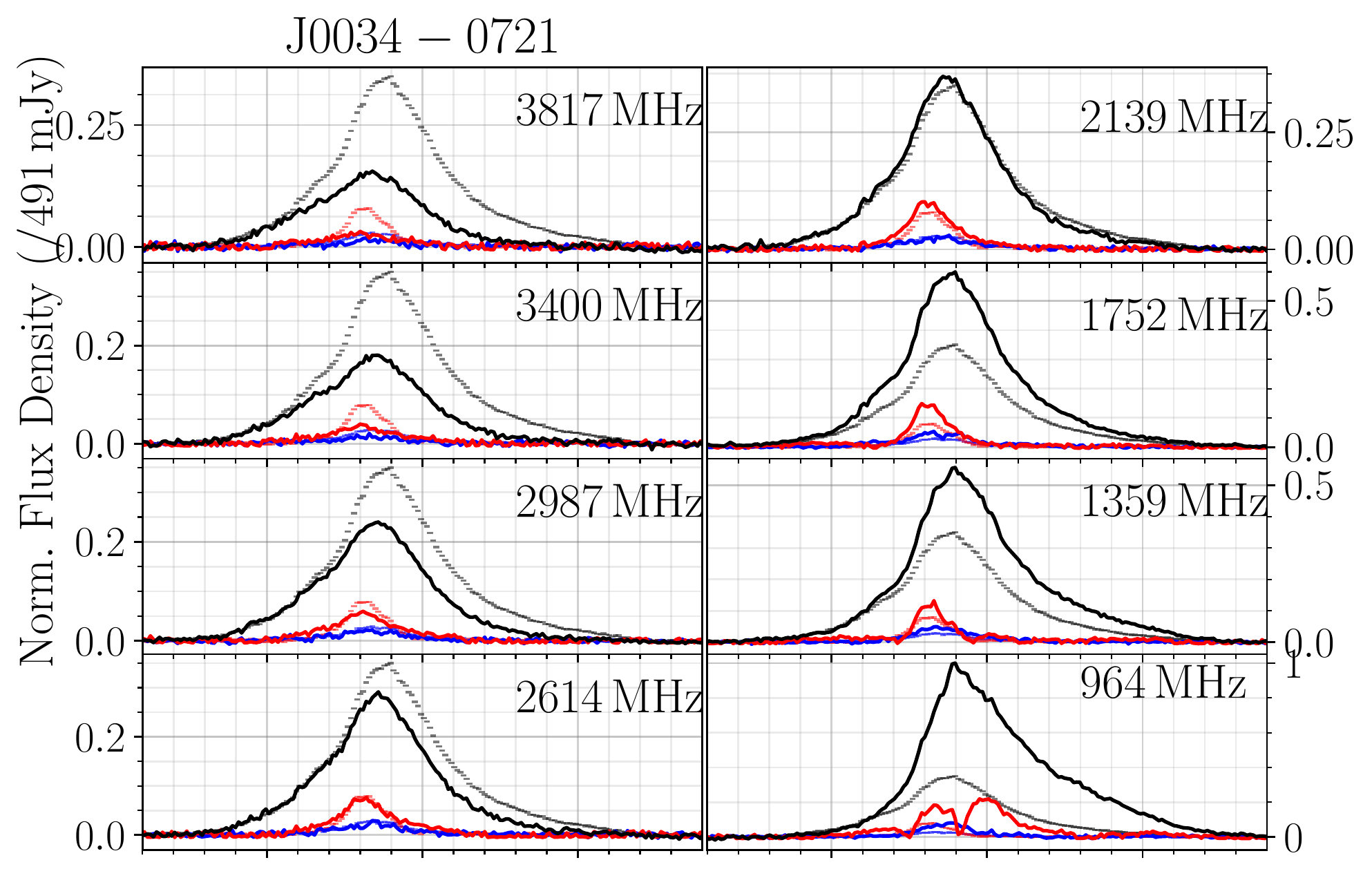}
	\includegraphics[width=\columnwidth]{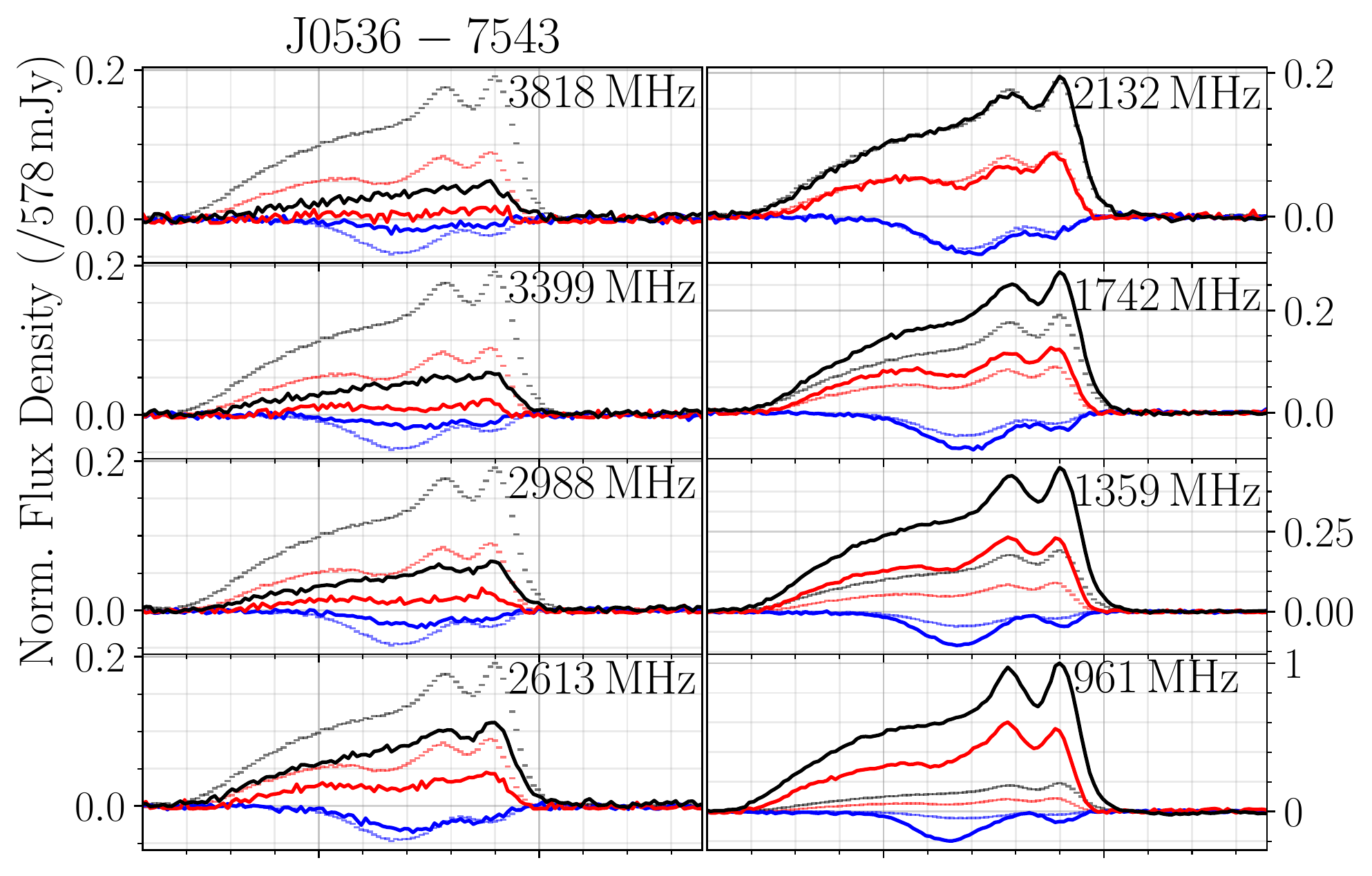}

	\includegraphics[width=\columnwidth]{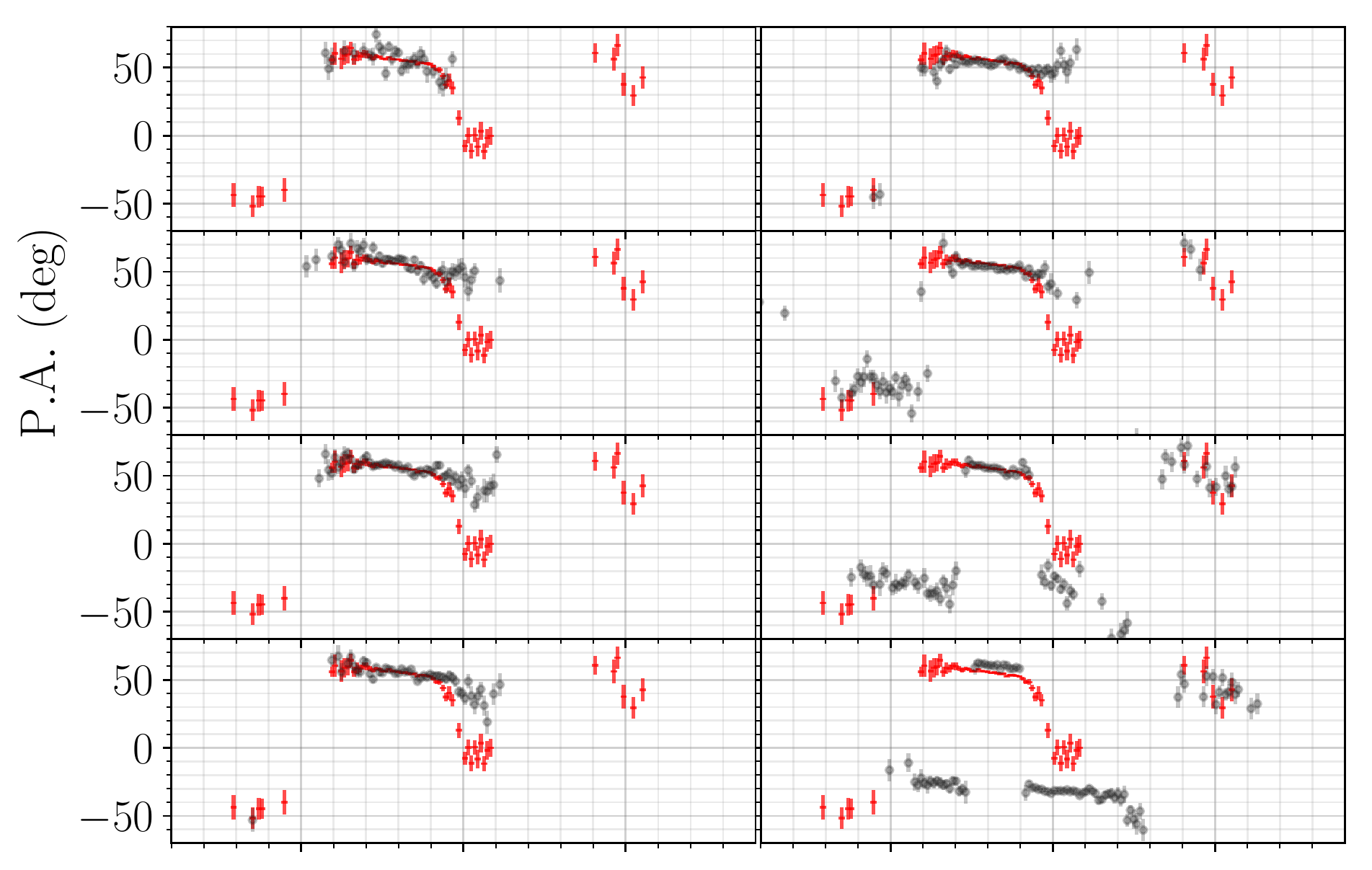}
	\includegraphics[width=\columnwidth]{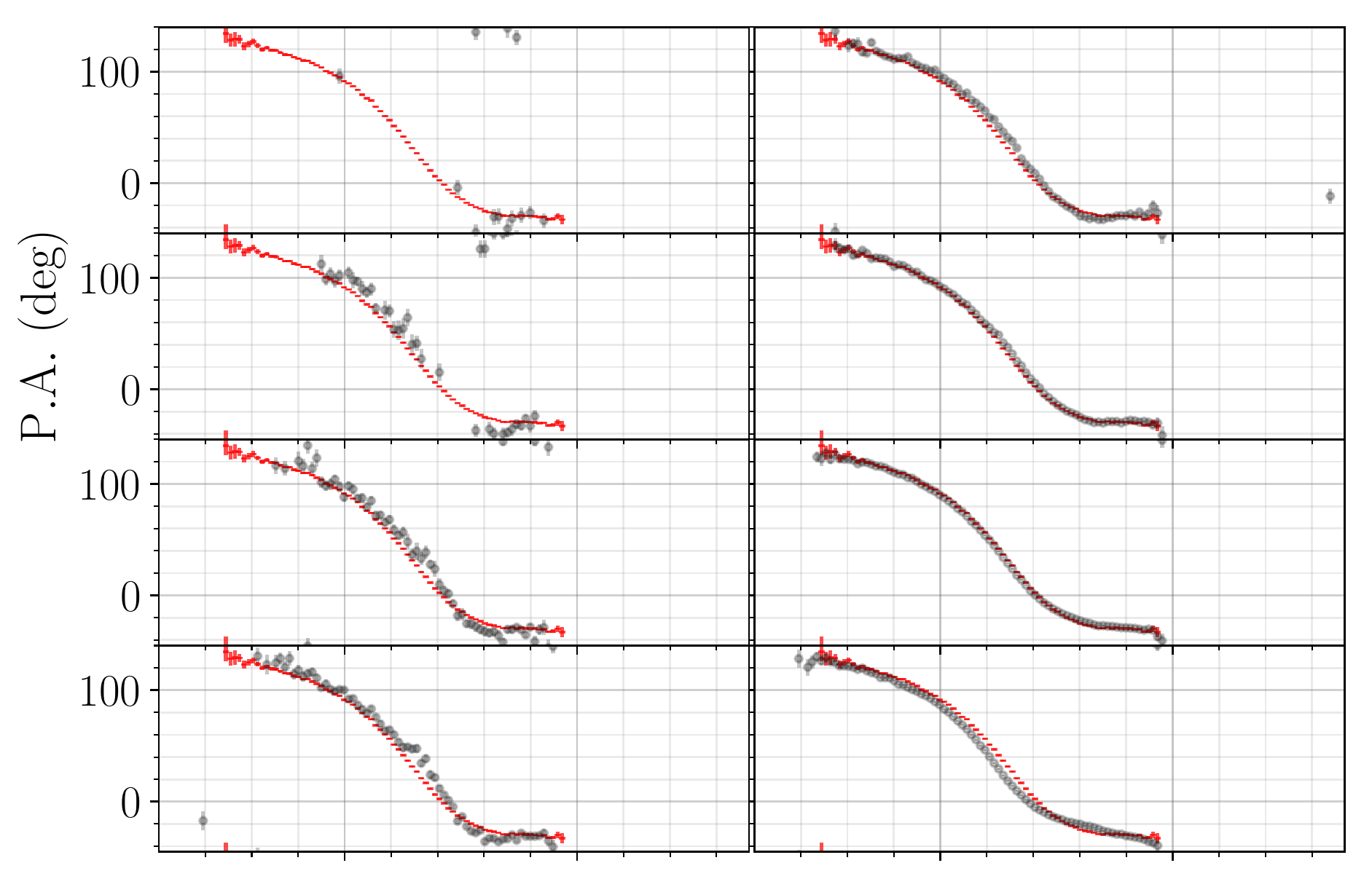}

	\includegraphics[width=\columnwidth]{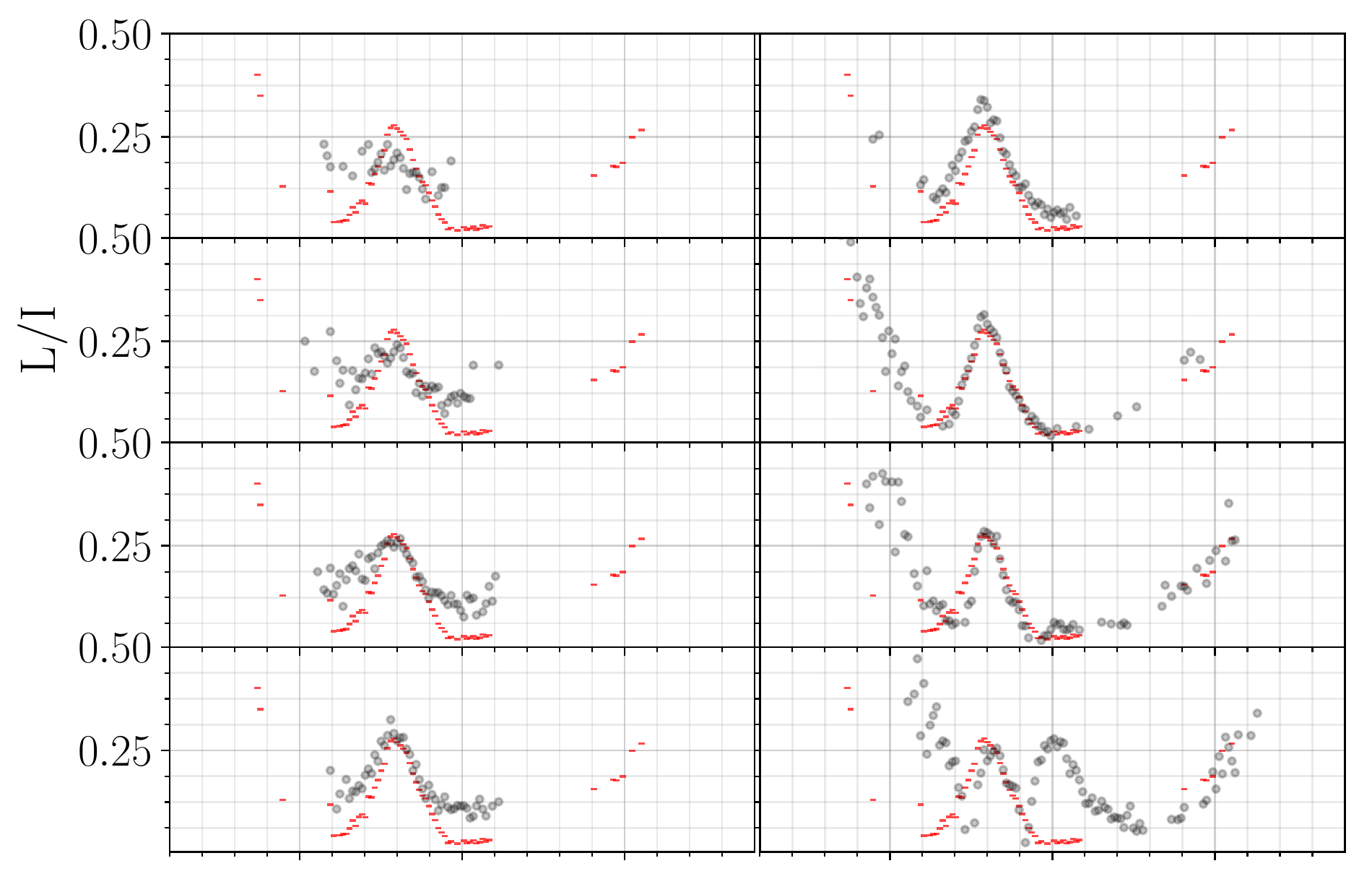}
	\includegraphics[width=\columnwidth]{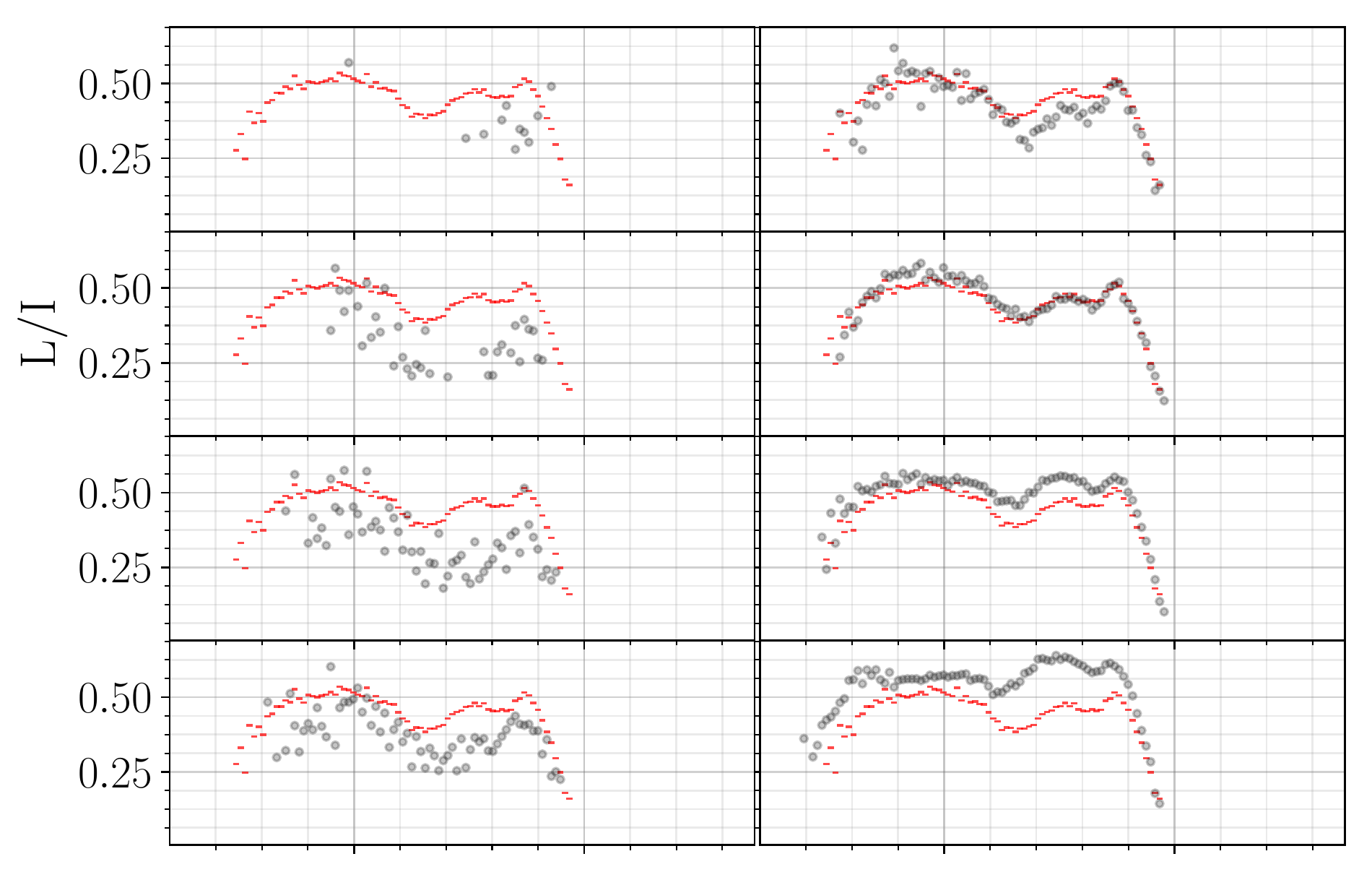}

	\includegraphics[width=\columnwidth]{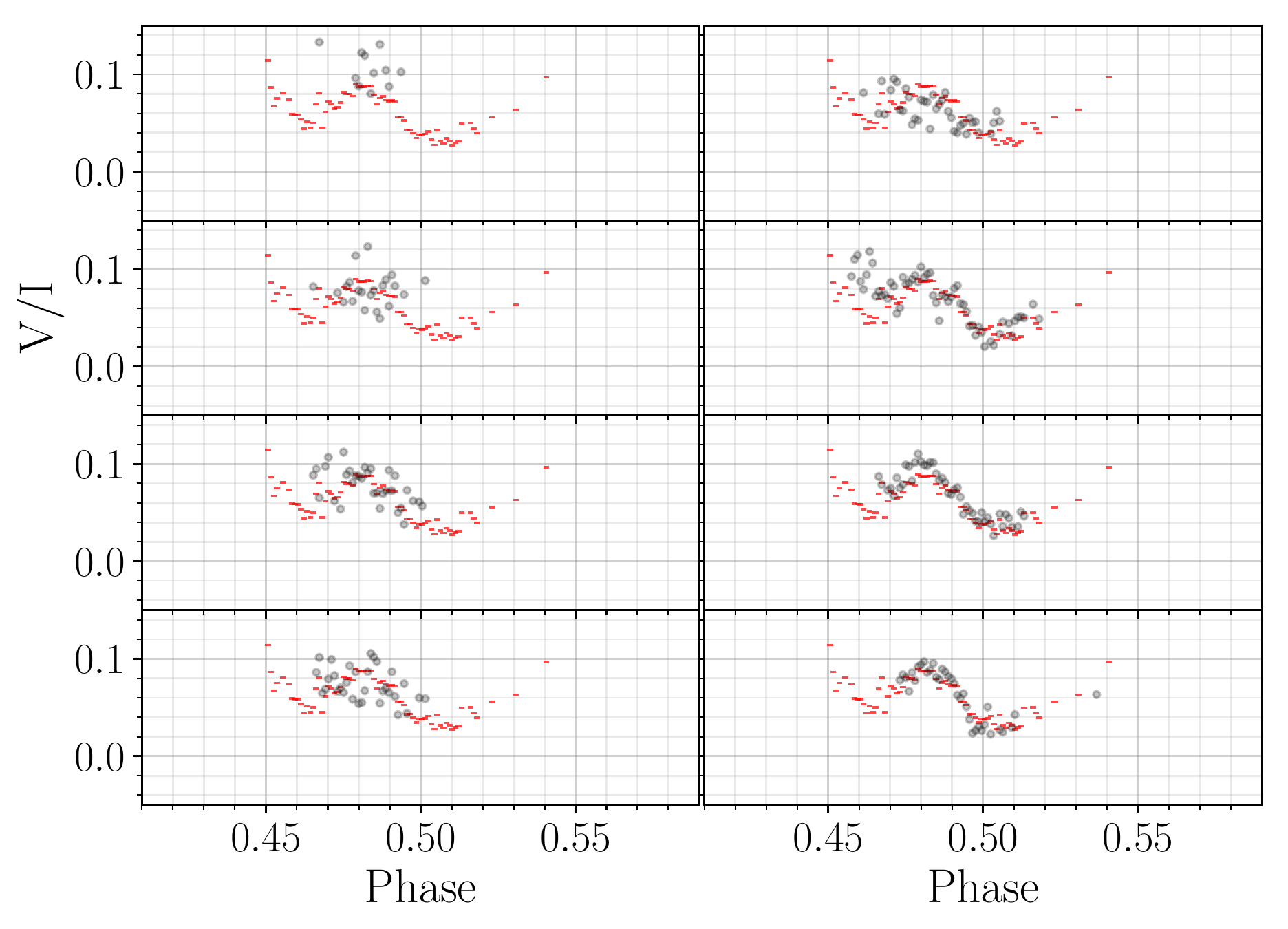}
	\includegraphics[width=\columnwidth]{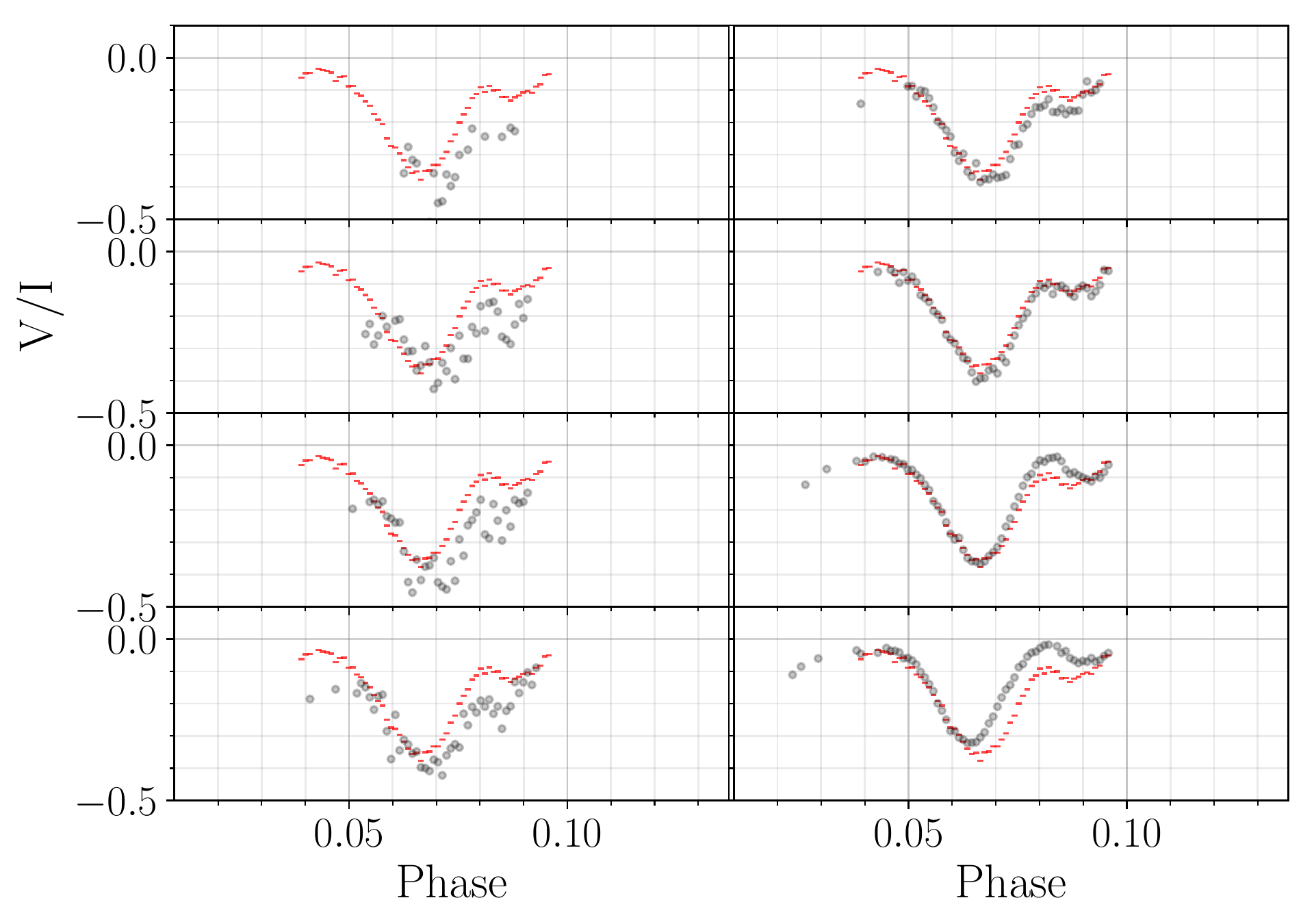}

    \caption{Plots shown are described in Figure \ref{fig:profs}.}
    \label{fig:profs1}
\end{figure*}

Figure \ref{fig:profs1} shows the polarization pulse profile for each pulsar. 
Two examples are shown in the main text in Figure \ref{fig:profs}.

\begin{figure*}
\includegraphics[width=0.693\columnwidth]{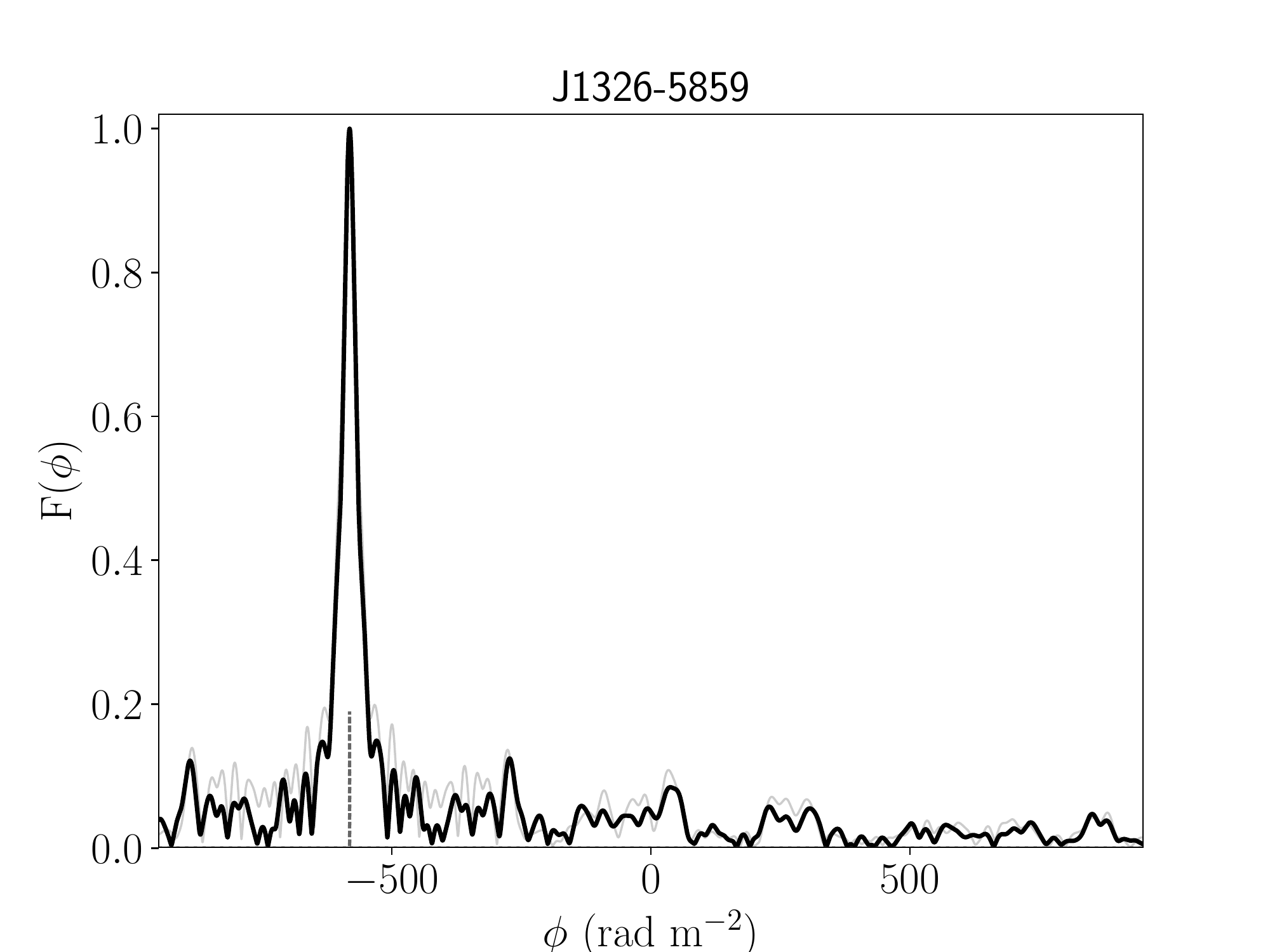}
\includegraphics[width=0.693\columnwidth]{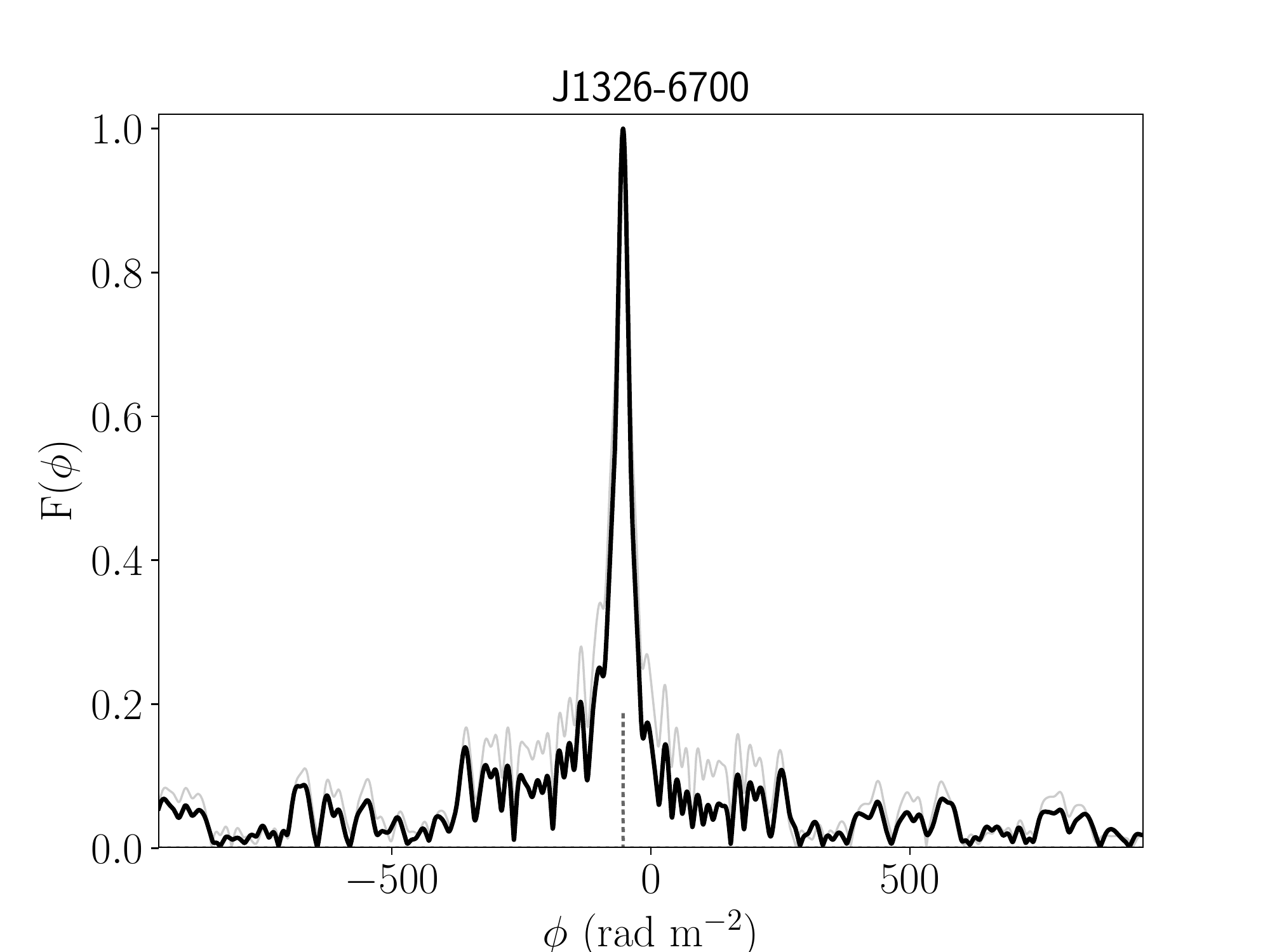}
\includegraphics[width=0.693\columnwidth]{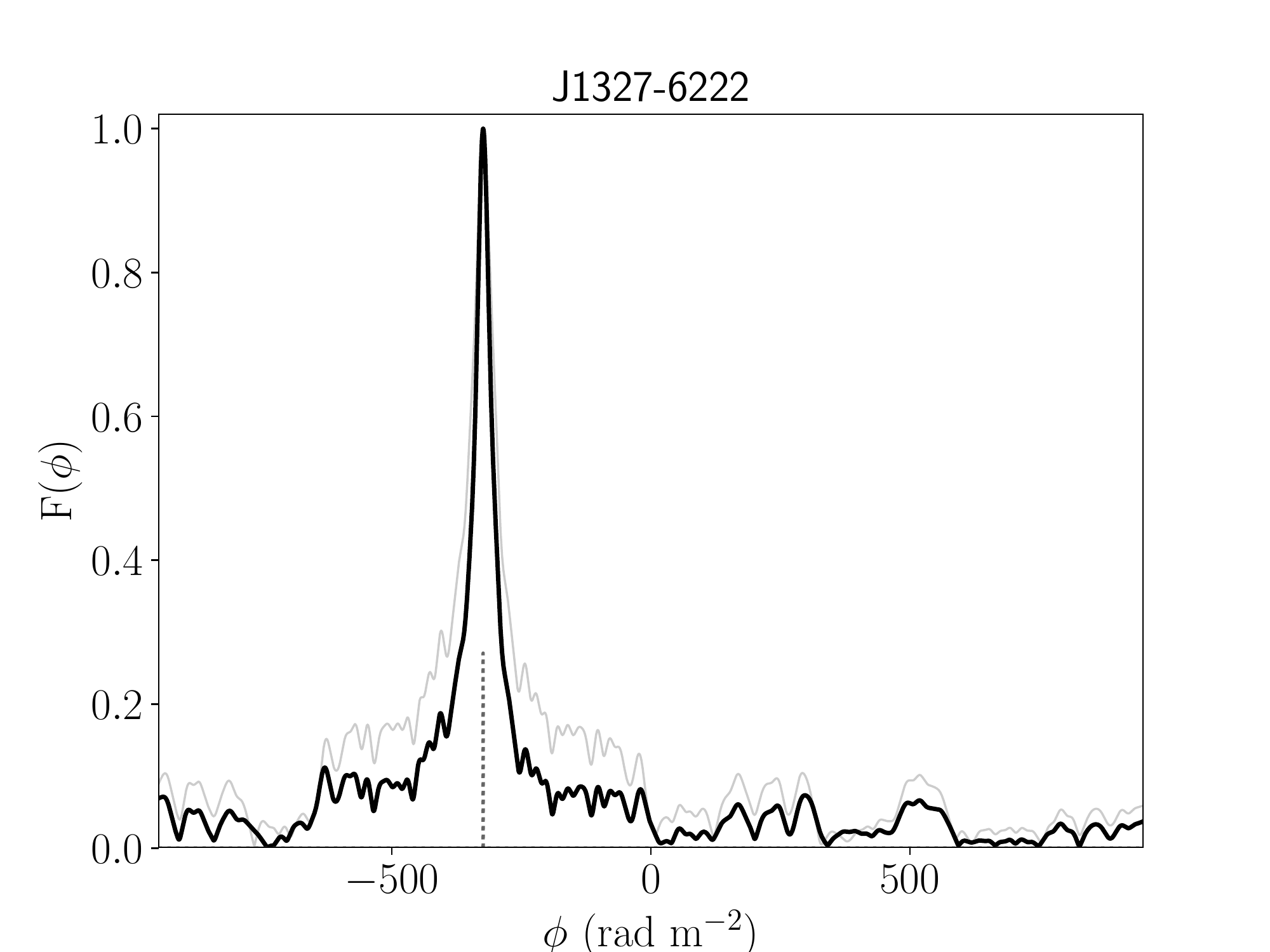}

\includegraphics[width=0.693\columnwidth]{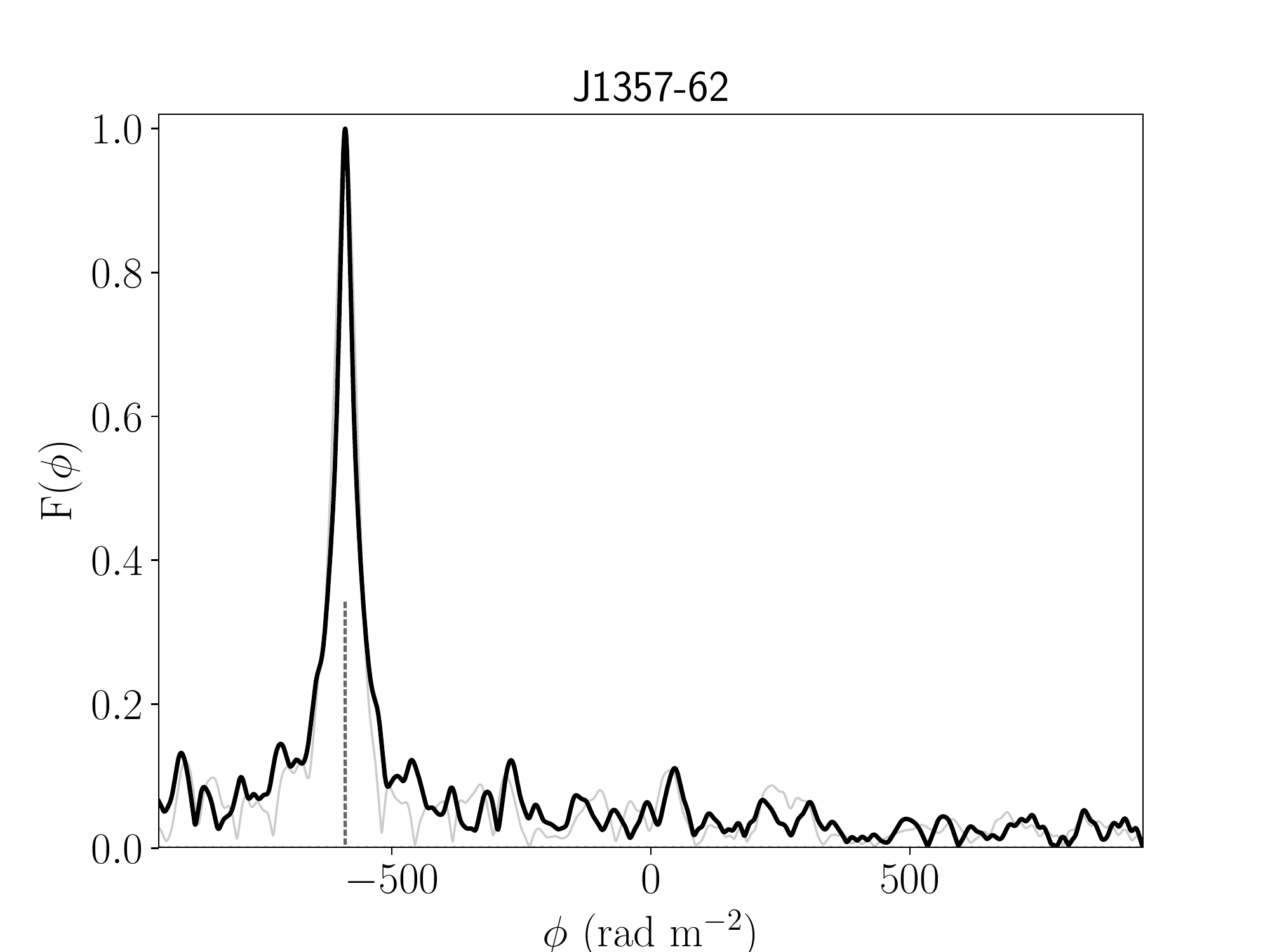}
\includegraphics[width=0.693\columnwidth]{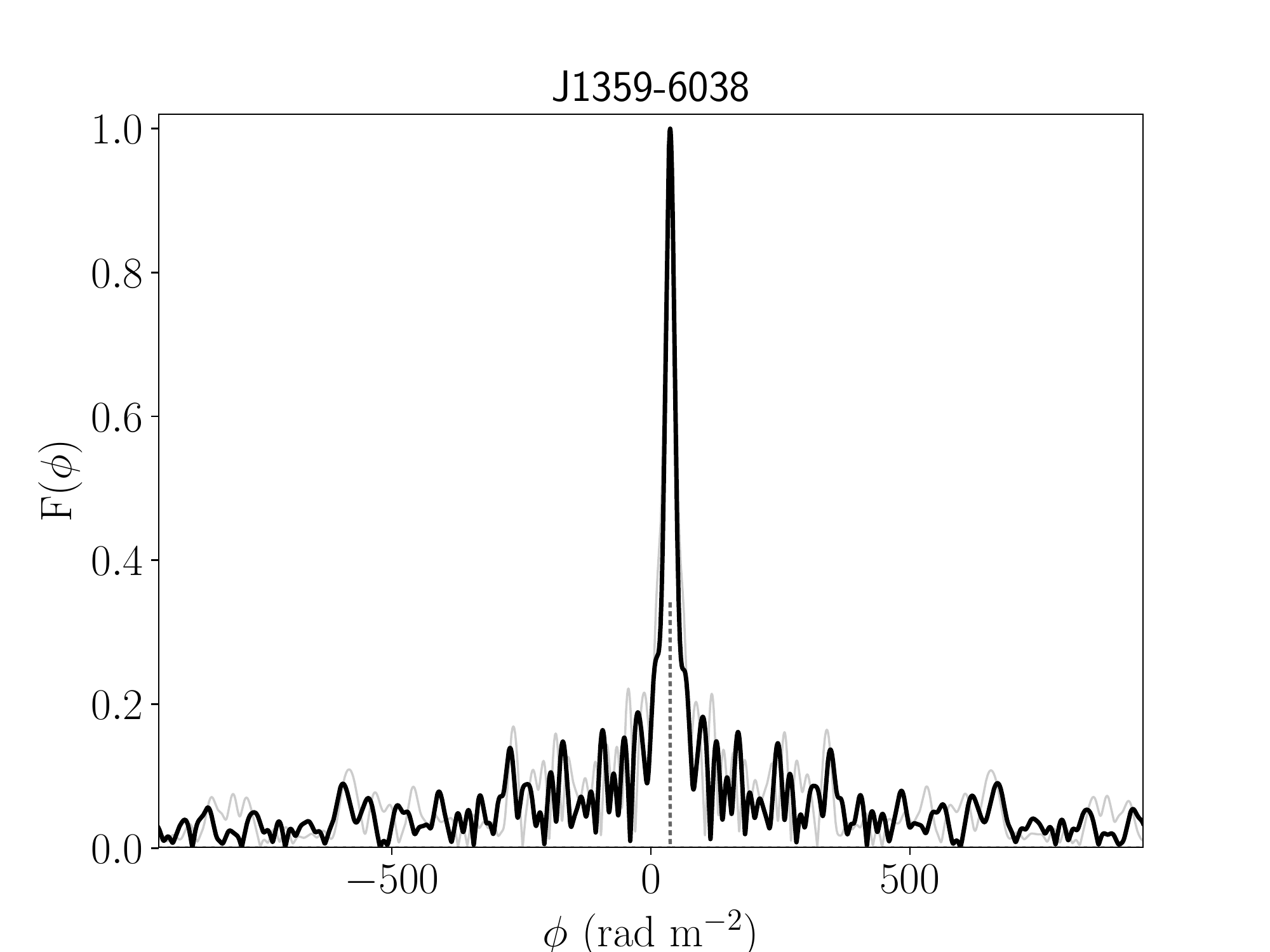}
\includegraphics[width=0.693\columnwidth]{FDFs/J1430-6623.pdf}

\includegraphics[width=0.693\columnwidth]{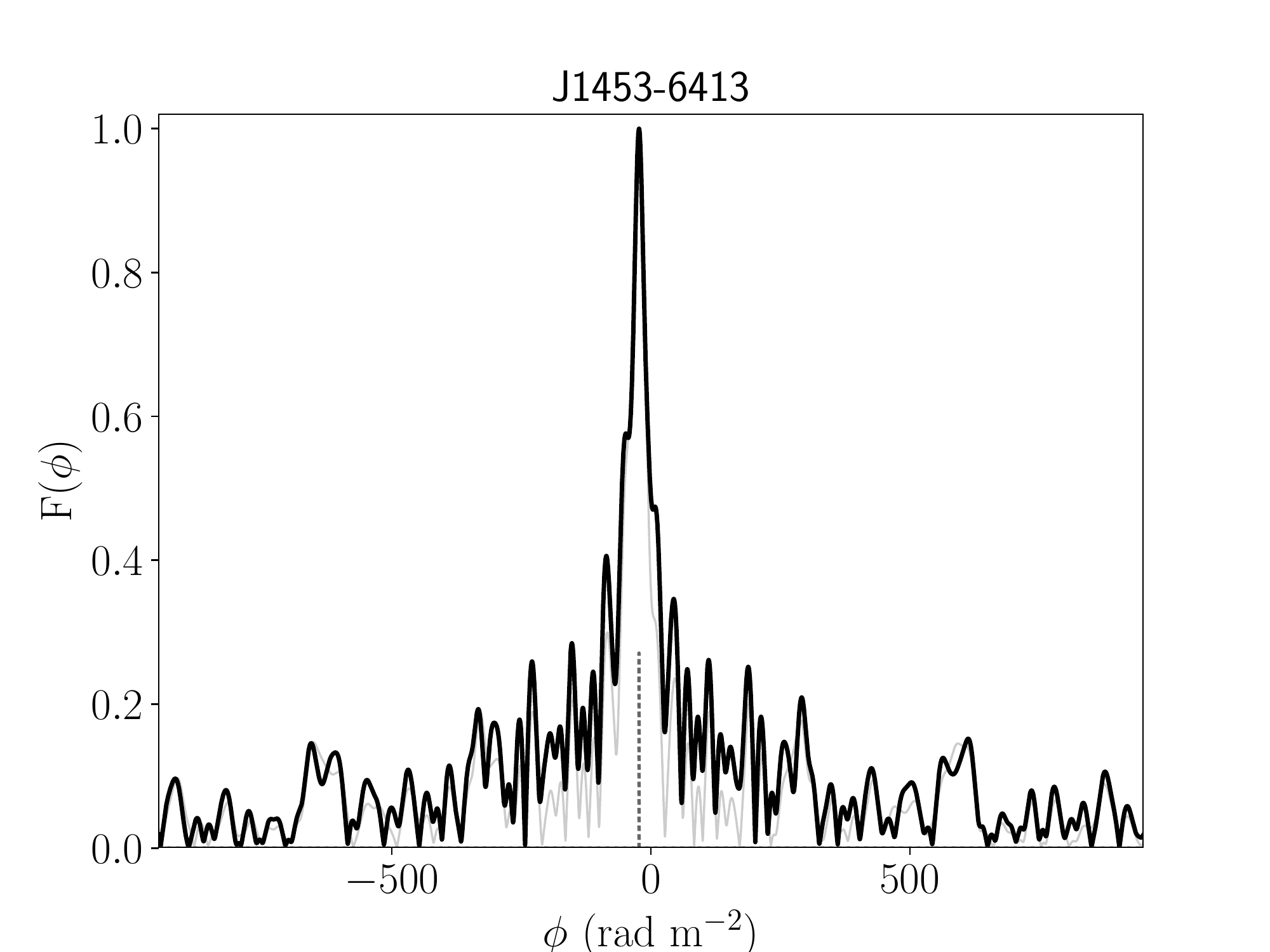}
\includegraphics[width=0.693\columnwidth]{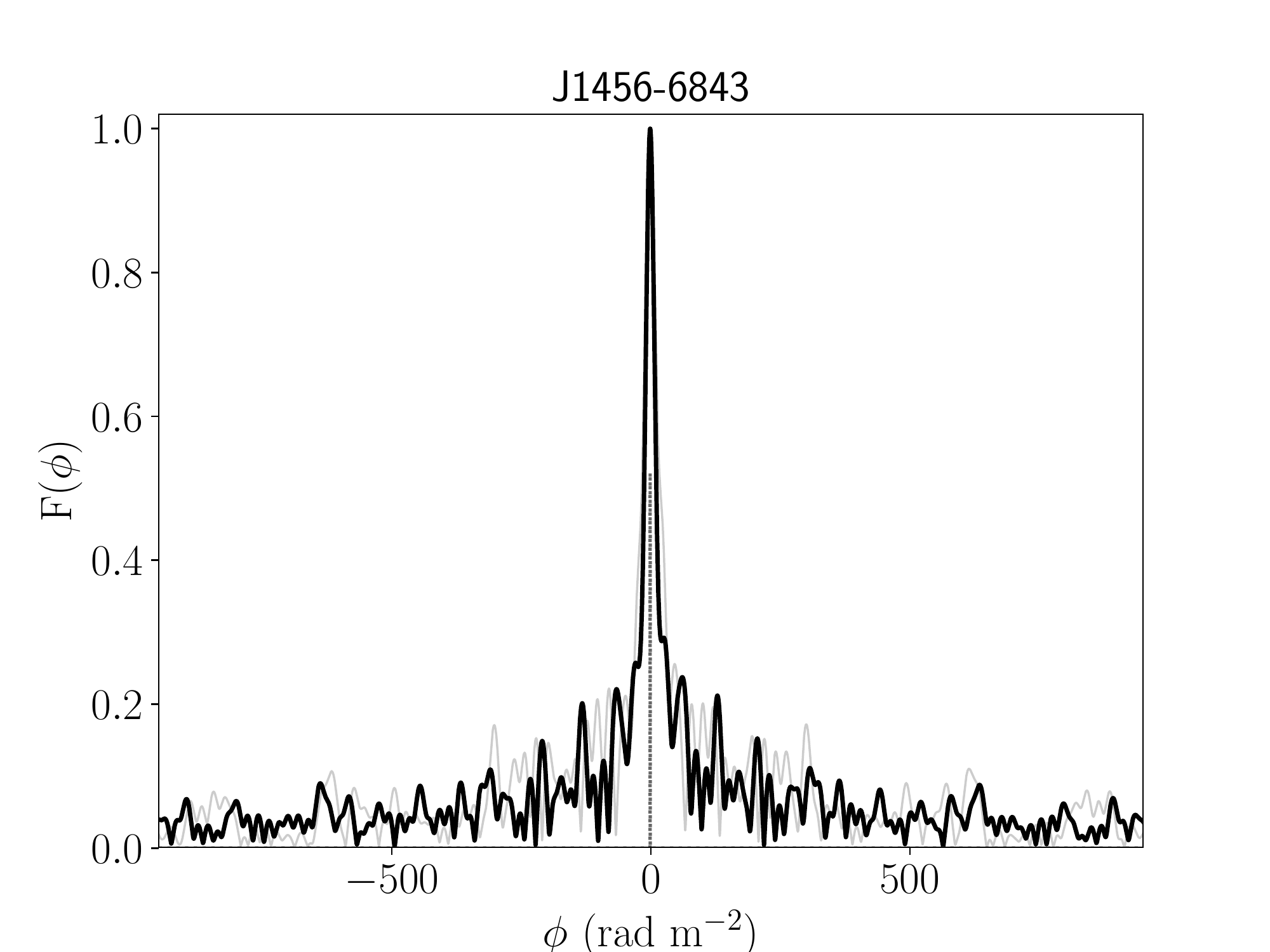}
\includegraphics[width=0.693\columnwidth]{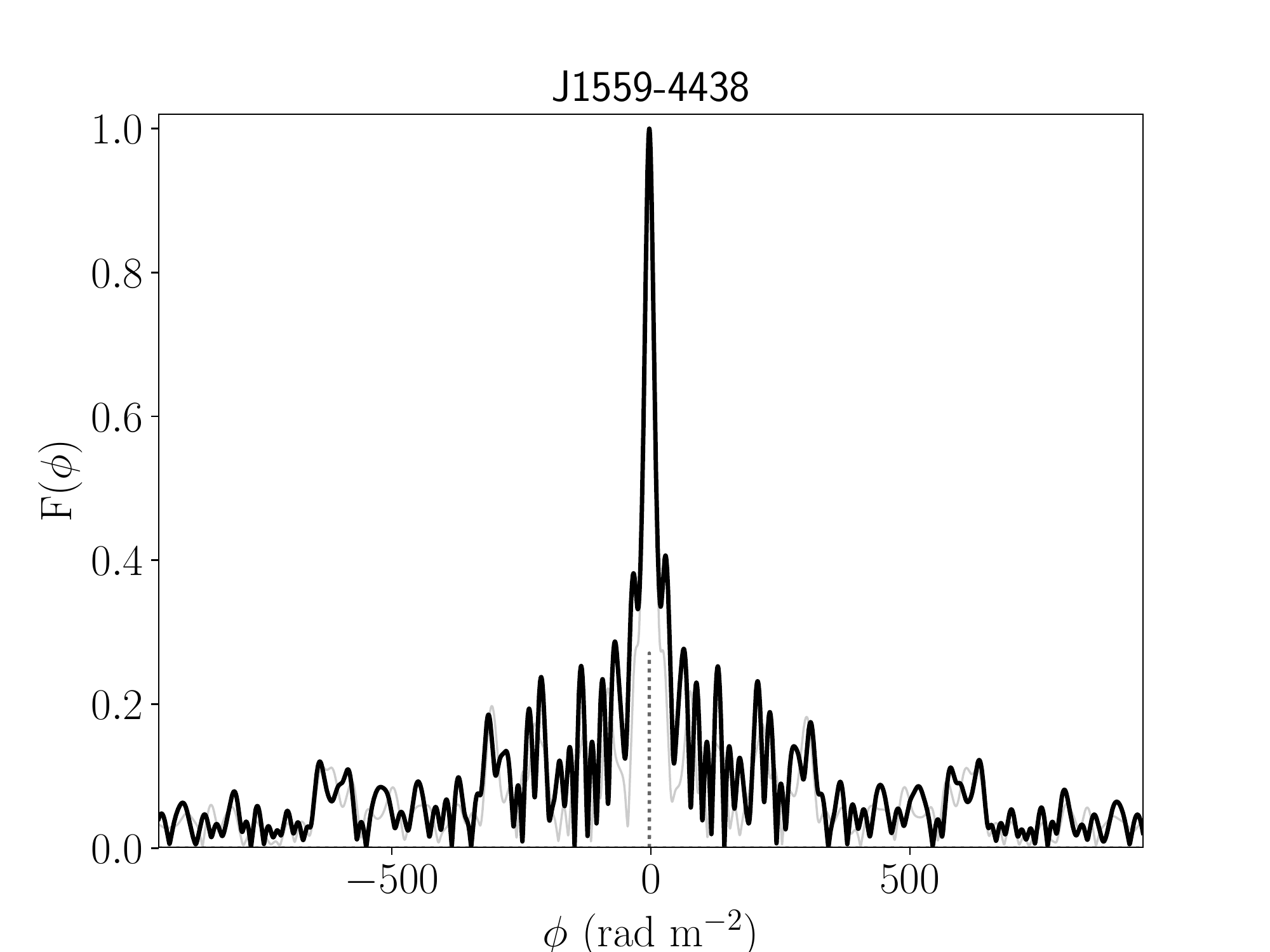}

\includegraphics[width=0.693\columnwidth]{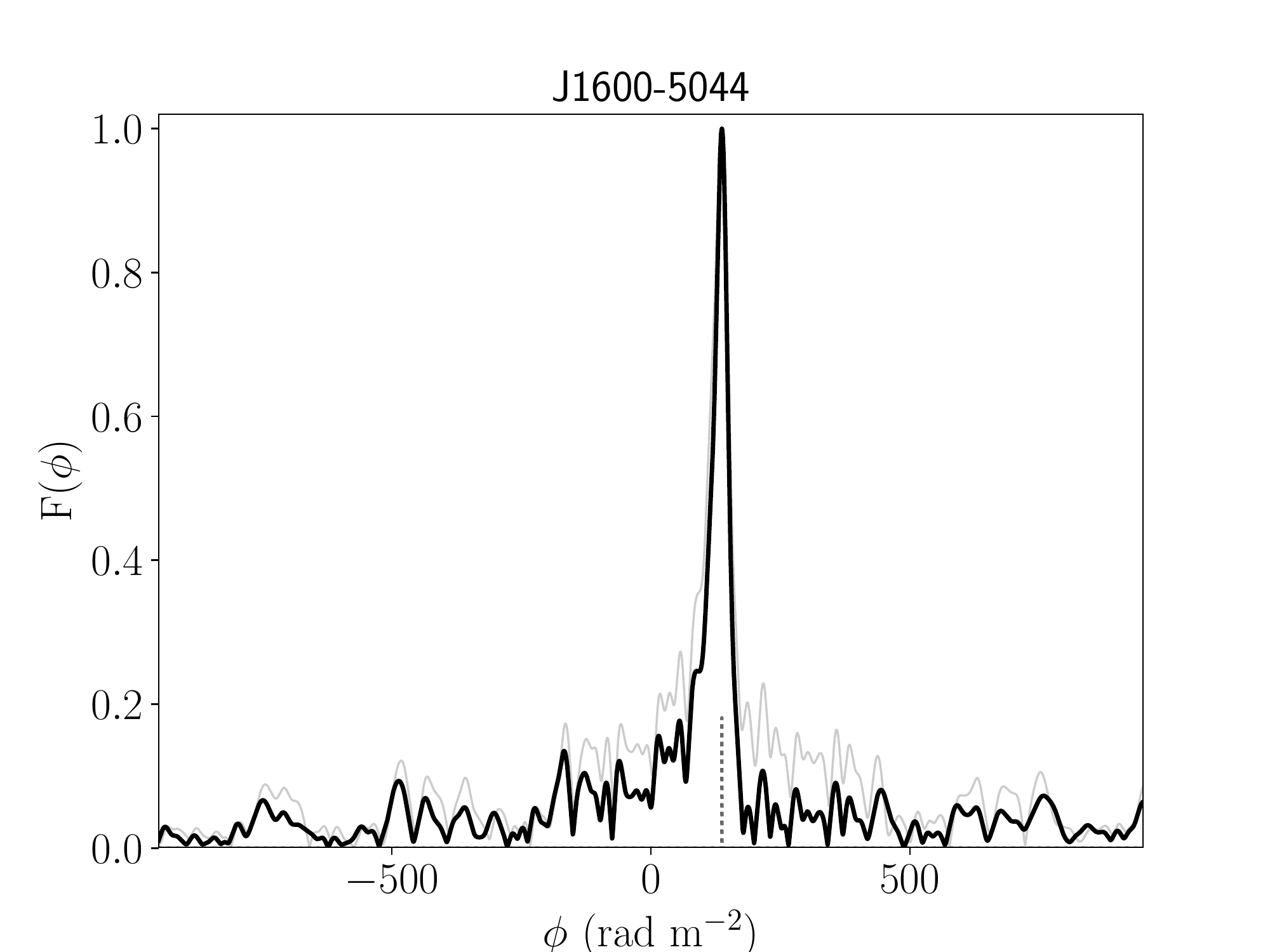}
\includegraphics[width=0.693\columnwidth]{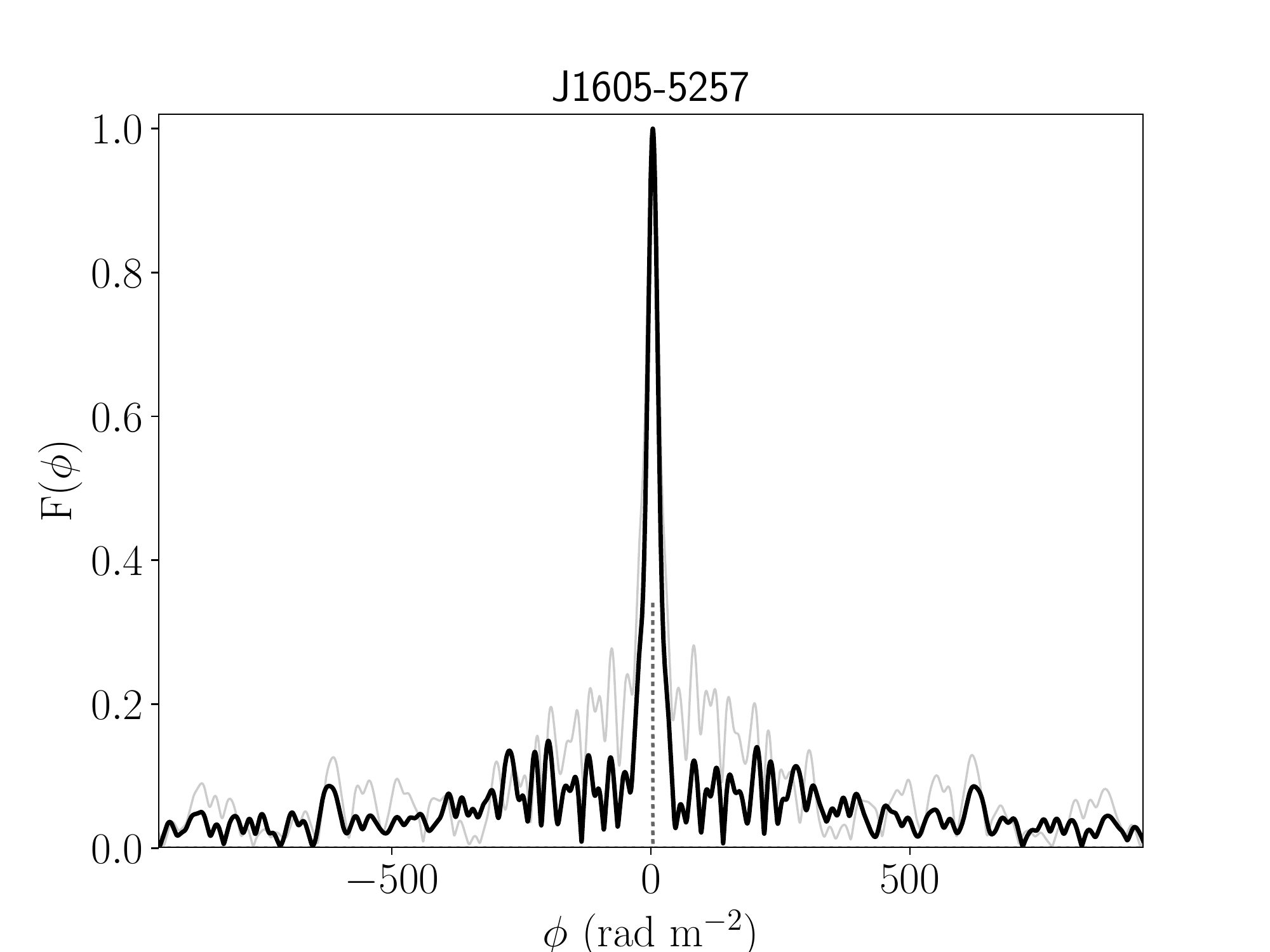}
\includegraphics[width=0.693\columnwidth]{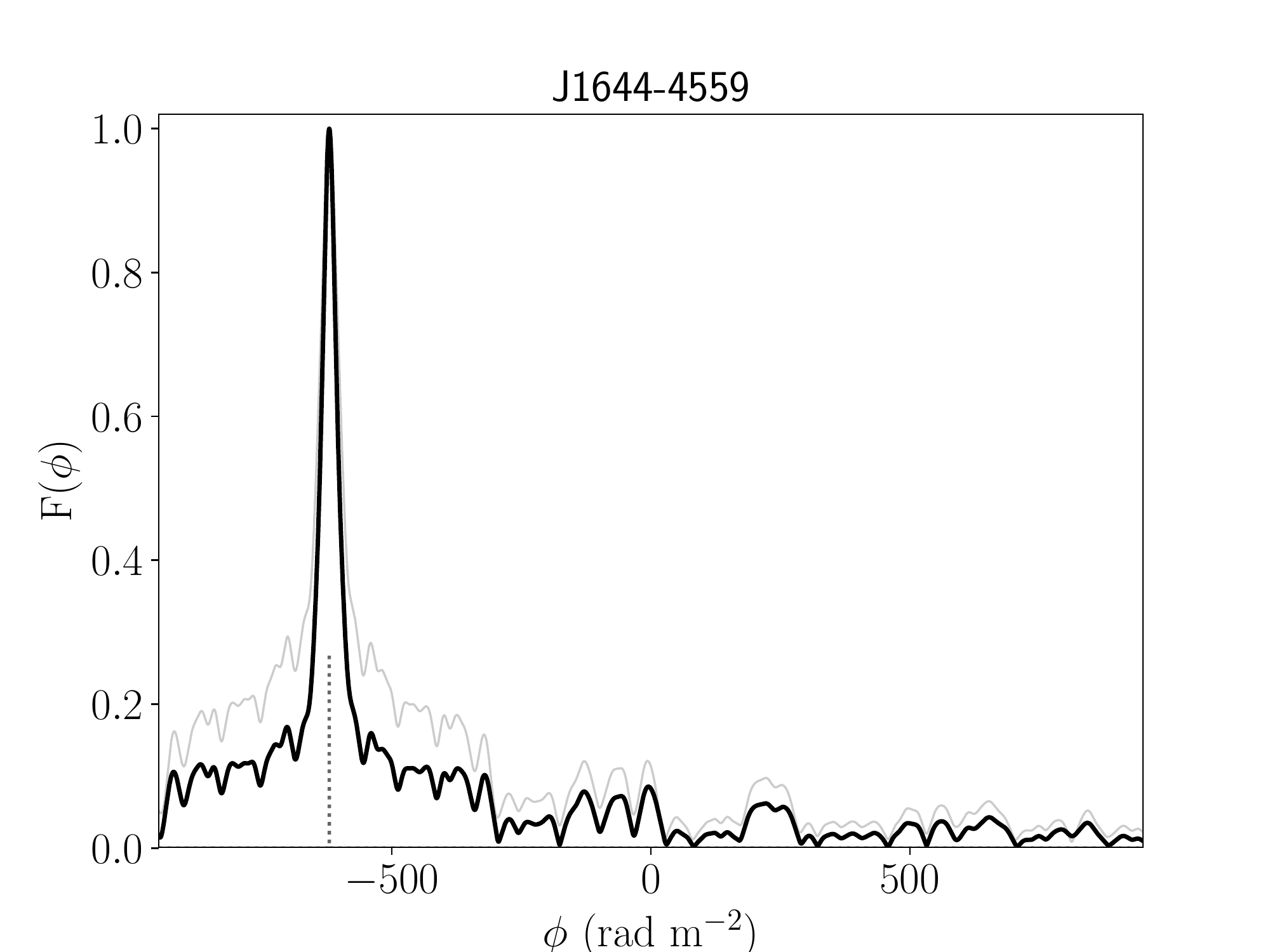}

\includegraphics[width=0.693\columnwidth]{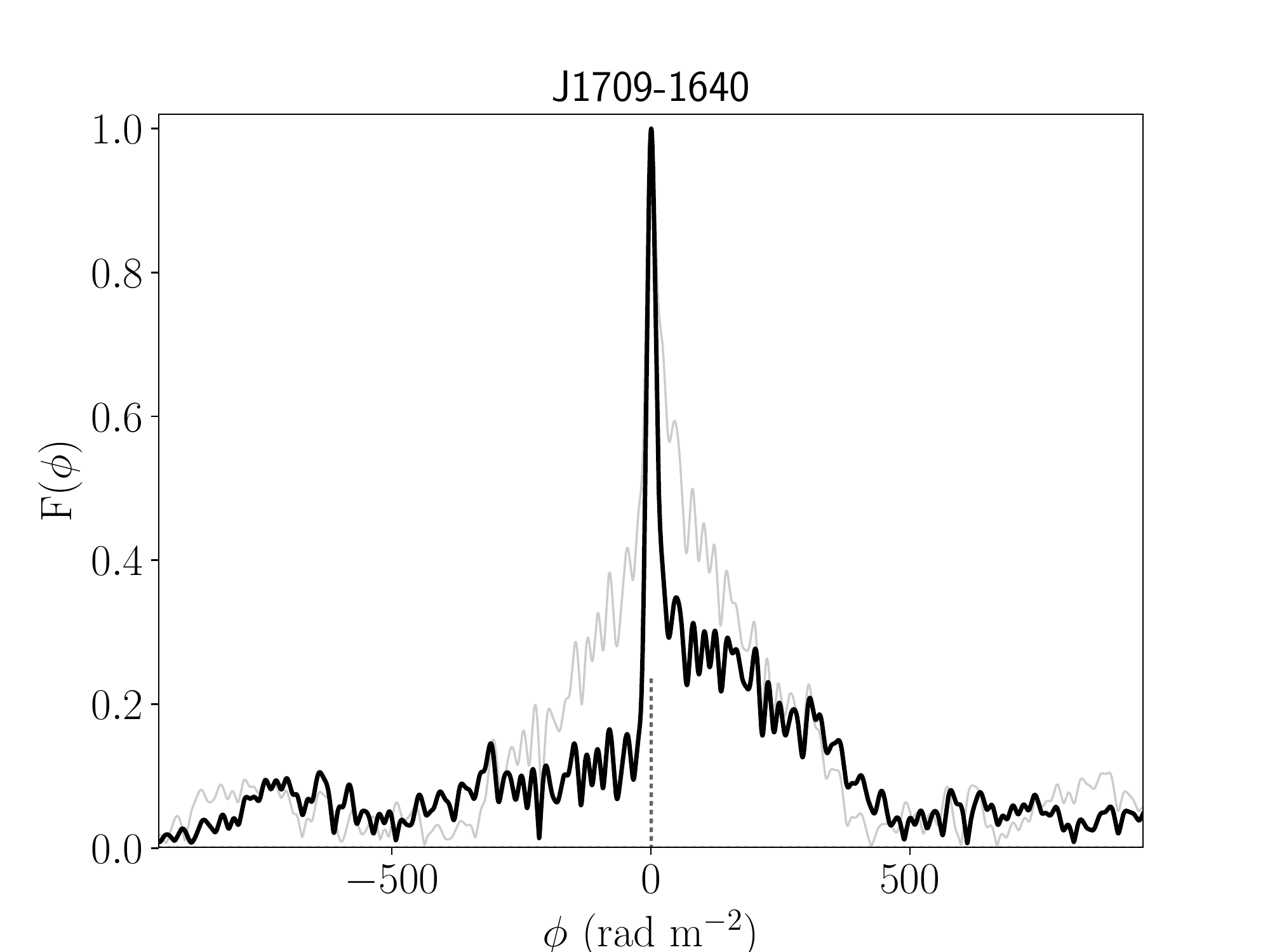}
\includegraphics[width=0.693\columnwidth]{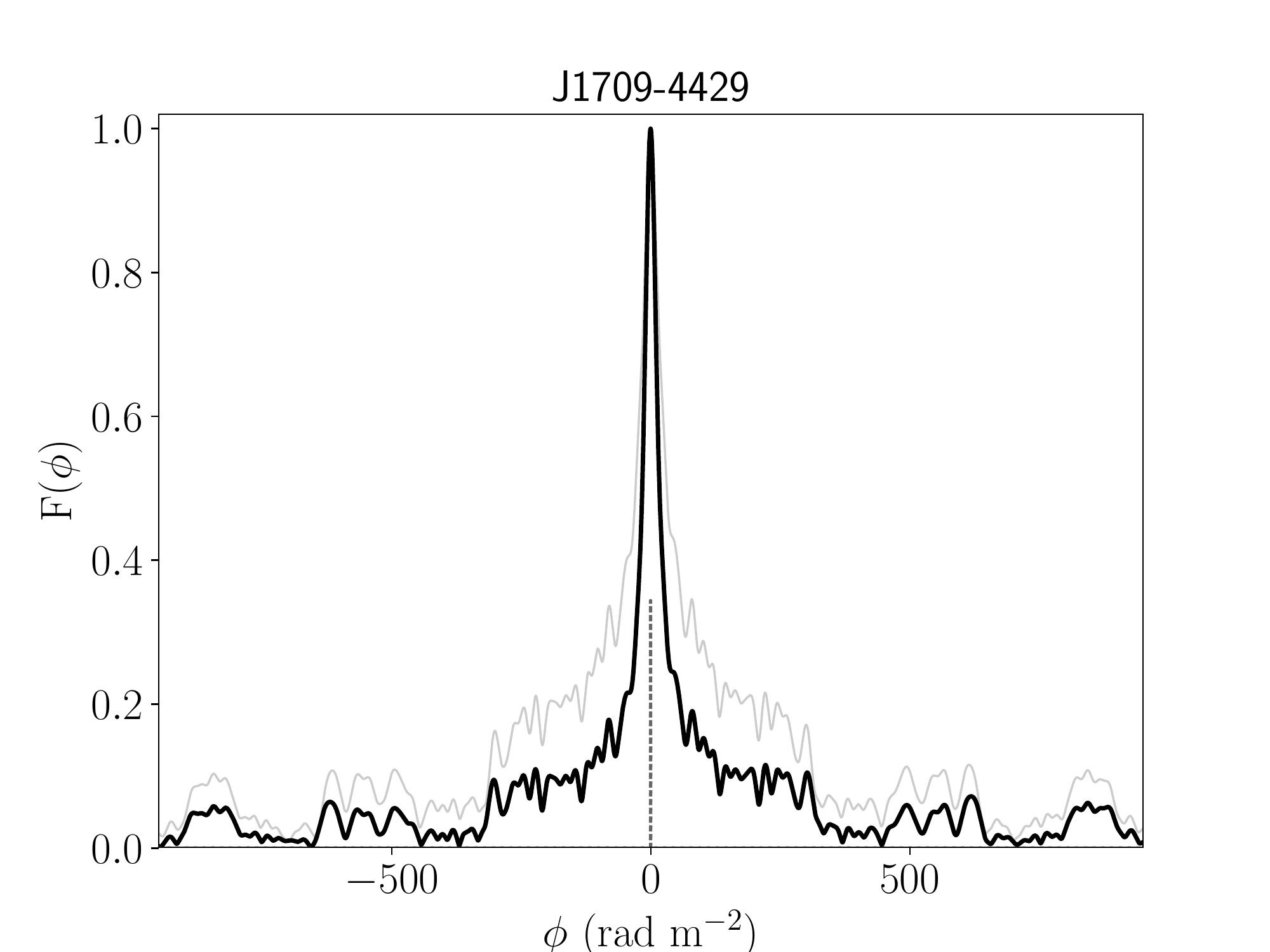}
\includegraphics[width=0.693\columnwidth]{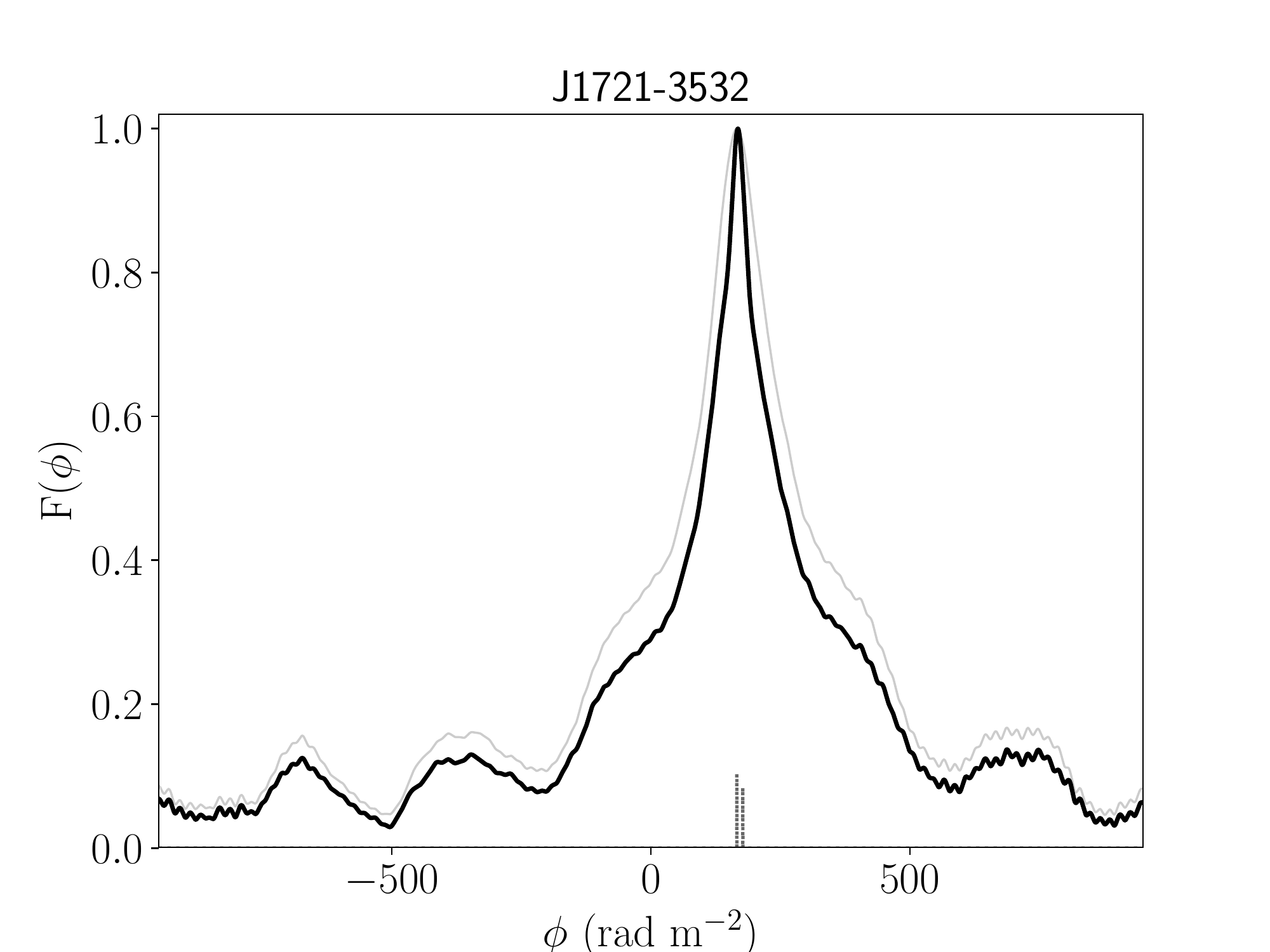}

    \caption{...Figure \ref{fig:FDFs1} continued...}
\end{figure*}

\begin{figure*}
\includegraphics[width=0.693\columnwidth]{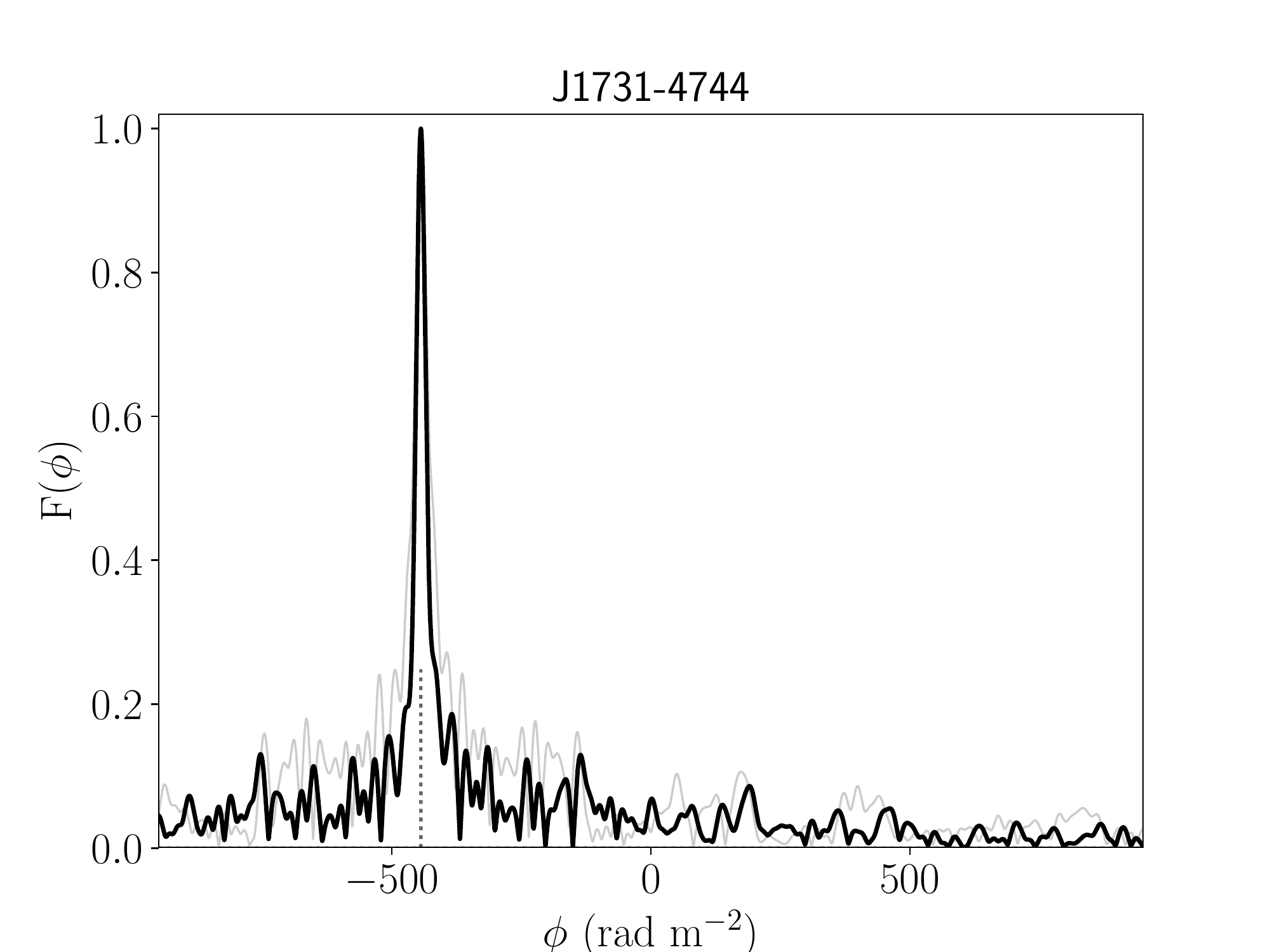}
\includegraphics[width=0.693\columnwidth]{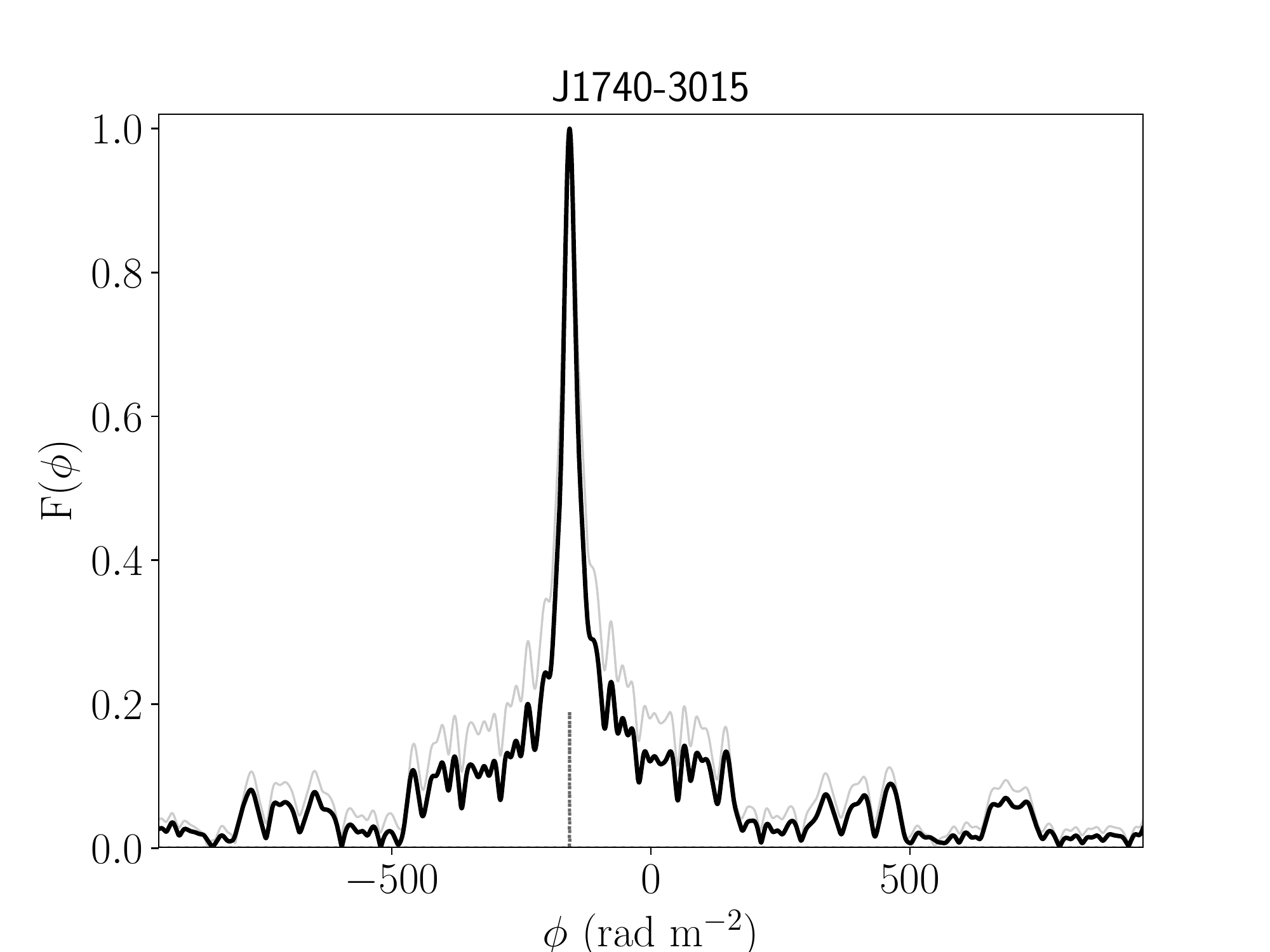}
\includegraphics[width=0.693\columnwidth]{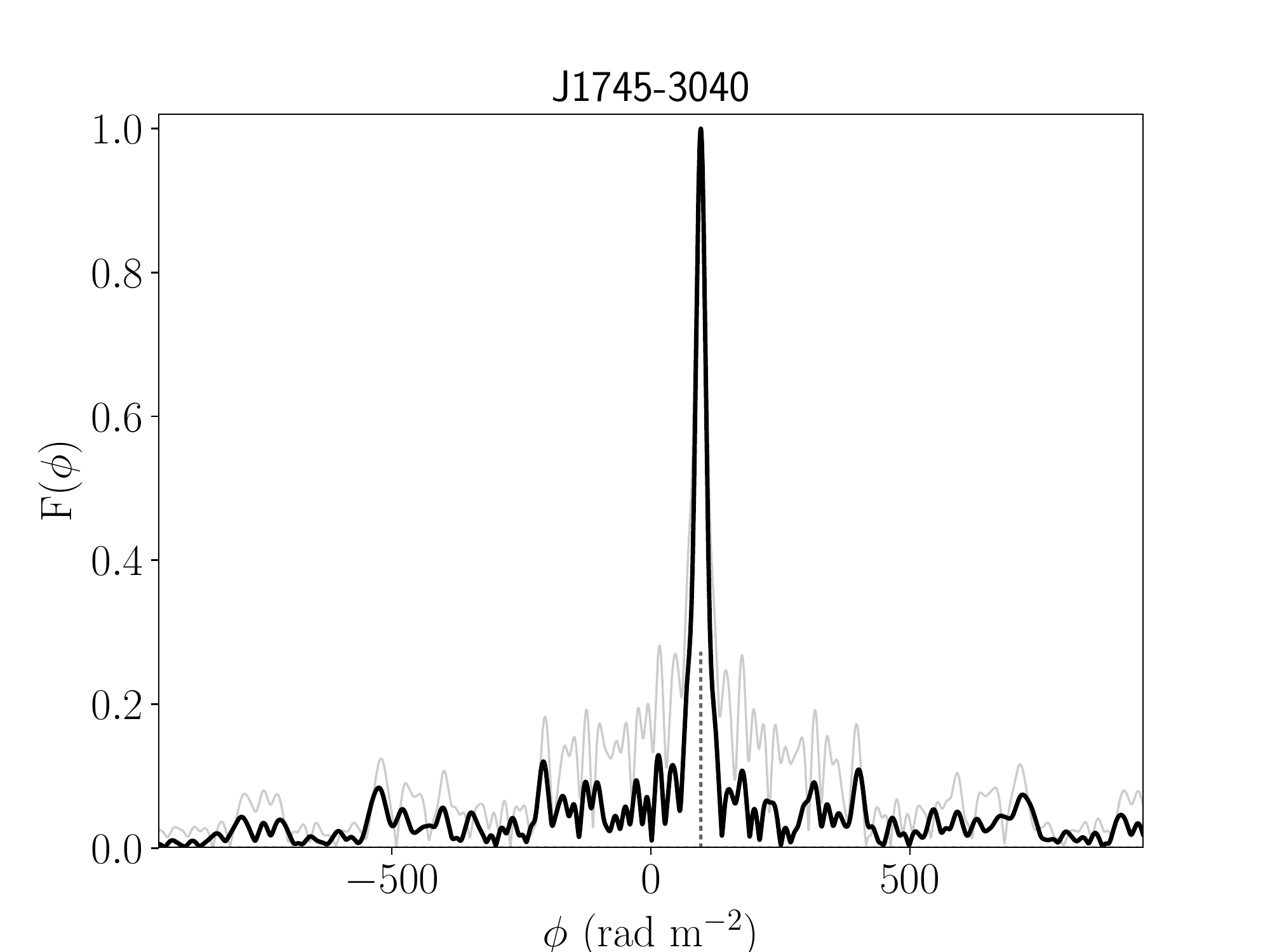}

\includegraphics[width=0.693\columnwidth]{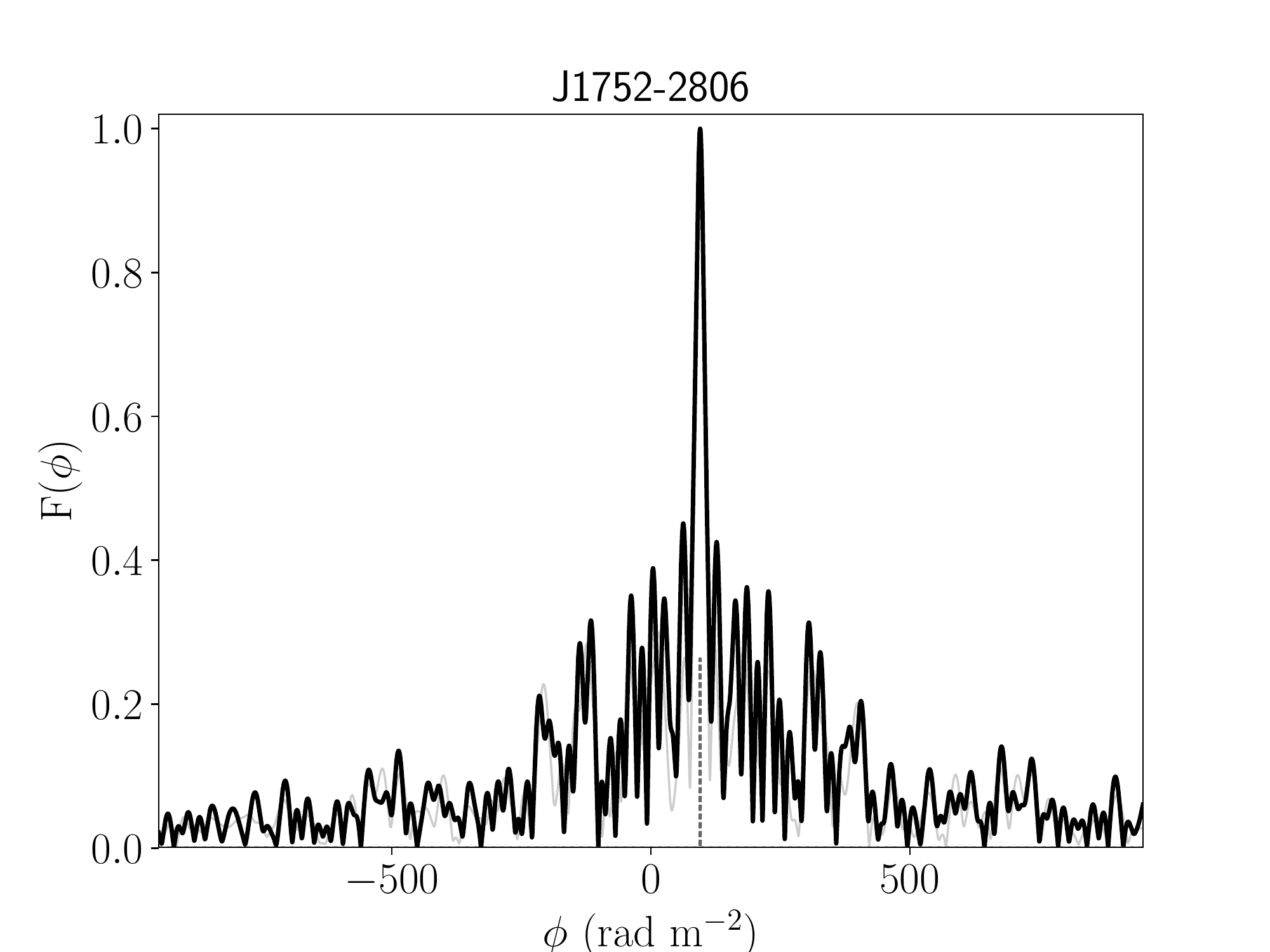}
\includegraphics[width=0.693\columnwidth]{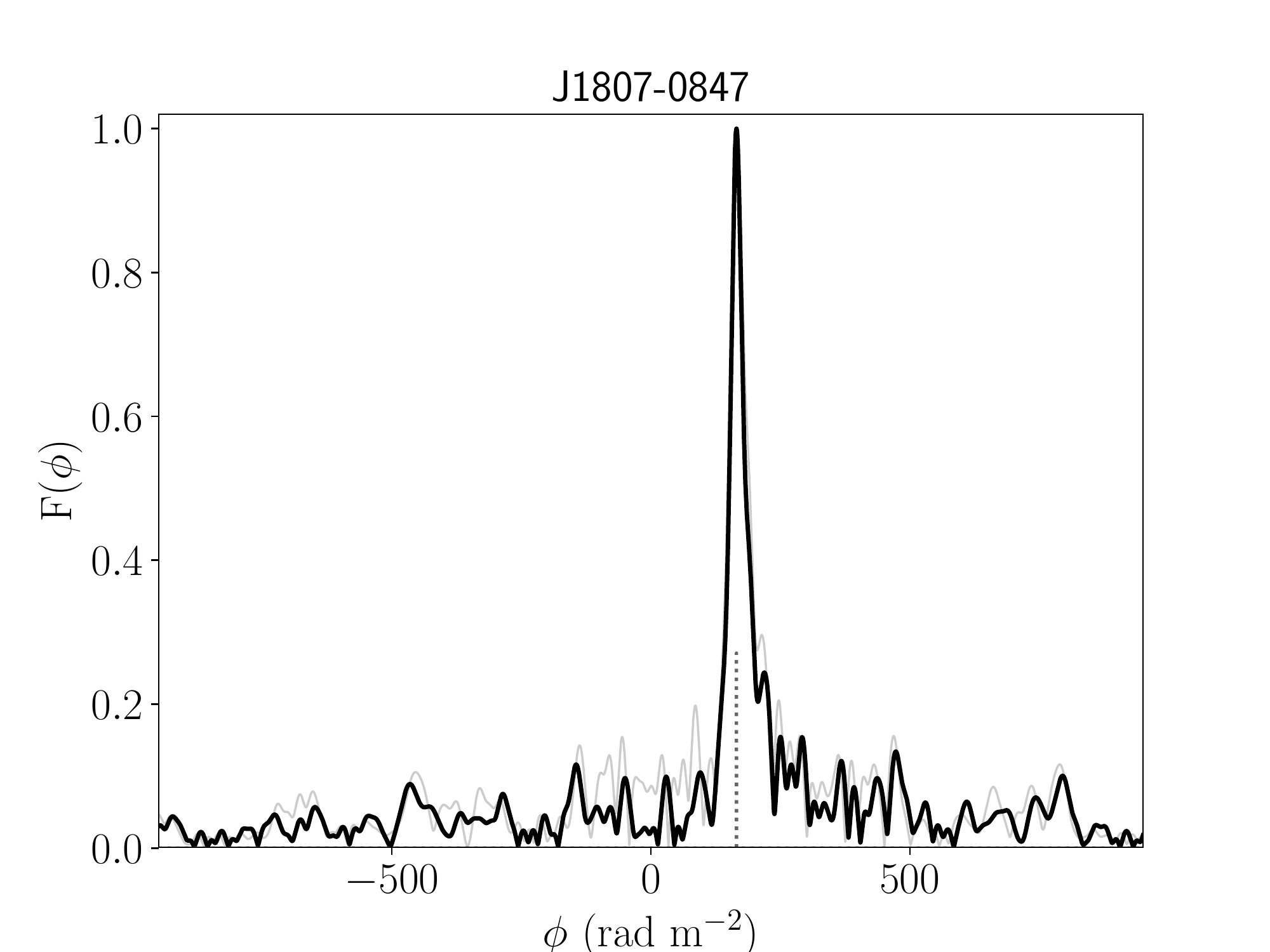}
\includegraphics[width=0.693\columnwidth]{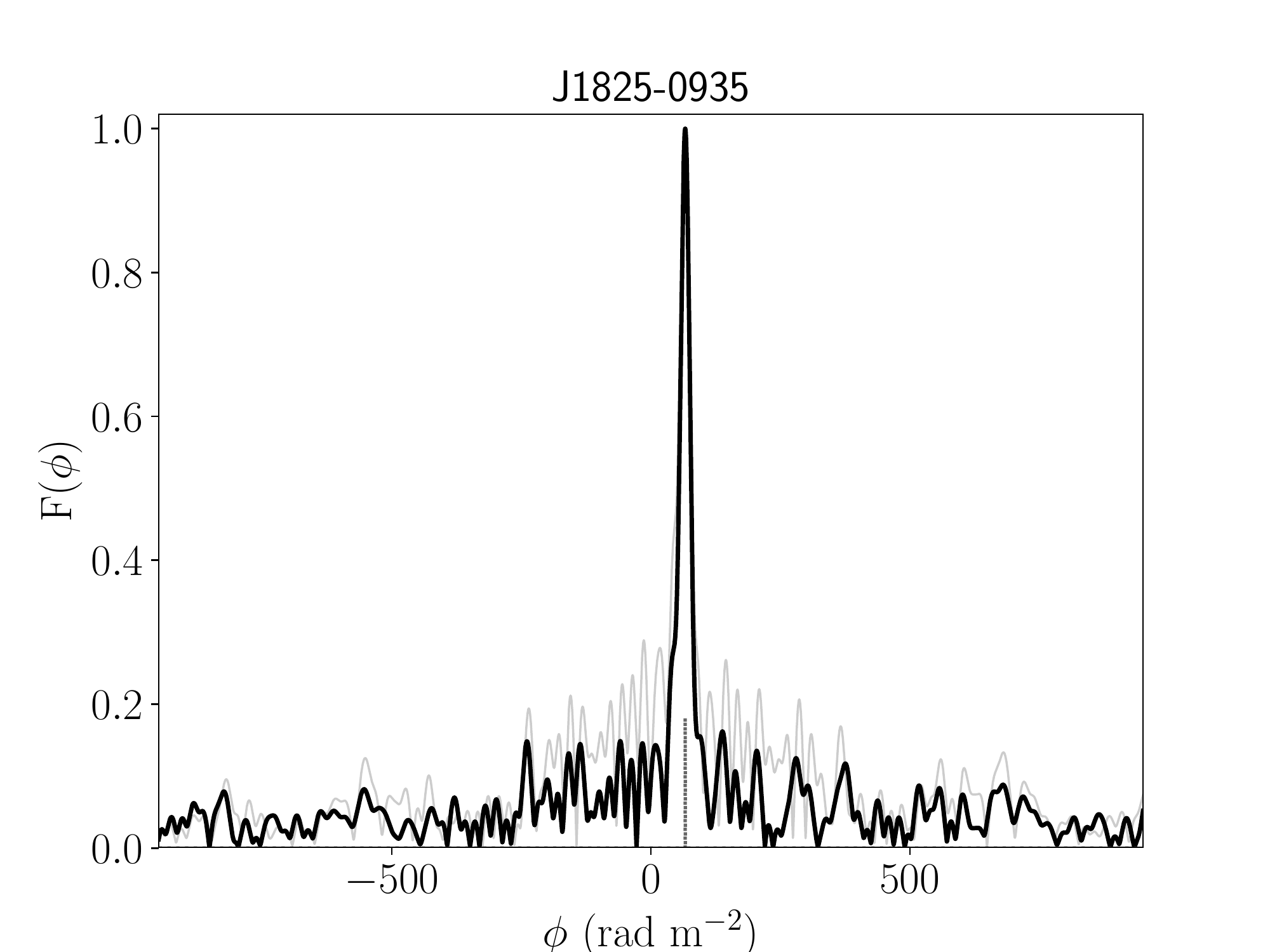}

\includegraphics[width=0.693\columnwidth]{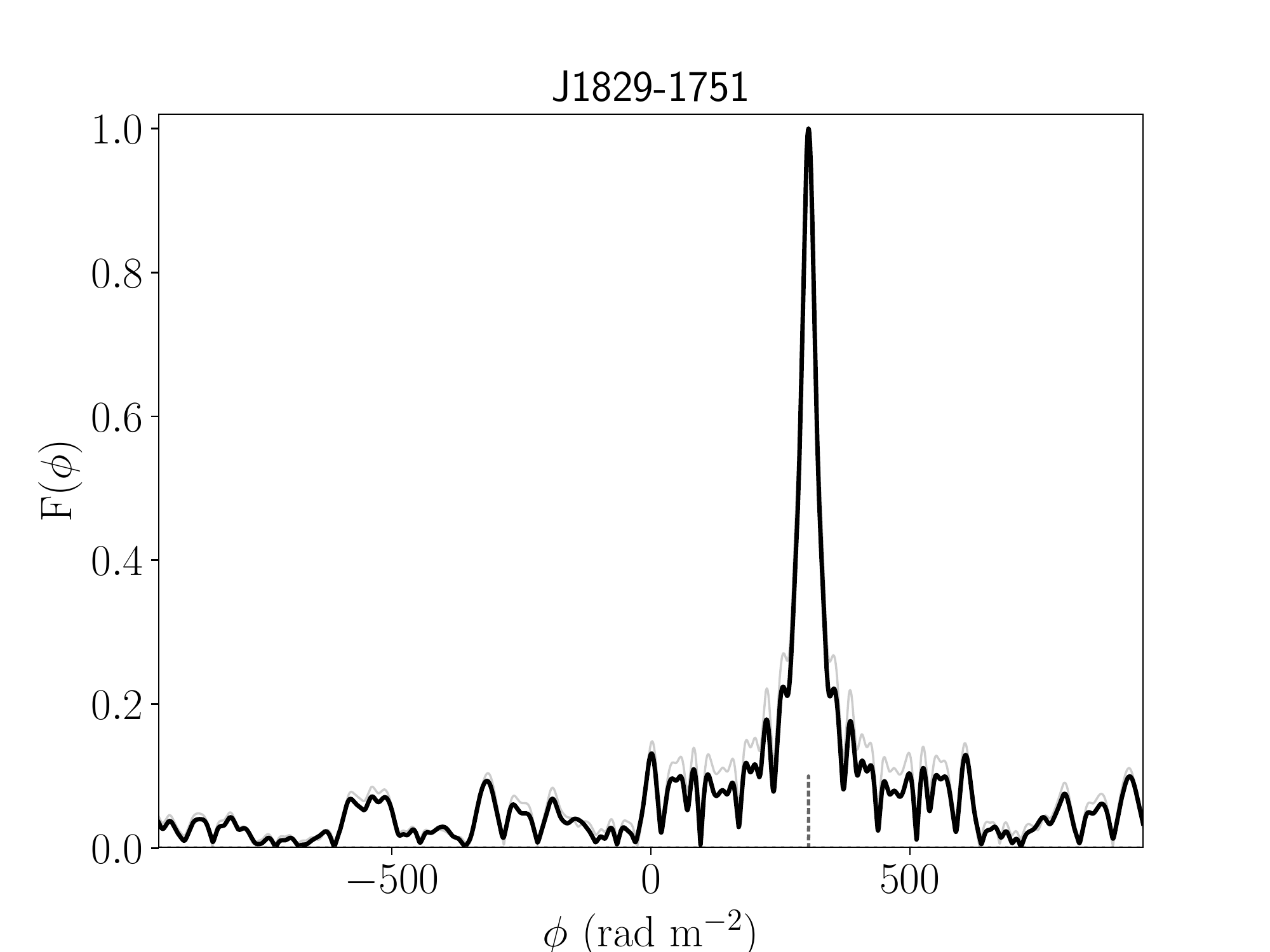}
\includegraphics[width=0.693\columnwidth]{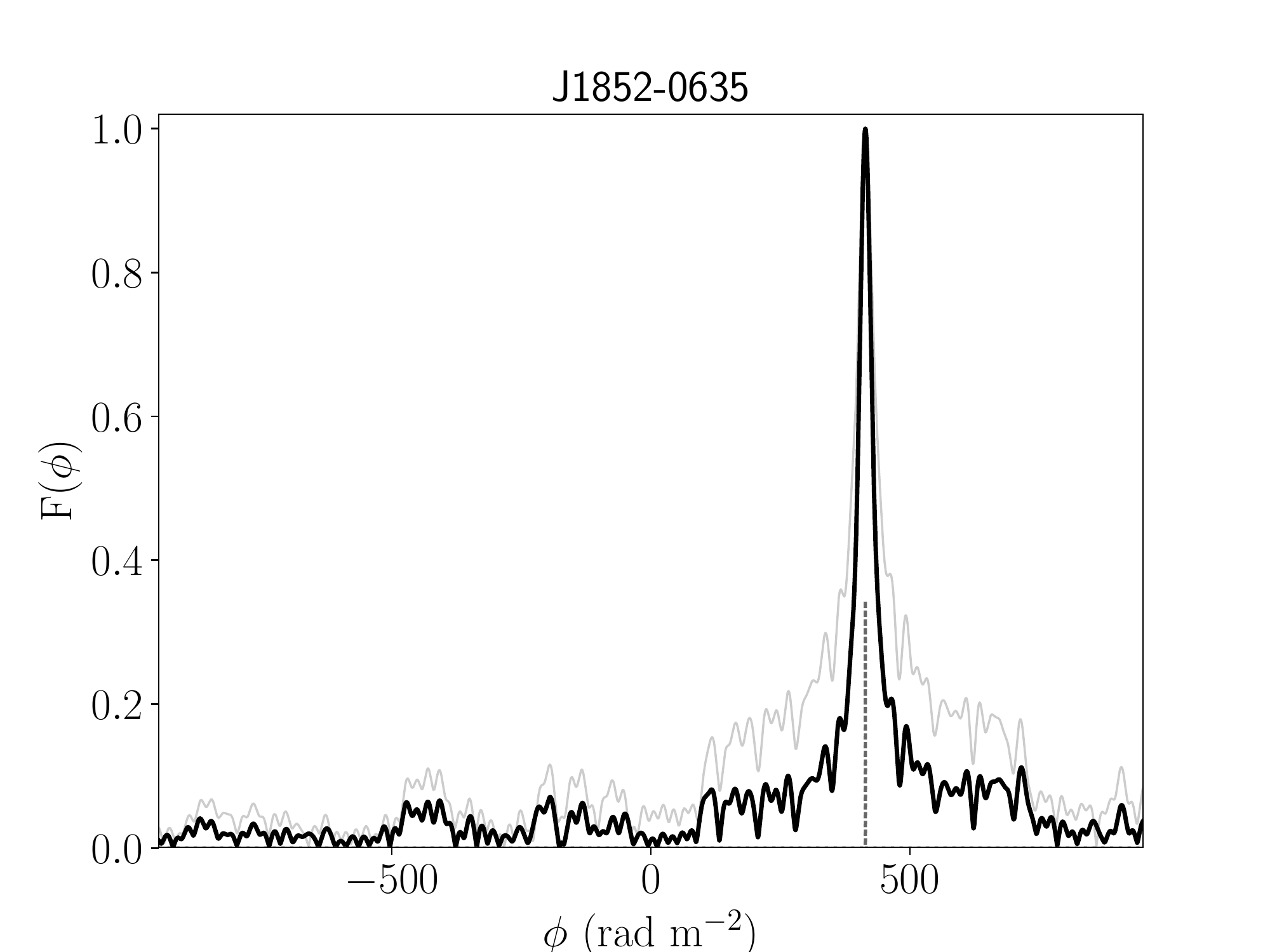}
\includegraphics[width=0.693\columnwidth]{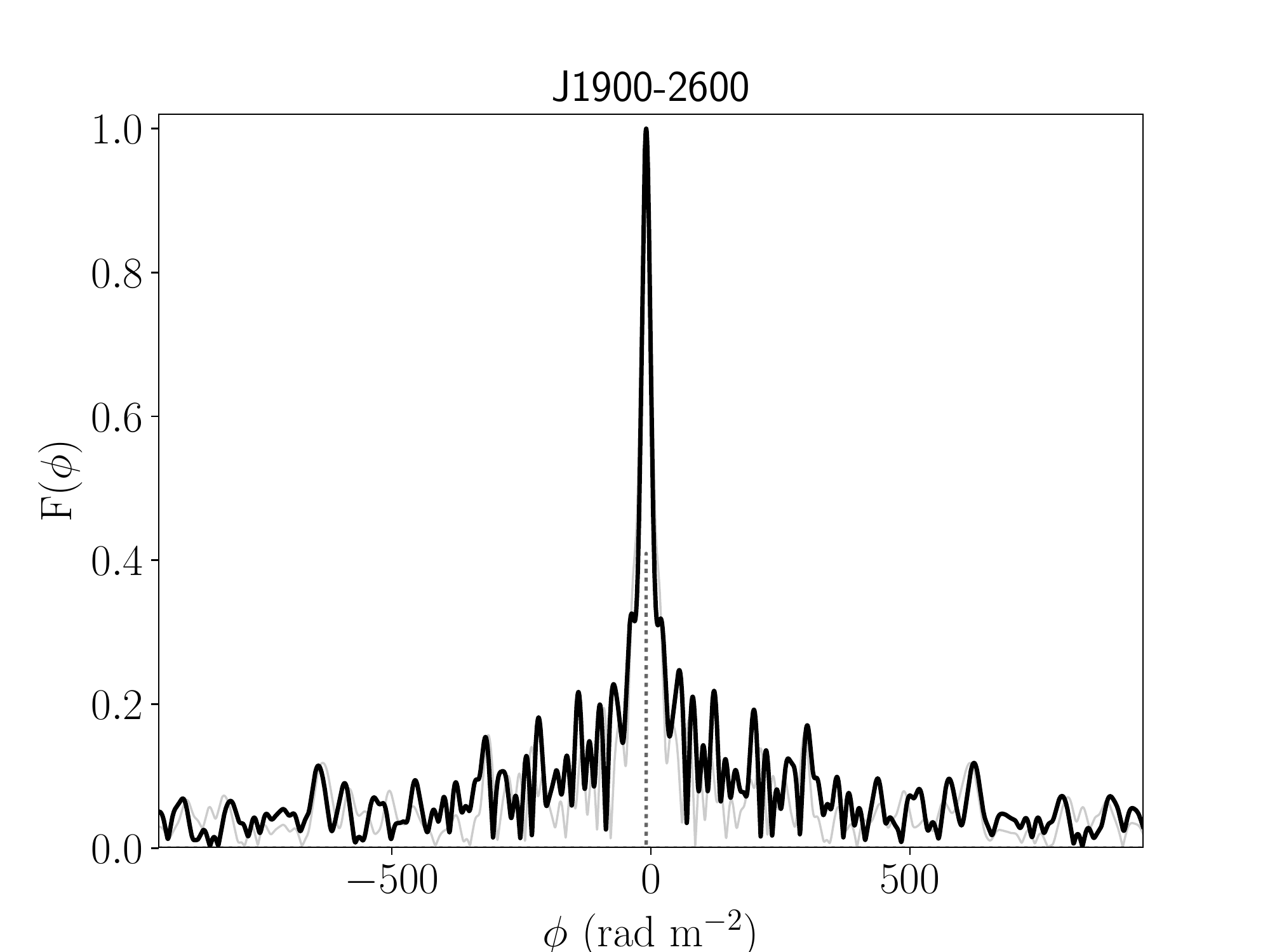}

\includegraphics[width=0.693\columnwidth]{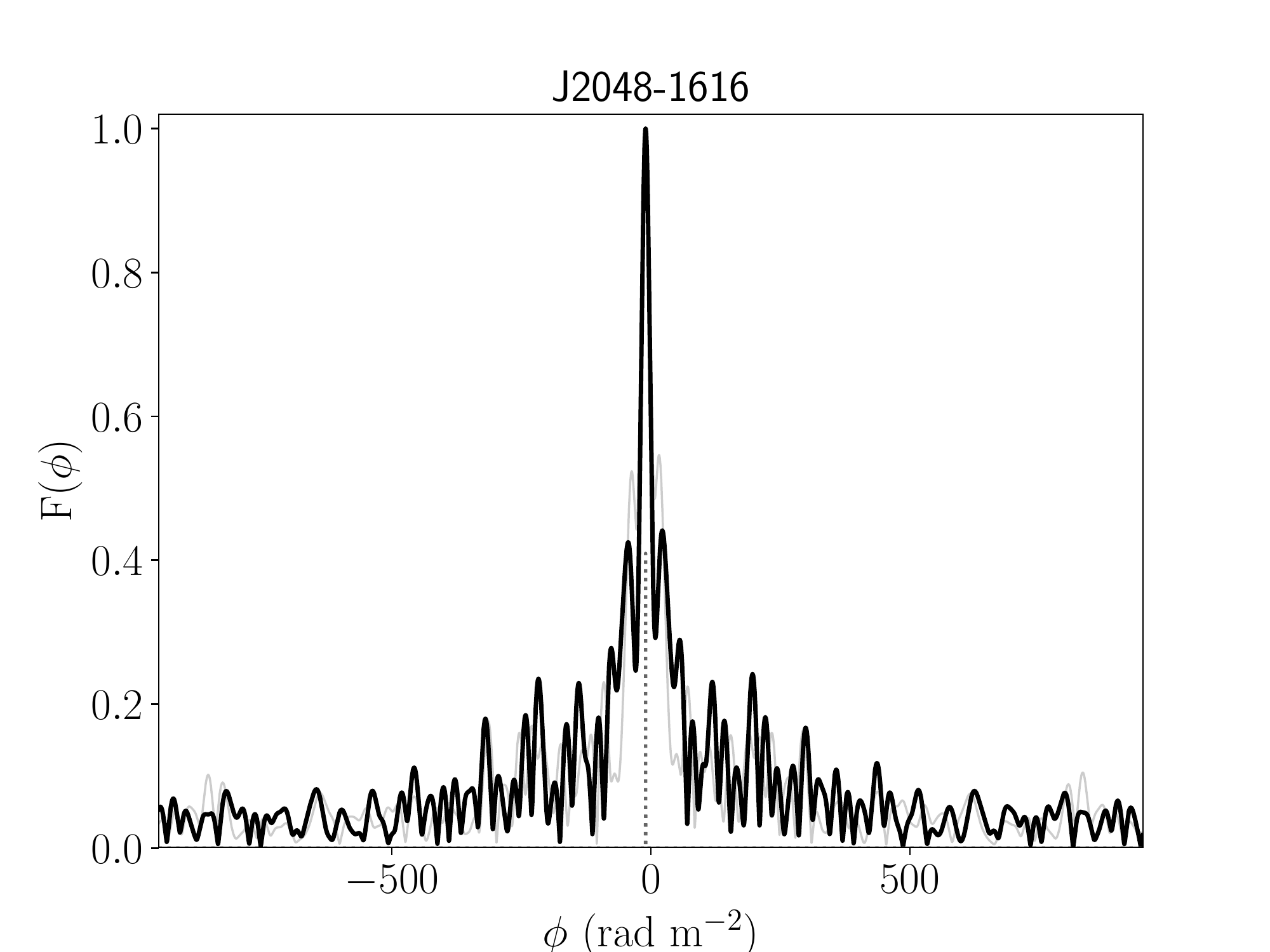}
	
    \caption{...Figure \ref{fig:FDFs1} continued.}
\end{figure*}


\begin{figure*} 
	\includegraphics[width=\columnwidth]{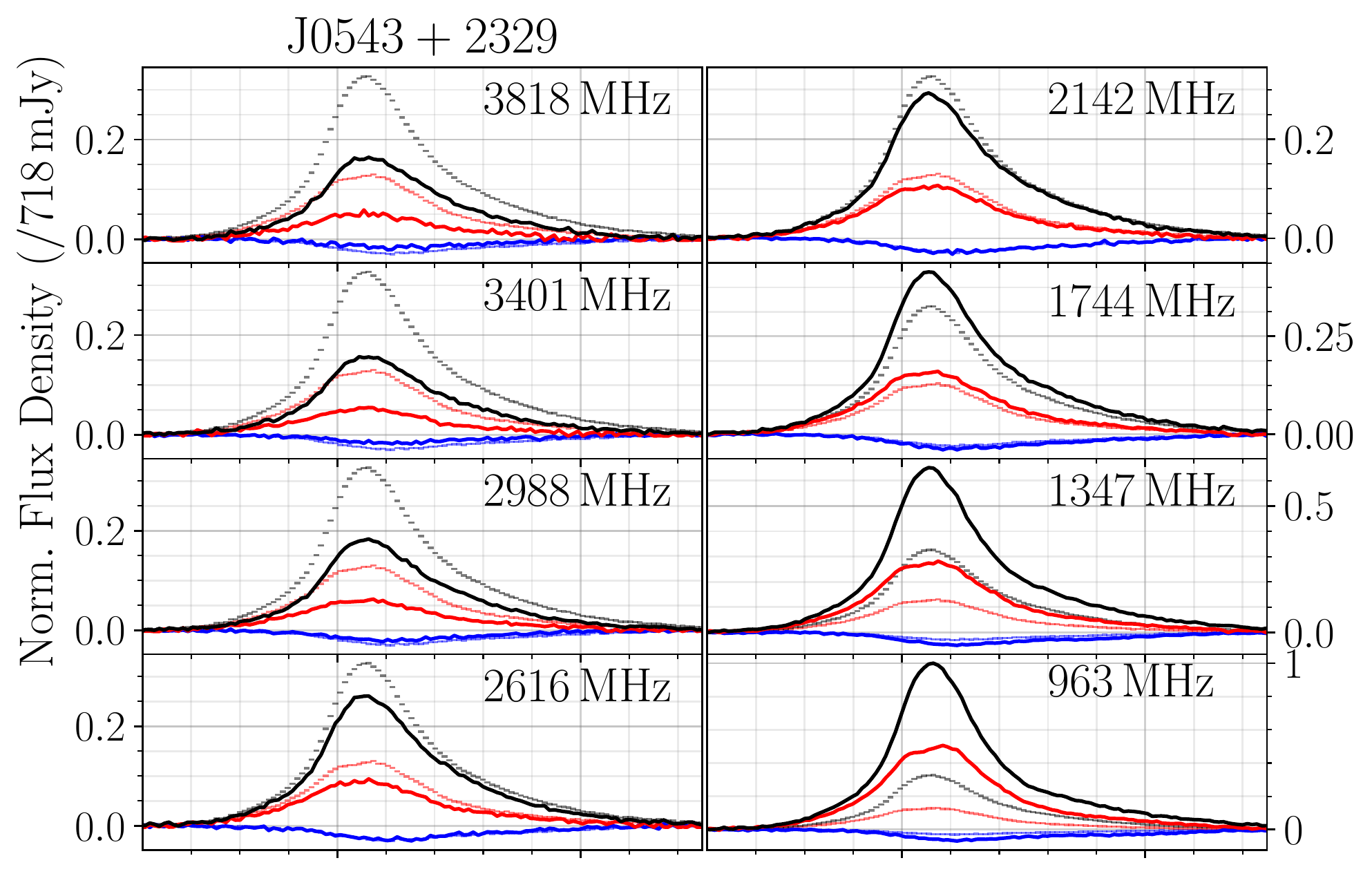}
	\includegraphics[width=\columnwidth]{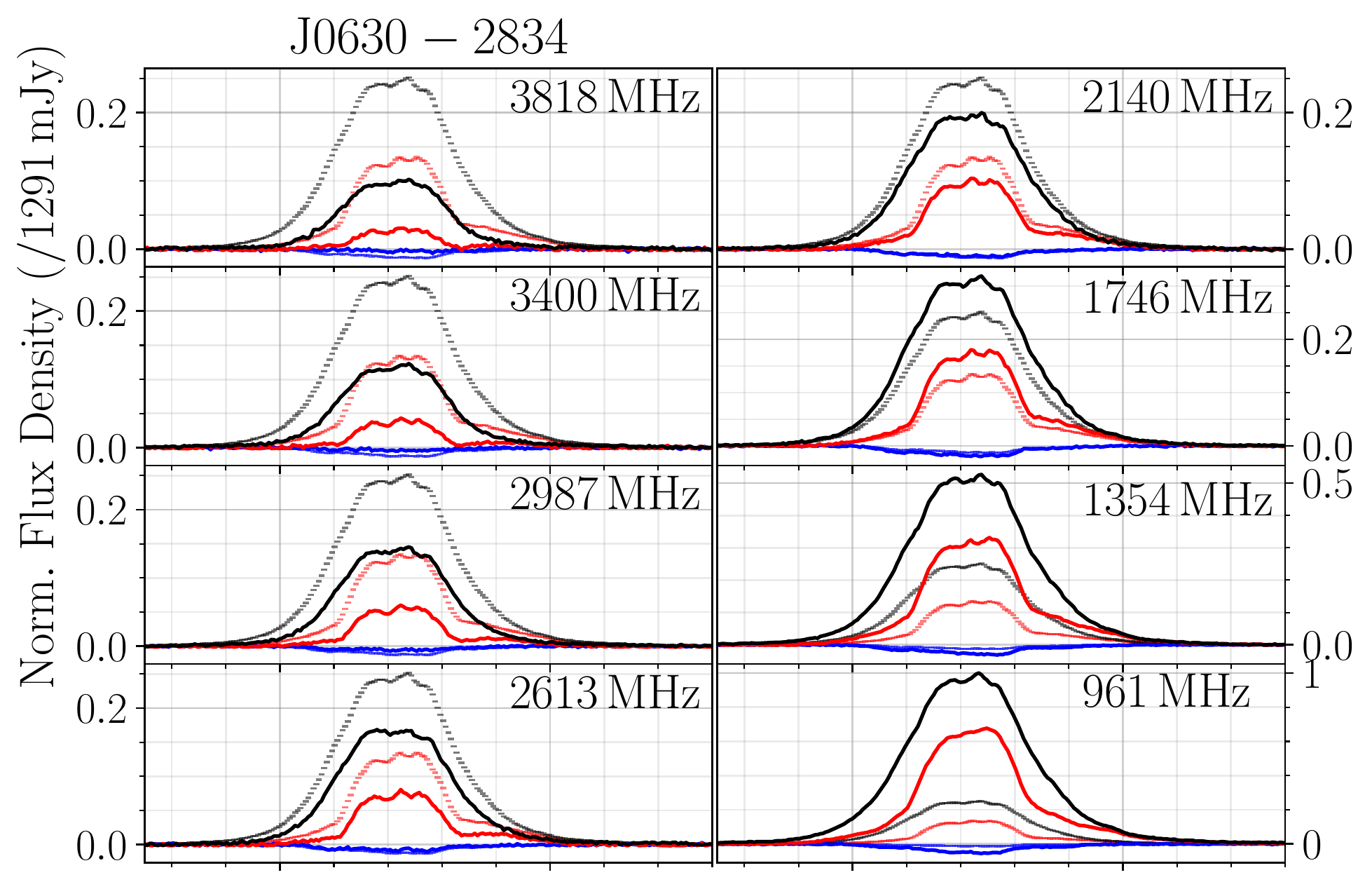}

	\includegraphics[width=\columnwidth]{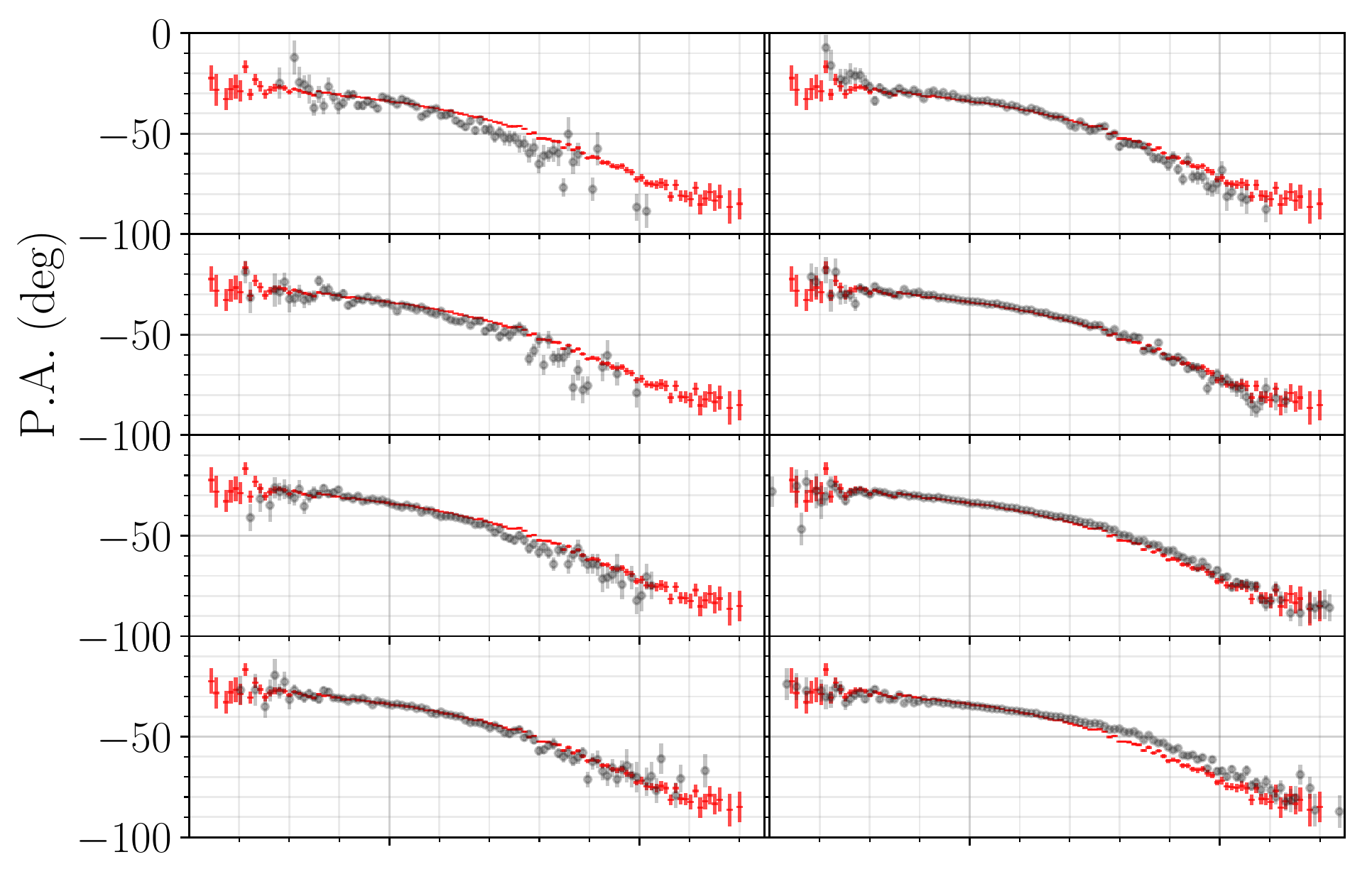}
	\includegraphics[width=\columnwidth]{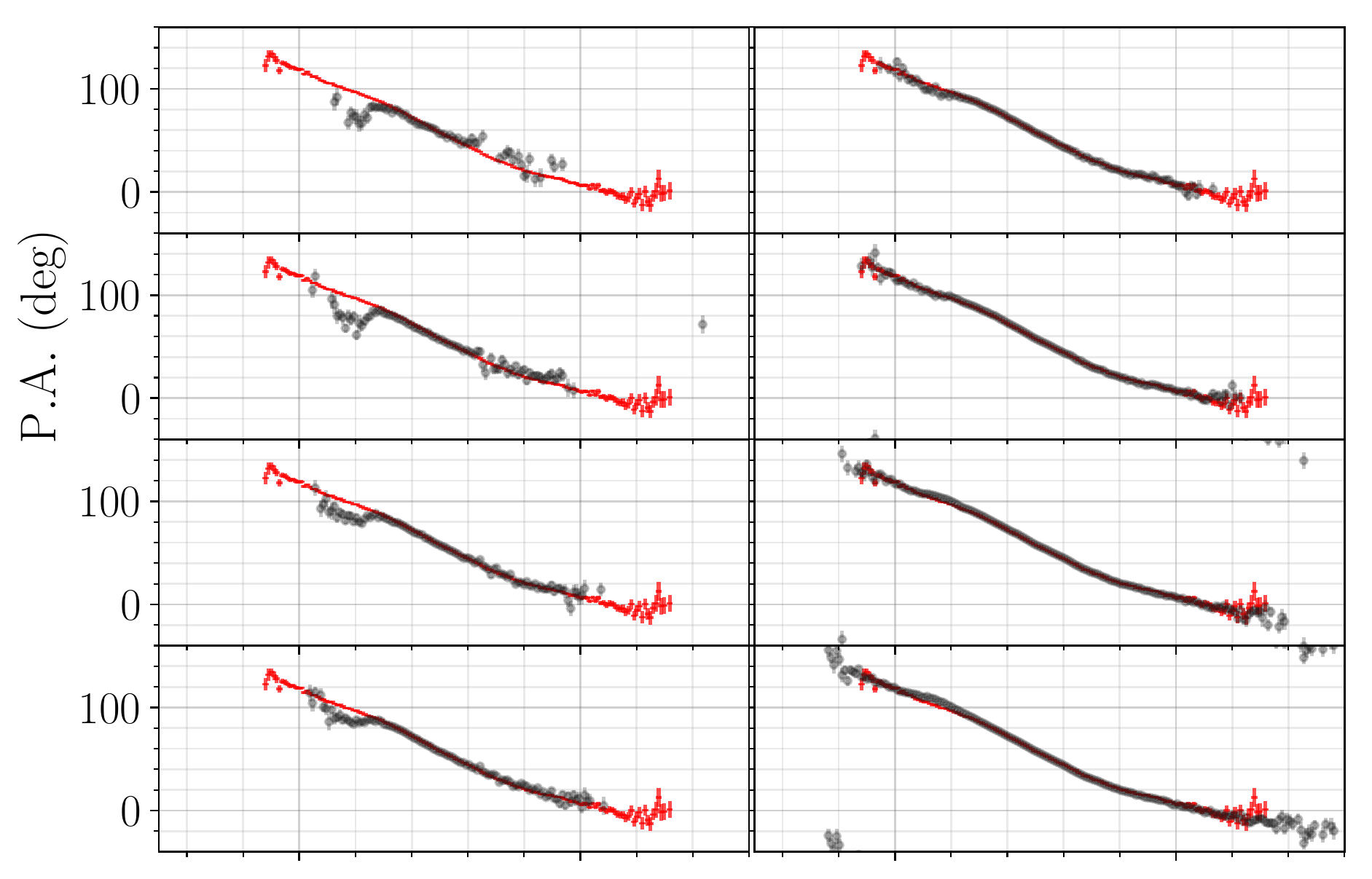}

	\includegraphics[width=\columnwidth]{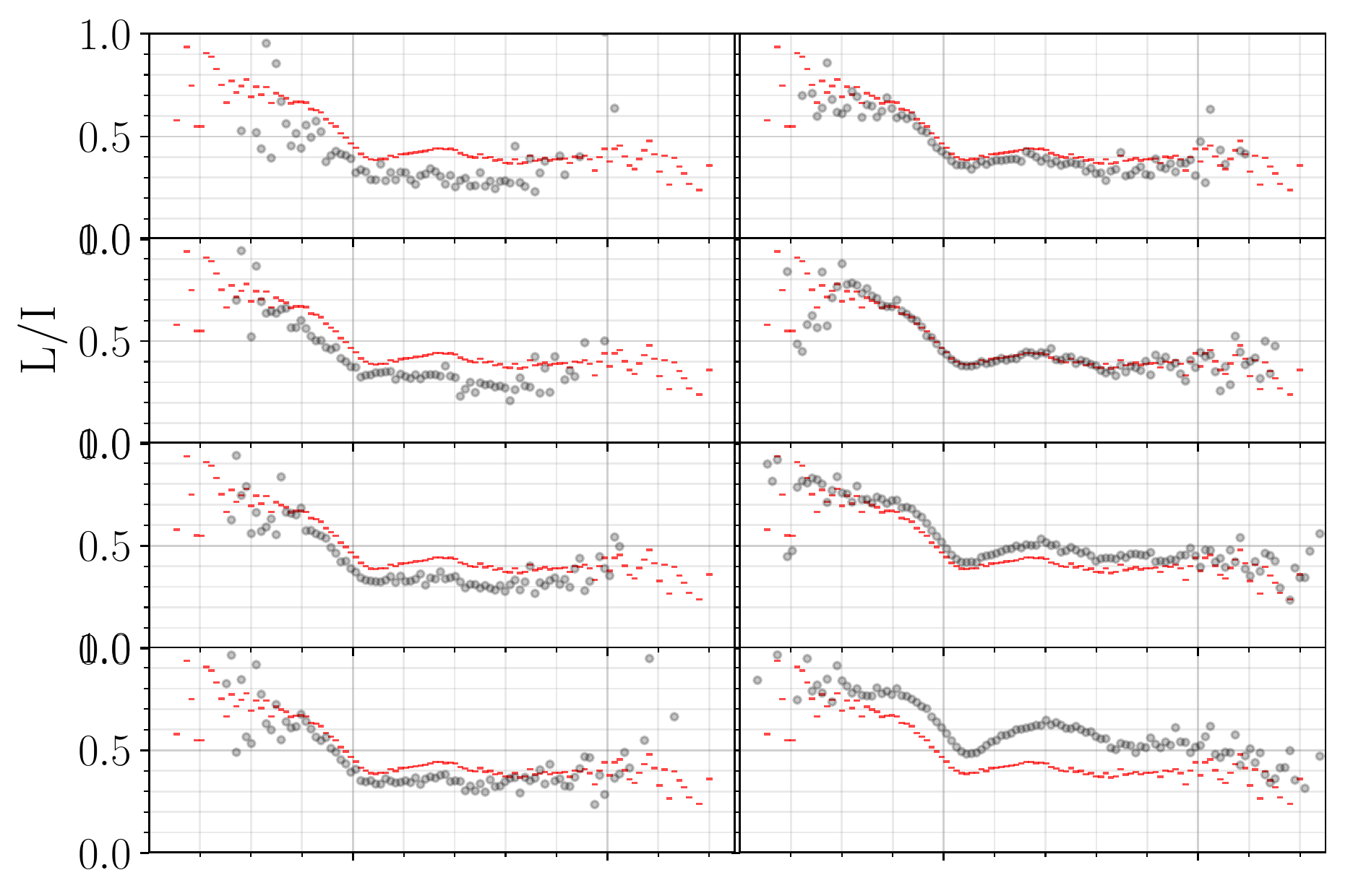}
	\includegraphics[width=\columnwidth]{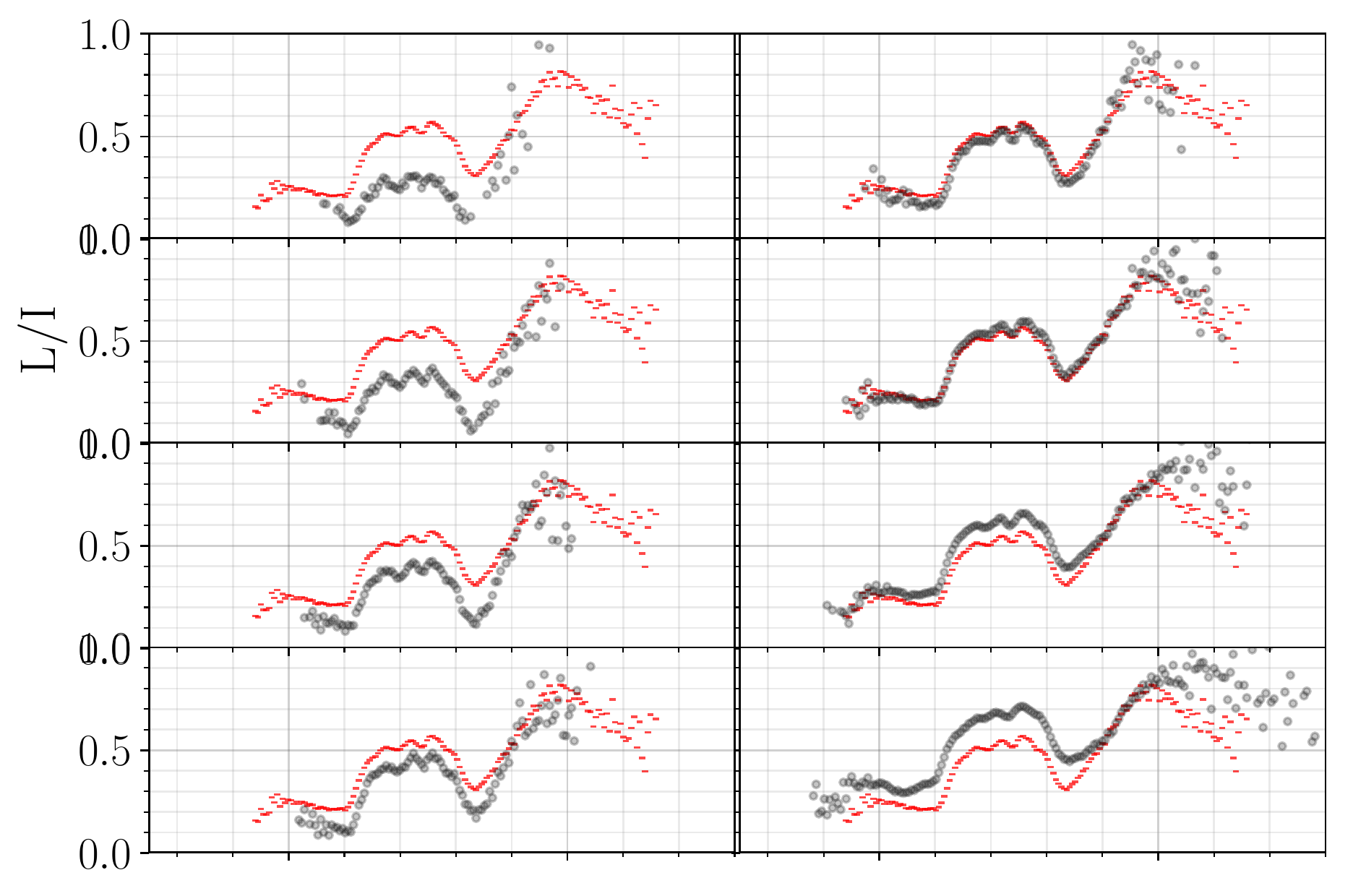}

	\includegraphics[width=\columnwidth]{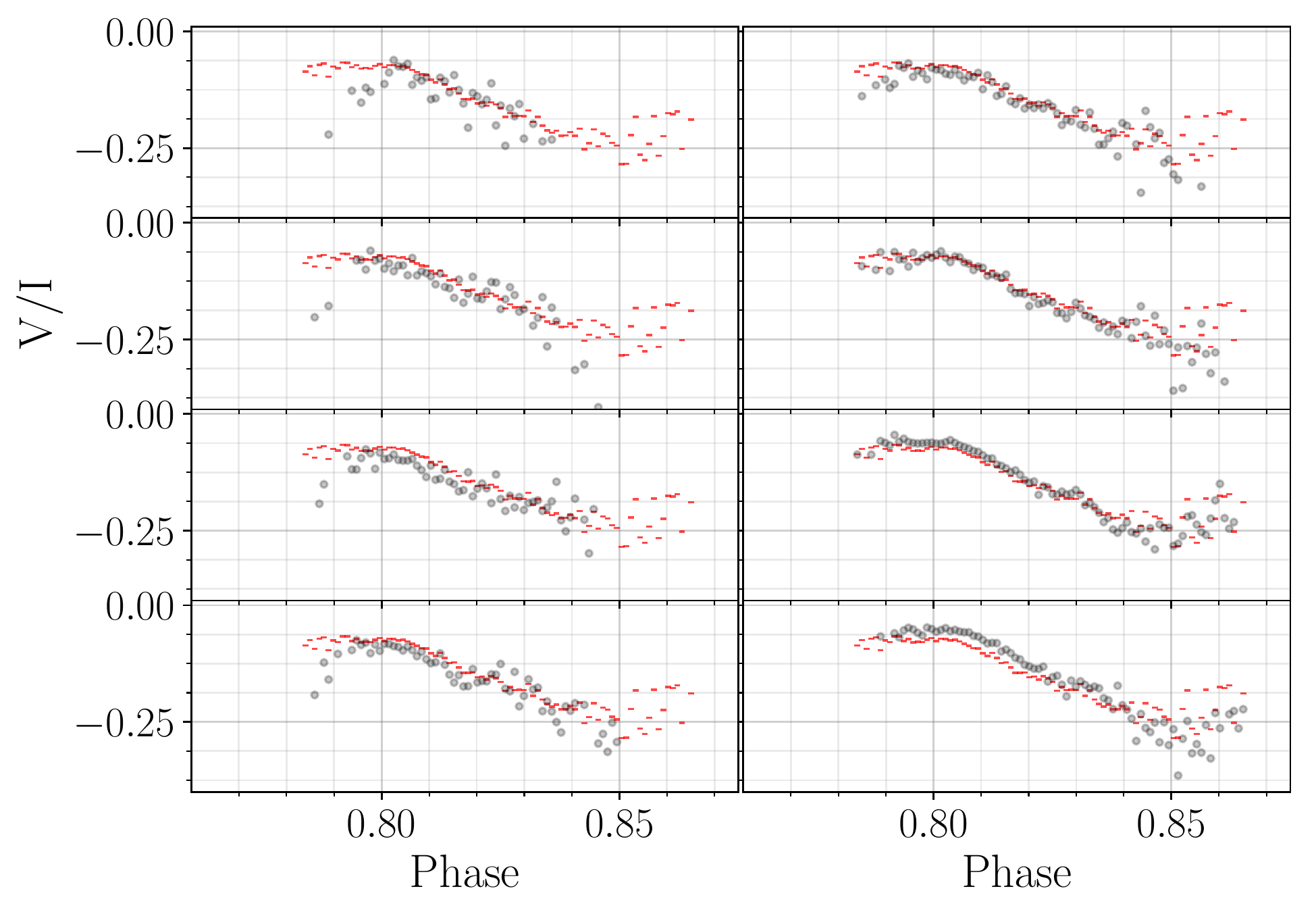}
	\includegraphics[width=\columnwidth]{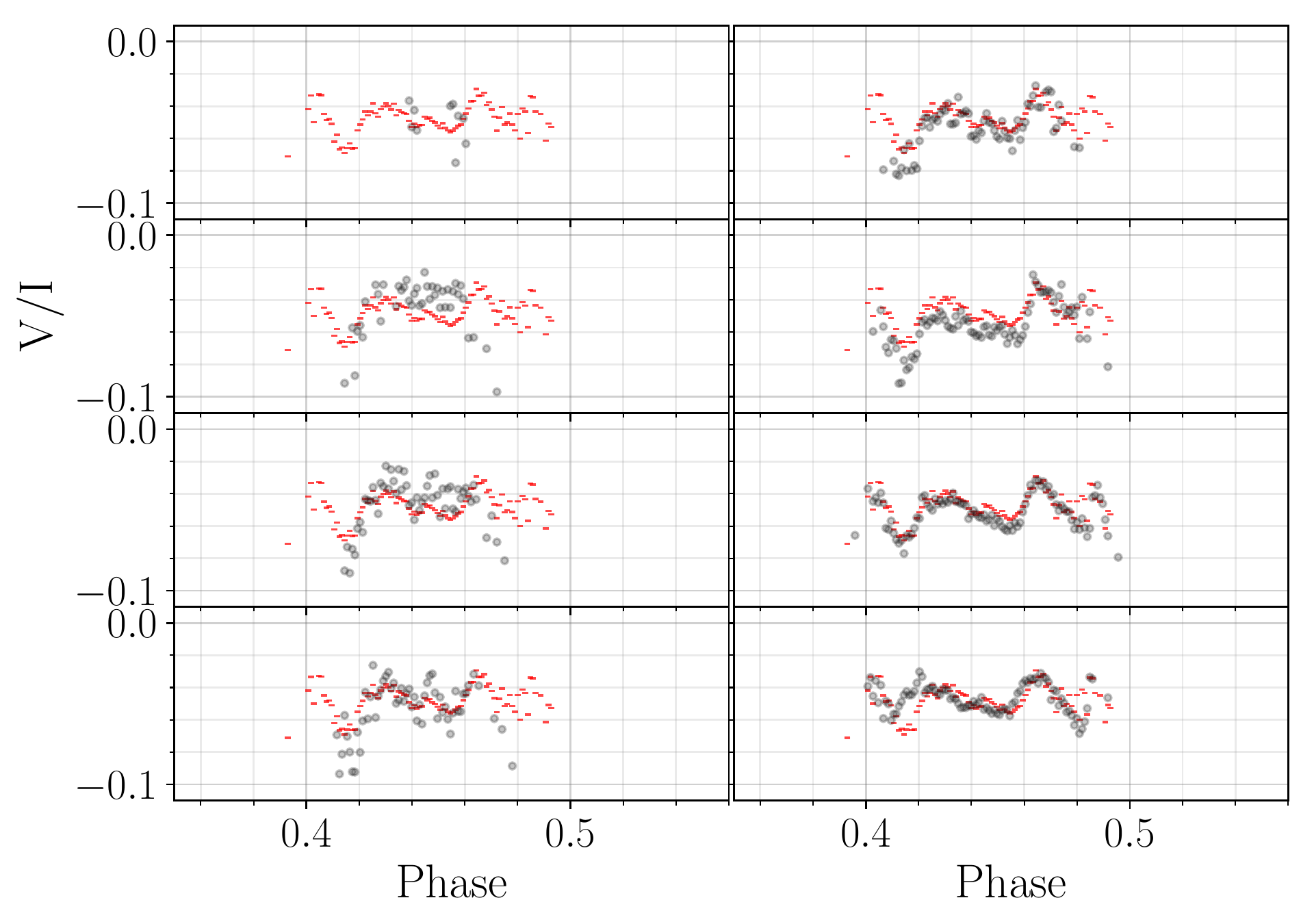}

    \caption{...Figure \ref{fig:profs1} continued...}
\end{figure*}

\begin{figure*} 
	\includegraphics[width=\columnwidth]{Profs/J0738-4042_profs.pdf}
	\includegraphics[width=\columnwidth]{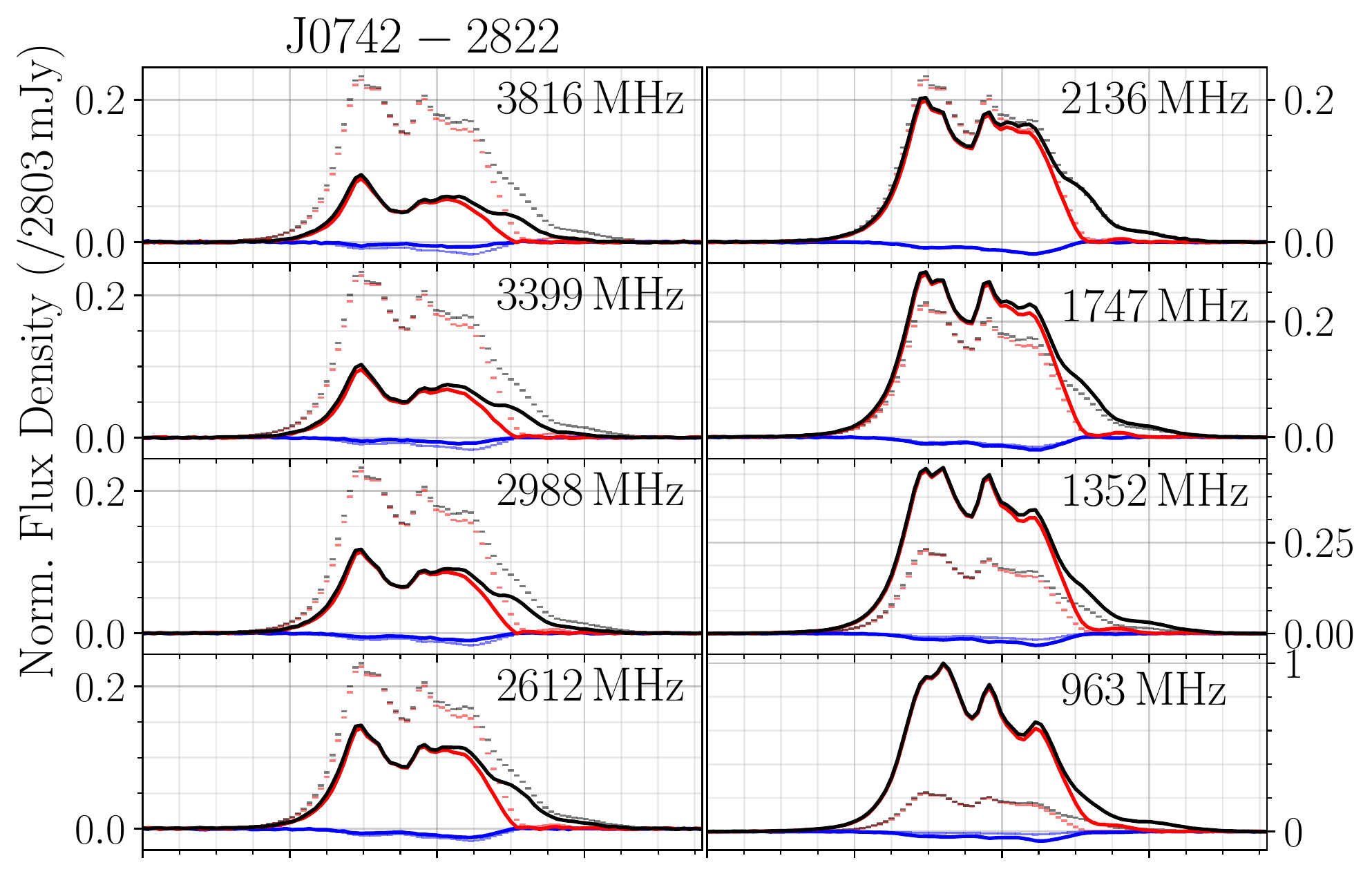}

	\includegraphics[width=\columnwidth]{Profs/J0738-4042_PAs.pdf}
	\includegraphics[width=\columnwidth]{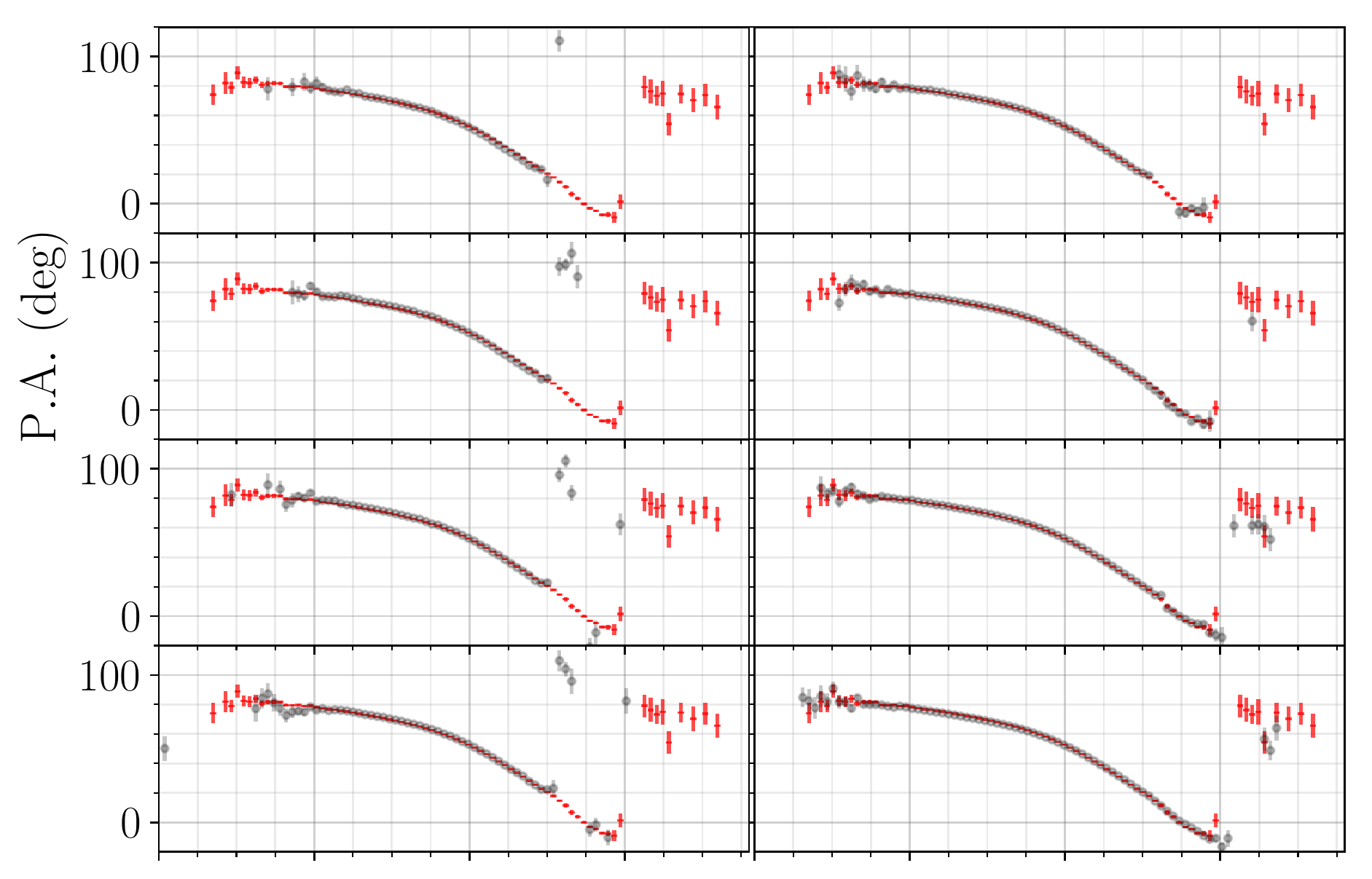}

	\includegraphics[width=\columnwidth]{Profs/J0738-4042_L.pdf}
	\includegraphics[width=\columnwidth]{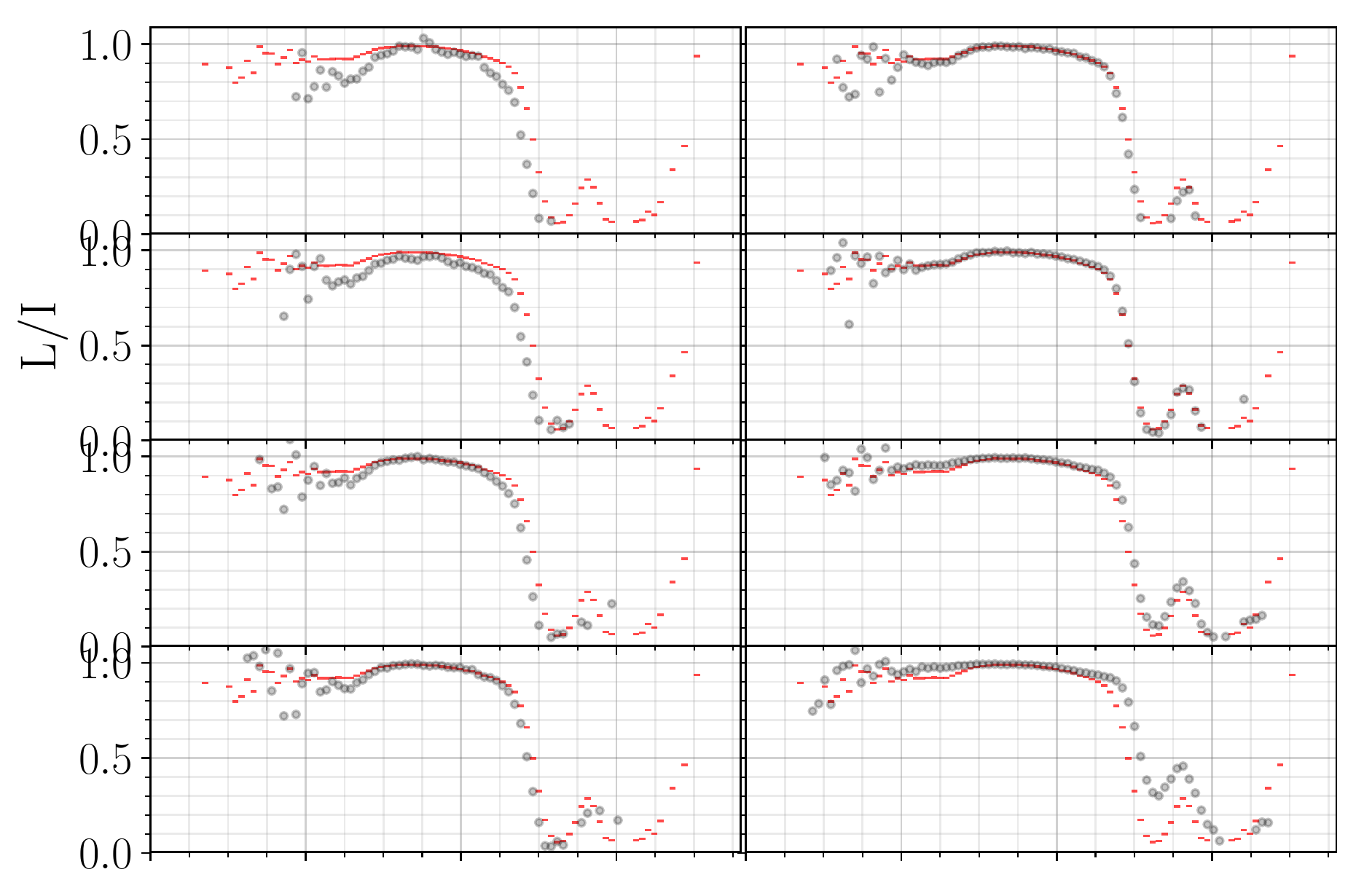}

	\includegraphics[width=\columnwidth]{Profs/J0738-4042_V.pdf}
	\includegraphics[width=\columnwidth]{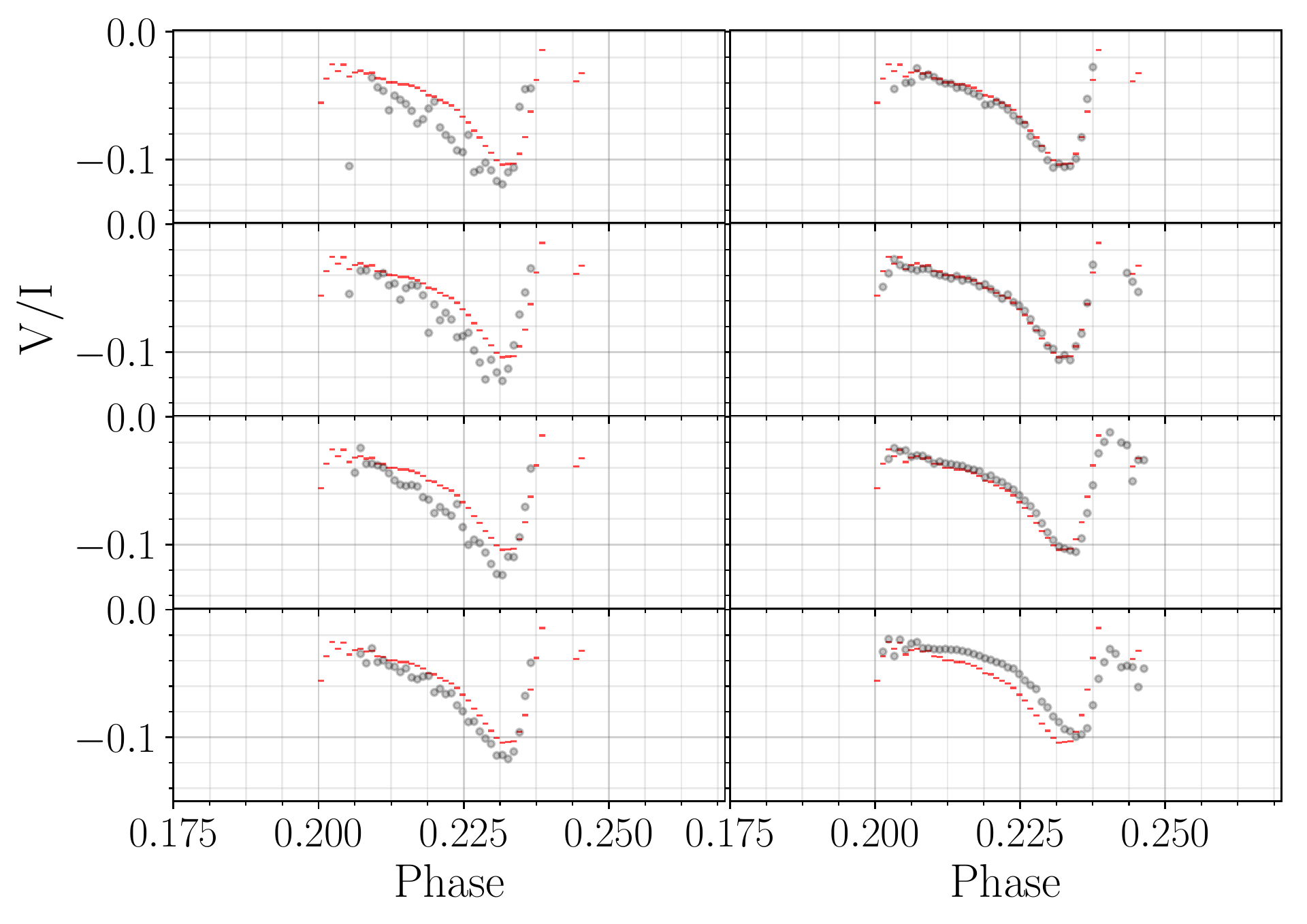}

    \caption{...Figure \ref{fig:profs1} continued...}
\end{figure*}

\begin{figure*} 
	\includegraphics[width=\columnwidth]{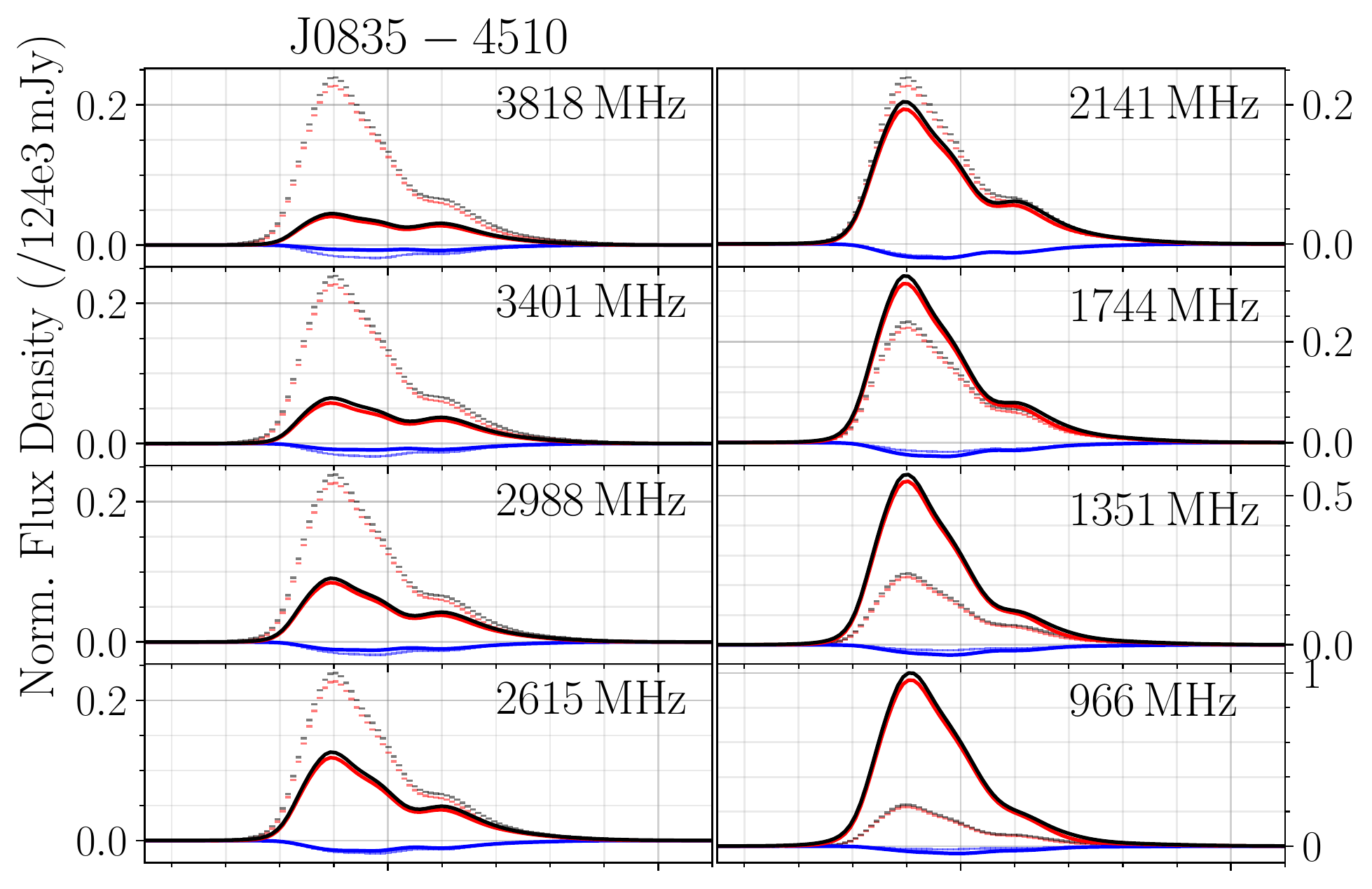}
	\includegraphics[width=\columnwidth]{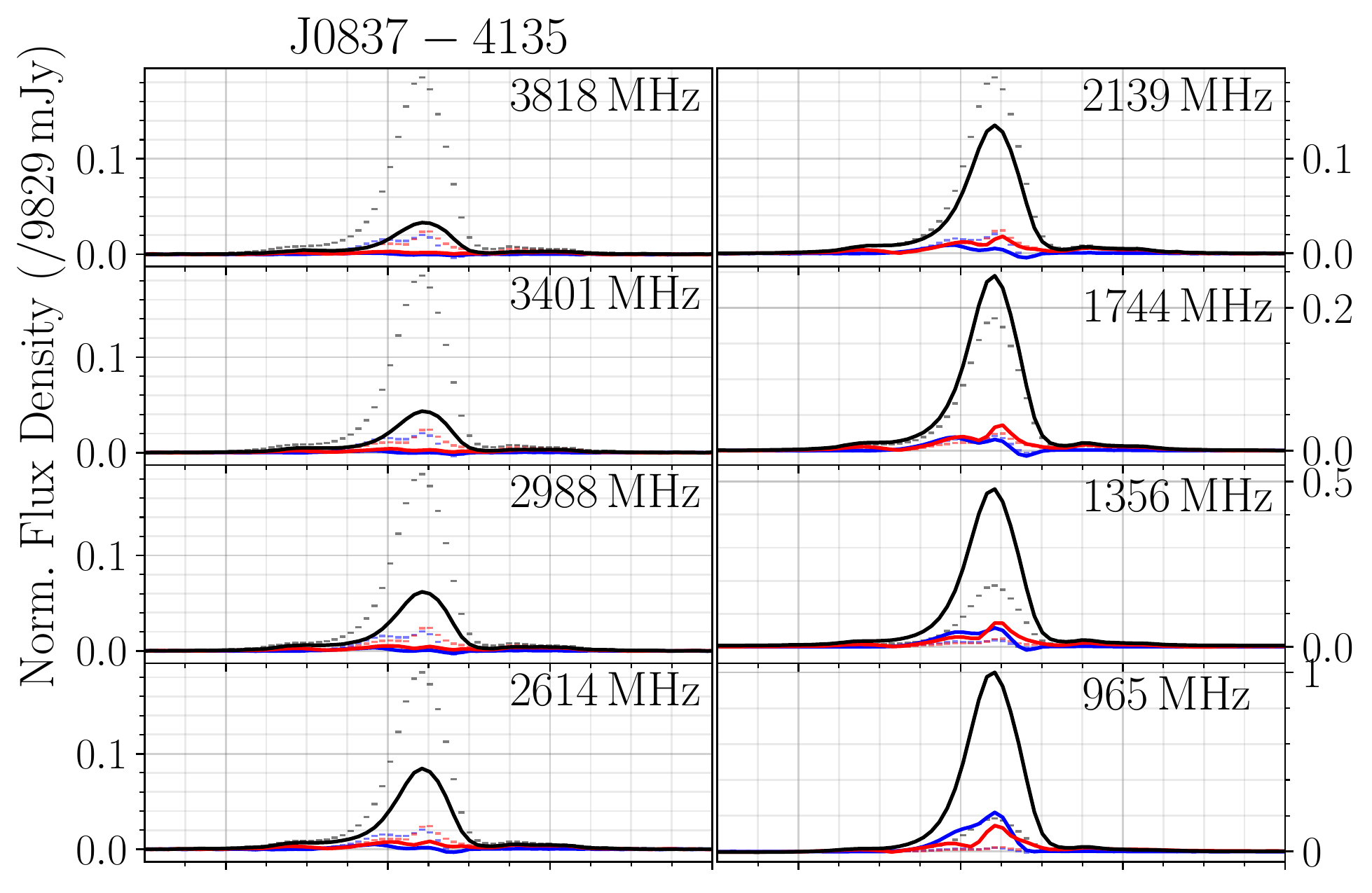}

	\includegraphics[width=\columnwidth]{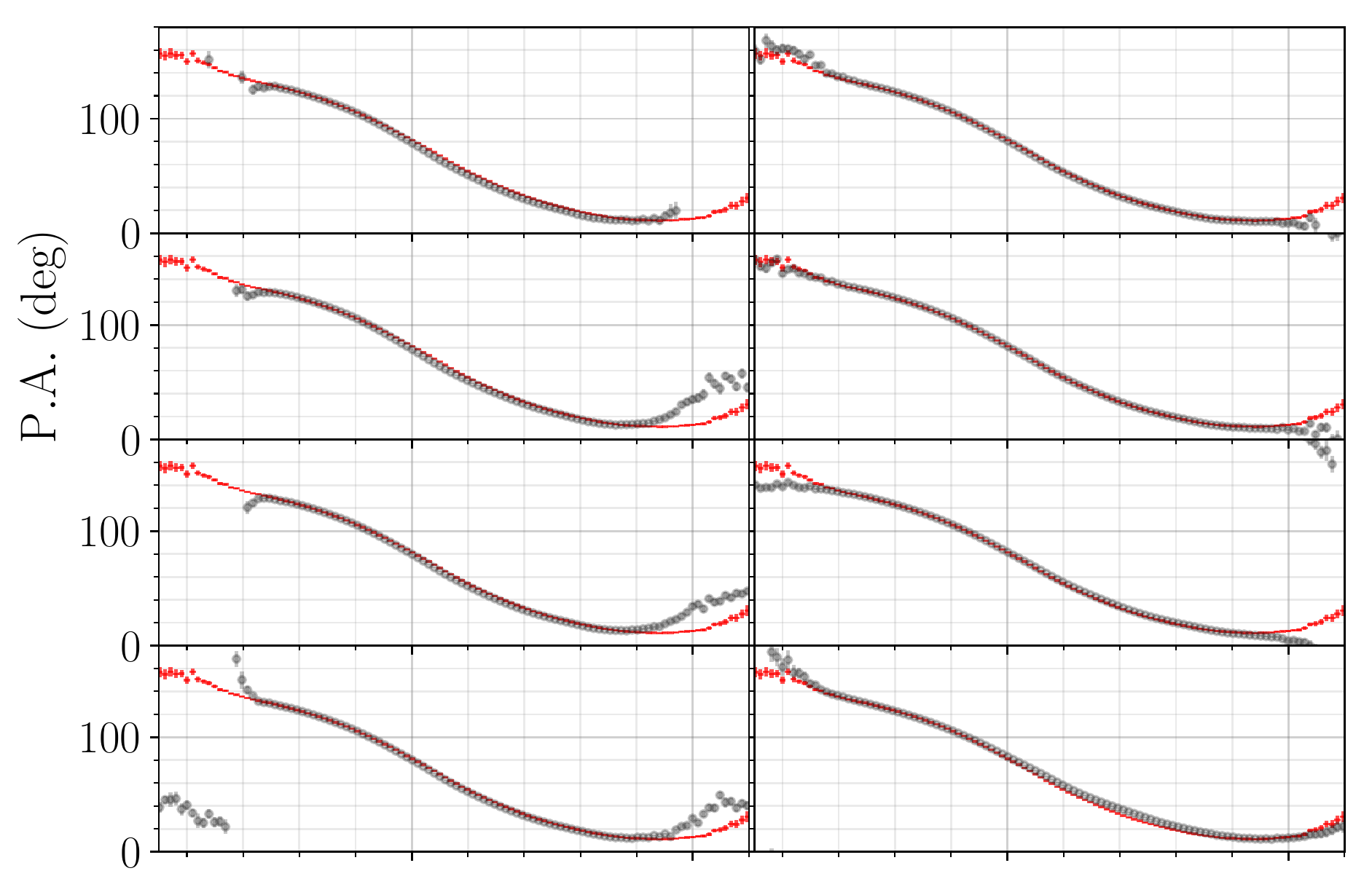}
	\includegraphics[width=\columnwidth]{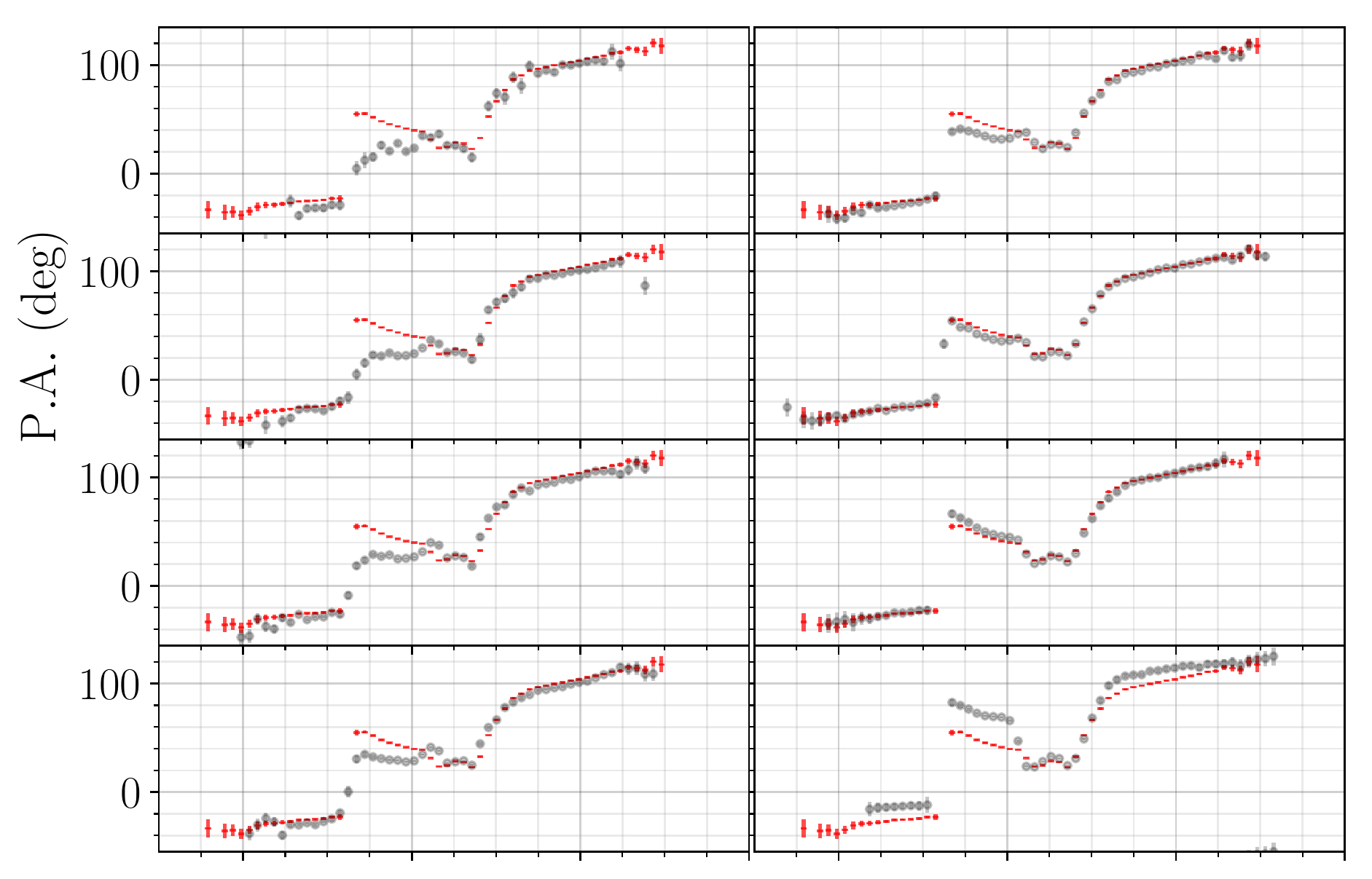}

	\includegraphics[width=\columnwidth]{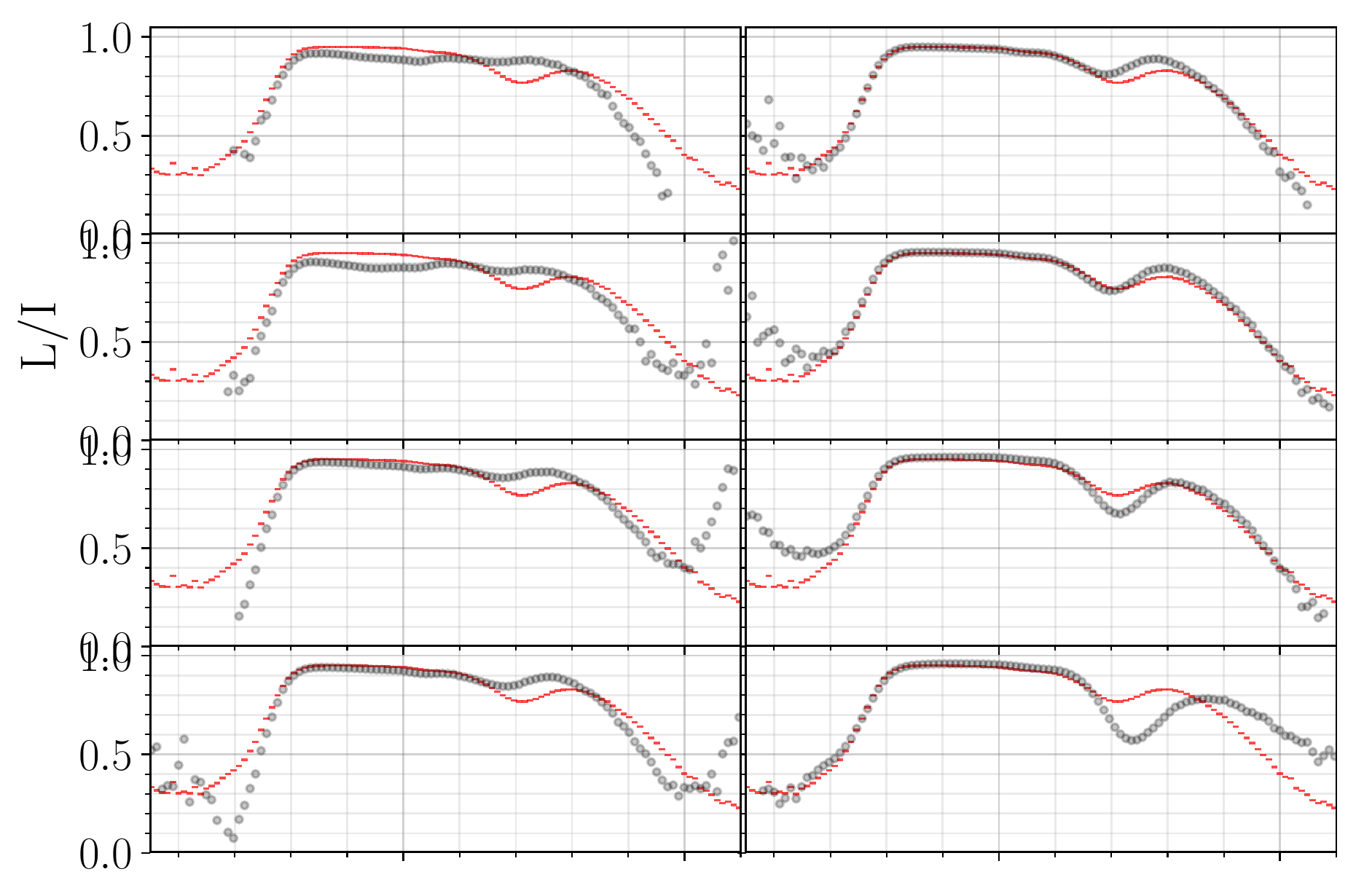}
	\includegraphics[width=\columnwidth]{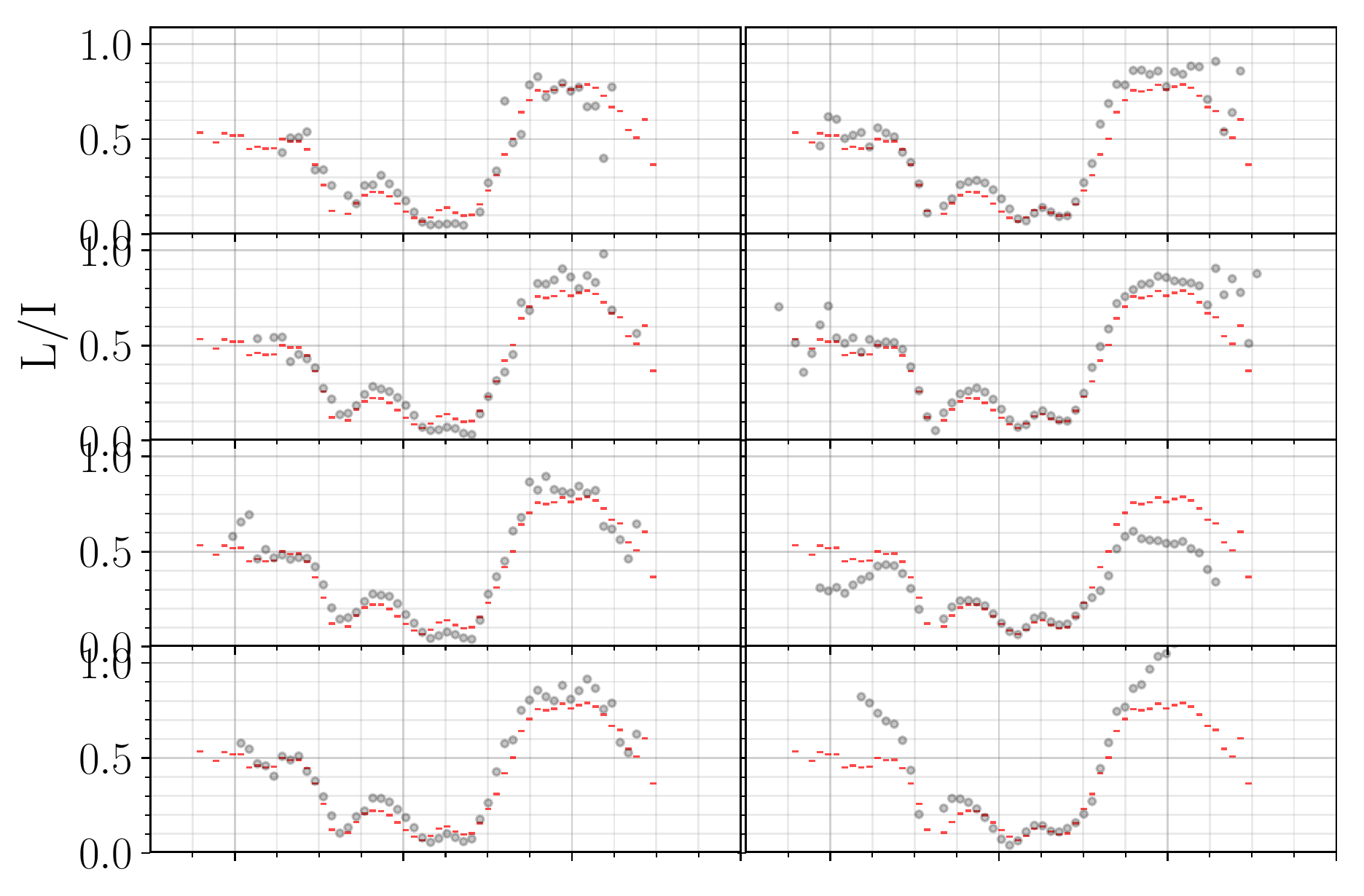}

	\includegraphics[width=\columnwidth]{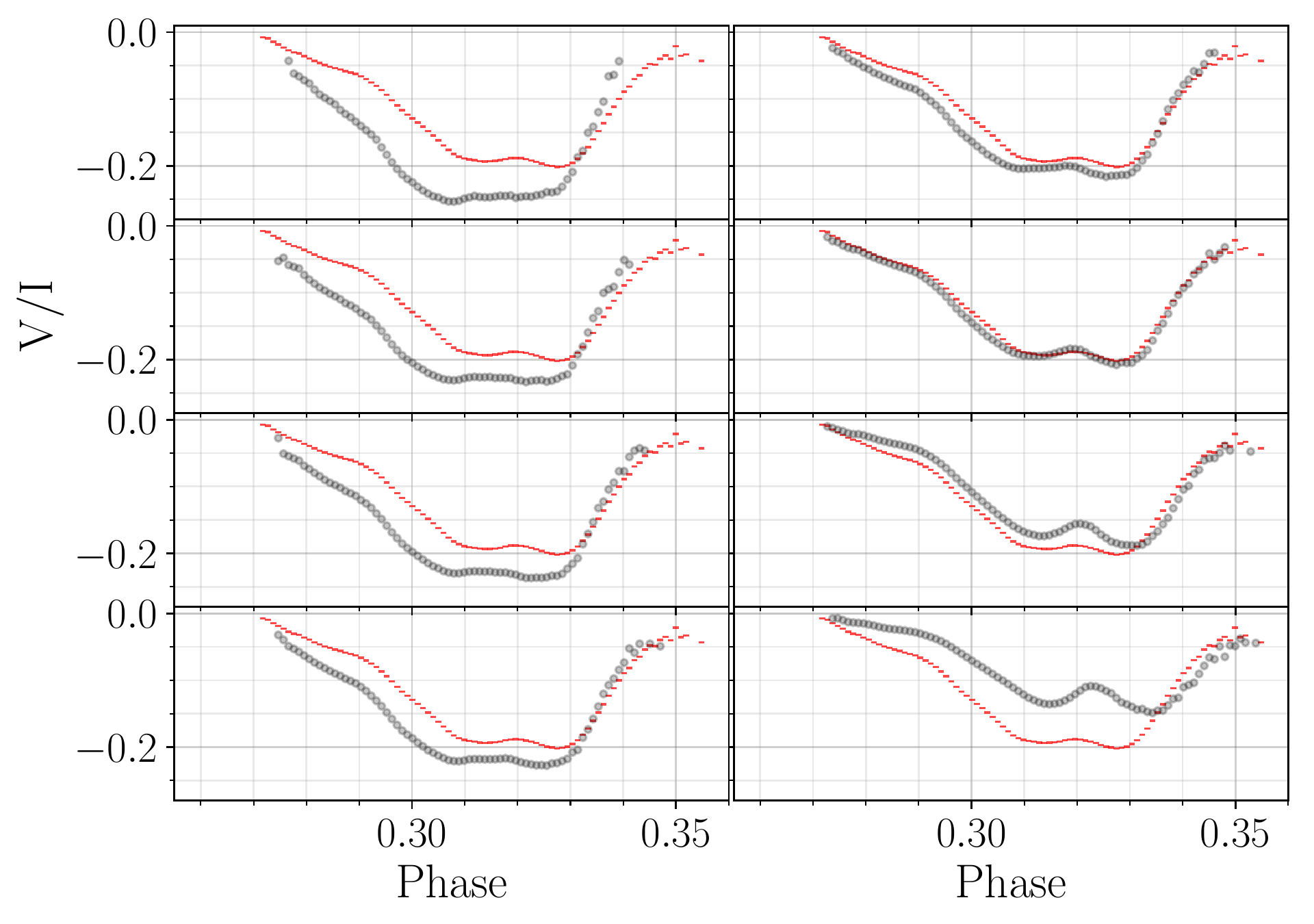}
	\includegraphics[width=\columnwidth]{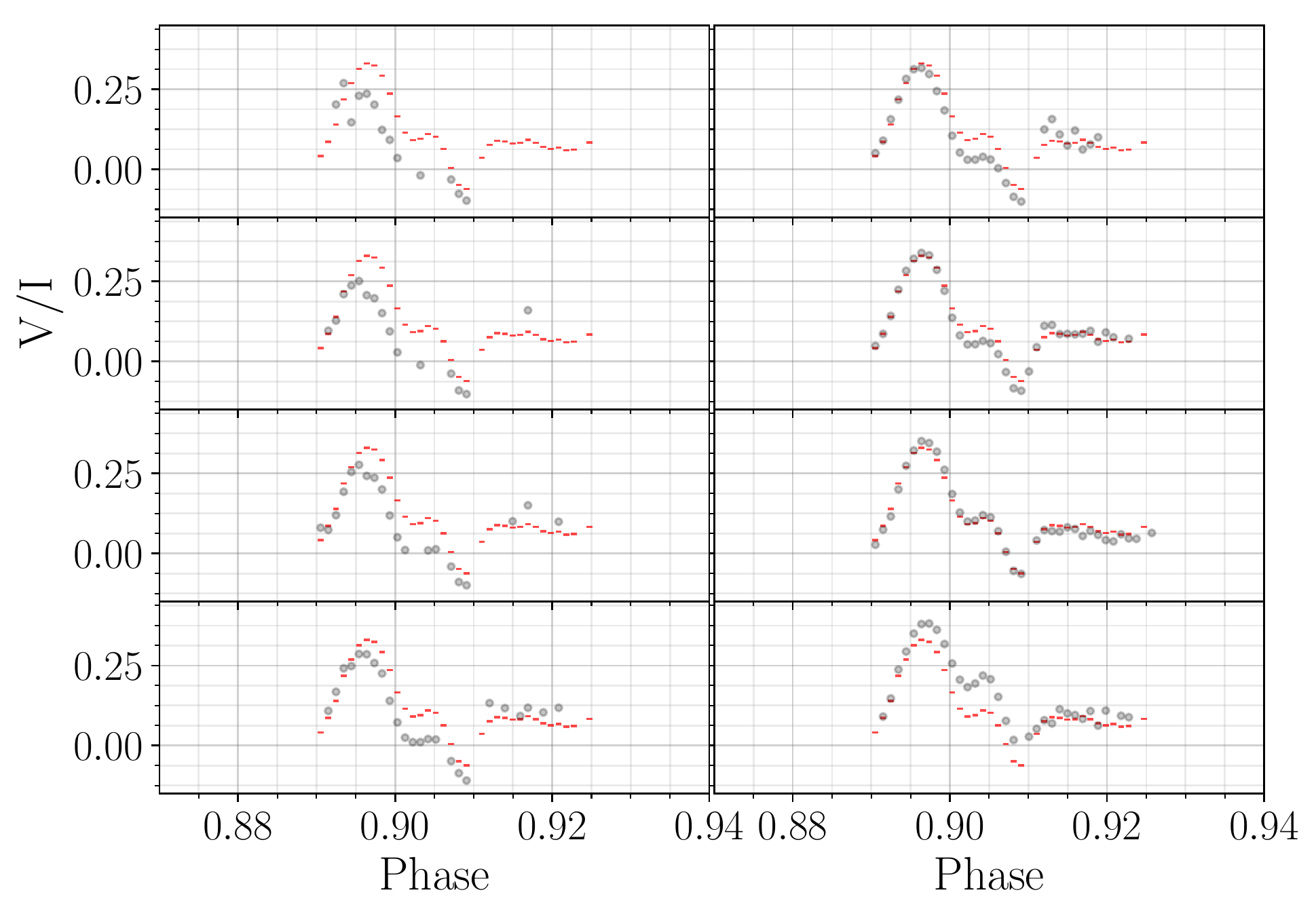}

    \caption{...Figure \ref{fig:profs1} continued...}
\end{figure*}

\begin{figure*} 
	\includegraphics[width=\columnwidth]{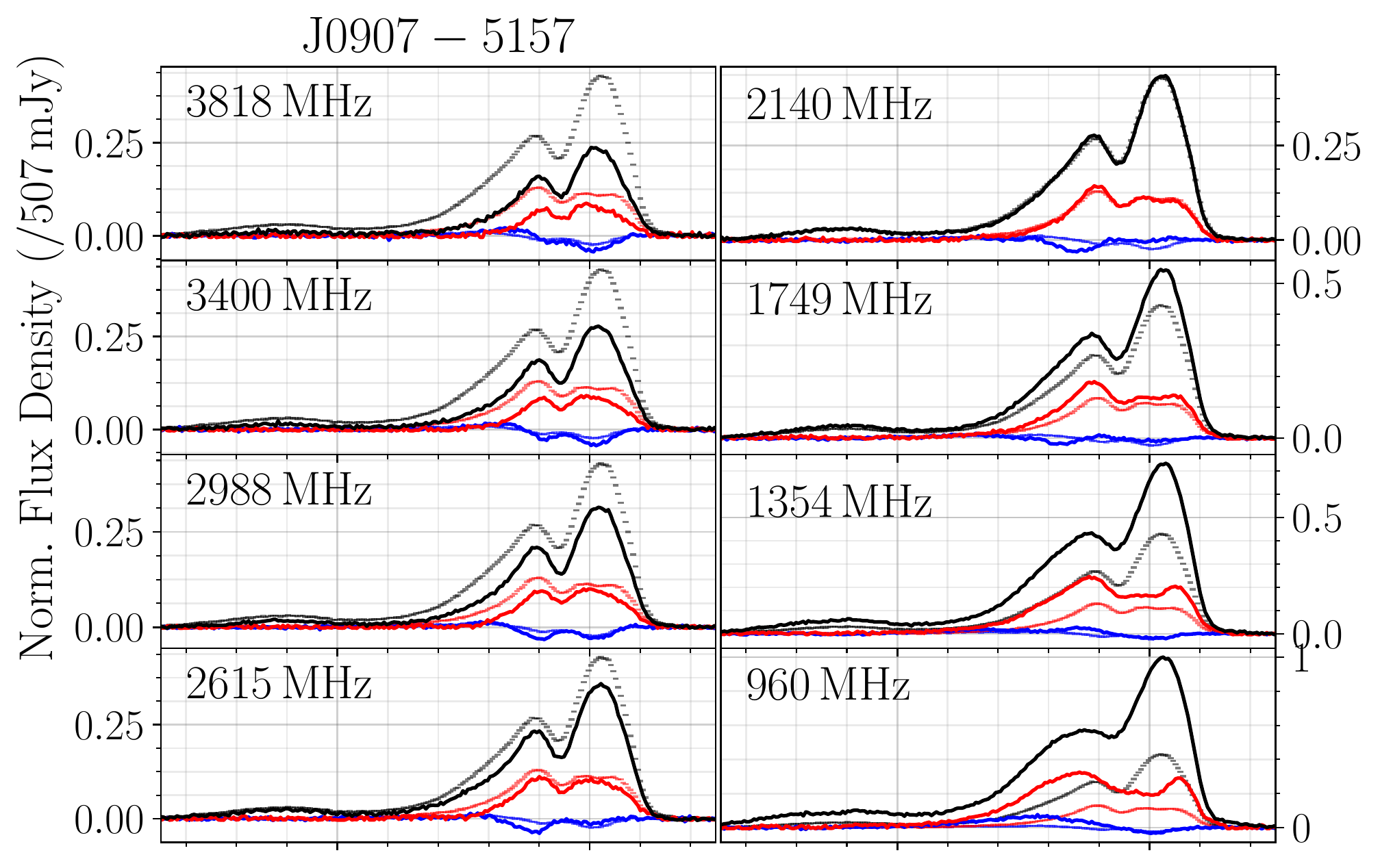}
	\includegraphics[width=\columnwidth]{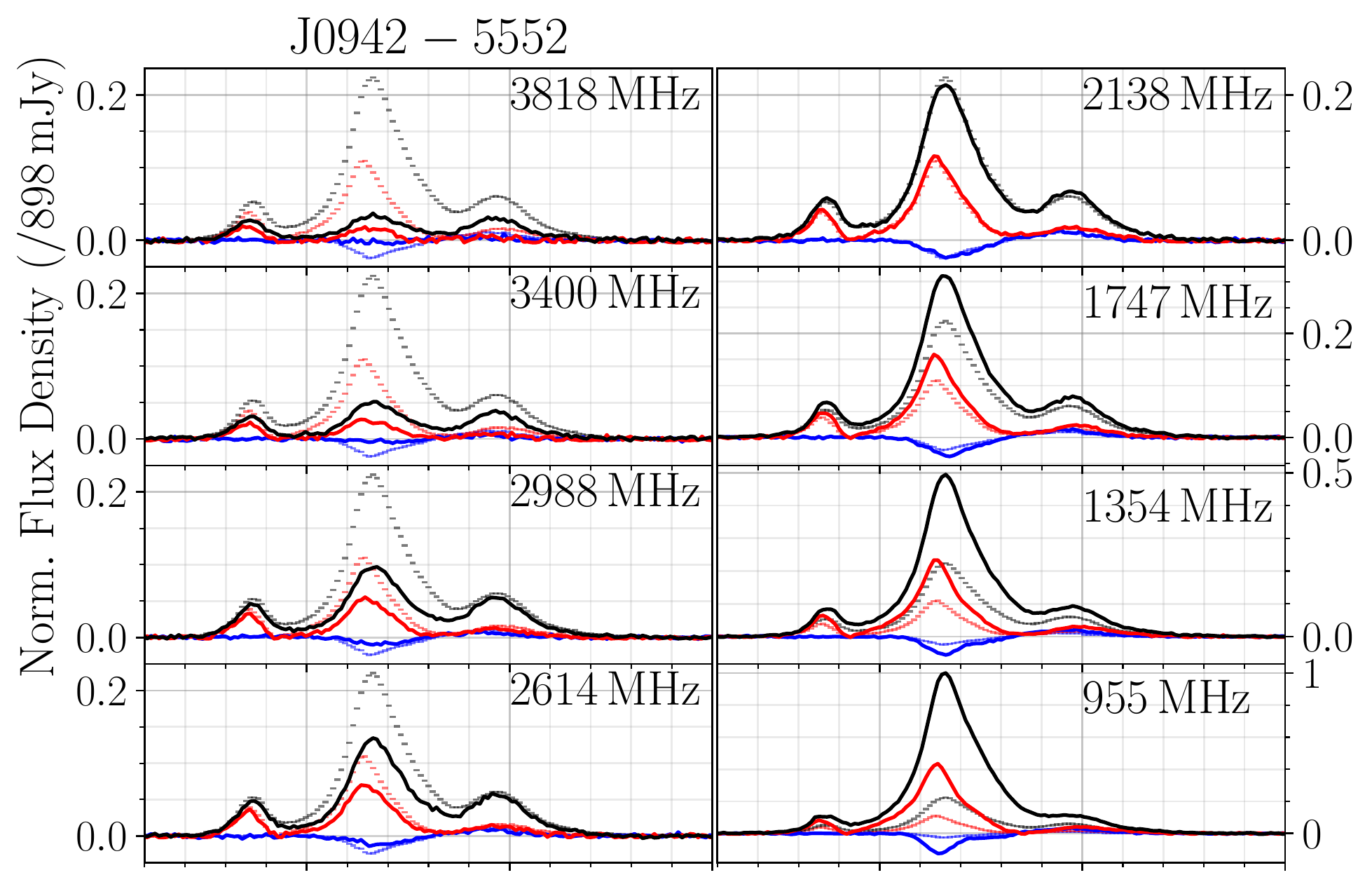}

	\includegraphics[width=\columnwidth]{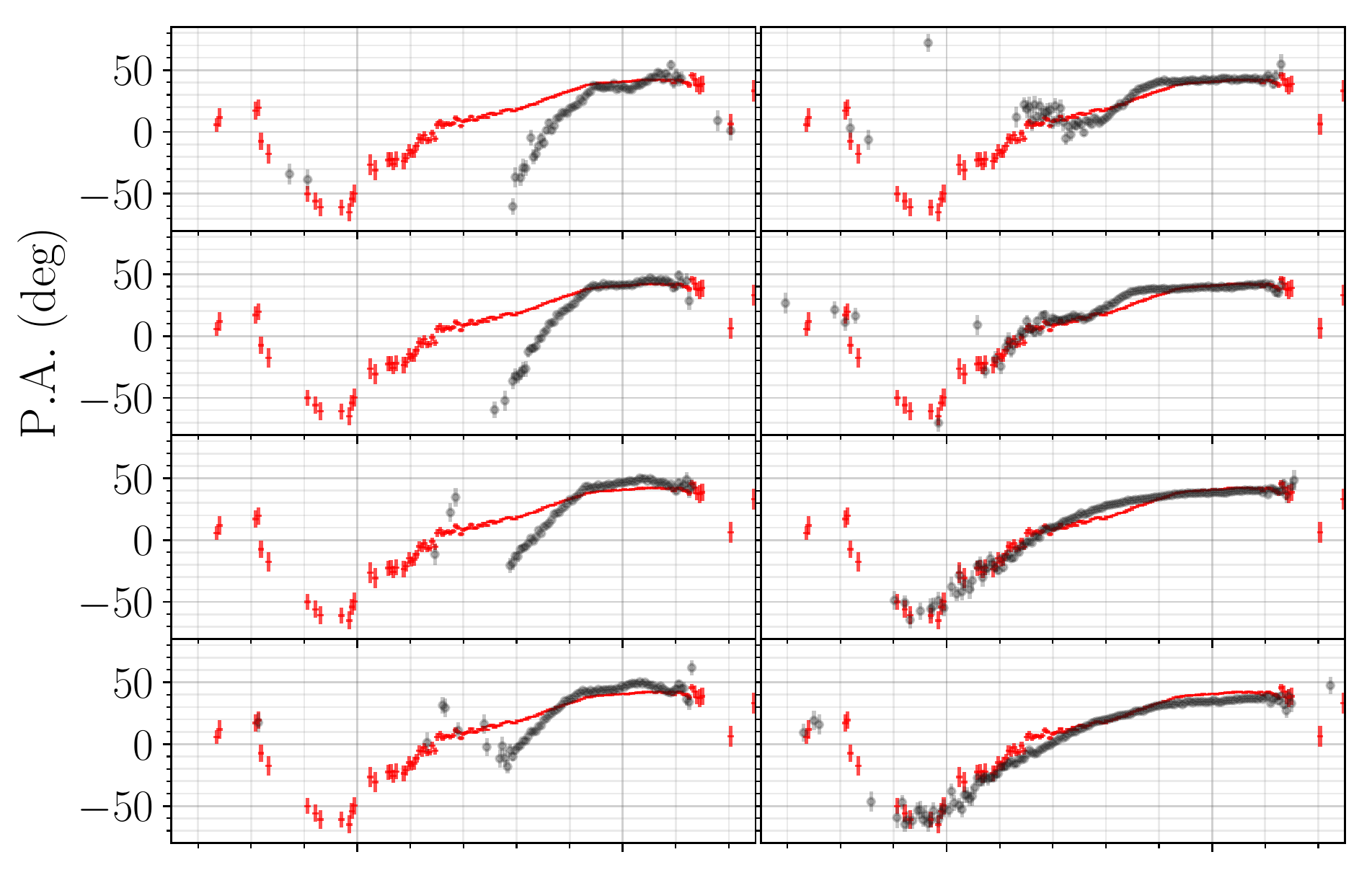}
	\includegraphics[width=\columnwidth]{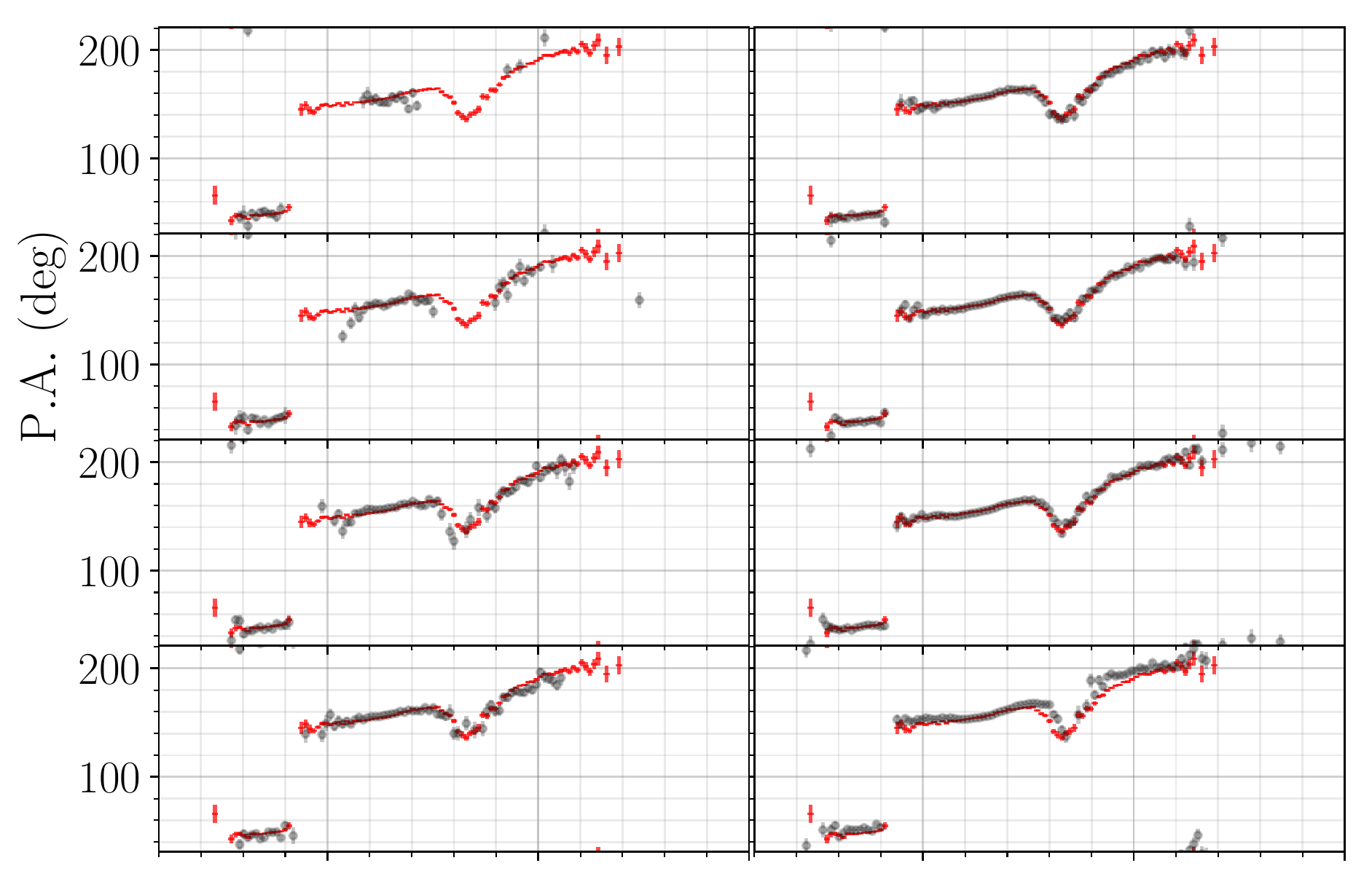}

	\includegraphics[width=\columnwidth]{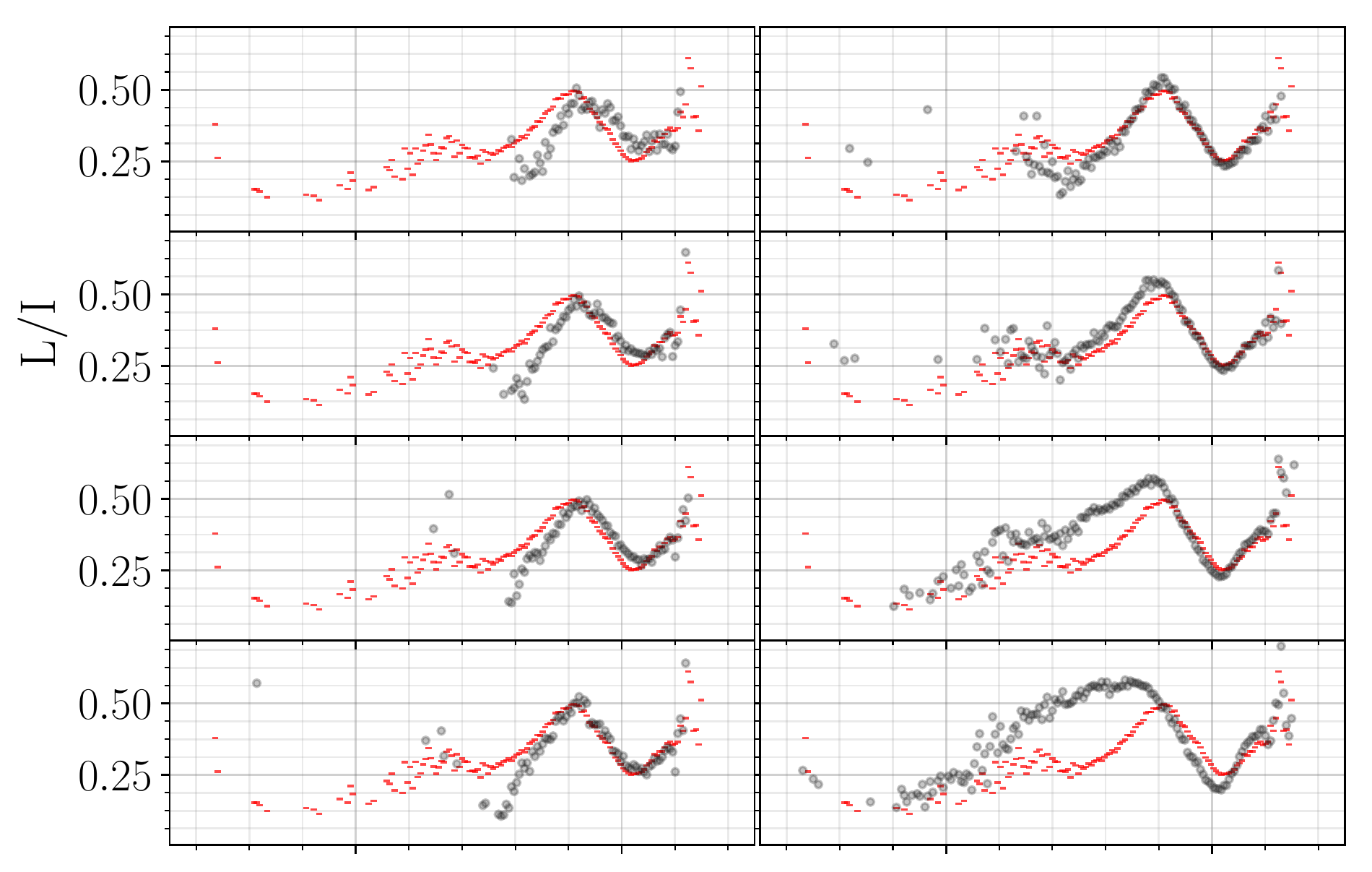}
	\includegraphics[width=\columnwidth]{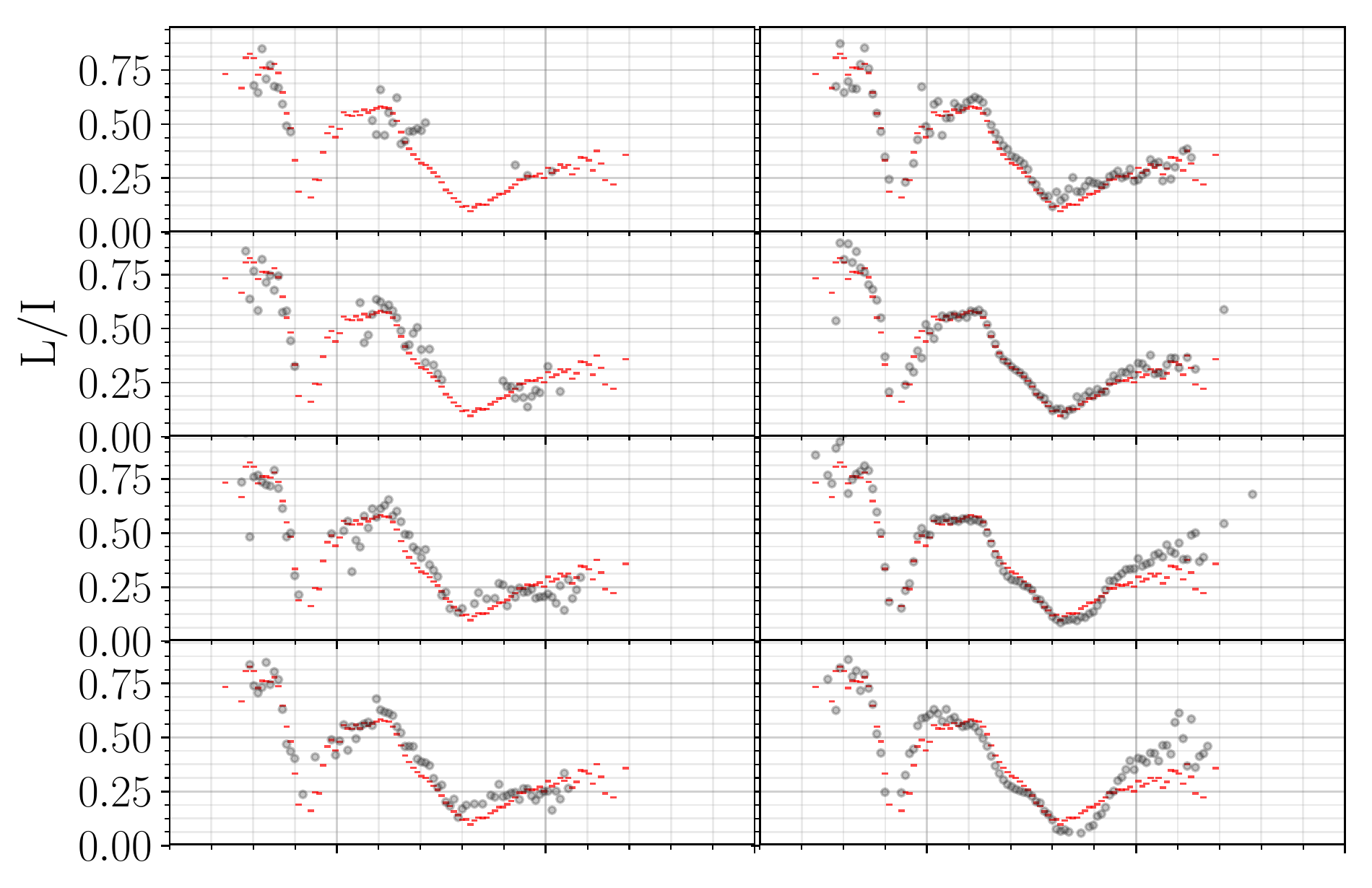}

	\includegraphics[width=\columnwidth]{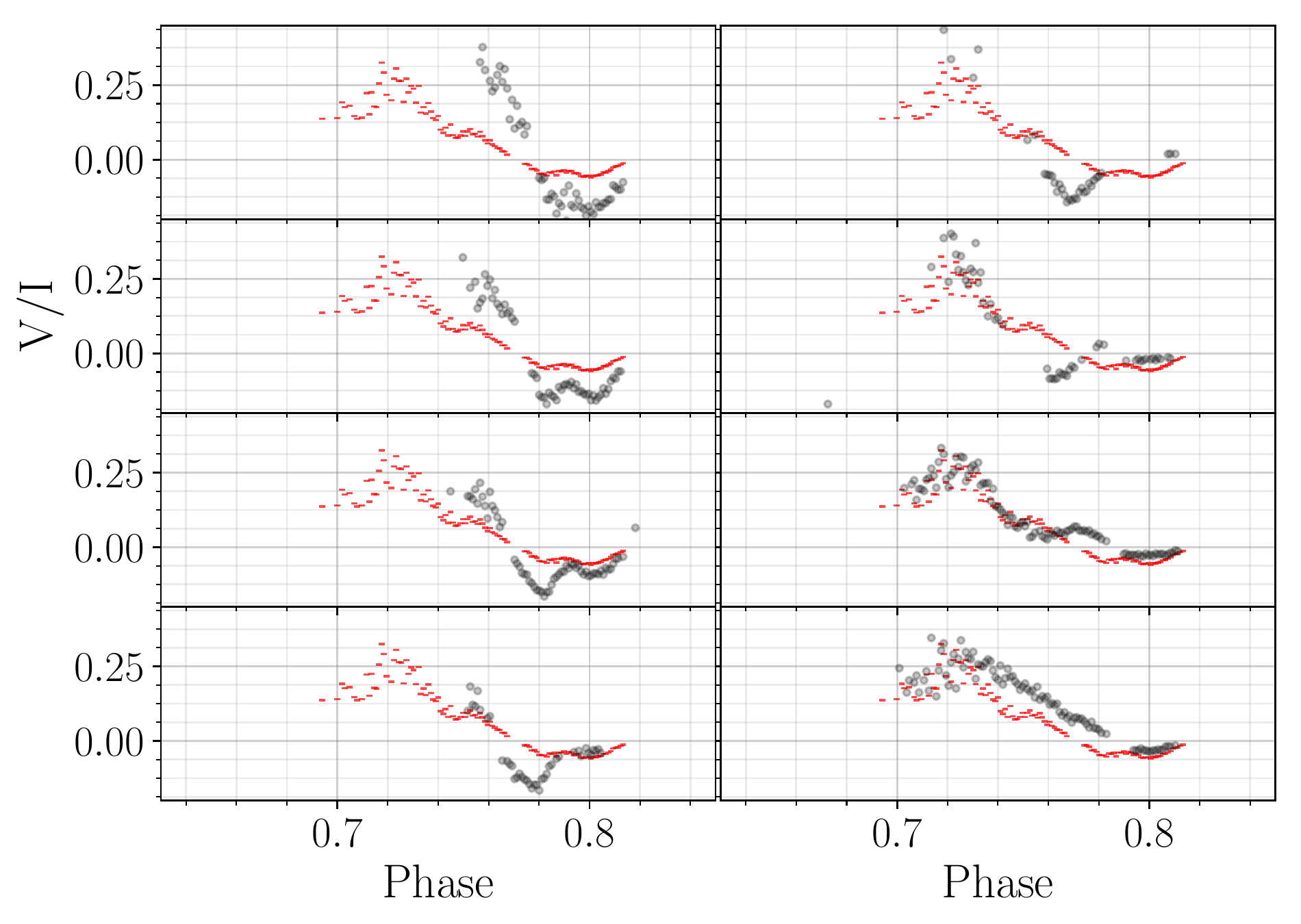}
	\includegraphics[width=\columnwidth]{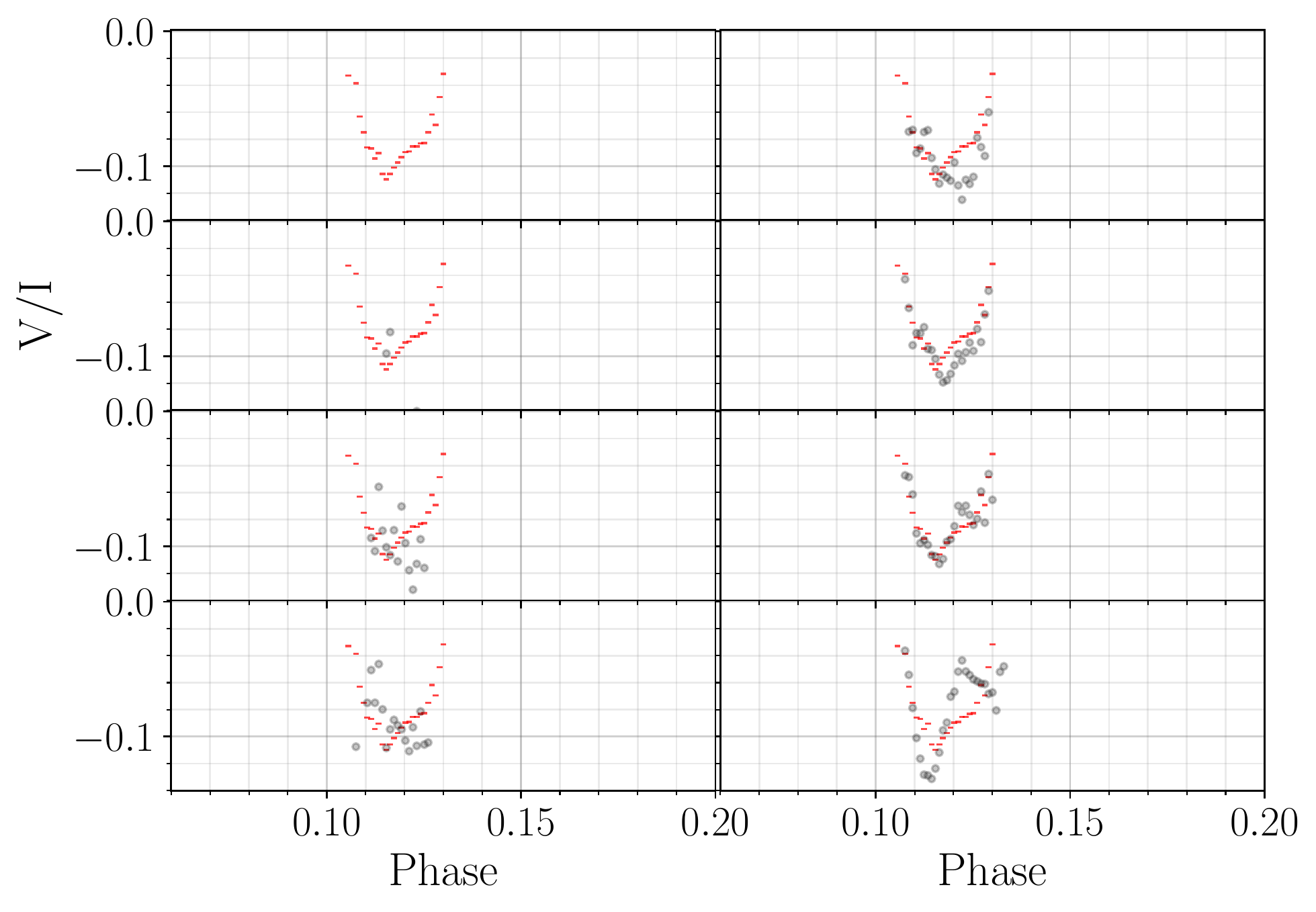}

    \caption{...Figure \ref{fig:profs1} continued...}
\end{figure*}

\begin{figure*} 
	\includegraphics[width=\columnwidth]{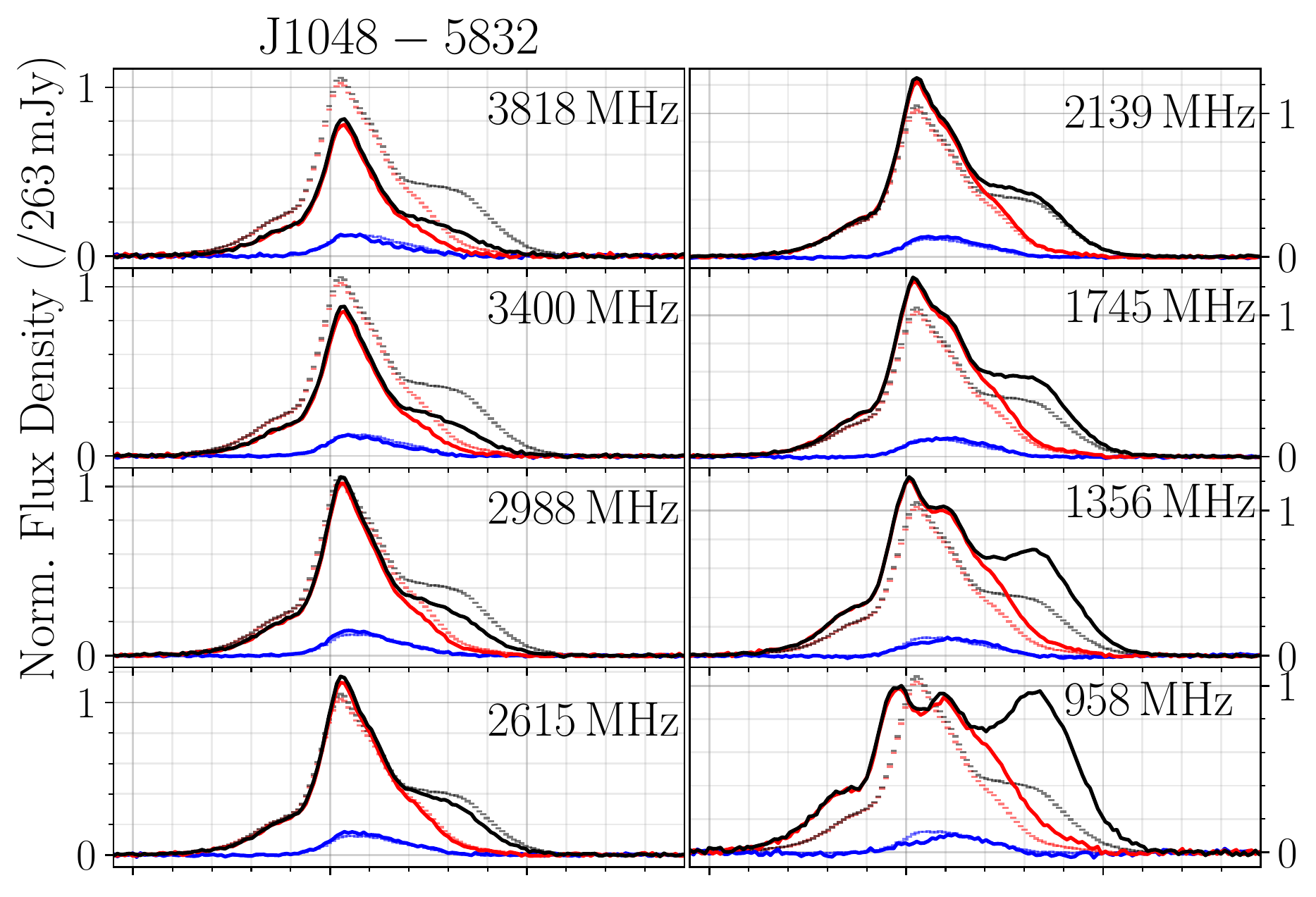}
	\includegraphics[width=\columnwidth]{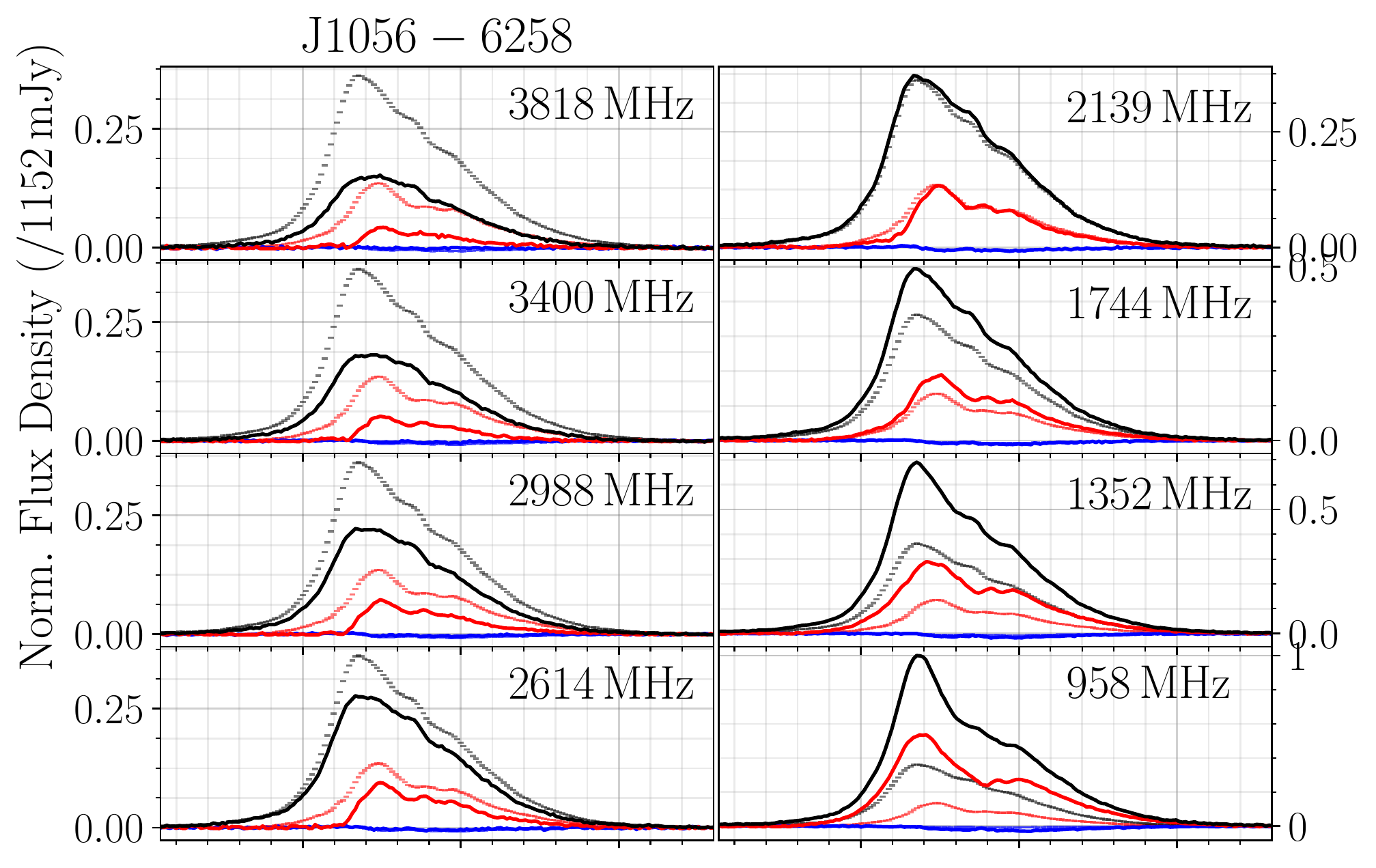}

	\includegraphics[width=\columnwidth]{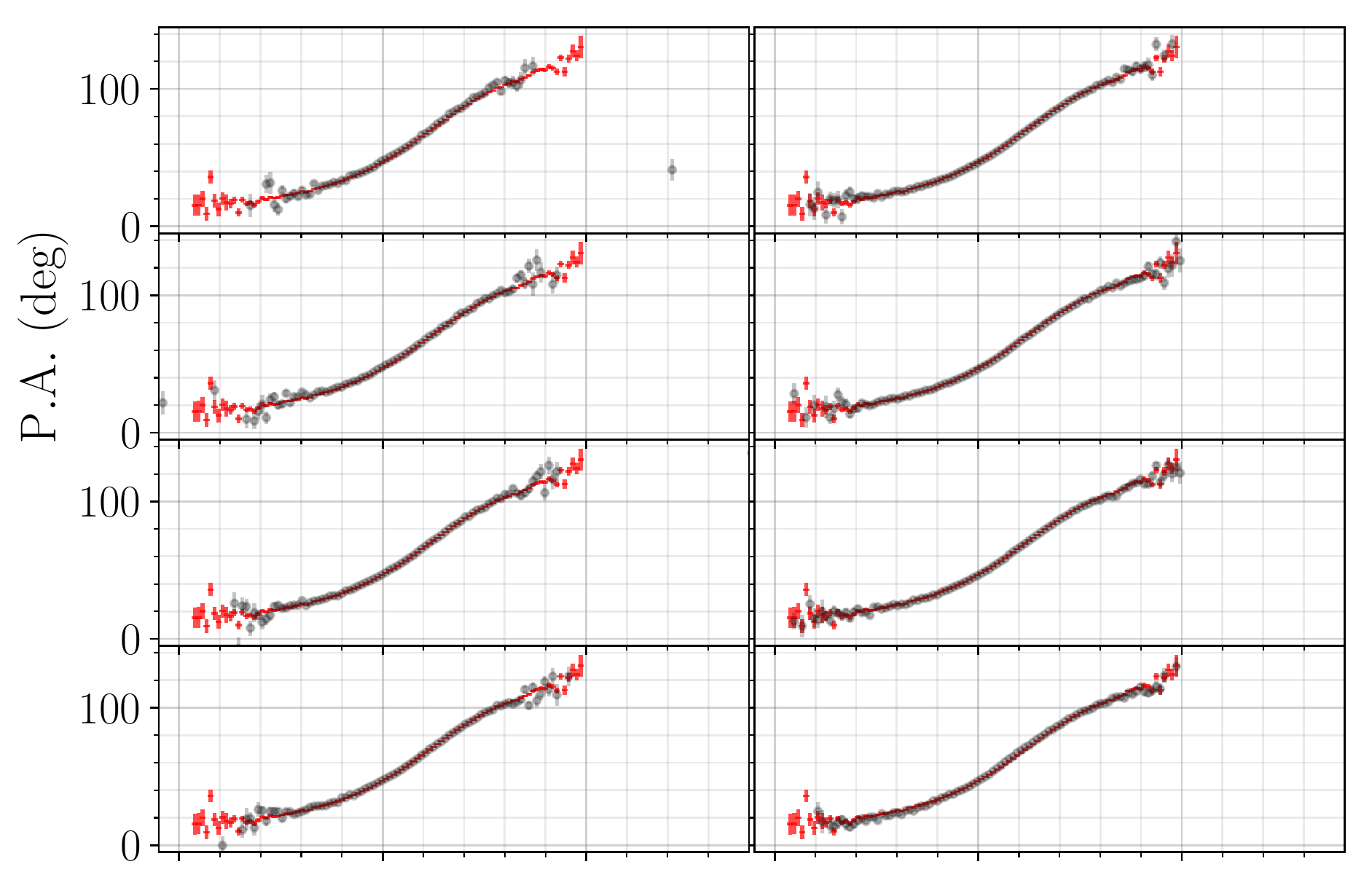}
	\includegraphics[width=\columnwidth]{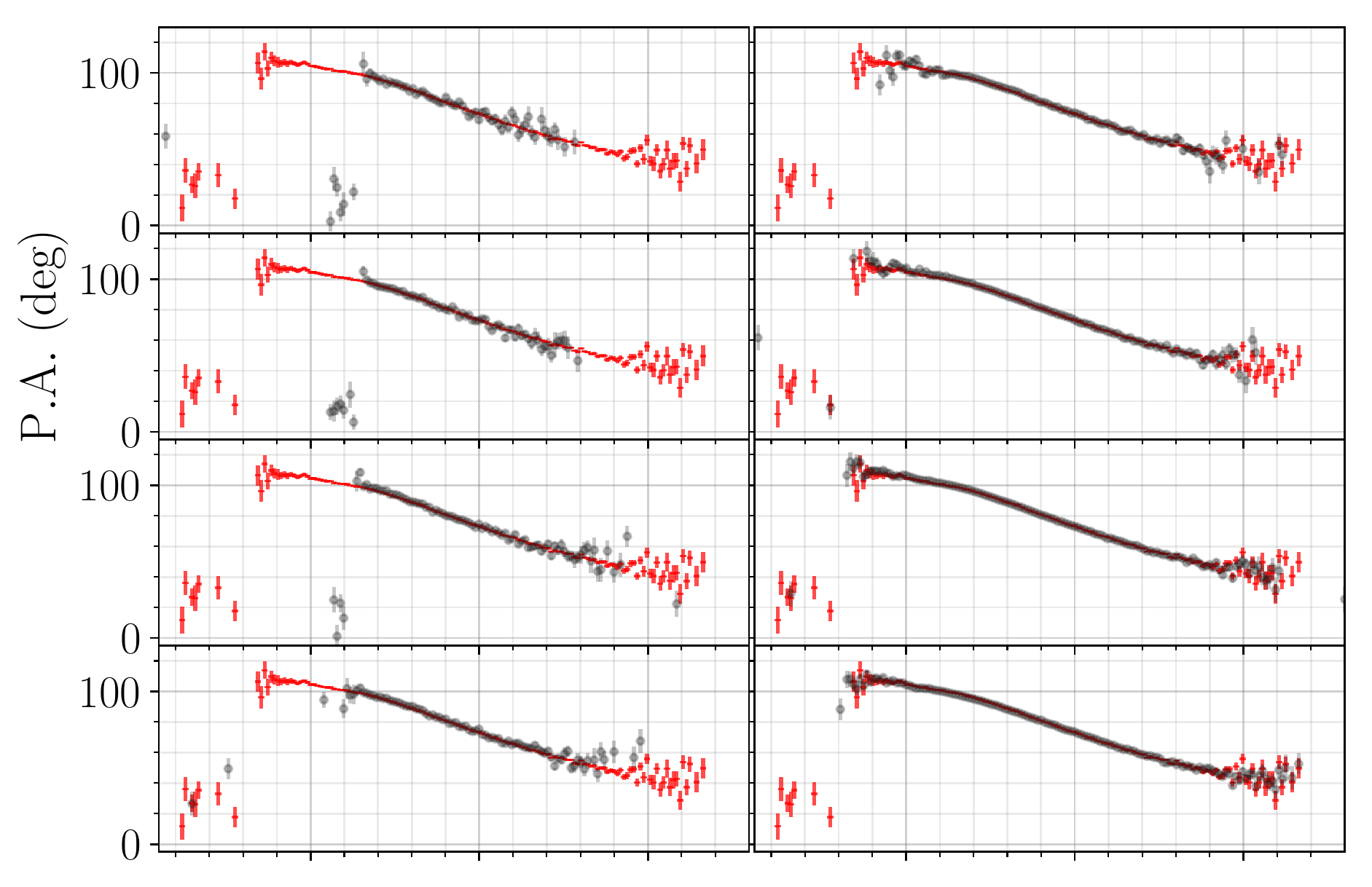}

	\includegraphics[width=\columnwidth]{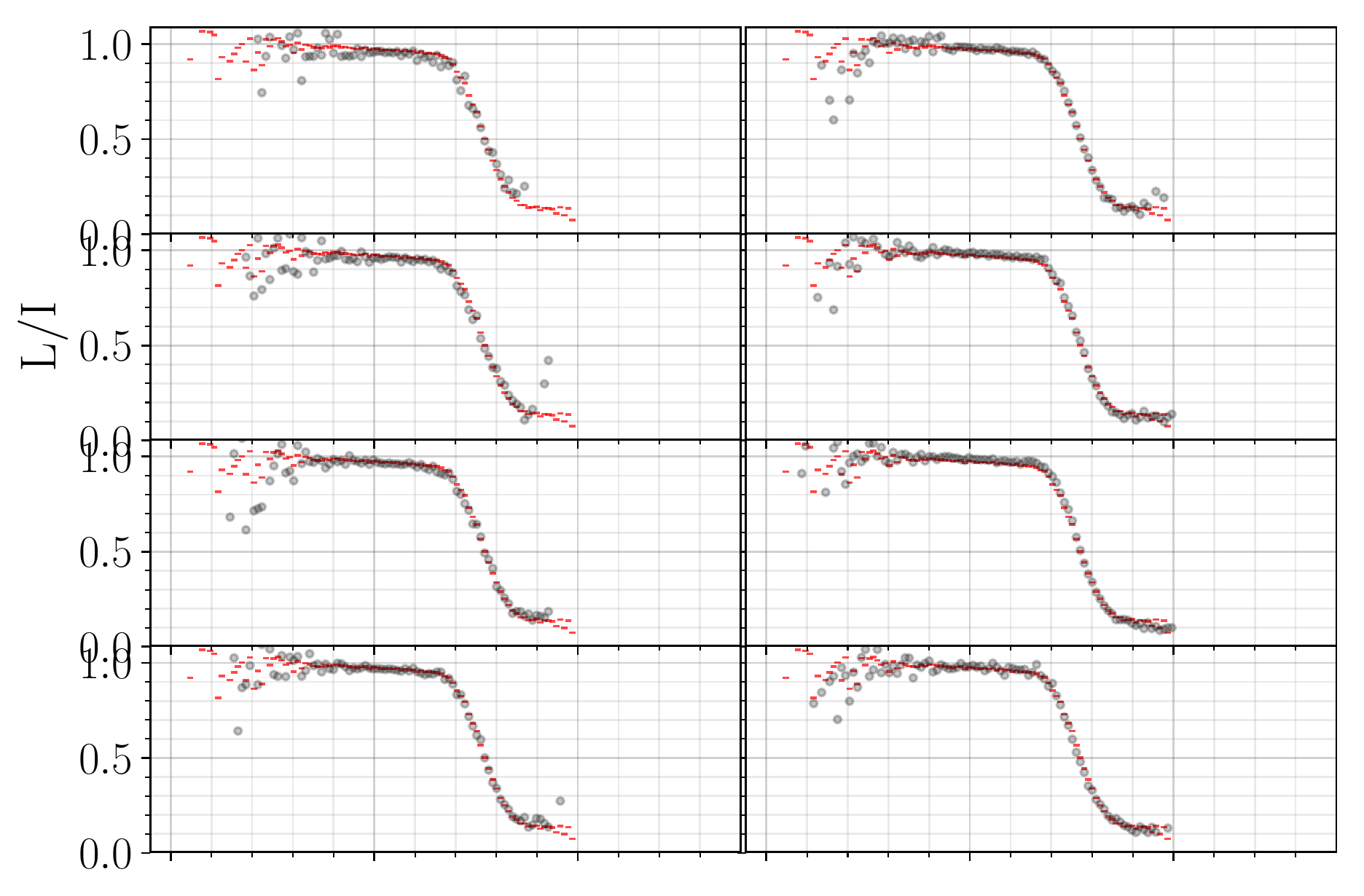}
	\includegraphics[width=\columnwidth]{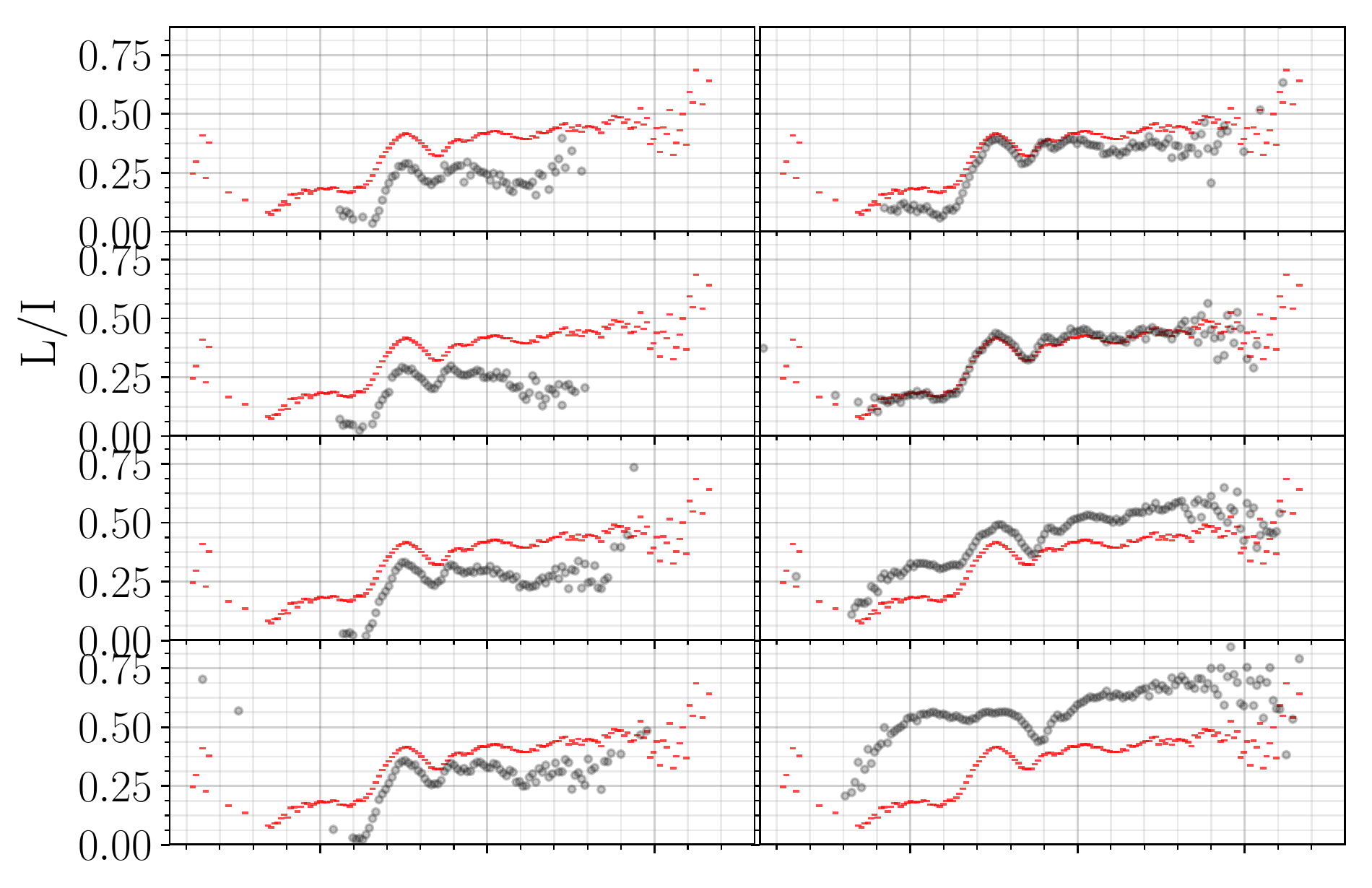}

	\includegraphics[width=\columnwidth]{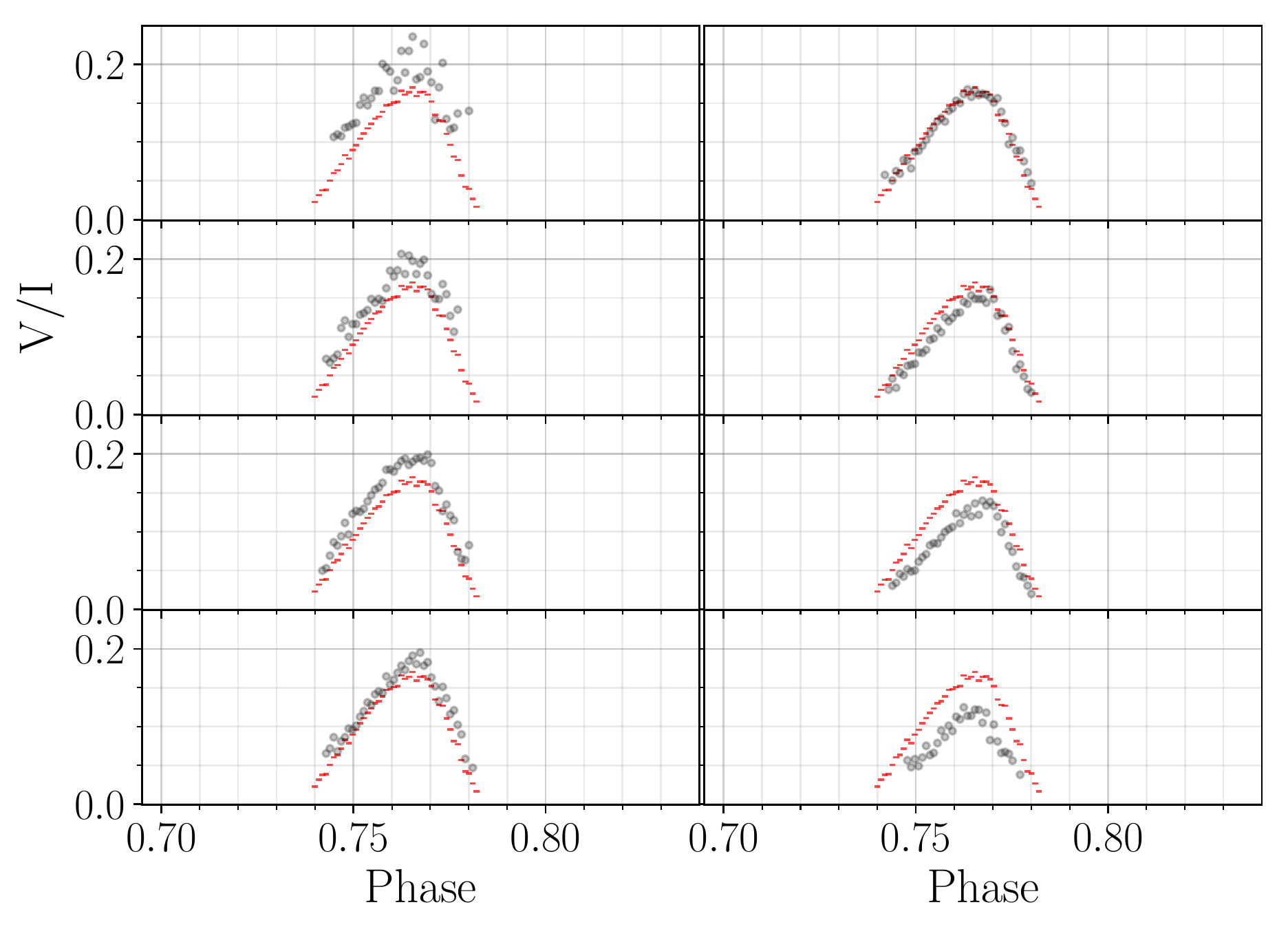}
	\includegraphics[width=\columnwidth]{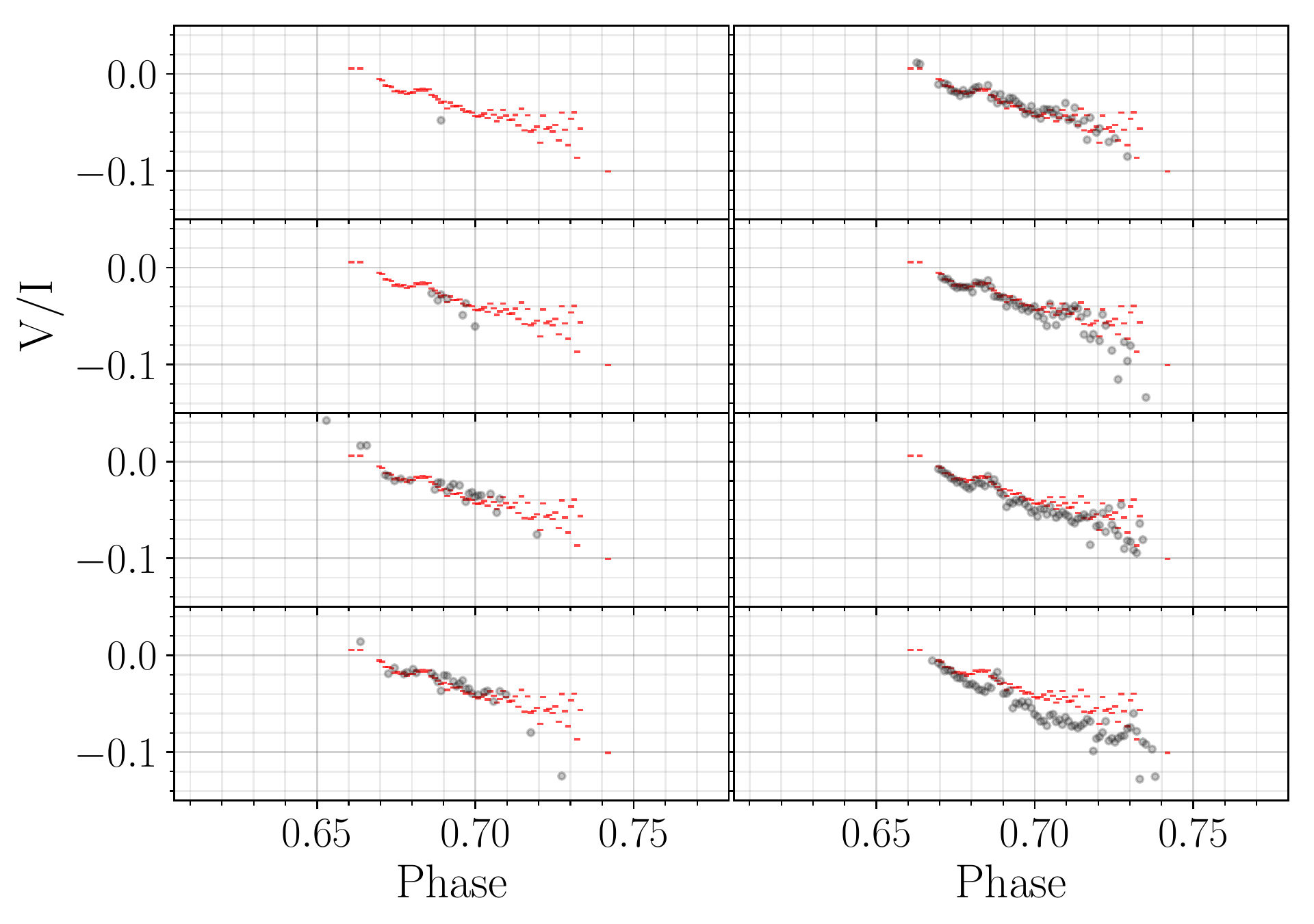}

    \caption{...Figure \ref{fig:profs1} continued...}
\end{figure*}

\begin{figure*} 
	\includegraphics[width=\columnwidth]{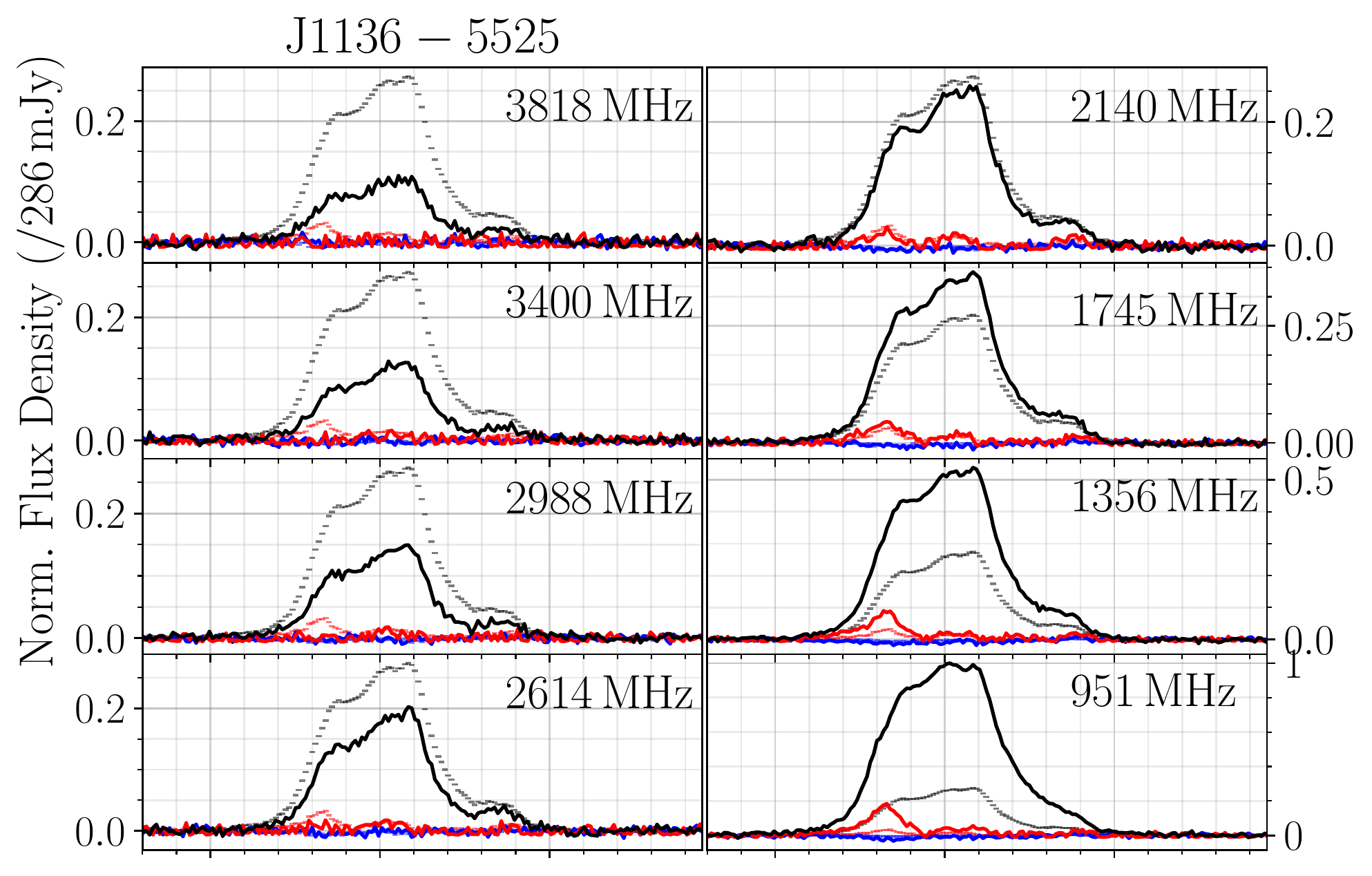}
	\includegraphics[width=\columnwidth]{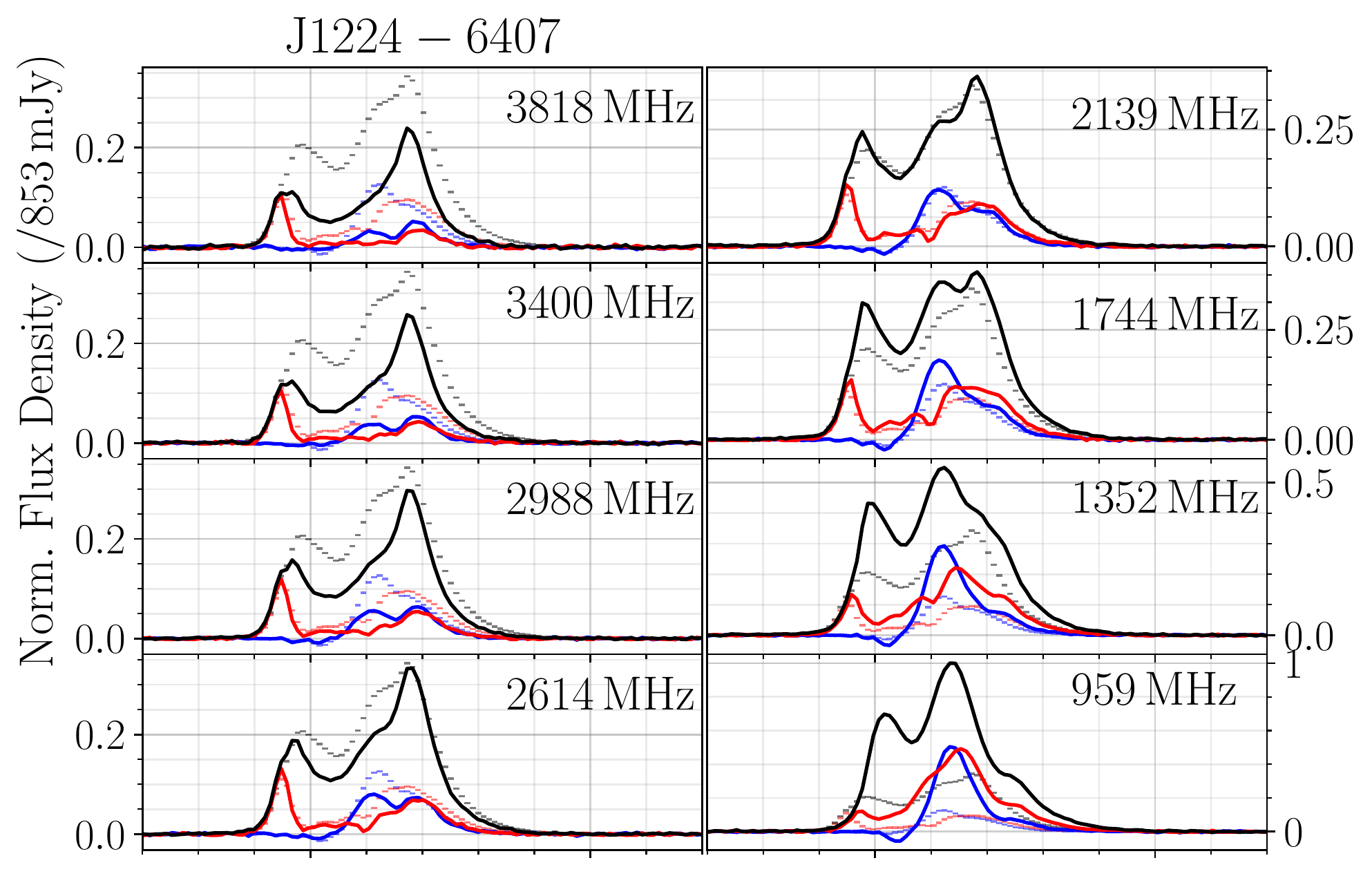}

	\includegraphics[width=\columnwidth]{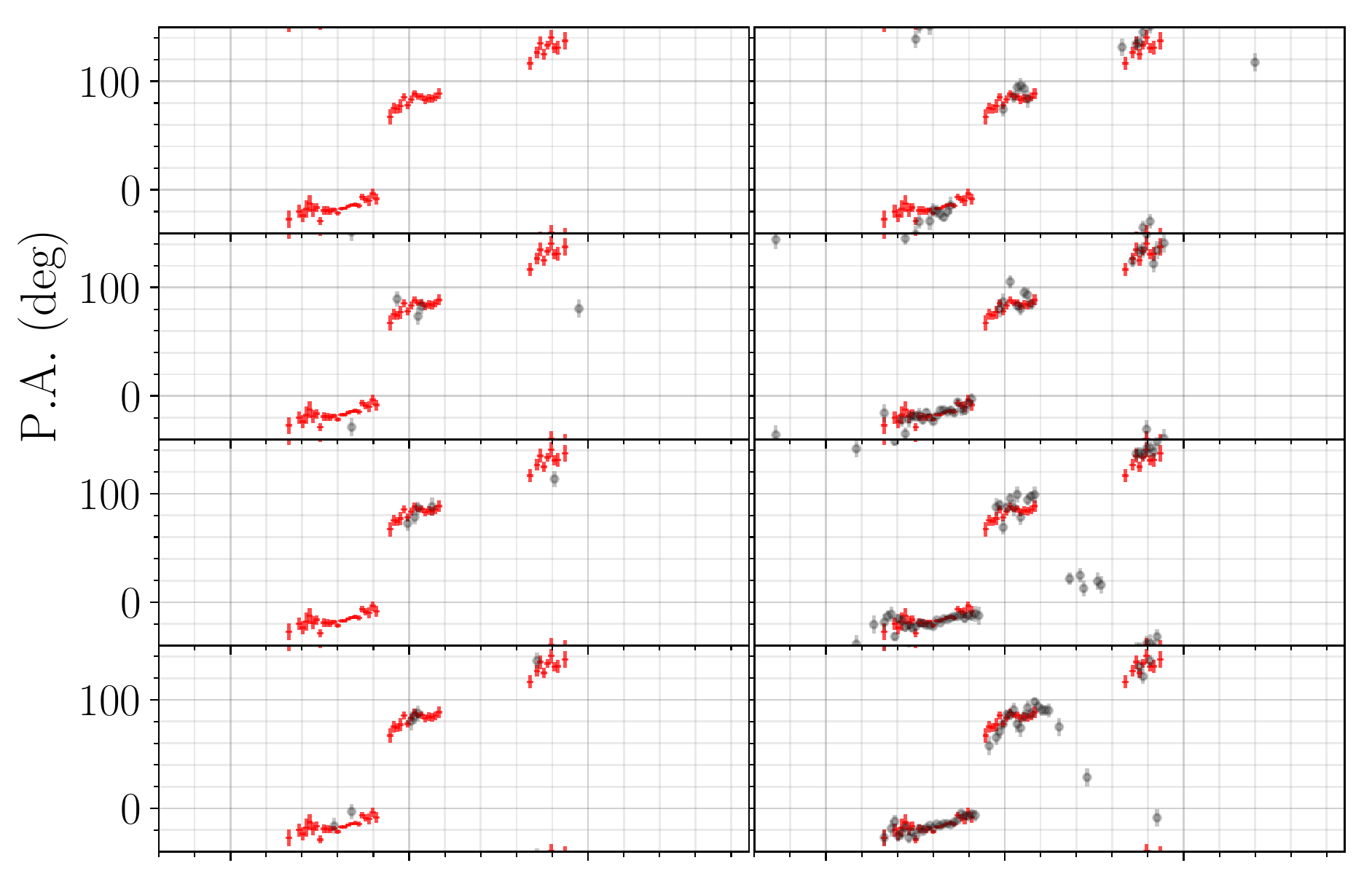}
	\includegraphics[width=\columnwidth]{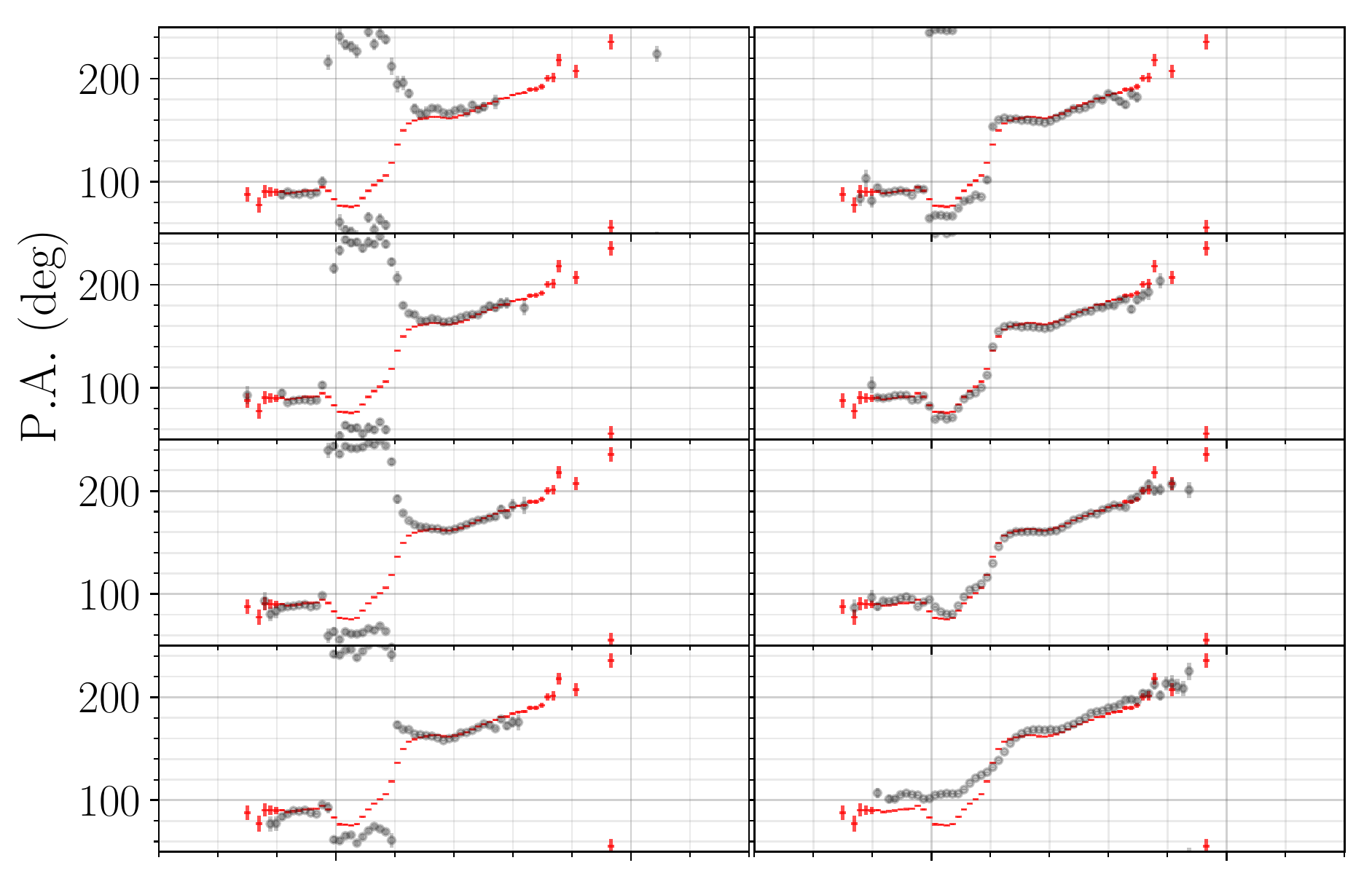}

	\includegraphics[width=\columnwidth]{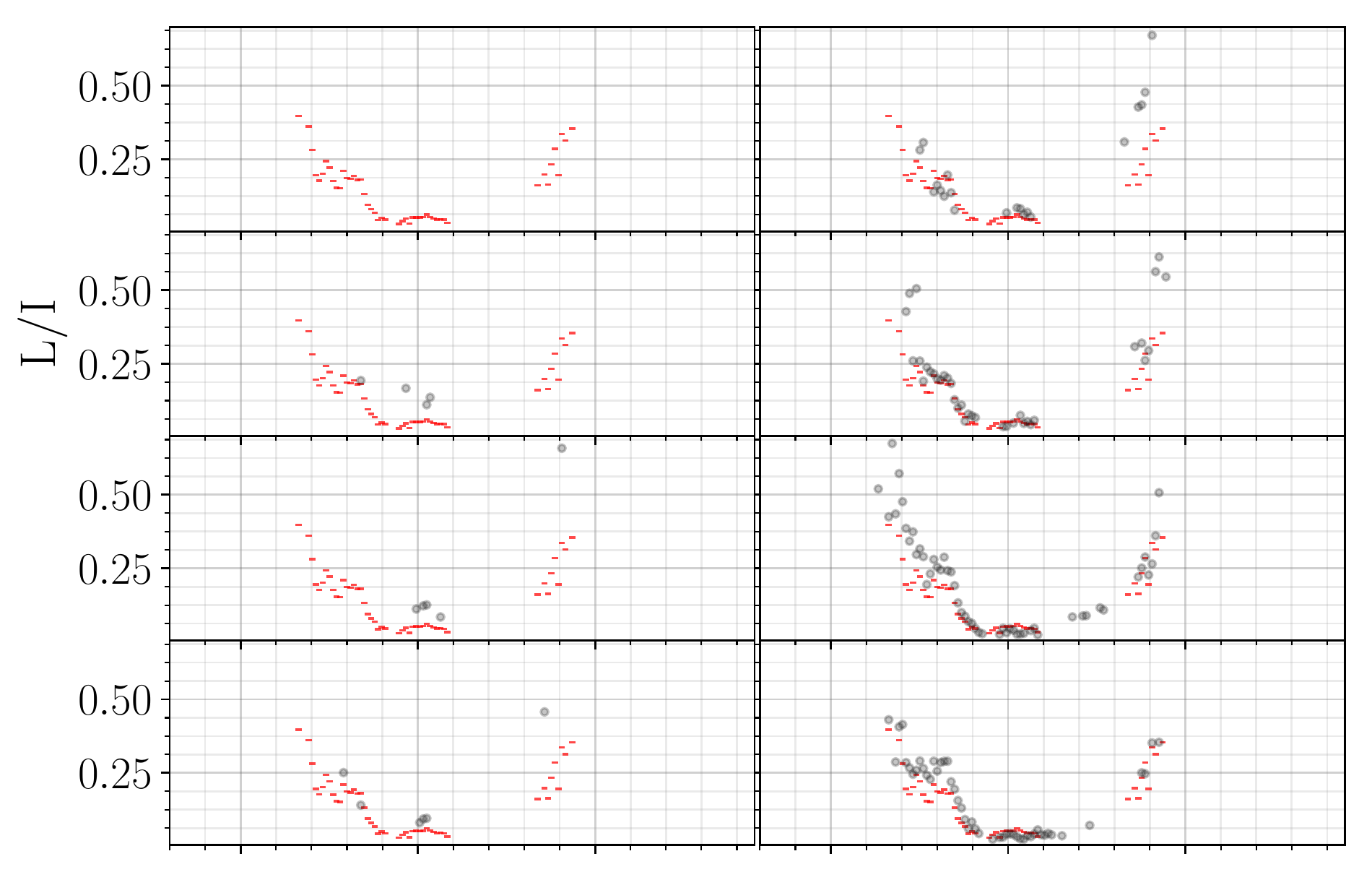}
	\includegraphics[width=\columnwidth]{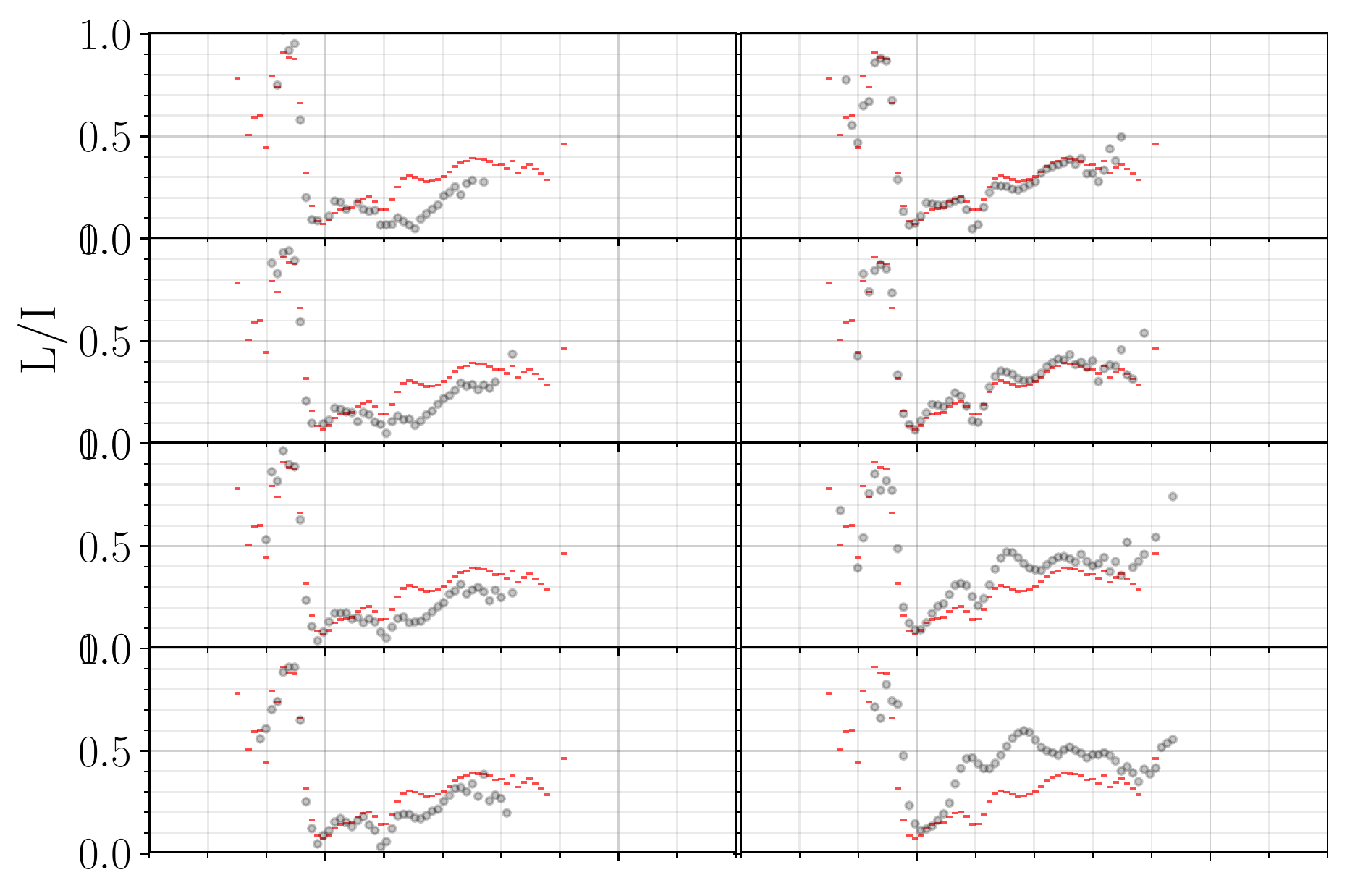}

	\includegraphics[width=\columnwidth]{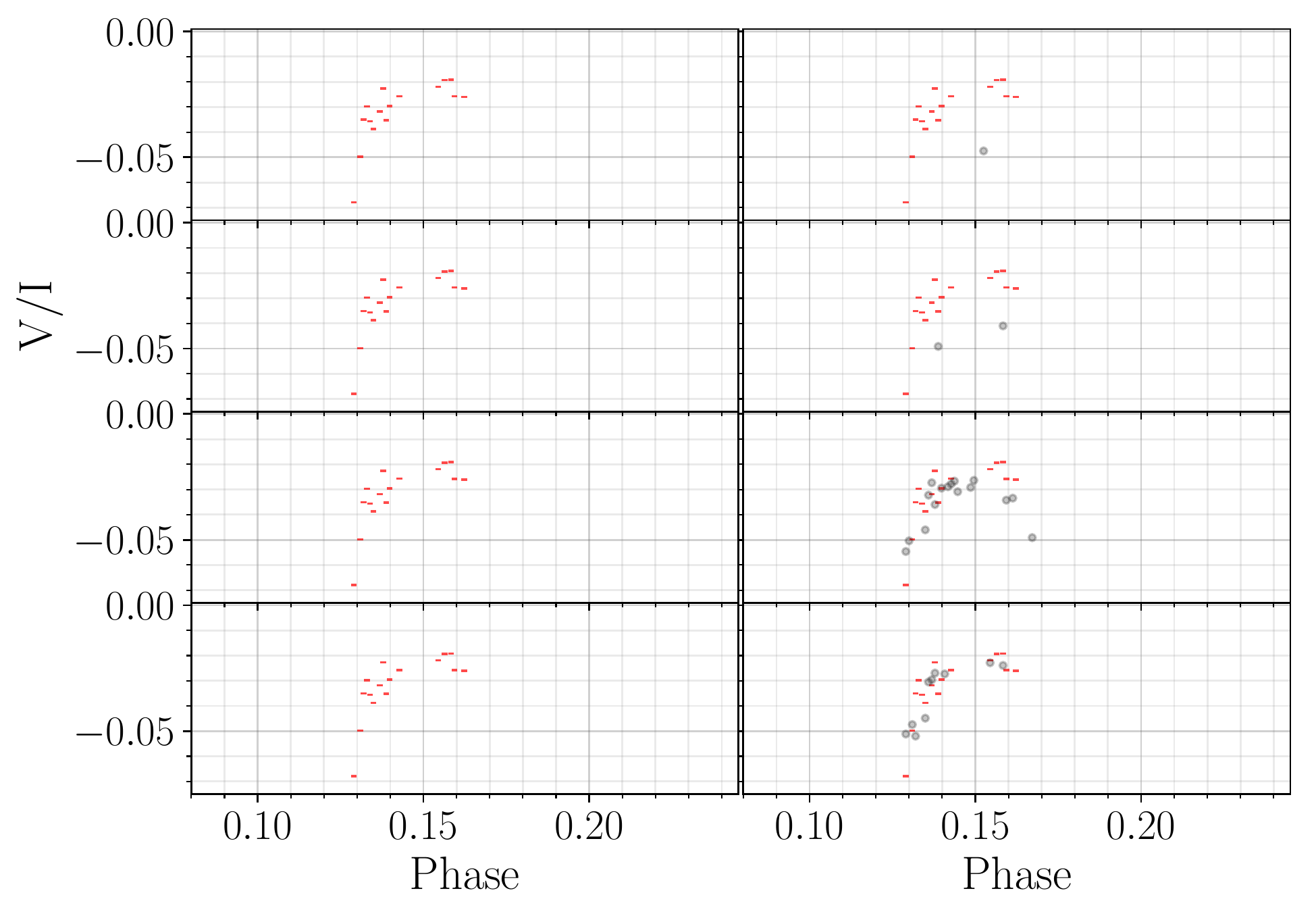}
	\includegraphics[width=\columnwidth]{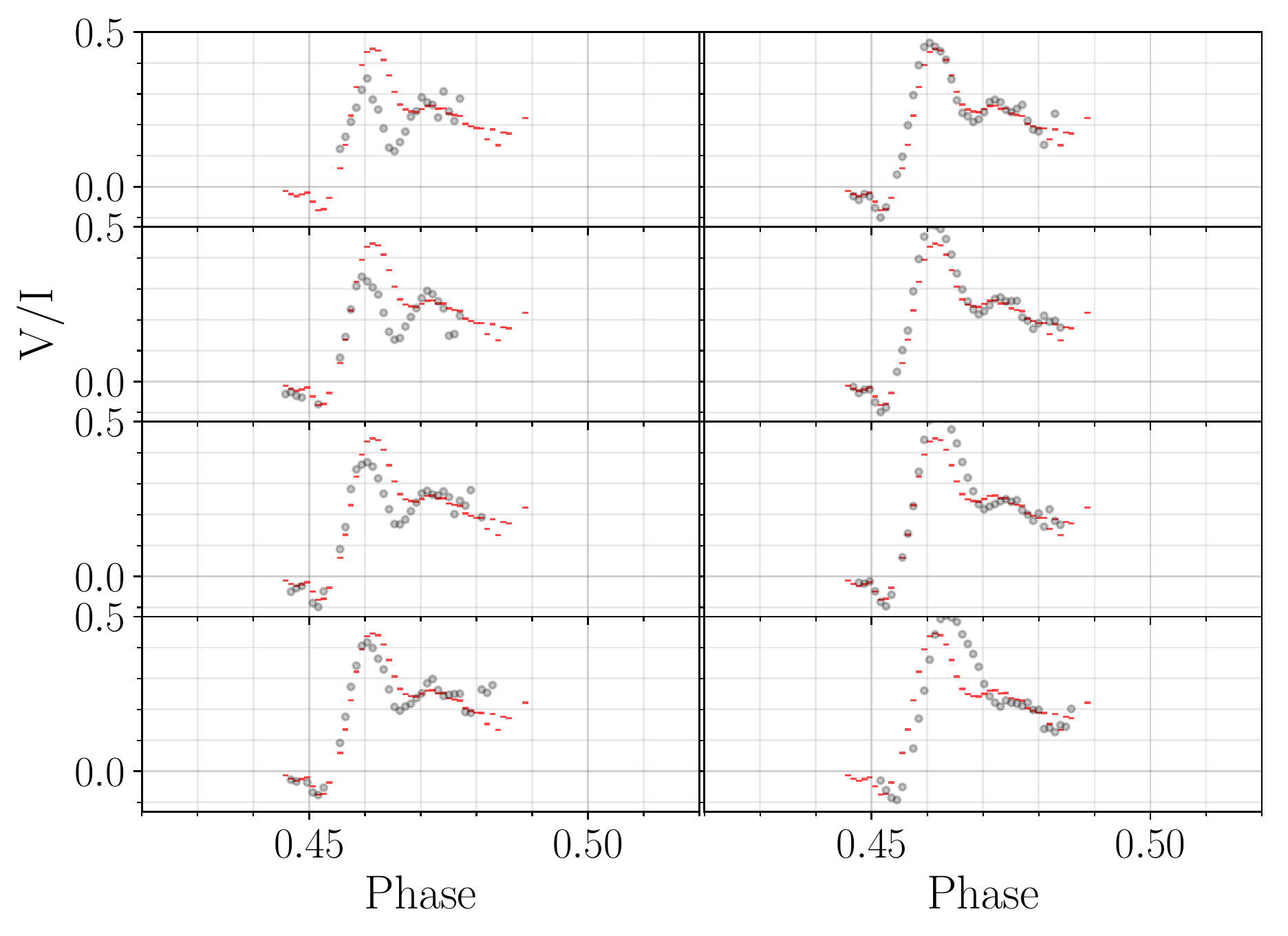}

    \caption{...Figure \ref{fig:profs1} continued...}
\end{figure*}

\begin{figure*} 
	\includegraphics[width=\columnwidth]{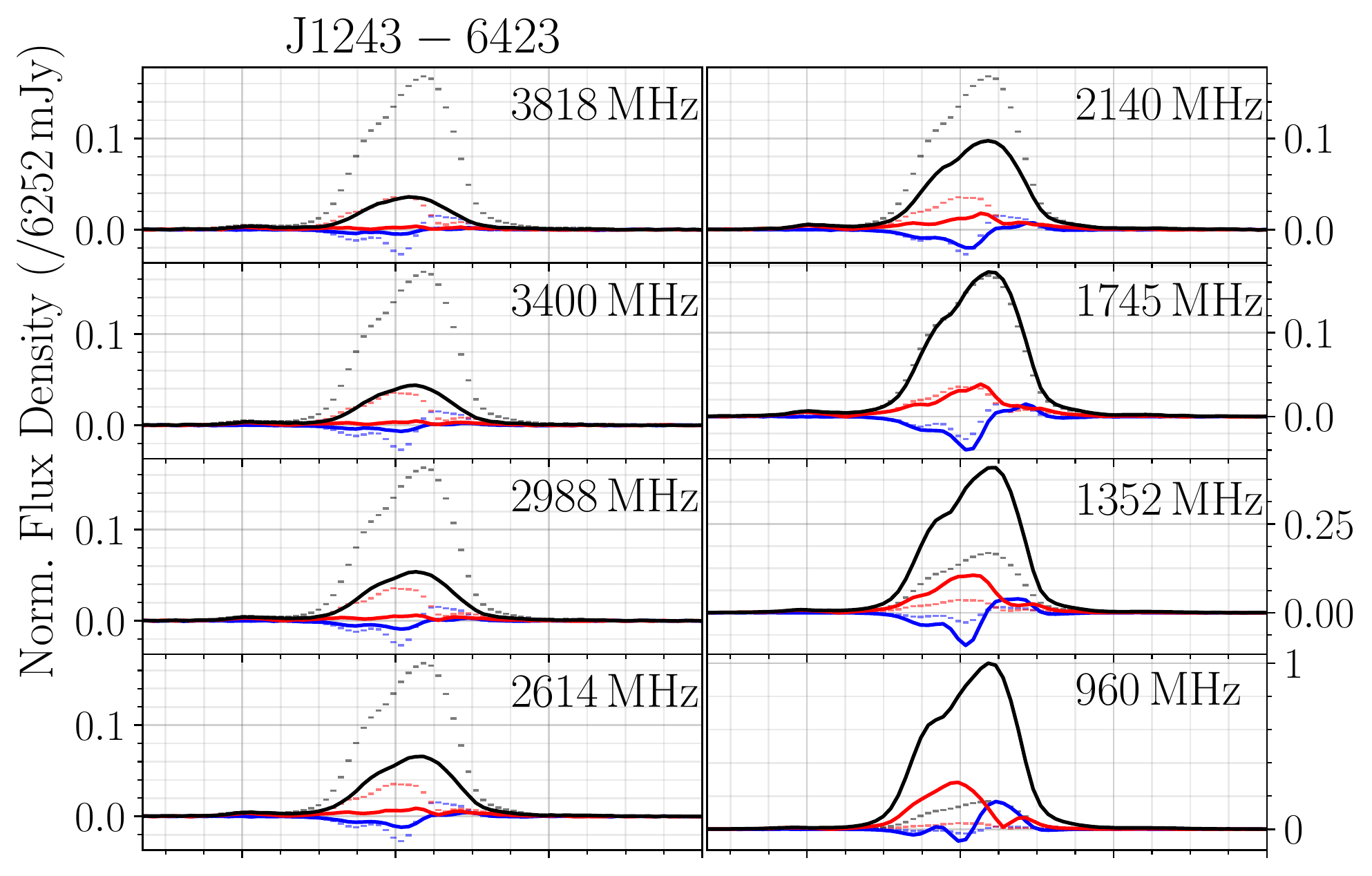}
	\includegraphics[width=\columnwidth]{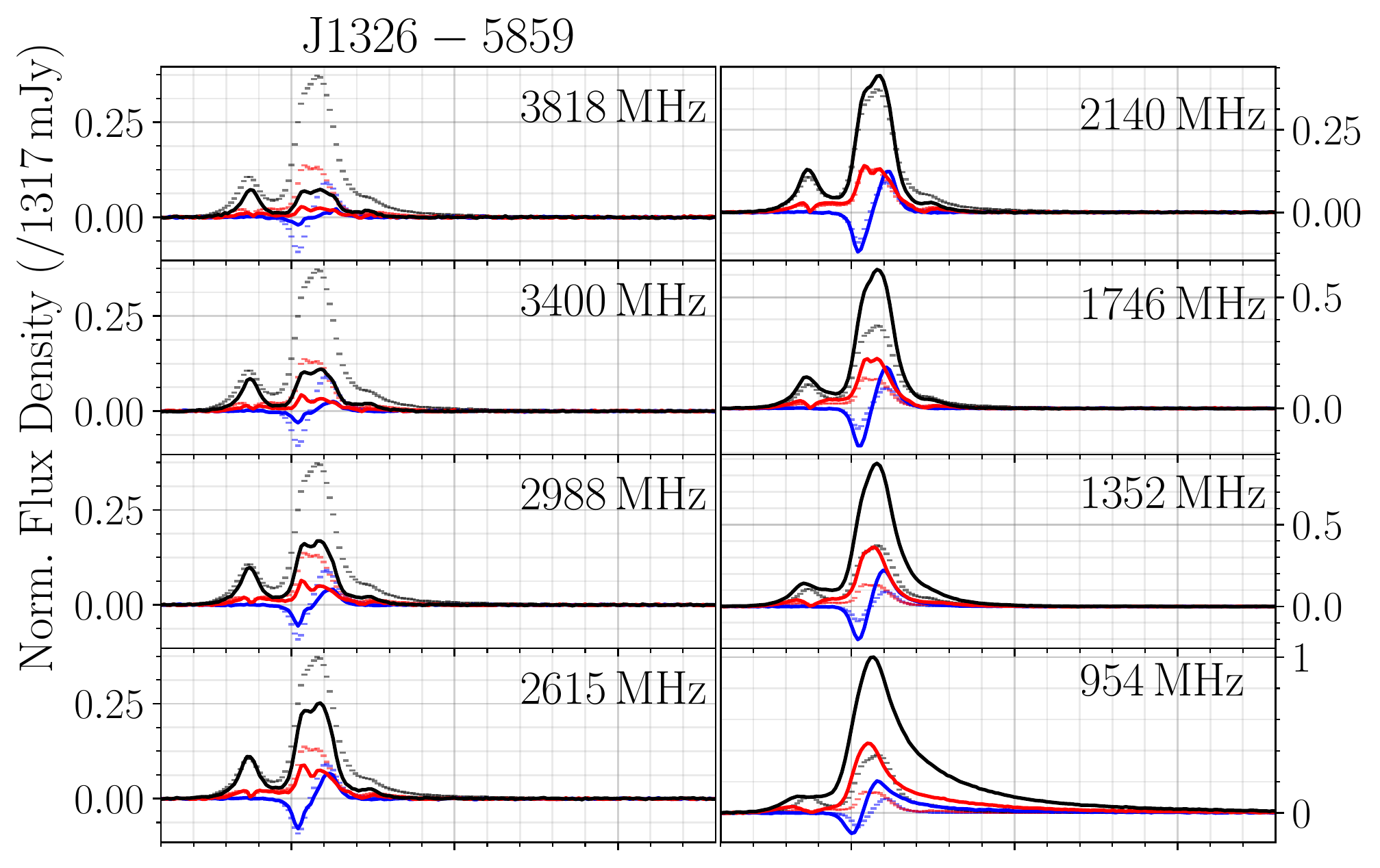}

	\includegraphics[width=\columnwidth]{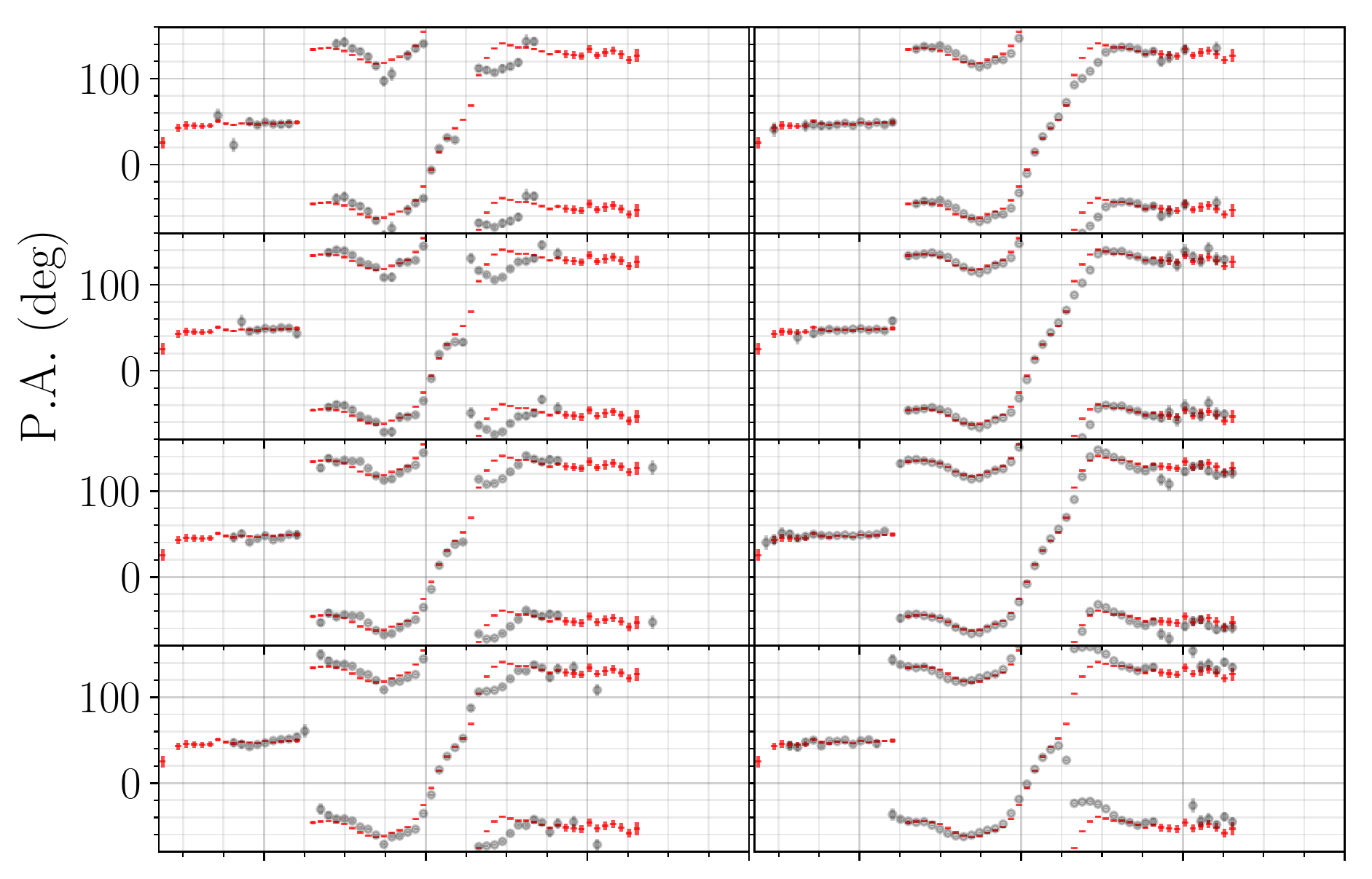}
	\includegraphics[width=\columnwidth]{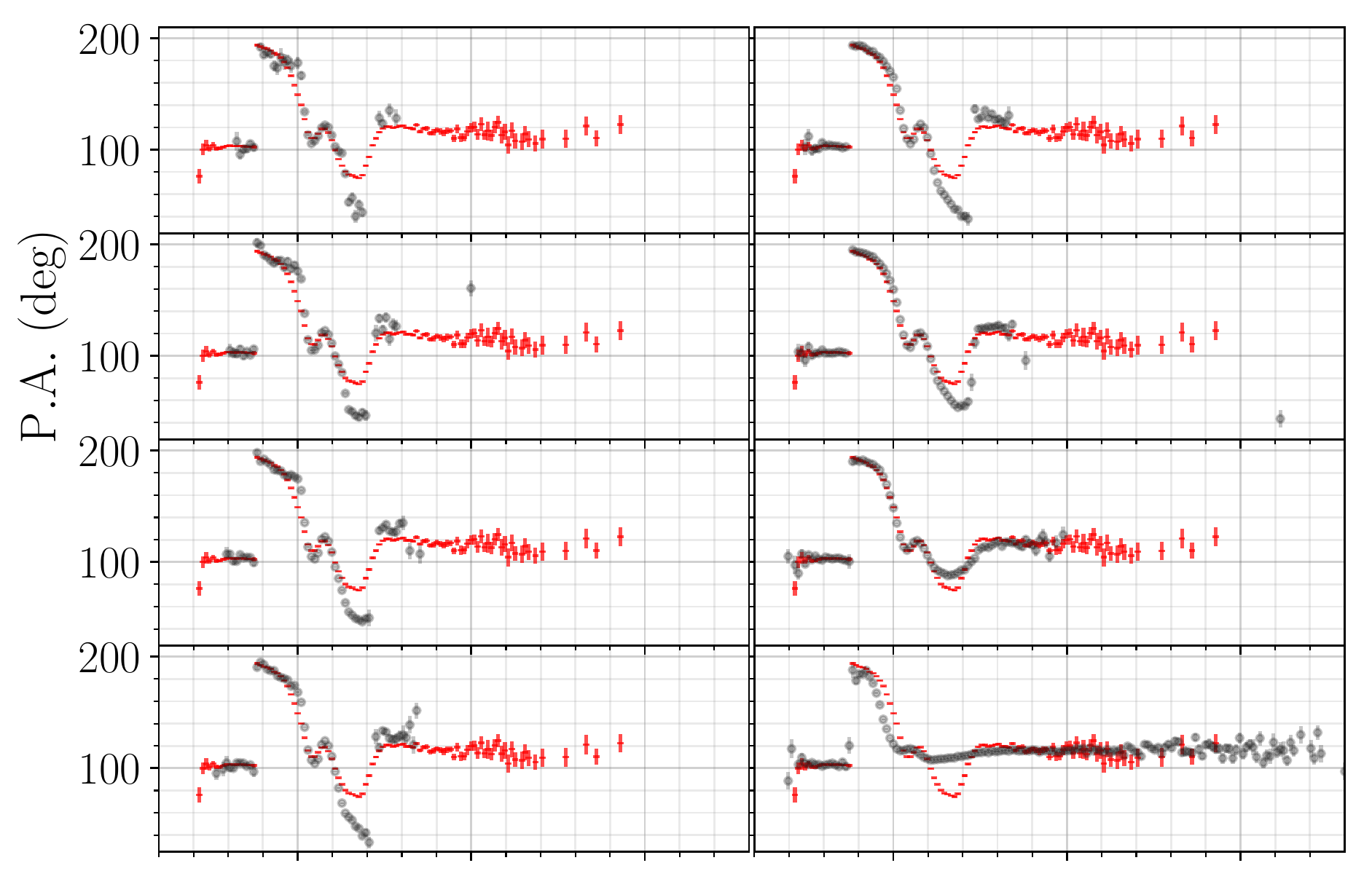}

	\includegraphics[width=\columnwidth]{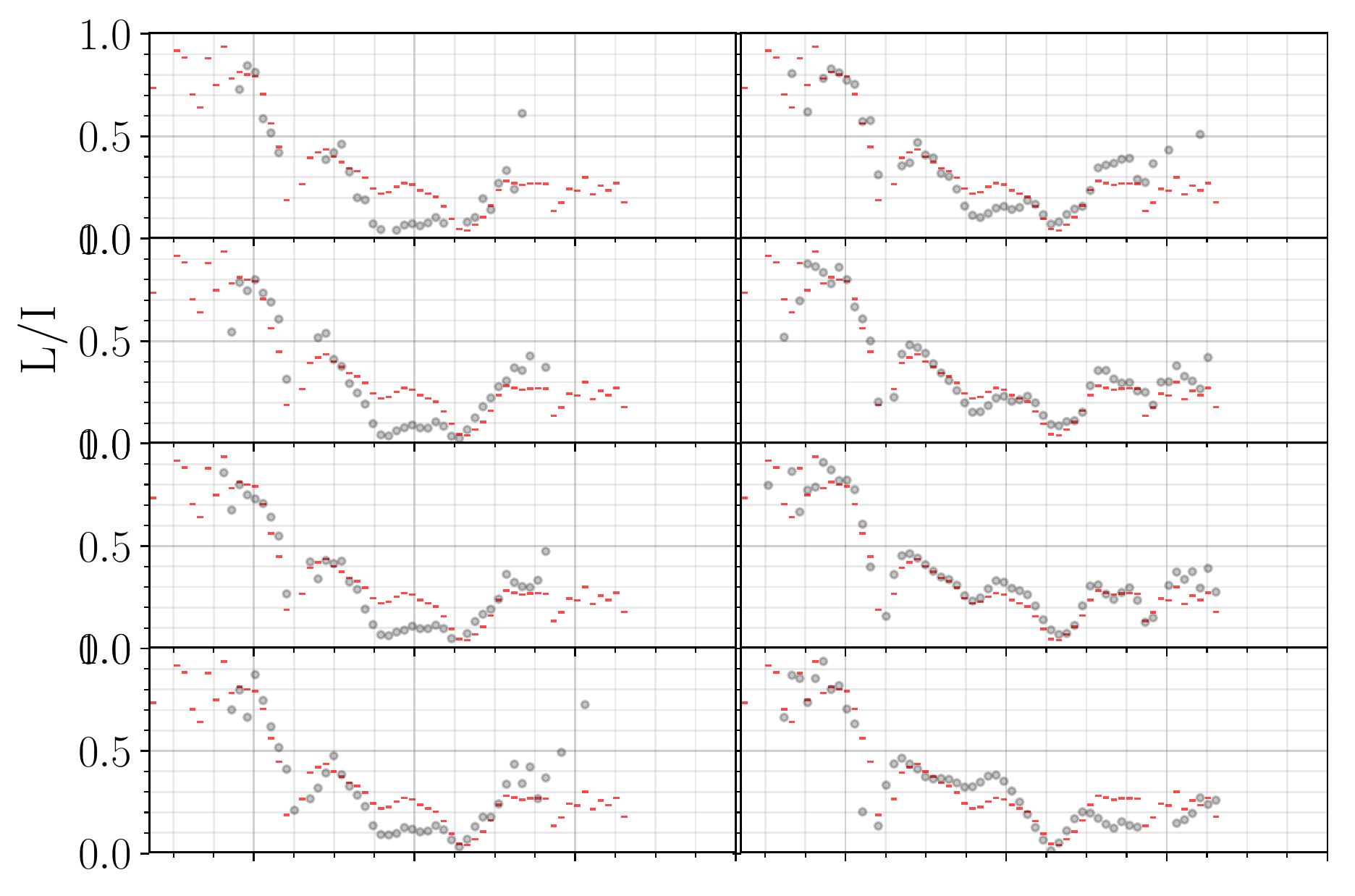}
	\includegraphics[width=\columnwidth]{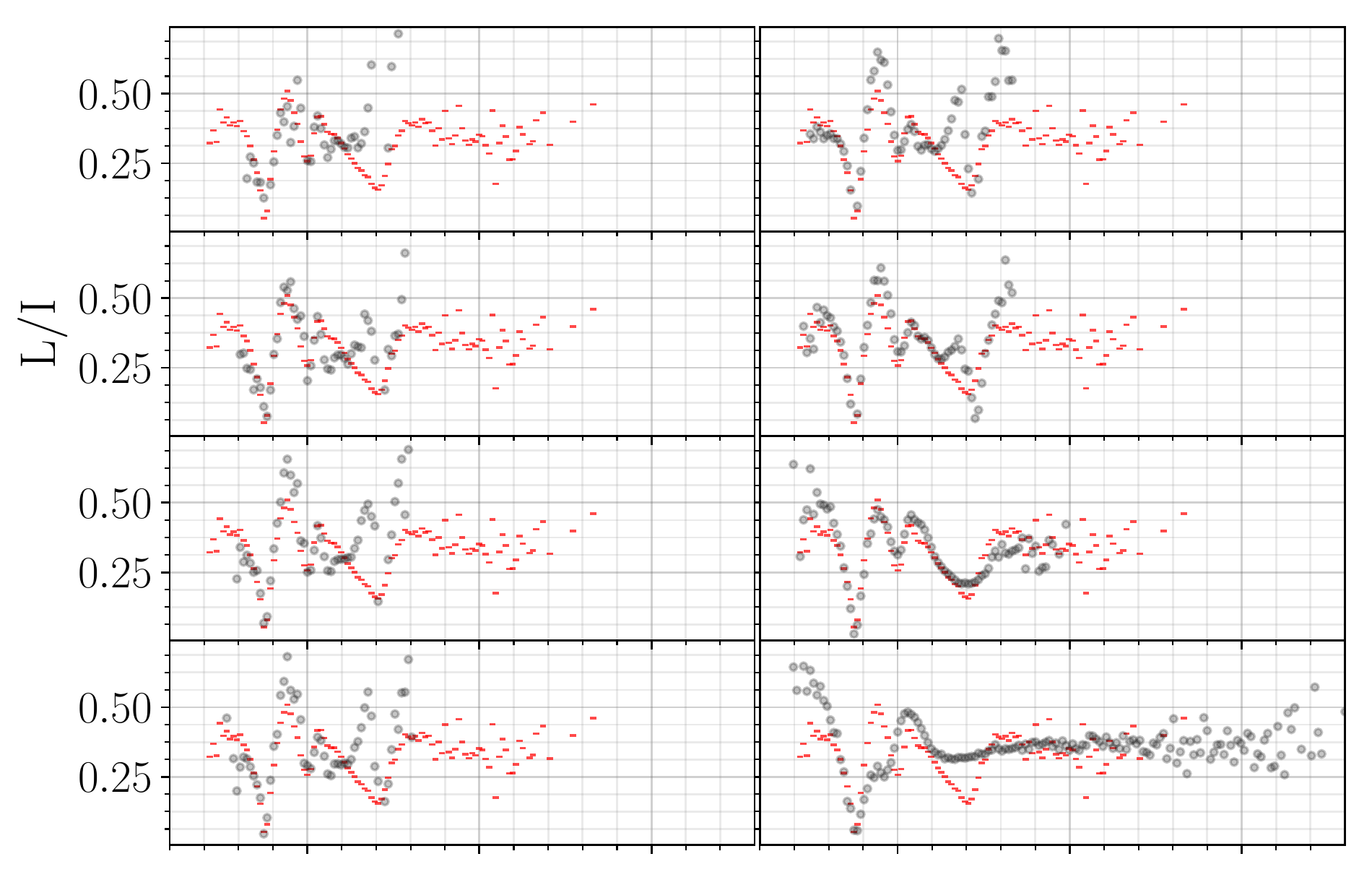}

	\includegraphics[width=\columnwidth]{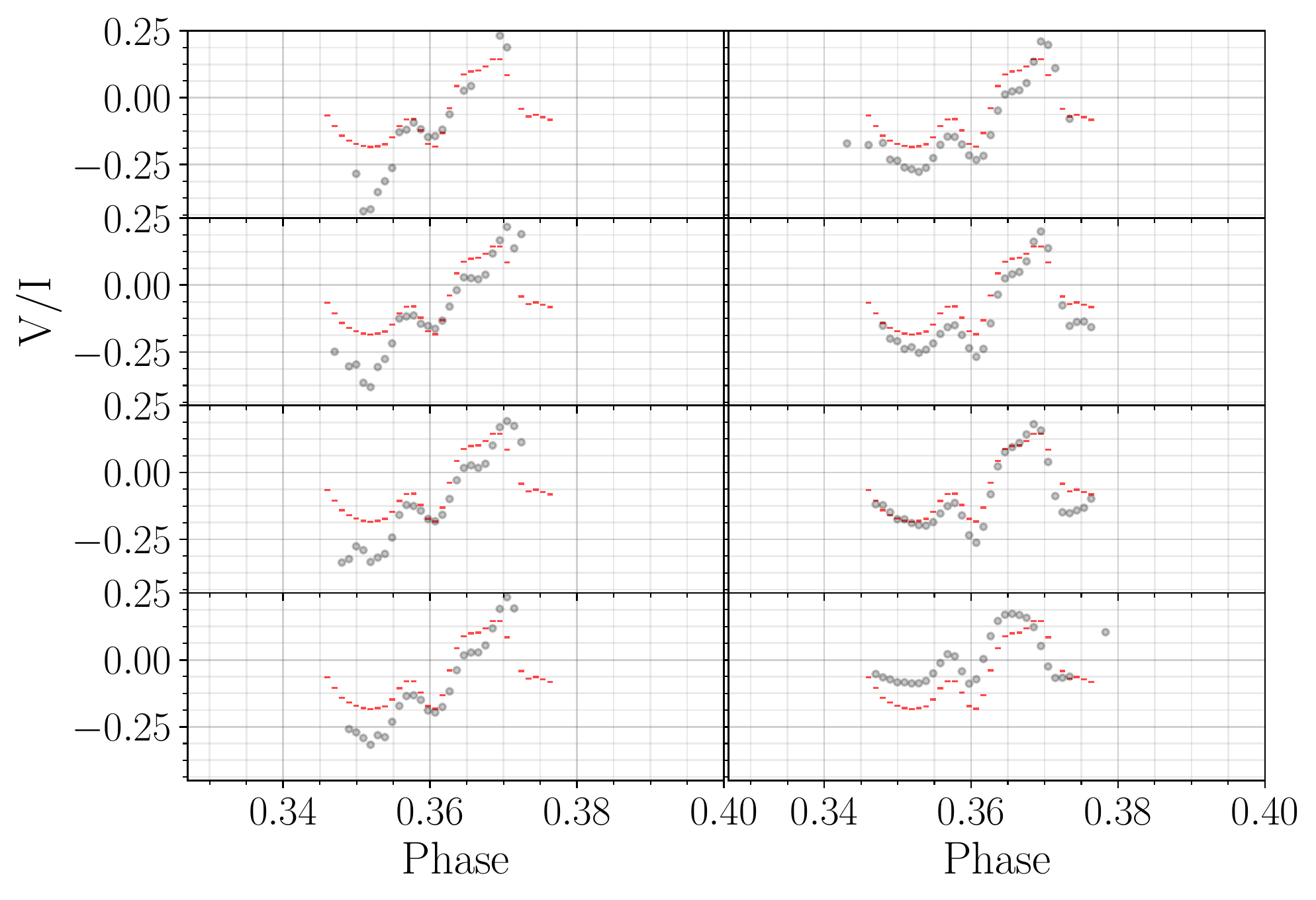}
	\includegraphics[width=\columnwidth]{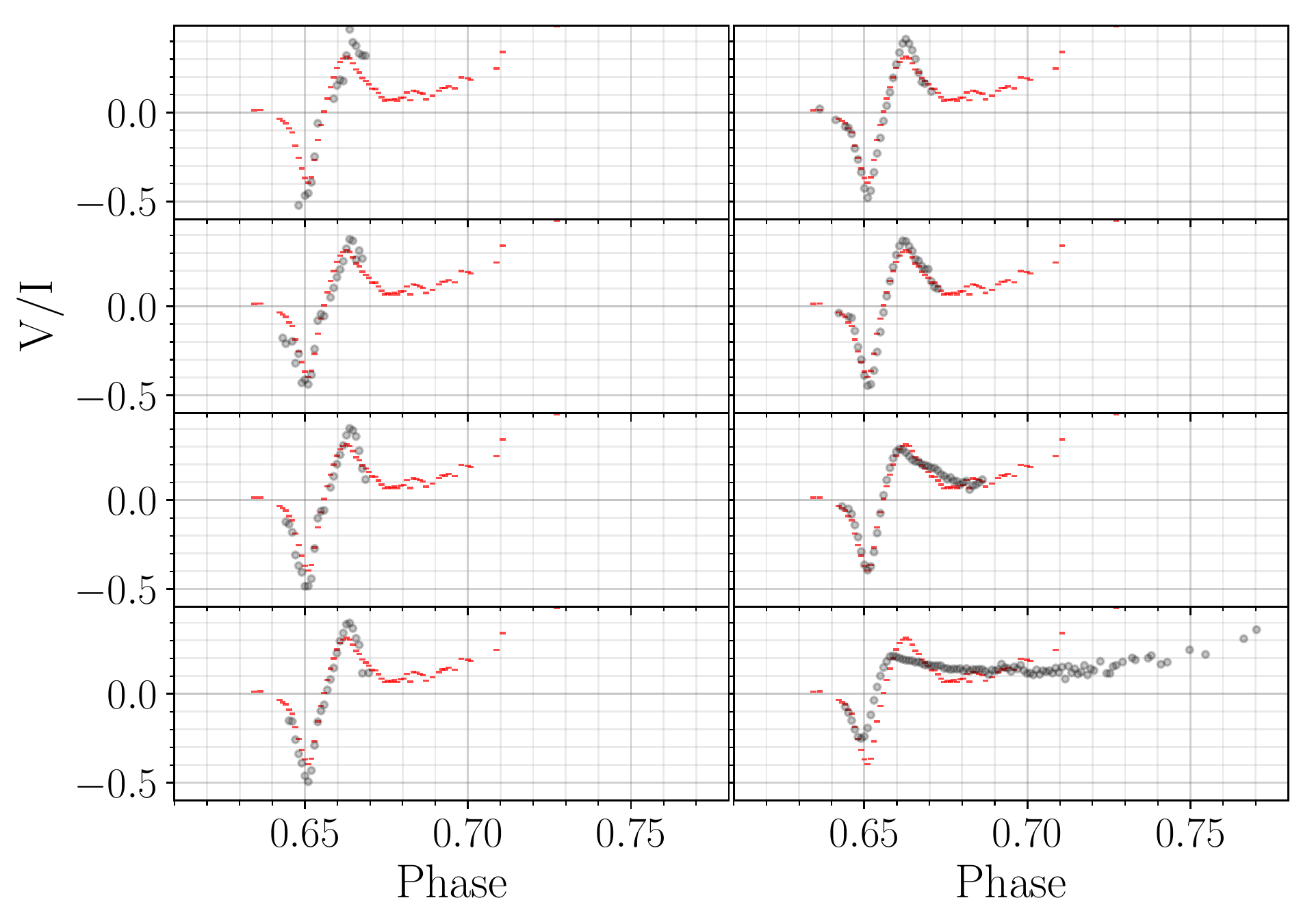}

    \caption{...Figure \ref{fig:profs1} continued...}
\end{figure*}

\begin{figure*} 
	\includegraphics[width=\columnwidth]{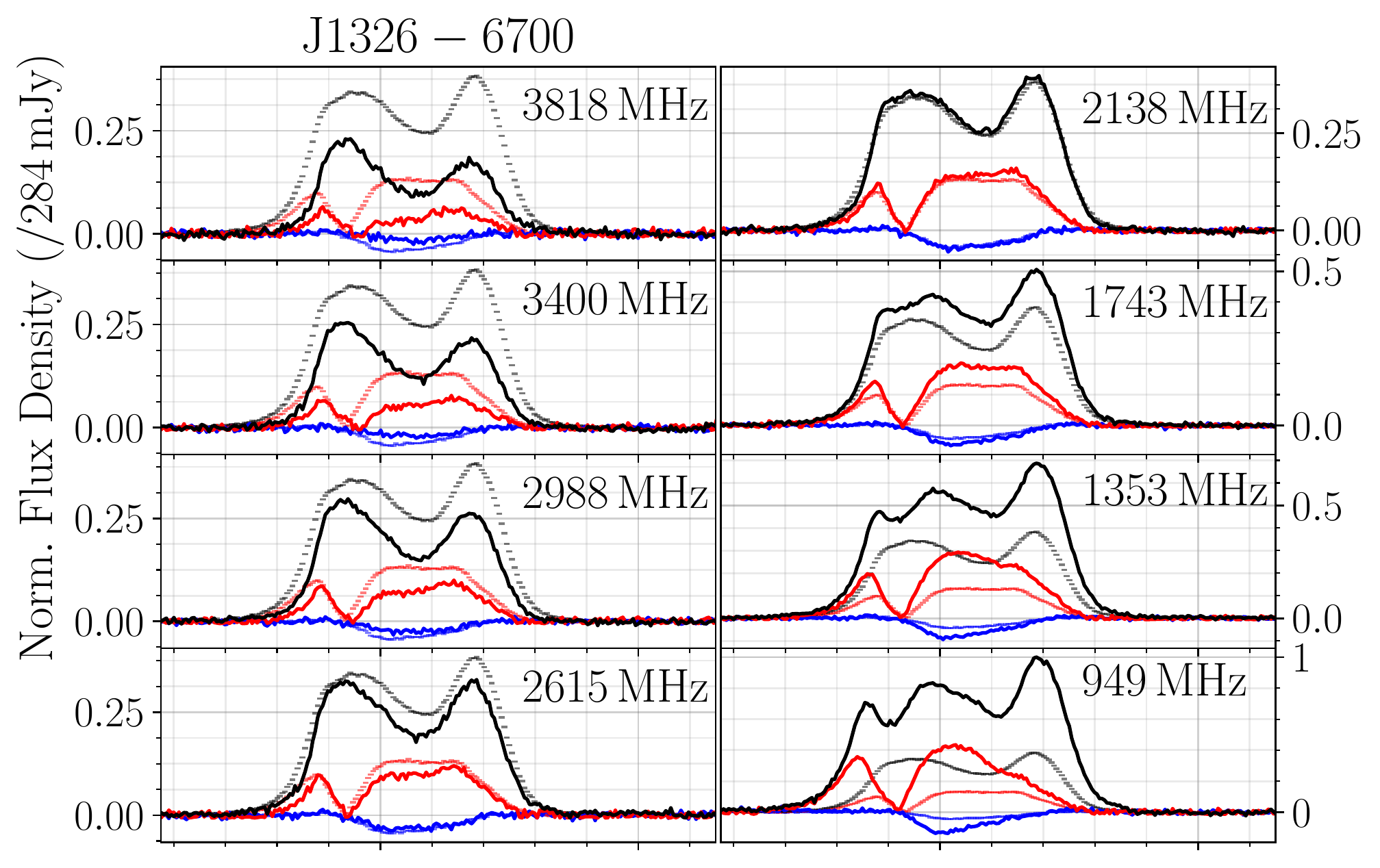}
	\includegraphics[width=\columnwidth]{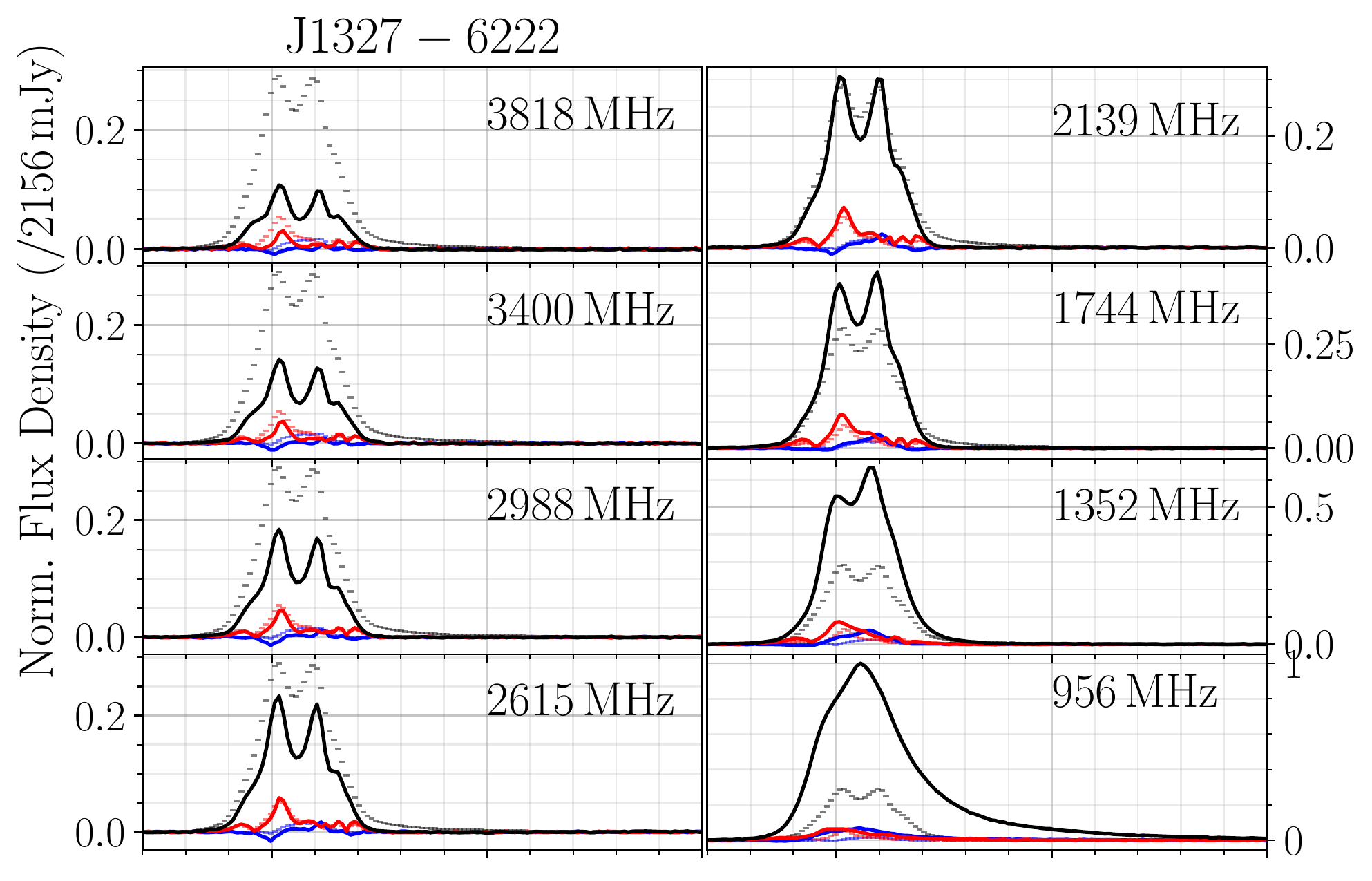}

	\includegraphics[width=\columnwidth]{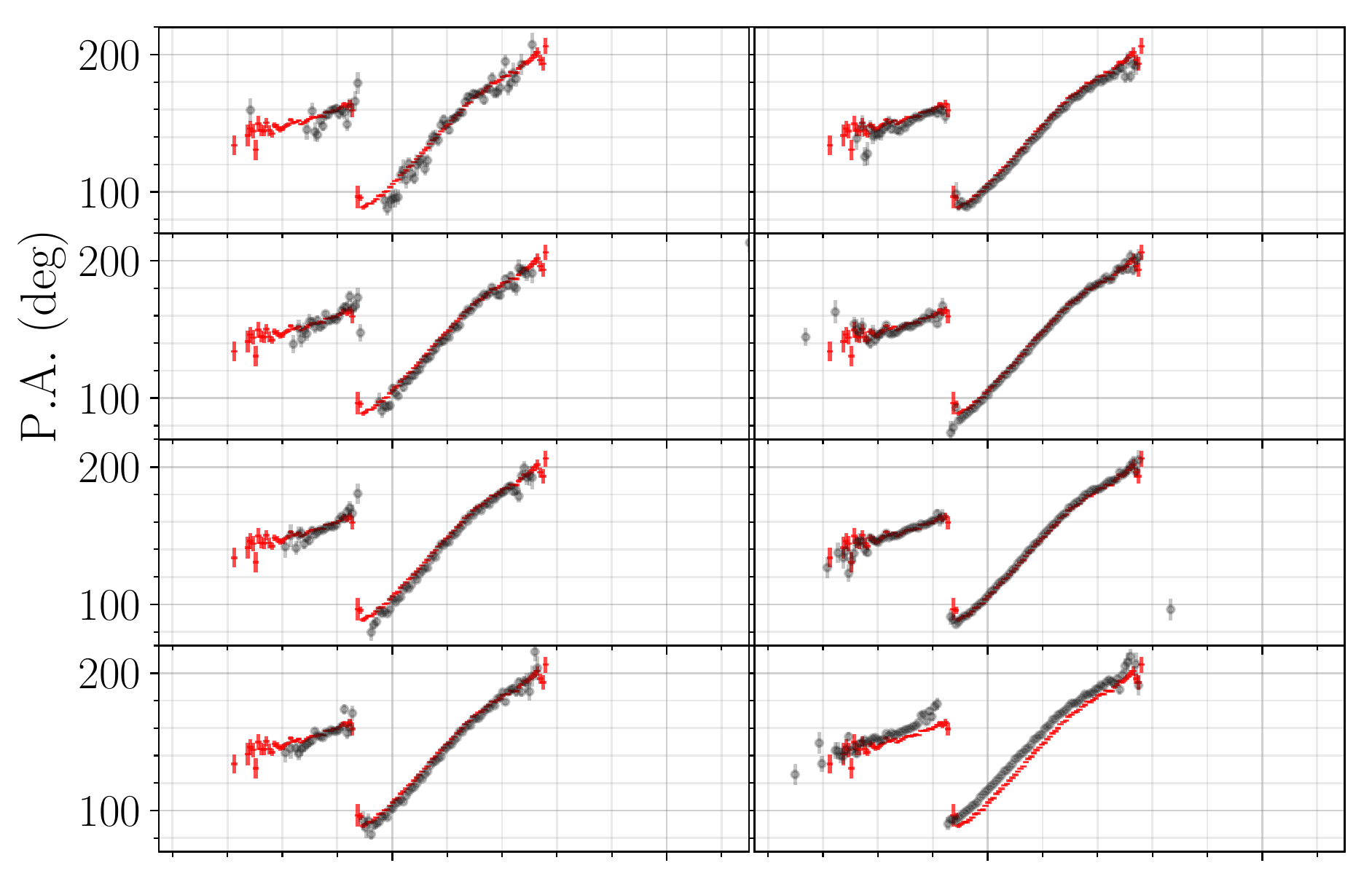}
	\includegraphics[width=\columnwidth]{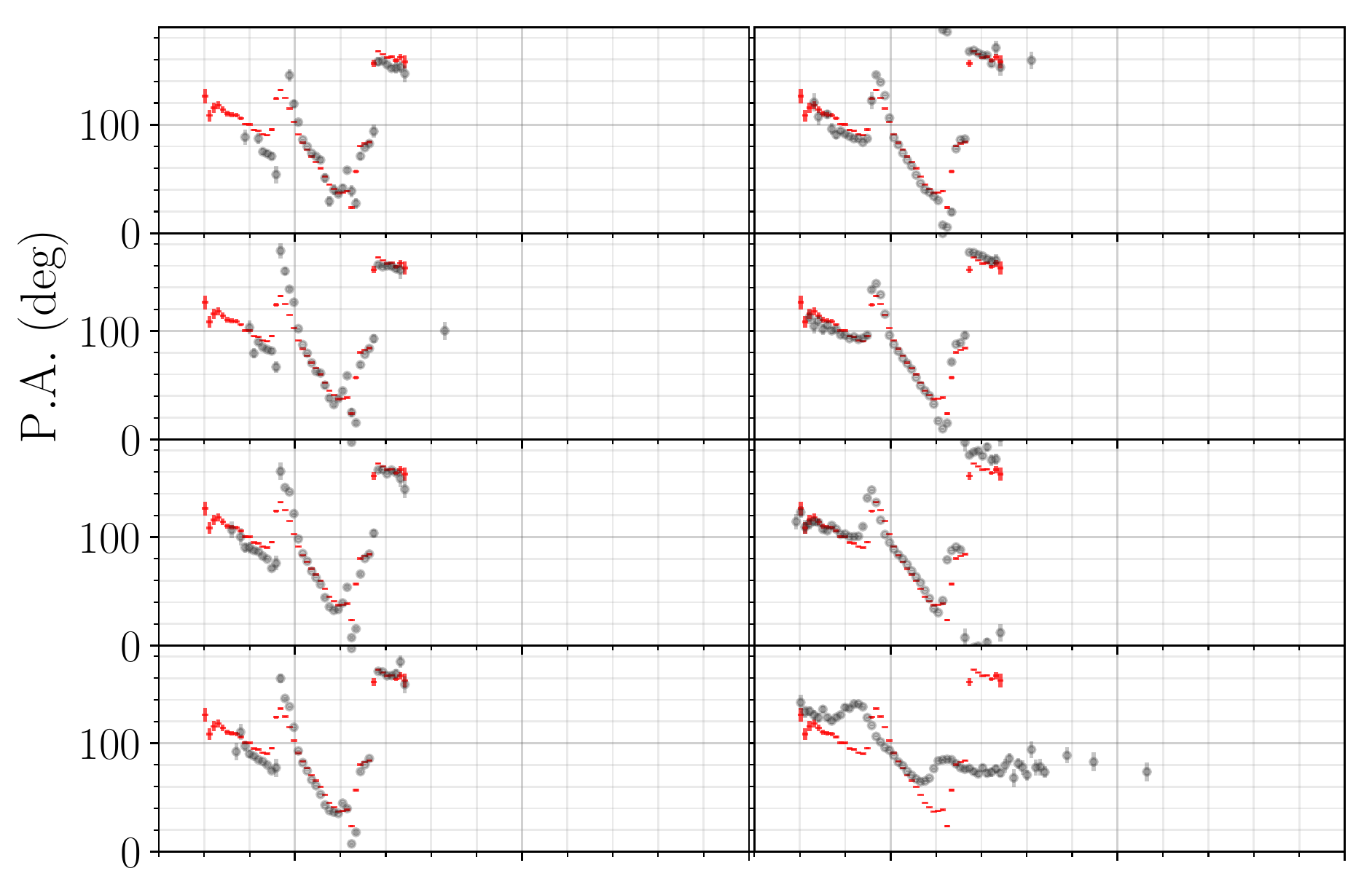}

	\includegraphics[width=\columnwidth]{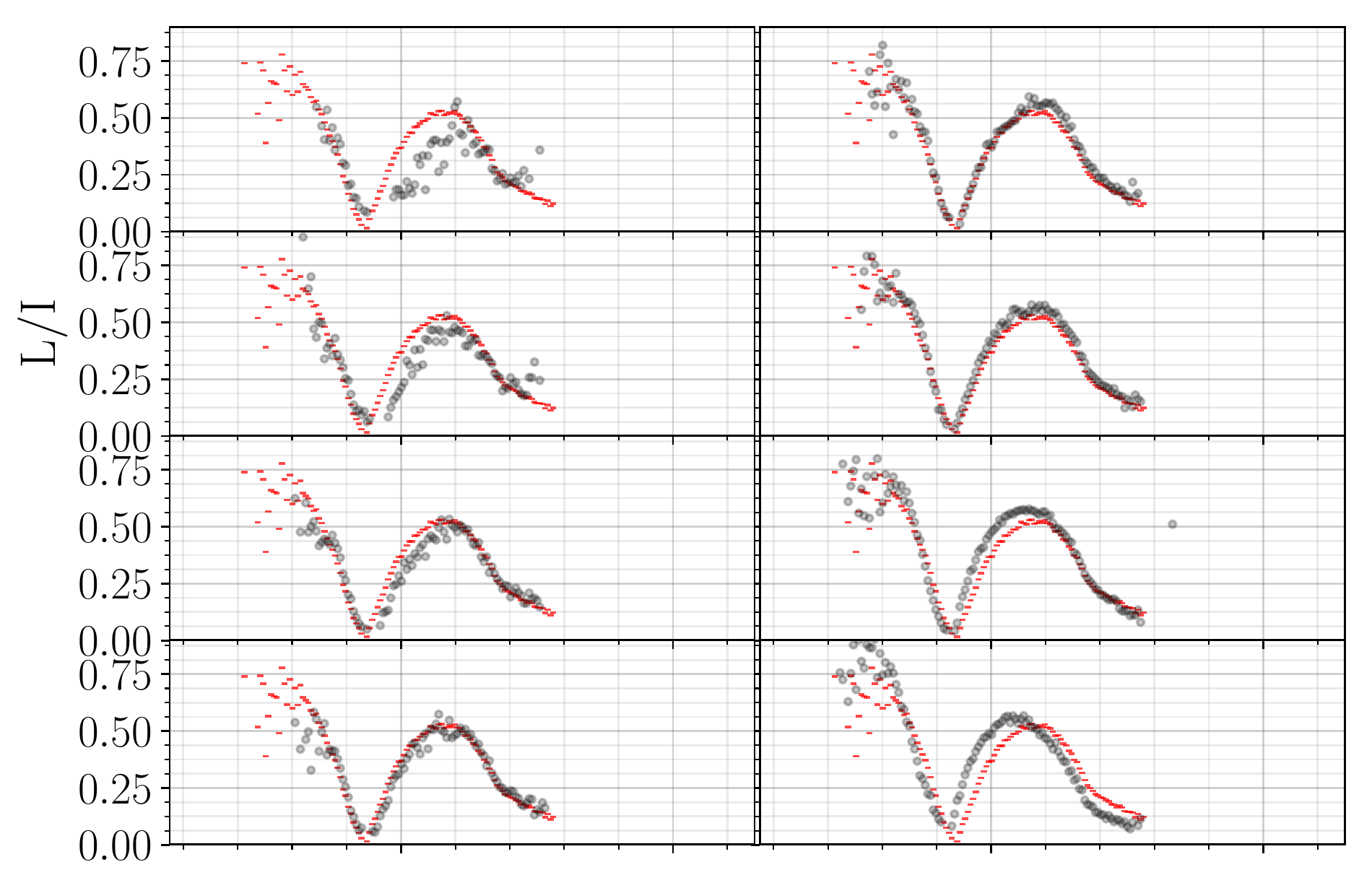}
	\includegraphics[width=\columnwidth]{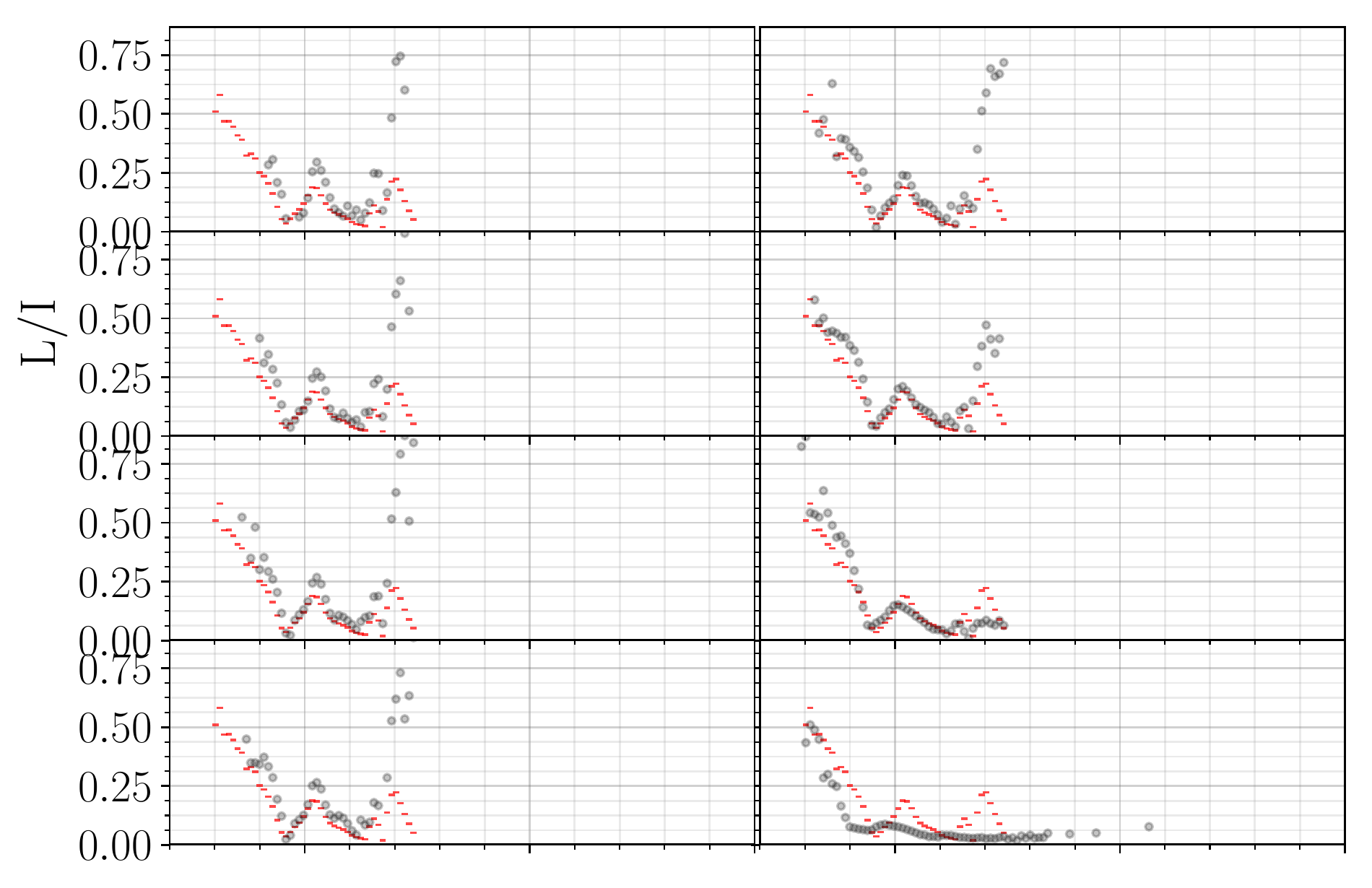}

	\includegraphics[width=\columnwidth]{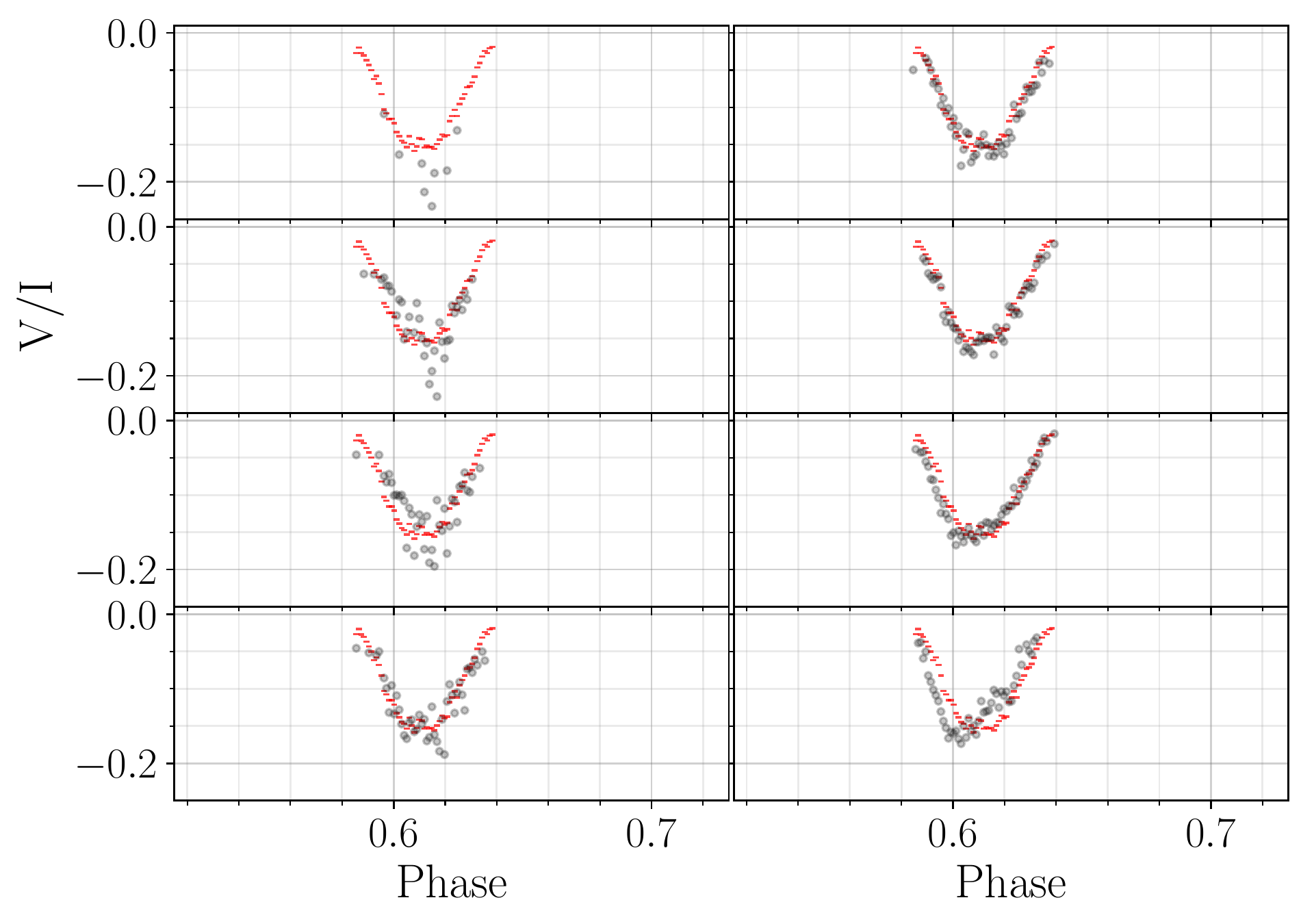}
	\includegraphics[width=\columnwidth]{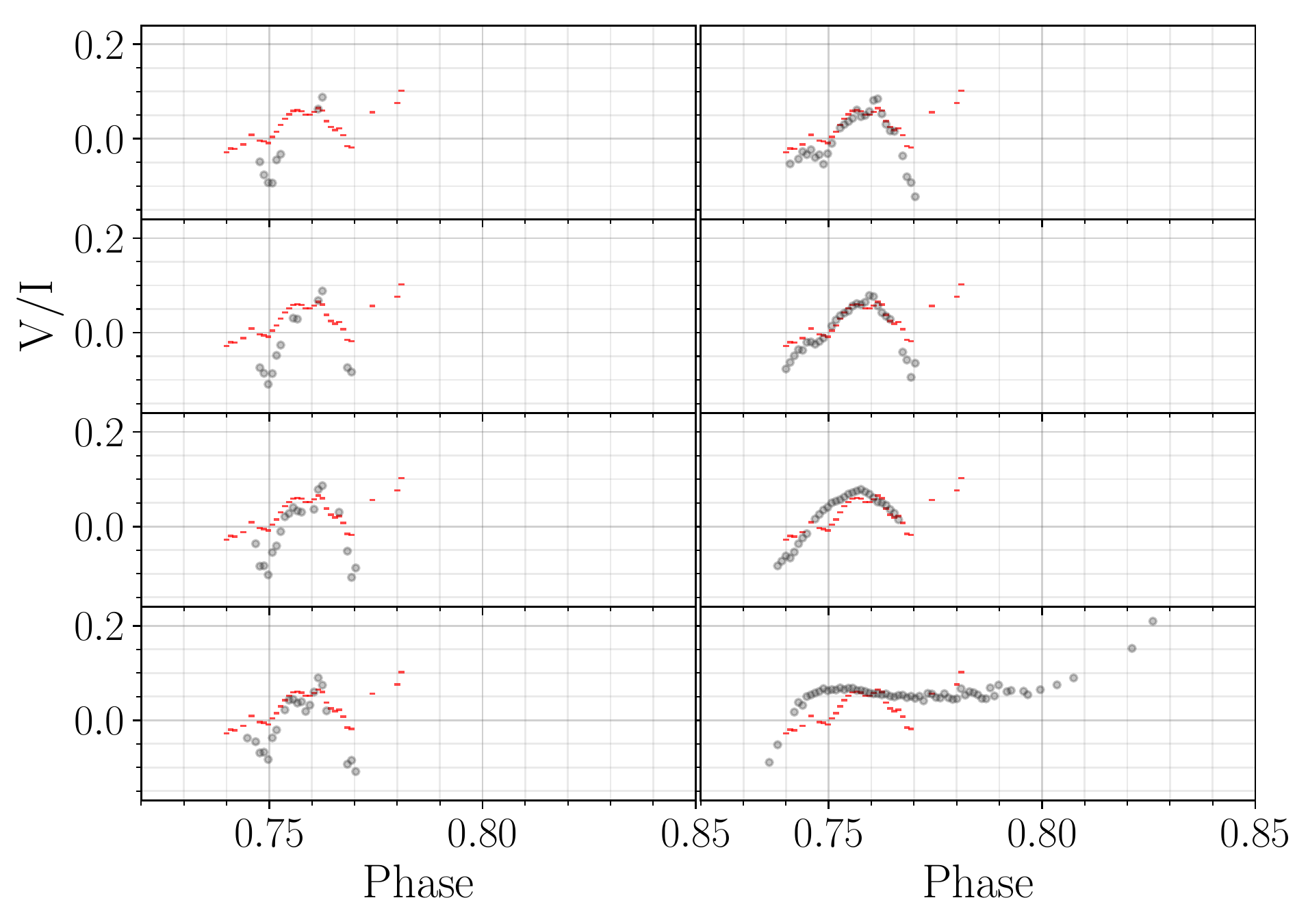}

    \caption{...Figure \ref{fig:profs1} continued..}
\end{figure*}

\begin{figure*} 
	\includegraphics[width=\columnwidth]{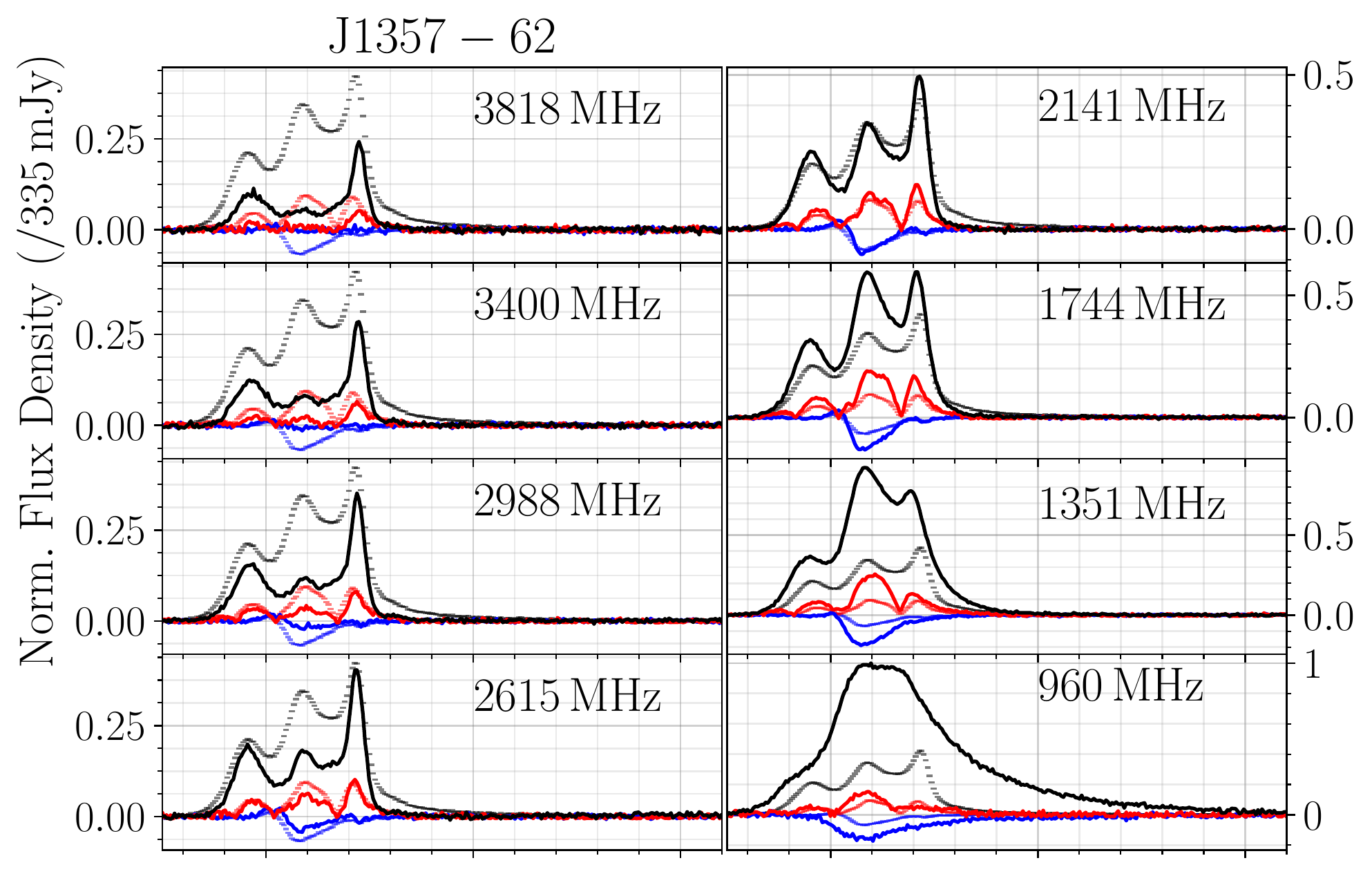}
	\includegraphics[width=\columnwidth]{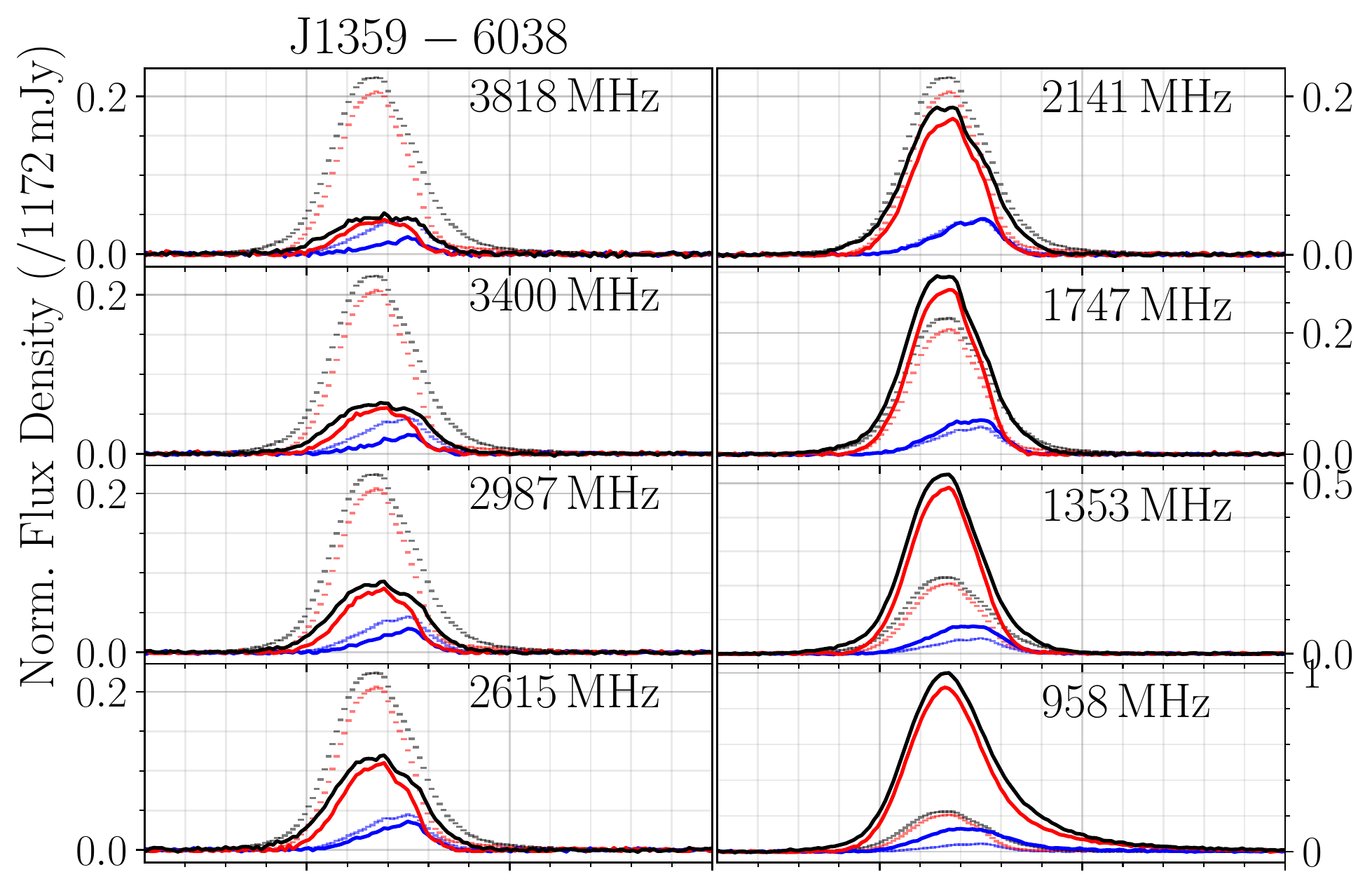}

	\includegraphics[width=\columnwidth]{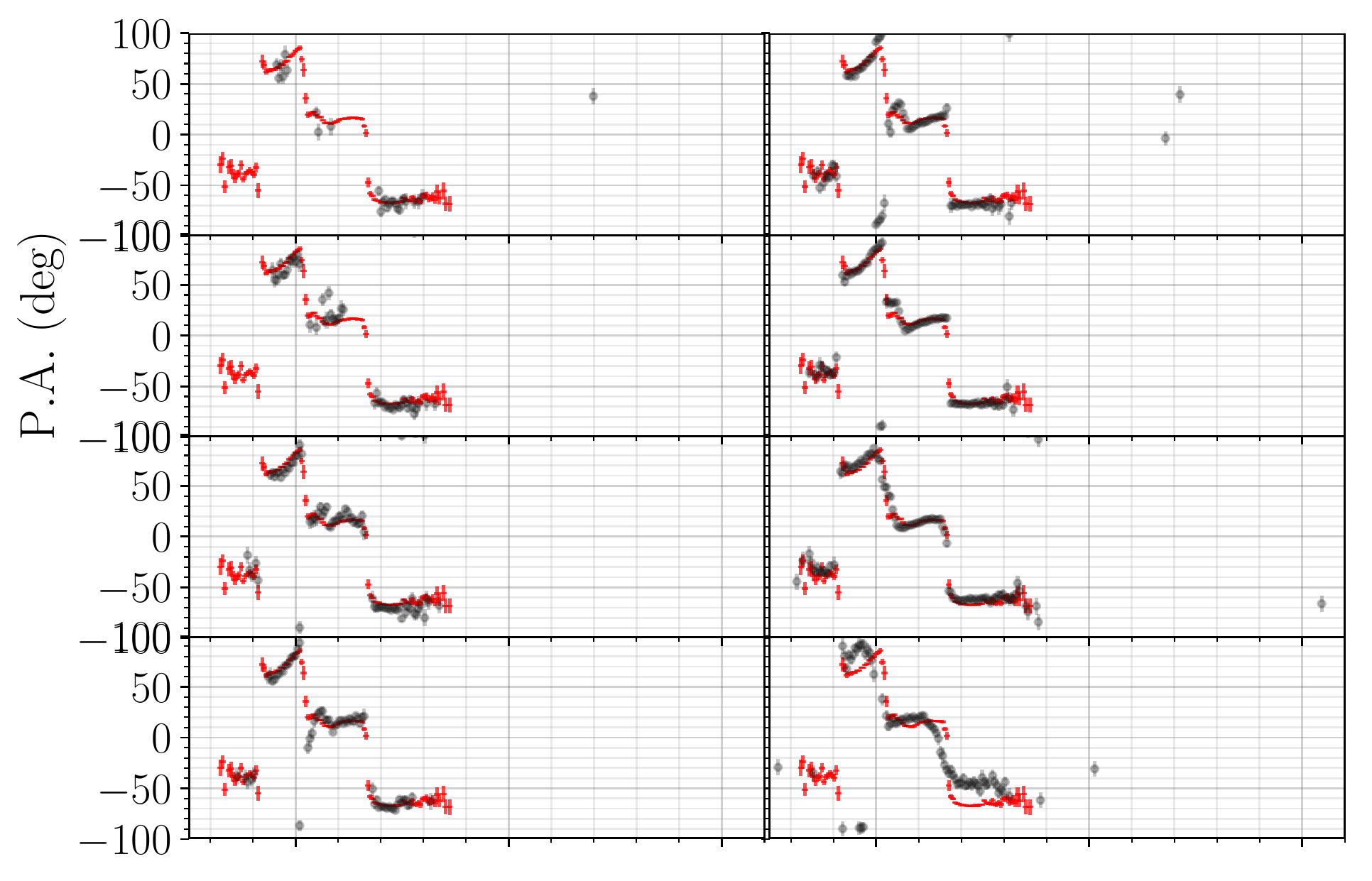}
	\includegraphics[width=\columnwidth]{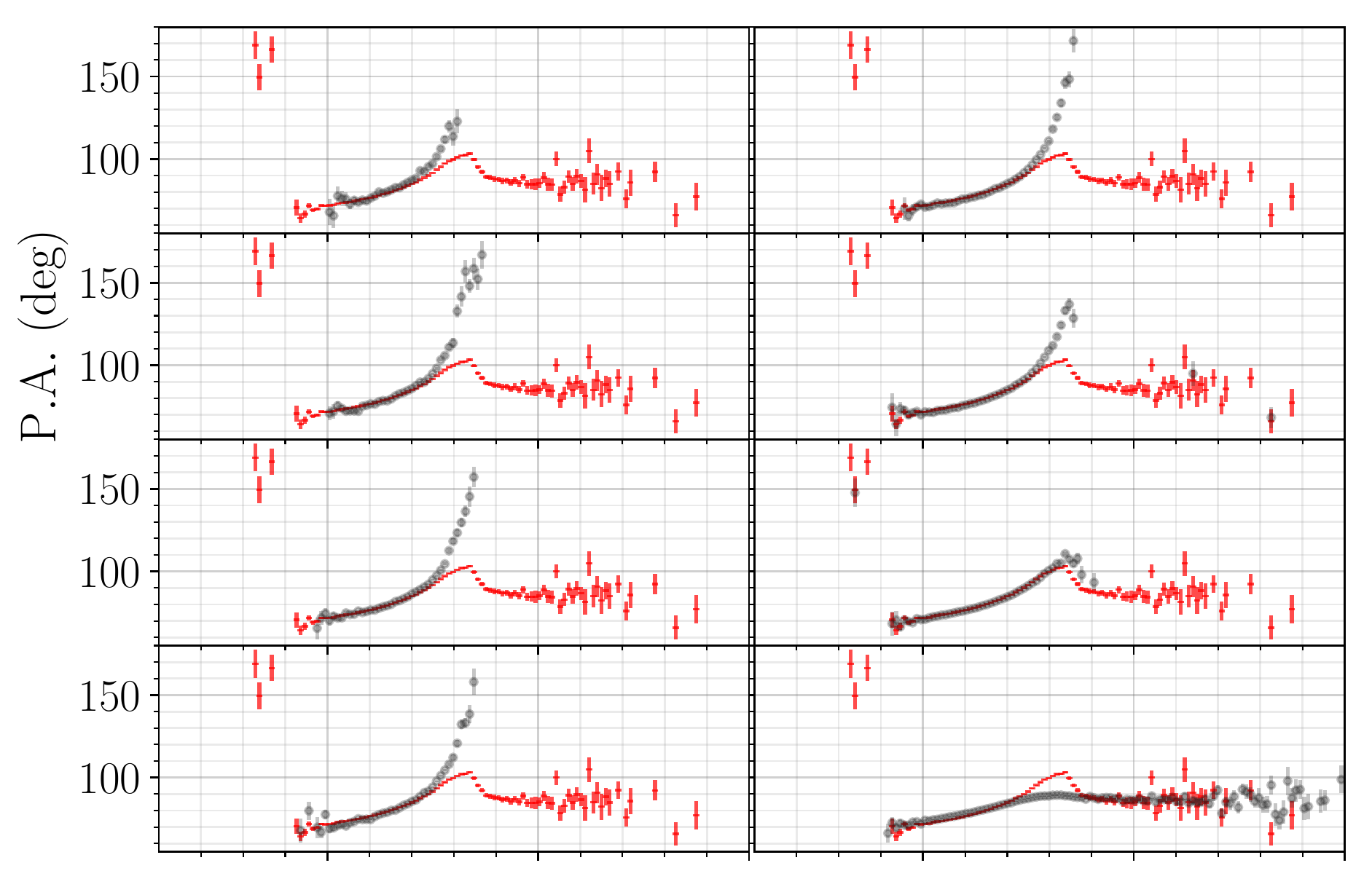}

	\includegraphics[width=\columnwidth]{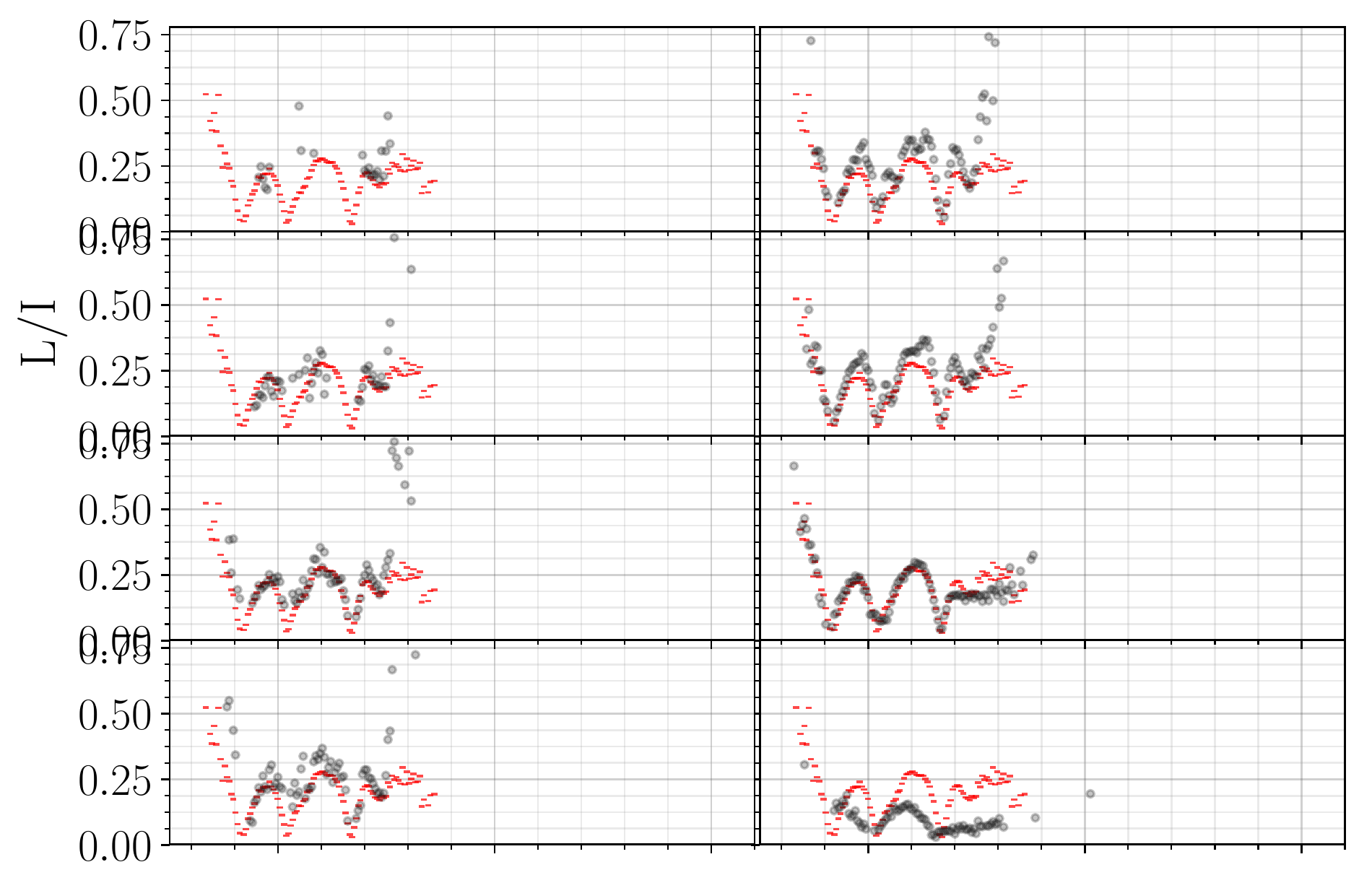}
	\includegraphics[width=\columnwidth]{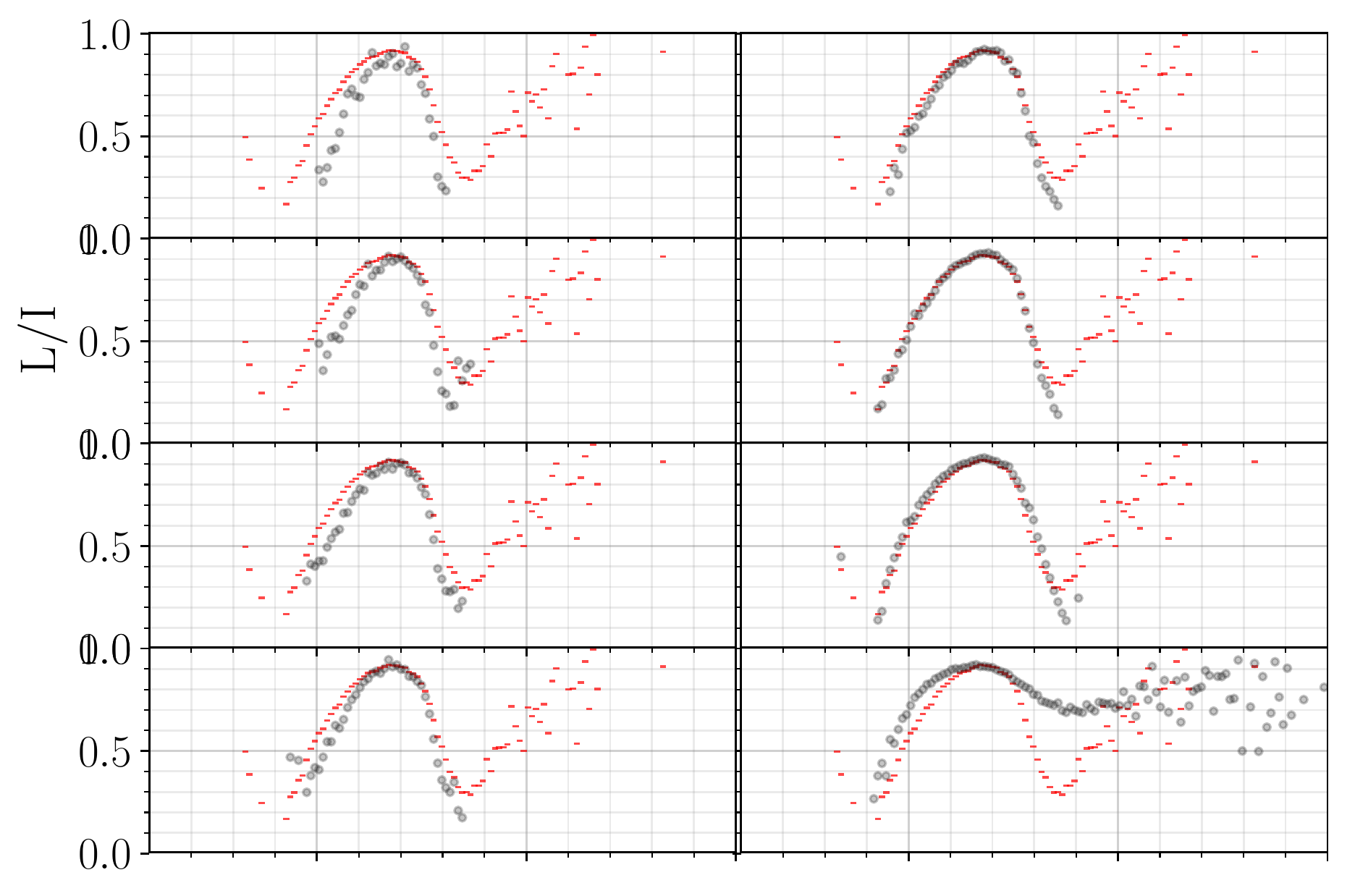}

	\includegraphics[width=\columnwidth]{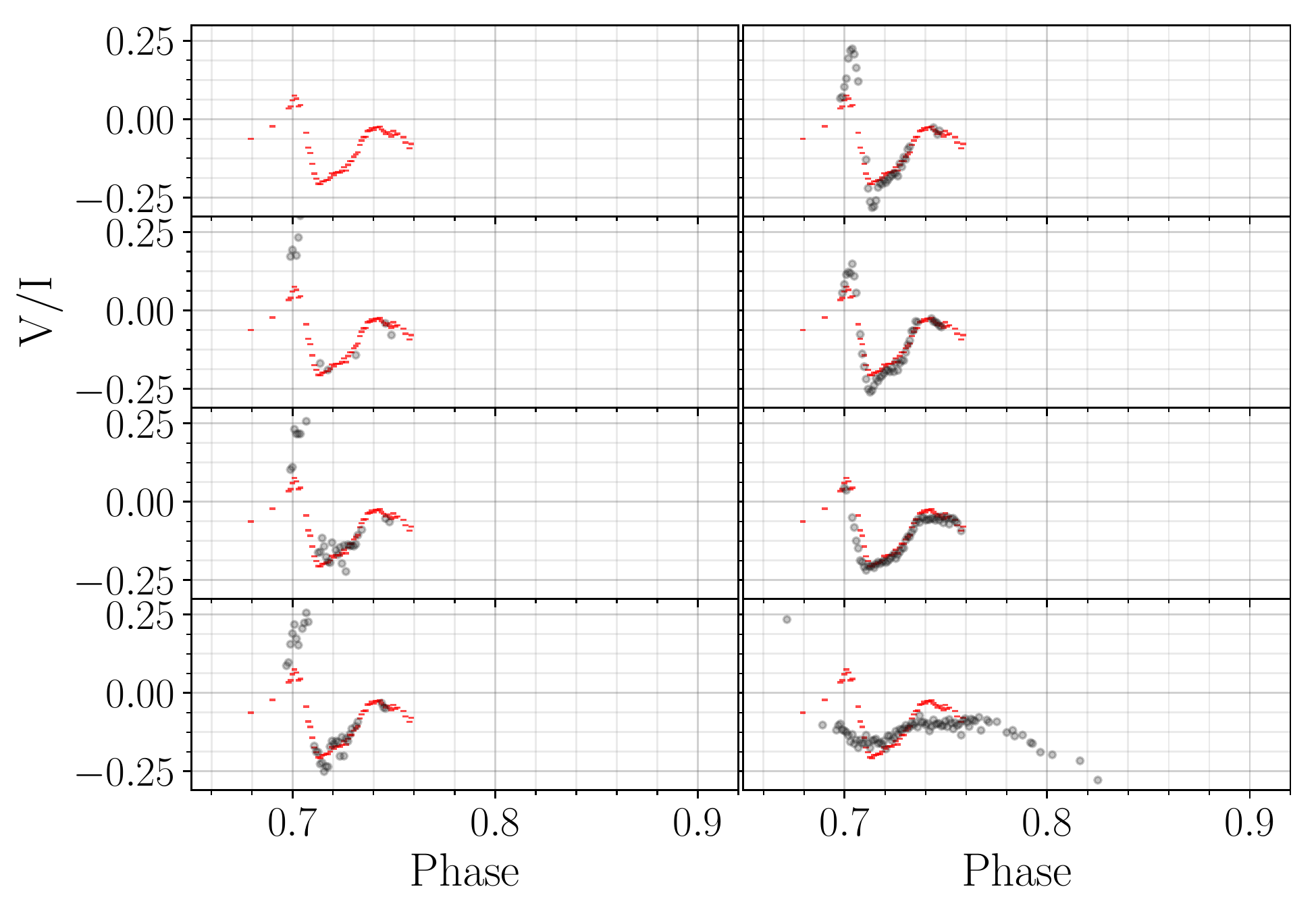}
	\includegraphics[width=\columnwidth]{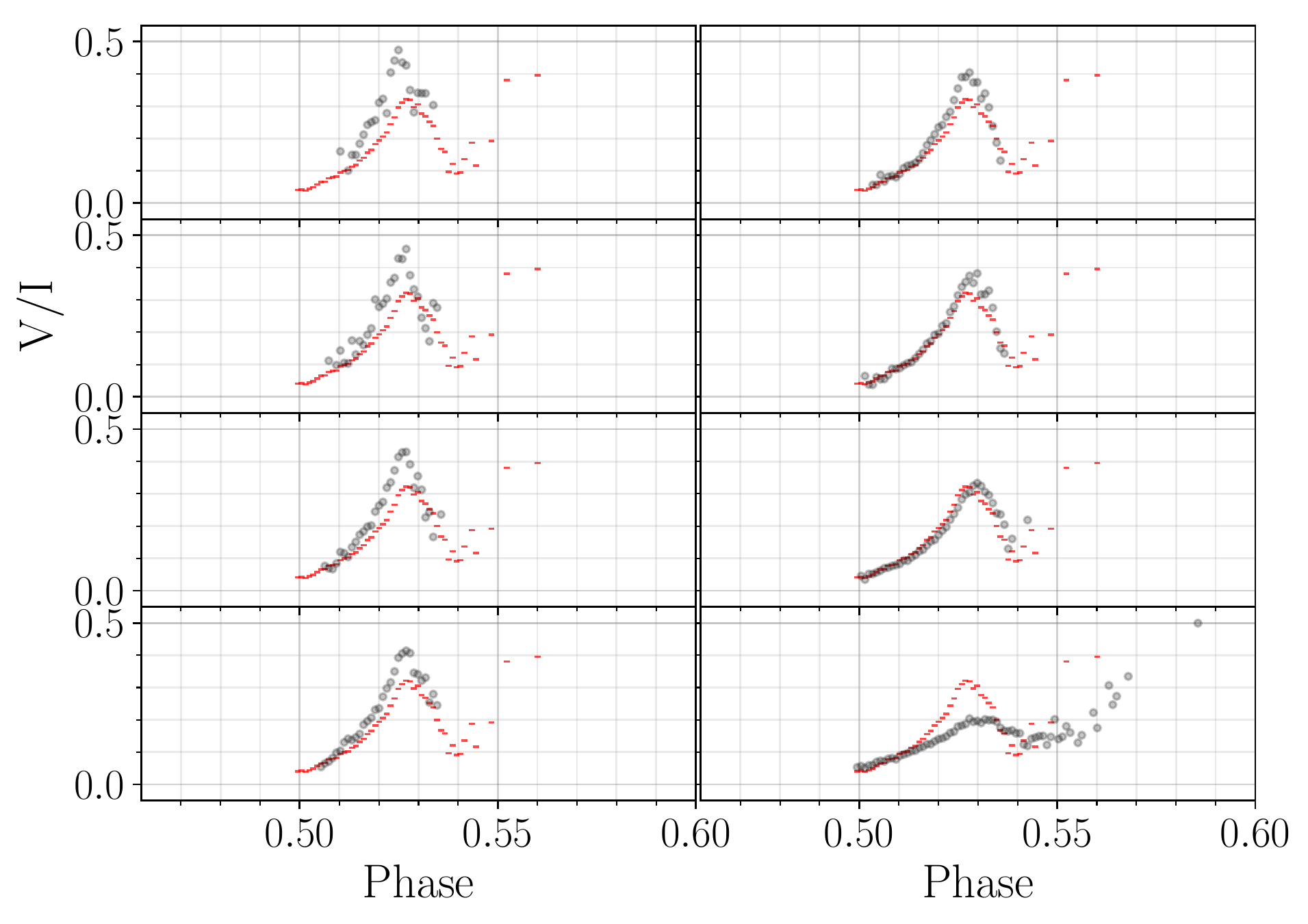}

    \caption{...Figure \ref{fig:profs1} continued...}
\end{figure*}

\begin{figure*} 
	\includegraphics[width=\columnwidth]{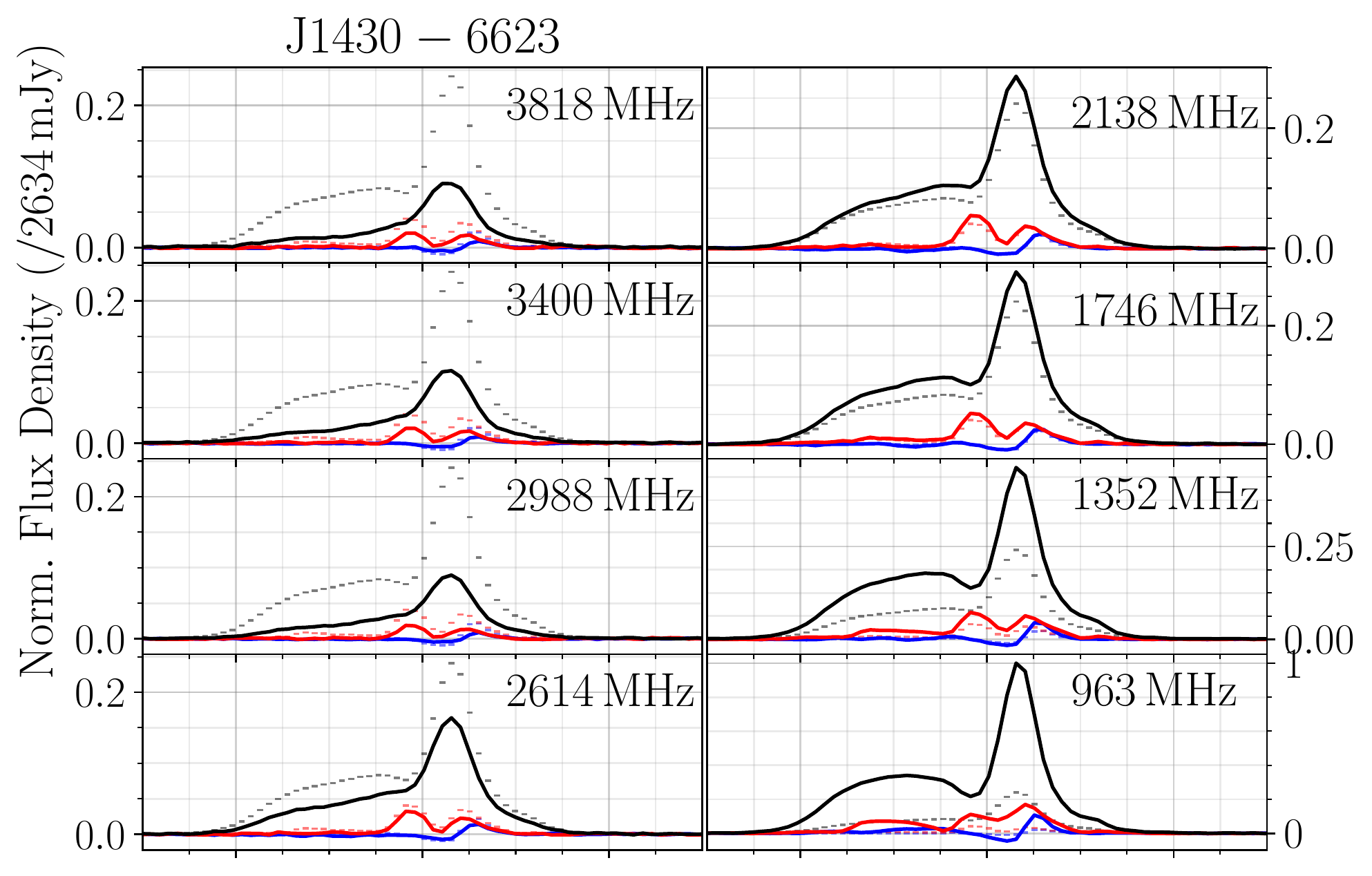}
	\includegraphics[width=\columnwidth]{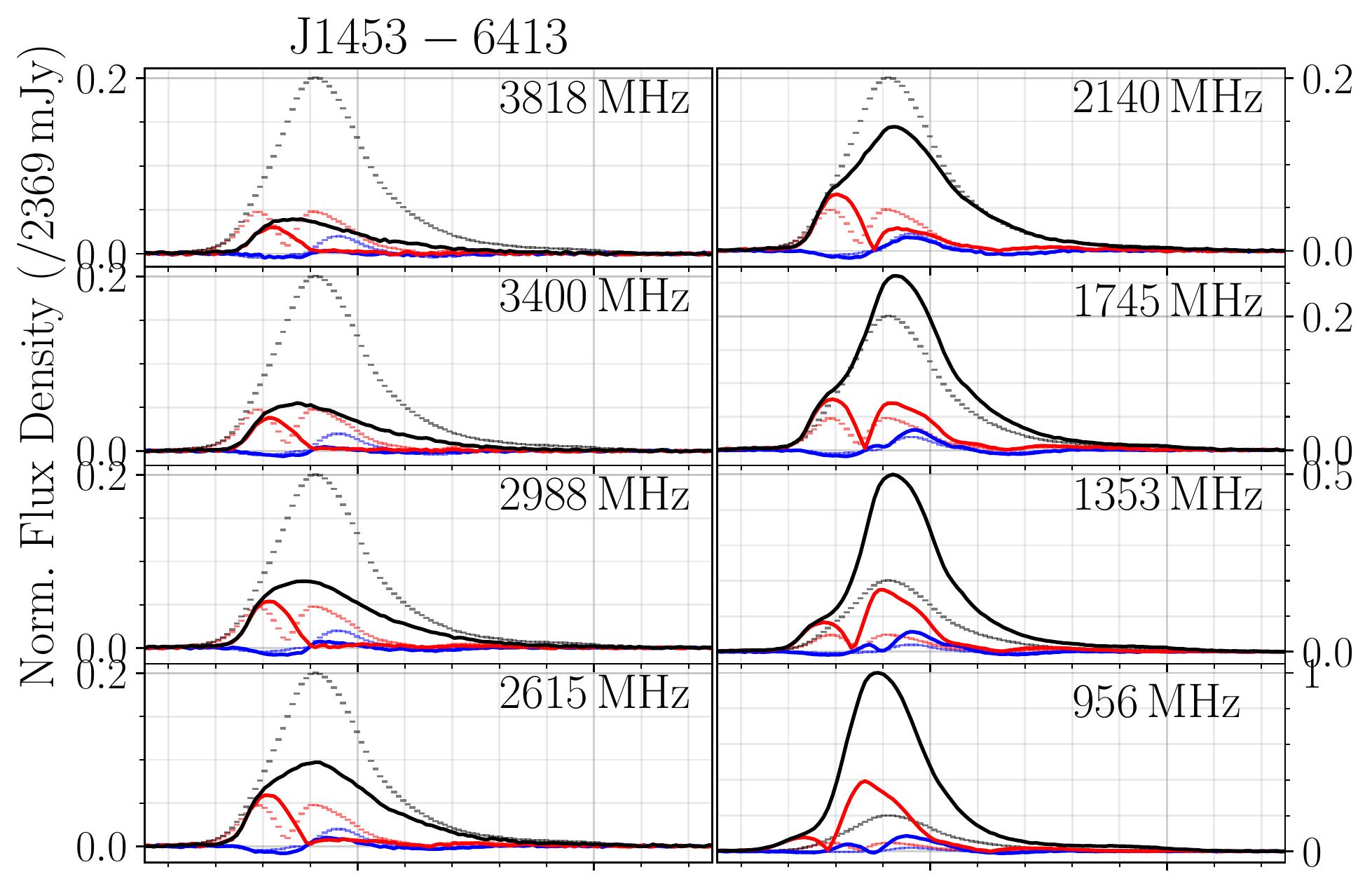}

	\includegraphics[width=\columnwidth]{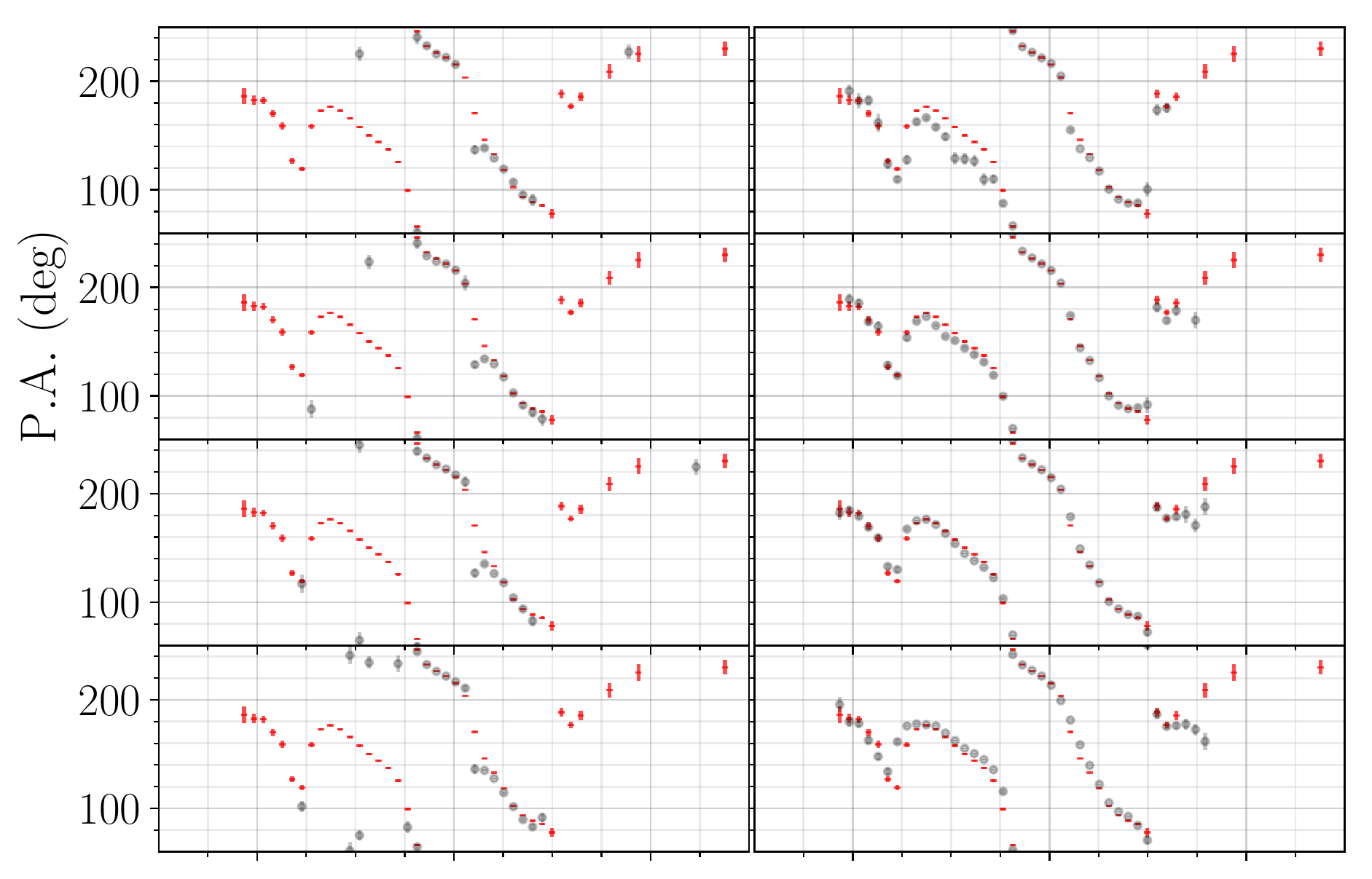}
	\includegraphics[width=\columnwidth]{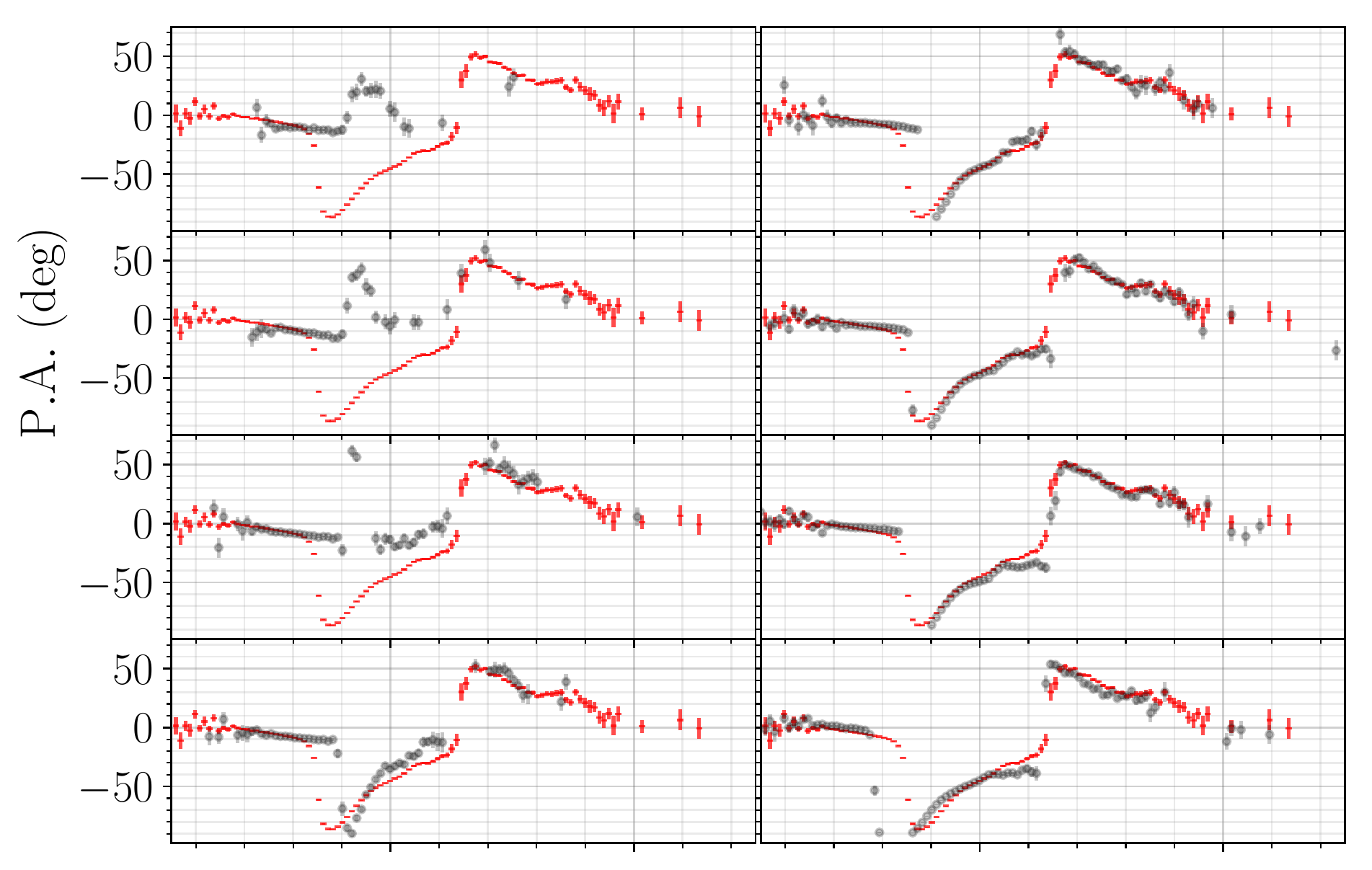}

	\includegraphics[width=\columnwidth]{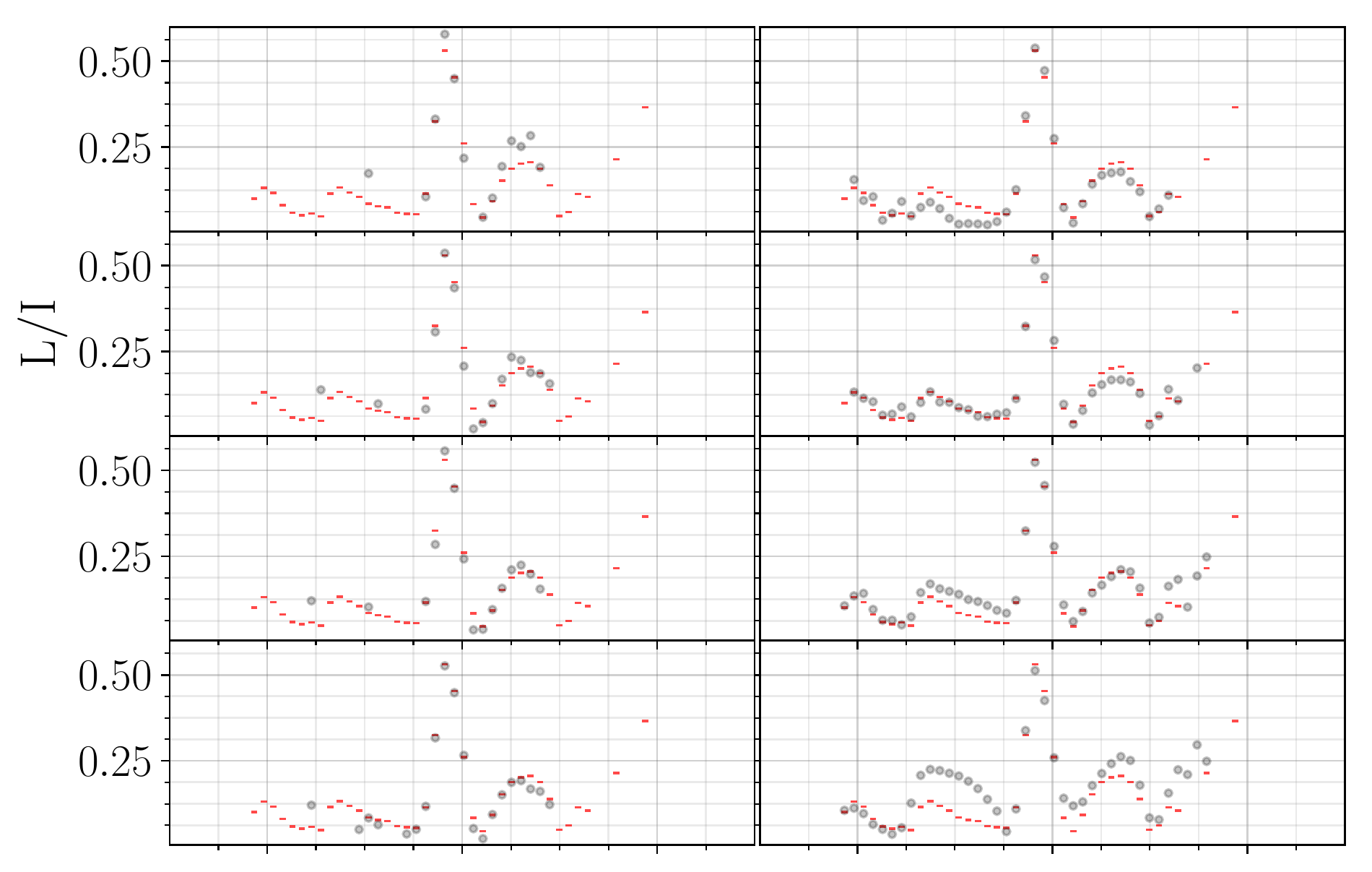}
	\includegraphics[width=\columnwidth]{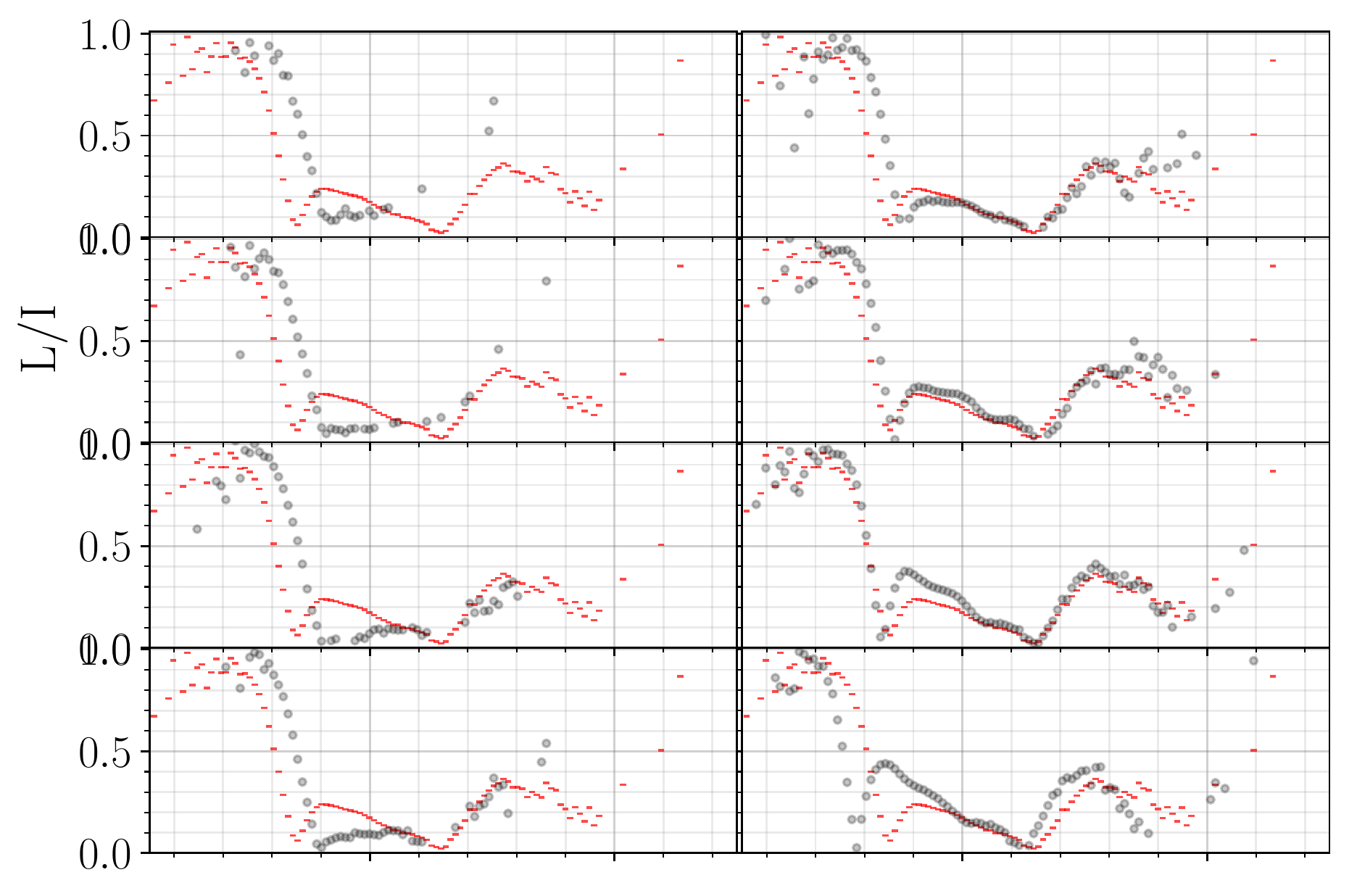}

	\includegraphics[width=\columnwidth]{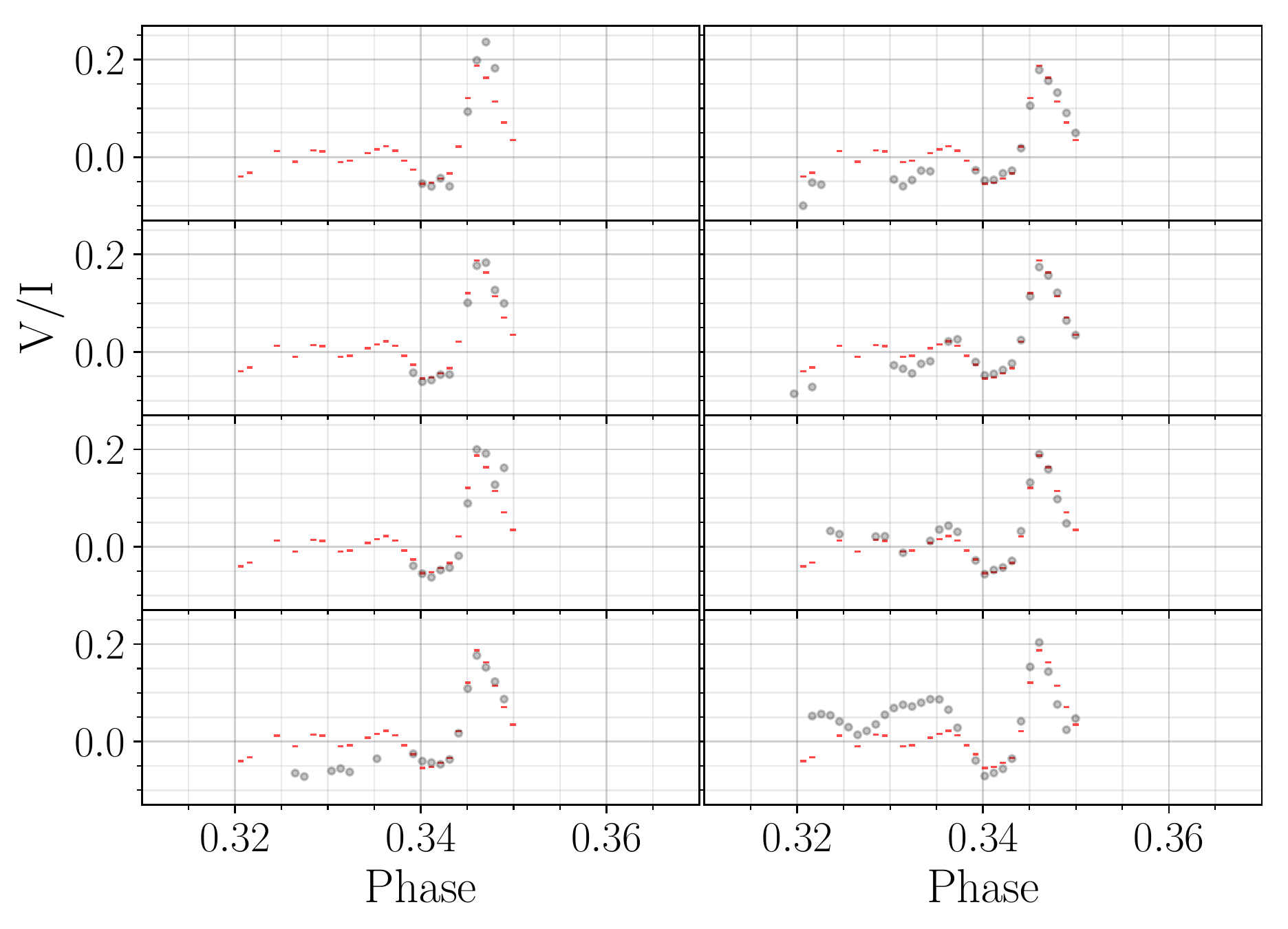}
	\includegraphics[width=\columnwidth]{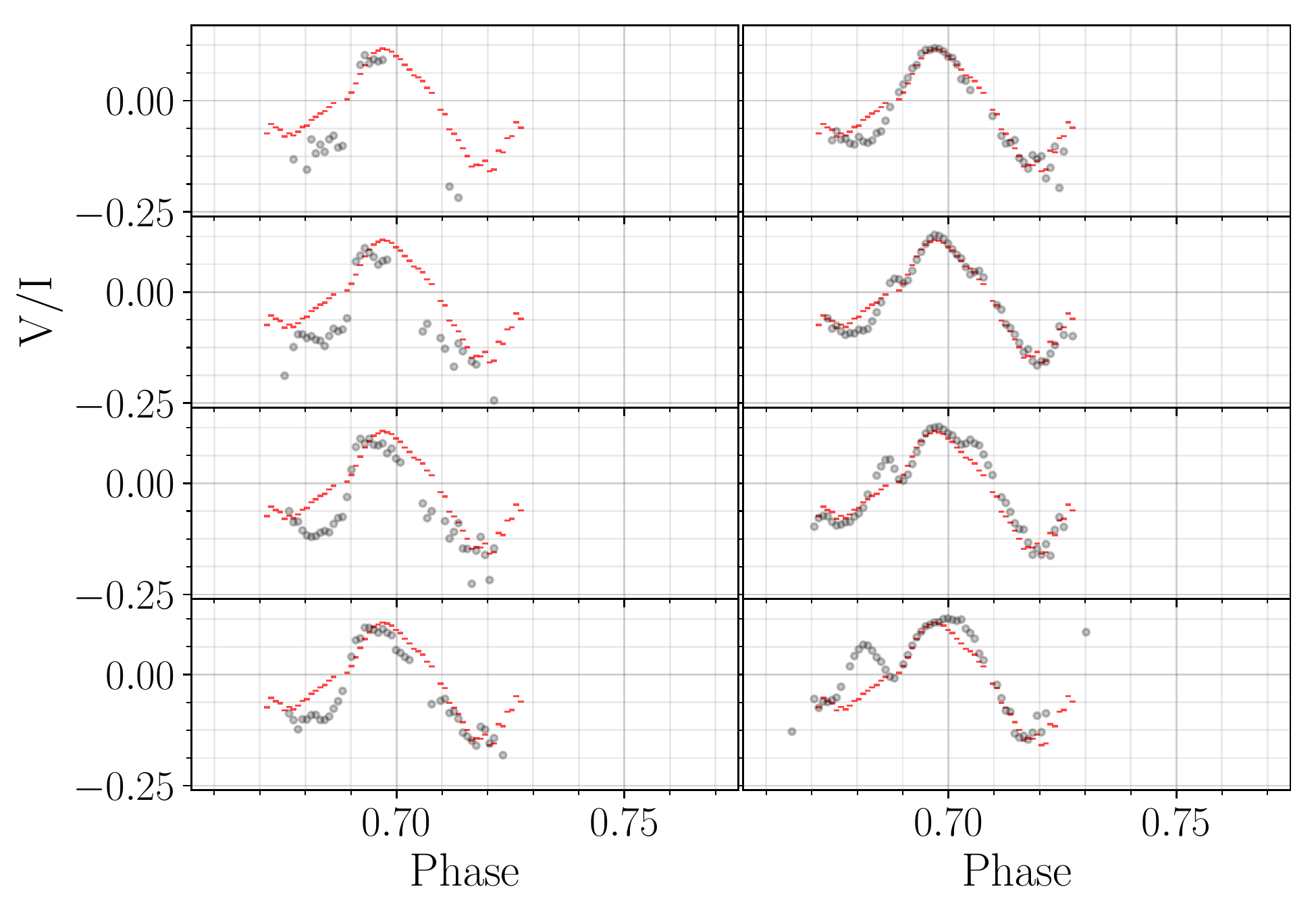}

    \caption{...Figure \ref{fig:profs1} continued...}
\end{figure*}

\begin{figure*} 
	\includegraphics[width=\columnwidth]{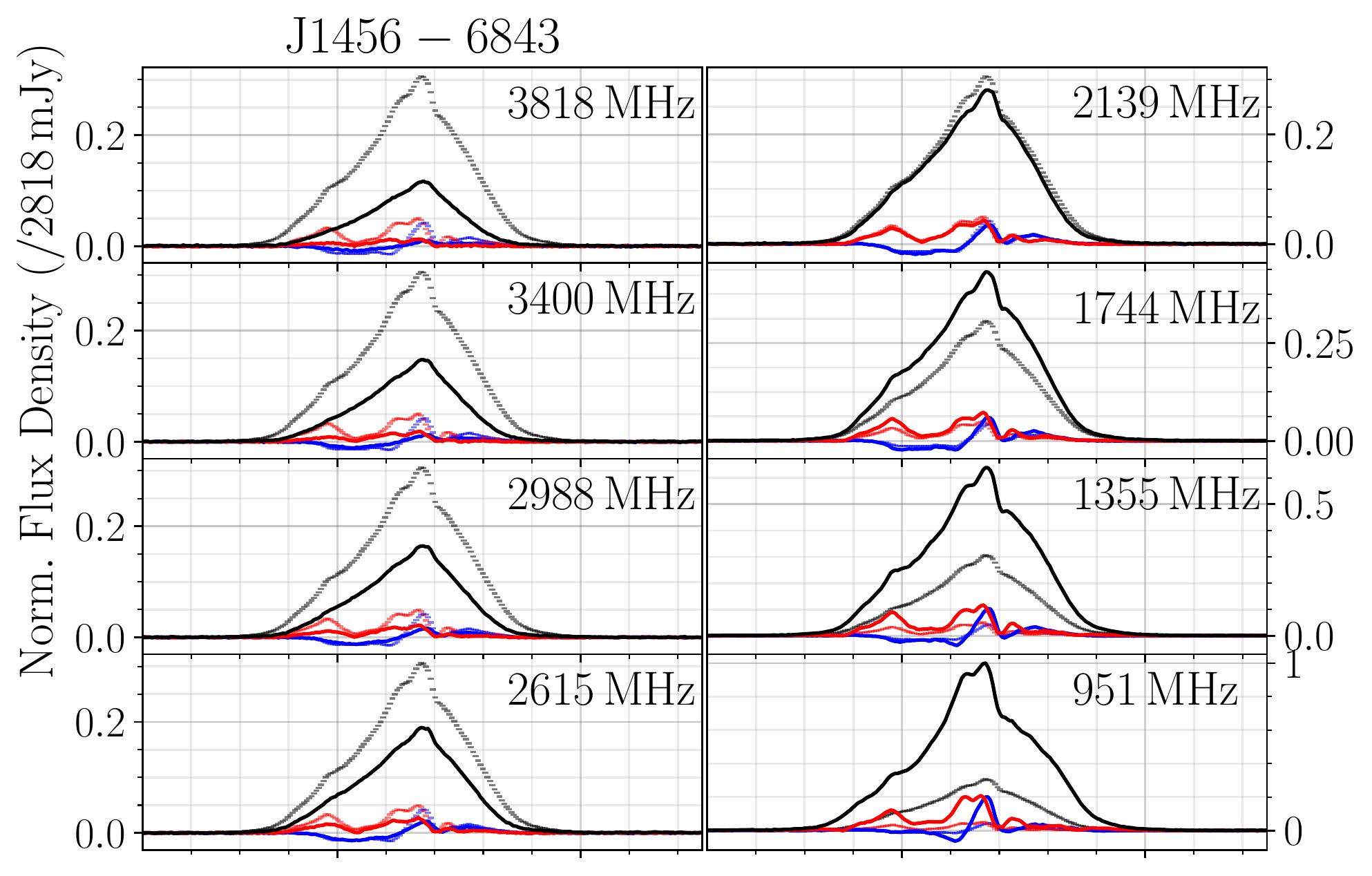}
	\includegraphics[width=\columnwidth]{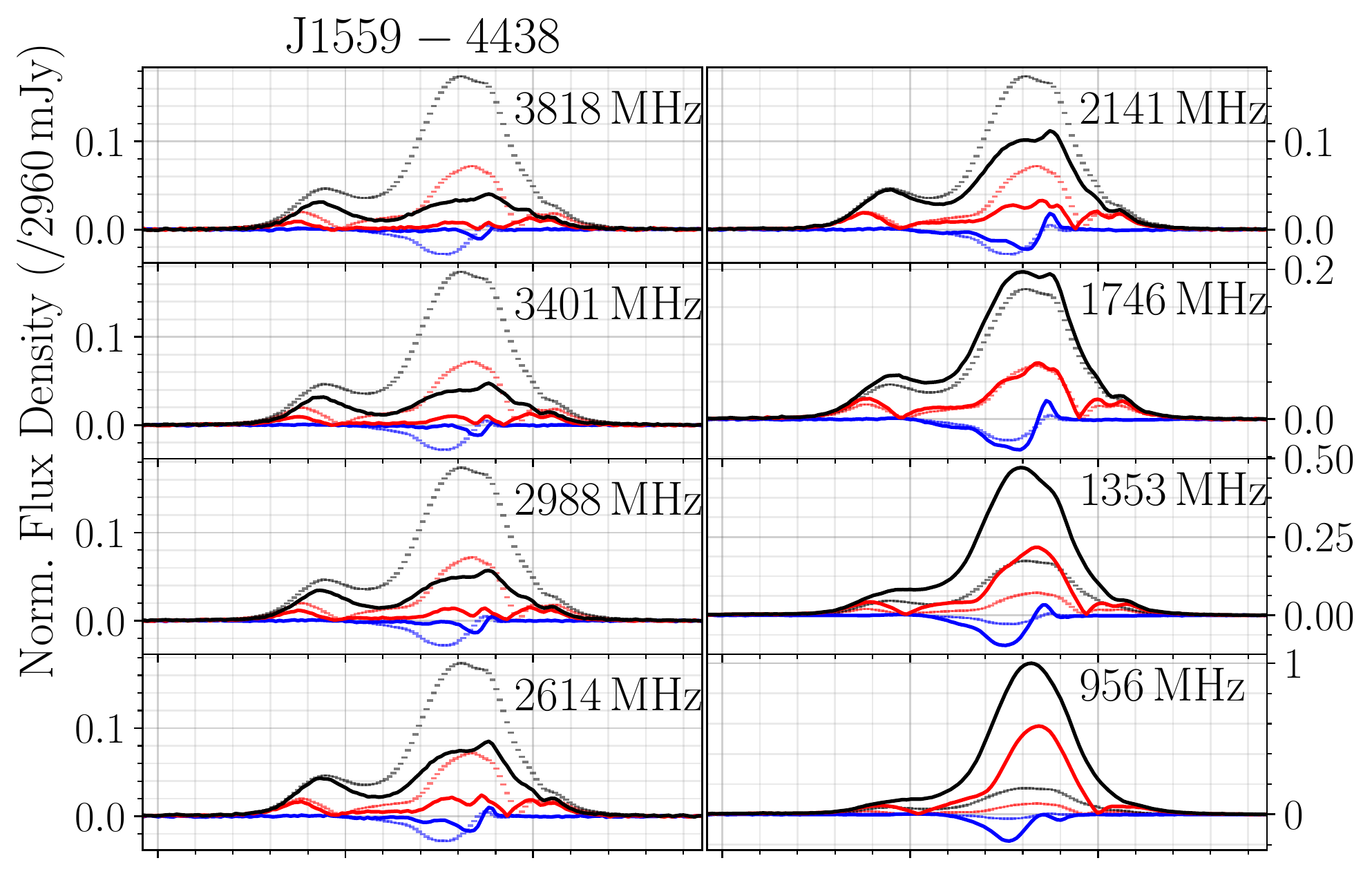}

	\includegraphics[width=\columnwidth]{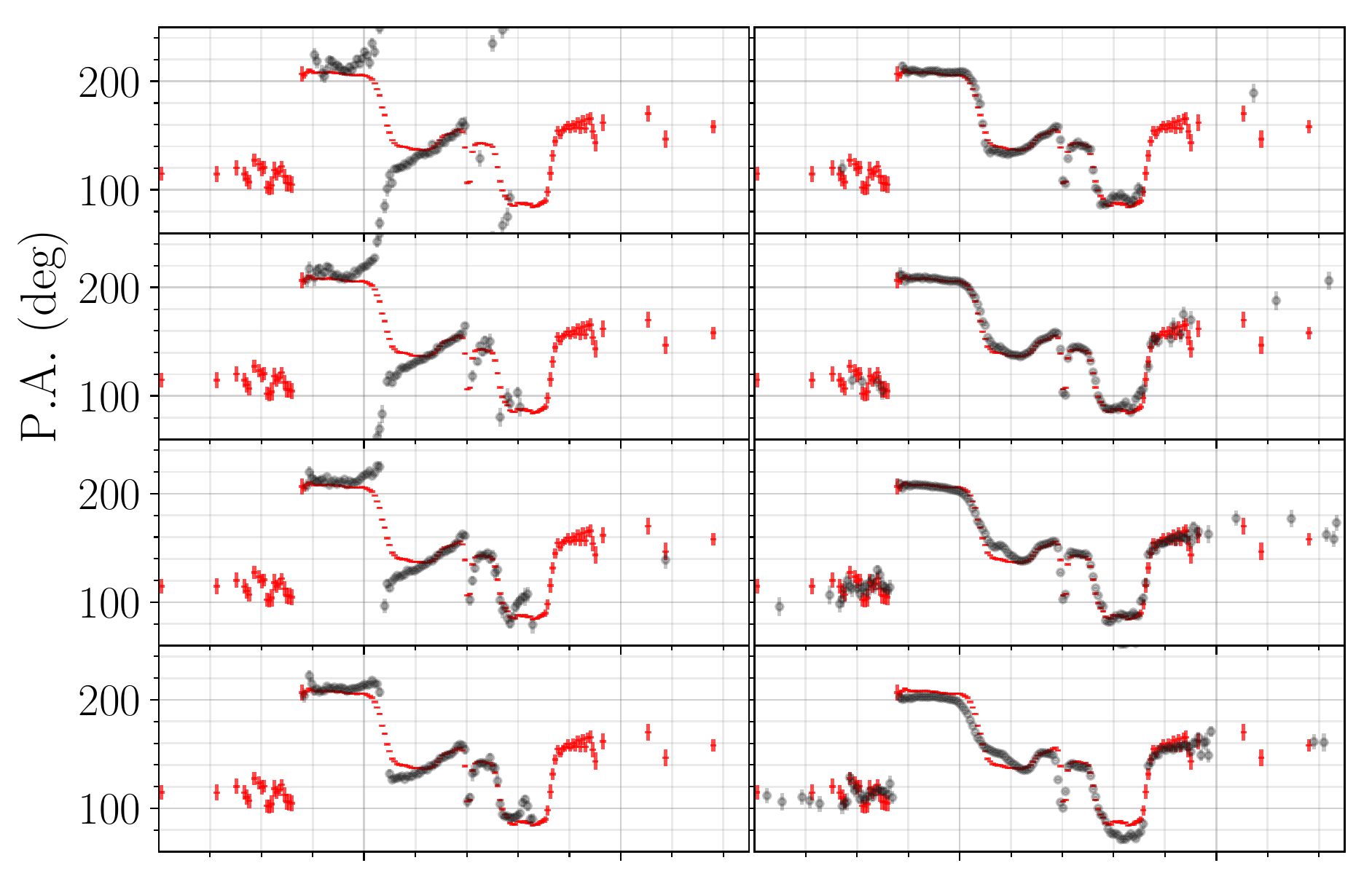}
	\includegraphics[width=\columnwidth]{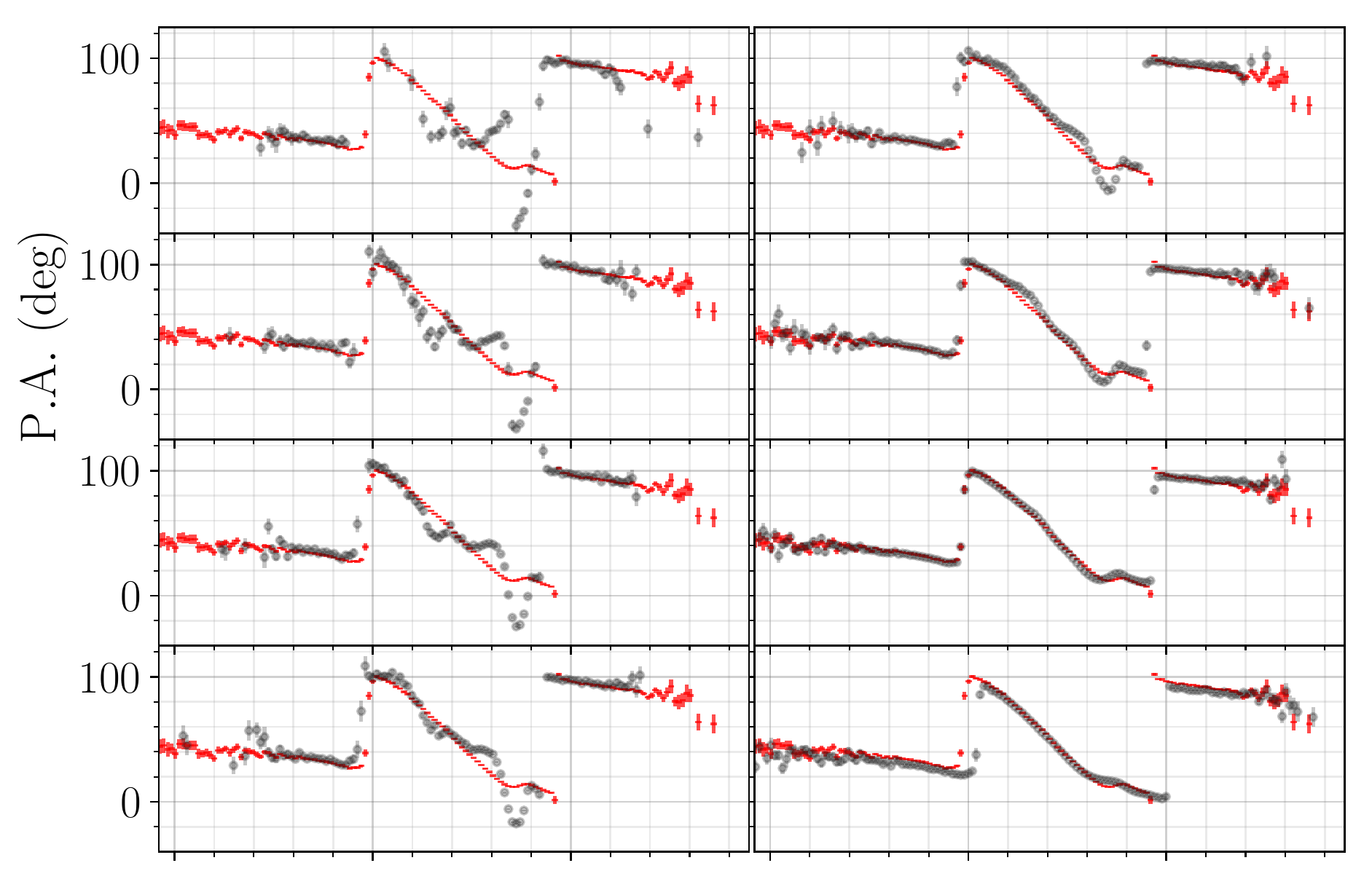}

	\includegraphics[width=\columnwidth]{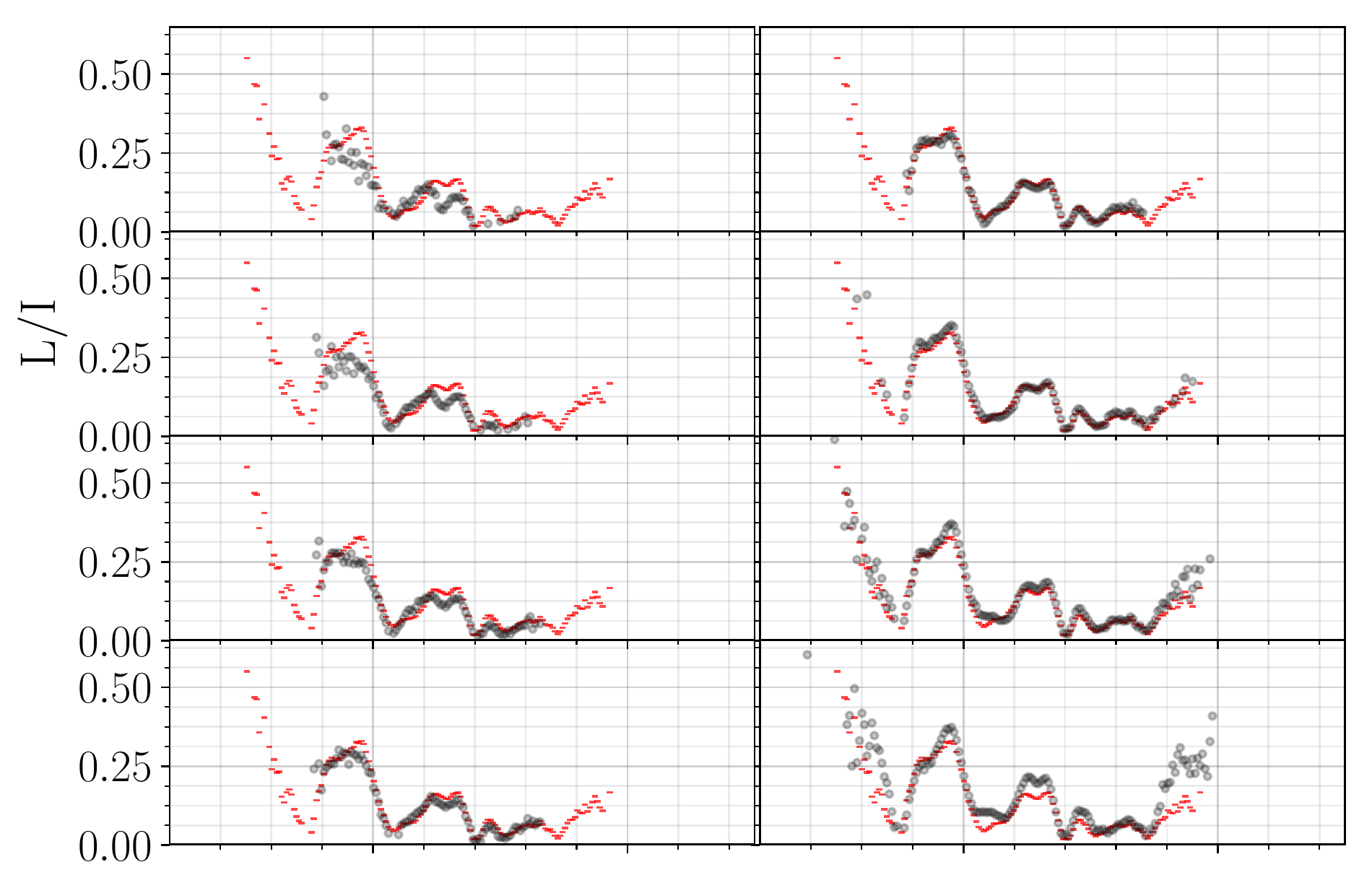}
	\includegraphics[width=\columnwidth]{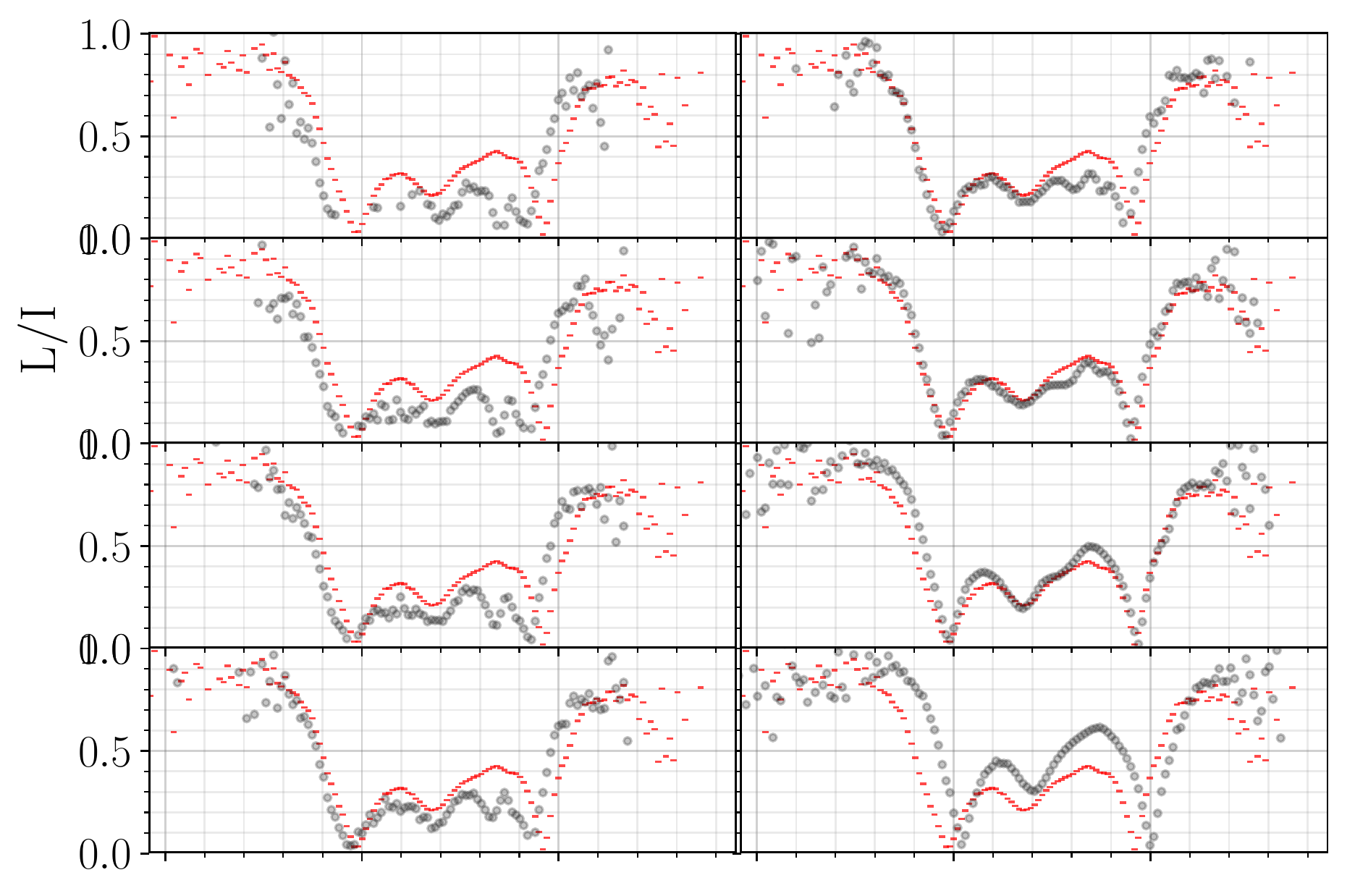}

	\includegraphics[width=\columnwidth]{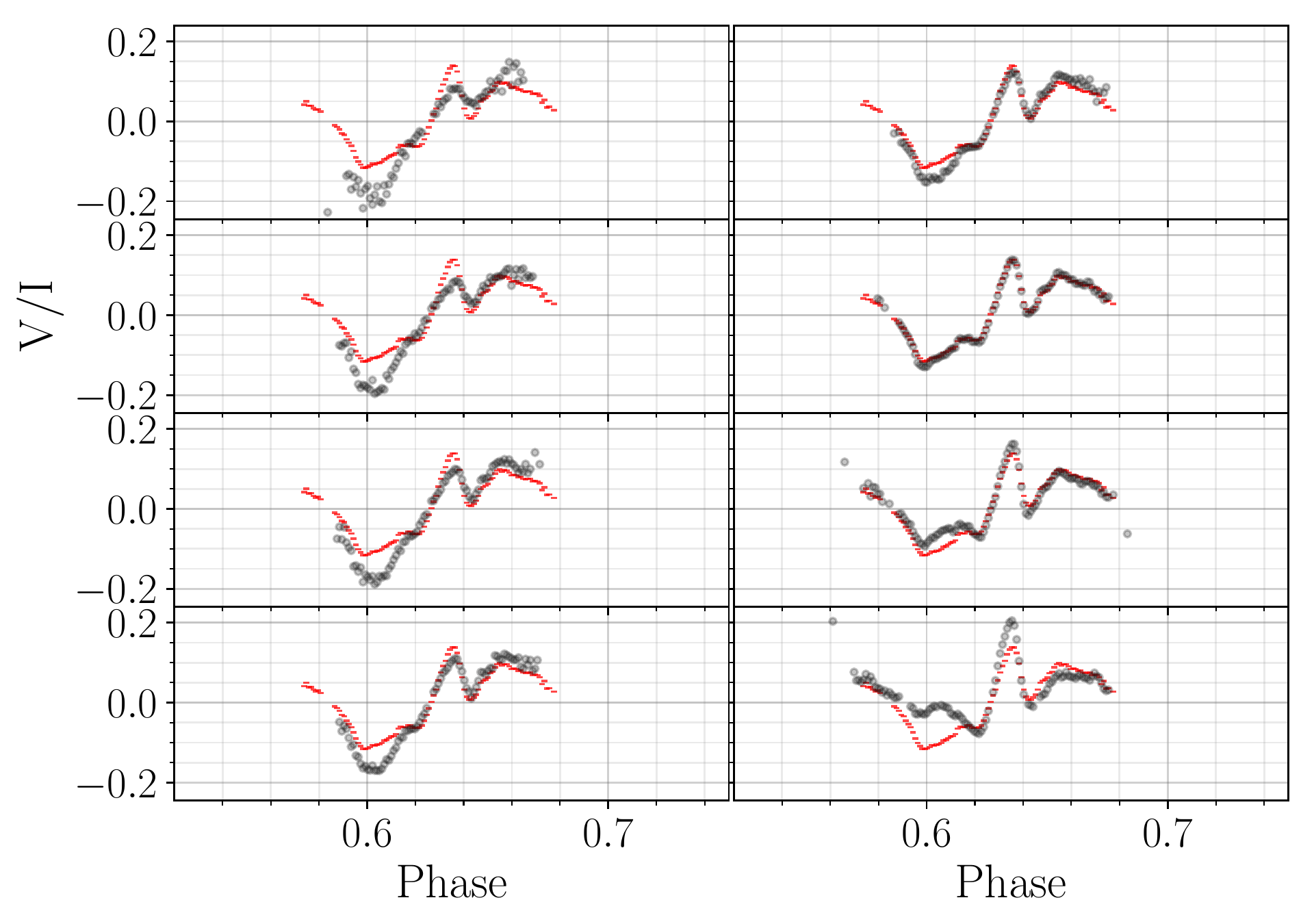}
	\includegraphics[width=\columnwidth]{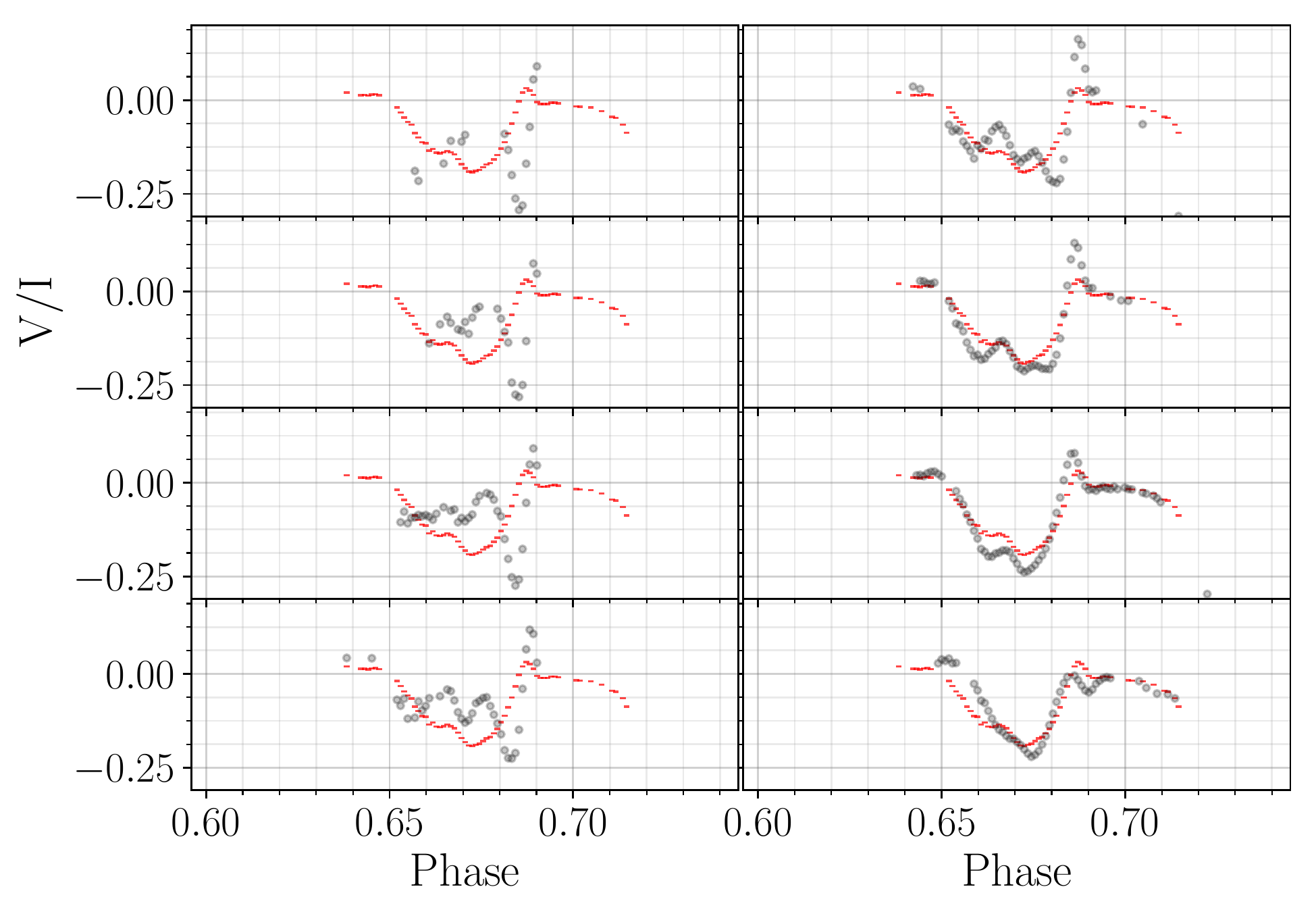}

    \caption{...Figure \ref{fig:profs1} continued...}
\end{figure*}

\begin{figure*} 
	\includegraphics[width=\columnwidth]{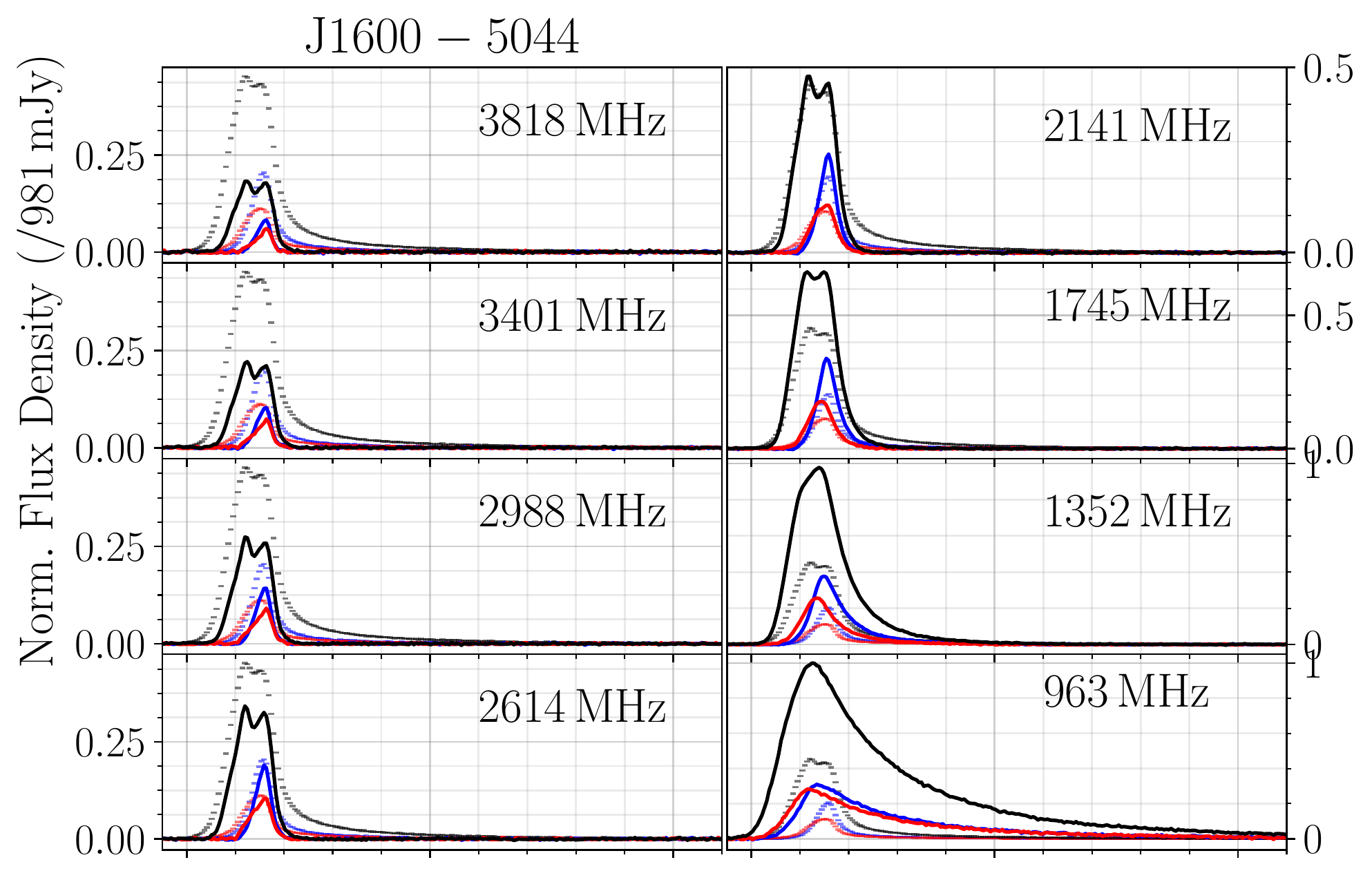}
	\includegraphics[width=\columnwidth]{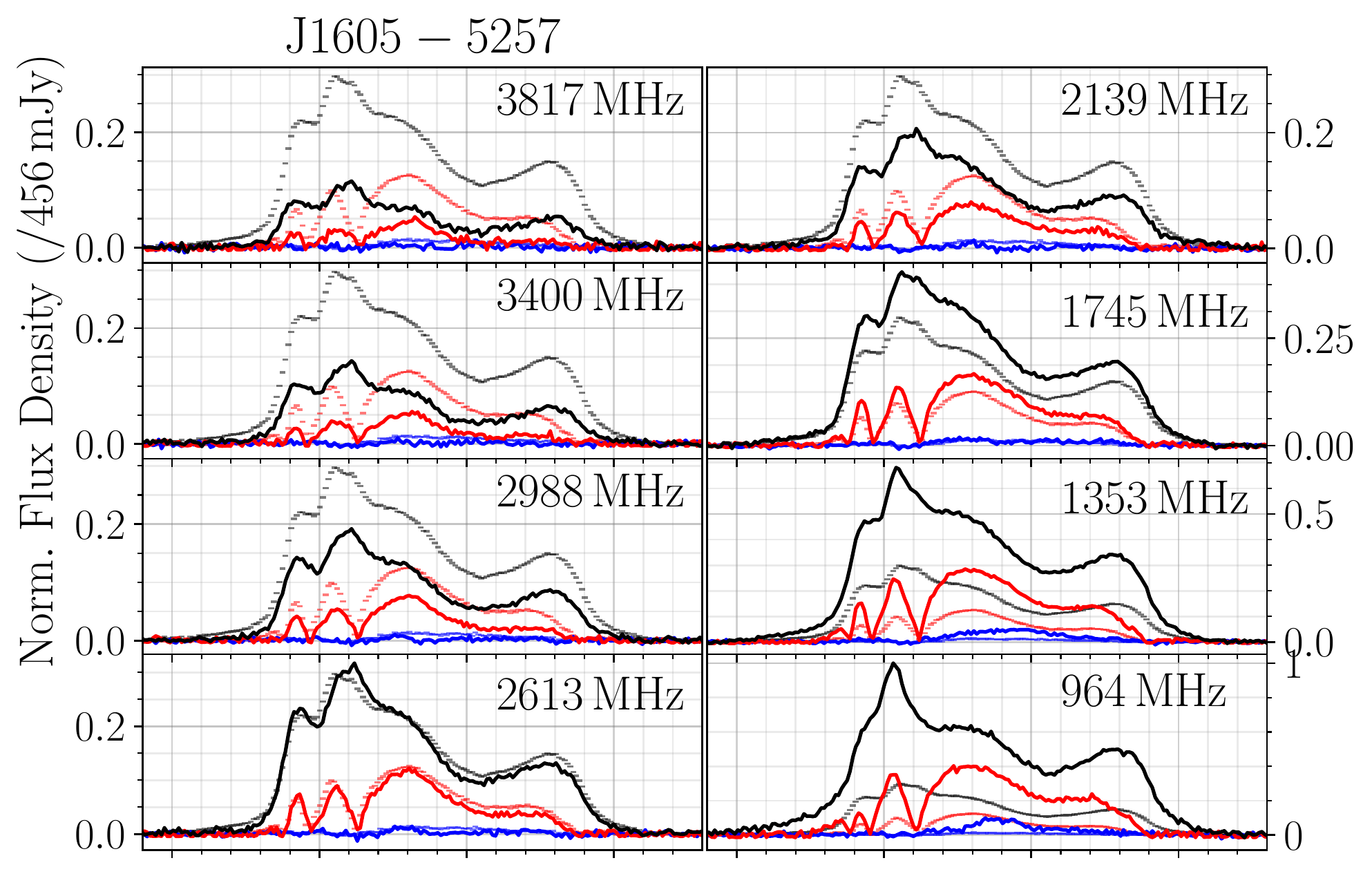}

	\includegraphics[width=\columnwidth]{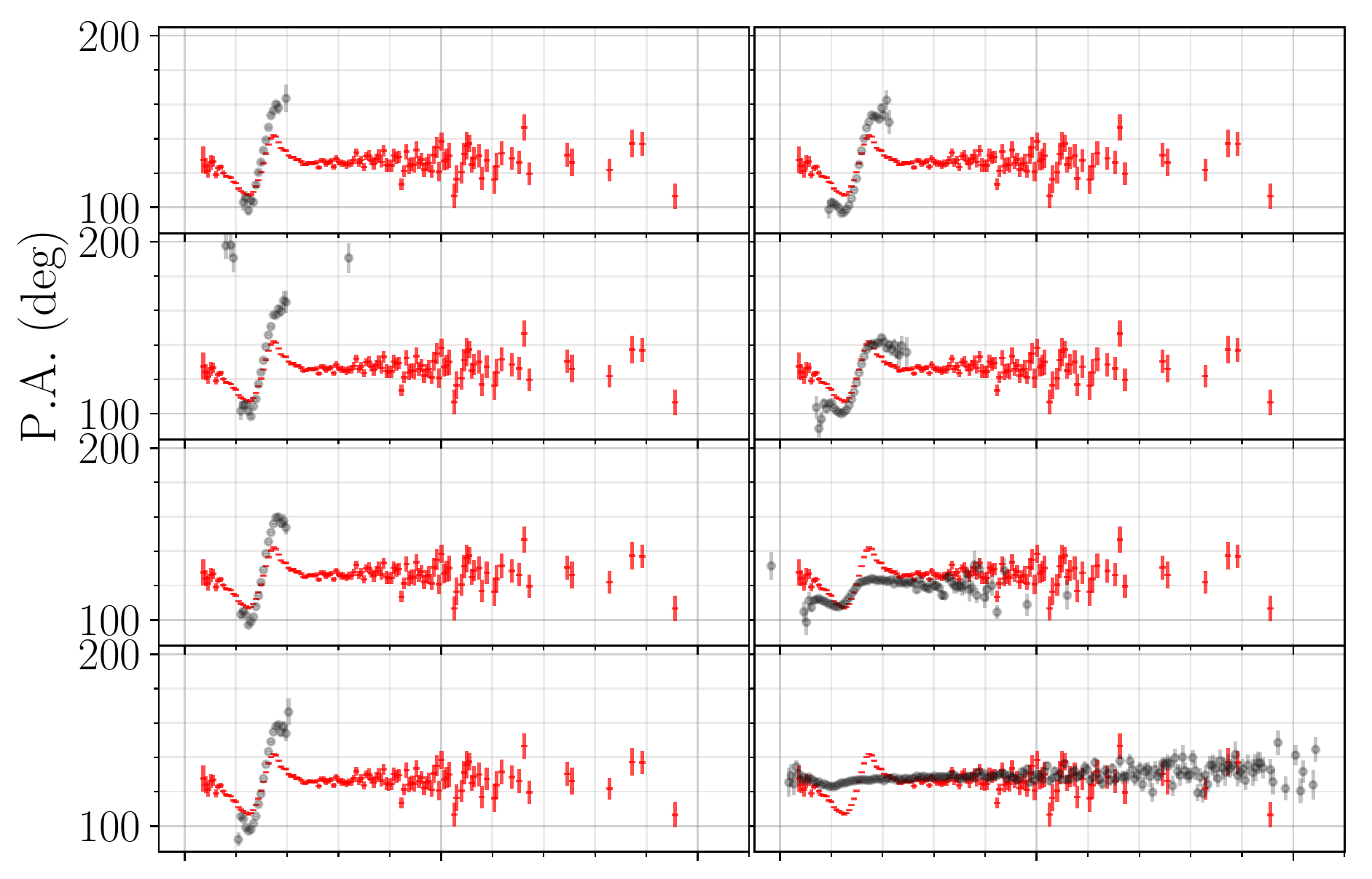}
	\includegraphics[width=\columnwidth]{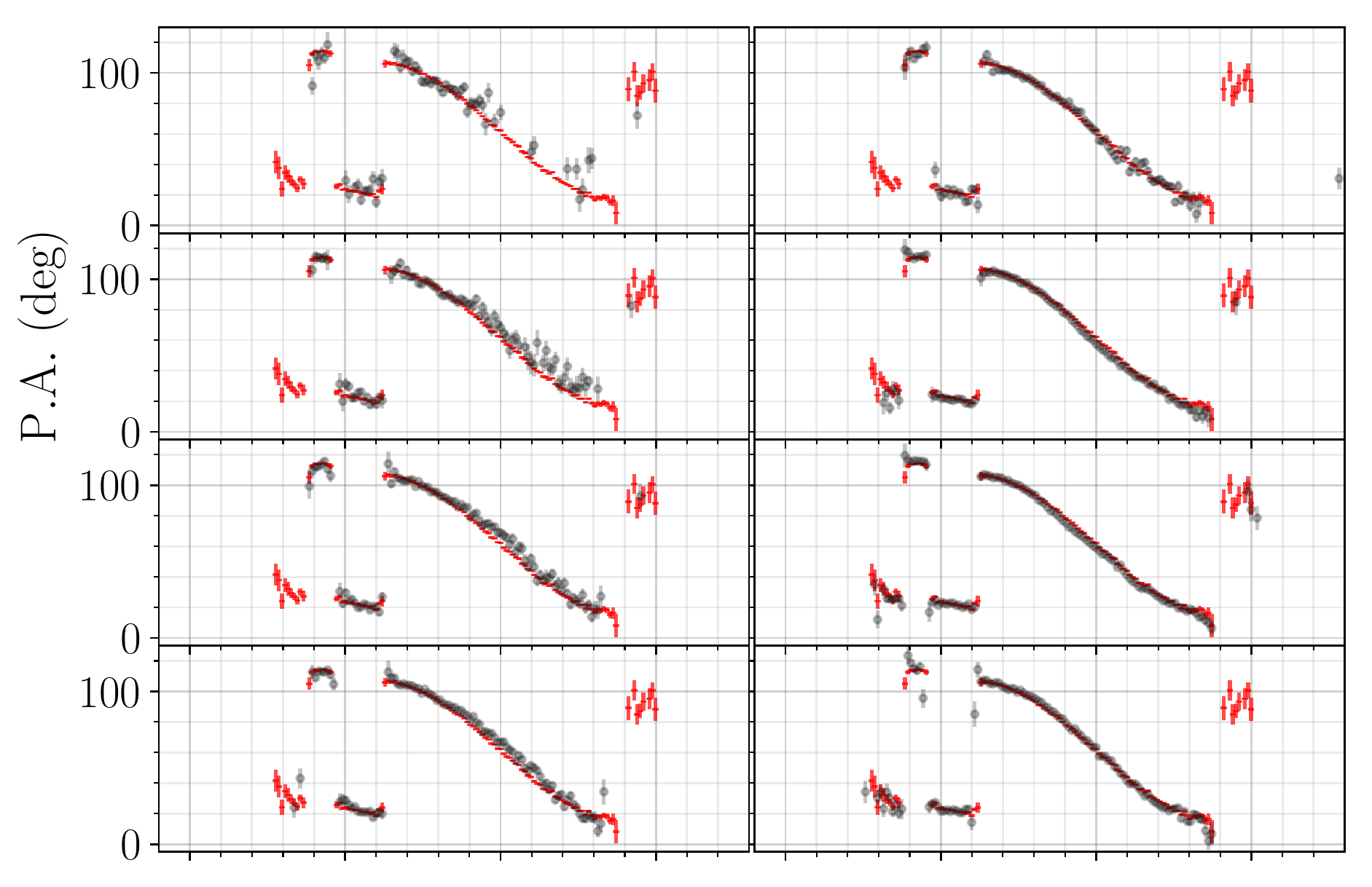}

	\includegraphics[width=\columnwidth]{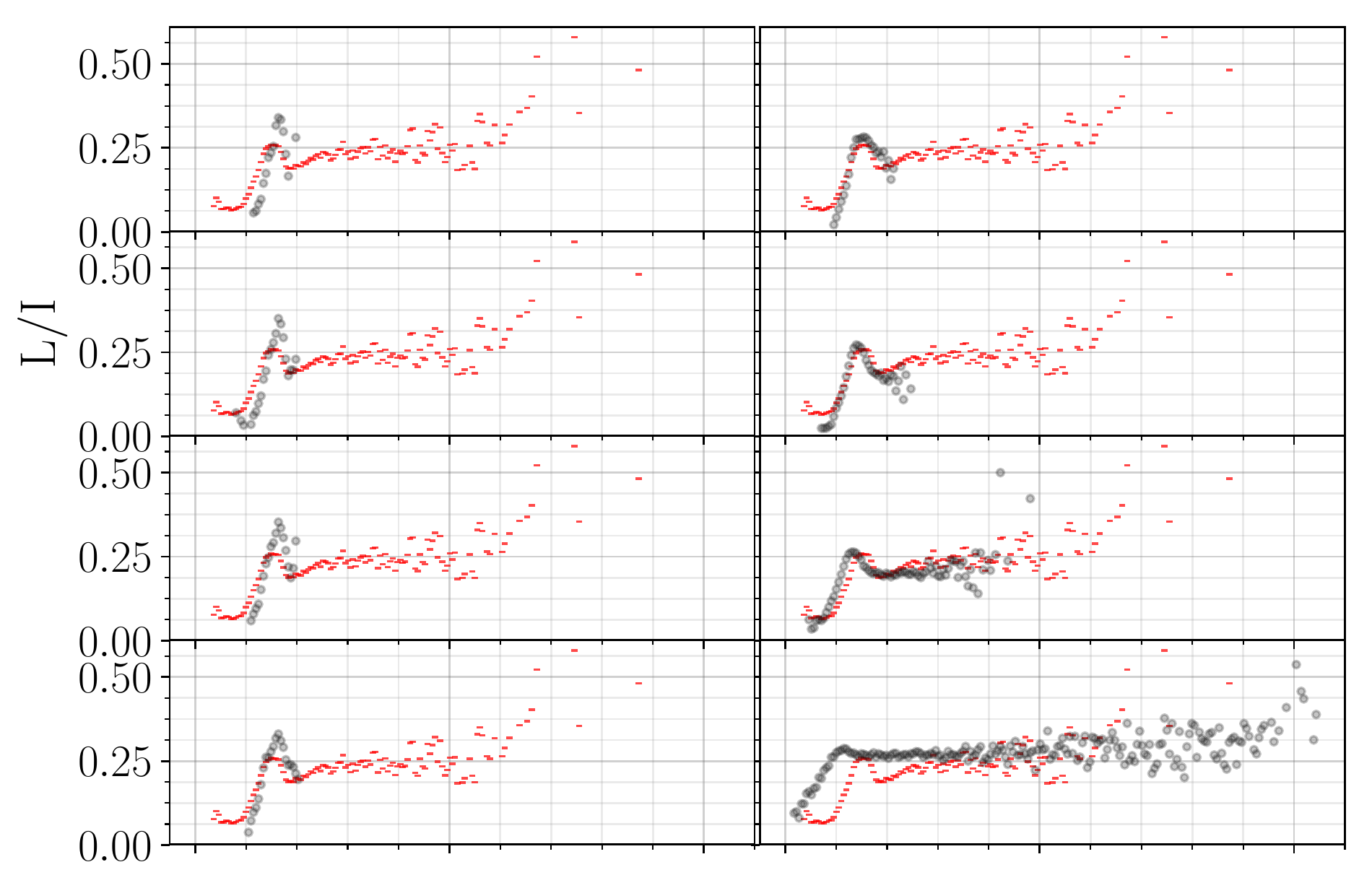}
	\includegraphics[width=\columnwidth]{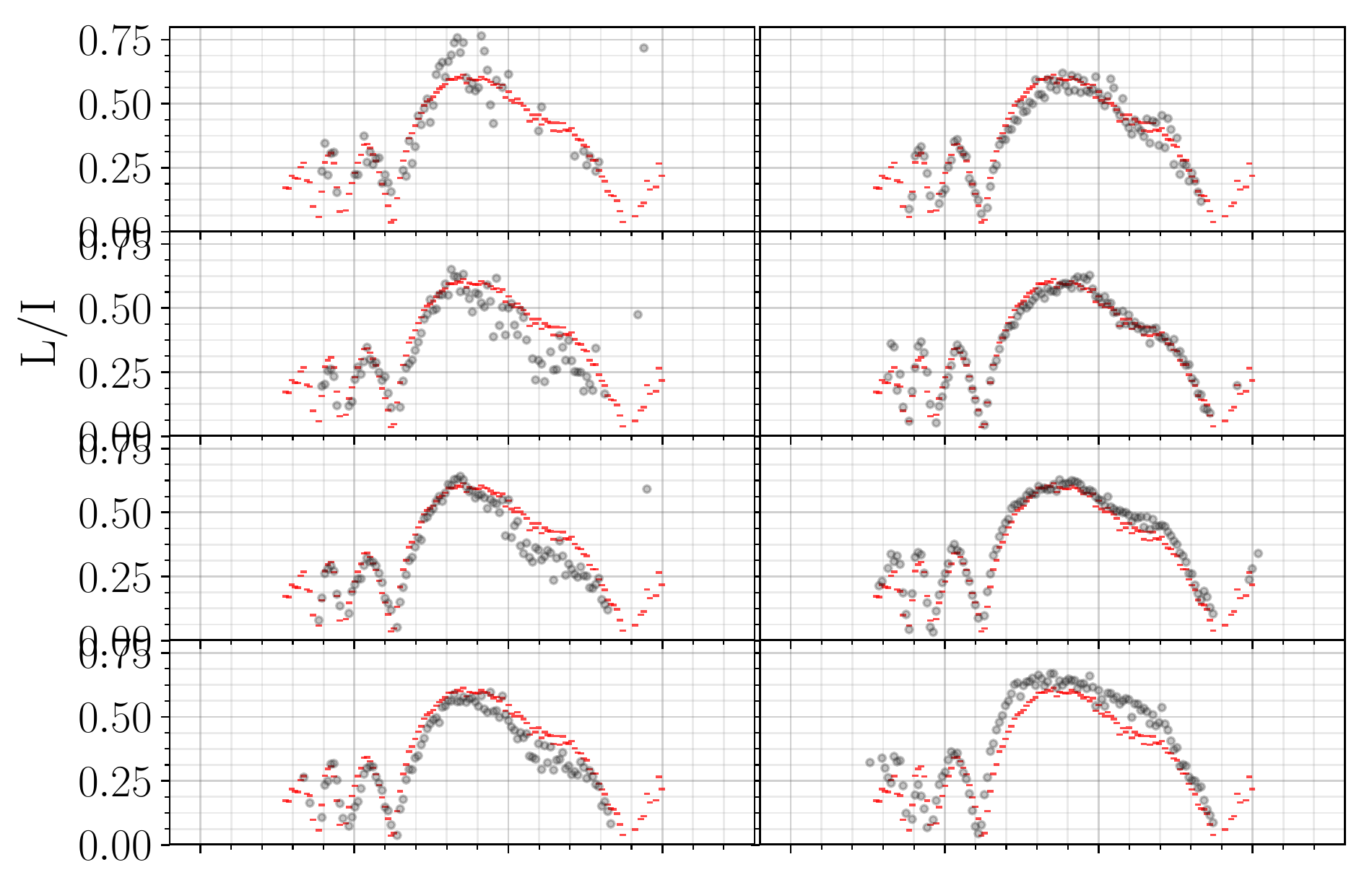}

	\includegraphics[width=\columnwidth]{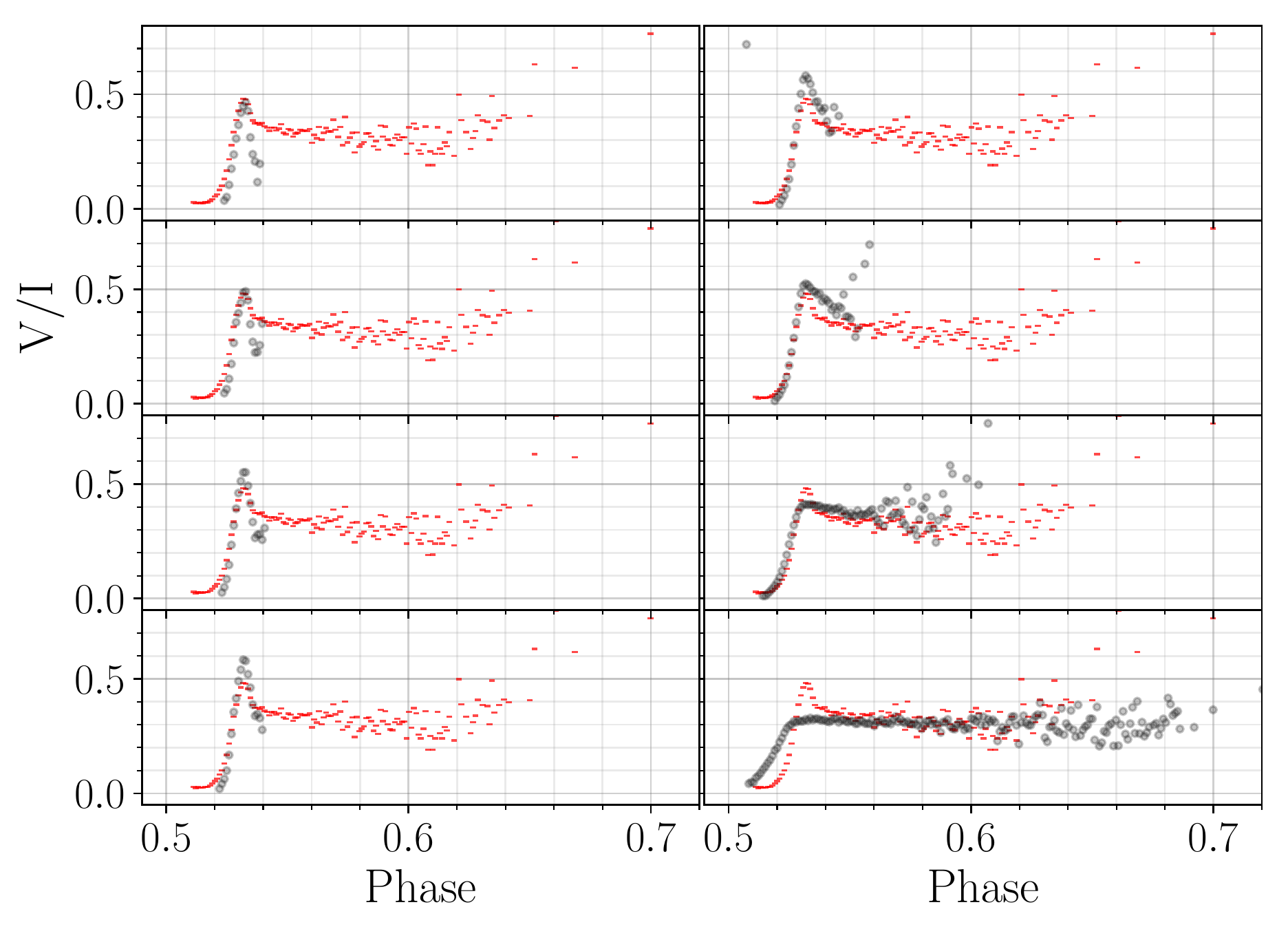}
	\includegraphics[width=\columnwidth]{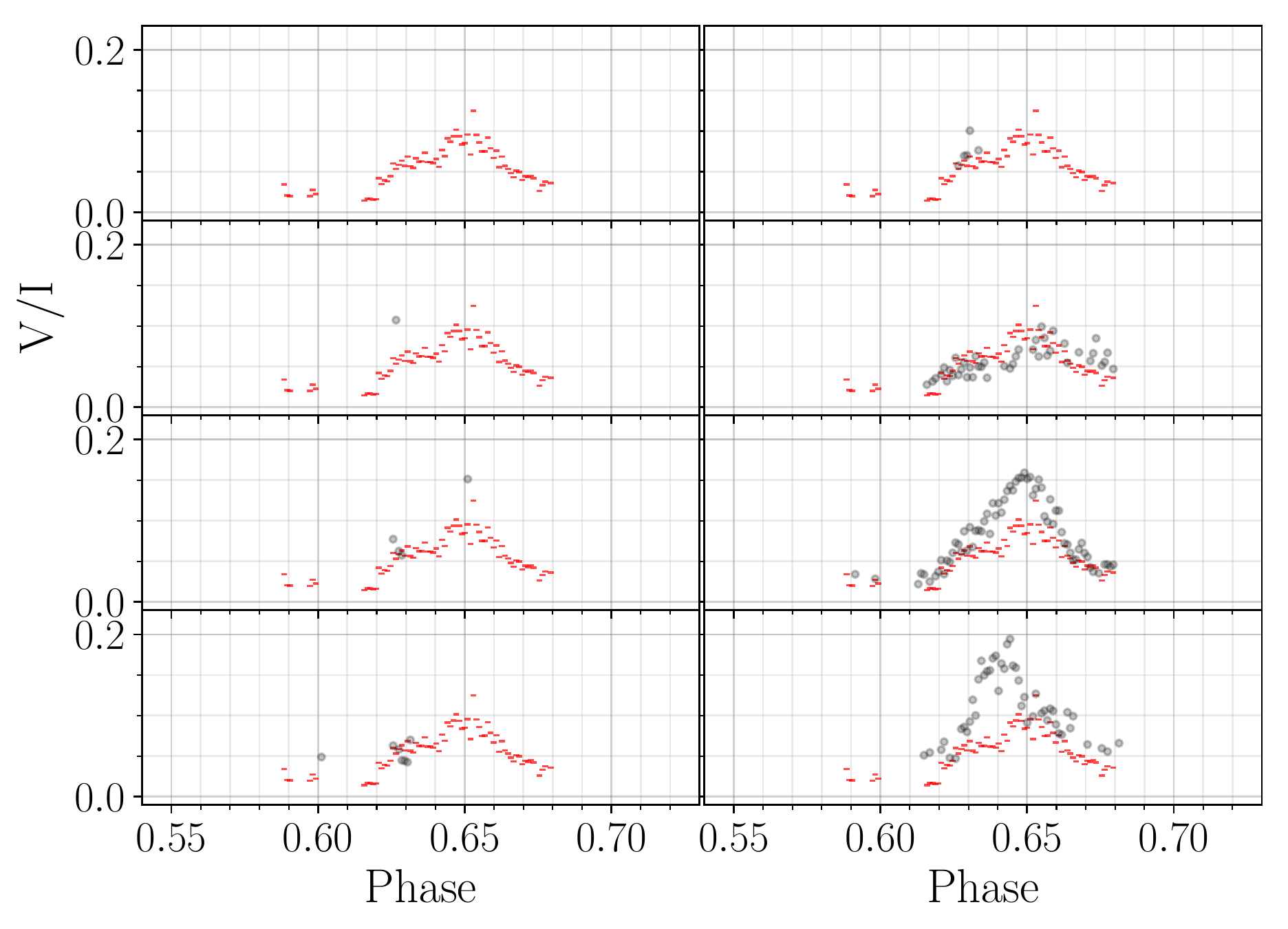}

    \caption{...Figure \ref{fig:profs1} continued...}
\end{figure*}

\begin{figure*} 
	\includegraphics[width=\columnwidth]{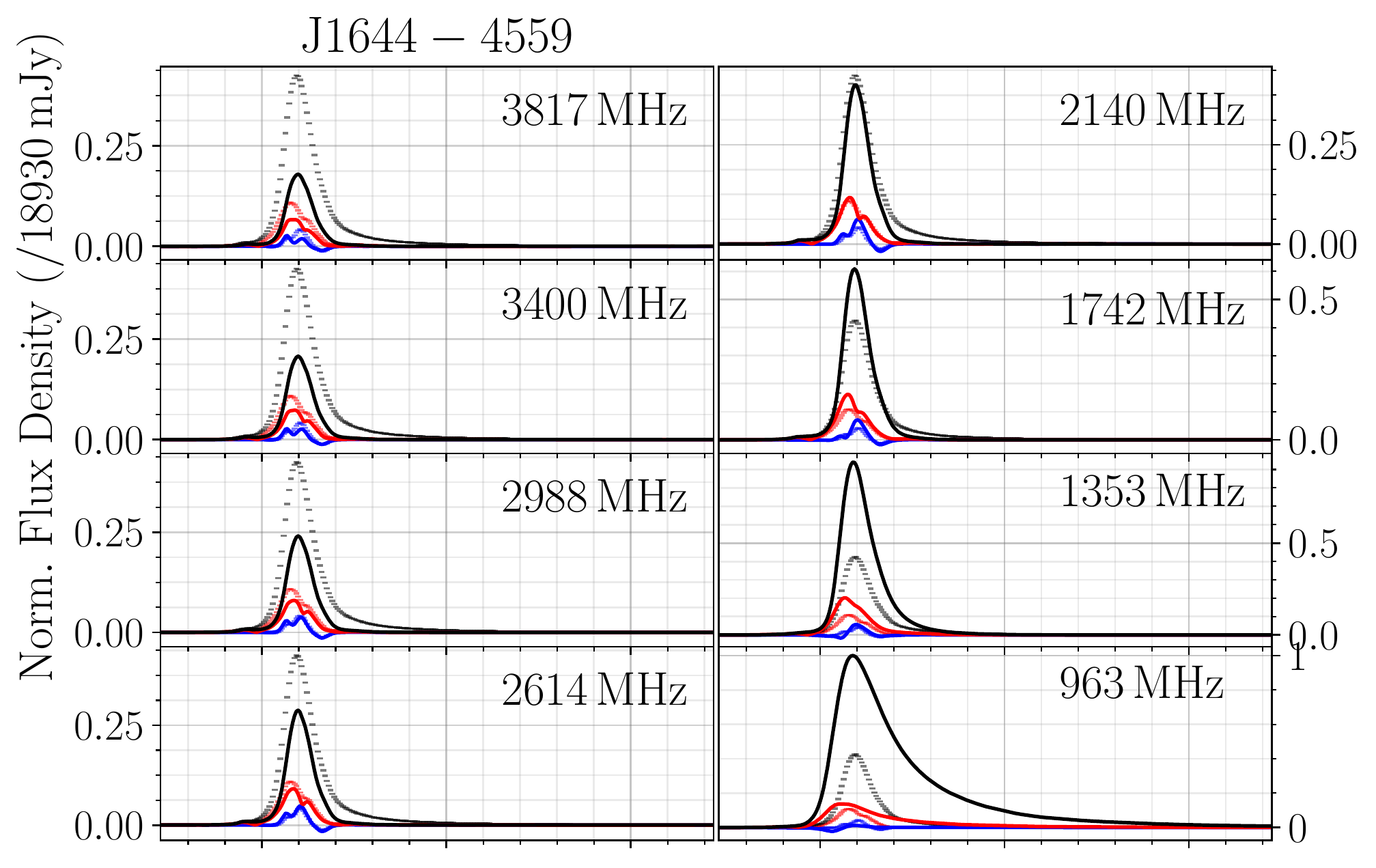}
	\includegraphics[width=\columnwidth]{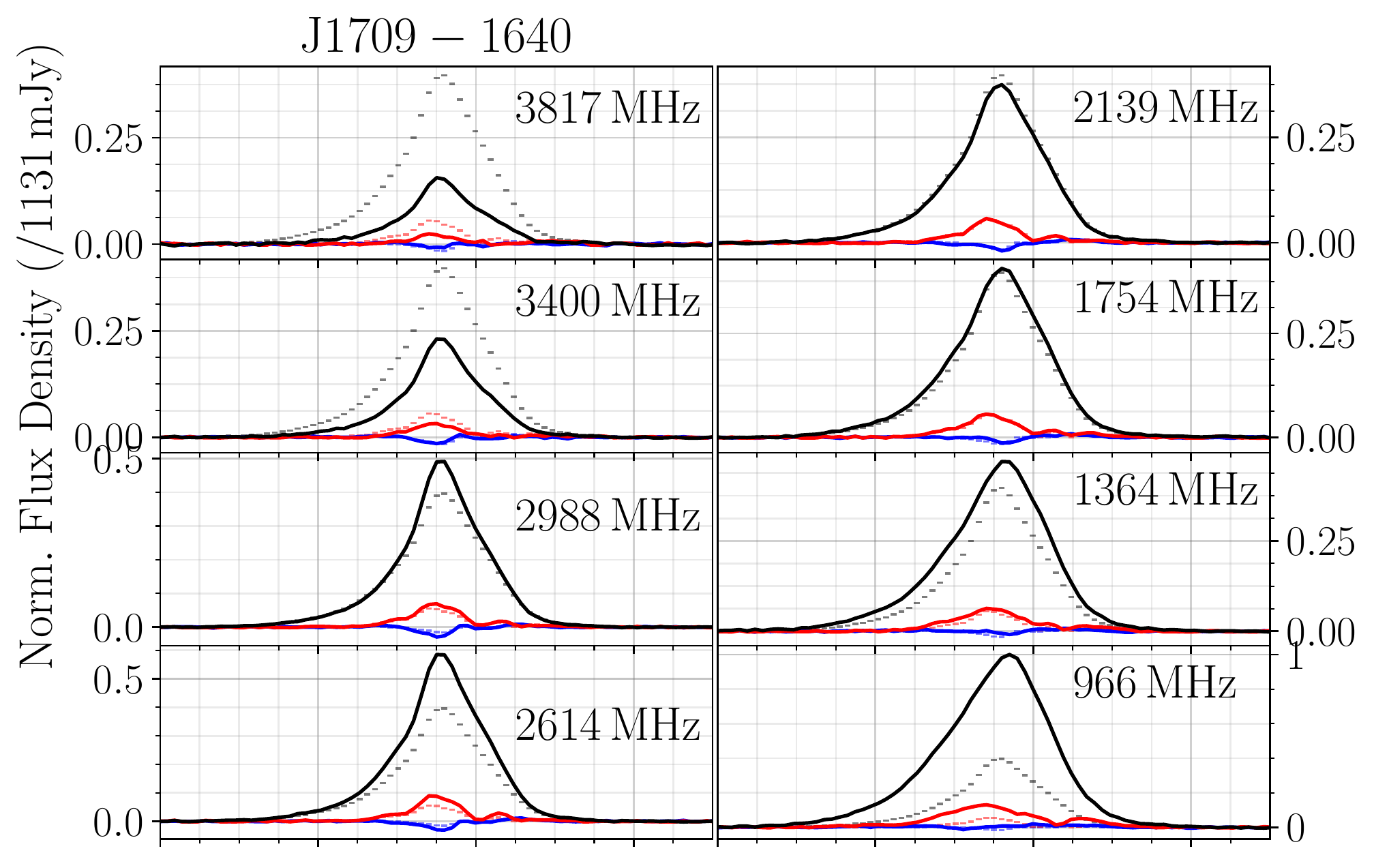}

	\includegraphics[width=\columnwidth]{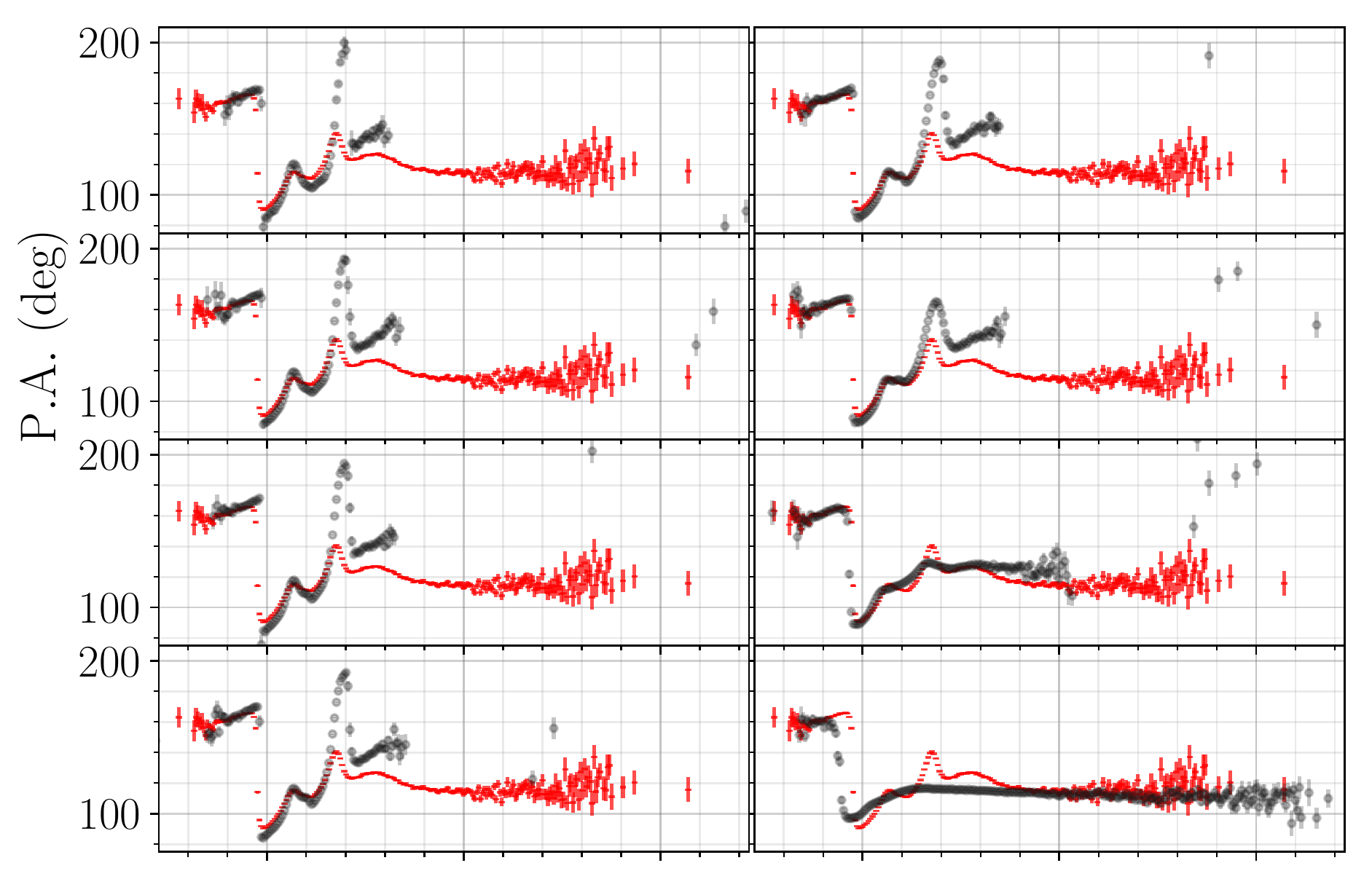}
	\includegraphics[width=\columnwidth]{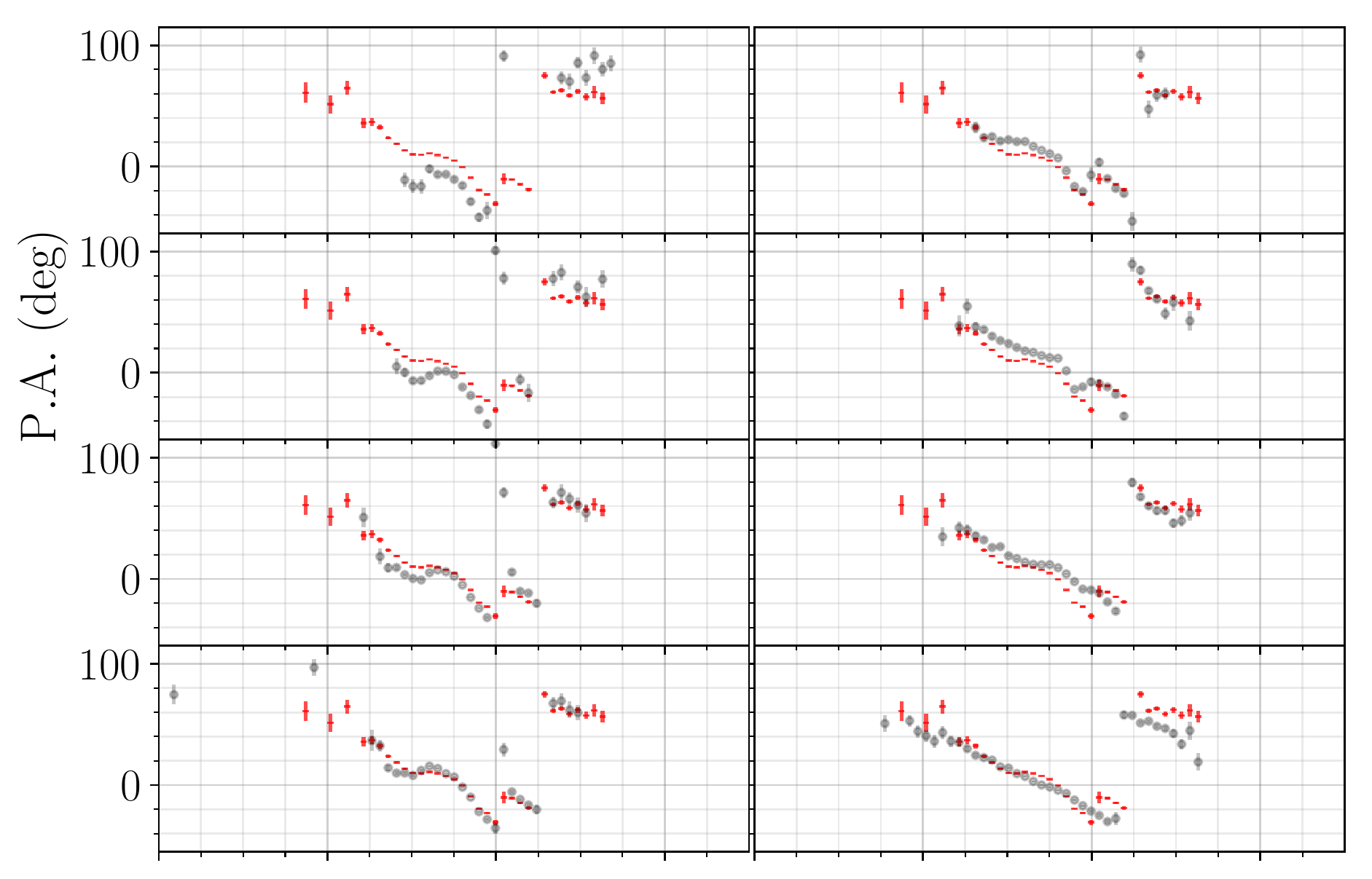}

	\includegraphics[width=\columnwidth]{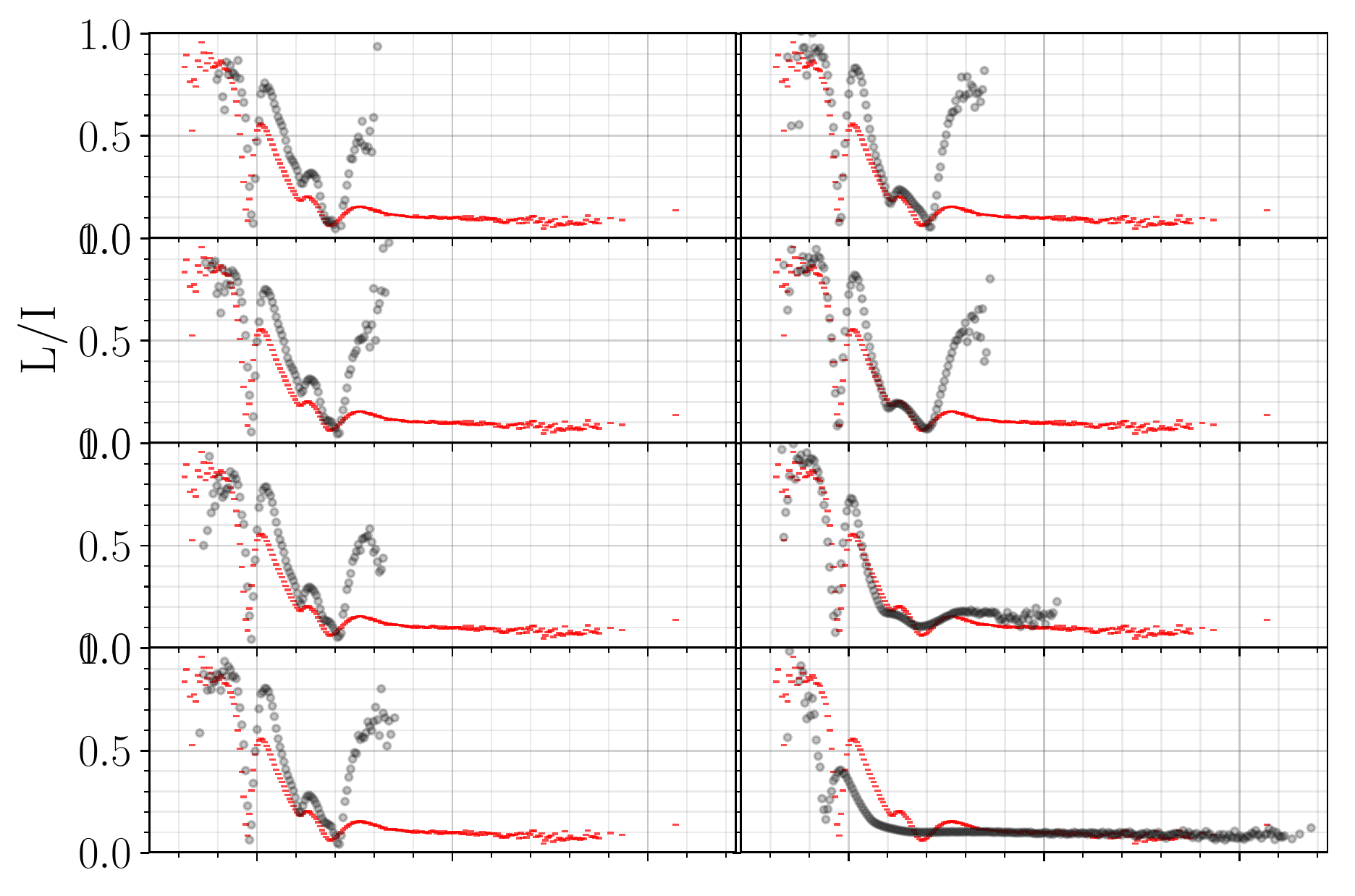}
	\includegraphics[width=\columnwidth]{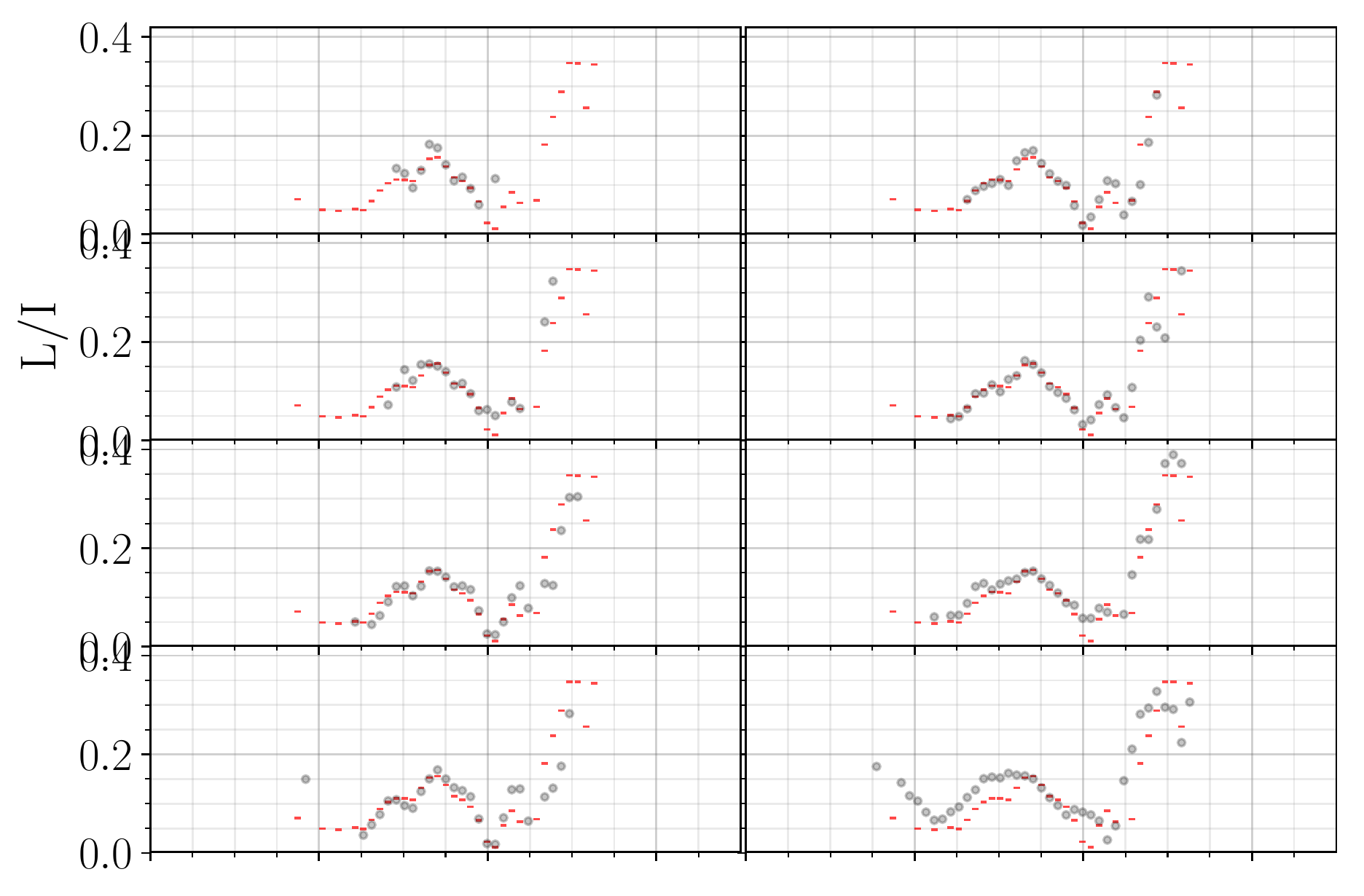}

	\includegraphics[width=\columnwidth]{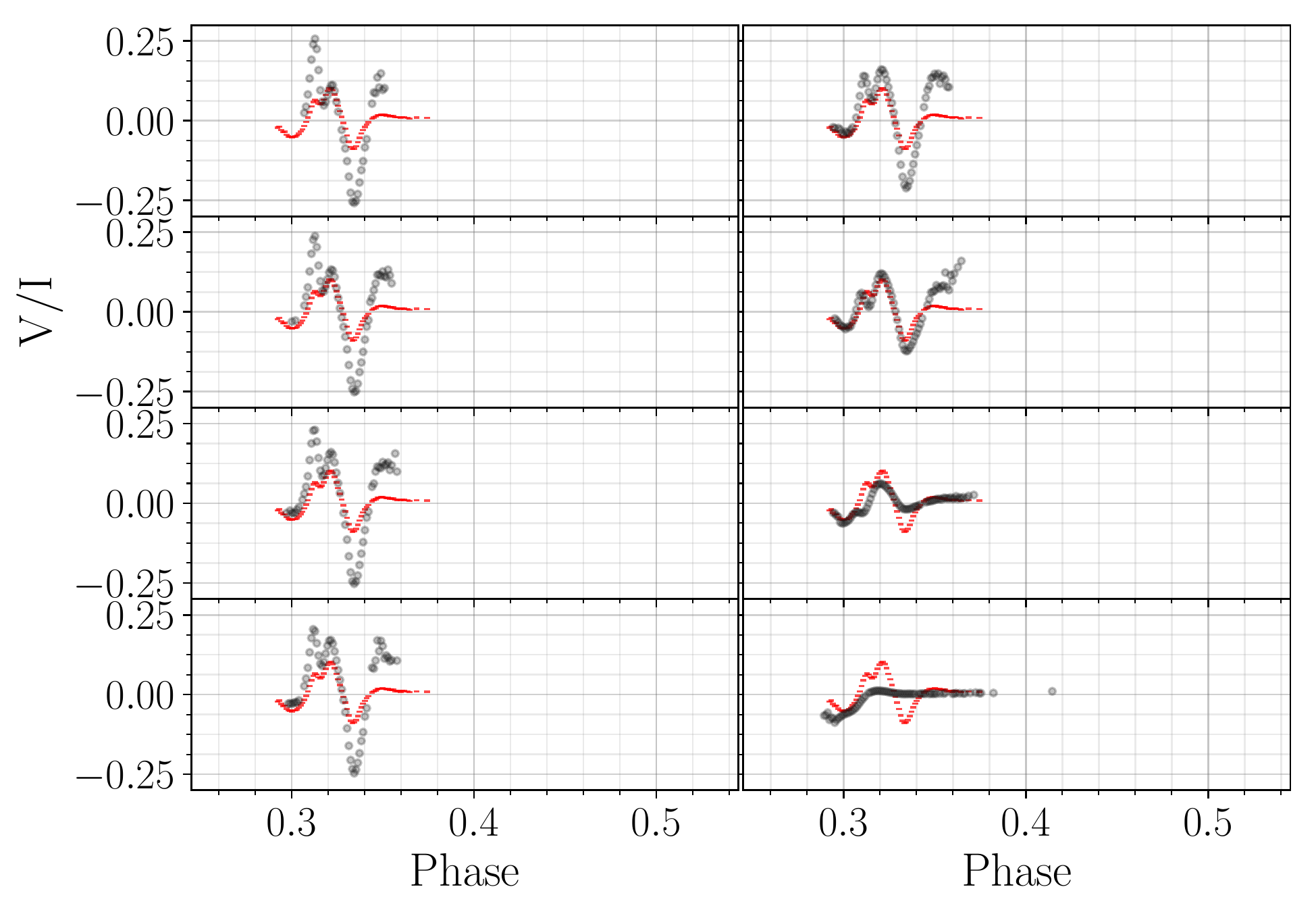}
	\includegraphics[width=\columnwidth]{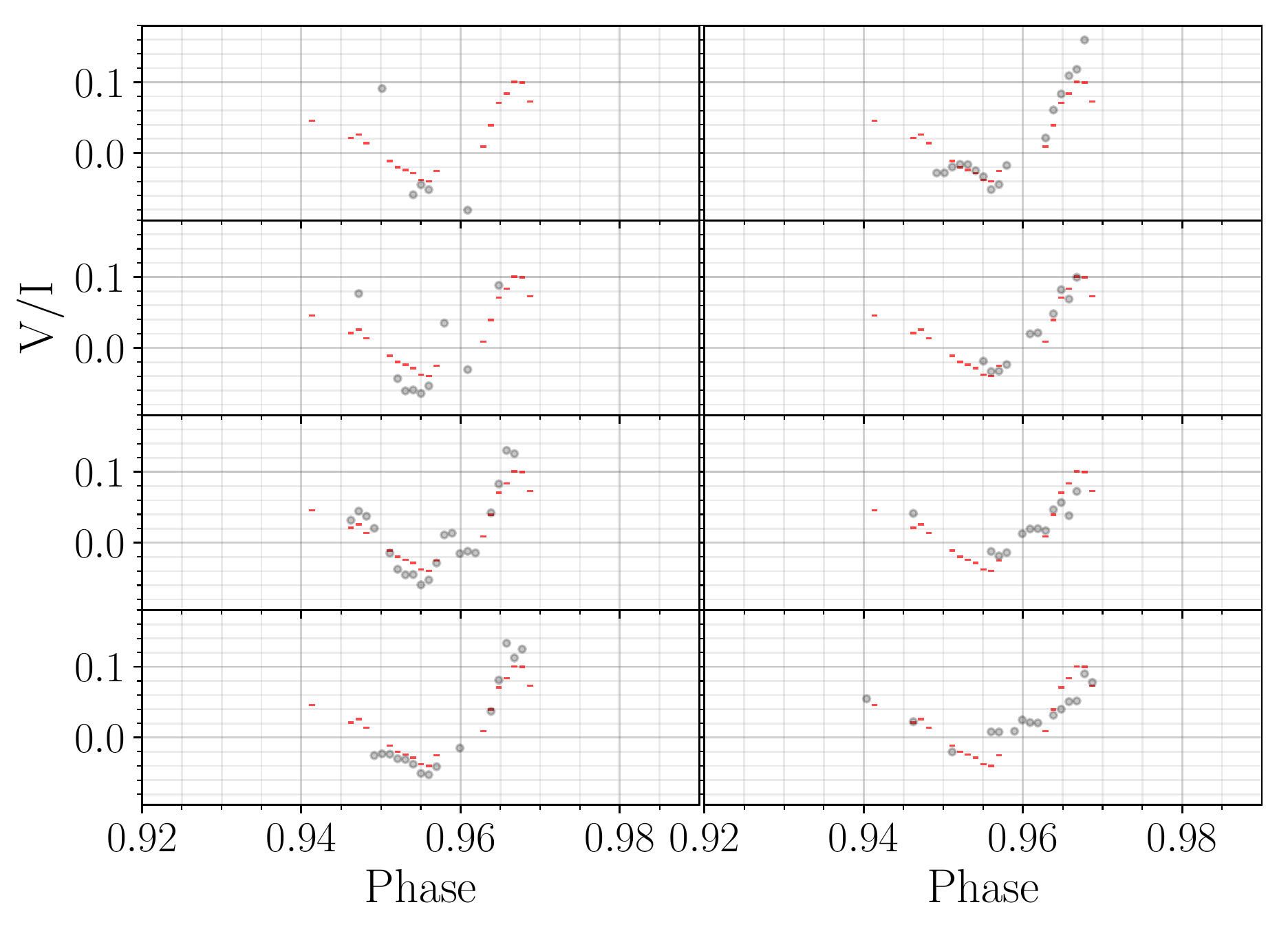}

    \caption{...Figure \ref{fig:profs1} continued...}
\end{figure*}

\begin{figure*} 
	\includegraphics[width=\columnwidth]{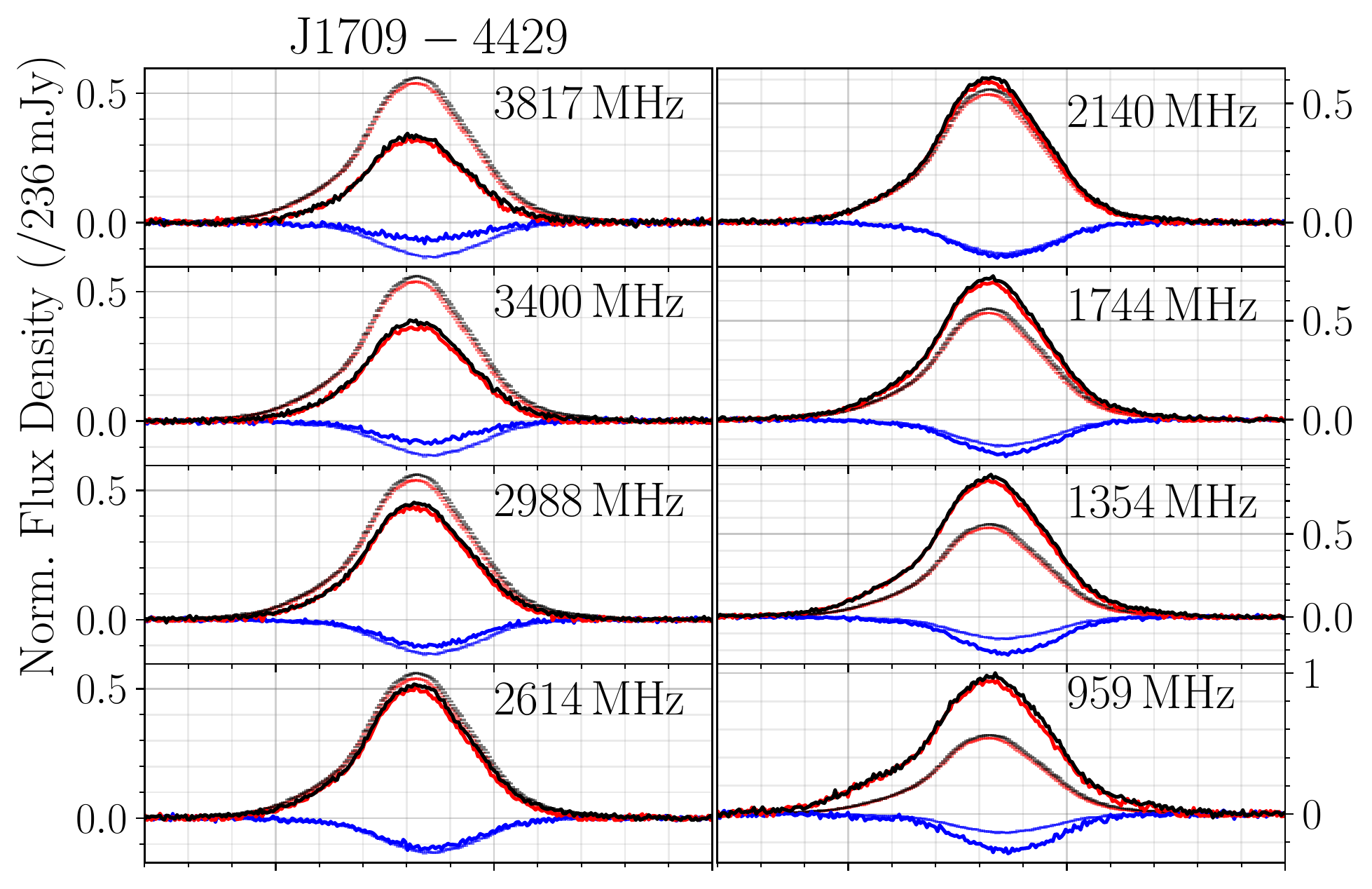}
	\includegraphics[width=\columnwidth]{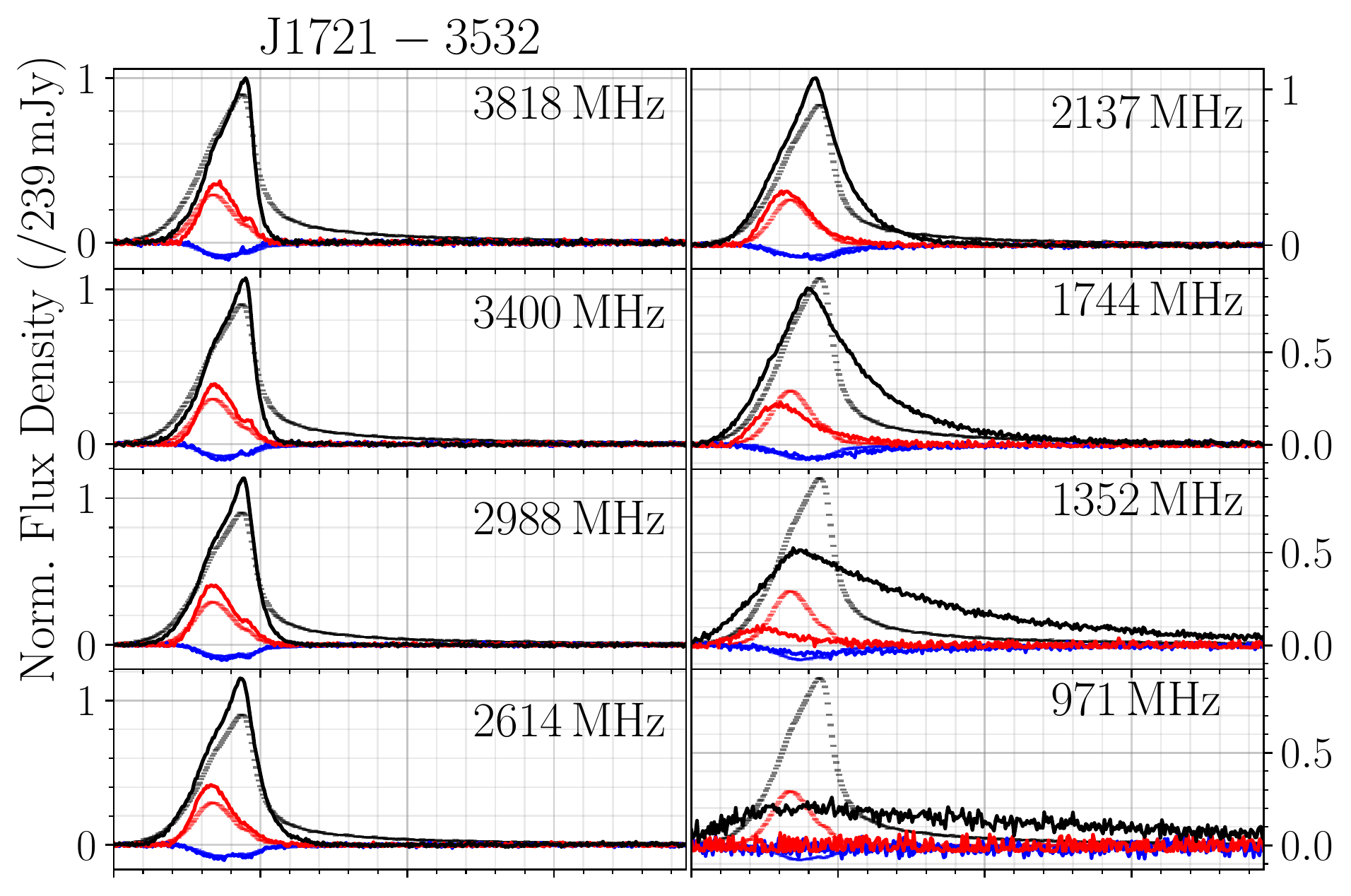}

	\includegraphics[width=\columnwidth]{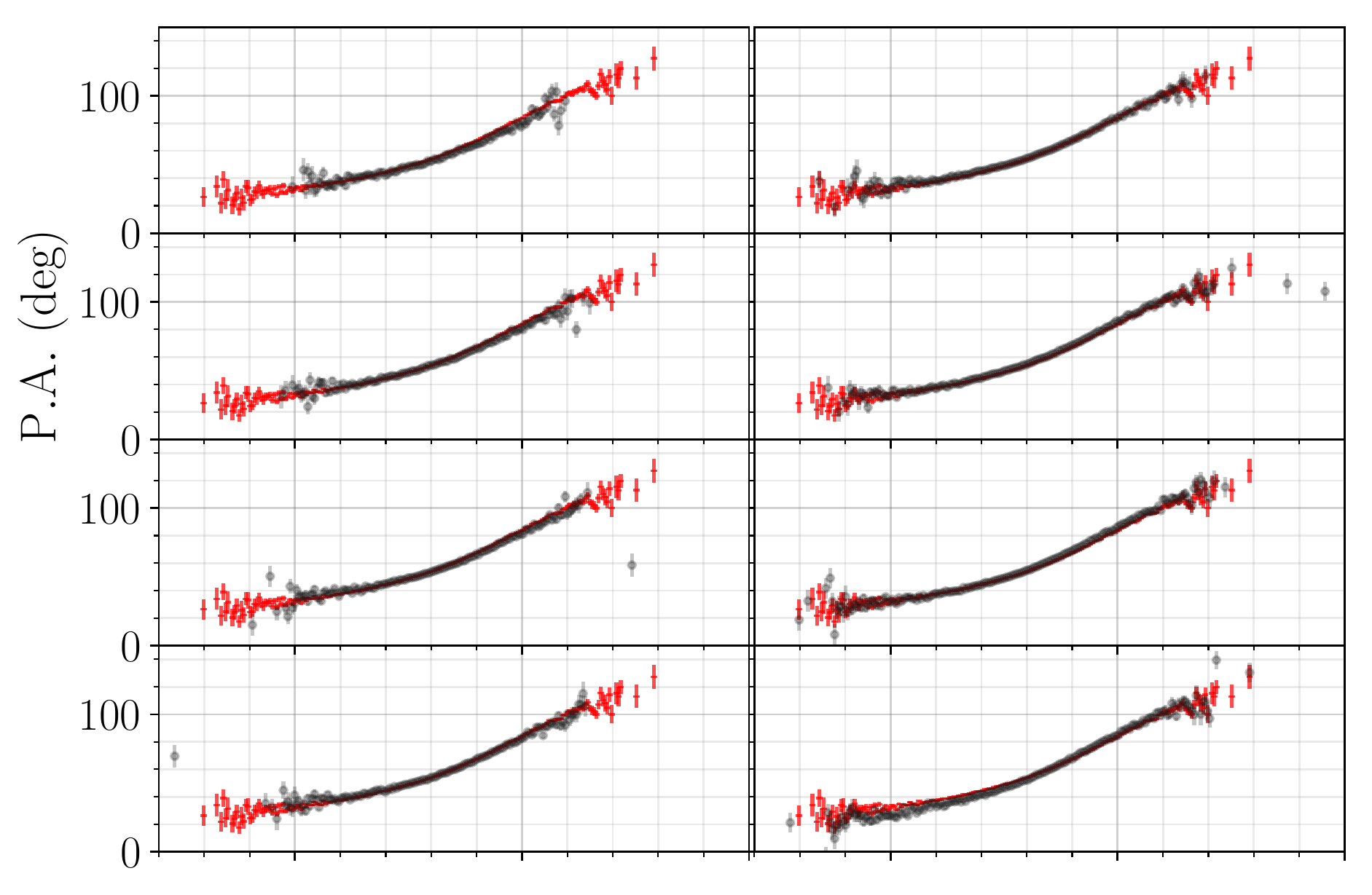}
	\includegraphics[width=\columnwidth]{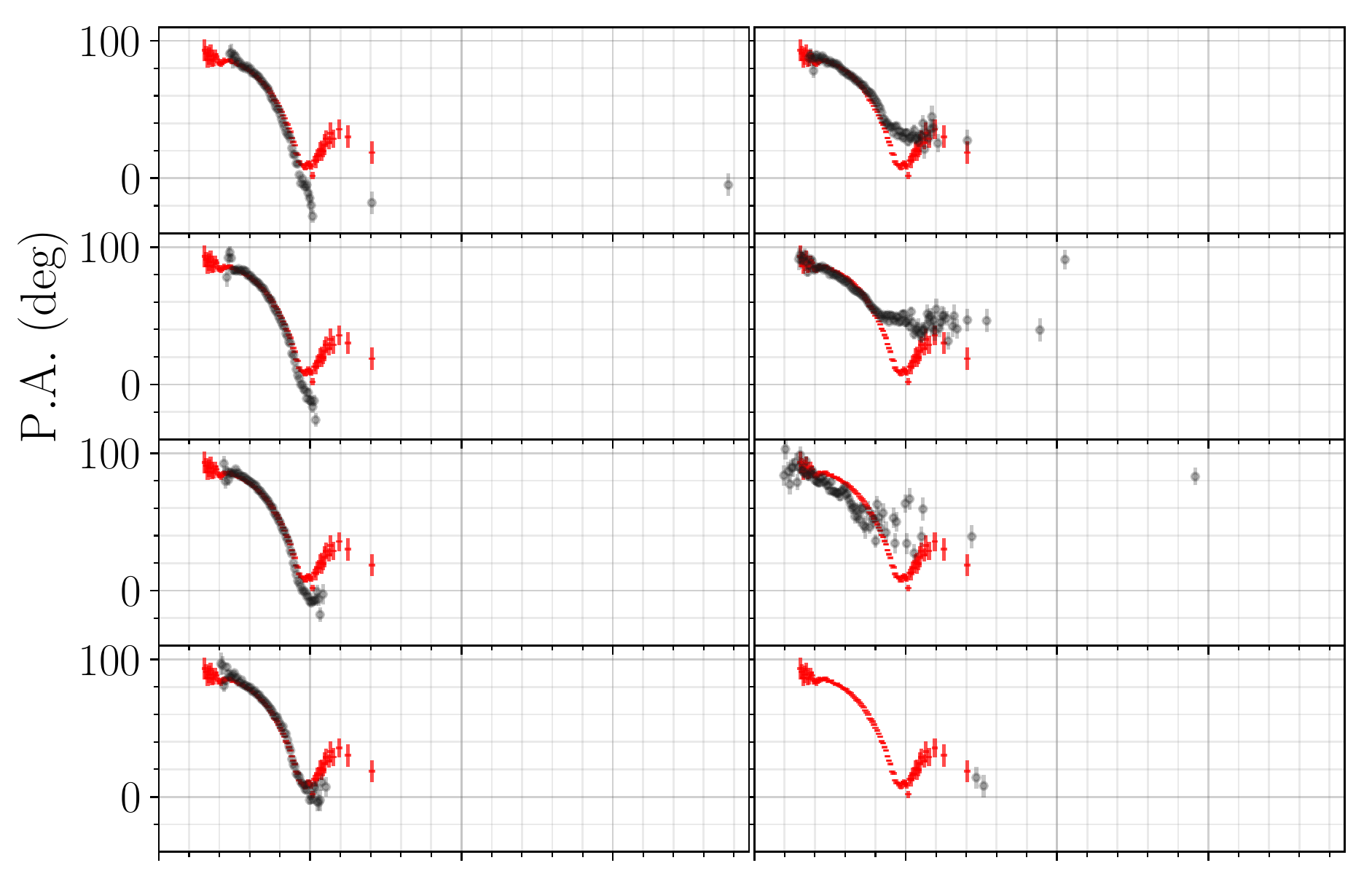}

	\includegraphics[width=\columnwidth]{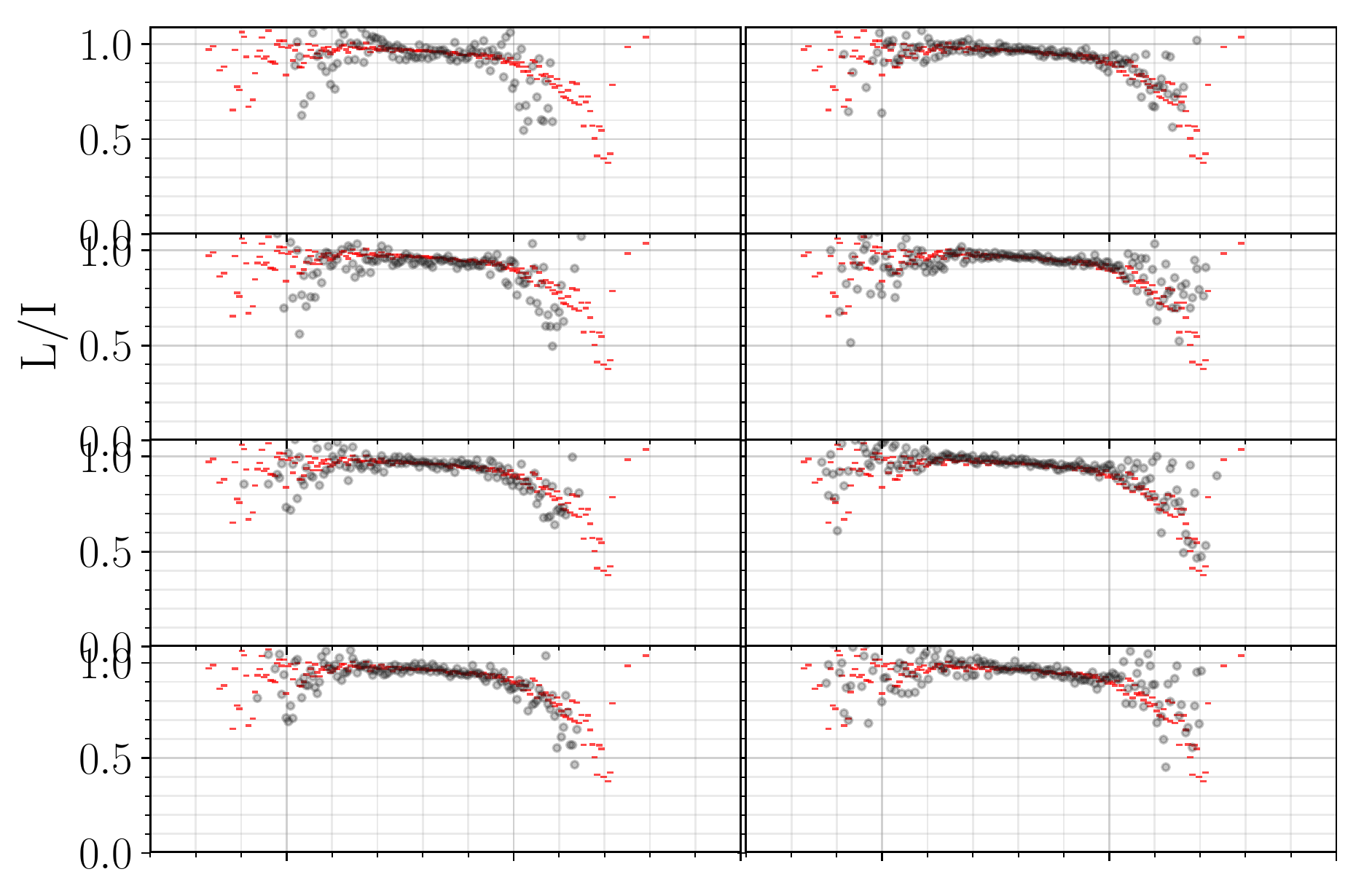}
	\includegraphics[width=\columnwidth]{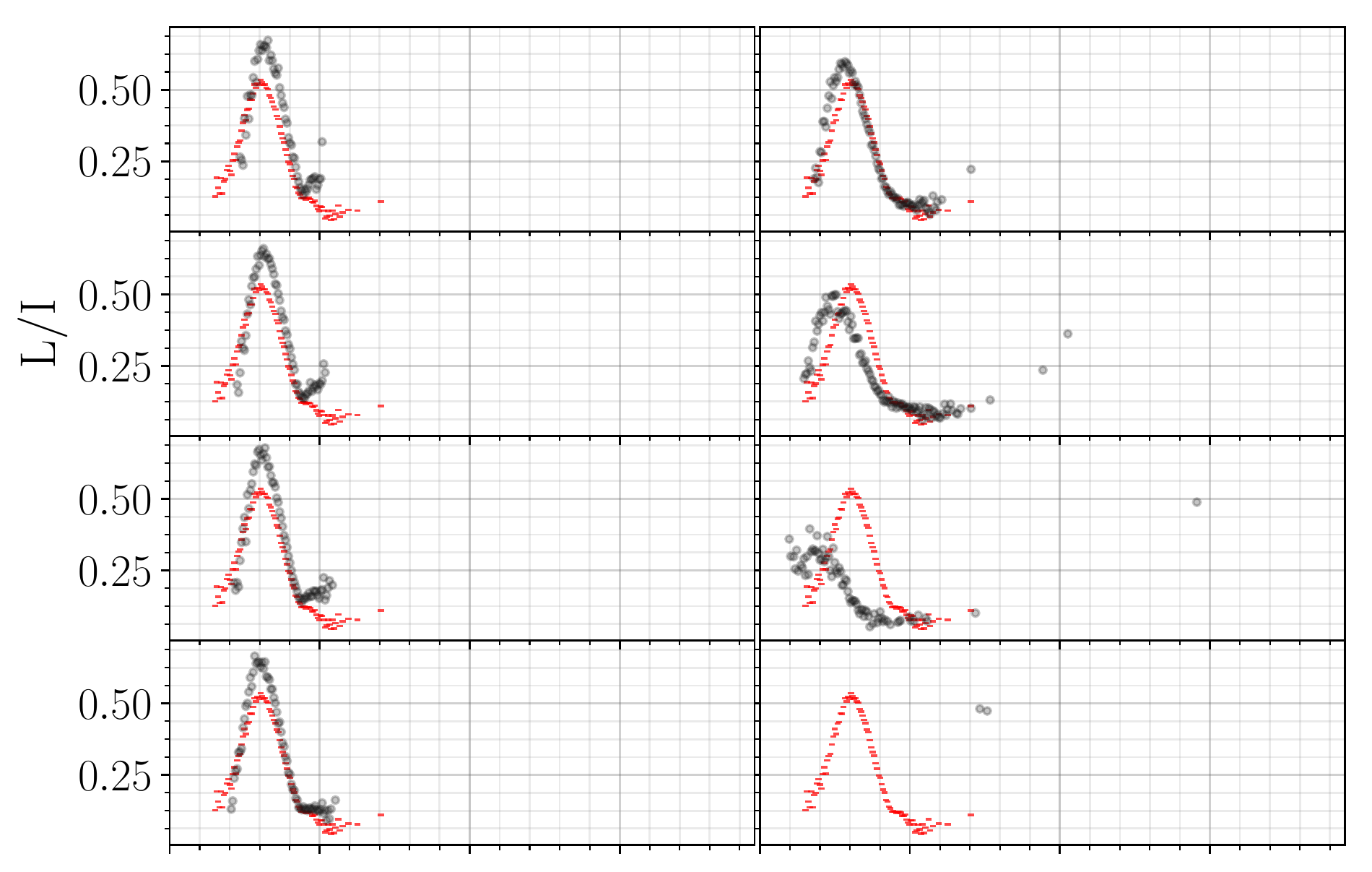}

	\includegraphics[width=\columnwidth]{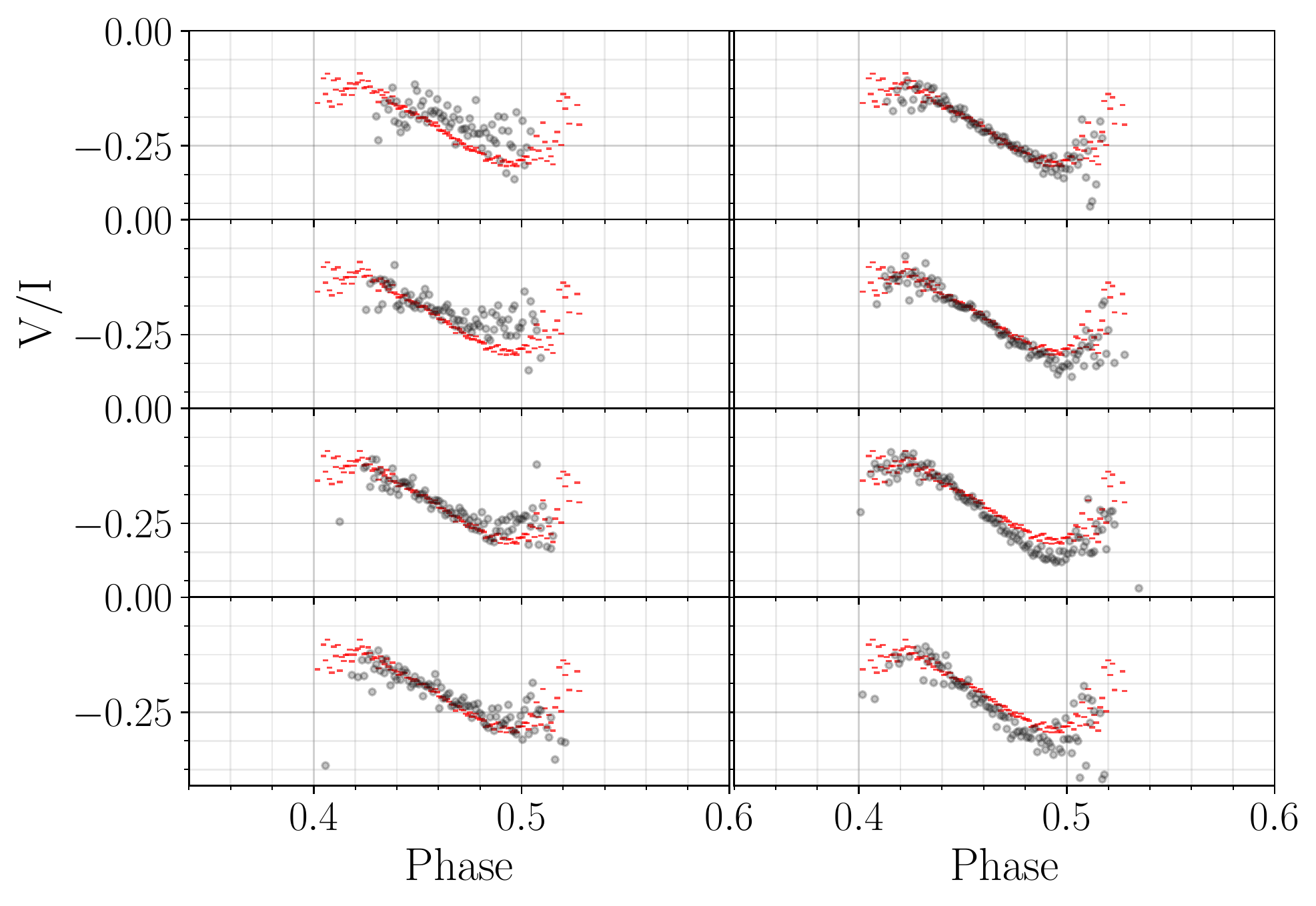}
	\includegraphics[width=\columnwidth]{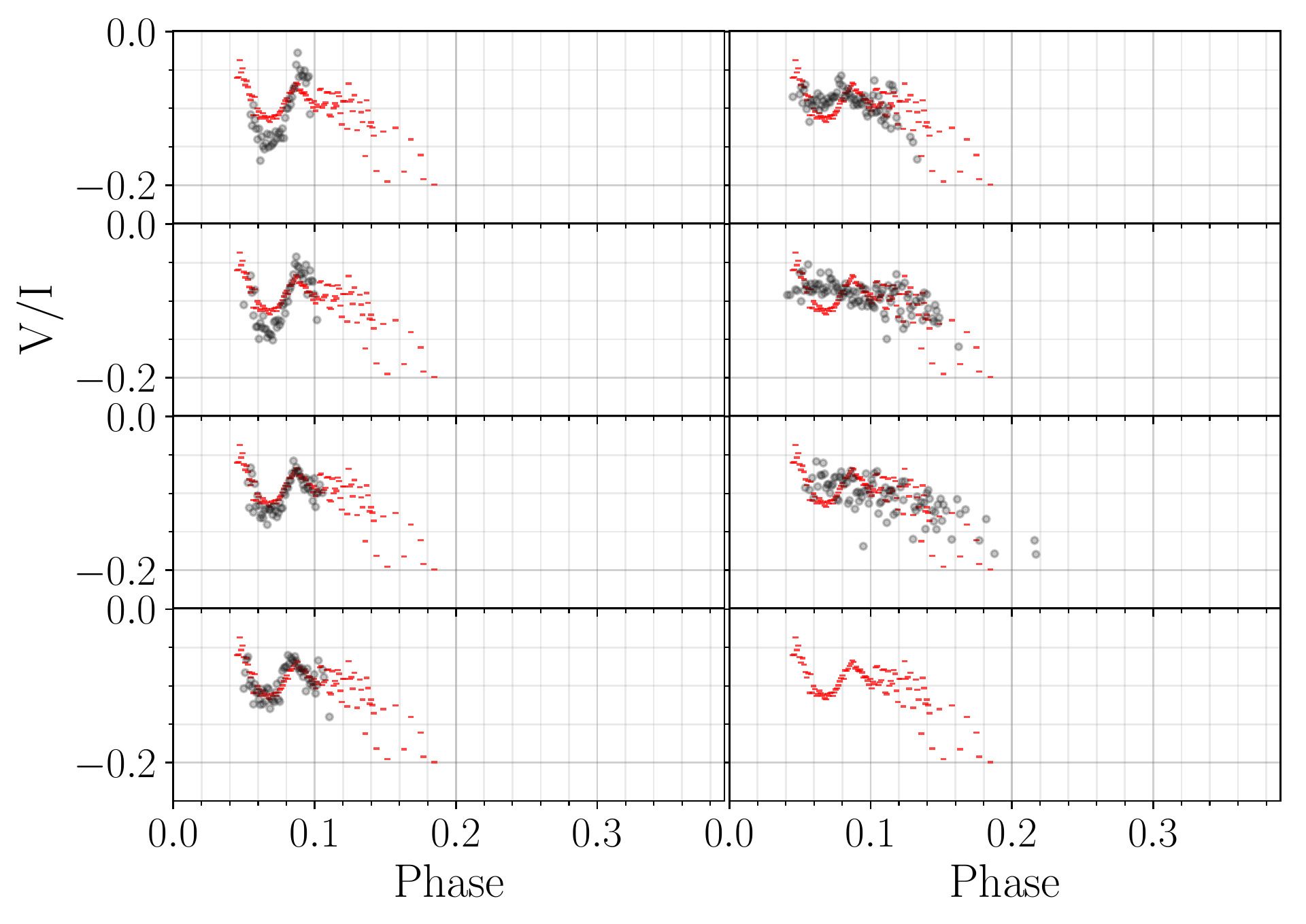}

    \caption{...Figure \ref{fig:profs1} continued...}
\end{figure*}

\begin{figure*} 
	\includegraphics[width=\columnwidth]{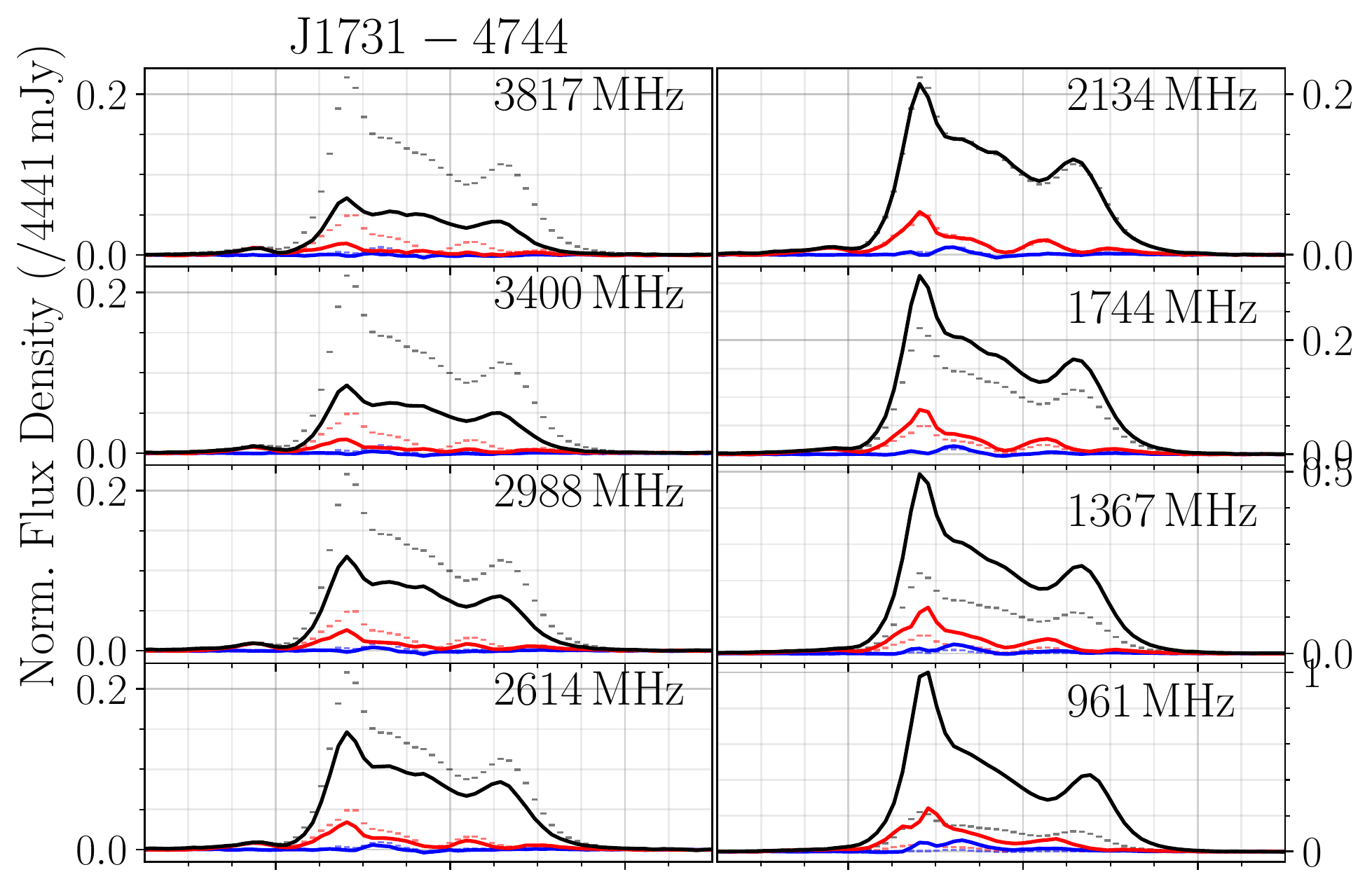}
	\includegraphics[width=\columnwidth]{Profs/J1740-3015_profs.pdf}

	\includegraphics[width=\columnwidth]{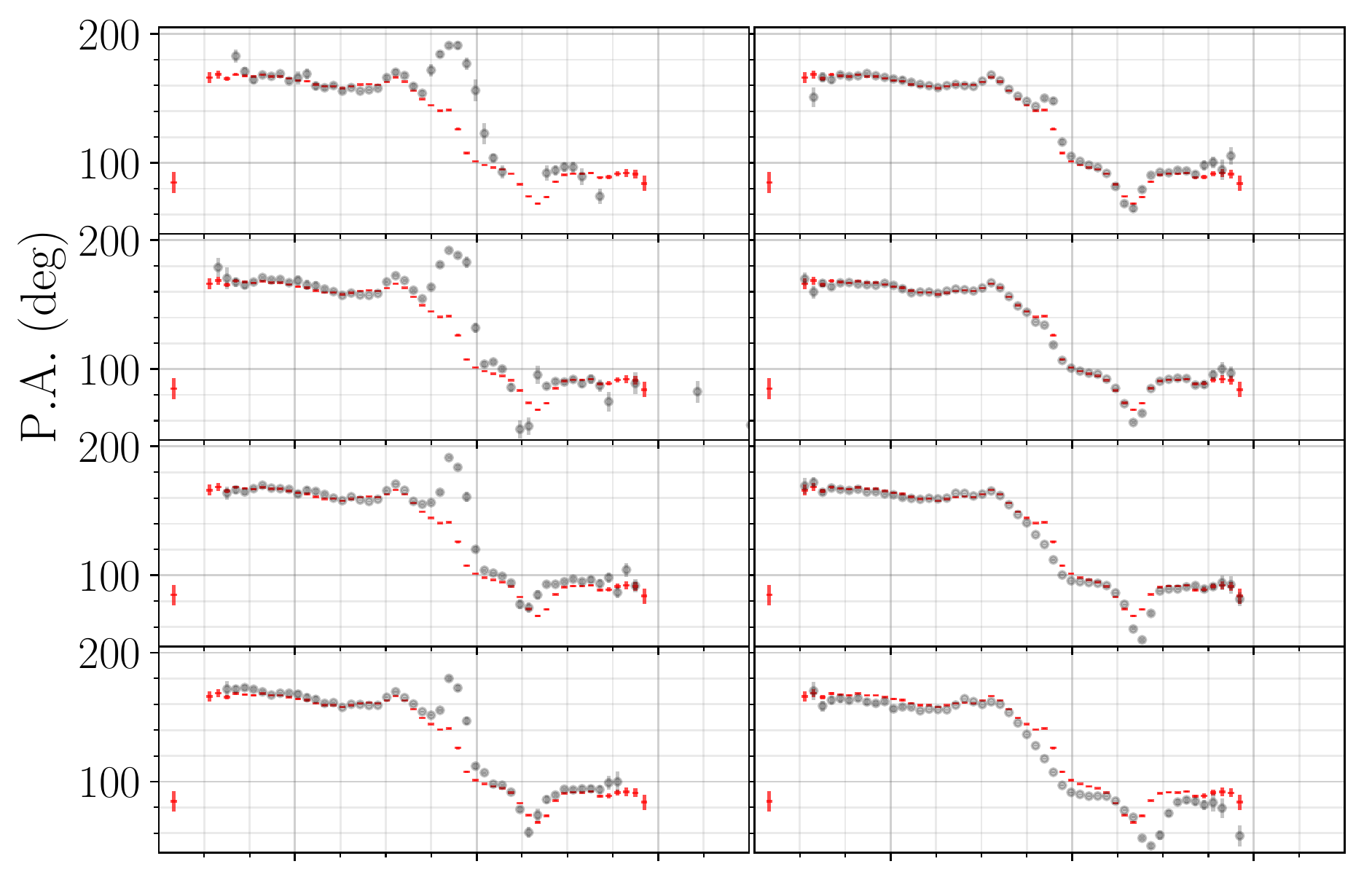}
	\includegraphics[width=\columnwidth]{Profs/J1740-3015_PAs.pdf}

	\includegraphics[width=\columnwidth]{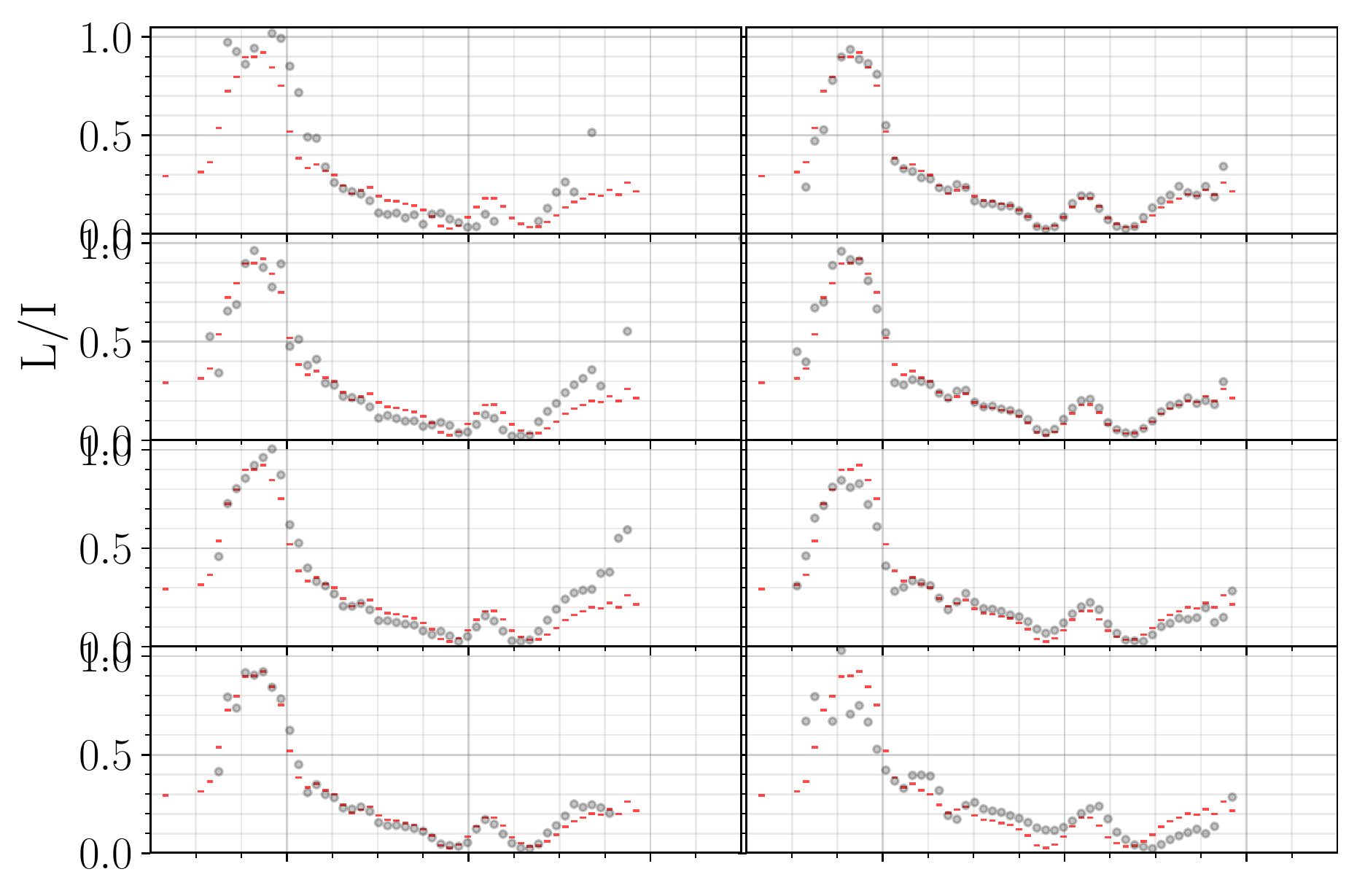}
	\includegraphics[width=\columnwidth]{Profs/J1740-3015_L.pdf}

	\includegraphics[width=\columnwidth]{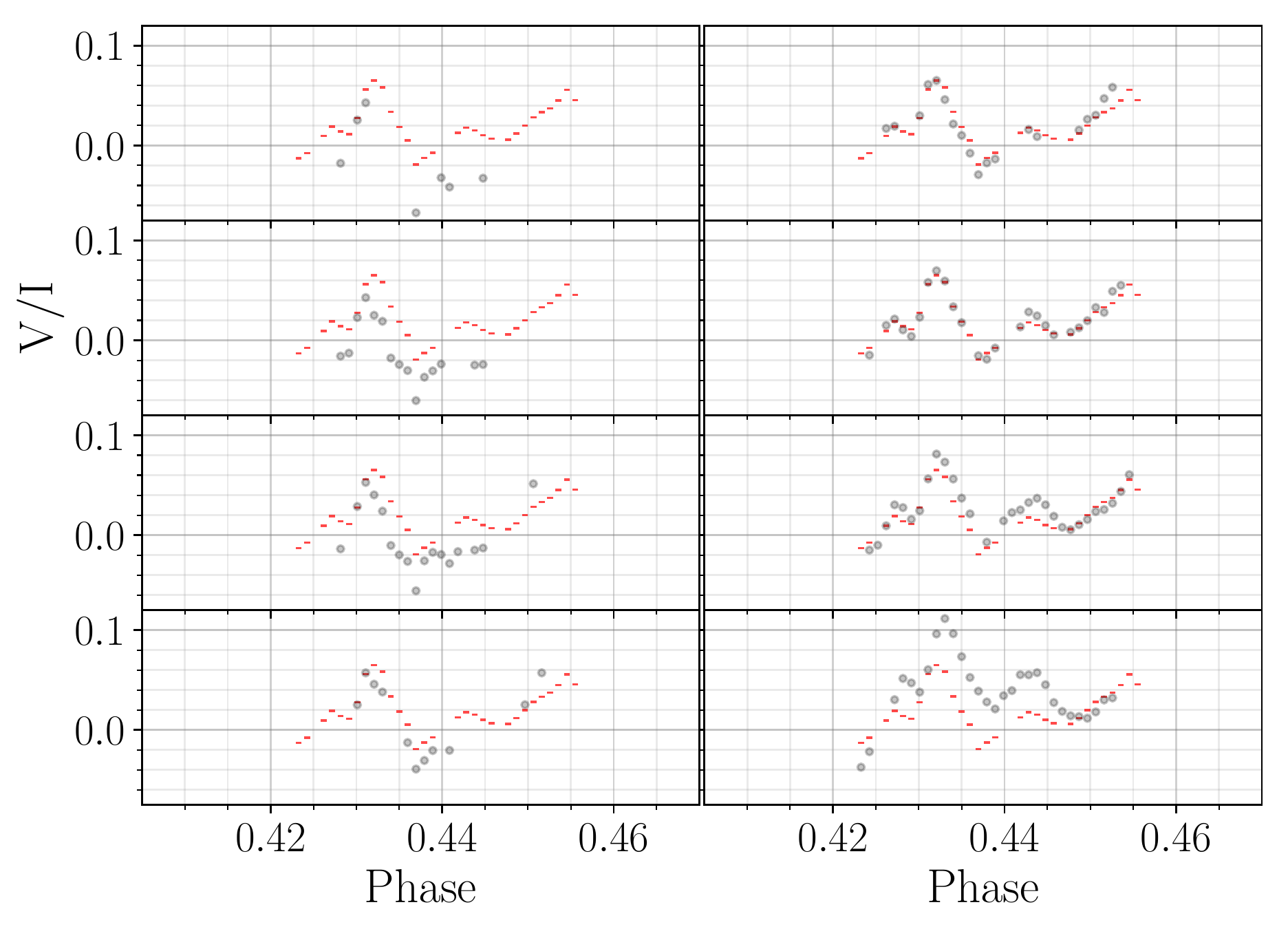}
	\includegraphics[width=\columnwidth]{Profs/J1740-3015_V.pdf}

    \caption{...Figure \ref{fig:profs1} continued...}
\end{figure*}

\begin{figure*} 
	\includegraphics[width=\columnwidth]{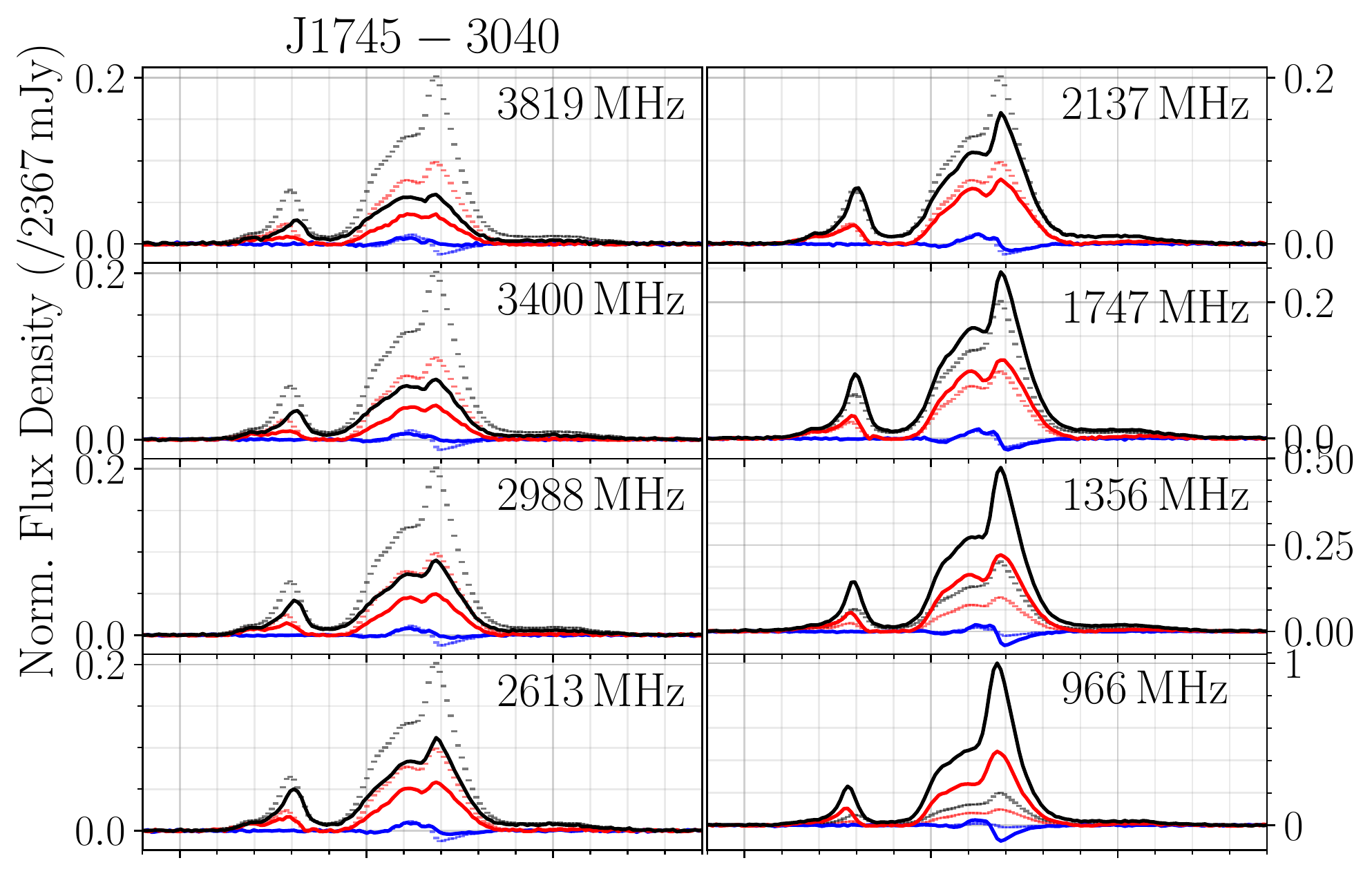}
	\includegraphics[width=\columnwidth]{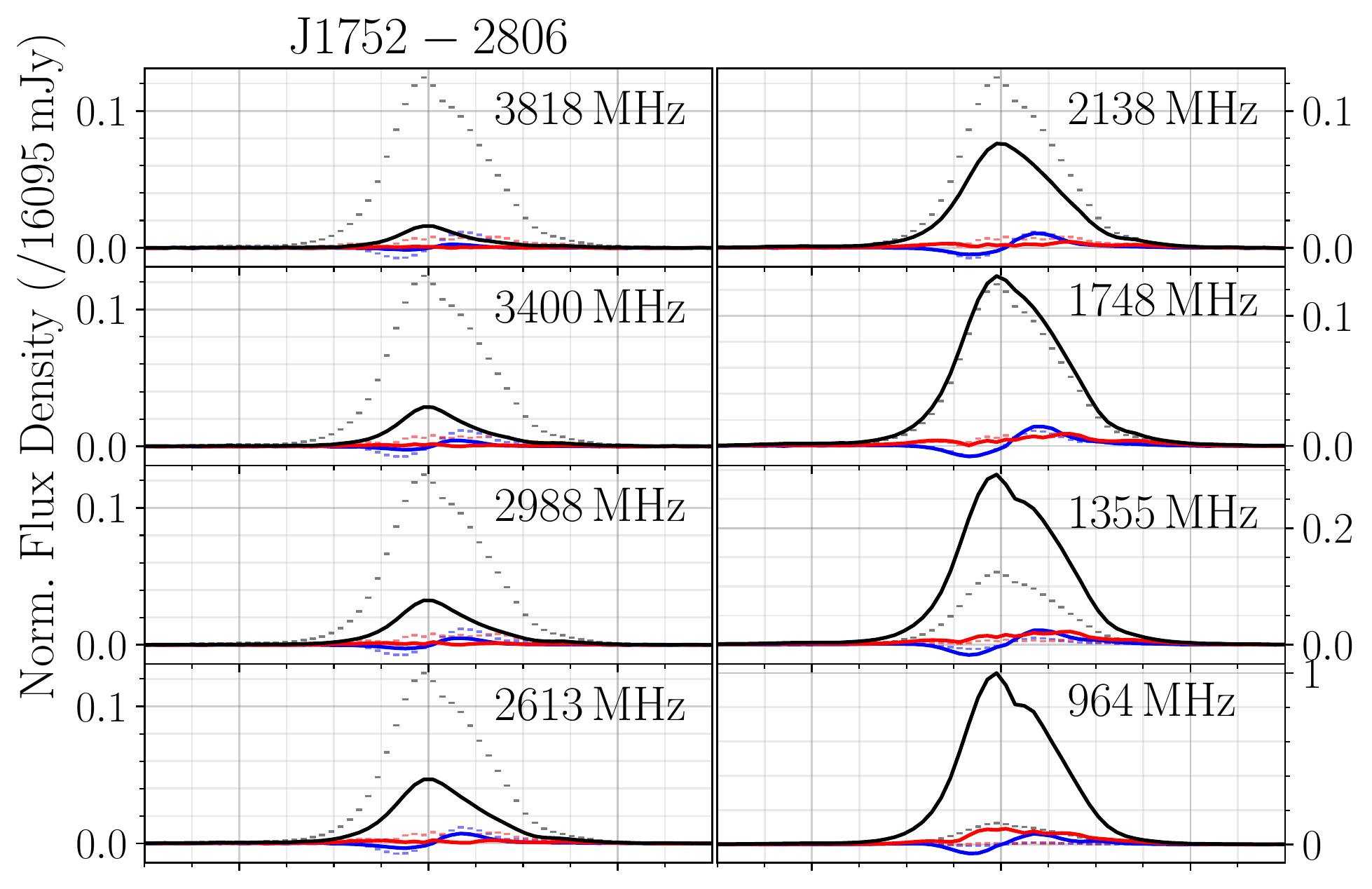}

	\includegraphics[width=\columnwidth]{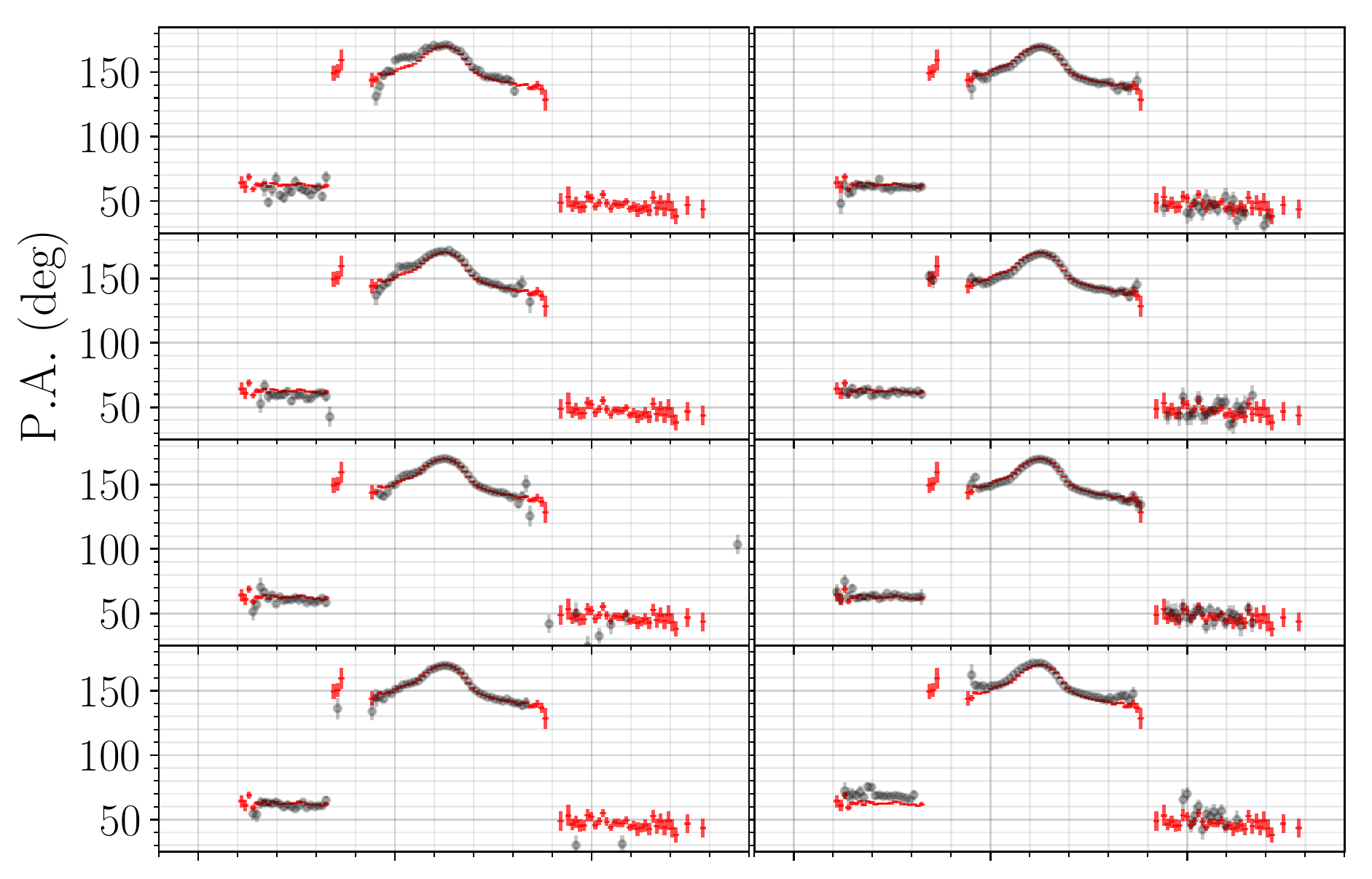}
	\includegraphics[width=\columnwidth]{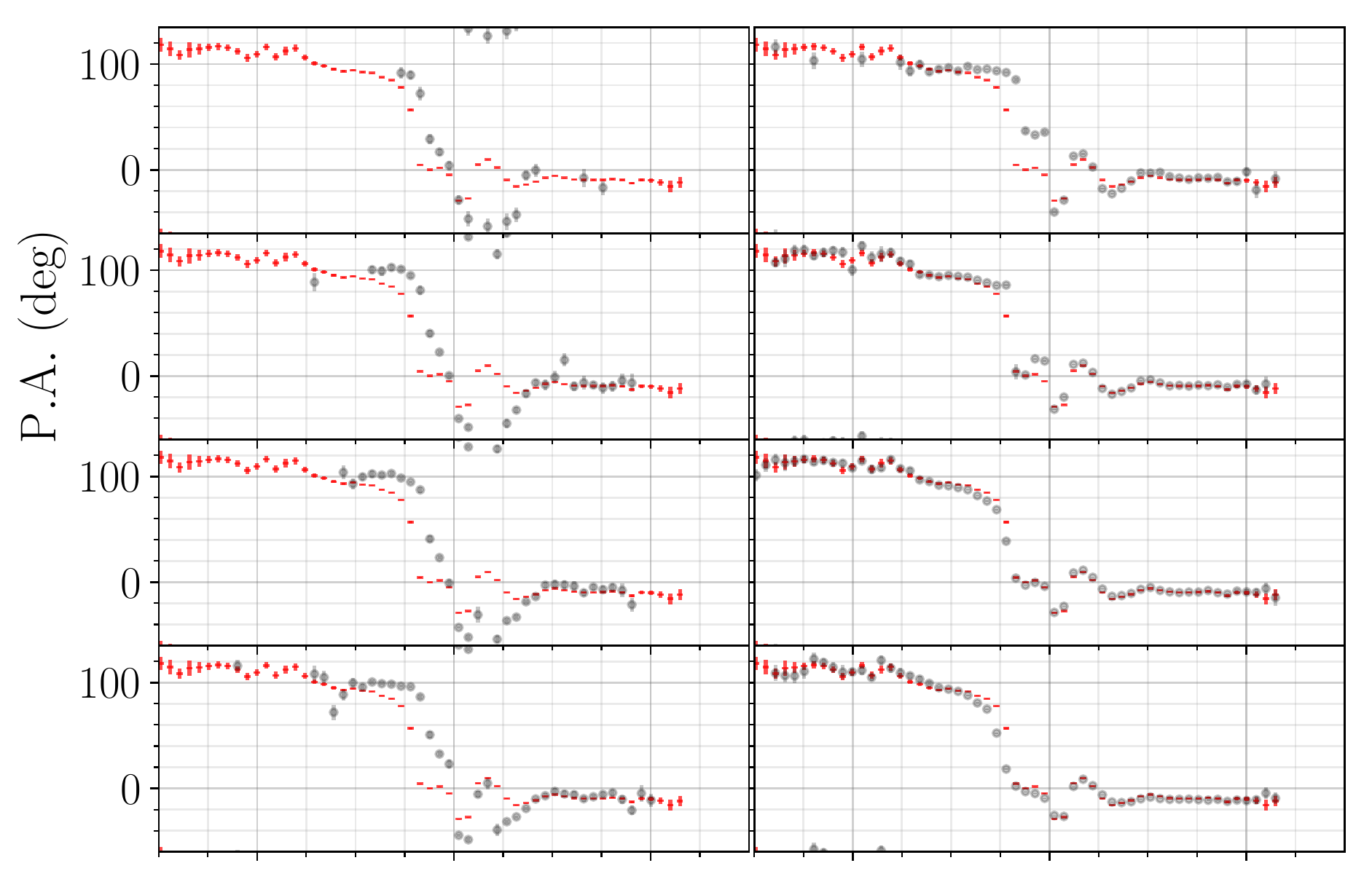}

	\includegraphics[width=\columnwidth]{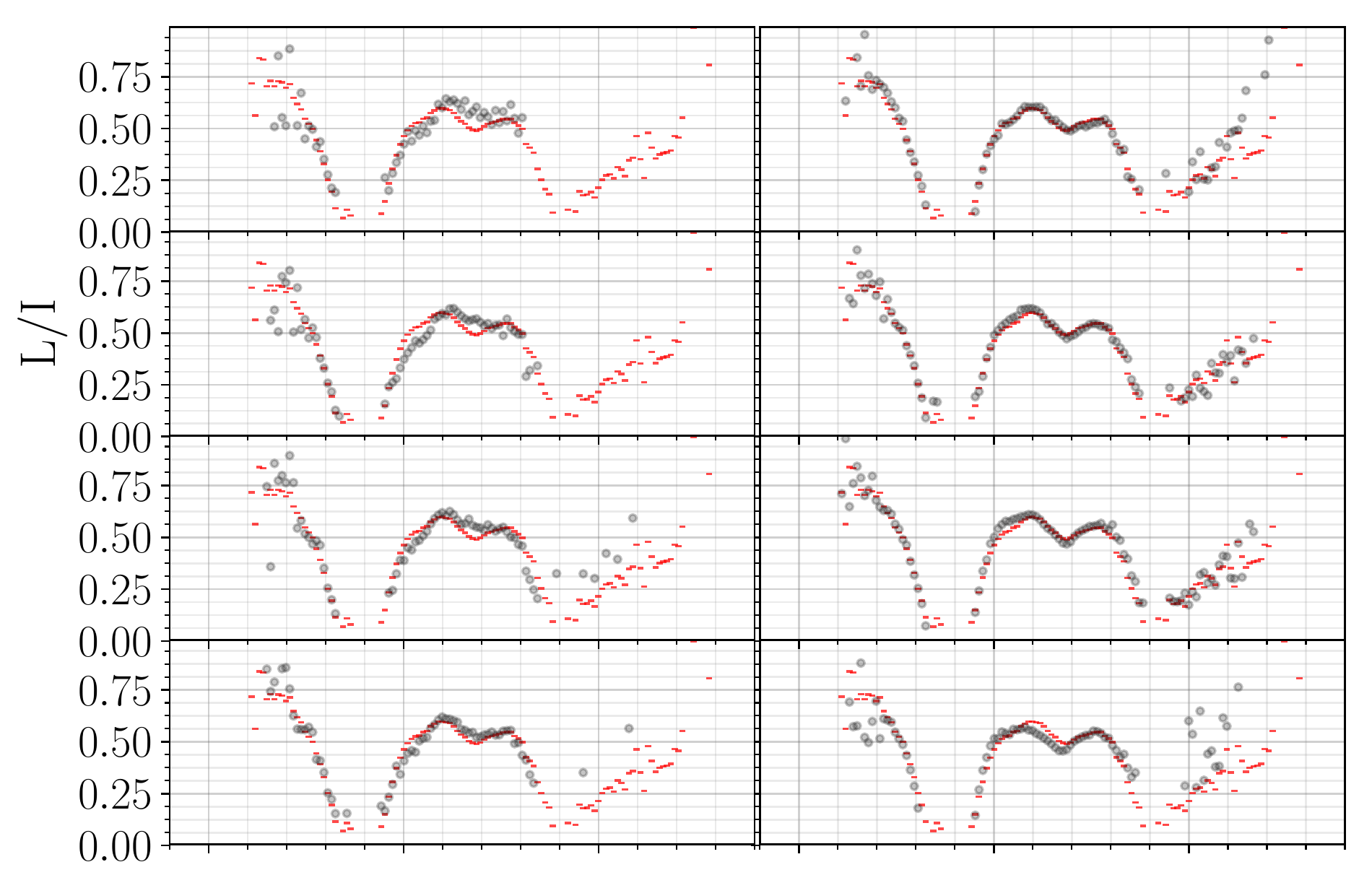}
	\includegraphics[width=\columnwidth]{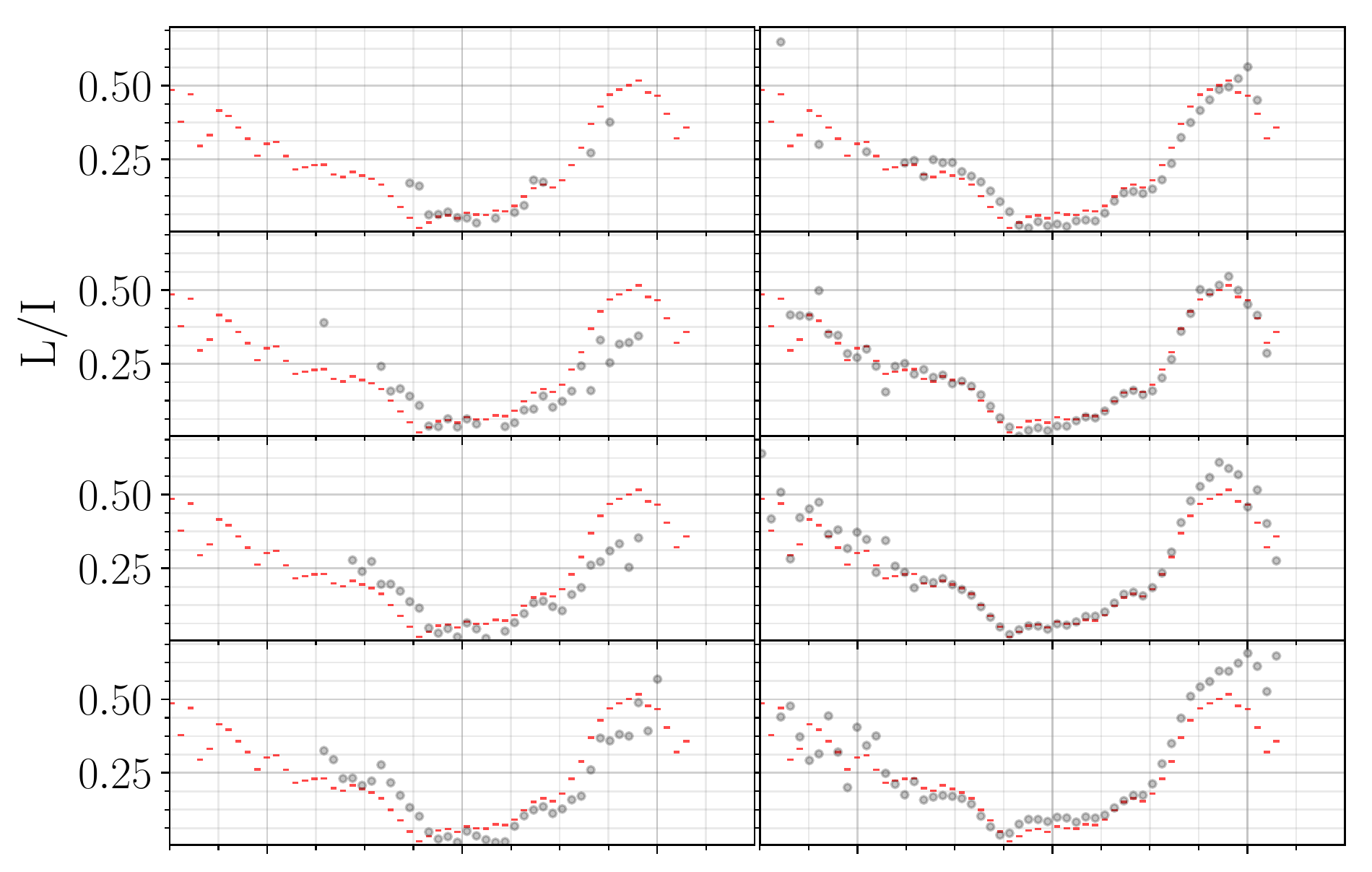}

	\includegraphics[width=\columnwidth]{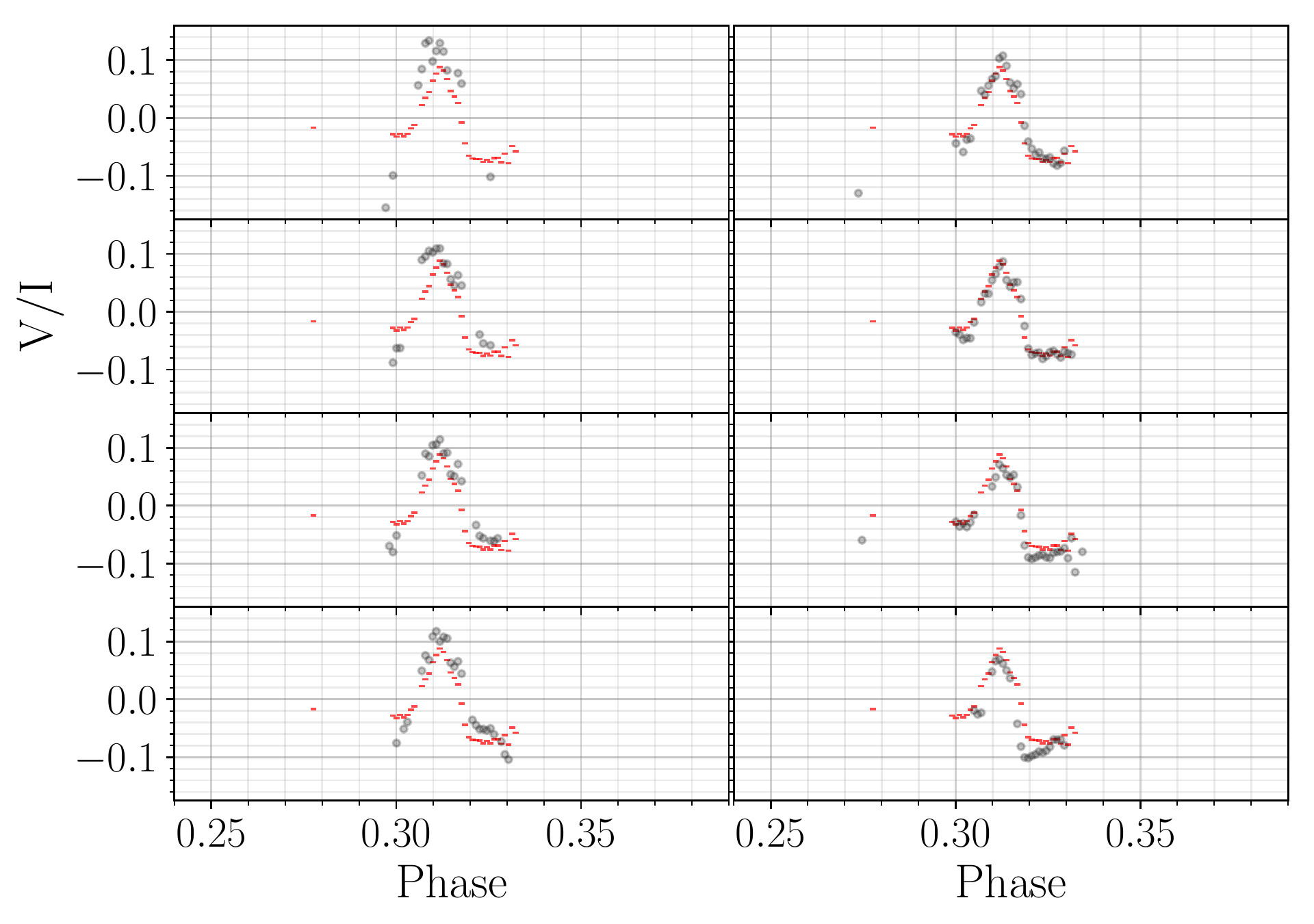}
	\includegraphics[width=\columnwidth]{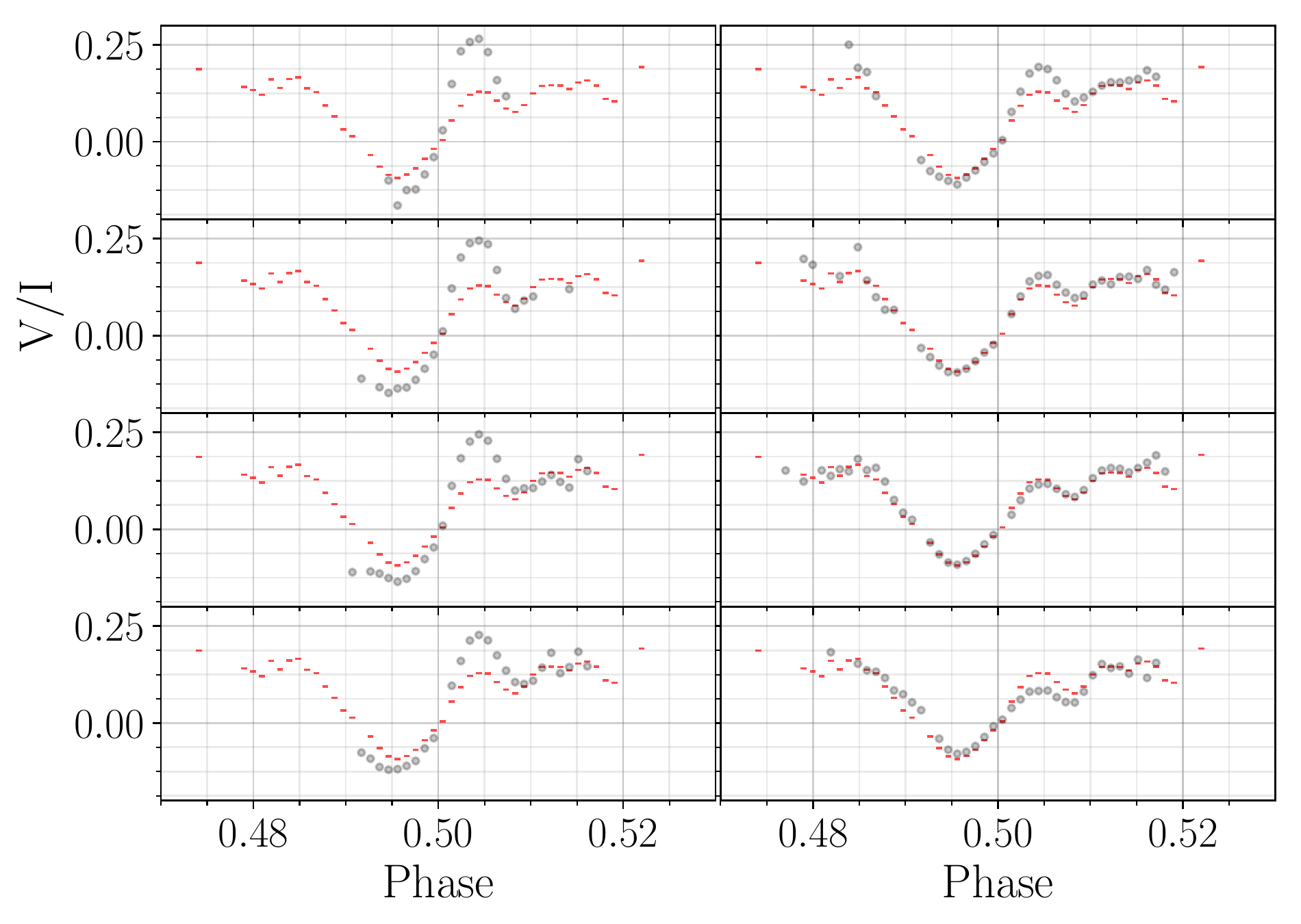}

    \caption{...Figure \ref{fig:profs1} continued...}
\end{figure*}

\begin{figure*} 
	\includegraphics[width=\columnwidth]{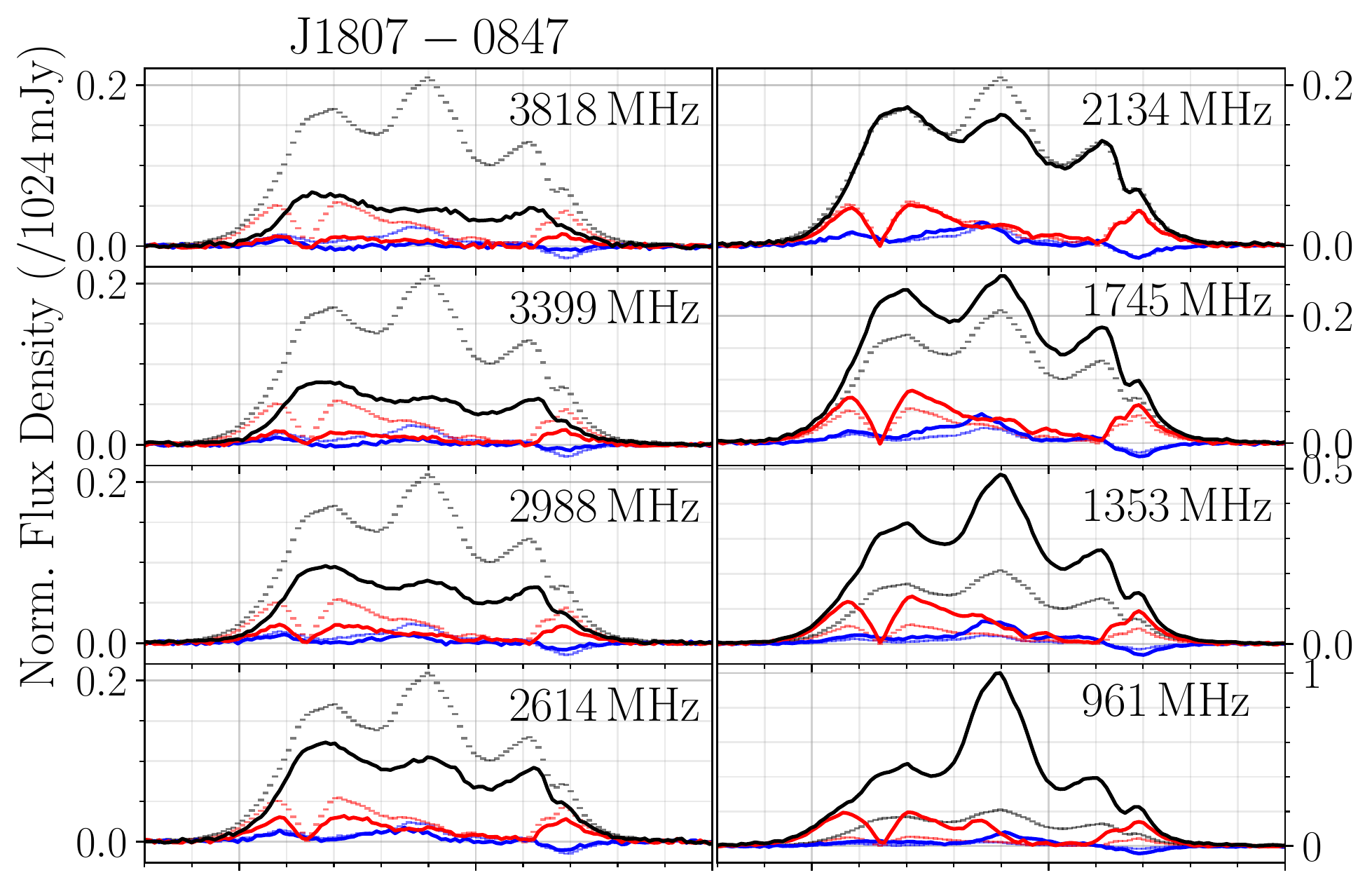}

	\includegraphics[width=\columnwidth]{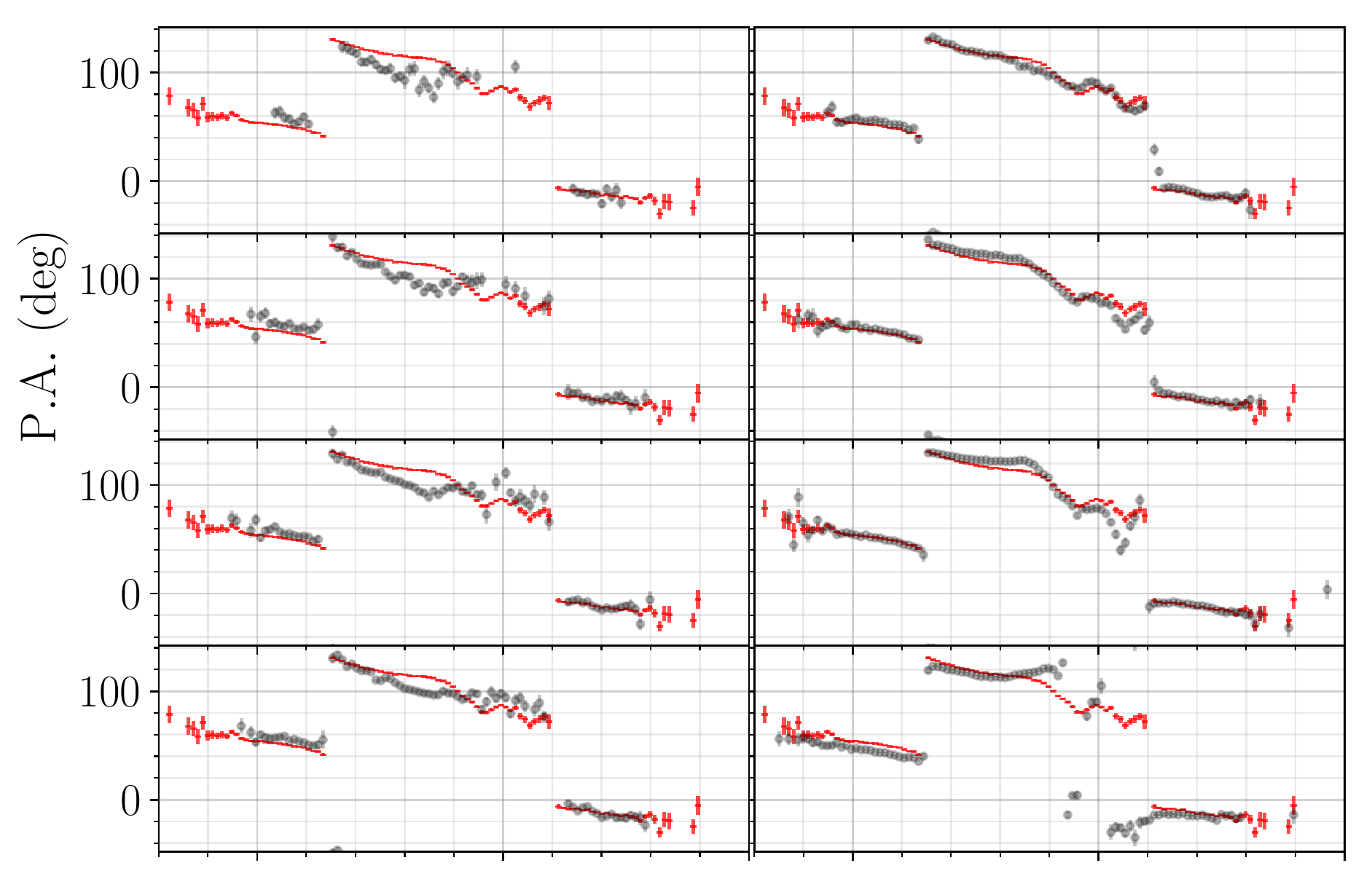}

	\includegraphics[width=\columnwidth]{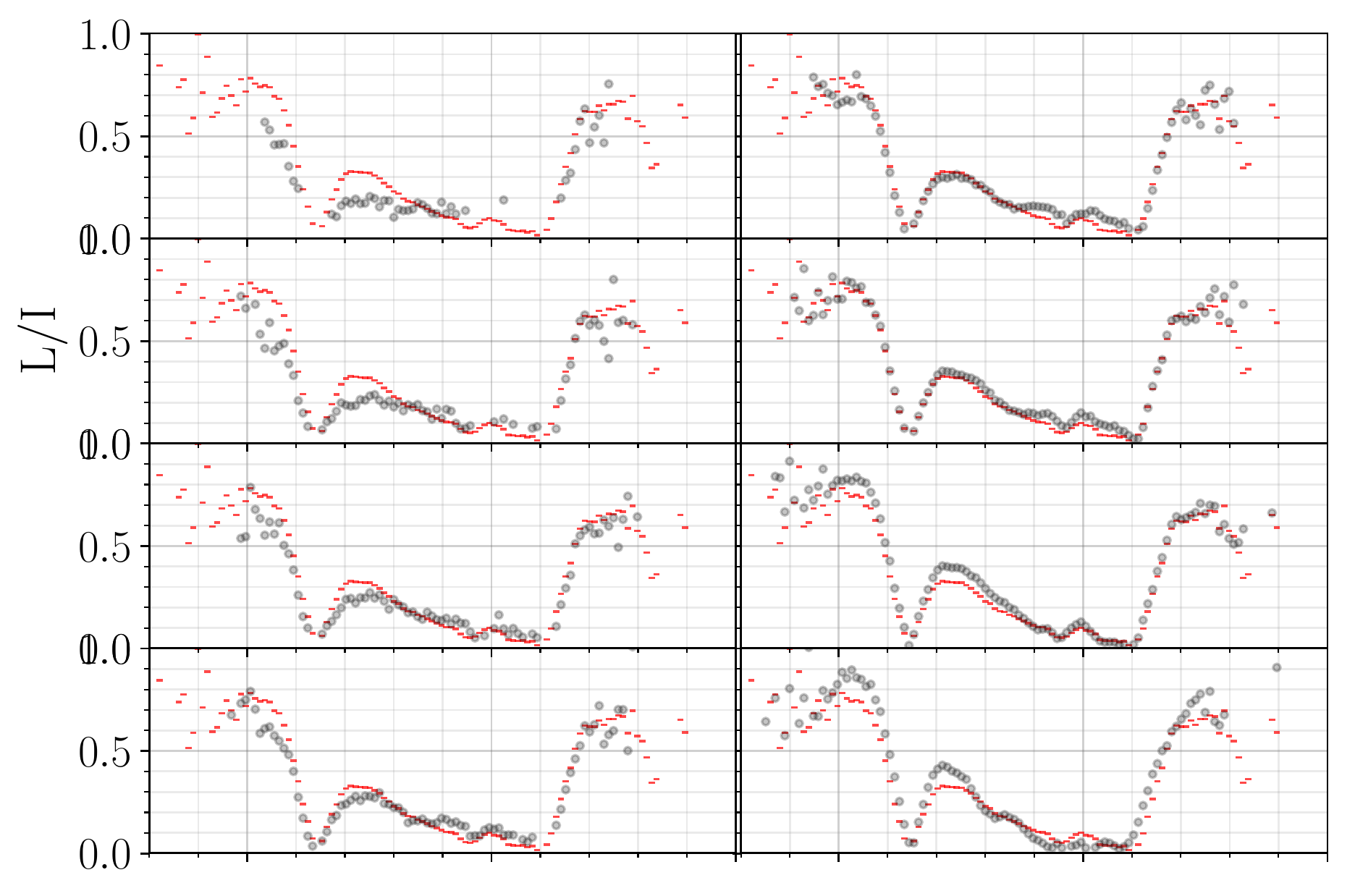}

	\includegraphics[width=\columnwidth]{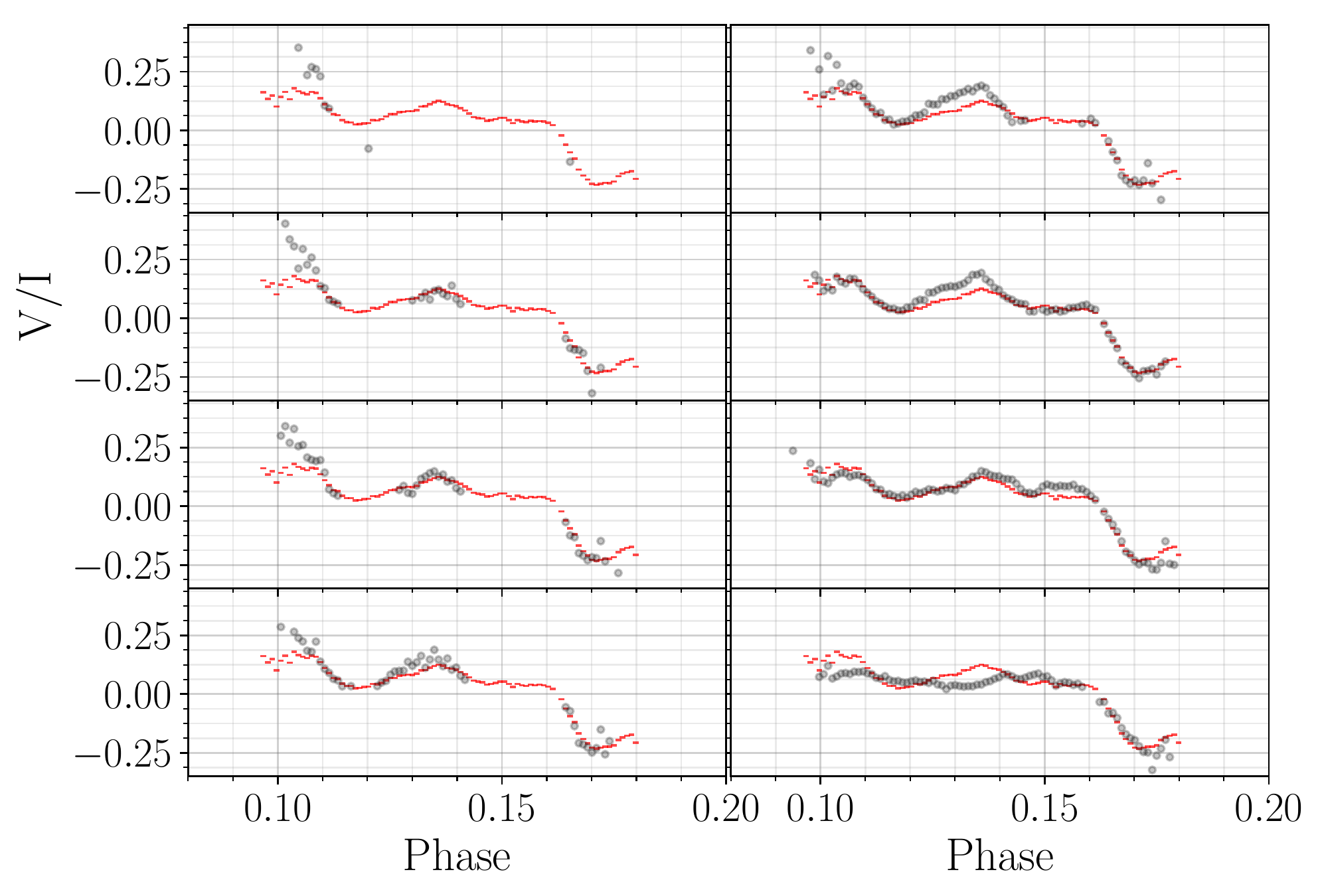}

    \caption{...Figure \ref{fig:profs1} continued...}
\end{figure*}

\begin{figure*} 
	\includegraphics[width=\columnwidth]{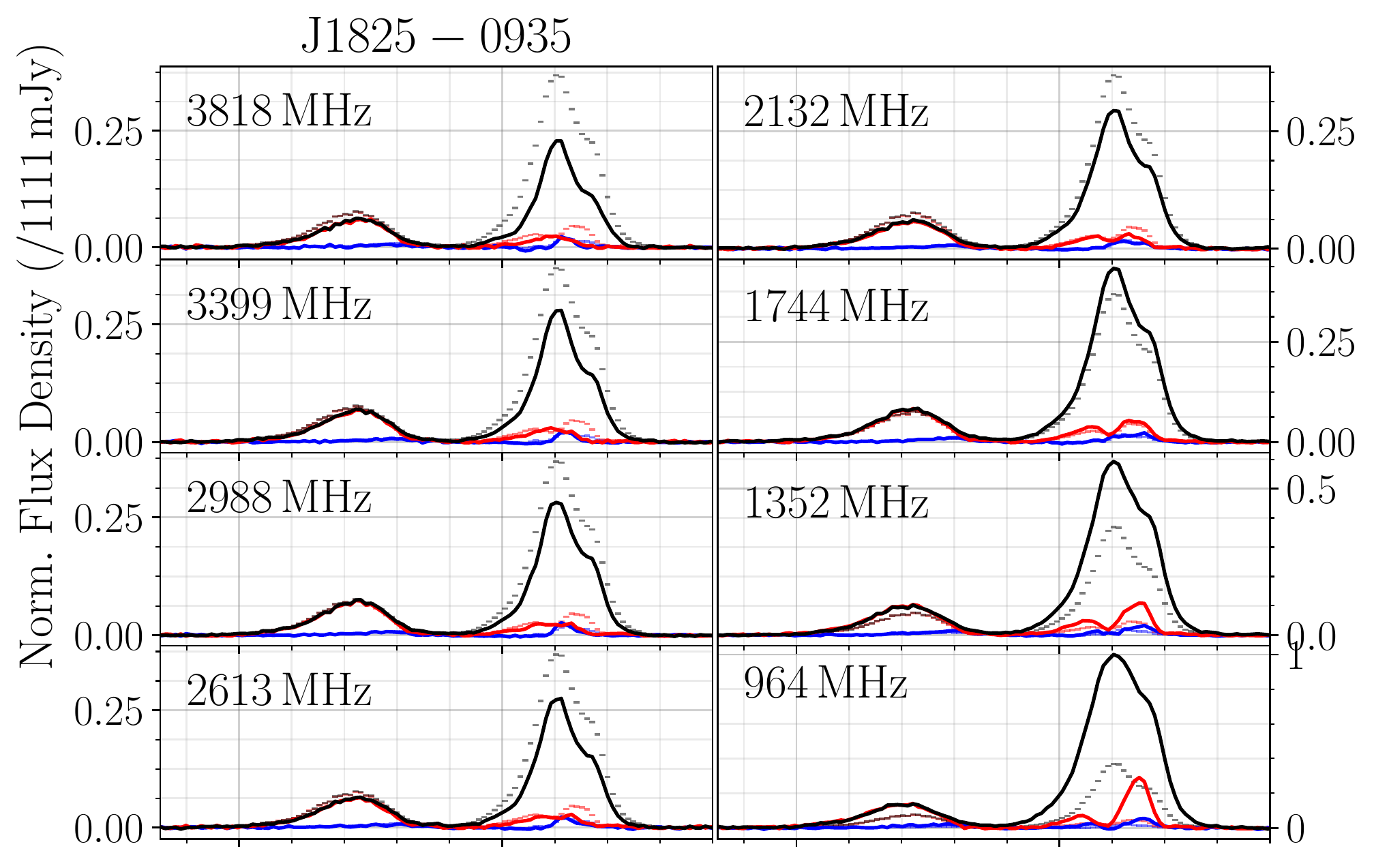}
	\includegraphics[width=\columnwidth]{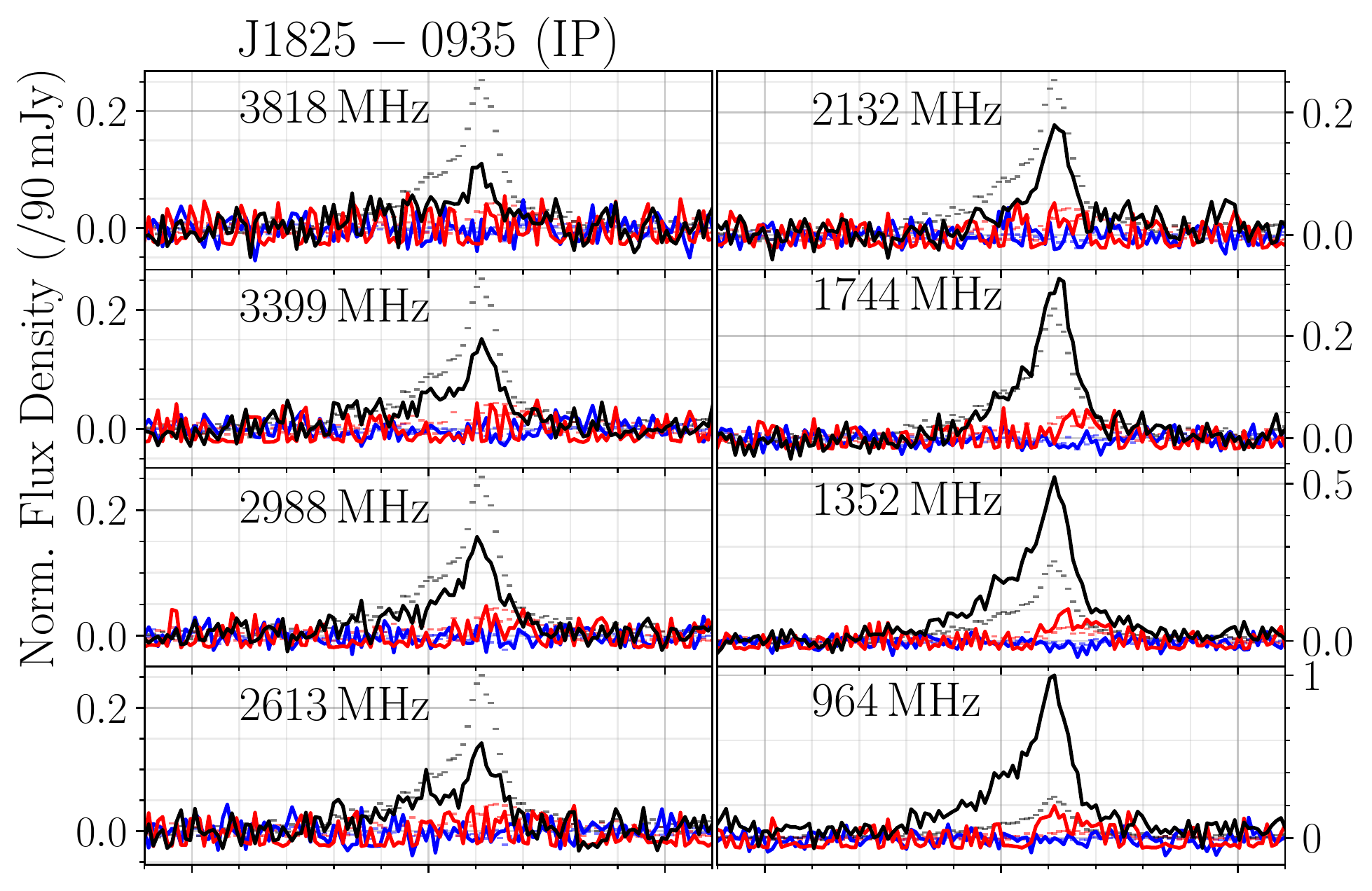}

	\includegraphics[width=\columnwidth]{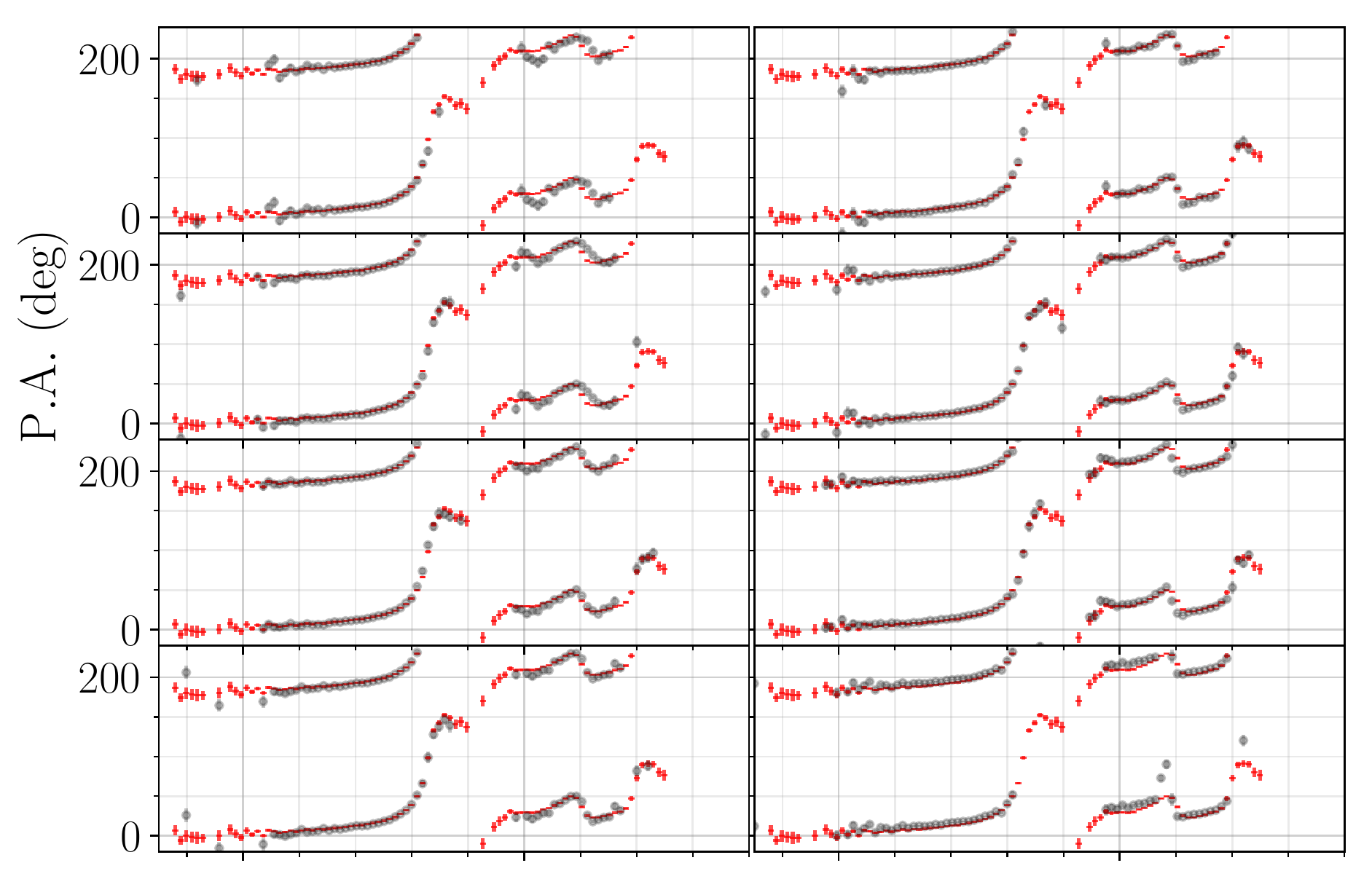}
	\includegraphics[width=\columnwidth]{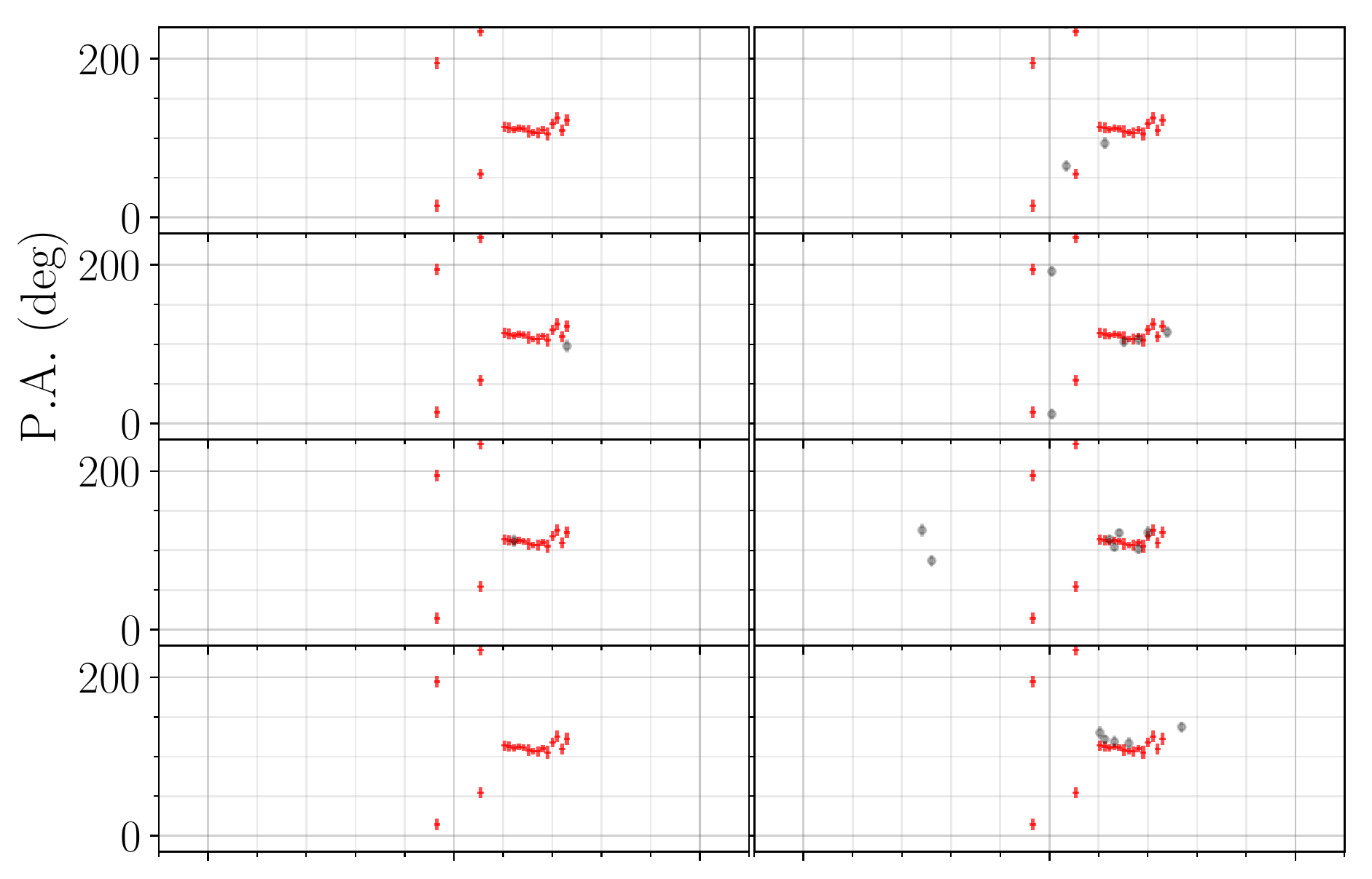}

	\includegraphics[width=\columnwidth]{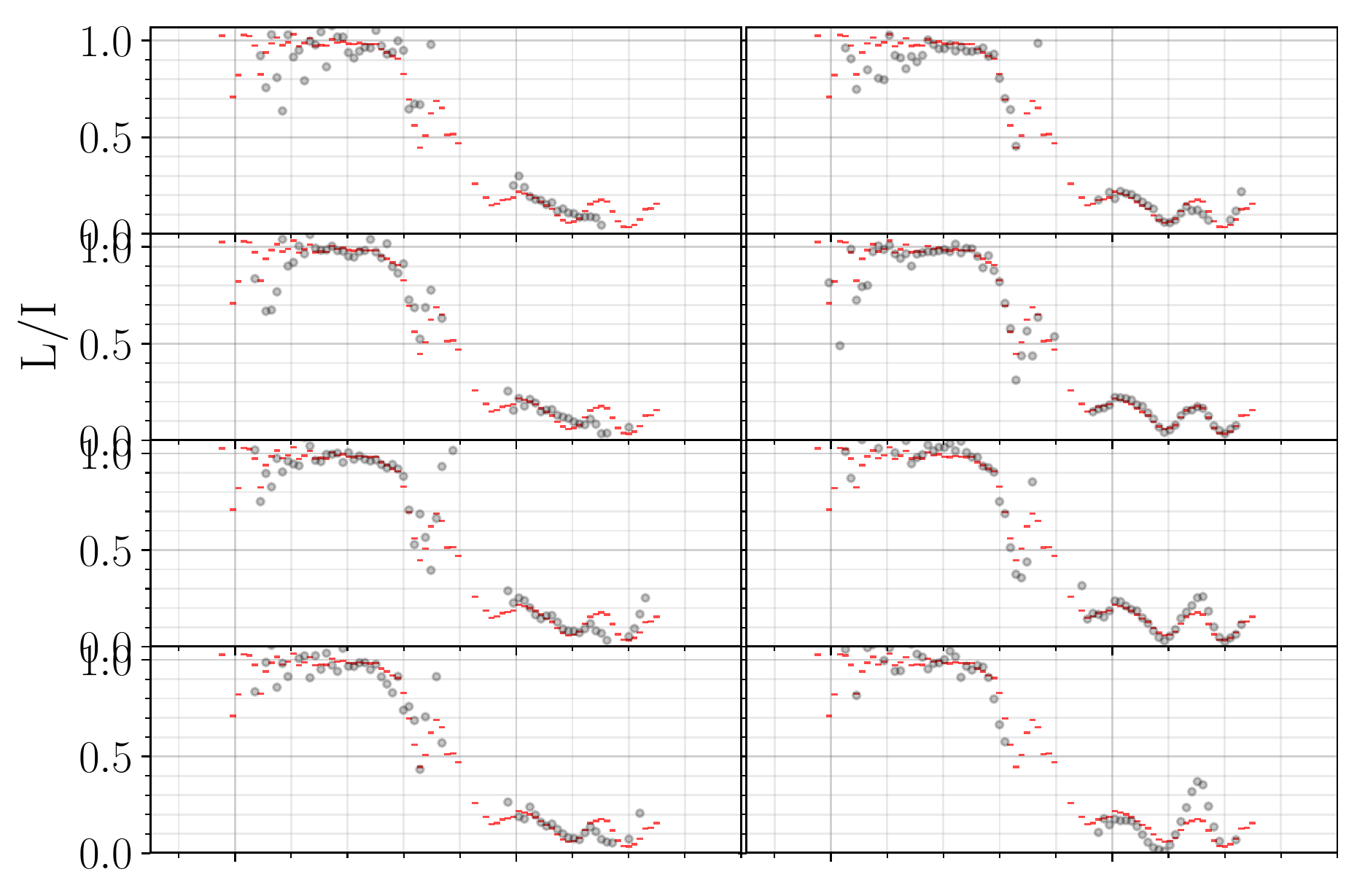}
	\includegraphics[width=\columnwidth]{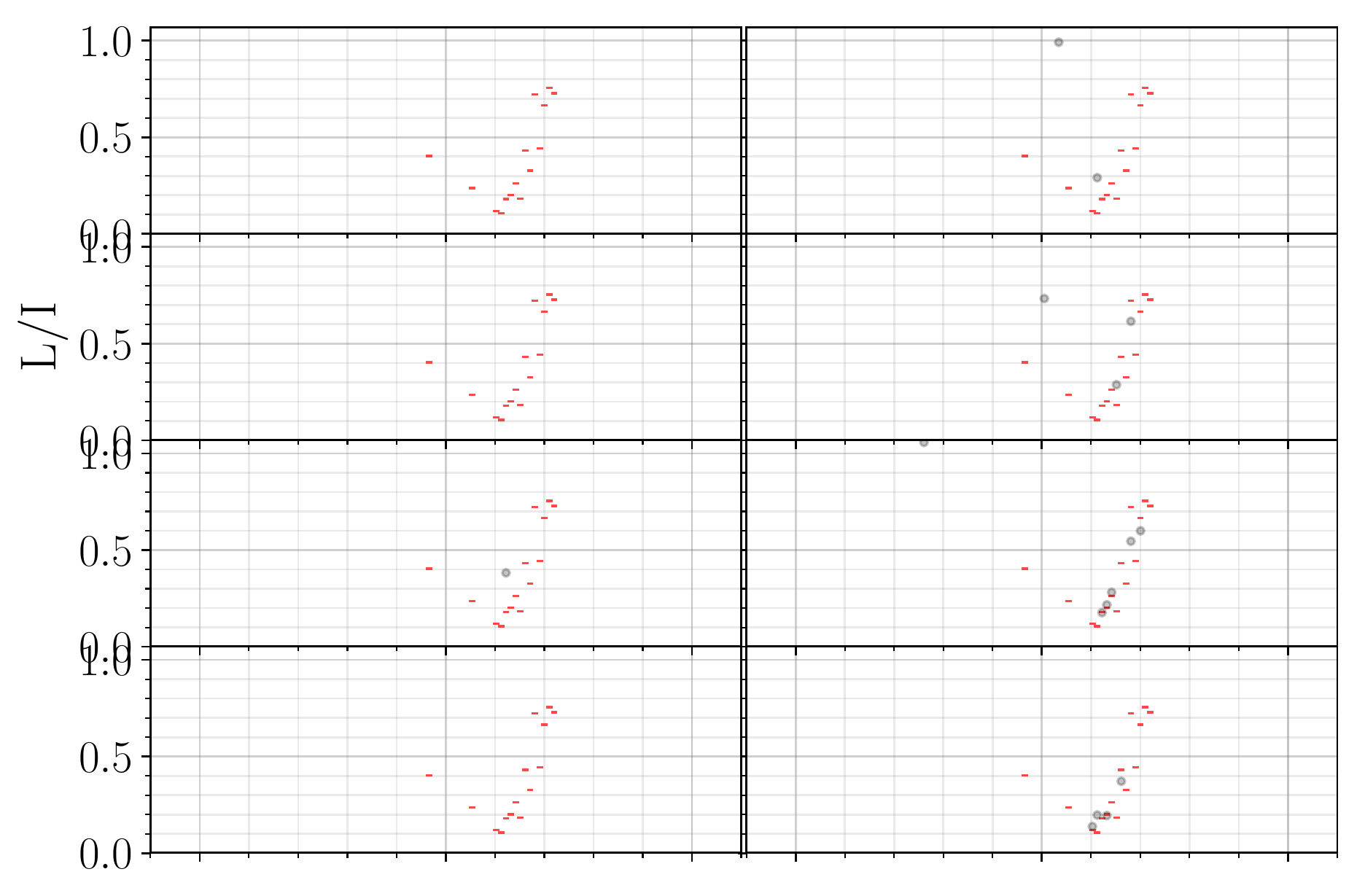}

	\includegraphics[width=\columnwidth]{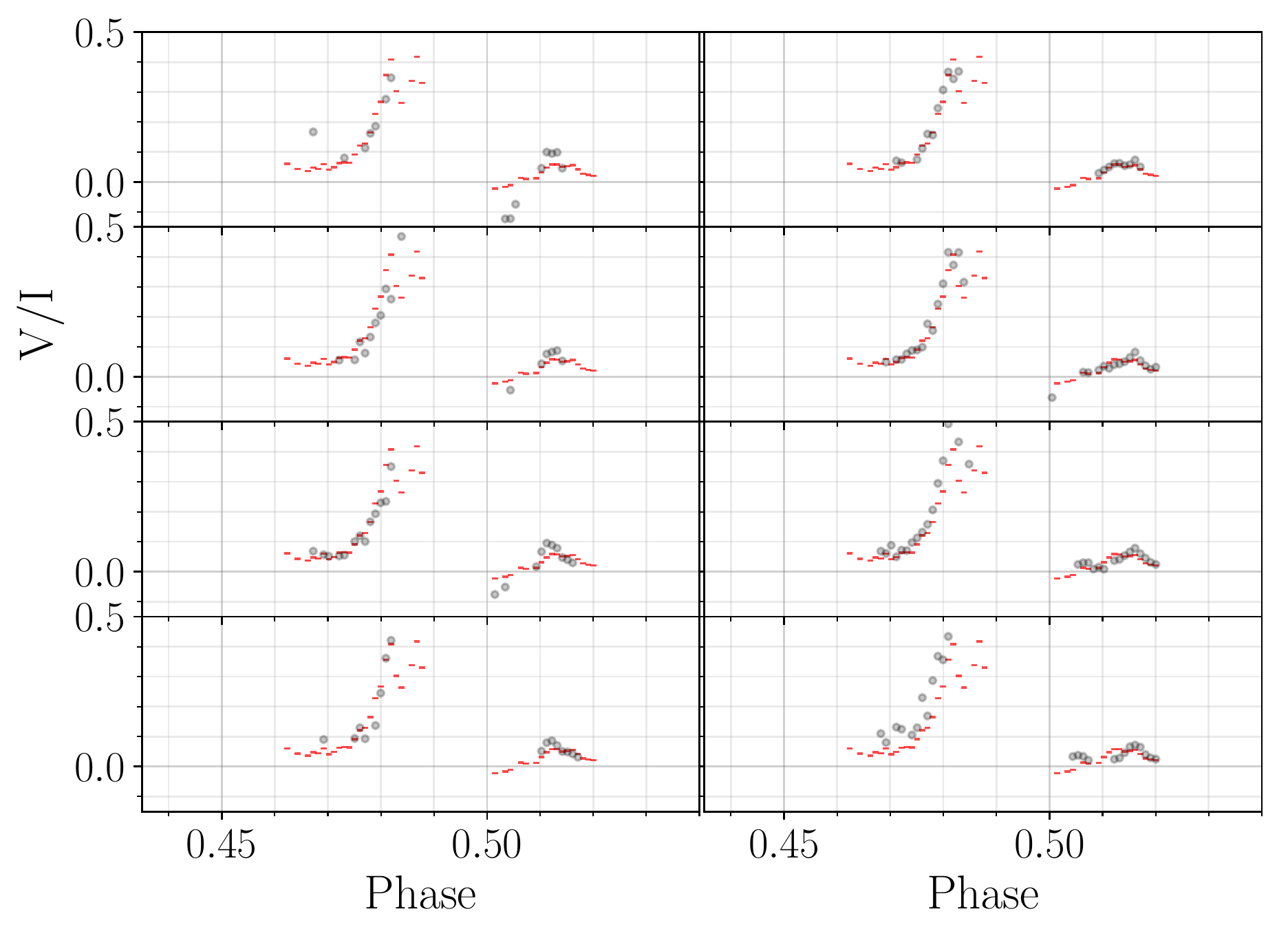}
	\includegraphics[width=\columnwidth]{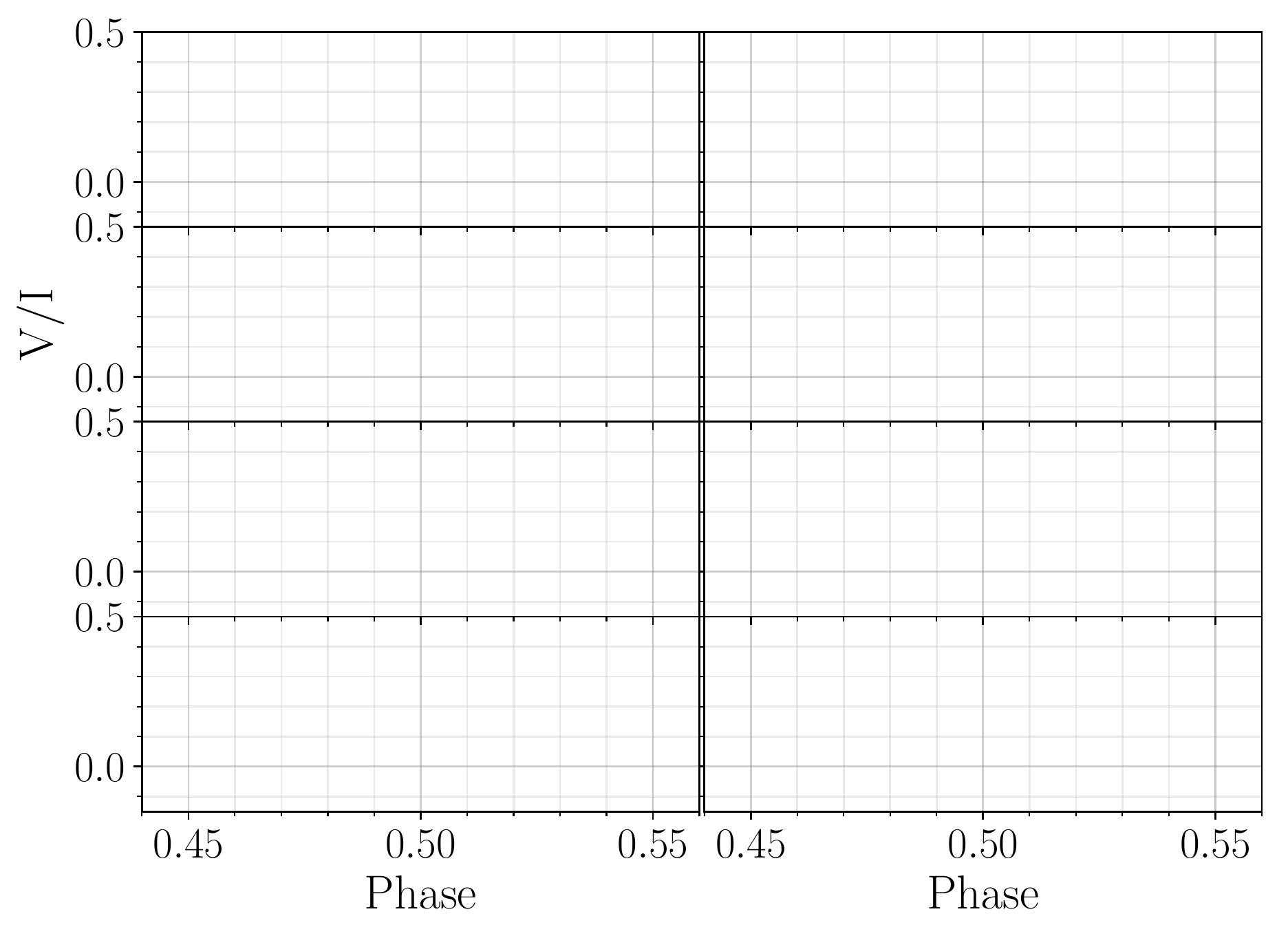}

    \caption{...Figure \ref{fig:profs1} continued...}
\end{figure*}

\begin{figure*} 

	\includegraphics[width=\columnwidth]{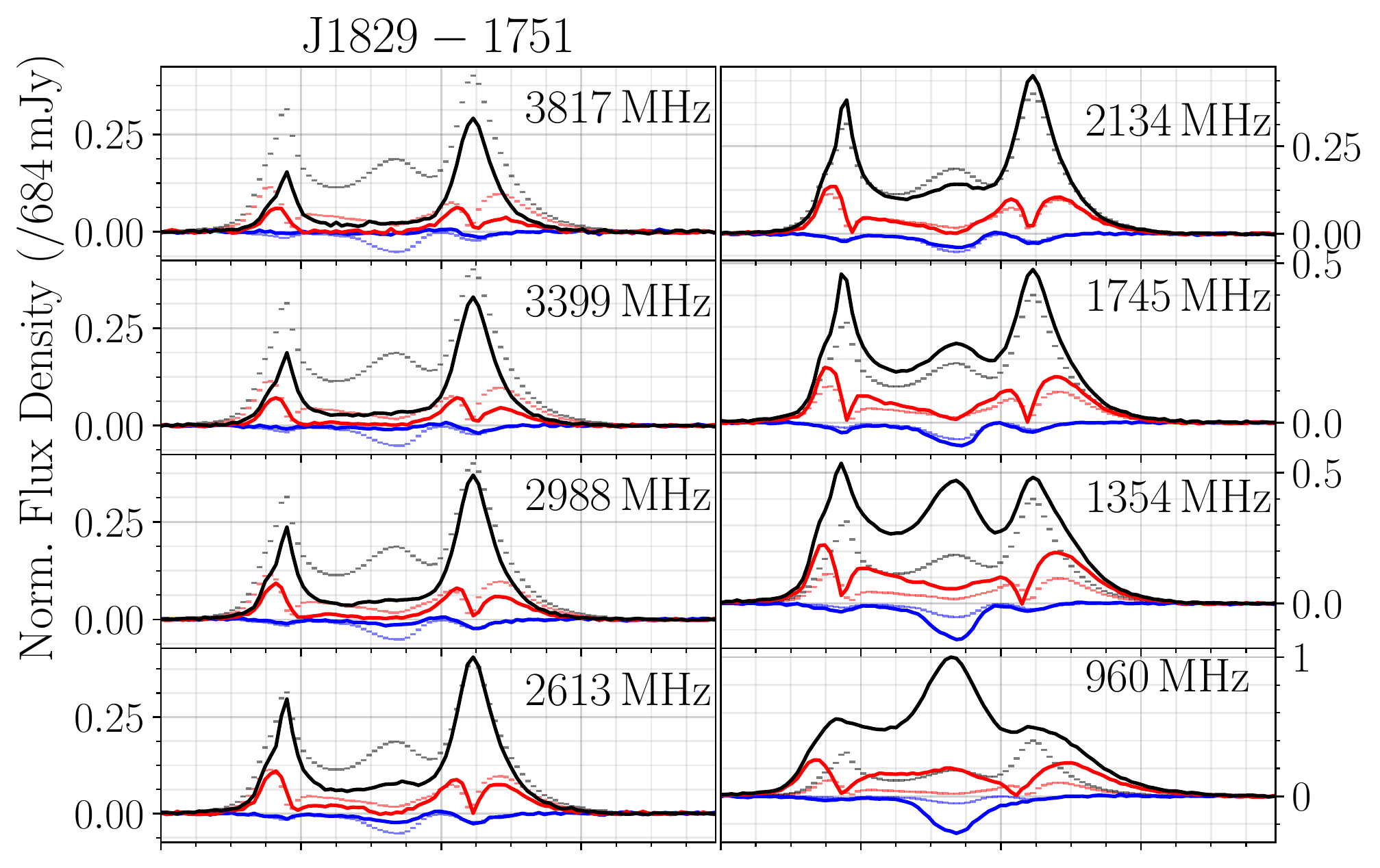}
	\includegraphics[width=\columnwidth]{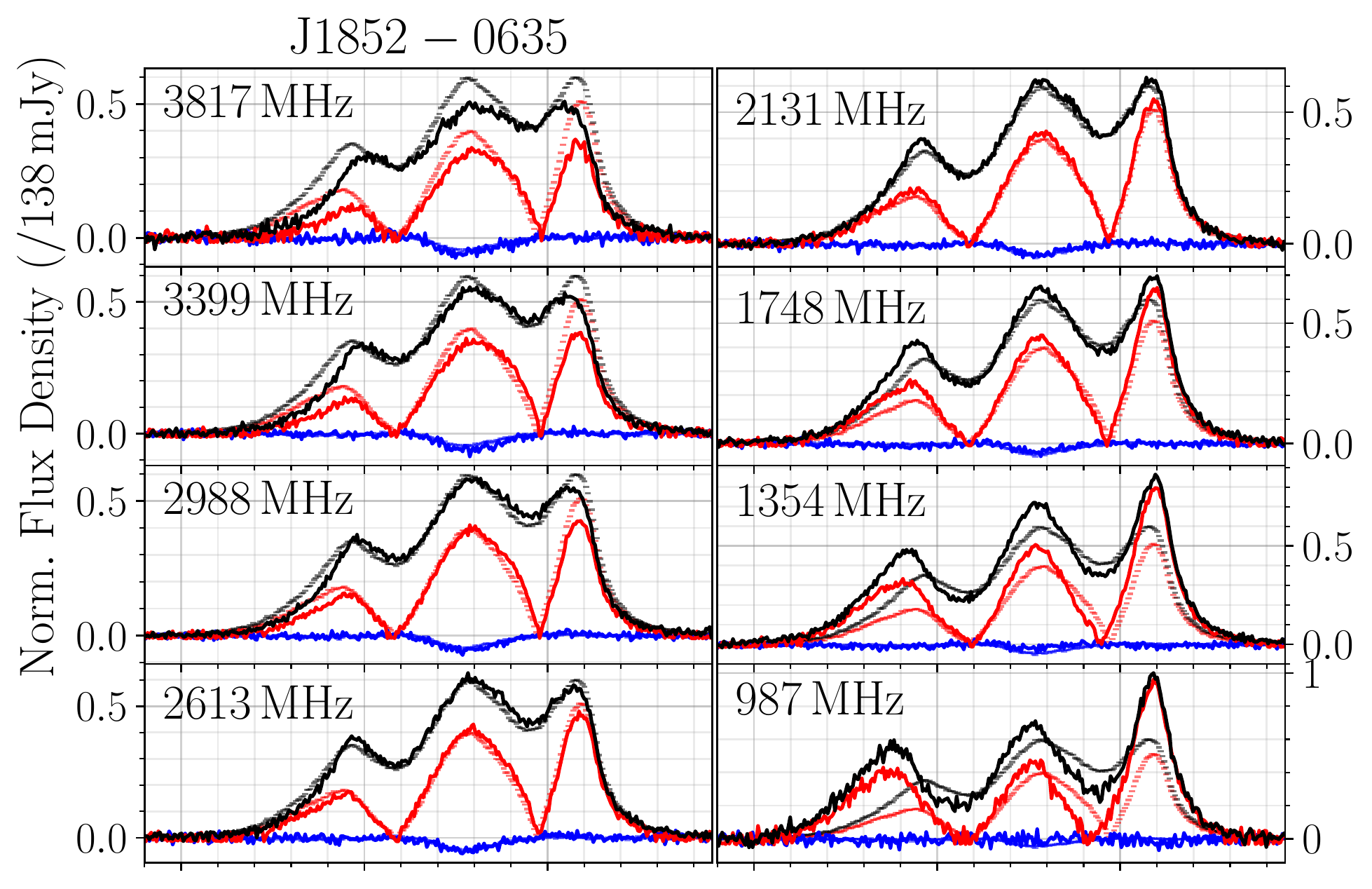}

	\includegraphics[width=\columnwidth]{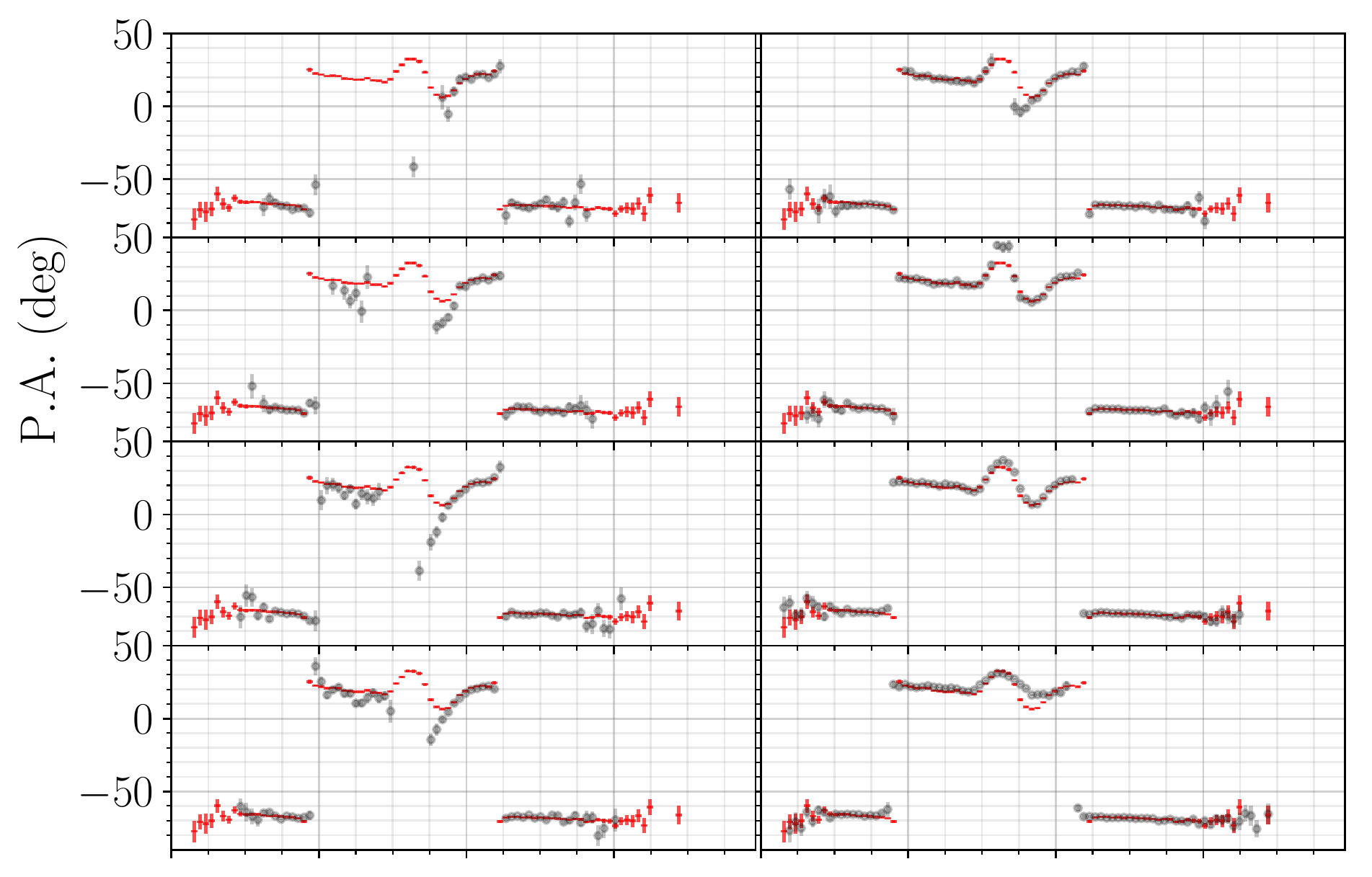}
	\includegraphics[width=\columnwidth]{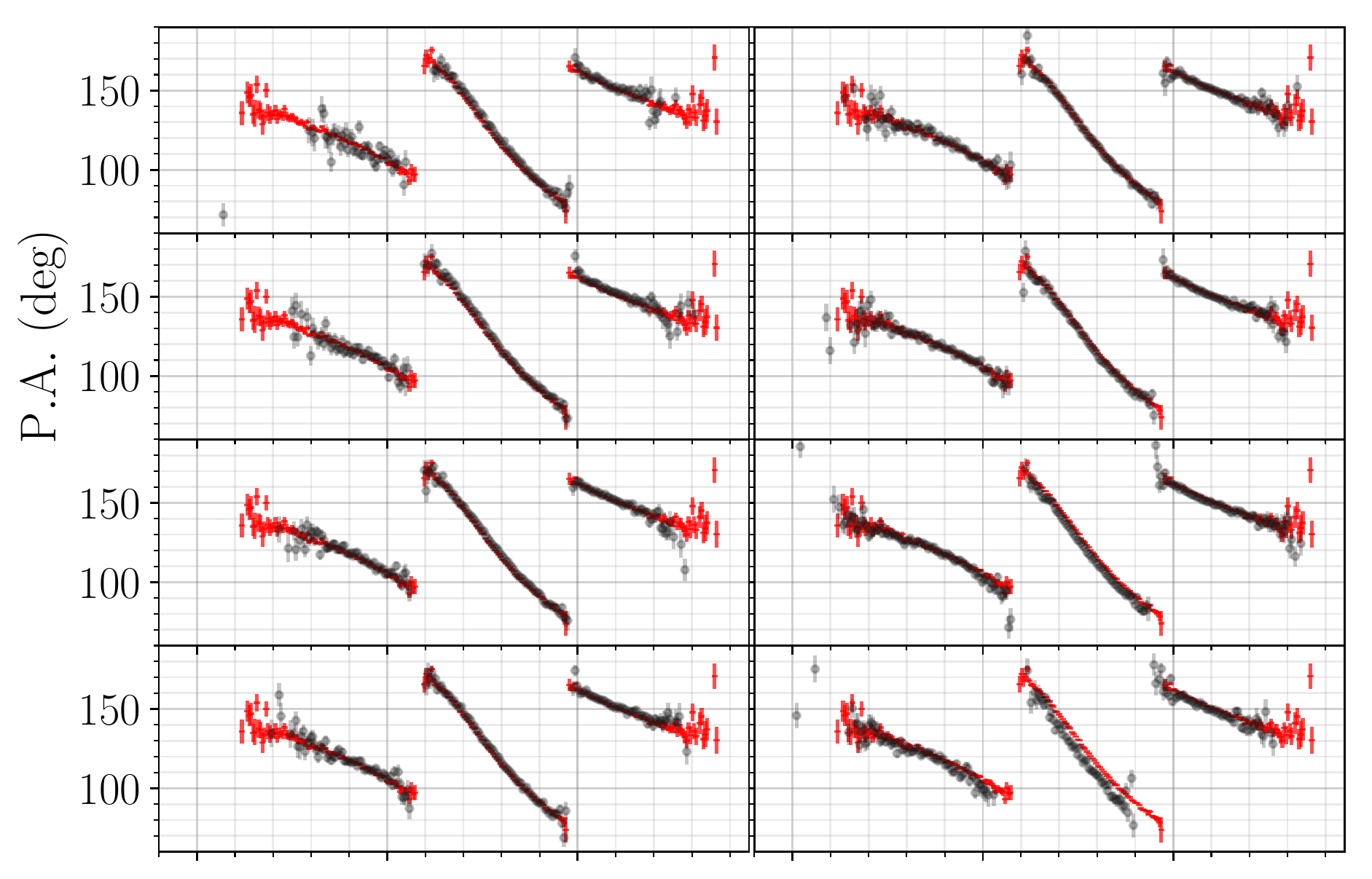}

	\includegraphics[width=\columnwidth]{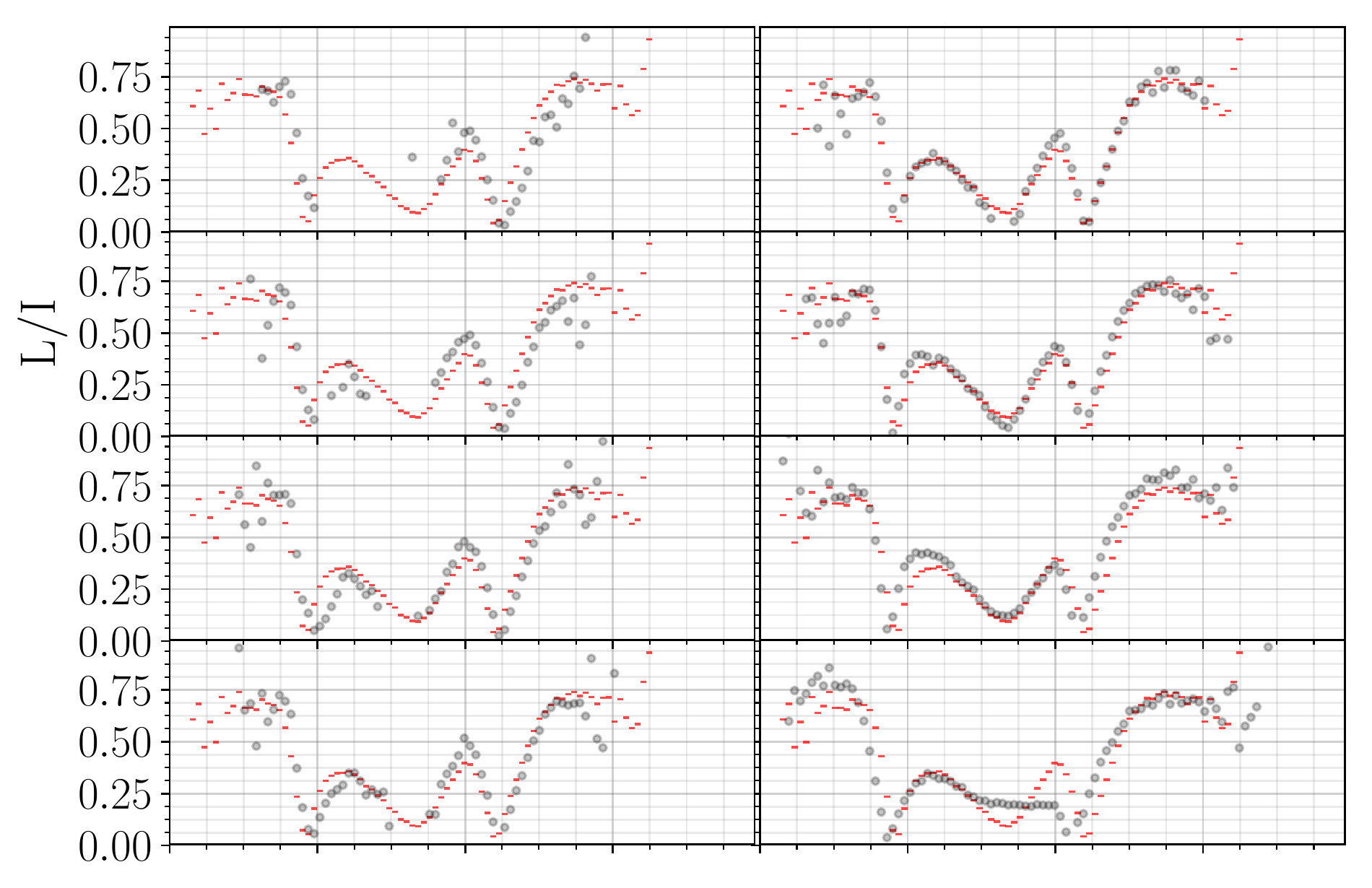}
	\includegraphics[width=\columnwidth]{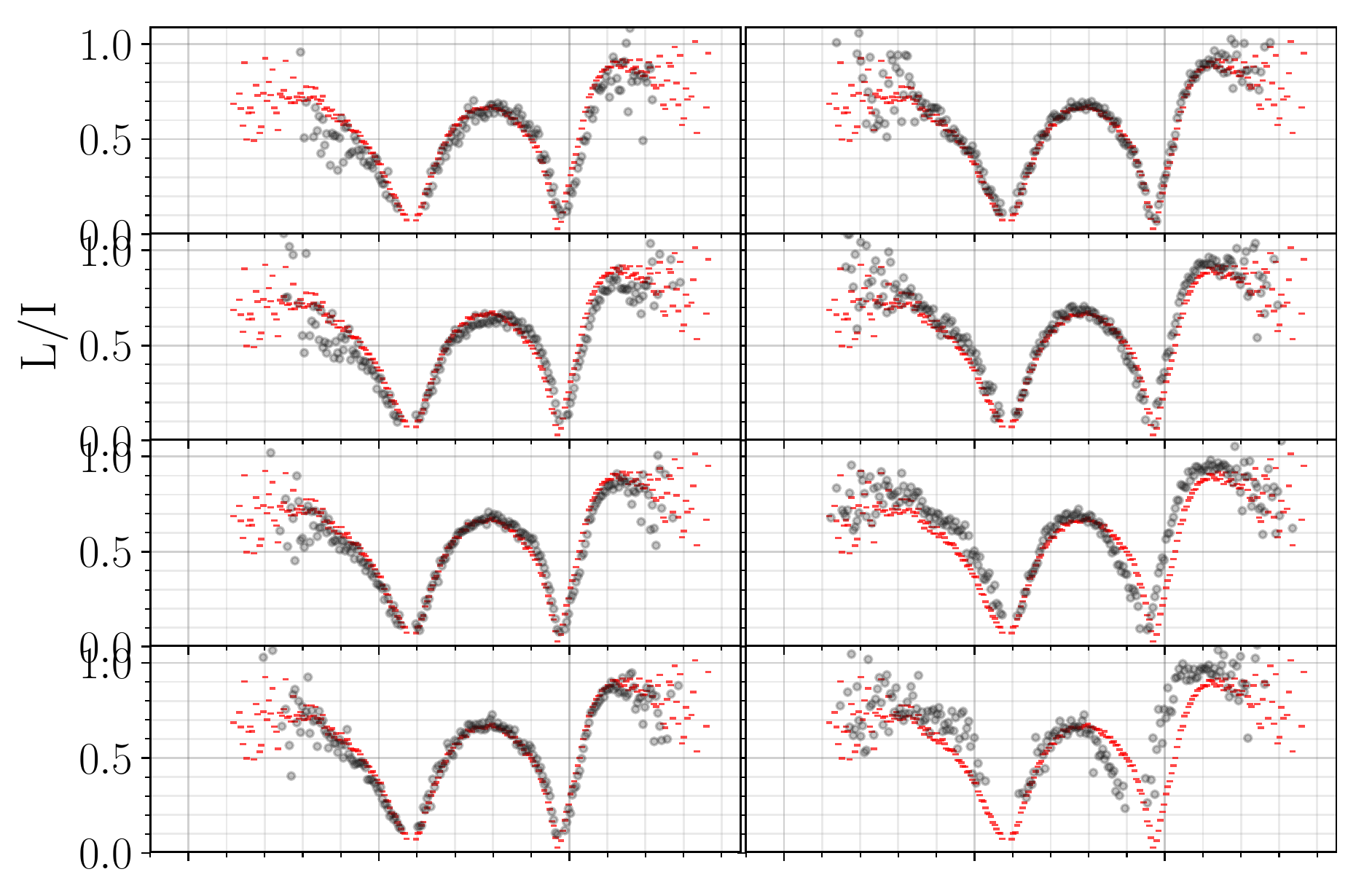}

	\includegraphics[width=\columnwidth]{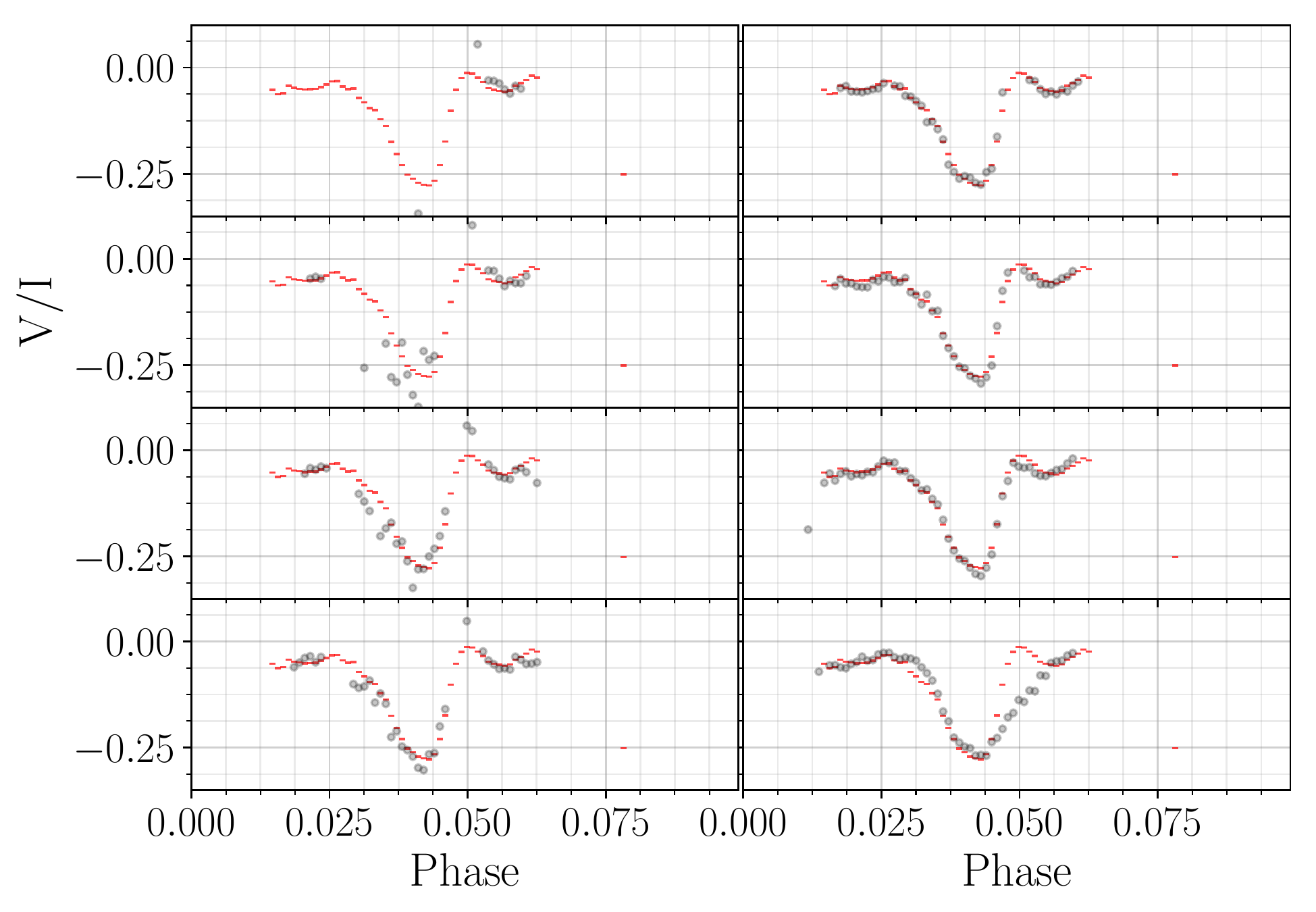}
	\includegraphics[width=\columnwidth]{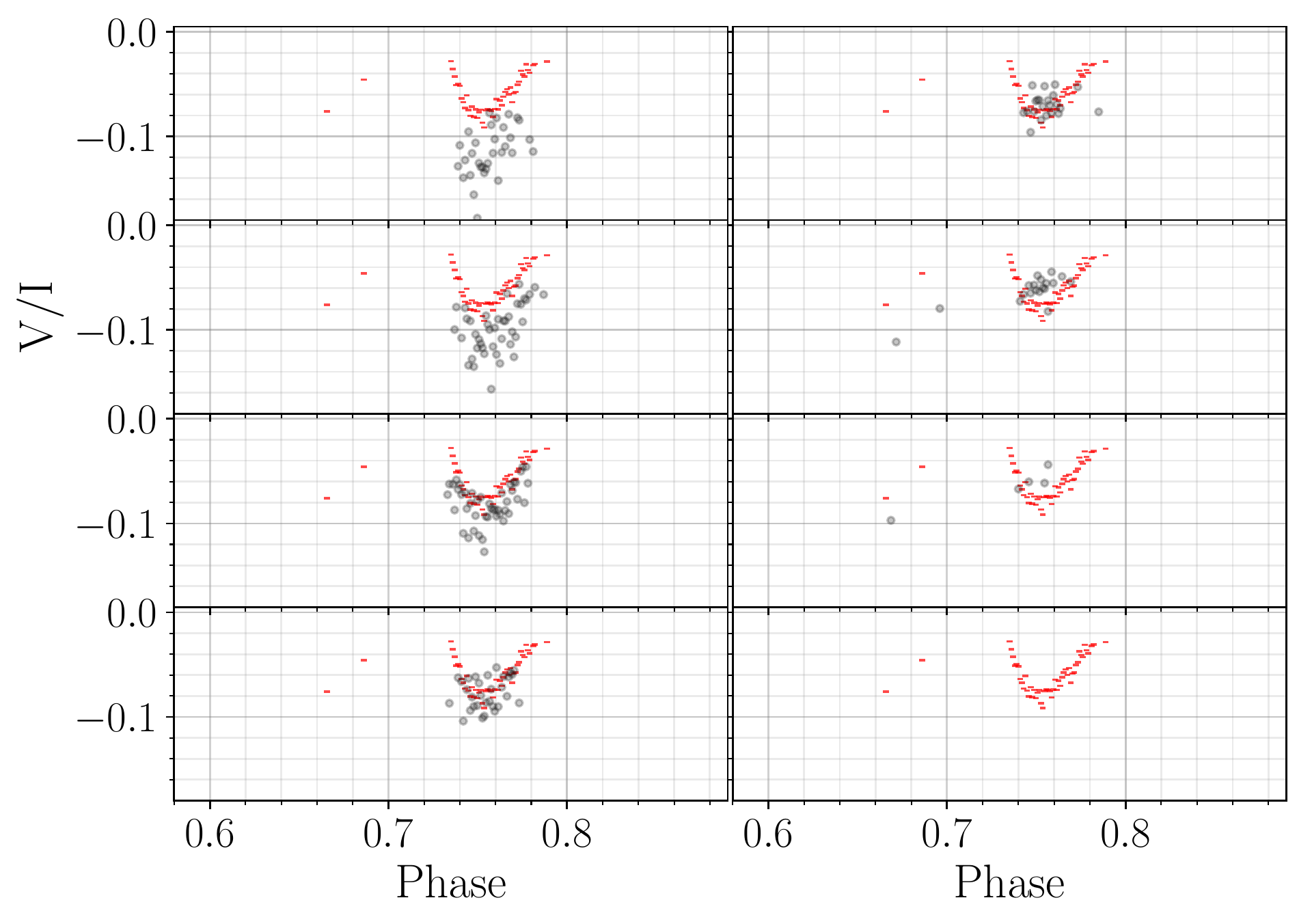}

    \caption{...Figure \ref{fig:profs1} continued.}
\end{figure*}

\begin{figure*} 

	\includegraphics[width=\columnwidth]{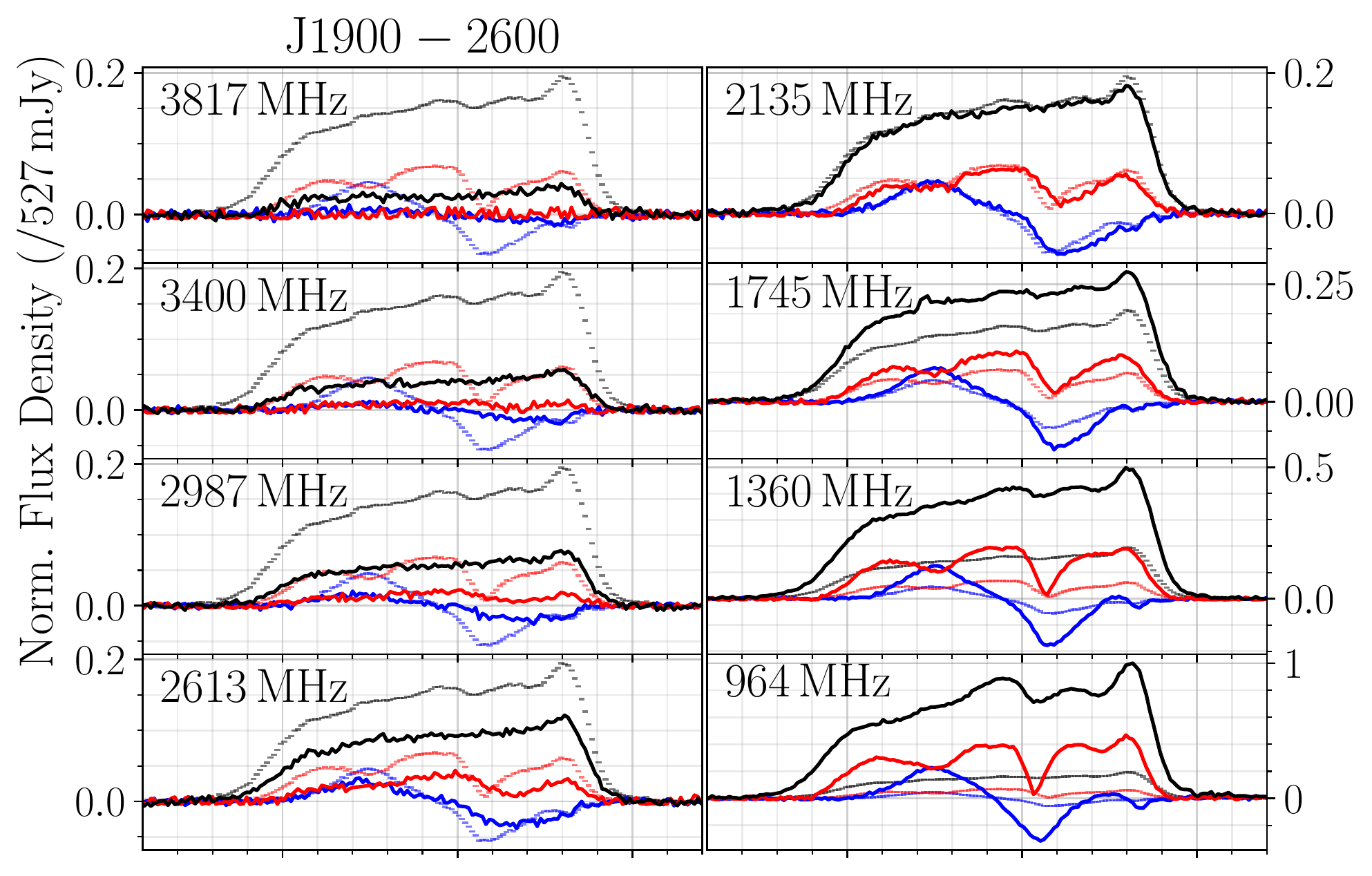}
	\includegraphics[width=\columnwidth]{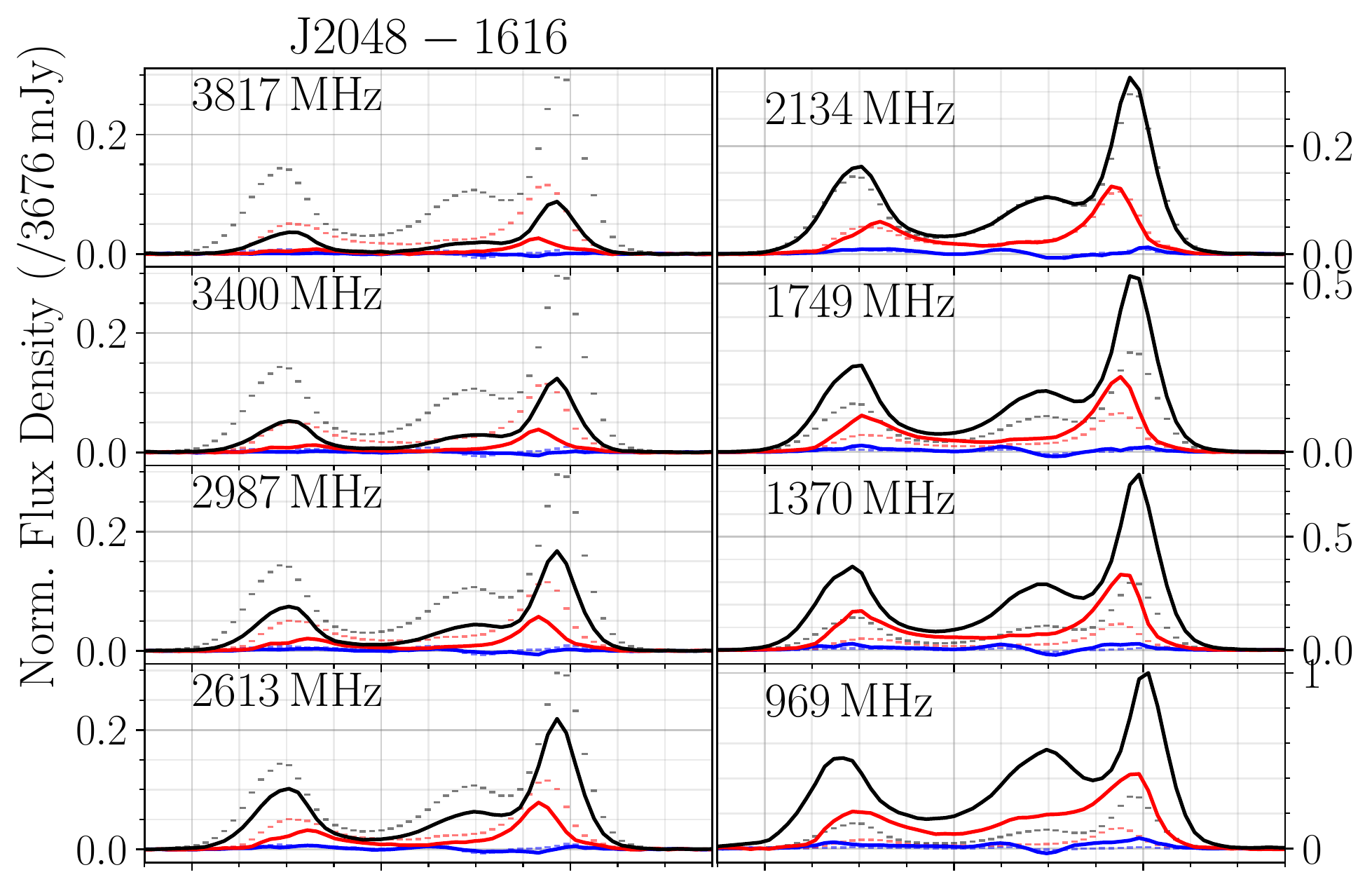}

	\includegraphics[width=\columnwidth]{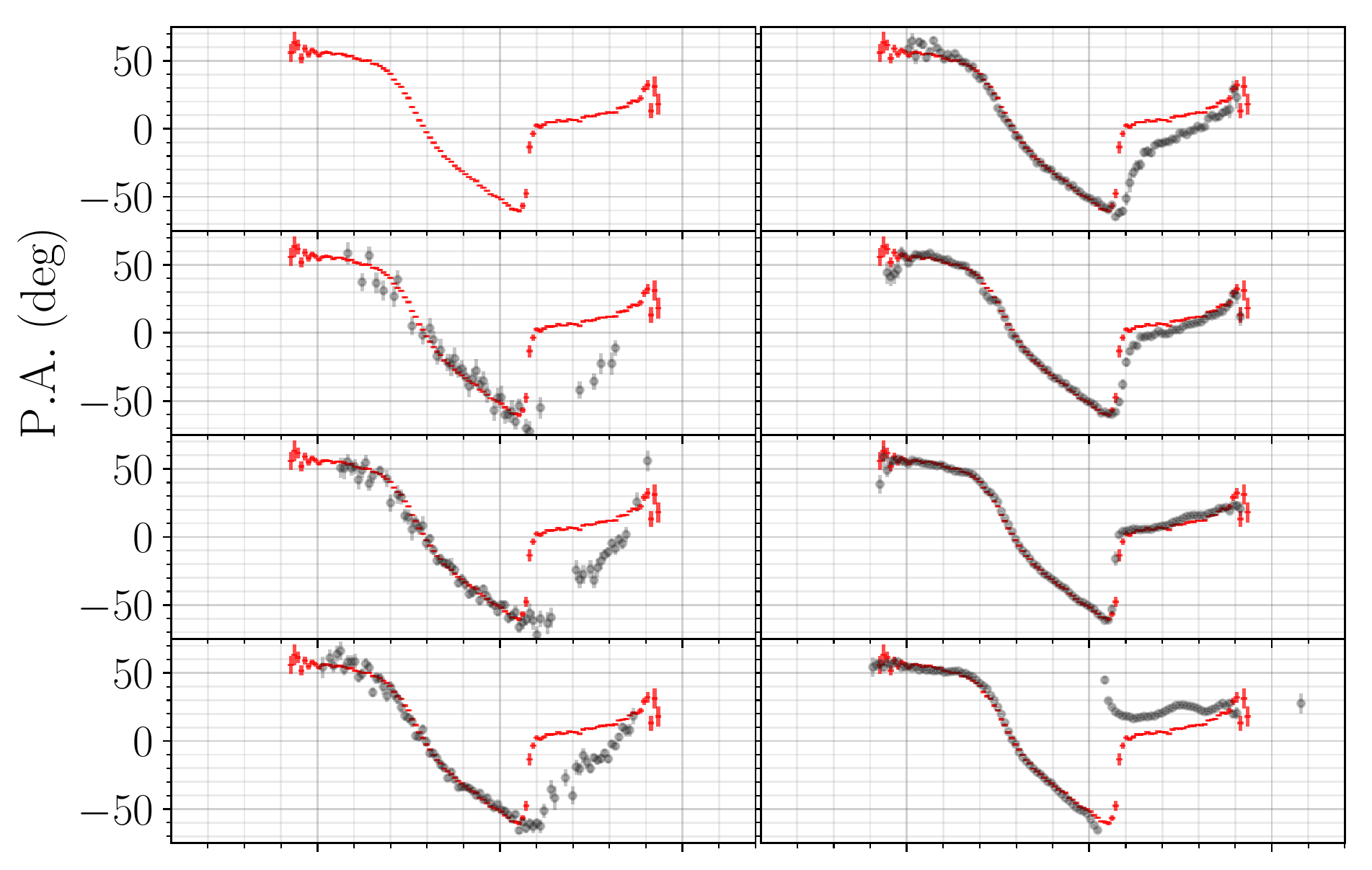}
	\includegraphics[width=\columnwidth]{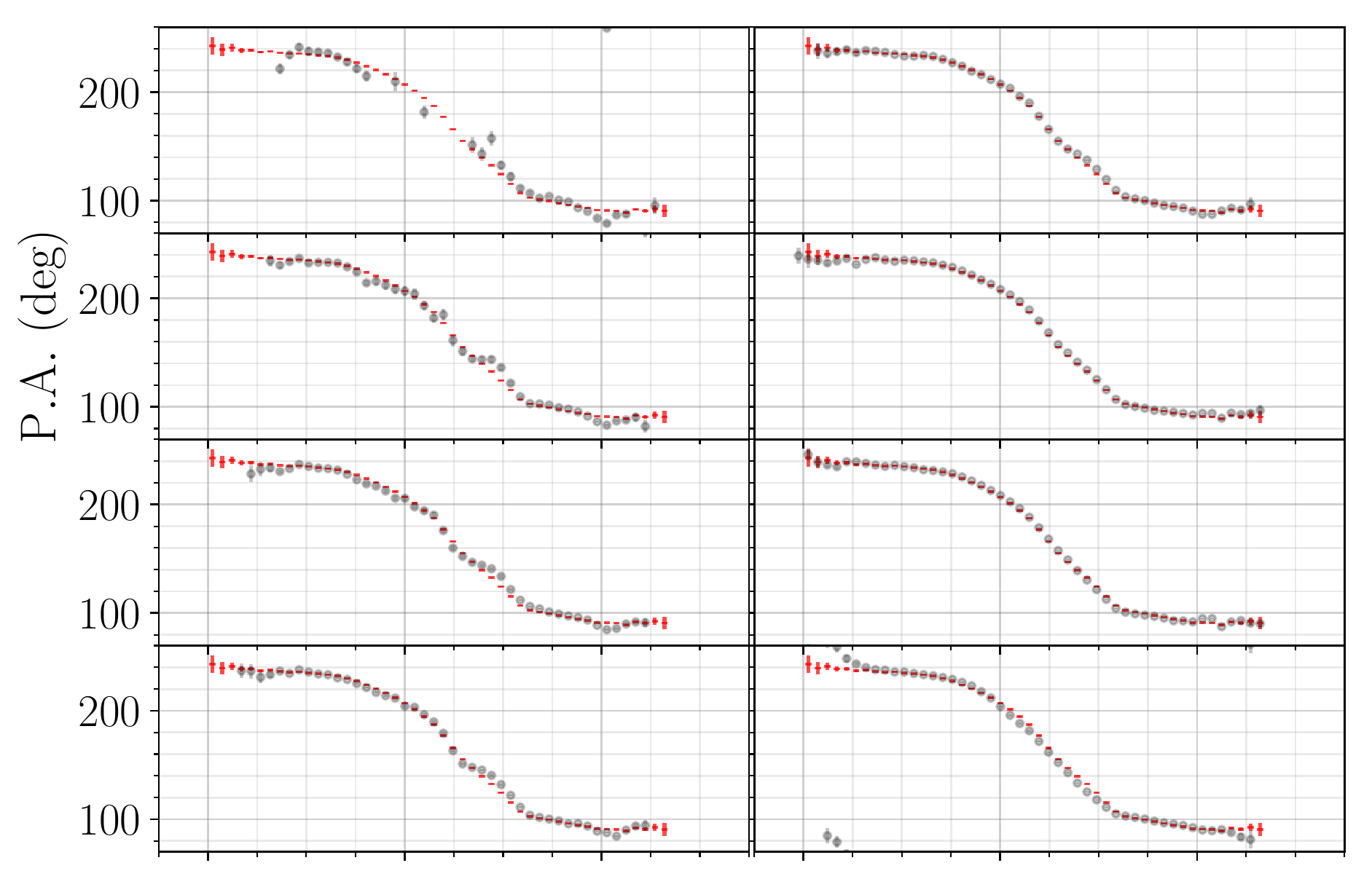}

	\includegraphics[width=\columnwidth]{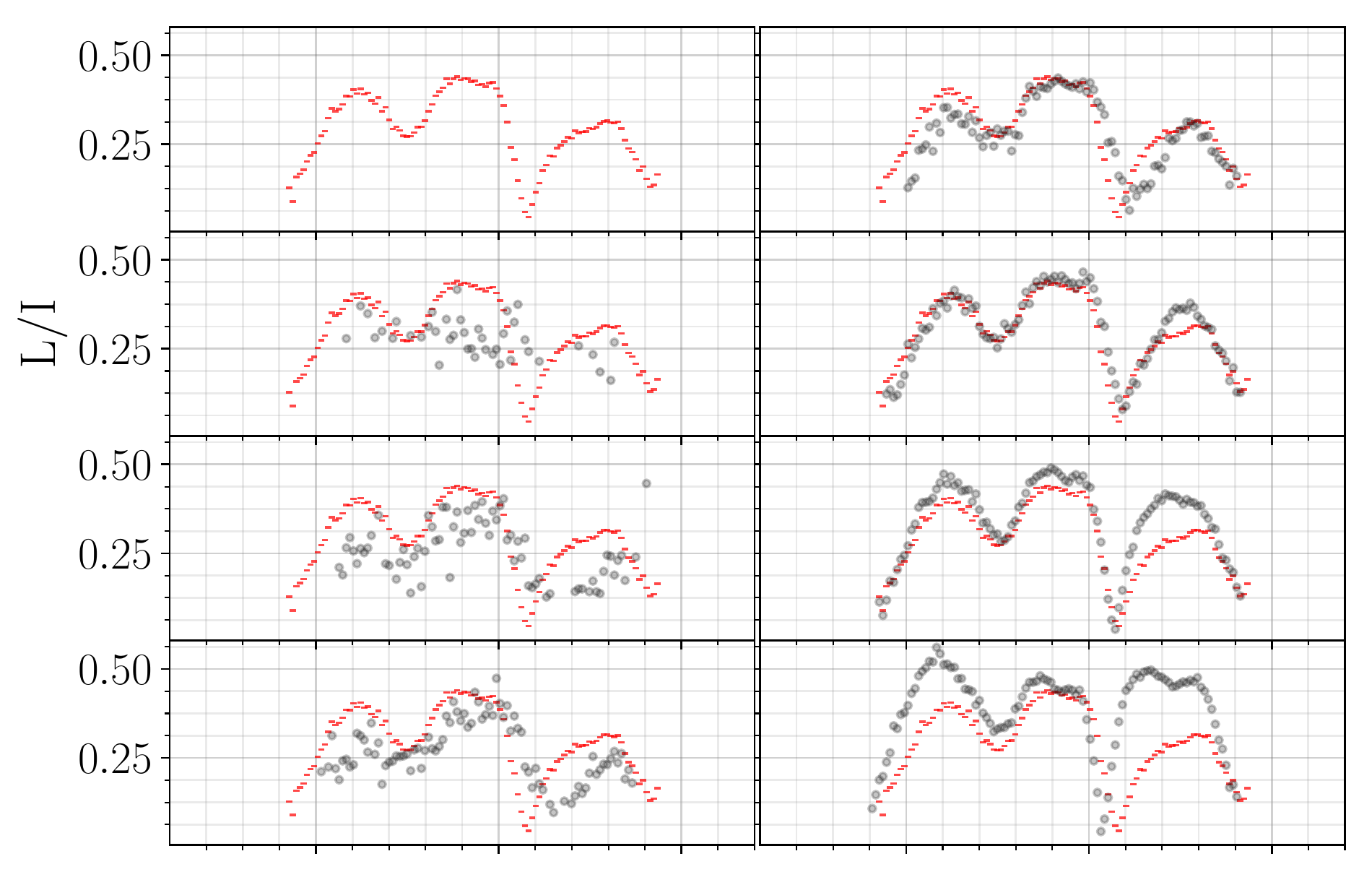}
	\includegraphics[width=\columnwidth]{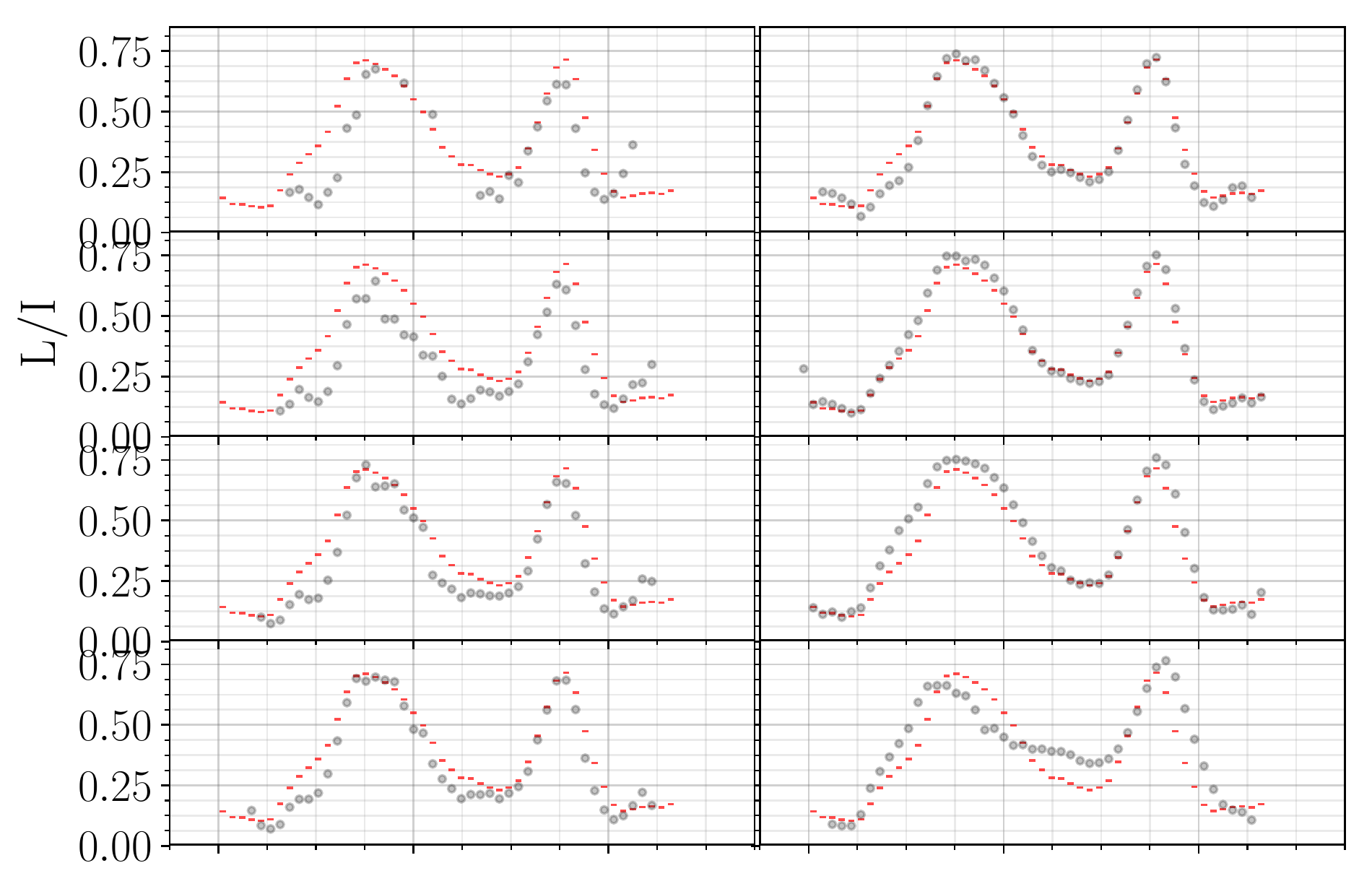}

	\includegraphics[width=\columnwidth]{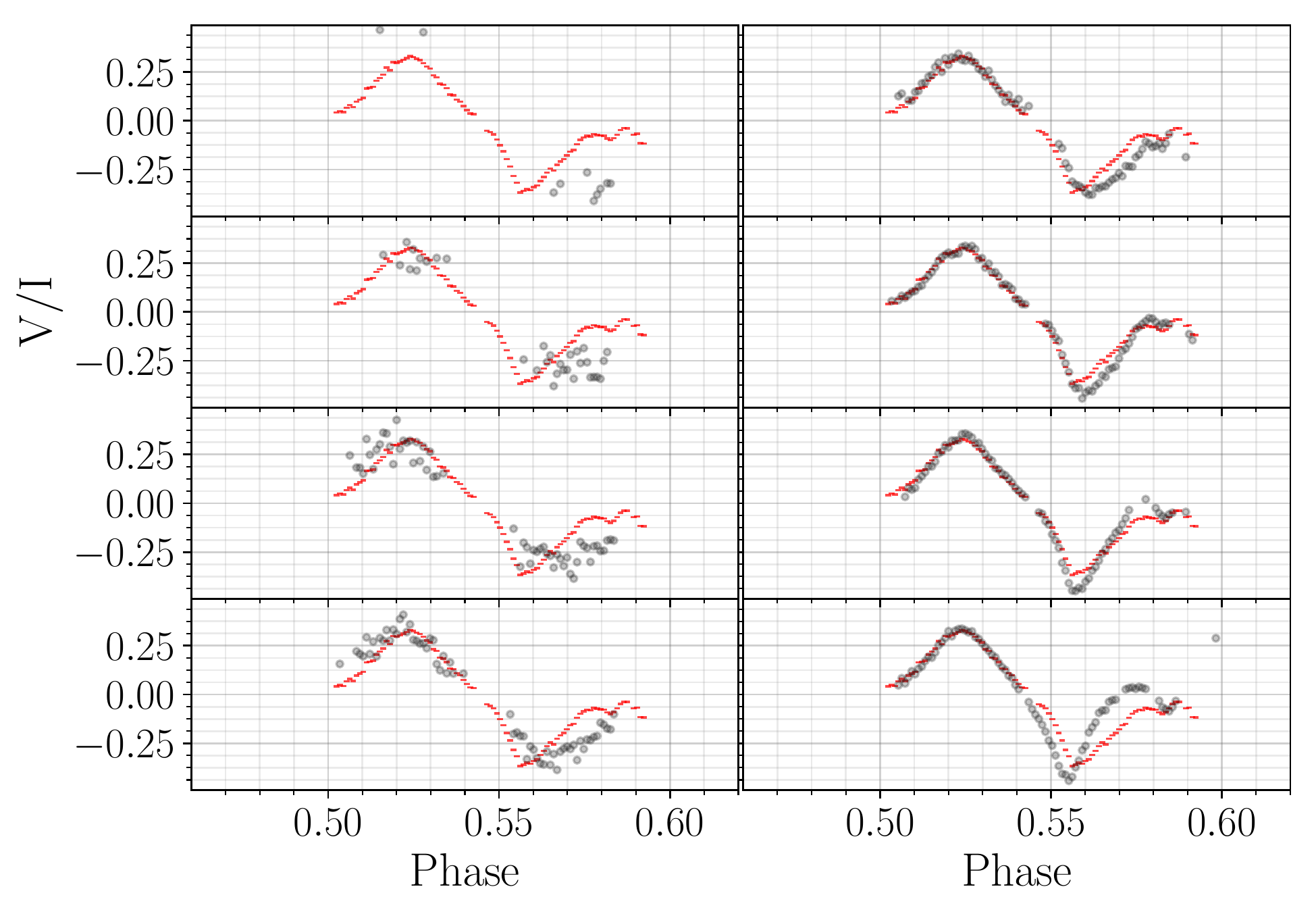}
	\includegraphics[width=\columnwidth]{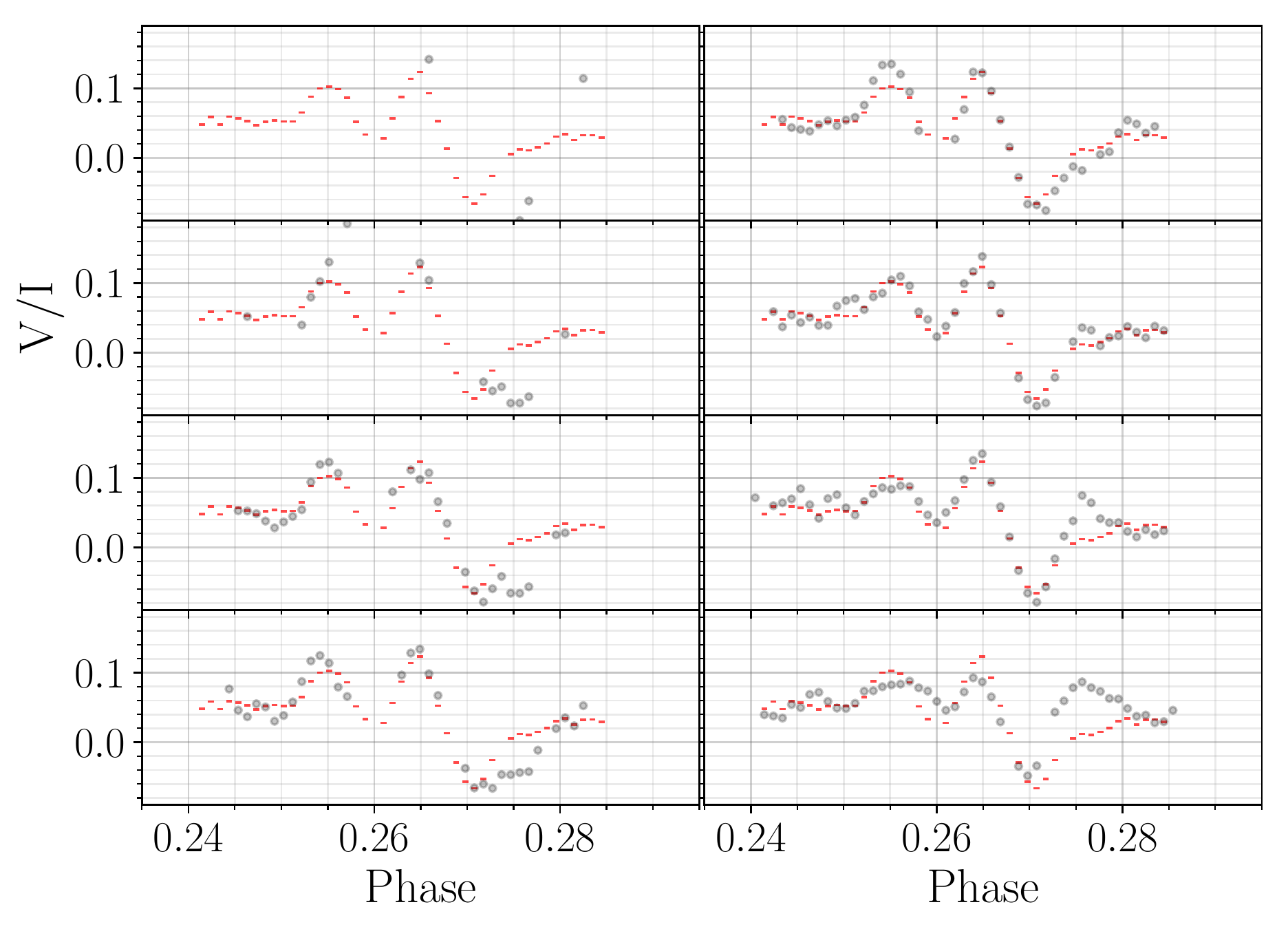}

    \caption{...Figure \ref{fig:profs1} continued.}
\end{figure*}


\bsp	
\label{lastpage}
\end{document}